\documentclass[12pt]{article}
\usepackage{amssymb,amsmath}
\usepackage[mathscr]{eucal}
\usepackage[dvips]{color}
\usepackage{graphicx}
\usepackage{psfrag}

\usepackage{hyperref}
\usepackage[numbers,sort&compress]{natbib}
\usepackage{hypernat}
\hypersetup{linkbordercolor=.8 .8 1,
            citebordercolor=.7 1 .7}

\DeclareFontFamily{U}{rsf}{}
\DeclareFontShape{U}{rsf}{m}{n}{
  <5> <6> rsfs5 <7> <8> <9> rsfs7 <10-> rsfs10}{}
\DeclareMathAlphabet\Scr{U}{rsf}{m}{n}

\catcode`\@=11
\def\@citex[#1]#2{%
\if@filesw \immediate \write \@auxout {\string \citation {#2}}\fi
\@tempcntb\m@ne \let\@h@ld\relax \def\@citea{}%
\@cite{%
  \@for \@citeb:=#2\do {%
    \@ifundefined {b@\@citeb}%
      {\@h@ld\@citea\@tempcntb\m@ne{\bf ?}%
      \@warning {Citation `\@citeb ' on page \thepage \space undefined}}%
      {\@tempcnta\@tempcntb \advance\@tempcnta\@ne%
      \@tempcntb\number\csname b@\@citeb \endcsname \relax%
      \ifnum\@tempcnta=\@tempcntb 
        \ifx\@h@ld\relax%
          \edef \@h@ld{\@citea\csname b@\@citeb\endcsname}%
        \else%
          \edef\@h@ld{\ifmmode{-}\else--\fi\csname b@\@citeb\endcsname}%
        \fi%
      \else
        \@h@ld\@citea\csname b@\@citeb \endcsname%
        \let\@h@ld\relax%
      \fi}%
    \def\@citea{,\penalty\@highpenalty\,}%
  }\@h@ld
}{#1}}

\def\@citeb#1#2{{[#1]\if@tempswa , #2\fi}}
\def\@citeu#1#2{{$^{#1}$\if@tempswa , #2\fi }}
\def\@citep#1#2{{#1\if@tempswa , #2\fi}}

\def\bcites{         
        \catcode`\@=11
        \let\@cite=\@citeb
        \catcode`\@=12
}

\def\upcites{         
        \catcode`\@=11
        \let\@cite=\@citeu
        \catcode`\@=12
}

\def\plaincites{      
        \catcode`\@=11
        \let\@cite=\@citep
        \catcode`\@=12
}

\newcount\hour
\newcount\minute
\newtoks\amorpm
\hour=\time\divide\hour by 60
\minute=\time{\multiply\hour by 60 \global\advance\minute by-\hour}
\edef\standardtime{{\ifnum\hour<12 \global\amorpm={am}%
        \else\global\amorpm={pm}\advance\hour by-12 \fi
        \ifnum\hour=0 \hour=12 \fi
        \number\hour:\ifnum\minute<10 0\fi\number\minute\the\amorpm}}
\edef\militarytime{\number\hour:\ifnum\minute<10 0\fi\number\minute}

\def\draftlabel#1{{\@bsphack\if@filesw {\let\thepage\relax
   \xdef\@gtempa{\write\@auxout{\string
      \newlabel{#1}{{\@currentlabel}{\thepage}}}}}\@gtempa
   \if@nobreak \ifvmode\nobreak\fi\fi\fi\@esphack}
        \gdef\@eqnlabel{#1}}
\def\@eqnlabel{}
\def\@vacuum{}
\def\marginnote#1{}
\def\draftmarginnote#1{\marginpar{\raggedright\scriptsize\tt#1}}
\def\draft{
        \pagestyle{plain}
        \overfullrule=2pt
        \oddsidemargin -.5truein
        \def\@oddhead{\sl \phantom{\today\quad\militarytime} \hfil
        \smash{\Large\sl DRAFT} \hfil \today\quad\militarytime}
        \let\@evenhead\@oddhead
        \let\label=\draftlabel
        \let\marginnote=\draftmarginnote
        \def\ps@empty{\let\@mkboth\@gobbletwo
        \def\@oddfoot{\hfil \smash{\Large\sl DRAFT} \hfil}
        \let\@evenfoot\@oddhead}
        \def\@eqnnum{(\theequation)\rlap{\kern\marginparsep\tt\@eqnlabel}%
        \global\let\@eqnlabel\@vacuum}  }

\def\section{\@startsection {section}{1}{\z@}{3.ex plus 1ex minus
 .2ex}{2.ex plus .2ex}{\Large\bf}}
\def\subsection{\@startsection{subsection}{2}{\z@}{2.75ex plus 1ex minus
 .2ex}{1.5ex plus .2ex}{\large\bf}}        
\def\subsubsection{\@startsection{subsubsection}{2}{\z@}{2.75ex plus 1ex minus
 .2ex}{1.5ex plus .2ex}{\large\bf}}

\def\abstract{\if@twocolumn
\section*{Abstract}
\else 
\begin{center}
{\bf Abstract\vspace{-.5em}\vspace{0pt}}
\end{center}
\quotation
\fi}

\catcode`\@=12

\newcommand{\beq}{\begin{equation}}
\newcommand{\eeq}{\end{equation}}
\newcommand{\beqa}{\begin{eqnarray}}
\newcommand{\eeqa}{\end{eqnarray}}
\newcommand{\dd}{{\rm d}}
\newcommand{\N}{{\bf N}}

\newcommand{\Z}{{\bf Z}}
\newcommand{\ZZ}{{\bf Z}}
\newcommand{\Q}{{\bf Q}}

\newcommand{\R}{{\bf R}}

\newcommand{\C}{{\bf C}}
\newcommand{\CC}{{\bf C}}
\newcommand{\PP}{{\bf P}}
\newcommand{\e}{\,{\rm e}}
\newcommand{\CP}{{\CC\PP}}

\newcommand{\be}{\begin{equation}}
\newcommand{\ee}{\end{equation}}
\newcommand{\bea}{\begin{eqnarray}}
\newcommand{\eea}{\end{eqnarray}}

\def\to{\rightarrow}

\def\lae{\mathrel{\mathop{\smash{\lower .5 ex \hbox{$\stackrel<\sim$}}}}}
\def\lae{\mathrel{\mathop{\smash{\lower .5 ex \hbox{$\stackrel>\sim$}}}}}

\def\l:{\mathopen{:}\,}
\def\r:{\,\mathclose{:}}

\catcode`\@=11
\def\theequation{\arabic{equation}}
\catcode`\@=12

\bcites

\catcode`\@=11
\def\theequation{\thesection.\arabic{equation}}
\@addtoreset{equation}{section}
\@addtoreset{footnote}{section}
\@addtoreset{footnote}{subsection}
\catcode`\@=12

\newcommand{\opsi}{\overline{\psi}}
\newcommand{\oQ}{\overline{Q}}
\newcommand{\btheta}{\overline{\theta}}

\newcommand{\bepsilon}{\overline{\epsilon}}
\newcommand{\bPhi}{\overline{\Phi}}

\newcommand{\bphi}{\overline{\phi}}
\newcommand{\bpsi}{\overline{\psi}}

\newcommand{\bi}{{\overline{\imath}}}
\newcommand{\bj}{{\overline{\jmath}}}
\newcommand{\bl}{\overline{l}}
\newcommand{\bz}{\overline{z}}

\newcommand{\blambda}{\overline{\lambda}}
\newcommand{\bsigma}{\overline{\sigma}}
\newcommand{\bSigma}{\overline{\Sigma}}
\newcommand{\bareta}{\overline{\eta}}
\newcommand{\barxi}{\overline{\xi}}

\newcommand{\lrd}{\overleftarrow{\partial}\!\!\!\!\!
\overrightarrow{\vbox to 8.35pt{}}}
\newcommand{\lrD}{\overleftarrow{D}\!\!\!\!\!
\overrightarrow{\vbox to 8.2pt{}}}
\newcommand{\bD}{\overline{D}}

\newcommand{\vp}{\varphi}

\newcommand{\tpsi}{\widetilde{\psi}}
\newcommand{\Ker}{{\rm Ker}}
\newcommand{\nn}{\nonumber}
\newcommand{\s}{s}
\newcommand{\wtpsi}{\widetilde{\psi}}
\newcommand{\bartial}{\overline{\partial}}

\newcommand{\Hom}{{\rm Hom}}
\newcommand{\Ext}{{\rm Ext}}

\newcommand{\Ho}{{\mathscr H}}

\newcommand{\ring}{{\mathcal R}}
\newcommand{\sring}{{\mathcal S}}

\newcommand{\HomHo}{{\rm Hom}_{\displaystyle{}_{\bf Ho}}}
\newcommand{\HomD}{{\rm Hom}_{\displaystyle{}_{\bf D}}}
\newcommand{\HomR}{{\rm Hom}_{\displaystyle{}_{\ring}}}
\newcommand{\HomgrR}{{\rm Hom}_{\displaystyle{}_{{\rm gr}\mbox{-}\ring}}}
\newcommand{\HomHogrR}{{\rm Hom}_{\displaystyle{
}_{{\bf Ho}\mbox{-}{\rm gr}\mbox{-}\ring}}}
\newcommand{\HomHogrS}{{\rm Hom}_{\displaystyle{
}_{{\bf Ho}\mbox{-}{\rm gr}\mbox{-}\sring}}}

\newcommand{\g}{{\it g}}

\newcommand{\moduli}{\mathfrak{M}}
\newcommand{\surface}{{\mathbb S}}

\newcommand{\lsmB}{{\mathfrak B}}
\newcommand{\nlsmB}{{\mathcal B}}

\begin{document}

\newcommand{\bX}{\overline{X}}
\newcommand{\bx}{\overline{x}}
\newcommand{\id}{{\rm id}}
\newcommand{\wA}{{\mathcal A}}

\newcommand{\brane}{{\mathscr W}}
\newcommand{\wilson}{{\mathscr W}}
\newcommand{\cp}{{\mathscr V}}

\newcommand{\lpi}{\mbox{\large $\pi$}}
\newcommand{\Rp}{{\rm Re}}
\newcommand{\Ip}{{\rm Im}}

\begin{titlepage}

\hfill {\scriptsize DESY-07-154, CERN-PH-TH/2008-048}

\begin{center}

\vskip 2.5 cm
{\large \bf Phases Of ${\mathcal N}=2$ Theories In $1+1$ Dimensions With
Boundary }
\vskip 1 cm
{Manfred Herbst,$\!{}^{{}^{ u}}\,$ 
Kentaro Hori${}^{{}^{\, v}}$ and David Page${}^{{}^{\, v}}$}\\
\vskip 0.6cm
${}^{{}^{ u}}${\it Theory Division, Department of Physics, CERN,
CH--1211 Geneva 23, Switzerland}

${}^{{}^{ v}}${\it Department of Physics, 
University of Toronto, 60 St. George Street,
Toronto, Ontario, M5S 1A7, Canada}

\end{center}

\vskip 0.5 cm
\begin{abstract}
We study B-type D-branes in linear sigma models with Abelian gauge groups.
The most important finding is the grade restriction rule. It
classifies representations of the gauge group on the Chan-Paton factor,
which can be used to define a family of D-branes over a region of
the K\"ahler moduli space that connects special points
of different character.
As an application, we find a precise, transparent relation between D-branes 
in various geometric phases as well as free orbifold and Landau-Ginzburg
points.
The result reproduces and unifies many of the earlier mathematical
results on equivalences of D-brane categories, 
including the McKay correspondence and Orlov's construction.
\end{abstract}

\vfill March, 2008

\end{titlepage}

\newpage

\pagestyle{empty}

\tableofcontents

\section*{Notation}

The time coordinate of the worldsheet is 
denoted $t=\s^0$ and the space coordinate is $\s=\s^1$,
except in Section~\ref{sec:vacuum} where we use $x$ for
the space coordinate.
Light cone coordinates are $\s^{\pm}=t\pm \s=\s^0\pm \s^1$
and their derivatives are
$\partial_{\pm}={1\over 2}(\partial_0\pm\partial_1)$.
The worldsheet itself is denoted as ${\mathbb S}$.
Supercharges are denoted by bold face ${\bf Q}$. For example, 
$(2,2)$ generators are ${\bf Q}_{\pm}$, $\overline{\bf Q}_{\pm}$.
On the other hand, $Q$ is used for the holomorphic part of an open string 
tachyon profile.

\setcounter{page}{0}

\newpage

\pagestyle{plain}

\section{Introduction}
\label{sec:intro}

D-branes in Calabi-Yau manifolds play central roles in
string theory and related fields. 
They can be used to construct and study models of particle physics 
and cosmology 
with spontaneously broken or unbroken ${\mathcal N}=1$ supersymmetry.
They also determine extremal black holes in 
four-dimensional ${\mathcal N}=2$ supergravity
or BPS states in ${\mathcal N}=2$ field theory. 
A lot of efforts have been devoted to this subject in the past twelve years.
Most are done in large volume regimes where 
the $\alpha'$ corrections are small or negligible and the ten-dimensional
supergravity can be used. 
There are also studies of D-branes at special non-geometric backgrounds
with exactly solvable worldsheet conformal field theories,
such as Gepner models and free orbifolds.
Furthermore, some works probe singular points of the moduli
space where the worldsheet description breaks down.
Although there are still many things to be understood,
a body of solid knowledge is accumulating at special points
of the moduli space where convenient descriptions of the theory are available.

A natural and important problem is to connect the information
at the special points and to obtain a global picture of D-branes
over the entire moduli space of backgrounds.
For example, this will be necessary to understand the totality
of ${\mathcal N}=1$ vacua.
Some quantities are protected from quantum corrections
and are either constant or holomorphic as functions of the 
moduli fields. For those, we may be able to
glue together the information at special points
trivially or by analytic continuation.
Ramond-Ramond (RR) charges of D-branes are good examples 
--- the connection between large volume regimes
and Gepner points was successfully found in the seminal paper \cite{BDLR}.
However, most quantities do not possess such properties.
We would like to have at least some hint to study the vast unexplored regions.

For closed strings, linear sigma models \cite{phases}
provide an ultra-violet
description of the worldsheet theories over the entire moduli space
in a large class of examples.
They were used to find a simple and global picture of the
stringy moduli space that had been available only via mirror symmetry,
and they had also been used to derive mirror symmetry itself.
A natural idea is to use them also to study D-branes.
The main goal of the present paper is to construct an ultra-violet
description of the worldsheet theory with boundary, using linear sigma
models, that is valid in regions of the moduli space that encompass
various special points of different character.
Just as in the bulk theories, we would like to have a 
workable method of construction which is very explicit and transparent.

To be precise our focus will be on the description of B-type D-branes 
over the bulk of the K\"ahler moduli space.
The moduli space of $(2,2)$ superconformal field theories is a product of
two spaces, $\moduli_C$ and $\moduli_K$,
which are referred to as the ``complex structure moduli space''
and the ``K\"ahler moduli space'' according to their interpretation
in large volume regimes.
It is the K\"ahler moduli space that
has special points with various 
different descriptions.
Relevant D-branes are those preserving a half of the $(2,2)$ supersymmetry,
and there are two types: A-branes and B-branes.
In a large volume regime, A-branes are wrapped on Lagrangian submanifolds
and B-branes are wrapped on complex submanifolds.
They have chiral sectors that are protected from renormalization.
The chiral sector of B-branes depends holomorphically on $\moduli_C$
and is invariant under deformations in $\moduli_K$.
Correspondingly, the tree-level spacetime superpotential 
depends holomorphically
on the complex structure moduli fields but is independent of the
K\"ahler moduli fields.
Note that this does not mean that B-branes do not depend at all
on $\moduli_K$. The spacetime D-term potential
and stability of the branes depend
primarily on the K\"ahler moduli.
These facts make it feasible and yet interesting 
to study B-branes over the K\"ahler moduli space.

Linear sigma models are a simple class of (2,2) supersymmetric gauge theories.
In this paper, we only consider models with Abelian gauge groups.
Fayet-Iliopoulos (FI) parameters enter into the worldsheet D-term
equations and determine the pattern of gauge symmetry breaking and massless 
fields.
The pattern decomposes the space of FI parameters into domains
called ``phases''. 
On a ``phase boundary'', there is a classical vacuum configuration
with an unbroken continuous subgroup which is generically a $U(1)$.
The quantum gauge theory also depends on the theta angles,
which are interpreted as the background electric fields.
The K\"ahler moduli space $\moduli_K$
is thus spanned by the FI parameters as well as the theta angles.

A part of the data to specify a D-brane in a linear sigma model
is the representation of
the gauge group on the Chan-Paton space, that is, the charges
of the Chan-Paton vectors. 
The most important result of the
present paper is the {\it grade (or band) restriction rule}.
It provides the necessary and sufficient condition
so that a ``parallel family'' of D-branes can be 
defined over a region of the moduli space $\moduli_K$ 
which covers two adjacent phases and their phase boundary.
The condition is on
the Chan-Paton charges of the brane 
with respect to the unbroken $U(1)$ subgroup
at the phase boundary, and goes as follows.
Let ${\mathscr S}$ be the sum of all positive charges 
under that $U(1)$ of the bulk matter fields.
Then the condition on the Chan-Paton charge $q$ is that
\beq
-{{\mathscr S}\over 2}<{\theta\over 2\pi}+q<{{\mathscr S}\over 2}
\eeq
for any value of $\theta$ 
at the phase boundary in the region of
$\moduli_K$ under consideration.

The present work is strongly motivated by recent developments in mathematics.
The D-brane category, which has the same information
as the chiral sector of all possible boundary interactions 
in a fixed bulk theory, 
provides active areas of research in algebraic geometry and 
symplectic geometry, after M. Kontsevich's
homological mirror symmetry conjecture \cite{HMS}.
The category of B-branes in a large volume regime
is the derived category of the target space, while in an orbifold theory 
it is the derived category of objects with orbifold group action.
In a Landau-Ginzburg model it is the category of matrix factorizations 
of the superpotential.
The categories of B-branes of bulk theories that are related by K\"ahler
deformations must be equivalent, as a consequence of the invariance
of the chiral sector.
Mathematically, an equivalence of categories
is given by a pair of maps of objects and morphisms with certain
isomorphism conditions.
Recently, such equivalences of
D-brane categories were constructed.
One example is the categorical version of McKay correspondence \cite{BKR},
that is,
the equivalence of the derived category of a non-compact
toric Calabi-Yau manifold and the derived category for the orbifold
theory which sits at a different point of
the same K\"ahler moduli space.
Also, D. Orlov constructed 
equivalences between the derived category of the Calabi-Yau
hypersurface defined by a polynomial
and the category of matrix factorizations of the same polynomial \cite{Orlov}.
A natural question is whether these equivalences
are the ones relevant for physics.
Our work grew out of an attempt to answer this question.

The organization of the rest of the paper is as follows.

In Section~\ref{sec:branecat}, we describe B-type D-branes
 in non-linear sigma models, Landau-Ginzburg models, and their orbifolds.
We determine the condition of ${\mathcal N}=2_B$
supersymmetry and $U(1)$ R-symmetry
on the ${\mathcal N}=1$ invariant boundary interactions
given by Quillen's superconnections.
This leads to complexes of vector bundles as the data of D-branes in 
non-linear sigma models
and homogeneous matrix factorizations of the superpotential 
for Landau-Ginzburg models. 
We also study the chiral sector of each system and describe
the corresponding D-brane category.

In Section~\ref{sec:D-isom}, we look into 
D-term deformations and brane-antibrane annihilation, which are
operations that do not change the low energy behaviour of the boundary
interactions. 
We show that a quasi-isomorphism between complexes of 
vector bundles can be obtained by a chain of D-term deformations and
brane-antibrane annihilation. 
This clarifies the relevance of quasi-isomorphisms
in brane-antibrane systems, which was
discussed earlier in \cite{Douglas} from the spacetime point of view.
We also study relevant and marginal deformations of the bulk theory.
We study what happens to D-branes when a pair of bulk fields with F-term mass
are integrated out, and find the map of D-branes
from the high energy theory to the low energy theory
(we call it {\it Kn\"orrer's map}).
We end the section with the study of
marginal K\"ahler deformations which is the main subject of 
this paper. We determine the rule of D-brane transport along a path in
the K\"ahler moduli space, and show that it defines the notion 
of a ``flat connection'' for the ``bundle'' of D-branes over $\moduli_K$.

In Section~\ref{sec:LSM}, we review bulk linear sigma models
and make some new observations that 
play important r\^oles later in the paper.
In particular, we present a simple way to find the phase structure
and the symmetry breaking patterns
by plotting the charges of the fields in the space of FI parameters.
We also find a simple relation 
of the symmetry breaking patterns between adjacent phases.

In Section~\ref{sec:LSMbranes}, we classify ${\mathcal N}=2_B$ supersymmetric
boundary interactions in linear sigma models with $U(1)$ R-symmetry.
(Earlier works on this subject can be found in
\cite{HIV,GJS2,blin,HKLM}.)
We first introduce the Wilson line branes as the basic building blocks.
Their interactions with the required symmetry
 are determined by gauge invariant and homogeneous matrix factorizations 
of the superpotential. In a system with vanishing superpotential,
they are given by complexes of Wilson line branes.
We also describe the chiral sector of the theory with zero gauge coupling.

In Section~\ref{sec:vacuum}, we study the boundary contribution to
the energy and charge density of the ground state
of a class of matter systems. 
This is to find the low energy effective theory on the Coulomb branch
of the linear sigma model.
One of the most important findings is the presence of normalizable modes
localized near the boundary, that become zero modes in
a particular direction of the Coulomb branch.
In such a direction, the effective theory in terms of vector multiplet fields
becomes singular.
We also digress to study, for later purpose,
 the vacuum energy and charge of the open string system
in a massive Landau-Ginzburg model.

Section~\ref{sec:GRR} is the main part in which we derive the grade 
restriction rule. The key is the Lagrangian boundary condition on the
Coulomb branch. We first re-examine the condition for
A-branes in Landau-Ginzburg models and find the condition that
the boundary potential must be bounded below.
This is then applied to the effective theory on the Coulomb branch.
For a brane that violates the grade restriction rule,
the Lagrangian submanifold must rotate as the phase boundary is crossed. 
It cannot avoid overlaping with a part of
the singular line on which the effective description breaks down.
We expect a non-trivial effect from such an overlap.
On the other hand, for a brane satisfying the grade restriction rule,
the Lagrangian submanifold is stable 
and nothing special happens on the Coulomb branch
as the phase boundary is crossed.

In Section~\ref{sec:noncompact}, we apply the grade restriction rule
to models with vanishing superpotential. We first study the reduction of
the linear sigma model branes to the low energy theory.
Worldsheet D-term equations give rise to
a tachyon condensation pattern that depends on 
the respective phase in $\moduli_K$.
The change of the condensation pattern across phase boundaries 
fits perfectly with the grade restriction rule.
As an application, we derive the monodromy along a closed loop in
the K\"ahler moduli space $\moduli_K$ that encircles a singular point.
We find that the effect is to bind the brane that becomes massless
at the singular point, as expected from the spacetime picture
and mirror symmetry.
We also demonstrate the power of our construction
in several key examples,
including the flop of the resolved conifold and McKay correspondence.
We close the section  with a comment on D-brane transport 
through the center of the moduli space
where multiple phase boundaries meet.

Section~\ref{sec:math} is a mathematical digression in which we
introduce some important notions in plain words and
prove some key facts used in the previous section.
This also paves the way to discuss compact models in the next section
where we need elaborate commutative algebra at some point.
We make a number of mathematical statements that follow from
our construction.

In Section~\ref{sec:compactI}, we apply the grade restriction rule
to models with non-trivial superpotential.
The problem of D-brane transport itself is equally simple as in
the models without superpotential.
An extra complication shows up
 when the superpotential gives mass to some of the
bulk fields: we need to integrate them out to arrive at the low energy
theory. 
To this end, we apply the Kn\"orrer map developed in Section~\ref{sec:D-isom}
to find the low energy description of the D-branes.
We exhibit the D-brane transport in some examples, including the large volume
images of Recknagel-Schomerus branes in the quintic and a two-parameter model.
We also randomly pick some brane at the geometric regime 
and find its Landau-Ginzburg image.
We discuss monodromy and again find that the effect
is to bind a brane that becomes massless at the singular point.
We end by showing the relation of our work to that of Orlov \cite{Orlov}
in a class of models. We also include a review of relevant mathematical
backgrounds \cite{Eisenbud,Ragnar}.

We include an appendix which summarizes the supersymmetry
transformations of the bulk fields and the bulk Lagrangians,
in non-linear sigma models, Landau-Ginzburg models and linear sigma models.

\subsection{A guide to read the paper}

We tried to write this work in a self-contained manner, 
and as a result it turned out quite comprehensive. 
In the following flow chart we therefore suggest various routes 
through the paper. We believe though that the most comprehendible 
way of doing so is to read through all sections, indicated by the bold arrows.

Alternatively the reader may take short cuts along the dashed arrows 
without missing the most important  conceptual points in the shaded 
subsections.  Depending on interests the reader may proceed after 
Section~\ref{sec:GRR} with D-branes in non-compact or compact models.

A remark on Section~\ref{sec:vacuum} is in order. It provides the basis 
for discussing the grade restriction rule in Section~\ref{sec:GRR}. 
Together these two sections are the heart of the paper. 
However, for getting the main conceptual ideas Section~\ref{sec:vacuum} 
may be left out in a first reading.
\begin{center}
\psfrag{1} {\small\ref{sec:intro}}
\psfrag{2} {\small\ref{sec:branecat}}
\psfrag{3} {\small\ref{sec:D-isom}}
\psfrag{4} {\small\ref{sec:LSM}}
\psfrag{5} {\small\ref{sec:LSMbranes}}
\psfrag{6} {\small\ref{sec:vacuum}}
\psfrag{7} {\small\ref{sec:GRR}}
\psfrag{8} {\small\ref{sec:noncompact}}
\psfrag{9} {\small\ref{sec:math}}
\psfrag{10}{\small\ref{sec:compactI}}
\psfrag{A} {\small\ref{app:susy}}
\psfrag{2.1} {\small\ref{subsec:tachyon}}
\psfrag{2.2} {\small\ref{subsec:NLSM}}
\psfrag{2.3} {\small\ref{subsec:orbifolds}}
\psfrag{2.4} {\small\ref{subsec:LG}}
\psfrag{3.1} {\small\ref{subsec:Diso}}
\psfrag{3.2} {\small\ref{subsec:Koszul}}
\psfrag{3.3} {\small\ref{subsec:DisomLG}}
\psfrag{3.4} {\small\ref{subsec:Knorrer}}
\psfrag{3.5} {\small\ref{subsec:whatwedo}}
\psfrag{4.1} {\small\ref{subsec:LSMlagrangian}}
\psfrag{4.2} {\small\ref{subsec:LSMphases}}
\psfrag{4.3} {\small\ref{subsec:LSMsing}}
\psfrag{4.4} {\small\ref{subsec:LSMexamples}}
\psfrag{4.5} {\small\ref{subsec:LSMdel}}
\psfrag{5.1} {\small\ref{subsec:LSMvariation}}
\psfrag{5.2} {\small\ref{subsec:LSMwilson}}
\psfrag{5.3} {\small\ref{subsec:noncompactbranes}}
\psfrag{5.4} {\small\ref{subsec:compactbranes}}
\psfrag{5.5} {\small\ref{subsec:firstlookBC}}
\psfrag{5.6} {\small\ref{subsec:UVlimit}}
\psfrag{6.1} {\small\ref{subsec:largesigma}}
\psfrag{6.2} {\small\ref{subsec:modeexpand}}
\psfrag{6.3} {\small\ref{subsec:groundstate}}
\psfrag{6.4} {\small\ref{subsec:energy}}
\psfrag{6.5} {\small\ref{subsec:charge}}
\psfrag{6.6} {\small\ref{subsec:SUSYnote}}
\psfrag{6.7} {\small\ref{subsec:half}}
\psfrag{6.8} {\small\ref{subsec:SeveralMatter}}
\psfrag{6.9} {\small\ref{subsec:gaugedynamics}}
\psfrag{6.10}{\small\ref{subsec:massfromW}}
\psfrag{7.1} {\small\ref{subsec:ALG}}
\psfrag{7.2} {\small\ref{subsec:AsympC}}
\psfrag{7.3} {\small\ref{subsec:rule}}
\psfrag{7.4} {\small\ref{subsec:nature}}
\psfrag{7.5} {\small\ref{subsec:fullBC}}
\psfrag{8.1} {\small\ref{subsec:LEBC}}
\psfrag{8.2} {\small\ref{subsec:grrrevisited}}
\psfrag{8.3} {\small\ref{subsec:NCmonodromy}}
\psfrag{8.4} {\small\ref{subsec:moreex}}
\psfrag{8.5} {\small\ref{subsec:center}}
\psfrag{9.1} {\small\ref{subsec:Ccrossaction}}
\psfrag{9.2} {\small\ref{subsec:generaltoric}}
\psfrag{9.3} {\small\ref{subsec:complexes}}
\psfrag{9.4} {\small\ref{subsec:gradedcase}}
\psfrag{9.5} {\small\ref{subsec:mathGRR}}
\psfrag{9.6} {\small\ref{subsec:multigradedcase}}
\psfrag{9.7} {\small\ref{subsec:generalizations}}
\psfrag{10.1} {\small\ref{subsec:compactLEBC}}
\psfrag{10.2} {\small\ref{subsec:LGOpt}}
\psfrag{10.3} {\small\ref{subsec:LVphase}}
\psfrag{10.4} {\small\ref{subsec:Excompact}}
\psfrag{10.5} {\small\ref{subsec:CompactMonodromy}}
\psfrag{10.6} {\small\ref{subsec:relOrlov}}
%
\includegraphics[width=16cm]{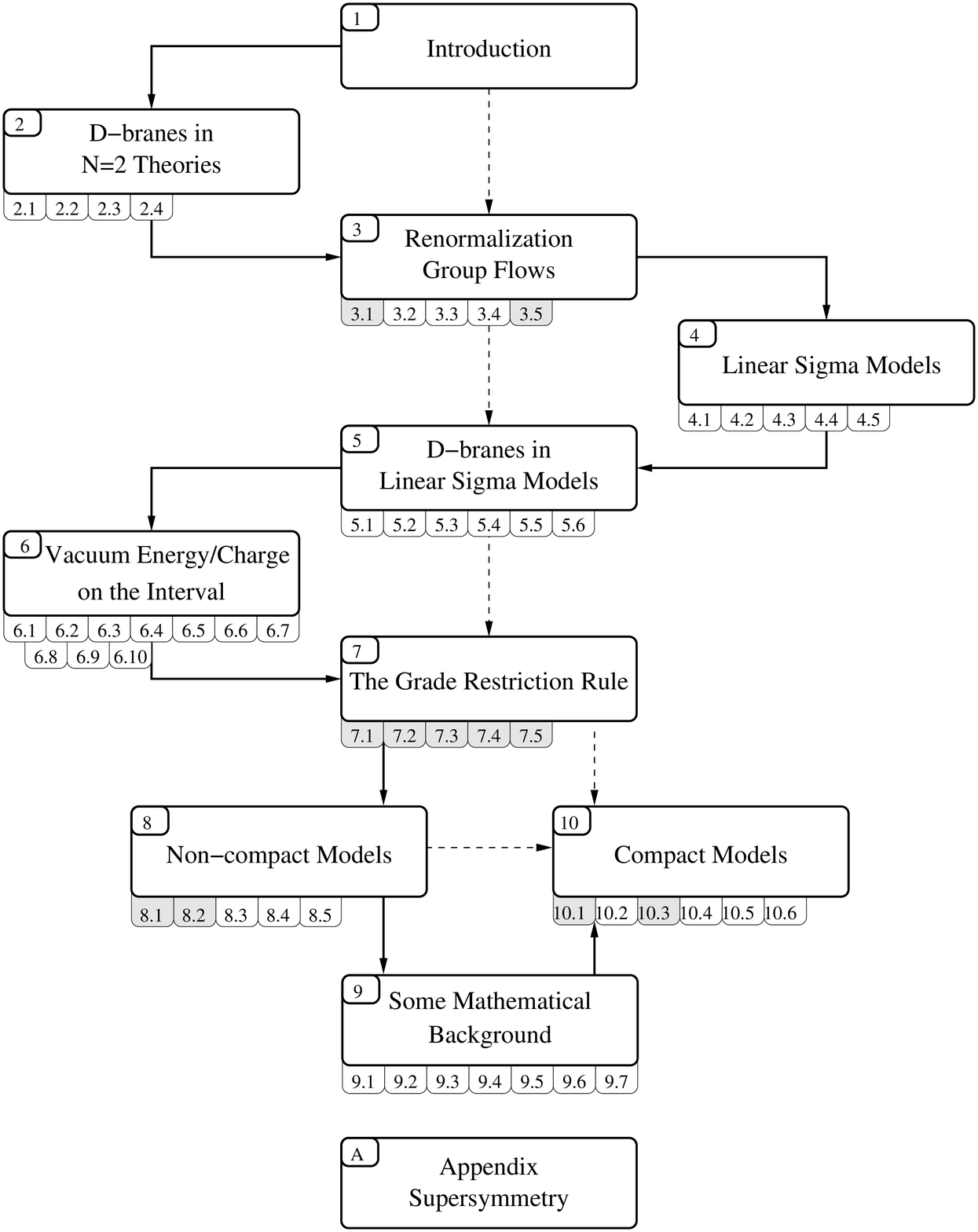}
\end{center}
\newpage

\section{D-branes In ${\mathcal N}=2$ Theories}
\label{sec:branecat}

\newcommand{\wtr}{\widetilde{r}}

In this section, we study D-branes 
in various $(2,2)$ supersymmetric field theories that
preserve a diagonal ${\mathcal N}=2$ supersymmetry.
In Sections \ref{subsec:NLSM}, \ref{subsec:orbifolds}
and \ref{subsec:LG}, we describe D-branes in non-linear sigma models,
in orbifolds and in Landau-Ginzburg models. We pay special
attention to R-symmetry and the ground state sector.
We start out in Section~\ref{subsec:tachyon} with the construction
of ${\mathcal N}=1$ supersymmetric boundary Lagrangians
based on open string tachyon profile,
which will be used throughout this paper.

\subsection{${\mathcal N}=1$ Supersymmetric Interactions}
\label{subsec:tachyon}

We first construct ${\mathcal N}=1$ supersymmetric boundary interactions
that are marginal or relevant.
For simplicity, we consider a string propagating
in flat Euclidean space $\R^n$, which is described by
real scalar fields $x^I$ and Majorana fermions $\psi_{\pm}^I$
($I=1,...,n$), with the Lagrangian density
\beq
{\mathcal L}=\sum_{I=1}^n\left(\,{1\over 2}(\partial_t x^I)^2
-{1\over 2}(\partial_{\s}x^I)^2
+{i\over 2}\psi_-^I(\partial_t+\partial_{\s})\psi_-^I
+{i\over 2}\psi_+^I(\partial_t-\partial_{\s})\psi_+^I\,\right).
\label{calL}
\eeq
The system has ${\mathcal N}=(1,1)$ supersymmetry
--- the action is invariant under the transformations
$\delta x^I=i\epsilon_-^1\psi_+^I-i\epsilon_+^1\psi_-^I,$
and
$\delta\psi_{\pm}^I=\mp\epsilon^1_{\mp}(\partial_t\pm\partial_{\s})x^I$.
If formulated on the worldsheet with boundary,
for instance on the strip $\surface=[0,L]\times \R$,
a diagonal ${\mathcal N}=1$ subalgebra can be preserved 
when a suitable boundary term is added.
In particular, the Lagrangian
\beq
L_{\rm bulk}=\int^L_0{\mathcal L}\,\dd\s
\,+\,\left[\,-{i\over 2}\sum_{I=1}^n\psi_+^I\psi_-^I\,\right]^L_0
\label{Lbulk}
\eeq
is invariant under the variation with
$\epsilon_-^1=-\epsilon_+^1=\epsilon_1$:
\beq
\delta x^I=i\epsilon_1\psi^I,\quad
\delta\psi^I=-2\epsilon_1\dot{x}^I,\quad
\delta\wtpsi^I=-2\epsilon_1\partial_{\s}x^I,
\label{N=1}
\eeq
where $\psi^I=\psi_+^I+\psi_-^I$ and $\wtpsi^I=\psi_+^I-\psi_-^I$. The
boundary term is needed so that $L_{\rm bulk}$ is supersymmetric
 without using equations of motion nor boundary conditions.
We shall assume this type of term (the ``standard boundary term'') throughout
this paper. 
In a curved background with metric $g_{IJ}$,
the standard boundary term is $-{i\over 2}g_{IJ}\psi_+^I\psi_-^J$.
If there is a closed B-field,
the action is supplemented by
\beq
\int_{\surface} {1\over 2}B_{IJ}(x)(\partial_tx^I\partial_{\s}x^J
-\partial_{\s}x^I\partial_tx^J)\dd^2\s
+{i\over 4}\int_{\partial\surface} B_{IJ}(x)\psi^I\psi^J\,\dd t.
\label{Bfield}
\eeq
There is another ${\mathcal N}=1$ subalgebra, $\epsilon_-^1=\epsilon_+^1$,
for which the standard boundary term has the opposite sign and the
boundary term for the B-field is ${i\over 4}B_{IJ}(x)\wtpsi^I\wtpsi^J$.

\subsubsection{Brane-Antibrane System And Open String Tachyons}

We recall that the vertex operators for the tachyon and
the massless vector boson take the following form
(see, for example, \cite{GSW})
\beqa
&&V_T=k\cdot \psi\,\e^{ik\cdot x},\quad (k^2=1),
\label{tachyon}
\\
&&V_A^{\epsilon}=\left(\epsilon\cdot \dot{x}-(\epsilon\cdot\psi)(k\cdot \psi)
\right)\e^{ik\cdot x},\quad (k^2=k\cdot \epsilon=0).
\label{vector}
\eeqa
We would like to find off-shell extension of these operators
and their finite  version
that can be included in the boundary Lagrangian. 
The off-shell and finite version of (\ref{vector}) is well-known:
\beq
\wA_t=\dot{x}^IA_I(x)-{i\over 4}F_{IJ}(x)\psi^I\psi^J,
\label{gauge}
\eeq
where $A_I$ is a $U(1)$ gauge field on $\R^n$ and $F_{IJ}$ is its field
strength
$F_{IJ}=\partial_IA_J-\partial_JA_I$.
The ${\mathcal N}=1$ variation of $\wA_t$
is a total time derivative, and hence it  can be added to the
boundary Lagrangian preserving the
 ${\mathcal N}=1$ supersymmetry.
Note that the full Lagrangian including the boundary term $-\wA_t$
 is invariant under the simultaneous 
shift by a one-form $\Lambda$:
\beq
B\to B+\dd \Lambda,\qquad
A\to A-\Lambda.
\label{shiftBA}
\eeq

Let us next consider the off-shell and finite version of
(\ref{tachyon}). We first note that $V_T$ is fermionic and one cannot 
include it in the boundary Lagrangian. This is how
the standard GSO projection eliminates the tachyon.
However, one can include it by introducing a $\Z_2$-graded 
Chan-Paton space, a vector space of the form
\beq
\cp=\cp^{\rm ev}\oplus \cp^{\rm od},
\eeq
where $\cp^{\rm ev}$ is the even (or bosonic) subspace
and $\cp^{\rm od}$ is the odd (or fermionic) subspace. We call
a linear map between such vector spaces {\it even} (resp. {\it odd})
when it maps even to even and odd to odd subspaces
(resp. even to odd and odd to even subspaces).
We define the action of the fermionic fields $\psi^I_{\pm}$ and
the fermionic parameters (such as $\epsilon_1$)
so that
they anticommutes with all odd linear maps of $\cp$.
Then, one can consider the following off-shell version of (\ref{tachyon})
\beq
\wA^{(1)}_t={i\over 2}\psi^I\partial_I{\bf T}(x)
\label{trial}
\eeq
where ${\bf T}(x)$ is an odd endomorphism of $\cp$, mapping
$\cp^{\rm ev}$ to $\cp^{\rm od}$ and $\cp^{\rm od}$ to $\cp^{\rm ev}$.
This $\wA_t^{(1)}$ is bosonic and can be included in the 
boundary Lagrangian.
Let us see if it is ${\mathcal N}=1$ invariant:
\beq
\delta \wA_t^{(1)}={i\over 2}\left(-2\epsilon_1\dot{x}^I\right)
\partial_I{\bf T}+{i\over 2}\psi^Ii\epsilon_1\psi^I\partial_I\partial_J
{\bf T}=-i\epsilon_1\dot{\bf T}.
\label{fails}
\eeq
This is a total derivative and it appears good at first sight.
However, since the Chan-Paton space has rank larger than one,
the term $\wA_t$ is placed in the
{\it path-ordered} exponential
$$
U(t_f,t_i)
={\rm P}\exp\left(-i\int_{t_i}^{t_f}\wA_t\dd t\right).
$$
A variation
$$
\delta U(t_f,t_i)
=-i\int_{t_i}^{t_f} U(t_f,t)\delta \wA_t
U(t,t_i)\,\dd t.
$$
is a symmetry only when
the whole integrand is a total derivative
${\dd\over \dd t}[U(t_f,t)X(t)U(t,t_i)]$,
which holds when $\delta\wA_t$
is a total {\it covariant} derivative
\beq
\delta\wA_t=\dot{X}+i[\wA_t,X]=:{\mathcal D}_tX.
\label{totcov}
\eeq
For the first trial (\ref{trial}),
this almost holds with
$X=-i\epsilon_1{\bf T}$, see (\ref{fails}), but there is an error term
$-i[\wA_t^{(1)},X]=-\epsilon_1[\wA_t^{(1)},{\bf T}]$. 
Thus, we would like to
modify $\wA_t^{(1)}$ by a term whose variation cancels it.
Note that
\beqa
\epsilon_1[\wA_t^{(1)},{\bf T}]
&=&\epsilon_1\left({i\over 2}\psi^I\partial_I{\bf T}\cdot{\bf T}
-{\bf T}\cdot{i\over 2}\psi^I\partial_I{\bf T}\right)
\nn\\&=&{i\over 2}\epsilon_1\psi^I\left(
\partial_I{\bf T}\cdot{\bf T}+{\bf T}\cdot\partial_I{\bf T}\right)=
{i\over 2}\epsilon_1\psi^I\partial_I({\bf T}^2)
=\delta\left({1\over 2}{\bf T}^2\right),
\nn
\eeqa
where we have used the anticommutativity
${\bf T}\psi^I=-\psi^I{\bf T}$ in the second equality.
Thus, if we modify the first trial to
$\wA_t=\wA_t^{(1)}+{1\over 2}{\bf T}^2$,
we have
$\delta\wA_t
=\dot{X}+i[\wA_t^{(1)},X]$.
Fortunately, ${1\over 2}{\bf T}^2$ commutes with
$X=-i\epsilon_1{\bf T}$,
and hence we have
$[\wA_t^{(1)},X]=[\wA_t,X]$
which means that $\delta \wA_t=\dot{X}+i[\wA_t,X]$.
The symmetry condition (\ref{totcov}) holds.
In this way, we find that
\beq
\wA_t={i\over 2}\psi^I\partial_I{\bf T}(x)
+{1\over 2}{\bf T}(x)^2
\label{AT}
\eeq
provides the ${\mathcal N}=1$ supersymmetric Wilson line
${\rm P}\exp\left(-i\int\wA_t\dd t\right)$.
The expression (\ref{AT}) is the off-shell and finite version of
the tachyon vertex operator (\ref{tachyon}).
Note that it requires a $\Z_2$-graded Chan-Paton space
$\cp=\cp^{\rm ev}\oplus \cp^{\rm od}$.
The standard interpretation is that $\cp^{\rm ev}$ and $\cp^{\rm od}$ are
the Chan-Paton spaces corresponding to branes and antibranes respectively.

When the Chan-Paton space has rank larger than one,
$F_{IJ}$ in (\ref{gauge})
must be the full curvature $F_{IJ}=\partial_IA_J-\partial_JA_I+i[A_I,A_J]$.
The quadratic part $i[A_I,A_J]$ is needed
for the same reason as the tachyon Lagrangian needs ${1\over 2}{\bf T}^2$.
We can also combine (\ref{gauge}) and (\ref{AT}).
Let $E=E^{\rm ev}\oplus E^{\rm od}$ be a 
$\Z_2$-graded vector bundle over $\R^n$.
For an odd endomorphism ${\bf T}$ of $E$ and an even gauge connection
$A$ of $E$, we have an invariant interaction
\beq
\wA_t=\dot{x}^IA_I(x)-{i\over 4}F_{IJ}(x)\psi^I\psi^J
+{i\over 2}\psi^ID_I{\bf T}(x)+{1\over 2}{\bf T}(x)^2,
\label{SWL}
\eeq
where $F_{IJ}$ is the curvature of
$A_I$ and $D_I{\bf T}$ is the ordinary covariant derivative
$\partial_I{\bf T}+i[A_I,{\bf T}]$.
$E^{\rm ev}$ is the Chan-Paton vector bundle suppoerted by branes
and $E^{\rm od}$ is the one supported by antibranes.
The expression (\ref{SWL}) was first obtained in \cite{blin,TTU,KraLar}
using boundary fermions.
The part other than $\dot{x}^IA_I$ is the curvature of
Quillen's superconnection
\cite{quillen}, and provides a concise expression for the 
Ramond-Ramond charge of a brane-antibrane system \cite{TTU,KraLar}.

We obtained boundary interactions that are off-shell
extensions of the ones generated by tachyon and gauge boson vertex operators,
which are relevant and marginal operators of the free theory.
Here we comment on those generated by a class of
operators with higher dimensions that have higher powers in 
the fermions $\psi$. (Other possibilities are those with
higher derivatives $\ddot{x}$, $\dd^3 x/\dd t^3$,...,
$\dot{\psi}$, $\ddot{\psi}$,..., which we do not
discuss here.) 
The idea is to allow ${\bf T}$ to depend not only on $x$ but also on
$\psi$. It turns out that modifying (\ref{AT})
by giving $\psi$-dependence to ${\bf T}$ and adding one simple term
does the job. Namely, for
\beq
\wA_t=-\dot{x}^I{\partial \over\partial \psi^I}{\bf T}(x,\psi)
+{i\over 2}\psi^I{\partial\over \partial x^I}{\bf T}(x,\psi)
+{1\over 2}{\bf T}(x,\psi)^2,
\label{moregeneral}
\eeq
we find
$$
\delta \wA_t={\mathcal D}_t\left(-i\epsilon_1{\bf T}\right).
$$
In fact, the one including the non-Abelian gauge field (\ref{SWL})
can be regarded as a special case of this: For ${\bf T}(x,\psi)
={\bf T}(x)-\psi^IA_I(x)$ we find that (\ref{moregeneral})
reproduces (\ref{SWL}).

\subsubsection*{\it The Landau-Ginzburg Case}

Finally, we comment on the case where the bulk theory has a superpotential
term
\beq
{\mathcal L}_h=-{1\over 2}\sum_{I=1}^n(\partial_I h(x))^2
-i\psi^I_+\psi^J_-\partial_I\partial_Jh(x).
\eeq
In this case, the ${\mathcal N}=1$ supersymmetry variation for
$\wtpsi^I$ is modified to
$\delta \wtpsi^I=-2\epsilon_1\partial_{\s}x^I-2\epsilon_1\partial_Ih(x)$
while the variation for $x^I$ and $\psi^I$ are intact.
Under this, the action varies as
\beq
\delta\left[\int_{\surface}\left({\mathcal L}+{\mathcal L}_h\right)
\dd^2\s+\int_{\partial\surface}\left[-{i\over 2}
\sum_{I=1}^n\psi_+^I\psi_-^I\right]
\dd t\right]
=\int_{\partial\surface}\left[
i\epsilon_1\psi^I\partial_Ih(x)\right]\dd t.
\label{varN=1LG}
\eeq
The right hand side is the same as the variation of
$\int h(x)\,\dd t$, and thus the action is 
${\mathcal N}=1$ invariant
provided that the following boundary term is added to the Lagrangian
\beq
\Delta L
=\Bigl[\,-h(x)\,\Bigr]_{\partial\surface}.
\label{DelLN=1LG}
\eeq
Since the variation of $x^I$ and $\psi^I$ are not modified, the
boundary interactions considered above, such as
(\ref{AT}) and (\ref{SWL}), remain invariant.

We next consider various $(2,2)$ supersymmetric bulk theories
and determine the condition for the gauge connection $A$ and 
the tachyon ${\bf T}$ to preserve
a diagonal ${\mathcal N}=2$ supersymmetry of B-type,
or ${\mathcal N}=2_B$ supersymmetry. The latter is a symmetry generated by
a linear combination of $\overline{\bf Q}_+$ and 
$\overline{\bf Q}_-$ and its complex conjugate, such as
\beq
{\bf Q}=\overline{\bf Q}_++\overline{\bf Q}_-,
\eeq
and its complex conjugate, ${\bf Q}^{\dag}={\bf Q}_++{\bf Q}_-$.
These supercharges obey the anticommutation relations 
${\bf Q}^2={\bf Q}^{\dag 2}=0$ and
$\{{\bf Q},{\bf Q}^{\dag}\}=2H$, where $H$ is the Hamiltonian.

\subsection{Non-Linear Sigma Models: Complex Of Vector Bundles}
\label{subsec:NLSM}

Let us first consider the non-linear sigma model on a K\"ahler manifold
$(X,g)$.
If formulated on the worldsheet with boundary,
a diagonal ${\mathcal N}=2$ subalgebra of the $(2,2)$ supersymmetry
can be preserved.
If we add the standard boundary term as in (\ref{Lbulk}),
\beq
L_{\rm bdry}^{(0)}=\left[-{i\over 2}g_{i\bj}
(\psi^i_+\bpsi_-^{\bj}+\bpsi_+^{\bj}\psi_-^i)\right]_{\partial\surface},
\label{L0bdry}
\eeq
the action is invariant under the ${\mathcal N}=2_B$
supersymmetry
\beq
\delta x^i=\epsilon\psi^i,\quad
\delta\psi^i=-2i\bepsilon\dot{x}^i,\quad
\delta\wtpsi^i=-2i\bepsilon\partial_{\s}x^i
-\epsilon\Gamma^i_{jk}\psi^j_+\psi^k_-.
\label{N=2}
\eeq
In this expression, $i,j,k,...$ are indices of complex coordinates,
$\epsilon$ is a complex variation parameter,
$\epsilon=i(\epsilon_1+i\epsilon_2)$, and $\bepsilon$ is its complex
conjugate, $\bepsilon=-i(\epsilon_1-i\epsilon_2)$.
The B-field term (\ref{Bfield}) is ${\mathcal N}=2_B$ invariant if
$B$ obeys $B(Jv,Jw)=B(v,w)$, that is, if $B$ is a $(1,1)$-form.
We stress again that the ${\mathcal N}=2_B$ invariance holds 
using neither equations of motion nor boundary conditions, provided that
the standard boundary term is added.

\subsubsection{Condition Of ${\mathcal N}=2$ Supersymmetry}

We now determine the condition on
the superconnection $(A,{\bf T})$ of a $\Z_2$ graded vector bundle
$E=E^{\rm ev}\oplus E^{\rm od}$ so that the corresponding
boundary interaction (\ref{SWL}) has ${\mathcal N}=2_B$ supersymmetry.
Let us first provide the answer. The condition on the gauge connection
$A$ is that its curvature is a $(1,1)$-form, namely
vanishing of the $(0,2)$ (and therefore $(2,0)$) components
\beq
F_A^{(0,2)}=0.
\label{11curv}
\eeq
Holomorphic structures are then defined on $E^{\rm ev}$ and $E^{\rm od}$
by the Cauchy-Riemann operator $\bartial_A$, which obeys
the integrability condition $(\bartial_A)^2=F_A^{(0,2)}=0$.
The condition on the tachyon ${\bf T}$ is that it can be decomposed as
\beq
{\bf T}=iQ-iQ^{\dag},
\label{Q+Qdag}
\eeq
where $Q$ is holomorphic
\beq
D_{\bi}Q=0,
\label{hol}
\eeq
and its square is proportional to the identity
\beq
Q^2=c\cdot {\rm id}_E,
\label{uni}
\eeq
in which $c$ is a numerical (field independent) constant.
The condition of the gauge connection (\ref{11curv}) is well-known \cite{CSOS},
and the proof is omitted here. 
The condition for ${\bf T}$ was derived in \cite{blin} for
those based on boundary fermions. We provide a general derivation below.

We start for simplicity with the case with trival gauge connection $A=0$.
We first note that
the additional supersymmetry variation can be expressed in terms
of the real coordiantes as
\beq
\delta_2 x^I=i\epsilon_2 J^I_{\,\,K}\psi^K,\qquad
\delta_2\psi^I=2\epsilon_2J^I_{\,\,K}\dot{x}^K,
\label{additional}
\eeq
where $J$ is the complex structure of $X$.
According to this, $\wA_t$ varies as
$$
\delta_2\wA_t=i\epsilon_2\dot{x}^KJ^I_{\,\,K}\partial_I{\bf T}
+\cdots
$$
where the ellipsis refers to terms without time derivative.
In order for this to be a total time derivative,
we would like $J^I_{\,\,K}\partial_I{\bf T}$ to be
$\partial_KY$ for some $Y$ which should be a linear recombination of
${\bf T}$. Since $J$ is a complex structure,
$J^2=-1$, the only possibility is that ${\bf T}$ is a sum of
two terms ${\bf T}_+$ and ${\bf T}_-$ such that
$J^I_{\,\,K}\partial_I{\bf T}_{\pm}=\pm i\partial_K{\bf T}_{\pm}$, for which
$Y=i{\bf T}_+-i{\bf T}_-$ does the job.
Writing ${\bf T}_+=iQ$ and using the hermiticity of ${\bf T}$,
one can write ${\bf T}=iQ-iQ^{\dag}$
where $\partial_{\bi}Q=0$.
The boundary interaction then takes the form
$$
\wA_t=-{1\over 2}\psi^i\partial_iQ
+{1\over 2}\bpsi^{\bj}\partial_{\bj}Q^{\dag}
+{1\over 2}\{Q,Q^{\dag}\}-{1\over 2}Q^2-{1\over 2}Q^{\dag 2}.
$$
Let us compute the full ${\mathcal N}=2_B$ variation of
$\wA_t$. As in the ${\mathcal N}=1$ case, we would like the variation 
to be the total covariant derivative 
${\mathcal D}_tZ=\dot{Z}+i[\wA_t,Z]$ for some $Z$:
\beqa
\delta\wA_t&=&
i\bepsilon\dot{x}^i\partial_iQ+{1\over 2}\{Q,-\bepsilon\bpsi^{\bj}
\partial_{\bj}Q^{\dag}\}-{1\over 2}\epsilon\psi^i\partial_iQ^2
\,+\,h.c.
\nn\\
&=&i\left({\dd\over \dd t}(\bepsilon Q)+i\left[
{1\over 2}\bpsi^{\bj}\partial_{\bj}Q^{\dag},\bepsilon Q\right]\right)
-{1\over 2}\epsilon\psi^i\partial_iQ^2
\,+\,h.c.
\nn\\
&=&i\left({\dd\over \dd t}(\bepsilon Q)+i[\wA_t,\bepsilon Q]\right)
+\left[-{1\over 2}\psi^i\partial_iQ+{1\over 2}\{Q,Q^{\dag}\}
-{1\over 2}Q^2-{1\over 2}Q^{\dag 2},\bepsilon Q\right]\nn\\
&&-{1\over 2}\epsilon\psi^i\partial_iQ^2
\,+\,h.c.
\nn\\
&=&i{\mathcal D}_t(\bepsilon Q)
-{1\over 2}(\bepsilon+\epsilon)\psi^i\partial_iQ^2
-{1\over 2}(\bepsilon+\epsilon)[Q^2,Q^{\dag}]\,+\,h.c.
\nn\\
&=&i{\mathcal D}_t(\bepsilon Q+\epsilon Q^{\dag})
+2{\rm Re}\left(\epsilon_2\psi^i\partial_iQ^2-[\epsilon_2
Q^{\dag},Q^2]\right).
\label{N=2VarAt}
\eeqa
We would like the second term to vanish so that
we are left with $i{\mathcal D}_t(\bepsilon Q+\epsilon Q^{\dag})$.
This is so if and only if $Q^2$ is field independent and 
proportional to the identity matrix.
When the gauge connection is non-trivial $A\ne 0$, 
the above consideration goes through with only a slight modification,
and we obtain the
conditions (\ref{Q+Qdag}), (\ref{hol}) and (\ref{uni}).

If we require $Q^2=c\cdot{\rm id}_E$ with a constant $c$,
the term $-{1\over 2}Q^2-{1\over 2}Q^{\dag 2}$ in $\wA_t$ is a constant
multiple of the identity and we may omit it.
From now on, we use the following version of the boundary interaction
without $-{\rm Re}(Q^2)$
\beq
{\mathcal A}_t=A_I\dot{x}^I-{i\over 4}F_{IJ}\psi^I\psi^J
-{1\over 2}\psi^i D_iQ
+{1\over 2}\opsi^{\bi}D_{\bi}Q^{\dag}
+{1\over 2}\{Q,Q^{\dag}\}.
\label{newA}
\eeq
Under the holomorphicity conditions (\ref{11curv}) and (\ref{hol}),
its ${\mathcal N}=2_B$ supersymmetry variation with time dependent parameters 
$\epsilon(t),\bepsilon(t)$ is given by
\beqa
\delta{\mathcal A}_t&=&
i{\mathcal D}_t\Bigl(\bepsilon (Q-\bpsi^{\bj}A_{\bj})
+\epsilon (Q^{\dag}-\psi^iA_i)\Bigr)
\nn\\
&&-{\rm Re}\Bigl(\,
\bepsilon \psi^iD_i Q^2-[\bepsilon Q^{\dag},Q^2]\,\Bigr)
\nn\\
&&
-i\Bigl(\dot{\bepsilon}Q+\dot{\epsilon}Q^{\dag}\Bigr).
\label{deltaA1}
\eeqa
where ${\mathcal D}_t$ is the covariant derivative
with respect to the new ${\mathcal A}_t$ in (\ref{newA}).
Under the additional condition (\ref{uni}), the variantion
$\delta{\mathcal A}_t$ is a total covariant derivative
up to $\dot{\epsilon}$ terms.
The $\dot{\bepsilon}$, $\dot{\epsilon}$ 
terms in (\ref{deltaA1}) show that $Q$ and $Q^{\dag}$
provide the boundary contribution to the supercharges ${\bf Q}$
and ${\bf Q}^{\dag}$.
On the strip $\surface=[0,L]\times\R$ 
with boundary interactions $(E_1,A_1,Q_1)$
and $(E_2,A_2,Q_2)$,
the supercharge ${\bf Q}$ found by
the standard Noether procedure is given by
$
{\bf Q}={\bf Q}_{\rm bulk}+{\bf Q}_{\rm bdry},
$
\beqa
&&
{\bf Q}_{\rm bulk}=\int_0^L\dd\s\left\{
g_{i\bj}(\bpsi_+^{\bj}+\bpsi_-^{\bj})\partial_tx^i
+g_{i\bj}(\bpsi_+^{\bj}-\bpsi_+^{\bj})\partial_{\s}x^i\right\}
\label{Qbulk}
\\
&&
{\bf Q}_{\rm bdry}=-iQ_2\Bigl|_{\s=L}\,+\,iQ_1\Bigl|_{\s=0}.
\label{Qbdry}
\eeqa
The boundary part acts only on the Chan-Paton factor,
$$
i{\bf Q}_{\rm bdry}(\Psi_{\rm CP}\otimes\Psi_{\rm internal})
=\left(Q_2\bigl|_{L}\Psi_{\rm CP}
-(-1)^{|\Psi_{\rm CP}|}\Psi_{\rm CP}Q_1\bigl|_{0}\right)
\otimes\Psi_{\rm internal},
$$
where $(-1)^{|\Psi_{\rm CP}|}$ is $+1$ or $-1$
depending on whether $\Psi_{\rm CP}$ is even or odd.
The supercharge ${\bf Q}$ squares to zero
as required by the ${\mathcal N}=2$ supersymmetry algebra,
as long as $Q_1^2$ and $Q_2^2$ are the same constants.

\subsubsection*{\it More General Interactions}

Let us next consider the tachyon profile ${\bf T}$
that depends also
on the fermions $\psi,\bpsi$.
We have seen that the boundary interaction
$$
\wA_t=-\dot{x}^I{\partial \over\partial \psi^I}{\bf T}
+{i\over 2}\psi^I{\partial\over \partial x^I}{\bf T}
+{1\over 2}{\bf T}^2
$$
is ${\mathcal N}=1$ supersymmetric.
We would like to see the condition for ${\mathcal N}=2$ invariance.
Assuming again the form ${\bf T}=iQ-iQ^{\dag}$, where
$Q$ is independent of $\psi^i$'s,
\beq
{\partial\over\partial \psi^i}Q=0,
\label{psiindep}
\eeq
we find the following supersymmetry variation
of the above $\wA_t$
\beqa
\delta \wA_t&=&{\mathcal D}_t\left(i\bepsilon Q+i\epsilon Q^{\dag}\right)
\nn\\
&&-{i\over 2}(\bepsilon+\epsilon)\delta_{\overline{\bf Q}}
\Bigl(\bpsi^{\bj}\partial_{\bj}Q+Q^2
\Bigr)
+{i\over 2}(\bepsilon+\epsilon)
\delta_{\bf Q}\Bigl(-\psi^i\partial_iQ^{\dag}+Q^{\dag 2}
\Bigr)\nn\\
&&-i\dot{\bepsilon}\left(Q-\bpsi^{\bi}{\partial\over \partial\bpsi^{\bi}}
Q\right)
-i\dot{\epsilon}\left(Q^{\dag}-\psi^i{\partial\over \partial\psi^i}
Q^{\dag}\right)
\eeqa
where
\beqa
&&i\delta_{\bf Q}Y:=\bpsi^{\bj}\partial_{\bj}Y+QY-(-1)^{|Y|}YQ
+2i\dot{x}^i{\partial\over\partial\psi^i}Y,
\\
&&i\delta_{\overline{\bf Q}}Y
:=\psi^i\partial_iY-Q^{\dag}Y+(-1)^{|Y|}YQ^{\dag}+2i\dot{x}^{\bi}
{\partial\over \partial\bpsi^{\bi}}Y.
\eeqa
We see that the interaction preserves the ${\mathcal N}=2$ supersymmetry
if and only if $Q$ satisfies, in addition to (\ref{psiindep}),
\beq
\bpsi^{\bj}\partial_{\bj}Q+Q^2=c\cdot {\rm id}_E
\label{genN2eq}
\eeq
where $c$ is a field independent constant.
Under these conditions, the boundary interaction takes the following form
up to an additive constant
\beq
\wA_t=-i\dot{x}^{\bj}{\partial\over\partial\bpsi^{\bj}}Q
+i\dot{x}^i{\partial\over\partial\psi^i}Q^{\dag}
-{1\over 2}\psi^i\partial_iQ+{1\over 2}\bpsi^{\bj}\partial_{\bj}Q^{\dag}
+{1\over 2}\{Q,Q^{\dag}\}.
\label{moregenN2}
\eeq
For the case where $Q(x,\bpsi)$ is at most linear in
 $\bpsi$, i.e.,
$Q(x,\bpsi)=Q(x)-\bpsi^{\bj}A_{\bj}(x)$, the equation 
(\ref{genN2eq}) splits into two equations,
$D_{\bj}Q(x)=0$ and $Q(x)^2=c\cdot {\rm id}_E$, which are nothing but
the condition obtained previously.

For the most part of this paper,
we will not consider such higher dimensional boundary interactions,
except for Section~\ref{sec:D-isom} where we discuss
D-term deformations.

\subsubsection{R-Symmetry}
\label{subsubsec:R-symmetry}

The bulk non-linear sigma model always has vector $U(1)$ R-symmetry
that acts trivially on the target space coordinates.
If preserved by the boundary interaction, the bulk vector $U(1)$ R-symmetry
 becomes an R-symmetry of the ${\mathcal N}=2_B$ superalgebra
under which the supercharge ${\bf Q}$ 
transforms as ${\bf Q}\to\lambda{\bf Q}$ for some phase $\lambda=\e^{i\alpha}$.
We restrict our attention to D-branes with such $U(1)$ R-symmetry.

Since the holomorphic part of the tachyon
$Q$ enters into the supercharge ${\bf Q}$, which has R-charge $1$,
$Q$ must also have R-charge $1$. Namely, the Chan-Paton bundle
$E$ must admit a $U(1)$ action $R(\e^{i\alpha}):E\to E$
such that
\beq
R(\lambda)Q(x) R(\lambda)^{-1}=\lambda Q(x).
\label{Rcharge1}
\eeq
This in particular requires $c=0$ in (\ref{uni}), so that
\beq
Q^2=0.
\label{nil}
\eeq
We of course require that the connection $A$ is invariant under
the same action
\beq
R(\lambda)A_I(x)R(\lambda)^{-1}=A_I(x).
\eeq
Let us denote the subbundle of $E$ of R-charge $j$ by $E^j$. That is,
$E^j$ is the $R(\lambda)=\lambda^j$ eigenbundle.
Each $E^j$ has a holomorphic structure determined by the connection
$A$ restricted to $E^j$. 
We denote the corresponding holomorphic vector bundle by
${\mathcal E}^j$. Then, by (\ref{Rcharge1})
the $Q$ action on ${\mathcal E}=\oplus_j{\mathcal E}^j$
is decomposed as
\beq
\cdots\stackrel{Q}{\longrightarrow}
{\mathcal E}^{j-1}\stackrel{Q}{\longrightarrow}
{\mathcal E}^{j}\stackrel{Q}{\longrightarrow}
{\mathcal E}^{j+1}\stackrel{Q}{\longrightarrow}\cdots.
\label{complex}
\eeq
The condition (\ref{nil}) means that this is a {\it complex} of holomorphic
vector bundles where $Q$ plays the r\^ole of a boundary operator.
We shall sometimes denote this complex by
${\mathcal C}={\mathcal C}({\mathcal E},Q)$.
We may assume that the R-charges $j$ are all integers, or equivalently,
that $\lambda\mapsto R(\lambda)$ is an honest $U(1)$ action
--- one can always redefine
$R(\lambda)$ by multiplying some phase $\lambda^{\delta_j}$
in each irreducible factor.
Since $Q$ is odd, the mod $2$ reduction of the grading by such integral
R-charges
matches with or is opposite to the original $\Z_2$-grading,
$E=E^{\rm ev}\oplus E^{\rm od}$, or mixture of the two cases.
We choose $R(\e^{i\alpha})$ so that they match:
\beq
E^{\rm ev}=\bigoplus_{j:\,\,{\rm even}}E^j,\quad
E^{\rm od}=\bigoplus_{j:\,\,{\rm odd}}E^j.
\label{mod2red}
\eeq
The reason is the charge integrality that is present
in the bulk non-linear sigma model, i.e., any bulk operator has integral
vector R-charge which agrees modulo 2 with the statistics of the
operator. 
By the requirement (\ref{mod2red})
we extend this charge integrality to
the boundary.

Let us consider two branes with such R-symmetry,
say, ${\mathcal B}_1=(E_1,A_1,Q_1)$ with symmetry $R_1$
and ${\mathcal B}_2=(E_2,A_2,Q_2)$ with symmetry $R_2$.
Then, there is an action of R-symmetry on the space of open string
states ${\mathcal H}({\mathcal B}_1,{\mathcal B}_2)$.
The action on the Chan-Paton factor is governed by $R_1$ and $R_2$;
\beq
{\bf R}(\lambda):\Psi_{\rm CP}\otimes \Psi_{\rm internal}
\longmapsto R_2(\lambda)\Psi_{\rm CP}R_1(\lambda)^{-1}\otimes
R_{\rm internal}(\lambda)\Psi_{\rm internal}.
\eeq
This introduces a grading on the space of open string states,
${\mathcal H}({\mathcal B}_1,{\mathcal B}_2)=\oplus_{p}
{\mathcal H}^p({\mathcal B}_1,{\mathcal B}_2)$,
with
\beq
{\mathcal H}^p({\mathcal B}_1,{\mathcal B}_2)
=\left\{\Psi\in {\mathcal H}({\mathcal B}_1,{\mathcal B}_2)
\,\,
\Bigl|\,\,{\bf R}(\lambda)\Psi=\lambda^p\Psi
\,\right\}.
\label{Hgrade}
\eeq
The R-charges $p$ of states are not necessarily
integers since the R-action on the internal part may not be integral.

There is an ambiguity in the choice of $R(\lambda)$ ---
the uniform shift $j\to j+2m$ does not violate
the condition (\ref{Rcharge1}) and (\ref{mod2red}). 
This shift of R-charges does not change the physical property
of the brane. However, for a given action
$R_{\rm internal}(\lambda)$ on the internal part,
different shifts of $R_1$ and $R_2$ will shift the grading 
of the space of states  (\ref{Hgrade}).
We sometimes keep this grading as a part of
the information of the D-brane. We call such branes with additional
information {\it graded D-branes}.

Let us briefly comment on the more general interaction (\ref{moregenN2})
for $\bpsi$-dependent $Q$ that obeys the equation (\ref{genN2eq}).
The R-symmetry condition is
$$
R(\lambda)Q(x,\lambda\bpsi)R(\lambda)^{-1}
=\lambda Q(x,\bpsi).
$$
This again requires $c=0$ in (\ref{genN2eq}).
If we write $Q=Q^0+Q^1+\cdots$ where $Q^m$ is the part that has
power $m$ in $\bpsi$, then the condition is
$R(\lambda)Q^mR(\lambda)^{-1}=\lambda^{1-m}Q^m$.
Thus, if we replace $\bpsi^{\bi}$ by the one form
$\dd x^{\bi}$ one can interpret $Q^m$ as a $(0,m)$ form that sends
$E^j$ to $E^{j+1-m}$. Then the supersymmetry condition (\ref{genN2eq})
becomes
\beq
\bartial Q+Q^2=0.
\label{moregenN3}
\eeq
for the total sum $Q=\sum_{m=0}^nQ^m$ of forms.
This data for D-branes is called a {\it twisted complex}.
It is found from the point of view of string field theory
in \cite{calin,emanuel} as a physical realization of an object of the
``enhanced triangulated category'' of Bondal and Kapranov \cite{BonKap}.
Here we showed its realization as an ordinary
${\mathcal N}=2$ supersymmetric boundary interaction.
This generalization is not necessary when $X$ is algebraic,
in the sense that will be explained at the end of Section~\ref{subsub:chiral}
and in Section~\ref{sec:D-isom}.

If the target space $X$ is a Calabi-Yau manifold, the bulk theory
flows to a non-trivial fixed point in the infra-red limit,
where vector and axial R-symmetries of the classical Lagrangian
become parts of the $(2,2)$ superconformal symmetry.
If it has a large volume limit,
the correct R-symmetries are the ones such that
the target space coordinates have zero R-charges. 
This is so even if $X$ has a non-trivial $U(1)$ symmetry by which the
R-symmetries could be modified \cite{somecomments}.
The superconformal field theory obtained this way is of special type ---
the R-charges in the NS-NS sector are all integers
and reduce modulo 2 to the $\Z_2$-grading that determines the
spin and statistics.
Also, there exist chiral {\it spectral flow operators} 
${\mathcal O}_{{1\over 2},0}$ and
${\mathcal O}_{0,{1\over 2}}$
in the NS-R and R-NS sectors, which are responsible
for spacetime supersymmetry \cite{FMS}
in the context of string compactifications.
In superconformal field theories
of this type, {\it BPS D-branes} are those such that the
two spectral flow operators are related by 
${\mathcal O}_{{1\over 2},0}=\e^{i\vp}{\mathcal O}_{0,{1\over 2}}$
for some phase $\e^{i\vp}$ \cite{OOY}. 
In the context of string theory,
this phase determines the
spacetime supersymmetry preserved by the D-branes.
For an open string stretched bewteen BPS branes with phases 
$\e^{i\vp_1}$ and $\e^{i\vp_2}$,
the R-charges of the states in the NS-sector
are $\vp_1-\vp_2$ plus integers \cite{Douglas}.
The brane defined by a boundary interaction
$(E,A,Q)$ may or may not flow to such a BPS D-brane.
If it does,
then the R-symmetry of $(E,A,Q)$ is expected to become a part of
the superconformal symmetry of the infra-red brane.
Often the brane $(E,A,Q)$ flows to the decoupled sum of several BPS branes
with different phases $\e^{i\vp}$.
In that case the R-symmetry of $(E,A,Q)$ may not correspond to
the infra-red R-symmetry.

\subsubsection{Chiral Sector}
\label{subsub:chiral}

In a supersymmetric field theory with a supercharge ${\bf Q}$ that
squares to zero,
such as $4d$ ${\mathcal N}=1$ and $2d$ $(2,2)$ theories,
the {\it chiral ring} is defined as the ring of ${\bf Q}$-cohomology
classes of local operators. It carries an important information of the theory
that is protected from renormalization.
This is the case also in $2d$ $(2,2)$ theories with
boundary interactions that preserve ${\mathcal N}=2$ supersymmetry.
In this context, a local operator is inserted on the boundary of the
worldsheet, say, at the point $z=0$ of the upper-half plane ${\rm Im}z\geq 0$.
Note that a boundary condition or a boundary interaction
must be specified.
Suppose the boundary interactions on the left ($z<0$) and 
the right ($z>0$) of the insertion point are
${\mathcal B}_1$ and ${\mathcal B}_2$ respectively.
Then the space of
${\bf Q}$-cohomology classes of operators is denoted by
$\Ho({\mathcal B}_1,{\mathcal B}_2)$.
If the two are the same brane,
the space $\Ho({\mathcal B},{\mathcal B})$ by itself
forms a ring by the product of operators.
For different branes, the product is of the form
$\Ho({\mathcal B}_1,{\mathcal B}_2)\times\Ho({\mathcal B}_2,{\mathcal B}_3)
\to \Ho({\mathcal B}_1,{\mathcal B}_3)$.
If we fix
a set of branes $\{{\mathcal B}_i\}_{i\in I}$,
the direct sum of spaces
$\Ho({\mathcal B}_i,{\mathcal B}_j)$ ($i,j\in I$) forms a ring
--- we may call it the chiral ring corresponding to
the set $\{{\mathcal B}_i\}_{i\in I}$.
If we consider all possible branes, 
it would be mathematically more appropriate
to use the language of category  --- objects are D-branes ${\mathcal B}$
and morphisms are elements of $\Ho({\mathcal B},{\mathcal B}')$.
This is how the {\it D-brane category} (in the chiral sector) is defined.
In this paper, however, we shall loosely refer to
 elements of $\Ho({\mathcal B},{\mathcal B}')$ as
 {\it chiral ring elements}.

If the bulk theory has an axial R-symmetry with integral R-charges,
such as a Calabi-Yau sigma model,
one can use B-twist to obtain a topological field theory in which
chiral ring elements play the r\^ole of physical observables.
In this context the D-brane category in the chiral sector
is called ``the category of topological D-branes''.
Also, the B-twist can be used to find a 
one-to-one correspondence between the chiral ring elements
in the ${\mathcal B}_1$-${\mathcal B}_2$ sector and
supersymmetric ground states of the open string stretched from
${\mathcal B}_1$ to ${\mathcal B}_2$.
Therefore,
the terms ``chiral ring elements'' and ``supersymmetric ground states''
can be used interchangeably in such a case.

In what follows, we determine the space
$\Ho({\mathcal B}_1,{\mathcal B}_2)$
for a pair of D-branes, ${\mathcal B}_1=(E_1,A_1,Q_1)$ and
${\mathcal B}_2=(E_2,A_2,Q_2)$, in the non-linear sigma model
on a K\"ahler manifold $X$.
We first realize it as a Dolbeault type cohomology,
and then try to translate it into a purely holomorphic description.
This leads us to the derived category of $X$.

The fields $\psi_{\pm}^i$ and $\partial_{z}x^i$, $\partial_{\bz}x^i$
are ${\bf Q}$-partners of each other and thus can be eliminated.
Also,
$\bpsi^{\bi}_+-\bpsi^{\bi}_-$ are set equal to zero by the boundary condition.
Thus, we may work only with the zero modes of
$x^i$, $x^{\bi}$ and $\bpsi^{\bi}=\bpsi_+^{\bi}+\bpsi_-^{\bi}$.
In this zero mode sector,\footnote{The meaning of ``zero mode'' may require
clarification.
It means ``constant (or more precisely {\it parallel}) mode on the
flat worldsheet with straight boundary (at which the operators are 
inserted)''.
It should not be confused with the ``zero mode'' in the
open string NS sector to which the space of local operators naturally
corresponds to: there is in fact no parallel
mode for spinors in the NS sector. 
However, when B-twist is possible, it literally corresponds to
``zero mode'' in the open string Ramond sector.}
the space of operators is identified as the space of
antiholomorphic 
forms with values in the bundle of linear maps $Hom(E_1,E_2)$;
\beq
{\mathcal H}_{\rm zero}({\mathcal B}_1,{\mathcal B}_2)
=\bigoplus_{i=1}^n\Omega^{0,i}(X,Hom(E_1,E_2)).
\label{Hzero}
\eeq
On this space, $\bpsi^{\bj}$ is represented as
the one-form $\dd \overline{x}^{\bj}$ while
$g_{i\bj}\dot{x}^i$ acts as the differentiation $-i\partial_{\bj}+A_{\bj}$.
Therefore the supercharge ${\bf Q}$, as shown in (\ref{Qbulk}) and
(\ref{Qbdry}),
is represented as
\beq
i{\bf Q}^{\rm zero}\phi=\bartial_{A_{1,2}}\phi+Q_2\phi-(-1)^{|\phi|}\phi Q_1.
\eeq
$\bartial_{A_{1,2}}$ is the Cauchy-Riemann operator
determined by the connections $A_1$ and $A_2$, which is expressed locally as
$\bartial_{A_{1,2}}\phi
=\dd z^{\bj}\wedge(\partial_{\bj}\phi+iA_{2,\bj}\phi-i\phi A_{1,\bj})$.
The $\Z_2$-grading $(-1)^{|\phi|}$ is the combination of
the one for $Hom(E_1,E_2)$ and the one by the form degree.
If both ${\mathcal B}_1$ and ${\mathcal B}_2$ have R-symmetry,
 the space (\ref{Hzero}) has a finer grading.
It is graded by the sum of the form degree and the grading of $Hom(E_1,E_2)$
determined by the R-charges of the bundles $E_1$ and $E_2$,
where the degree $j$ elements of $Hom(E_1,E_2)$
increase the R-charge by $j$,
$Hom^j(E_1,E_2)=\oplus_{j'}Hom(E_1^{j'},E_2^{j'+j})$.
The R-charge $p$ subspace of (\ref{Hzero}) hence is
\beq
{\mathcal H}_{\rm zero}^p({\mathcal B}_1,{\mathcal B}_2)
=\bigoplus_{i+j=p}\Omega^{0,i}(X,Hom^j(E_1,E_2)).
\label{Hzerop}
\eeq
The supercharge ${\bf Q}^{\rm zero}$ is a degree one operator 
that squares to zero, and hence defines a $\Z$-graded complex.
The space of chiral ring elements is
isomorphic to the cohomology group of this complex
\beq
\Ho^p({\mathcal B}_1,{\mathcal B}_2)
\cong H^p_{{\bf Q}^{\rm zero}}
({\mathcal H}_{\rm zero}^{\bullet}({\mathcal B}_1,{\mathcal B}_2)).
\label{cohomo}
\eeq
The ring structure $\Ho({\mathcal B}_1,{\mathcal B}_2)\times
\Ho({\mathcal B}_2,{\mathcal B}_3)\to\Ho({\mathcal B}_1,{\mathcal B}_3)$
is simply realized as the wedge product of
forms combined with the composition of homomorphisms.
In particular, it preserves the grading by the R-symmetry.

Note that the degree $p$ here is not necessarily the same as the R-charge of
the quantum fields that
includes the contribution from the internal part (such as the
sum of the zero point charges).
For distinction, we call it the {\it R-degree}
or ``R-charge in the zero mode approximation''.
If ${\mathcal B}_1$ and ${\mathcal B}_2$ flow to BPS D-branes
with
phases $\e^{i\vp_1}$ and $\e^{i\vp_2}$,
the true R-charge is $p+\vp_1-\vp_2$ for a suitably chosen
integral part of $\vp_i$.

\subsubsection*{\it Flat Space $X=\C^n$ --- Homotopy Category}

As the simplest example, 
let us consider the Euclidean space $X=\C^n$
and complexes based on vector bundles with trivial flat connections.
Namely, we consider complexes of the form (\ref{complex})
where ${\mathcal E}^j$ are all trivial bundles ${\mathcal O}^{\oplus k_j}$
over $\C^n$.
A brane is represented by an odd square matrix $Q(x)$
which is holomorphic in the complex coordinates $x=(x^1,...,x^n)$ 
and squares to zero. For two branes of this kind,
${\mathcal B}_1=({\mathcal E}_1,Q_1)$ and 
${\mathcal B}_2=({\mathcal E}_2,Q_2)$,
the space of chiral ring elements is realized as
the cohomology group (\ref{cohomo}) but there is a more convenient
realization. Let 
$\Hom({\mathcal E}_1,{\mathcal E}_2)$ 
be the space of global holomorphic bundle
maps from ${\mathcal E}_1$ to ${\mathcal E}_2$, which is
a graded subspace of the $i=0$ part of (\ref{Hzerop}).
The operator ${\bf Q}^{\rm zero}$ acts on this subspace as
$$
i{\bf Q}^{\it hol}\phi=Q_2\phi-(-1)^{|\phi|}\phi Q_1
$$
and defines a complex
$\cdots\to \Hom^p({\mathcal E}_1,{\mathcal E}_2)\to
\Hom^{p+1}({\mathcal E}_1,{\mathcal E}_2)\to\cdots$.
We claim that the space (\ref{cohomo}) is isomorphic to
the cohomology group of this complex,
\beq
\Ho^p({\mathcal B}_1,{\mathcal B}_2)
\cong
H^p_{{\bf Q}^{\it hol}}(\Hom^{\bullet}
({\mathcal E}_1,{\mathcal E}_2)).
\label{cohomo2}
\eeq
This can be shown as follows.
Let us pick a degree zero element
$\phi\in {\mathcal H}_{\rm zero}^0({\mathcal B}_1,{\mathcal B}_2)$ 
annihilated by ${\bf Q}^{\rm zero}$. 
We decompose it with respect to its 
form-degree, $\phi=\phi^0+\phi^1+\cdots+\phi^n$ where
$\phi^i$ belongs to $\Omega^{0,i}(\C^n,Hom^{-i}(E_1,E_2))$.
Then the equation ${\bf Q}^{\rm zero}\phi=0$ can be decomposed as follows
\beqa
\bartial \phi^{n-1}+Q_2\phi^n-\phi^nQ_1=0,&&\nn\\
\bartial \phi^{n-2}+Q_2\phi^{n-1}-\phi^{n-1}Q_1=0,&&\nn\\
\vdots~~~~ &&\label{decompo}\\
\bartial \phi^0+Q_2\phi^1-\phi^1Q_1=0,&&\nn\\
Q_2\phi^0-\phi^0Q_1=0.&&
\nn
\eeqa
Since any $(0,n)$ form on $\C^n$ 
is $\bartial$-exact, there is an $(0,n-1)$-form
$\beta^{n-1}$ such that $\phi^n=\bartial\beta^{n-1}$.
Then the first equation means that $\phi^{n-1}-Q_2\beta^{n-1}
-\beta^{n-1}Q_1$ is $\bartial$-closed. Hence it
can be written as $\bartial\beta^{n-2}$ for some
$(0,n-2)$-form $\beta^{n-2}$.
Then the second equation means that $\phi^{n-2}-Q_2\beta^{n-2}
-\beta^{n-2}Q_1$ is $\bartial$-closed and hence can be written
as $\bartial\beta^{n-3}$ for some
$(0,n-3)$-form $\beta^{n-3}$.
Repeating this procedure, using the fact that
any $\bartial$-closed $(0,i)$-form on $\C^n$ is $\bartial$-exact if $i>0$,
we recursively
find a sequence of forms, $\beta^i\in \Omega^{0,i}(Hom^{-i-1}(E_1,E_2))$
such that
$\phi^i-Q_2\beta^{i}-\beta^{i}Q_1=\bartial\beta^{i-1}$
at every $i=n-1,n-2,...,1$.
The last two equations of (\ref{decompo}) mean that
$\phi^0-Q_2\beta^0-\beta^0Q_1=:\widetilde{\phi}$ is a holomorphic
$0$-form that is ${\bf Q}^{\it hol}$-closed.
Summarizing, we found
\beq
\phi=\widetilde{\phi}+i{\bf Q}^{\rm zero}\beta,\qquad
\bartial\widetilde{\phi}={\bf Q}^{\it hol}\widetilde{\phi}=0,
\eeq
where $\beta=\beta^0+\beta^1+\cdots+\beta^{n-1}$.
Namely, every ${\bf Q}^{\rm zero}$-closed element
has a holomorphic representative.
This proves the claim (\ref{cohomo2}) for the case $p=0$.
Proof for higher $p$ is similar.
The only non-trivial property of $\C^n$ 
we have used in the above argument
is that any $\bartial$-closed form of positive degree
is $\bartial$-exact.
This holds more generally in a class of spaces 
called {\it Stein manifolds}.

Let us restate the claim (\ref{cohomo2})
in a more conventional mathematical language.
For this we need to introduce some terminology.
Here everything is stated in the context of complexes of trivial vector 
bundles where maps are holomorphic bundle maps, but the
terminology can be applied straightforwardly to more general context.
By definition, a {\it cochain map} of a complex 
$\cdots\to {\mathcal E}^j\to{\mathcal E}^{j+1}\to\cdots$
to another $\cdots\to {\mathcal F}^j\to{\mathcal F}^{j+1}\to\cdots$
is a sequence of maps ${\mathcal E}^j\to
{\mathcal F}^j$ such that the following diagram commutes
\beq
\begin{array}{ccccccc}
\cdots\to&{\mathcal E}^{j-1}&\longrightarrow&
{\mathcal E}^j&\longrightarrow&{\mathcal E}^{j+1}
&\to\cdots\\
&\downarrow&&\downarrow&&\downarrow&\\
\cdots\to&{\mathcal F}^{j-1}&\longrightarrow&
{\mathcal F}^j&\longrightarrow&{\mathcal F}^{j+1}
&\to\cdots
\end{array}
\nn
\eeq
A {\it homotopy} between these complexes
is a cochain map of the form $hQ_{\mathcal E}+Q_{\mathcal F}h$
where $h$ is a degree $-1$ map of ${\mathcal E}$ to
${\mathcal F}$, that is, a collection of maps
${\mathcal E}^j\to{\mathcal F}^{j-1}$.
For a complex ${\mathcal C}={\mathcal C}({\mathcal E},Q)$, we denote by
${\mathcal C}[p]$ the same complex ${\mathcal C}$ shifted $p$ steps
to the left with the coboundary operator given by $(-1)^pQ$.
In these terms, the claim (\ref{cohomo2}) means that the
space of degree $p$ chiral ring elements is isomorphic to the space
of cochain maps ${\mathcal C}_1\to {\mathcal C}_2[p]$ modulo
homotopies, where ${\mathcal C}_1$ and ${\mathcal C}_2$
are the complexes associated with $({\mathcal E}_1,Q_1)$
and $({\mathcal E}_2,Q_2)$. Such a space is simply denoted by
$\HomHo({\mathcal C}_1,{\mathcal C}_2[p])$. Thus, we may say
\beq
\Ho^p({\mathcal B}_1,{\mathcal B}_2)
\cong \HomHo({\mathcal C}_1,{\mathcal C}_2[p]).
\label{homotopy}
\eeq
The ring structure is given by composition of cochain maps.
The D-brane category we obtained, where the
objects are complexes of trivial vector bundles over $\C^n$
and morphisms are
cochain maps modulo homotopies (\ref{homotopy}), is what is known as
{\it homotopy category} of the category of complexes of vector bundles
over $\C^n$.
This is in fact the same as the derived category
of $\C^n$, which we will discuss momentarily in more general context.
This point will be discussed further in Section~\ref{sec:math}.

\subsubsection*{\it Extensions}

We would like to have a holomorphic or algebraic description of
the chiral ring, like (\ref{homotopy}),
in more general spaces.
Here we present such a description for branes which are vector bundles, 
namely complexes where ${\mathcal E}^j$ is non-zero only for $j=0$.
Let $(E_1,A_1)$ and $(E_2,A_2)$ be vector bundles with 
connection which determine holomorphic vector bundles
${\mathcal E}_1$ and ${\mathcal E}_2$.
The space (\ref{cohomo}) is simply
the Dolbeault cohomology group which
is linearly isomorphic to the \v Cech cohomology group
of (the sheaf of holomorphic sections of)
the holomorphic bundle
 ${\mathcal Hom}({\mathcal E}_1,{\mathcal E}_2)\cong 
{\mathcal E}_1^*\otimes {\mathcal E}_2$;
\beq
\Ho^p({\mathcal B}_1,{\mathcal B}_2)
\cong H^{0,p}_{\bartial_{A_{1,2}}}(X,Hom(E_1,E_2))
\cong H^p(X,{\mathcal Hom}({\mathcal E}_1,{\mathcal E}_2)).
\eeq
For $p=0$, this is equal to the space of global holomorphic sections
of ${\mathcal Hom}({\mathcal E}_1,{\mathcal E}_2)$, namely the space 
of global holomorphic bundles maps,
\beq
\Ho^0({\mathcal B}_1,{\mathcal B}_2)
\cong\Hom({\mathcal E}_1,{\mathcal E}_2).
\eeq
Indeed this is nothing but the answer obtained in (\ref{homotopy}).
However, for $p>0$, (\ref{homotopy}) would tell us 
that there is no chiral ring elements in the present situation,
where both ${\mathcal B}_1$ and ${\mathcal B}_2$ are vector bundles. 
That is, however, not true in general since we may have
$H^{0,p}_{\bartial_{A_{1,2}}}(X,Hom(E_1,E_2))\ne 0$
for $p>0$.

The $p=1$ subspace $\Ho^1({\mathcal B}_1,{\mathcal B}_2)$
has the following algebraic characterization.
Let $\phi\in \Omega^{0,1}(X,Hom(E_1,E_2))$ represent an element of
$H^{0,1}_{\bartial_{A_{1,2}}}(X,Hom(E_1,E_2))$. It obeys
$\bartial_{A_{1,2}}\phi=0$. Then one can define a holomorphic structure
${\mathcal F}_{\phi}$ on $E_1\oplus E_2$ by the operator
\beq
\bartial_{A_{1,2},\phi}=\left(\begin{array}{cc}
\bartial_{A_1}&0\\
\phi&\bartial_{A_2}
\end{array}\right).
\label{A12phi}
\eeq
Indeed it squares to zero under the condition $\bartial_{A_{1,2}}\phi=0$.
The exact sequence
$0\to E_2\to E_1\oplus E_2\to E_1\to 0$ given by
the trivial maps $e_2\mapsto (0,e_2)$ and
$(e_1,e_2)\mapsto e_1$ defines an exact sequence of holomorphic bundles
\beq
0\longrightarrow{\mathcal E}_2\longrightarrow {\mathcal F}_{\phi}
\longrightarrow {\mathcal E}_1\longrightarrow 0.
\label{extension}
\eeq
Such a sequence is called an {\it extension} of
${\mathcal E}_1$ by ${\mathcal E}_2$.
If $\phi$ is shifted by
$\bartial_{A_{1,2}}$-exact form, $\phi\to \phi+\bartial_{A_{1,2}}\beta$,
where $\beta\in \Gamma(X,Hom(E_1,E_2))$,
there is an {\it isomorphism} of the extension (\ref{extension})
to the new one. Namely, there is a cochain map from
(\ref{extension}) to the new one, which is identity at 
${\mathcal E}_1$ and ${\mathcal E}_2$. 
The map in the middle is necessarily an isomorphism
--- explicitly it is given by $(e_1,e_2)\to (e_1,e_2-\beta e_1)$.
The set of isomorphism classes of extensions, denoted by
$\Ext^1({\mathcal E}_1,{\mathcal E}_2)$,
is thus bijective to the cohomology group
$H^{0,1}_{\bartial_{A_{1,2}}}(X,Hom(E_1,E_2))$.
Thus, the space of $p=1$ 
chiral ring elements is 
\beq
\Ho^1({\mathcal B}_1,{\mathcal B}_2)
\cong \Ext^1({\mathcal E}_1,{\mathcal E}_2).
\label{ext1}
\eeq
A structure of 
complex vector space can be defined 
on the set $\Ext^1({\mathcal E}_1,{\mathcal E}_2)$
and (\ref{ext1}) is a linear isomorphism.
An extension (\ref{extension}) is a zero element
if $\phi$ itself is exact, $\phi=\bartial_{A_{1,2}}\beta$.
In that case, there is a holomorphic map
${\mathcal E}_1\to {\mathcal F}_{\phi}$ 
(resp. ${\mathcal F}_{\phi}\to{\mathcal E}_2$)
which gives 
the identity on ${\mathcal E}_1$ 
(resp. on ${\mathcal E}_2$) 
if it is followed by
the map ${\mathcal F}_{\phi}\to {\mathcal E}_1$ of (\ref{extension})
(resp. preceded by 
the map ${\mathcal E}_2\to {\mathcal F}_{\phi}$ of (\ref{extension})).
Explicitly, the map is given by $e_1\to (e_1,-\beta e_1)$
(resp. $(e_1,e_2)\mapsto e_2+\beta e_1$).
Such a map is called a {\it splitting}, and an exact sequence with
a splitting is called {\it split exact}. An exact sequence represents the zero 
element of the group $\Ext^1({\mathcal E}_1,{\mathcal E}_2)$ when it is 
split exact.

For higher $p$, algebraic characterization is not as simply derived as
in the $p=1$ case. We just quote the fact that
$H^p(X,{\mathcal Hom}({\mathcal E}_1,{\mathcal E}_2))$
is linearly isomorphic to the group of equivalence classes of $p$-extensions
of ${\mathcal E}_1$ by ${\mathcal E}_2$,
$\Ext^p({\mathcal E}_1,{\mathcal E}_2)$.
A $p$-extension of ${\mathcal E}_1$ by ${\mathcal E}_2$
is an exact sequence of the form
$$
0\longrightarrow
{\mathcal E}_2\longrightarrow {\mathcal F}_1
\longrightarrow \cdots\longrightarrow 
{\mathcal F}_p
\longrightarrow {\mathcal E}_1\longrightarrow 0.
$$
Equivalences of $p$-extensions are defined in a similar but 
slightly more involved way than in the $p=1$ case. 
The  extension defines the zero element of the group
$\Ext^p({\mathcal E}_1,{\mathcal E}_2)$
if there is a splitting
${\mathcal E}_1\to{\mathcal F}_p$ or
${\mathcal F}_1\to {\mathcal E}_2$.
We refer the reader to \cite{HAbooks} as well as
Section~\ref{sec:math} for more details.

\subsubsection*{\it Derived Category}

What is the algebraic description of the space of chiral ring elements
(\ref{cohomo})? 
We have seen partial answers: For $X=\C^n$ it is the space of
homotopy classes of cochain maps (\ref{homotopy}). For vector bundles it is the
extension group $\Ext^p({\mathcal E}_1, {\mathcal E}_2)$.
In general, the answer is provided by the formalism of 
the derived category\footnote{Possible relevance of the derived category
for brane-antibrane systems was first emphasized in \cite{Sharpe}.}
---
It is the space of morphisms of the derived category ${\bf D}$:
\beq
\Ho^p({\mathcal B}_1,{\mathcal B}_2)
\cong \HomD({\mathcal C}_1,{\mathcal C}_2[p]).
\label{D-hom}
\eeq
Here ${\mathcal C}_a$ is the complex of vector bundles
associated with ${\mathcal B}_a=({\mathcal E}_a,Q_a)$
and ${\mathcal C}_2[p]$ is the complex
${\mathcal C}_2$ shifted $p$-steps to the left.
Let us sketch the definition of the right hand side of (\ref{D-hom}).

First we need to make a technical remark.
A disadvantage in working with complexes of vector bundles is that the
kernel and cokernel of a bundle map are not in general vector bundles. 
To remedy this problem, we introduce some mathematical
objects called {\it sheaves (of ${\mathcal O}_X$-modules)}
as generalization of vector bundles. 
Basically, we consider the space of local holomorphic 
sections of a vector bundle as a module over the ring of 
local holomorphic functions, and we generalize 
it by including any module of that ring.
A map of sheaves is given by linear maps of the modules defined locally which
satisfy certain compatibility condition.
Section~\ref{sec:math} provides a slightly more detailed
explanation.
The main point is that we can freely talk about the kernel and cokernel
of a map of sheaves. 
We emphasize that this generalization is just to describe $\HomD(-,-)$.
Our branes are still complexes of vector bundles.

Of most importance is the notion of a {\it quasi-isomorphism}.
A quasi-isomorphism from a complex of sheaves 
$\cdots\to {\mathcal E}^j\to{\mathcal E}^{j+1}\to\cdots$
to another one $\cdots\to {\mathcal F}^j\to{\mathcal F}^{j+1}\to\cdots$
is a cochain map that descends to an isomorphism at the cohomology level.
Namely, each map ${\mathcal E}^j\to{\mathcal F}^j$
induces an isomoprphism from the cohomology sheaf
 $\Ker({\mathcal E}^j\to{\mathcal E}^{j+1})/
{\rm Im}({\mathcal E}^{j-1}\to{\mathcal E}^j)$
to the cohomology sheaf $\Ker({\mathcal F}^j\to{\mathcal F}^{j+1})/
{\rm Im}({\mathcal F}^{j-1}\to{\mathcal F}^j)$.
Note that a shift of a cochain map by a homotopy 
does not change the map at the cohomology level. Thus,
a chain map that is homotpy equivalent to a quasi-isomorphism
is again a quasi-isomorphism.

Let us now describe the space of morphisms in the derived category ${\bf D}$.
The homotopy classes of cochain maps define morphisms.
In addition, we include
the formal inverses of the homotopy classes of quasi-isomorphisms.
Namely, if
$q:{\mathcal C}_1\to {\mathcal C}_2$ is a quasi-isomorphism,
we include its formal inverse $q^{-1}$ as a morphism from
${\mathcal C}_2$ to ${\mathcal C}_1$ which
obeys the property that $q\circ q^{-1}={\rm id}_{{\mathcal C}_2}$
and $q^{-1}\circ q={\rm id}_{{\mathcal C}_1}$.
Then $\HomD({\mathcal C}_1,{\mathcal C}_2)$
is defined as the set of all sequences of such extended morphisms
starting from ${\mathcal C}_1$ and ending at ${\mathcal C}_2$,
modulo the obvious identification
$$
\Bigl({\mathcal C}_1\to\cdots\to{\mathcal C}\stackrel{f}{\to}
{\mathcal C}'\stackrel{g}{\to}{\mathcal C}''\to\cdots\to
{\mathcal C}_2\Bigr)\,\,\,
\equiv \,\,\,
\Bigl({\mathcal C}_1\to\cdots\to
{\mathcal C}\stackrel{g\circ f}{\longrightarrow}{\mathcal C}''
\to\cdots\to{\mathcal C}_2\Bigr).
$$
One can show that each morphism from ${\mathcal C}_1$ to
${\mathcal C}_2$ have presentations of the following forms
$$
{\mathcal C}_1\stackrel{q^{-1}}{\longrightarrow}{\mathcal C}
\stackrel{f}{\longrightarrow}
{\mathcal C}_2,\qquad
{\rm and}\qquad
{\mathcal C}_1\stackrel{g}{\longrightarrow}{\mathcal C}'
\stackrel{{q'}^{-1}}{\longrightarrow}
{\mathcal C}_2,
$$
where $f$ and $g$
are (the homotopy classes of) ordinary cochain maps and
$q^{-1}$ and ${q'}^{-1}$ are the inverses of 
(the homotopy classes of) quasi-isomorphisms.

If there is a quasi-isomorphism $q:{\mathcal C}_1\to{\mathcal C}_2$, then
for any complex ${\mathcal C}$ there are linear isomorphisms,
$\HomD({\mathcal C},{\mathcal C}_1)\cong \HomD({\mathcal C},{\mathcal C}_2)$
and
$\HomD({\mathcal C}_1,{\mathcal C})\cong \HomD({\mathcal C}_2,{\mathcal C})$,
given by composition with
$q$ or $q^{-1}$ on the left and  on the right.
This means that two objects related by a chain of quasi-isomorphisms
are isomorphic in the derived category.

Let us consider a complex ${\mathcal C}$ given by an exact sequence
\beq
\cdots \to {\mathcal E}^i\stackrel{Q_i}{\longrightarrow}{\mathcal E}^{i+1}
\stackrel{Q_{i+1}}{\longrightarrow}{\mathcal E}^{i+2}\to\cdots
\label{exact1}
\eeq
By definition of exactness, $\Ker(Q_i)={\rm Im}(Q_{i-1})$,
the cohomology sheaves are all
zero. Hence the zero map from ${\mathcal C}$
to the zero sequence 
$$
\cdots \longrightarrow 0\longrightarrow 0\longrightarrow 0\longrightarrow 0
\longrightarrow \cdots
$$
(or back) is a quasi-isomorphism.
In particular, the morphism space to and from any complex ${\mathcal C}'$
vanishes, 
$\HomD({\mathcal C},{\mathcal C}')=\HomD({\mathcal C}',{\mathcal C})=0$.
Namely, ${\mathcal C}$ is a zero object in the derived category. 
Another important observation is that
the exact complex ${\mathcal C}$ can be broken
at any position $j$ to obtain a pair of quasi-isomorphic complexes
\beq
\begin{array}{ccccccccccc}
\cdots\to&{\mathcal E}^{j-2}&\stackrel{-Q_{j-2}}{\longrightarrow}
&{\mathcal E}^{j-1}&\stackrel{-Q_{j-1}}{\longrightarrow}&
{\mathcal E}^j&\longrightarrow&0&\longrightarrow&0&\to\cdots\\
&\downarrow&&\downarrow&&\,\,\,\downarrow {{}^{Q_j}}\!\!\!\!\!\!
&&\downarrow&&\downarrow&\\
\cdots\to&0&\longrightarrow&0&\longrightarrow&{\mathcal E}^{j+1}&
\stackrel{Q_{j+1}}{\longrightarrow}&
{\mathcal E}^{j+2}&\stackrel{Q_{j+2}}{\longrightarrow}&
{\mathcal E}^{j+3}&\to\cdots
\end{array}
\label{qisex}
\eeq
That this is a quasi-isomorphism can be see easily by the exactness
of ${\mathcal C}$.
Conversely, if this is a quasi-isomorphism, the complex 
(\ref{exact1}) is exact.

Let us comment on how (\ref{D-hom}) may be related to
the results obtained earlier for special cases.
First, for $X=\C^n$ we must show that the derived 
category is equivalent
 to the homotopy category. It is equal to the statement 
that a quasi-isomorphism is a chain isomorphism up to homotopy.
In particular, there is no non-trivial extension between trivial
vector bundles.
This point has a particular importance
in our paper and will be explained in Section~\ref{sec:math}.
Next, let us consider vector bundles ${\mathcal E}_1$ and ${\mathcal E}_2$
on a more general space $X$
and ask whether $\HomD({\mathcal E}_1,{\mathcal E}_2[p])$
is isomorphic to the extension group $\Ext^p({\mathcal E}_1,{\mathcal E}_2)$.
Let us demonstrate the map from the latter to the former for the $p=1$ case.
Take an element of $\Ext^1({\mathcal E}_1,{\mathcal E}_2)$
represented by an exact sequence
$$
0\longrightarrow{\mathcal E}_2\stackrel{a}{\longrightarrow}
{\mathcal F}\stackrel{b}{\longrightarrow}{\mathcal E}_1\longrightarrow
0.
$$
Then we find the following morphism ${\mathcal E}_1\to {\mathcal E}_2[1]$
in the derived category
$$
\begin{array}{ccccccccc}
\to&0&\longrightarrow&0&\longrightarrow&{\mathcal E}_1&\longrightarrow&0&\to\\
&\uparrow&&\uparrow&&\,\,\,\uparrow \!{}_a\!\!&&\uparrow&\\
\to&0&\longrightarrow&
{\mathcal E}_2&\stackrel{b}{\longrightarrow}&{\mathcal F}&\longrightarrow&
0&\to\\
&\downarrow&&\,\,\downarrow \!{}^{\rm id}\!\!&&\downarrow&&\downarrow&\\
\to&0&\longrightarrow&{\mathcal E}_2&\longrightarrow&0&\longrightarrow&0&\to
\end{array}
$$
From the first line to the second
is the inverse of a quasi-isomorphism of the type (\ref{qisex}).
Alternatively, we could consider the morphism with
$0\to {\mathcal F}\to{\mathcal E}_1\to 0$ in the second line and
an inverse quasi-isomorphism from the second line to the third, but that is
homotopy equivalent to the above one.
The generalization to $p>1$ is obvious.

Let us end the description of the derived category with another
technical remark.\footnote{We thank A. Bondal for everything that is said 
in this remark.}
A general sheaf of ${\mathcal O}_X$-modules is in some way too general
and, if possible, we would like to work with something close to vector bundles.
This motivates us to consider {\it coherent sheaves}. 
A coherent sheaf is realized locally as the cokernel
of a map of sheaves of holomorphic sections of vector bundles.
When $X$ is algebraic, namely, if it is covered by
open affine varieties,
then it is known that the intermediate complexes of sheaves that appear in
the definition of $\HomD({\mathcal C}_1,{\mathcal C}_2)$ can be taken from
complexes of coherent sheaves.
Also, any complex of coherent sheaves is known to be quasi-isomorphic to
a complex of vector bundles.
Thus, if $X$ is algebraic, we can describe D-branes entirely by complexes of
coherent sheaves, and the category ${\bf D}$ can be identified as
what is known as the derived category of coherent sheaves of $X$.
If $X$ is not algebraic, we may need more general sheaves than coherent
sheaves in the intermediate complexes to define
$\HomD({\mathcal C}_1,{\mathcal C}_2)$.
As the D-branes themselves, it is more natural to consider
complexes of sheaves whose cohomology sheaves are coherent.
They are slightly more general than complexes of vector bundles, but
are not more general than twisted complexes (\ref{moregenN3}). Namely,
a twisted complex determines a complex of sheaves with coherent cohomology
sheaves and any complex with coherent cohomologies 
is quasi-isomorphic to a complex coming from a twisted complex 
\cite{BoRo}.
(There is also a work which studies this point \cite{Block}.) 
In technical terms, the relevant category ${\bf D}$ is
the full subcategory of the derived category of sheaves of
${\mathcal O}_X$-modules consisting of complexes with coherent cohomologies.
When $X$ is algebraic, this subcategory is equivalent to
the derived category of coherent sheaves.
In the rest of this paper, 
we only consider algebraic $X$'s where we can entirely work with
complexes of coherent sheaves. 
Fortunately enough, Calabi-Yau manifolds with $h^{2,0}=0$ are all algebraic.

\subsection{Orbifolds}
\label{subsec:orbifolds}

When a quantum field theory in $1+1$ dimensions
has a finite group of symmetries,
one may consider gauging it. 
This is the operation known as {\it orbifold}.
We remove all the states and operators that are not invariant
under the group action, and
at the same time, we include field configurations on a circle
with preiodocity twisted by group elements, thus adding new sectors
(twisted sectors) to the space of states and operators \cite{DHVW}.
D-branes in an orbifold theory are simply boundary conditions and interactions
that are invariant under the orbifold group action.
The orbifold action on the Chan-Paton vectors must be included
as a part of the data,
in order to specify the action on the open string states so that
one can select only the invariant states.

For the non-linear sigma model on a Riemannian manifold $(X,g)$,
a typical orbifold is associated with a group $\Gamma$
of isometries of $(X,g)$. A D-brane of the type
$(E,A,{\bf T})$ is $\Gamma$-invariant when there is a lift of the 
$\Gamma$-action to the vector bundle $E$ that preserves 
the $\Z_2$-grading, the
gauge field $A$ and the tachyon profile ${\bf T}$.
In particular, 
there is an even linear bundle map $\rho(\gamma):E\to E$ over 
each element $\gamma\in\Gamma$ such that
the pull back connection of $A$ is $A$ itself and
$$
\rho(\gamma )^{-1}{\bf T}(\gamma x)\rho(\gamma)={\bf T}(x).
$$
Note that the maps $\rho(\gamma)$ encode
the information of the $\Gamma$-action on Chan-Paton factors:
For the open string stretched from a brane
$(E_1,A_1,{\bf T}_1,\rho_1)$ to another brane $(E_2,A_2,{\bf T}_2,\rho_2)$,
the $\gamma\in\Gamma$ action on open string wavefunctions
is
$$
\gamma:\Psi(x)\longmapsto
\rho_2(\gamma)^{-1}\Psi(\gamma x)\rho_1(\gamma).
$$
We are interested only in wavefunctions that are invariant under this
orbifold action.

If $(X,g)$ is a K\"ahler manifold and if the isometry group $\Gamma$
preserves also the complex structure, 
the orbifold theory has $(2,2)$ supersymmetry.
An ${\mathcal N}=2_B$ invariant D-brane
in the orbifold theory
can be provided by the usual data $(E,A,Q)$ together with
a $\Gamma$-action,
$\rho(\gamma):E\to E$, with the obvious invariance condition
including
$$
\rho(\gamma )^{-1}Q(\gamma x)\rho(\gamma)=Q(x).
$$
A $U(1)$ R-symmetry of such a brane is 
an R-symmetry $R$ of $(E,A,Q)$ that preserves
the $\Gamma$-invariance condition of open string wavefunctions.
It is easy to see that it requires
$$
R(\lambda)\rho(\gamma)=c_{\lambda,\gamma}\cdot \rho(\gamma)R(\lambda)
$$
where $c_{\lambda,\gamma}$ is a complex number which is independent
of the brane.
Setting $\lambda=1$ or $\gamma=1$
we find $c_{1,\gamma}=c_{\alpha,1}=1$. Also, by group 
property $R(\lambda_1\lambda_2)=R(\lambda_1)R(\lambda_2)$
and $\rho(\gamma_1\gamma_2)=\rho(\gamma_1)\rho(\gamma_2)$, we find
$c_{\lambda_1\lambda_2,\gamma}=c_{\lambda_1,\gamma}c_{\lambda_2,\gamma}$
and
$c_{\lambda,\gamma_1\gamma_2}=c_{\lambda,\gamma_1}c_{\lambda,\gamma_2}$.
By the first equation, one can write $c_{\lambda,\gamma}=
\lambda^{f(\gamma)}$ for some complex valued function
$f$ of $\Gamma$
and then the second equation says
$f(\gamma_1\gamma_2)=f(\gamma_1)+f(\gamma_2)$.
Also, $c_{\alpha,1}=1$ means $f(1)=0$.
Since $\Gamma$ is a finite group, say with order $d$,
we have $\gamma^d=1$ for any $\gamma\in \Gamma$. 
Thus, we find
$0=f(1)=f(\gamma^d)=d\cdot f(\gamma)$, namely, $f(\gamma)=0$ for any
$\gamma\in \Gamma$. This means $c_{\lambda,\gamma}=1$ for any $\lambda$
and $\gamma$.
Thus, we conclude that the R-symmetry in the orbifold theory must satisfy
$$
R(\lambda)\rho(\gamma)=\rho(\gamma)R(\lambda).
$$
We find that the data to specify an R-graded D-brane in the
orbifold theory is the quintuple $(E,A,Q,R,\rho)$ and corresponds to
a complex of $\Gamma$-equivariant vector bundles.

Of particular importance is the orbifold of the Euclidean space $X=\C^n$
with a finite group $\Gamma$ of linear transformations.
For each representation of the orbifold group,
$\rho:\Gamma\to GL(V)$, we have a D-brane
associated with the trival vector bundle with fibre $V$ and
 trivial gauge connection $A=0$ with the natural $\Gamma$-action.
We denote it by ${\mathcal O}(\rho)$.
If the group is isomorphic to the cyclic group $\Gamma\cong
\Z_d$, its irreducible representation
is  one-dimensional
and is specified by a mod $d$ integer,
$m\in\Z_d$, with $\rho_m:l\in \Z_d\mapsto \e^{2\pi iml}\in U(1)$.
We often write ${\mathcal O}(\rho_m)$ simply by
${\mathcal O}(m)$.
Note that all these branes ${\mathcal O}(\rho)$ are extending in the
entire space $\C^n/\Gamma$.
A general D-brane may be ``represented'' as a complex of these branes
in the sense that is specified in the next section.
For example, in Section~\ref{subsec:Koszul} we will find
such a representation for D0-branes stuck at the orbifold fixed point
$0\in \C^n$. These are known as the {\it fractional branes} 
\cite{Quiver,fractional,EmJau}.

The closed string sector of the orbifold theory on $\C^n/\Gamma$
has a charge 
integrality --- any NS-NS operator has integral vector R-charge
that matches modulo 2 to the statistics.
(The axial R-charges are not necessarily integral unless the Calabi-Yau
condition $\det\gamma=1$ is met.)
In order to preserve this integrality, 
we require the R-charge of the even ({\it resp}. odd)
 Chan-Paton vector to be even ({\it resp}. odd) integer.

\subsection{Landau-Ginzburg Models: Matrix Factorizations}
\label{subsec:LG}

Let us consider the $(2,2)$ supersymmetric
Landau-Ginzburg model
of $N$ variables $X_1,...,X_N$
with a polynomial superpotential $W(X)=W(X_1,...,X_N)$.
The Lagrangian density is given by
\beqa
{\mathcal L}&=&\int\sum_{i=1}^N\overline{X_i}X_i\,\dd^4\theta\,
+\,\left({i\over 2}\int W(X)\,\dd^2\theta\,\,+\,c.c.\,\right)
\,+\,\mbox{total derivative}
\nn\\
&=&\sum_{i=1}^N\left(\,|\partial_t x_i|^2-|\partial_{\s}x_i|^2
+i\bpsi_{i-}(\lrd_{\!\!\! t}+\lrd_{\!\!\!\s})\psi_{i-}
+i\bpsi_{i+}(\lrd_{\!\!\! t}-\lrd_{\!\!\!\s})\psi_{i+}
-{1\over 4}\left|{\partial W(x)\over \partial x_i}\right|^2\,\right)
\nn\\
&&+\left(-{i\over 2}\sum_{i,j=1}^N\psi_{i+}\psi_{j-}{\partial^2 W(x)
\over\partial x_i\partial x_j}
+\,\, c.c\,\right),
\eeqa
where the auxiliary field is eliminated in the second equality.
We would like to formulate this theory on
the worldsheet $\surface$ with boundary, and
find boundary interactions that respect
the ${\mathcal N}=2_B$ supersymmetry
\beq
\delta x_i=\epsilon \psi_i,
\qquad
\delta\psi_{i\pm}=-i\bepsilon(\partial_0\pm\partial_1)x_i
\pm {i\over 2}\epsilon 
{\partial \overline{W(x)}\over\partial \overline{x_i}}.
\eeq
With the addition of the standard boundary term
\beq
L_{\rm bdry}^{(0)}
=\left[-{i\over 2}\sum_{i=1}^N
(\psi_{i+}\bpsi_{i-}+\bpsi_{i+}\psi_{i-})
\right]_{\partial\surface},
\eeq
the bulk action varies as
\beq
\delta\left[\int_{\surface}{\mathcal L}\,\dd^2\s
+\int_{\partial\surface}L_{\rm bdry}^{(0)}\dd t\,\right]
=-{\rm Re}\int_{\partial\surface}
\sum_{i=1}^N\left(\bepsilon\psi^i{\partial W(x)\over\partial x_i}\right)
\,\dd t.
\label{Warner}
\eeq
This is known as the ``Warner term'' after \cite{Warner}.
(See \cite{blin} for a simple superfield derivation.)

The main task is to cancel the Warner term (\ref{Warner})
by adding a suitable boundary term to the action \cite{HKLM,KapLi1}.
Let us try our friend
\beq
{\mathcal A}_t=-{1\over 2}\sum_{i=1}^N\psi_i{\partial\over\partial x_i}Q(x)
+{1\over 2}\sum_{i=1}^N\bpsi_i{\partial \over
\partial \overline{x}_i}Q(x)^{\dag}
+{1\over 2}\{Q(x),Q(x)^{\dag}\},
\label{SWLLG}
\eeq
where $Q(x)$ is an odd operator on a 
$\Z_2$-graded vector space $\cp$ which depends holomorphically on $x$.
Its ${\mathcal N}=2_B$ variation, computed in (\ref{deltaA1}), reads
\beq
\delta{\mathcal A}_t\,=\,
-{\rm Re}\left\{\,\sum_{i=1}^N
\left(\bepsilon \psi_i{\partial \over\partial x_i}Q^2\right)
-[\bepsilon Q^{\dag},Q^2]
\,\right\}
+i{\mathcal D}_t\Bigl(\bepsilon Q+\epsilon Q^{\dag}\Bigr)
-i\Bigl(\dot{\bepsilon}Q+\dot{\epsilon}Q^{\dag}\Bigr).
\label{delALG}
\eeq
The first term cancels the Warner term (\ref{Warner})
provided that $Q$ satisfies\footnote{$Q^2=(W+c)\cdot{\rm id}_V$ 
for some constant $c$ is also allowed.
However, since we will consider branes with R-symmetry, which require
$c=0$, we restrict our attention to those with $c=0$.}
\beq
Q^2=W\cdot{\rm id}_{\cp}.
\label{mf}
\eeq
The remaining terms in (\ref{delALG}) are as in non-linear sigma models:
The second term, where ${\mathcal D}_tX=\dot{X}+i[{\mathcal A}_t,X]$,
leads to a total derivative when inserted between the
Wilson lines 
 $P\exp\left(-i\int_{\partial\surface}{\mathcal A}_t\dd t\right)$
with the right time-ordering.
The third term shows that $Q$ and $Q^{\dag}$ enter into the 
${\mathcal N}=2_B$ supercharges ${\bf Q}$ and
${\bf Q}^{\dag}$. For the open string
stretched between $(\cp_1,Q_1)$ and $(\cp_2,Q_2)$, the supercharge
${\bf Q}$ is expressed as the sum ${\bf Q}_{\rm bulk}+
{\bf Q}_{\rm bdry}$ where
\beqa
&&
{\bf Q}_{\rm bulk}=\int_0^L\dd\s\sum_{i=1}^N\left(
\overline{\psi_i}\partial_tx_i
+\overline{\widetilde{\psi}}_i\partial_{\s}x_i
+\widetilde{\psi}_i{\partial W\over\partial x_i}\right)
\nn\\
&&
{\bf Q}_{\rm bdry}=-iQ_2\Bigl|_{\s=L}\,+\,iQ_1\Bigl|_{\s=0}.
\nn
\eeqa
The canonical commutation relation yields
the supersymmetry relation ${\bf Q}^2=0$; 
${\bf Q}^2_{\rm bulk}$ gives a boundary
term $W|_L-W|_0$ that is canced by ${\bf Q}_{\rm bdry}^2$
as a consequence of (\ref{mf}).

An odd holomorphic operator $Q(x)$
of a $\Z_2$-graded vector space $\cp=\cp^{\rm ev}\oplus \cp^{\rm od}$
can be represented by a holomorphic matrix of the form
$$
Q(x)=\left(\begin{array}{cc}
0&f(x)\\
g(x)&0
\end{array}\right).
$$
It satisfies the condition (\ref{mf}) if and only if the even and the odd
parts of $\cp$ have the same rank, say $r$, and
\beq
f(x)g(x)=g(x)f(x)=W(x){\bf 1}_r.
\label{mf2}
\eeq
Such a matrix $Q$, or a pair of matrices $(f,g)$, is called a 
{\it matrix factorization} of $W$.

Let us briefly comment on the relation of this to the 
tachyon ${\bf T}$ in
${\mathcal N}=1$ Landau-Ginzburg model with
the real superpotential $h(x)$.
By comparison of the potential and Yukawa-type terms, we see that
the ${\mathcal N}=(2,2)$ Landau-Ginzburg model with
superpotential $W$ is equal to
the ${\mathcal N}=(1,1)$ Landau-Ginzburg model
with $h={W\over 2}+{\overline{W}\over 2}={\rm Re}(W)$.
The variation (\ref{Warner}) is of course the same as (\ref{varN=1LG}) 
as far as the ${\mathcal N}=1$ part is concerned.
We have seen that the system is ${\mathcal N}=1$
supersymmetric with the additional boundary term 
(\ref{DelLN=1LG}),
\beq
\Delta L_{\rm bdry}^{(0)}
=-\left[{1\over 2}W+{1\over 2}\overline{W}\right]_{\partial\surface}.
\label{DLLG}
\eeq
Also, the boundary interaction of the form (\ref{AT}) is
${\mathcal N}=1$ invariant by itself for any tachyon profile ${\bf T}(x)$.
If we include it with ${\bf T}=iQ(x)-iQ(x)^{\dag}$,
we obtain precisely the above system (\ref{SWLLG})
provided $Q(x)$ obeys (\ref{mf}).
Indeed the interaction (\ref{AT})
includes extra terms $-{1\over 2}Q^2-{1\over 2}Q^{\dag 2}$
compared to (\ref{SWLLG}) but those are cancelled by
the term $-\Delta L_{\rm bdry}^{(0)}$
from (\ref{DLLG}) provided $Q^2=W\cdot {\rm id}$.

\subsubsection*{\it Chiral Sector}

The space of chiral ring elements for a pair of branes
${\mathcal B}_1=(\cp_1,Q_1)$ and
${\mathcal B}_2=(\cp_2,Q_2)$ can be studied, as before,
by the zero mode approximation where the supercharge $i{\bf Q}$
acts on the space $\Omega^{0,\bullet}(\C^N,Hom(\cp_1,\cp_2))$
as a Dolbeault-like operator
$\bartial\phi+Q_2\phi-(-1)^{|\phi|}\phi Q_1$.
Since any $\bartial$-closed form of positive degree
is $\bartial$-exact on $\C^N$, 
one can further truncate to the subspace
consisting of  holomorphic functions of $\C^N$ with 
values in $Hom(\cp_1,\cp_2)$.
Thus, we have a relation similar to (\ref{cohomo2}).
In this paper, as we will explain momentarily, we consider 
matrix factorizations that are polynomials of $x_1,.., x_N$.
Then, one can further truncate to the space of
polynomial functions of $x_1,...,x_N$ with values in
$Hom(\cp_1,\cp_2)$.
In this situation, it turns out to be
 convenient to use algebraic terminology
associated with the polynomial ring $\ring=\C[x_1,...,x_N]$.
For each $\Z_2$-graded Chan-Paton space $\cp=\cp^{\rm ev}\oplus \cp^{\rm od}$
 we introduce
the $\Z_2$-graded $\ring$-module $M=\cp\otimes_{\C} \ring=M^{\rm ev}\oplus
M^{\rm od}$.
The space of polynomial functions with values in $Hom(V_1,V_2)$
is equal to the space $\HomR(M_1,M_2)$
of homomorphisms of the $\ring$-module $M_1$ to
the $\ring$-module $M_2$. It is of course $\Z_2$-graded,
$\HomR^{\rm ev}(M_1,M_2)=\HomR(M_1^{\rm ev},M_2^{\rm ev})
\oplus \HomR(M_1^{\rm od},M_2^{\rm od})$
and $\HomR^{\rm od}(M_1,M_2)=\HomR(M_1^{\rm ev},M_2^{\rm od})
\oplus \HomR(M_1^{\rm ev},M_2^{\rm od})$.
The relation analogous to (\ref{cohomo2}) can be written as
\beq
\Ho^{p}({\mathcal B}_1,{\mathcal B}_2)
\cong H^p_{{\bf Q}^{\it pol}}(\HomR^{\bullet}(M_1,M_2))
\qquad 
p={\rm ev}/{\rm od},
\eeq
where ${\bf Q}^{\it pol}$ is given by
$$
i{\bf Q}^{\it pol}\phi=Q_2\phi-(-1)^{|\phi|}\phi Q_1.
$$
There is an alternative way to describe these cohomology groups.
Let us introduce an infinite sequence of maps which is 2-periodic
\beq
{\mathcal C}_{Q}:\qquad\cdots
\stackrel{f}{\longrightarrow}M^{\rm ev}\stackrel{g}{\longrightarrow}M^{\rm od}
\stackrel{f}{\longrightarrow}M^{\rm ev}\stackrel{g}{\longrightarrow}M^{\rm od}
\stackrel{f}{\longrightarrow}
\cdots\qquad
\label{TAC1}
\eeq
Over the ring
$B=\C[x_1,\ldots,x_N]/(W)$, where any multiple of $W$ is regarded zero,
this is a complex of $B$-modules
due to the matrix factorization condition
$f\cdot g=g\cdot f=W{\rm id}$.
In fact it is exact everywhere. 
(We will show this later in
Section~\ref{subsec:LVphase} where we will revisit such infinite
complexes.)
Such a complex is called a
{\it totally acyclic complex} of $B$-modules.
Then, the space of chiral ring elements for a pair of branes 
is isomorphic to the space of
cochain maps from 
${\mathcal C}_{Q_1}$ 
to ${\mathcal C}_{Q_2}$ modulo homotopies,
\beq
\Ho^{p}({\mathcal B}_1,{\mathcal B}_2)
\cong \HomHo({\mathcal C}_{Q_1},{\mathcal C}_{Q_1}[p])
\qquad 
\mbox{for~$
p={\rm ev}/{\rm od}$},
\label{Htac}
\eeq
where ${\mathcal C}[{\rm ev}]$ is ${\mathcal C}$ itself while
${\mathcal C}[{\rm od}]$ is the complex ${\mathcal C}$ shifted
by one with $Q$ replaced by $-Q$.

\subsubsection*{\it R-Symmetry}

If the superpotential $W(x_1,...,x_N)$ is quasi-homogeneous of degree
$(d_1,...,d_N)$, such as the Fermat polynomial
$W=x_1^{d_1}+\cdots+x_N^{d_N}$,
the bulk theory has
vector $U(1)$ R-symmetry where $x_i$ has R-charge $2/d_i$:
\beq
W(\lambda^{2/d_1}x_1,...,\lambda^{2/d_N}x_N)=\lambda^2W(x_1,...,x_N).
\label{Whomog}
\eeq
The bulk LG model is believed to flow in the infra-red limit to
a $(2,2)$ superconformal field theory with central charge
$\hat{c}=\sum_{i=1}^N(1-2/d_i)$, where this R-symmetry
becomes a part of the superconformal algebra. The R-charges of
NS-NS states are in general not integral.

This vector R-symmetry is preserved by the D-brane $(\cp,Q)$ if
the matrix factorization is quasi-homogeneous \cite{HW1}. Namely,
\beq
R(\lambda)Q(\lambda^{2/d_1}x_1,...,\lambda^{2/d_N}x_N)
R(\lambda)^{-1}=\lambda Q(x_1,...,x_N).
\label{Qhomog}
\eeq
$R(\lambda)$ is a one parameter group of
 operators on $\cp$ --- it depends on
$\lambda=\e^{i\alpha}$ in such a way that $R(\lambda_1\lambda_2)
=R(\lambda_1)R(\lambda_2)$.
We do not require that it
 is invariant under $\alpha\to\alpha+2\pi$,
i.e., $R(\lambda)$ may contain fractional powers of $\lambda$.
Notice that the condition (\ref{Qhomog}) requires that
$Q(x)$ must be a polynomial in $x_1,...,x_N$.

For two R-symmetric branes, $(\cp_1,Q_1)$ with $R_1$
and $(\cp_2,Q_2)$ with $R_2$, the R-symmetry acts on the
open string states. 
 In the holomorphic sector,\footnote{Somewhat loosely, 
we use the terms ``local operators'' and ``open string states''
interchangeably.
Note that local operators naturally correspond to open string states
in the NS sector and, if B-twist is possible,
also to open string states in the Ramond sector.
} 
the action is
\beq
{\bf R}(\lambda):\phi(x_1,...,x_N)\longmapsto
R_2(\lambda)\phi(\lambda^{2/d_1}x_1,...
\lambda^{2/d_N}x_N)R_1(\lambda)^{-1}.
\label{Raction}
\eeq
It introduces a new grading in the space of open string states.
Note that a state of definite R-degree is necessarily a polynomial.
The R-degree is not in general integral.
In particular, the R-grading and the $\Z_2$-grading can be
completely independent.

As an example, let us consider the LG model of single variable $X$
with superpotential $W=X^d$. The bulk theory
flows to a
rational conformal field theory (RCFT)
called the A-type minimal model
at level $k=d-2$. 
An obvious matrix factorization is
$W=x^n\cdot x^{d-n}$, that is,
\beq
Q_n(x)=\left(\begin{array}{cc}
0&x^n\\
x^{d-n}&0
\end{array}\right).
\label{minimal}
\eeq
If $1\leq n\leq d-1$,
this brane is believed to flow to the conformally 
invariant boundary condition in the minimal model
known as the Cardy brane with $L=(n-1)$ 
(see for example \cite{Cardy} for
Cardy branes in RCFTs, \cite{MMS,BH2} for D-branes in
$(2,2)$ minimal models and \cite{BHLS,KapLi2} for relation to
LG branes).
It has the property (\ref{Qhomog}) with
\beq
R_n(\lambda)=\left(\begin{array}{cc}
\lambda^{{1\over 2}-{n\over d}}&0\\
0&\lambda^{-{1\over 2}+{n\over d}}
\end{array}\right).
\label{Rn}
\eeq
The $n$-dependence in the
overall phase of $R_n(\lambda)$ is chosen so that it agrees with
the R-symmetry of the infra-red fixed point \cite{bRG,HW1}.
Indeed, for the $Q_{n_1}$-$Q_{n_2}$ string, 
the space of chiral ring elements is spanned by
\beq
\phi^{\rm ev}_{j}(x)=\left(\begin{array}{cc}
x^{j-{n_1-n_2\over 2}}&0\\
0&x^{j+{n_1-n_2\over 2}}
\end{array}\right),\qquad
\phi^{\rm od}_j(x)=\left(\begin{array}{cc}
0&x^{{n_1+n_2\over 2}-j-1}\\
-x^{d-{n_1+n_2\over 2}-j-1}&0
\end{array}\right)
\eeq
where $j$ runs over ${|n_1-n_2|\over 2}, {|n_1-n_2|\over 2}+1,\ldots,
 {\rm min}\left\{
{n_1+n_2\over 2}-1,d-{n_1+n_2\over 2}-1\right\}$. 
With the choice (\ref{Rn}), the R-charges of these
elements are ${2j\over d}$ for $\phi^{\rm ev}_j$ and
$1-{2j+2\over d}$ for $\phi^{\rm od}_j$ which are the right R-charges
for open string NS-states between the corresponding minimal model branes
\cite{BH2}.

\subsubsection{Landau-Ginzburg Orbifold}
\label{subsub:LGO}

The Landau-Ginzburg model with quasi-homogeneous superpotential $W(x)$
of degree $(d_1,...,d_N)$ has symmetry
\beq
(x_1,...,x_N)\longmapsto (\e^{2\pi i/d_1}x_1,...,\e^{2\pi i/d_N}x_N).
\label{ds}
\eeq
This generates the cyclic group of order $d={\rm l.c.m.}(d_1,...,d_N)$, 
which we call $\Gamma_0$. 
Gauging this symmetry group, we obtain a
Landau-Ginzburg orbifold theory.
This orbifold theory has a charge integrality ---
the vector R-charges of NS-NS states
are integers whose mod 2 reduction
matches with the statistics of the corresponding
operators.
In particular, there is a one-to-one correspondence between
R-R ground states
and {\it ac}-primary operators.\footnote{If the degrees obey
the ``Calabi-Yau condition''
$1/d_1+\cdots +1/d_N=1$, then the axial R-charges are also
integral and there is a spectral flow between
R-R ground states
and {\it cc}-primary operators.
}

A  D-brane in this LG orbifold is specified by the triple $(\cp,Q,\rho)$
where $(\cp,Q)$ is a matrix factorization and $\rho$ is a representation of
the group $\Gamma_0$ such that 
\beq
\rho(\omega)^{-1}Q(\omega\cdot x)\rho(\omega)=Q(x),\qquad\forall
\omega\in \Gamma_0.
\label{oinv}
\eeq
$\rho$ determines the action of the orbifold group 
on the Chan-Paton factor in the theory before orbifolding, and the 
equation (\ref{oinv}) means that the tachyon profile is invariant
under that orbifold group action.
For the brane pair $(\cp_1,Q_1,\rho_1)$, 
$(\cp_2,Q_2,\rho_2)$, the open string states in the orbifold theory
are $\Gamma_0$-invariant states.
In the polynomial sector, the condition of $\Gamma_0$-invariance is
\beq
\rho_2(\omega)^{-1}\phi(\omega\cdot x)\rho_1(\omega)=\phi(x).
\label{oinvst}
\eeq
Since the bulk theory has vector R-symmetry with integrality,
we would like the branes to respect that as well \cite{HW1,jo}.
Namely, we would like that the branes are quasihomogeneous, that is,
equation (\ref{Qhomog}) is satisfied,
and that the R-grading in the NS sector is integral and reduces modulo
$2$ to the original $\Z_2$-grading.
The R-symmetry action (\ref{Raction}) preserves the $\Gamma_0$-invariance 
condition when $R(\lambda)\rho(\omega)
=c_{\lambda,\omega}\rho(\omega)R(\lambda)$
for a brane independent scalar $c_{\lambda,\omega}$.
As in the case of orbifolds of non-linear
sigma models, one can show $c_{\lambda,\omega}=1$.
Thus, we require
\beq
R(\lambda)\rho(\omega)=\rho(\omega)R(\lambda).
\eeq
The charge integrality requires that
$$
R_2(\e^{\pi i})\phi(\e^{2\pi i\over d_1}x_1,...,
\e^{2\pi i\over d_N}x_N)R_1(\e^{\pi i})^{-1}
=(-1)^{|\phi|}\phi(x_1,...,x_N).
$$
Using the orbifold invariance (\ref{oinvst}), 
this is equivalent to
\beq
R_2(\e^{\pi i})\rho_2(\e^{2\pi i\over d})
\phi(x)\rho_1(\e^{2\pi i\over d})^{-1}R_1(\e^{\pi i})^{-1}
=(-1)^{|\phi|}\phi(x),
\label{integrality}
\eeq
where $\omega=\e^{2\pi i\over d}$ is the generator (\ref{ds})
of $\Gamma_0$.
Let $\sigma_1$ and $\sigma_2$ be the $\Z_2$-grading
operators
on $\cp_1$ and $\cp_2$, acting as
$1$ on even elements and $-1$ on odd elements.
Then, we have $(-1)^{|\phi|}\phi
=\sigma_2\phi\sigma_1^{-1}$.
Thus the integrality condition (\ref{integrality}) is satisfied 
if $R_i(\e^{\pi i})\rho_i(\e^{2\pi i\over d})= c\cdot\sigma_i$ for $i=1,2$
for some common  constant $c$. We simply choose $c=1$.
Thus, we restrict our attention to branes
$(\cp,Q,\rho)$ with R-symmetry $R$ that obeys
\beq
R(\e^{\pi i})\rho(\e^{2\pi i\over d})=\sigma_{\cp},
\label{intmf}
\eeq
where $\sigma_V$ is the $\Z_2$-grading operator on $\cp$.
As in non-linear sigma models, there is an ambiguity $R(\lambda)\to
\lambda^2R(\lambda)$ that does not change the physical property of
the brane but changes the R-grading. We again call the brane with this 
additional data $(\cp,Q,\rho,R)$ a {\it graded D-brane}.
By the condition (\ref{intmf}), a graded D-brane 
in the theory with the orbifold group $\Gamma_0$
is specified simply by
$(\cp,Q,R)$ such that $R(\e^{\pi i})\sigma_{\cp}$ obeys
$(R(\e^{\pi i})\sigma_{\cp})^d=1$.
For this reason, in some literature, such as \cite{Orlov},
only the R-symmetry action is used to specify
a data for a graded D-brane in this class of LG orbifolds.

The superpotential $W$ may have a larger symmetry group
$\Gamma$ by which we can define the orbifold theory.
As long as $\Gamma$ includes $\Gamma_0$, the closed string
sector has a charge integrality, and we require that it is extended to
the boundary or open string sector.
The requirement is just like (\ref{intmf}), $R(\e^{\pi i})\rho(\omega_{-1})
=\sigma_V$, where $\omega_{-1}$ is the element of $\Gamma$ that acts on $x_i$ 
in the same way as the R-symmetry action for $\lambda=\e^{\pi i}$.
Such an element exists in $\Gamma$ since it includes $\Gamma_0$ by assumption.

\subsubsection*{\it Recknagel-Schomerus Branes}

As examples, let us consider tensor products 
of minimal model branes (\ref{minimal}) in 
the model with Fermat type superpotential
$W=X_1^{d_1}+\cdots+X_N^{d_N}$. They are known as {\it RS-branes}
after Recknagel and Schomerus who first studied these branes
in the framework of RCFT \cite{ReSc}.
They are most conveniently described
in terms of the Clifford algebra
\beq
\{\eta_i,\bareta_j\}=\delta_{i,j},
\quad
\{\eta_i,\eta_j\}=\{\bareta_i,\bareta_j\}=0.
\label{Cliffordalg}
\eeq
The latter is represented on the $2^N$-dimensional space $\cp_N$ 
(the Clifford module),
which is generated by a vector $|0\rangle$ annihilated by all $\eta_i$'s.
There are two $\Z_2$-gradings on $\cp_N$: the first is such that even and odd
multiples of $\bareta_i$'s on $|0\rangle$ are even and odd, and the second
is the opposite one.
The sum of minimal model branes is written as
\beq
Q_{{\bf L}}
=\sum_{i=1}^N \left(x_i^{L_i+1}\eta_i+x_i^{d_i-L_i-1}\bareta_i\right).
\eeq
This is invariant under the orbifold group $\Gamma_0\cong \Z_d$
as well as the R-symmetry, with the transformations
$\eta_i\to \omega^{-L_i-1}\eta_i$
and $\eta_i\to \lambda^{1-{2(L_i+1)\over d_i}}\eta_i$
(with conjugate action on $\bareta_i$'s).
The representations of the orbifold and R-symmetry groups on the 
Chan-Paton space are specified by the action on the vector $|0\rangle$
--- action on other vectors are determined by the transformations of
the $\bareta_i$'s.
The orbifold representation is labeled by a mod $d$ integer $\bar q$:
$$
\rho_{\bar q}(\omega):|0\rangle\mapsto \omega^{\bar q}|0\rangle.
$$
The condition of integrality (\ref{intmf}) requires that
the R-charge $R_{|0\rangle}$ of the state $|0\rangle$ must be
of the form
$$
R_{|0\rangle}
= -{2q\over d}+r
$$
where $q$ is an integer representing $\bar q$ mod $d$,
while $r$ is an even integer for 
the first
$\Z_2$-grading and an odd interger for the second one.
We shall denote the D-brane that corresponds to the above data 
by ${\mathcal B}_{{\bf L},q,r}$.
The label is a little redundant: obviously $({\bf L},q,r)\to
({\bf L}, q+d, r+2)$ does not change the brane. Also, 
the exchange $x_i^{L_i+1}\leftrightarrow x_i^{d_i-L-1}$
is compensated by the exchange 
$\eta_i\leftrightarrow\bareta_i$, and this leads to the 
identification of branes under
$L_i\to d_i-L_i-1$ (for one $i$), 
$q\to q-{d(L_i+1)\over d_i}$, $r\to r-1$.
The translation to the standard notation
for the RS-branes $B_{{\bf L},M,S}$ (see e.g. \cite{BHHW})
is 
$2q=M+\sum_{i=1}^N{d(L_i+1)\over d_i}
=\sum_{i=1}^N{d(M_i+L_i+1)\over d_i}$ (mod $2d$) and
$2r=S$ (mod $4$) up to an overall shift.
Matrix factorization realizations of these branes were first discussed
in \cite{BHHW,ADD}.

These branes are indecomposable except for the case where there are 
two or more $i$'s with $L_i+1={d_i\over 2}$ (which necessarily requires 
that $d_i$
is even for such $i$). 
In the latter case, it splits into a sum of
indecomposable ones known as {\it short orbit branes}
$\widehat{\mathcal B}^{(\pm )}_{{\bf L},q,r}$. For example,
if there are exactly two or three such $i$'s
the RS-brane
${\mathcal B}_{{\bf L},q,r}$
 splits into two:
\beq
{\mathcal B}_{{\bf L},q,r}\,\cong\, \widehat{\mathcal B}^{(+)}_{{\bf L},q,r}
\,\oplus \,\widehat{\mathcal B}^{(-)}_{{\bf L},q,r}.
\eeq
To define short orbit branes, we denote by ${\bf S}$ the set of
$i$'s such that $L_i+1={d_i\over 2}$
and consider
\beq
Q=\sum_{i\in {\bf S}}x_i^{d_i\over 2}\xi_i
+\sum_{j\not\in {\bf S}}\Bigl(x_j^{L_j+1}\eta_j+x_j^{d_j-L_j-1}\bareta_j\Bigr),
\label{shortorbitbrane}
\eeq
where $\xi_i$ are {\it real} Clifford generators
$$
\{\xi_i,\xi_j\}=2\delta_{i,j}, \qquad \{\xi_i,\eta_k\}=\{\xi_i,\bareta_k\}=0.
$$
If the number of elements $|{\bf S}|$ is odd, we introduce one extra
real Clifford generator $\xi_0$.
We choose an orientation in the space of $\xi_i$'s.
We then introduce {\it complex} generators
 $\eta_{i_1j_1}={1\over 2}(\xi_{i_1}+i\xi_{j_1})$,...,
$\eta_{i_sj_s}={1\over 2}(\xi_{i_s}+i\xi_{j_s})$, and their complex conjugates
 $\bareta_{i_1j_1}={1\over 2}(\xi_{i_1}-i\xi_{j_1})$,...,
$\bareta_{i_sj_s}={1\over 2}(\xi_{i_s}-i\xi_{j_s})$,
where 
$(\xi_{i_1},\xi_{j_1},...,\xi_{i_s},\xi_{j_s})$ is positively oriented.
They form a complex Clifford algebra together with $\eta_j$ and $\bareta_j$.
The expression (\ref{shortorbitbrane}) then becomes a matrix factorization
on the Clifford module.
As before we denote the element annihilated by all $\eta_{ij}$ and $\eta_k$
by $|0\rangle$. Defining the label $(q,r)$ in the same way as
the RS-branes, we have a brane in the LG orbifold which we denote
by $\widehat{\mathcal B}^{(+)}_{{\bf L},q,r}$. 
One can see that it depends only on the orientation of $\xi_i$'s.
The key point is that $\xi_i$ is invariant under the R-symmetry
and transforms as 
$\xi_i\to -\xi_i$ under the orbifold group generator 
($x_i^{d_i/2}$ already has R-charge $1$ and transforms by
sign under the orbifold generator).
The one associated
with the opposite orientation is denoted as
$\widehat{\mathcal B}^{(-)}_{{\bf L},q,r}$.
Thus, $(+)$ versus $(-)$ 
is defined with respect to the orientation of $\xi_i$'s.
Using the key point again, we find the relation
$\widehat{\mathcal B}^{(-)}_{{\bf L},q,r}
\cong\widehat{\mathcal B}^{(+)}_{{\bf L},q+{d\over 2},r+1}$. 
If $|{\bf S}|$ is odd, the $(+)$ and $(-)$ branes are isomorphic
and we shall simply write 
$\widehat{\mathcal B}_{{\bf L},q,r}$ for
$\widehat{\mathcal B}^{(+)}_{{\bf L},q,r}
\cong\widehat{\mathcal B}^{(-)}_{{\bf L},q,r}$.
For the case $|{\bf S}|=1$, it is the same as the standard RS-brane,
$\widehat{\mathcal B}_{{\bf L},q,r}\cong{\mathcal B}_{{\bf L},q,r}$.
A matrix factorization realization of
short orbit branes, which is equivalent to the above, 
was first noticed in \cite{CFG} based on the analysis of
R-R charge.
A derivation from the short orbit branes 
in minimal models will be given in 
Section~\ref{subsec:Knorrer}.

The Fermat potential $X_1^{d_1}+\cdots+X_N^{d_N}$ has a larger
symmetry group $\Z_{d_1}\times \cdots\times \Z_{d_N}$ and the brane
$Q_{\bf L}$ is invariant under all the elements of that group.
If we take the orbifold group $\Gamma$ to be larger than
$\Gamma_0\cong \Z_d$,
then we need more labels
to specify the orbifold action on the Chan-Paton factor.
For example, we may use $\vec{q}=(q_1,...,q_N)$
with
$$
\rho_{\vec{q}}(\omega)=\omega_1^{\bar q_1}\cdots\omega_N^{\bar q_N}
\quad\mbox{on $|0\rangle$},
$$
for the orbifold element $\omega$ which maps $x_i\mapsto \omega_ix_i$.
The translation to the standard notation is then $2q_i=M_i+L_i+1$
(again up to a universal shift).

\section{Renormalization Group Flows}
\label{sec:D-isom}

\newcommand{\ve}{\varepsilon}

Given a D-brane realized as a boundary Lagrangian,
we are interested in how it behaves at low energies, or equivalently,
at long distances on the worldsheet.
In this section, we discuss two classes of operations on D-branes
that do not change the low energy behaviour:
boundary D-term deformations and brane-antibrane annihilation.
We shall call them {\it D-isomorphisms}.
Our main interest in this paper is
the entire set of all possible D-branes
up to D-isomorphisms.
We start our discussion with
 D-branes in non-linear sigma models as described
in Section~\ref{subsec:NLSM}.
We show, in particular, that D-isomorphisms of D-branes are nothing
else but quasi-isomorphisms of complexes.
The argument is applicable to a more general class of theories
including those discussed in other parts of Section~\ref{sec:branecat}
as well as those for linear sigma models that we will study later
in this paper. As an example, we discuss
Landau-Ginzburg models in Section~\ref{subsec:DisomLG}.

Subsequently,
 in Section~\ref{subsec:Knorrer},
we study the effect on D-branes of bulk RG flows associated with 
integrating out fields with F-term masses.

We end with Section~\ref{subsec:whatwedo} where
we discuss parallel transport of B-type D-branes 
over marginal K\"ahler deformations of the bulk theory.

\subsection{D-Term Deformations And Brane-Antibrane Annihilation}
\label{subsec:Diso}

In ${\mathcal N}=2$ supersymmetric systems, there are
two kinds of invariant Lagrangians --- D-terms and F-terms.
In terms of superspace integrals 
these are respectively of the form
$\int V\dd\theta\dd\btheta$ and $\int W\dd\theta$ 
where $\theta$ and $\btheta$ are fermionic coordinates of the
${\mathcal N}=2$ superspace, $V$ is any superfield 
and $W$ is any chiral superfield.
For ${\mathcal N}=2_B$ supersymmetric
 D-banes $(E,A,Q)$ on a K\"ahler manifold $X$, 
the choice of a fibre metric of $E$ defines
the D-term and the choice of complex structure of $E$ together with
the holomorphic part $Q$ of the tachyon determines the F-term.
In other words, the F-term is fixed when the complex
of holomorphic vector bundles
${\mathcal C}={\mathcal C}({\mathcal E},Q)$
is fixed and then the D-term is fixed when the fibre metric
of each bundle (and therefore
the hermitian connection) is fixed.
This can be seen explicitily when the brane is realized using
fermionic boundary chiral superfields \cite{blin}.
As in the bulk theory (for example, with $4d$ ${\mathcal N}=1$ or
$2d$ $(2,2)$ supersymmetry) there is a non-renormalization theorem:
the F-term does not change under the renormalization group flow.
On the other hand,
the D-term does get renormalized and adjust itself to a unique form 
in the deep infra-red limit.
For this reason, one can say that 
deformations of the D-term do not affect the low energy behaviour
and it is the F-term which uniquely determines the infra-red fixed point.

Let us explicitly see how deformations of the fibre metric
correspond to D-term deformations.
For a brane $(E,A,{\bf T})$, we choose some local unitary frame
$\{e_a\}$ and local holomorphic frame $\{\ve_a\}$ of $E$,
which are related by, say
$\ve_a=e_bH^b_{\,\,a}$. The fibre metric of $E$
is represented with respect to the holomorphic frame by
$h(\ve_a,\ve_b)=\sum_{c}(H^c_{\,\,a})^*H^c_{\,\,b}=(H^{\dag}H)_{\bar{a}b}$.
Let ${\mathcal Q}$ be the matrix expression of the holomorphic part 
$Q$ of the tachyon with respect to the holomorphic frame.
With respect to the unitary frame, the connection and
the tachyon is expressed as
$$
iA=H^{\dag -1}\partial H^{\dag}-\bartial H H^{-1},\qquad
Q=H{\mathcal Q}H^{-1}, \quad Q^{\dag}=H^{\dag-1}{\mathcal Q}^{\dag}H^{\dag}.
$$
Since we are going to deform the fibre metric for a fixed holomorphic 
structure $({\mathcal E},{\mathcal Q})$, it is more convenient to
use the holomorphic frame. The expression reads
$$
iA=h^{-1}\partial h,\qquad
Q={\mathcal Q},\quad Q^{\dag}=h^{-1}{\mathcal Q}^{\dag}h.
$$
Let us now deform the fibre metric $h\to h+\delta h=h(1+\epsilon)$
 and see how the boundary interaction (\ref{newA}) changes. 
By a straightforward computation, we find
\beq
\delta {\mathcal A}_t
={\mathcal D}_t(-i\epsilon)
-{1\over 2}{\bf Q}^{\dag}{\bf Q}\epsilon
\eeq
where
\beq
i{\bf Q}\epsilon=\bpsi^{\bi}\partial_{\bi}\epsilon+[Q,\epsilon],
\qquad
i{\bf Q}^{\dag}\alpha=\psi^iD_i\alpha-\{Q^{\dag},\alpha\}.
\eeq
Up to a total covariant derivative, it is indeed
a D-term ${\bf Q}^{\dag}{\bf Q}(*)$.
This exercise lets us notice that D-term deformations are not limited to
deformations of the fibre metric.
Any deformation of the form 
\beq
{\bf Q}^{\dag}{\bf Q}\epsilon
\eeq
is a D-term deformation. If $\epsilon$ is a function of $x$ only,
it is equivalent up to total covariant derivative
 to a deformation of the fibre metric, as we have just seen.
However, we can take a more general $\epsilon$ such as
a matrix that depends also on bulk fermions. This generalization will
play an important r\^ole later on.

Let us next discuss brane-antibrane annihilation.
Ashoke Sen proposed in \cite{Ashoke}
that in a system of an equal number of 
coincident branes and antibranes 
in Type I or Type II string theory, the classical minimum of 
the tachyon potential has zero energy and can be identified with
the supersymmetric vacuum state corresponding to the spacetime without branes.
What is relevant for us is a version of his conjecture described in terms of
renormalization group flow of the worldsheet quantum field theory.
See, for example, \cite{HKM,wsRGs} for works in this direction.
The idea is very simple.
Let us look at the boundary interaction (\ref{AT}) or (\ref{SWL}).
It includes the boundary potential term
\beq
U(x)={1\over 2}{\bf T}(x)^2
\eeq
for a hermitian tachyon profile ${\bf T}(x)$.
It is natural to expect that, at low energies, $x$ would like to be
near the minimum of this potential.
In fact only the {\it zero} matters:
The dynamics concentrates at the locus where ${\bf T}(x)$
has zero eigenvalues. 
A block of ${\bf T}(x)$ that has non-zero eigenvalues everywhere
``can be ignored'' in the low energy dynamics.
These claims are understood as follows.
If $\det{\bf T}(x)$ is nowhere vanishing, the potential 
$U(x)$ is strictly positive everywhere on the target space.
In such a case, the Euclidean path-integral weight,
$P\exp\left(-\int_{\partial \surface}U(x)\sqrt{h}\dd\tau\right)$,
vanishes as the worldsheet metric $h$ is sent to 
infinity. As a consequence, the partition function
and all (unnormalized) correlation functions vanish in the infra-red limit.
In other words, the renormalization group sends ${\bf T}(x)$ to infinity. 
From the spacetime point of view,
this is interpreted to mean that
the minimum of the tachyon potential is located at infinity.
If there is no topological obstruction, a tachyon configuration
stabilizes at the vacuum ${\bf T}\to \infty$.
If ${\bf T}(x)$ is of block-diagonal form,
${\rm diag}({\bf T}_0(x),{\bf T}_1(x),...)$,
and one of the blocks, say ${\bf T}_0(x)$, is everywhere invertible,
that part of the potential blows up 
and the corresponding weight
vanishes in the infra-red limit:
\beqa
P\exp\left(-\int_{\partial\surface}U(x)\sqrt{h}\dd\tau\right)
&=&\left(\begin{array}{ccc}
P\e^{-\int U_0(x)\sqrt{h}\dd\tau}&0&0\\
0&P\e^{-\int U_1(x)\sqrt{h}\dd\tau}&0\\
0&0&\ddots
\end{array}\right)
\nn\\
&\longrightarrow&
\left(\begin{array}{ccc}
{\bf 0}&0&0\\
0&P\e^{-\int U_1(x)\sqrt{h}\dd\tau}&0\\
0&0&\ddots
\end{array}\right).
\nn
\eeqa
As a result
the partition function and all correlation functions
receive contributions only from the remaining blocks.
In this sense a block with strictly positive boundary potential
can be ignored in the infra-red limit.
Namely, the full theory is infra-red equivalent to
the theory without such a block.
Let us describe the condition of postitivity ${\bf T}^2>0$ in the
${\mathcal N}=2$ supersymmetric system,
where the tachyon is expressed as ${\bf T}=iQ-iQ^{\dag}$.
Positivity of ${\bf T}^2=\{Q,Q^{\dag}\}$ is
equivalent to $\Ker Q\cap \Ker Q^{\dag}=\{0\}$.
Since $\Ker Q^{\dag}=({\rm Im} Q)^{\perp}$ and ${\rm Im}Q\subset \Ker Q$,
it simply means that $\Ker Q={\rm Im} Q$.
Namely, invertibility of
${\bf T}(x)$ at every point $x$
is equivalent to the statement that
the complex ${\mathcal C}({\mathcal E},Q)$ is exact.
Thus, {\it the D-brane corresponding to an exact complex
can be ignored in the infra-red limit.}
Recall that an exact complex is quasi-isomorphic to the zero complex
in the derived category.
Later in this subsection,
we will understand the relation of quasi-isomorphism and
brane-antibrane annihilation in more generality.

The operation of brane-antibrane annihilation is analogous to
integrating out massive fields in the bulk theory.
In the bulk, if there is a field of mass $m$ it is appropriate 
to integrate it out in the effective theory at energies below $m$.
Similarly,
in the boundary theory, if there is a block in the
Chan-Paton factor with everywhere invertible ${\bf T}_0$, it is appropriate to
``eliminate'' that factor at energies below $\sim |{\bf T}_0^2|$.
In the bulk, we know that integrating out a massive field
may induce a new term in the superpotential
when it is interacting with other fields.
For example, consider a Landau-Ginzburg model of two variables
$X$ and $Y$ with the superpotential
\beq
W=X^n+X^2Y+{m\over 2}Y^2.
\eeq
$Y$ has a mass $m$ but it is interacting with $X$ via the term $X^2Y$.
At energies below $m$ it is appropriate to integrate out the field
$Y$. This is done simply by solving the equation of motion for $Y$ 
and plugging the result back in, or equivalently, by completing the square
for $Y$ and eliminating the square \cite{ADS2}. 
In any case, the outcome is
\beq
W_{\rm low}=X^n-{1\over 2m}X^4.
\label{WIR}
\eeq
A term $-{1\over 2m}X^4$ has emerged in this process.

Likewise, the brane-antibrane annihilation may produce
a non-trivial effect, when
the eliminated sector is interacting with the rest of the system.
Suppose there is a block in the Chan-Paton factor with
everywhere invertible ${\bf T}_0$: 
$$
{\bf T}=\left(\begin{array}{c|c}
{\bf T}_0&*\\
\hline
*&\begin{array}{c}
\\
\\
\end{array}
{\bf T}'\,\,
\end{array}\right)
$$
This block is interacting with the other sector
if the off-diagonal parts, denoted by asterisks,
are non-zero.
We may consider erasing the off-diagonal parts 
 by the standard linear algebra operation
--- addition/subtraction of raws and columns including
the maximal rank ${\bf T}_0$.
If that is possible and if that is done, 
then the ${\bf T}_0(x)$ block is decoupled from the rest
and can be ignored.
But this may have induced new terms in the remaining part,
thus shifting ${\bf T}'$.
This operation is analogous to the process of
completing the square in the bulk theory. 
This is exactly what we will often do in this paper.
For illustration, let us consider the boundary interaction given by
the following complex:
\beq
\mbox{
\setlength{\unitlength}{0.7mm}
\begin{picture}(170,45)(-5,-2)
\put(-18,0){\small $a_{j-2}$}
\put(-23,4){\vector(1,0){20}}
\put(2,2){${\mathcal E}^{j-1}$}
\put(26,0){\small $a_{j-1}$}
\put(15,4){\vector(1,0){30}}
\put(15,8){\vector(3,2){30}}
\put(15,16){\small $c_{j-1}$}
\put(50,30){${\mathcal F}^{j}$}
\put(50,17){$\oplus$}
\put(50,2){${\mathcal E}^j$}
\put(72,35){\small $b_{j}$}
\put(60,32){\vector(1,0){30}}
\put(60,28){\vector(3,-2){30}}
\put(82,14){\small $m$}
\put(60,8){\vector(3,2){30}}
\put(63,15){\small $c_{j}$}
\put(60,4){\vector(1,0){30}}
\put(72,0){\small $a_{j}$}
\put(95,30){${\mathcal F}^{j+1}$}
\put(95,17){$\oplus$}
\put(95,2){${\mathcal E}^{j+1}$}
\put(116,35){\small $b_{j+1}$}
\put(107,32){\vector(1,0){30}}
\put(107,7){\vector(3,2){30}}
\put(123,15){\small $c_{j+1}$}
\put(142,30){${\mathcal F}^{j+2}$}
\put(161,35){\small $b_{j+2}$}
\put(156,32){\vector(1,0){20}}
\end{picture}
}
\label{complexD}
\eeq
where ${\mathcal F}^j\stackrel{m}{\longrightarrow}{\mathcal E}^{j+1}$
is an isomorphism of vector bundles, that is, there is an inverse
$$
m^{-1}:{\mathcal E}^{j+1}\longrightarrow {\mathcal F}^j.
$$
This invertible part is the analog of the massive field $Y$ in the bulk theory
and we would like to eliminate it. This sector is interacting with the rest
of the system via the terms $c_{j-1}$, $b_j$
$a_j$, and $c_{j+1}$.
Let us order the even and odd vector bundles (assuming $j$ is even)
as
$$
{\mathcal E}^{\rm ev}={\mathcal F}^j\oplus{\mathcal E}^j\oplus
{\mathcal F}^{j+2}\oplus\cdots,\qquad
{\mathcal E}^{\rm od}={\mathcal E}^{j+1}\oplus{\mathcal F}^{j+1}\oplus
{\mathcal E}^{j-1}\oplus\cdots
$$
Then, the operator $Q$ for the above complex is written as
$$
Q=\left(\begin{array}{cc}
0&f\\
g&0
\end{array}\right)
$$
where
$$
g=\left(\begin{array}{c|cccc}
m&a_j&0&0&\cdots\\
\hline
b_j&c_j&0&0&\cdots\\
0&0&0&*&\cdots\\
0&0&*&&\\
\vdots&\vdots&\vdots&&
\end{array}\right),\qquad
f=\left(\begin{array}{c|cccc}
0&0&c_{j-1}&0&\cdots\\
\hline
0&0&a_{j-1}&0&\cdots\\
c_{j+1}&b_{j+1}&0&0&\cdots\\
0&0&0&&\\
\vdots&\vdots&\vdots&&
\end{array}\right).
$$
We wish to erase the off-diagonal terms
by moving around the invertible map $m$. The following does the job
for $g$:
$$
g\longrightarrow
\left(\begin{array}{cccc}
1&0&&\\
\!\! -b_jm^{-1}\!\!&1&&\\
&&\ddots&\\
&&&1
\end{array}\right)g
\left(\begin{array}{cccc}
1&\!\!-m^{-1}a_j\!\!&&\\
0&1&&\\
&&\ddots&\\
&&&1
\end{array}\right)
=\left(\begin{array}{c|cccc}
m&0&0&0&\cdots\\
\hline
0&c_j-b_jm^{-1}a_j&0&0&\cdots\\
0&0&0&*&\cdots\\
0&0&*&&\\
\vdots&\vdots&\vdots&&
\end{array}\right).
$$
Note that the map $c_j$ is modified by $-b_jm^{-1}a_j$.
For $f$, the same basis change works as well,
$$
f\longrightarrow
\left(\begin{array}{cccc}
1&\!\!m^{-1}a_j\!\!&&\\
0&1&&\\
&&\ddots&\\
&&&1
\end{array}\right)
f
\left(\begin{array}{cccc}
1&0&&\\
\!\!b_jm^{-1}\!\!&1&&\\
&&\ddots&\\
&&&1
\end{array}\right)
=\left(\begin{array}{c|cccc}
0&0&{\bf 0}&0&\cdots\\
\hline
0&0&a_{j-1}&0&\cdots\\
{\bf 0}&b_{j+1}&0&0&\cdots\\
0&0&0&&\\
\vdots&\vdots&\vdots&&
\end{array}\right).
$$
The two off-diagonal entries are erased,
thanks to the equations
$mc_{j-1}+a_ja_{j-1}=0$ and
$c_{j+1}m+b_{j+1}b_j=0$ that are part of the condition that (\ref{complexD})
is a complex.
Changing the fibre metric so that the new frame is orthogonal,
which is a D-term deformation,
the sector ${\mathcal F}^j\stackrel{m}{\longrightarrow}
{\mathcal E}^{j+1}$ is decoupled from the rest and thus can be ignored.
The remaining part is, however, not the one obtained 
by just ignoring all the maps
involving ${\mathcal F}^j$ and ${\mathcal E}^{j+1}$.
The original
interaction between the two sectors has a non-trivial 
effect:
A zigzag map, $-b_jm^{-1}a_j$, is added to $c_j$.
This is the analog of the new term $-X^4/2m$ in (\ref{WIR})
that results from integrating out $Y$.
To conclude,
the boundary theory based on the complex (\ref{complexD})
is equivalent up to D-term deformations and brane-antibrane annihilation,
and in particular infra-red equivalent,
to the brane based on the complex
\beq
\mbox{
\setlength{\unitlength}{0.7mm}
\begin{picture}(170,14)(-10,-2)
\put(-18,5){\small $a_{j-2}$}
\put(-23,2){\vector(1,0){20}}
\put(2,0){${\mathcal E}^{j-1}$}
\put(26,5){\small $a_{j-1}$}
\put(15,2){\vector(1,0){30}}
\put(50,0){${\mathcal E}^{j}$}
\put(62,5){\small $c_{j}\!-\!b_jm^{-1}a_j$}
\put(60,2){\vector(1,0){30}}
\put(95,0){${\mathcal F}^{j+1}$}
\put(116,5){\small $b_{j+1}$}
\put(107,2){\vector(1,0){30}}
\put(142,0){${\mathcal F}^{j+2}$}
\put(161,5){\small $b_{j+2}$}
\put(156,2){\vector(1,0){20}}
\end{picture}
}
\label{complexD'}
\eeq
Note that, without the modification $c_j\to c_j-b_jm^{-1}a_j$, 
this (\ref{complexD'}) is not even a complex in general. In fact 
the new complex (\ref{complexD'}) is quasi-isomorphic to
the original one (\ref{complexD}).

Now, let us turn to the general relation of
quasi-isomorphisms to D-term deformations and brane-antibrane annihilation.
We start with introducing the cone construction.

\subsubsection{Binding D-Branes: Cone Construction}

{\it Cone} is an operation which ``binds''
two D-branes together using a chiral ring element. 
Let us choose two D-branes $({\mathcal E},Q_E)$ and $({\mathcal F},Q_F)$.
Let $\vp:{\mathcal E}\to {\mathcal F}$ be a degree zero 
bundle map obeying $Q_F\vp-\vp Q_E=0$, or equivalently, a cochain map
$\vp:{\mathcal C}_E\to {\mathcal C}_F$ of the associated
complexes.
Then, one can construct a new brane, called the cone of
$\vp$, which is denoted by $C(\vp)=
C(\vp:{\mathcal C}_E\to {\mathcal C}_F)$.
It is based on the graded vector bundle
${\mathcal E}_{C(\vp)}={\mathcal E}[1]\oplus {\mathcal F}$, and 
the holomorphic part of the tachyon is given by
\beq
Q_{C(\vp)}=\left(\begin{array}{cc}
-Q_E&0\\
\vp&Q_F
\end{array}\right).
\label{cone}
\eeq
The associated complex, the cone complex, looks like this:
\beq
\begin{array}{ccccccccc}
\cdots\to&{\mathcal E}^{j}&\longrightarrow
&{\mathcal E}^{j+1}&\longrightarrow
&{\mathcal E}^{j+2}&\longrightarrow
&{\mathcal E}^{j+3}&\to\cdots\\
\searrow&\oplus&\searrow&\oplus&\searrow&\oplus&\searrow&\oplus&\searrow
\\
\cdots\to&{\mathcal F}^{j-1}&\longrightarrow
&{\mathcal F}^{j}&\longrightarrow
&{\mathcal F}^{j+1}&\longrightarrow
&{\mathcal F}^{j+2}&\to\cdots  .
\end{array}
\label{conecomplex}
\eeq
The horizontal arrows on the first line are $-Q_E$,
the southeast arrows are $\vp$, and the horizontal arrows on
the second line are $Q_F$. A very important fact is:

\underline{If $\vp$ is a quasi-isomorphism, the cone complex is exact.}
Let us provide a proof of this statment since
it plays an important r\^ole in our paper.
Suppose $(e^{j+1},f^j)$ is in the kernel of the cone complex
(\ref{conecomplex})
at degree $j$,
$$
Q_E e^{j+1}=0,\qquad 
\vp e^{j+1}+Q_F f^j=0.
$$
The first equation means that $e^{j+1}$ represents an element of the
cohomology class of ${\mathcal C}_E$, and by
the second equation we see that
the cohomology class $[e^{j+1}]$ is mapped by
$\vp_*$
 to the zero element of the cohomology of 
${\mathcal C}_F$. 
By the fact that $\vp$ is a quasi-isomorphism,
this means that the class $[e^{j+1}]$ is zero, namely,
$e^{j+1}$ can be written as $-Q_E e_1^j$ for
some element $e_1^j$ of ${\mathcal E}^j$.
Using the second equation again,
we find $-\vp Q_Ee^j_1+Q_Ff^j=0$. Since $\vp$ is a cochain map,
$\vp Q_E=Q_F\vp$, this means that $Q_F(f^j-\vp e_1^j)=0$
and therefore we have a cohomology class
$[f^j-\vp e_1^j]$ of ${\mathcal C}_F$ at degree $j$.
Again using the fact that $\vp$ is a quasi-isomorphism,
this class can be written as
$\vp_*[e_2^j]$ for some element $e_2^j\in {\mathcal E}^j$
that obeys $Q_Ee_2^j=0$. In other words,
$[f^j-\vp(e_1^j+e_2^j)]=0$. Thus, 
$f^j-\vp(e_1^j+e_2^j)$ can be written as
$Q_Ff^{j-1}$ for some element $f^{j-1}$ of ${\mathcal F}^{j-1}$.
Writing $e^j=e_1^j+e_2^j$, we find
$$
e^{j+1}=-Q_E e^j,\qquad
f^j=\vp e^j+Q_F f^{j-1}.
$$
Namely, $(e^{j+1},f^j)$ is in the image of the complex
(\ref{conecomplex}) at degree $j$.
This proves that the cone complex is exact.
The converse,
\underline{$\vp$ is a quasi-isomorphism if its cone is exact},
also holds and can be proved easily.

We have already seen an example of this fact.
Recall that a cochain map (\ref{qisex}) can be obtained 
by breaking a complex (\ref{exact1}).
Its cone is the original complex itself, (\ref{exact1}),
 and it is exact if and only if
the map (\ref{qisex}) is a quasi-isomorphism.

\subsubsection{D-Isomorphisms Versus Quasi-Isomorphisms}

We claim that two quasi-isomorphic branes are related
by a chain of D-term deformations and brane-antibrane annihilation.
Let $A$ and $B$ be two D-branes and suppose
there is a quasi-isomorphism $\vp:{\mathcal C}_A\to {\mathcal C}_B$
between the corresponding complexes.
Let $C$ be the cone of $\vp$. 
As we have seen, the cone complex is
exact and hence can be ignored in the infra-red limit.
However, does this mean that $A$ and $B$ 
are equivalent at
low energies?
One possible way to show this is to use the following line of arguments:
\beq
A\cong A+(\overline{B}+B)_{{\rm id}_B}
\stackrel{?}{\cong} (A+\overline{B})_{\vp}+B
\cong B,
\label{outline}
\eeq
where $(\overline{B}+B)_{{\rm id}_B}$ 
is the brane-antibrane system with
 the tachyon ${\rm id}_B$ turned on, and 
$(A+\overline{B})_{\vp}$ 
is the cone $C$.\footnote{To be very precise
$C$ is the cone of $-\vp$ shifted by
one to the right and reversing the sign of the tachyon.
Anyway the complex for $C$ is
exact and thus $C$ can be ignored in the IR limit.}
The first and the last equivalence relations are associated with
brane-antibrane annihilation
--- both
$(\overline{B}+B)_{{\rm id}_B}$ and
of $C=(A+\overline{B})_{\vp}$ can be ignored in the infra-red limit.
But what about the equivalence in the middle? 
One may try to prove it by considering
an alternative brane based on the graded vector bundle
$\widetilde{\mathcal E}
={\mathcal E}_A\oplus {\mathcal E}_B[-1]\oplus {\mathcal E}_B$
where $\vp$ and ${\rm id}_B$ are turned on at the same time:
\beq
\widetilde{Q}=\left(\begin{array}{ccc}
Q_A&0&0\\
\vp&-Q_B&{\rm id}_B\\
0&0&Q_B
\end{array}\right).
\label{tilQ}
\eeq
We wish to find two similarity transformations of
$\widetilde{Q}$,
one erasing $\vp$ and the other erasing ${\rm id}_B$.
That would show that
$(\widetilde{\mathcal E},\widetilde{Q})$ 
is related to both 
$A+(\overline{B}+B)_{{\rm id}_B}$
and
$(A+\overline{B})_{\vp}+B$ 
by changes of the fibre metric.
It is easy to find a transformation of the first type:
\beq
\left(\begin{array}{ccc}
{\rm id}_A&0&0\\
0&{\rm id}_B&0\\
\vp&0&{\rm id}_B
\end{array}\right)
\left(\begin{array}{ccc}
Q_A&0&0\\
\vp&-Q_B&{\rm id}_B\\
0&0&Q_B
\end{array}\right)
\left(\begin{array}{ccc}
{\rm id}_A&0&0\\
0&{\rm id}_B&0\\
-\vp&0&{\rm id}_B
\end{array}\right)
=\left(\begin{array}{ccc}
Q_A&0&0\\
0&-Q_B&{\rm id}_B\\
0&0&Q_B
\end{array}\right),
\label{tilQQA}
\eeq
where we have used $Q_B\vp=\vp Q_A$ to see that the lower left corner is zero.
However, a transformation of the second kind does not always exist.
Let us try the similarity transformation $\widetilde{Q}\to 
M \widetilde{Q} M^{-1}$ of the following form,
\beq
M=\left(\begin{array}{ccc}
{\rm id}_A&0&\psi\\
0&{\rm id}_B&q\\
0&0&{\rm id}_B
\end{array}\right).
\label{simi}
\eeq
We find that it eliminates ${\rm id}_B$ in (\ref{tilQ})
if and only if 
\beqa
&&\psi Q_B=Q_A\psi,\\
&&\vp\psi={\rm id}_B+qQ_B+Q_Bq.
\eeqa
The first equation means that $\psi:{\mathcal C}_B\to {\mathcal C}_A$
is a cochain map, and the second is the condition that
$\vp$ and $\psi$ are inverse to each other at the
level of the cohomology sheaves. This is actually a very special
situation. For example, if 
$0\to {\mathcal E}\to {\mathcal F}\to {\mathcal G}\to 0$
is an exact sequence, then there is a quasi-isomorphism
of the type (\ref{qisex})
\beq
\begin{array}{cccccccccc}
{\mathcal C}_A:
&\cdots\to&0&\to&{\mathcal E}&\to&{\mathcal F}&\to&0&\to\cdots\\
\vp\downarrow\,\,\,\,\,\,\,\,\,
&  &\downarrow&&\downarrow&&\downarrow&&\downarrow&\\
{\mathcal C}_B:
&\cdots\to&0&\to&0&\to&{\mathcal G}&\to&0&\to\cdots
\end{array}
\label{egcase}
\eeq
However, there is no
inverse $\psi$ from ${\mathcal C}_B$ to ${\mathcal C}_A$ 
satisfying the above condition
unless the sequence $0\to {\mathcal E}\to {\mathcal F}\to {\mathcal G}\to 0$ 
is split-exact.
We conclude that $(\widetilde{\mathcal E},\widetilde{Q})$
is not in general holomorphically isomorphic to
the decoupled sum $C+B$.

We have thus seen that the proof is not straightforward.
It is this point where a more general D-term deformation
comes to the rescue.
Let us consider a one parameter family of theories
given by
\beq
Q_s=\left(\begin{array}{ccc}
Q_A&0&0\\
\vp&-Q_B&s\cdot{\rm id}_B\\
0&0&Q_B
\end{array}\right).
\label{Qs}
\eeq
At $s=0$ the theory is the decoupled sum of $C$ and
$B$. For any non-zero $s\ne 0$
the theory is equivalent up to deformations of fibre metric
to the decoupled sum of $A$ and $(\overline{B}+ B)_{{\rm id}_B}$.
We have seen that $Q_s$ with non-zero $s$, no matter how
small it is, is not in general related to $Q_0$
by a change of fibre metric.
Actually, turning on $s$ is a more general D-term deformation.
The variation of the boundary interaction
is given by
\beq
\left.{\partial\over \partial s}{\mathcal A}_t\right|_{s=0}
={1\over 2}\{Q_0^{\dag},\alpha\}=-{i\over 2}{\bf Q}^{\dag}\alpha
\label{varAt}
\eeq
where
$$
\alpha=\left(\begin{array}{ccc}
0&0&0\\
0&0&{\rm id}_B\\
0&0&0
\end{array}\right)
=:\left(\begin{array}{c|c}
\!\!\begin{array}{cc}
0&0\\
0&0
\end{array}&u\\
\hline
\!\!0\,\,\,\,\, 0&0
\end{array}\right).
$$
Note that $\alpha$ anticommutes with $Q_0$, or equivalently,
$u:{\mathcal E}_B\to {\mathcal E}_C$
satisfies $Q_Cu+uQ_B=0$.
In particular, $u$ determines a chiral ring element
 for the open string from $B$ to $C$.
However, since the cone complex ${\mathcal C}_C$ is exact,
the space of chiral ring elements is zero,
$$
\Ho^p(B,C)\cong
H^p_{{\bf Q}^{\rm zero}}({\mathcal H}_{\rm zero}^{\bullet}(B,C))
\cong \HomD({\mathcal C}_B,{\mathcal C}_C[p])=\{0\}.
$$
This in particular means that $u$ is ${\bf Q}^{\rm zero}$-exact, that is, 
there is some differential form
$\beta\in {\mathcal H}_{\rm zero}(B,C)
=\oplus_{i=0}^n\Omega^{0,i}(X,Hom^{-i}(E_B,E_C))$ such that
\beq
u=i{\bf Q}^{\rm zero}\beta.
\eeq
Note that $\beta$ is not necessarily a holomorphic zero-form;
that would be the case when $Q_0$ and $Q_{s\ne 0}$
are related by a similarity transformation.
As remarked earlier, that is a very special case.
The point is that, even if that fails, there is a differential form $\beta$
with higher degree components
such that $u=i{\bf Q}^{\rm zero}\beta$.

Let us digress for a moment to see how it works
in the example (\ref{egcase}) associated with
a non-split exact sequence
$0\to {\mathcal E}\to{\mathcal F}\to {\mathcal G}\to 0$.
We suppose that the holomorphic bundles
${\mathcal E}$ and ${\mathcal G}$ are realized as
the smooth bundles with hermitian connections,
$(E_2,\bartial_{A_2})$ and $(E_1,\bartial_{A_1})$,
and ${\mathcal F}$ is the extension by
a non-zero element
$\phi\in H^{0,1}_{\bartial_{A_{1,2}}}(X,Hom(E_1.E_2))$.
We recall that ${\mathcal F}$ corresponds to the vector bundle $E_1\oplus E_2$
with the connection $\bartial_{A_{1,2},\phi}$ given in (\ref{A12phi}),
and the maps ${\mathcal E}\to {\mathcal F}$ and
${\mathcal F}\to{\mathcal G}$ are
$e_2\mapsto (0,e_2)$ and $(e_1,e_2)\mapsto e_1$.
We would like to find $\beta=\beta^0+\beta^1$,
with $\beta^0\in \Omega^0(X,Hom(E_1,E_1\oplus E_2))$
and $\beta^1\in \Omega^{0,1}(X,Hom(E_1,E_2))$,
such that $u=i{\bf Q}^{\rm zero}\beta$. 
Here, $u$ is the identity
map of $E_1$ that sends
${\mathcal G}$ of $B$ to the rightmost ${\mathcal G}$ 
in the cone $C$.
\beq
\mbox{
\setlength{\unitlength}{0.7mm}
\begin{picture}(170,45)(-15,1)
\put(-28,30){$C:$}
\put(-12,32){\vector(1,0){18}}
\put(9,30){$0$}
\put(14,32){\vector(1,0){18}}
\put(34,30){$E_2$}
\put(42,40){\small $\left(\!\!\begin{array}{c}0\\\,{\rm id}_{{}_{E_2}}\!\!
\end{array}\!\!\right)$}
\put(42,32){\vector(1,0){20}}
\put(64,30){$E_1\oplus E_2$}
\put(86,38){\small $(\,{\rm id}_{{}_{E_1}},0\,)$}
\put(86,32){\vector(1,0){20}}
\put(108,30){$E_1$}
\put(117,32){\vector(1,0){18}}
\put(139,30){$0$}
\put(145,32){\vector(1,0){18}}

\put(46,14){$\beta^1$}
\put(49,22){\vector(-3,2){6}}
\put(50,16.8){$\ddots$}
\put(57,12.1){$\ddots$}

\put(68,18){$\beta^0$}
\put(74.8,23){\vector(0,1){3}}
\put(74,18.5){$\vdots$}
\put(74,12.5){$\vdots$}

\put(84,11.5){\vector(3,2){22}}
\put(94,13.5){$u={\rm id}_{E_1}$}

\put(-28,4){$B:$}
\put(-12,6){\vector(1,0){18}}
\put(9,4){$0$}
\put(14,6){\vector(1,0){18}}
\put(36,4){$0$}
\put(42,6){\vector(1,0){23}}
\put(72,4){$E_1$}
\put(83,6){\vector(1,0){23}}
\put(110,4){$0$}
\put(117,6){\vector(1,0){18}}
\put(139,4){$0$}
\put(145,6){\vector(1,0){18}}
\end{picture}
}
\label{u}
\eeq
In fact, $\beta^0(e_1)=(e_1,0)$ and $\beta^1(e_1)=-\phi(e_1)$
does the job:
\beqa
&&Q_C\beta^0={\rm id}_{E_1},
\nn\\
&&\bartial_{A_{1,2},\phi}\beta^0-\beta^0\bartial_{A_1}
=\left(\begin{array}{c}
\bartial_{A_1}\\
\phi
\end{array}\right)
-\left(\begin{array}{c}
\bartial_{A_1}\\
0
\end{array}\right)
=\left(\begin{array}{c}
0\\
\phi
\end{array}\right),
\nn\\
&&
Q_C\beta^1
=\left(\begin{array}{c}
0\\
\phi
\end{array}\right),
\nn\\
&&
\bartial_{A_2}\beta^1-\beta^1\bartial_{A_1}=0.
\nn
\eeqa
We indeed have $\bartial_A(\beta)+Q_C\beta-\beta Q_B={\rm id}_{E_1}$.
(End of digression.)

Let us place $\beta$ into the $3\times 3$ block matrix,
just as $u$ fits into $\alpha$, and replace the differential form
$\dd x^{\bi}$ by the fermion $\bpsi^{\bi}$. We denote the
resulting matrix by $\widetilde{\beta}$.
$$
\widetilde{\beta}:=
\left.\left(\begin{array}{c|c}
\!\!\begin{array}{cc}
0&0\\
0&0
\end{array}&\beta\\
\hline
\!\!0\,\,\,\,\, 0&0
\end{array}\right)\right|_{\displaystyle \dd x^{\bi}\to \bpsi^{\bi}}.
$$
Then $u=i{\bf Q}^{\rm zero}\beta$ means 
$\alpha=i{\bf Q}\widetilde{\beta}$.
Thus, we find that the variation of the boundary interaction
is a D-term
\beq
\left.{\partial\over \partial s}{\mathcal A}_t\right|_{s=0}
={1\over 2}{\bf Q}^{\dag}{\bf Q}\widetilde{\beta}.
\eeq
We have shown that turning on $s$ is indeed a D-term deformation.
Once $s$ is turned on, the system is equivalent
to the decoupled sum of $A$ and $(\overline{B}+B)_{\rm id}$.
In this way, the middle part of (\ref{outline}) is shown to hold.
This completes the proof that quasi-isomorphic D-branes
are related by a chain of D-term deformations and
brane-antibrane annihilation. In particular
we proved that quasi-isomorphic branes flow to the same 
infra-red fixed point.

Conversely, one can also show that
D-branes that are related by D-isomorphisms are related
by a chain of quasi-isomorphisms. Firstly,
brane-antibrane annihilation is a trivial example
of a quasi-isomorphism. Secondly, D-term deformations do not change
the D-brane category in the chiral sector \cite{book}.
On the other hand, the D-brane category in the chiral sector is
equivalent to the derived category, and
two complexes are isomorphic in the derived category if and only if
they are related by a chain of quasi-isomorphisms.
Thus, two D-branes related by a D-term deformation
are related by a chain of quasi-isomorphisms.

To summarize, we have seen that {\it D-isomorphisms are equivalent to
quasi-isomorphisms}.

When a B-twist to a topological field theory is possible,
our result can also be stated as follows: 
two D-branes are D-isomorphic if and only if they determine
isomorphic D-branes in topological field theory.
Thus, in a fixed closed string background,
our main target of study --- D-branes up to D-isomorphisms ---
is nothing but the isomorphism class of objects in
the category of topological D-branes. Nevertheless,
we decide not to use the term ``topological D-branes''
since our ultimate motivation is to study the full physics of D-branes,
rather than the property of D-branes in topological field theory
and topological strings. In particular, we pay attention to the
dependence of the boundary RG flow
on the bulk parameter corresponding to
the complexified K\"ahler class, whereas topological 
field/string theory is insensitive to such a dependence.

The relevance of quasi-isomorphisms concerning identification of
D-branes was argued in \cite{Douglas} employing the spacetime picture
(crossing symmetry). 
We have finally managed to clarify their precise 
relevance in the general case from the worldsheet view point.

\subsection{Lower-Dimesional Branes As Complexes Of Vector Bundles}
\label{subsec:Koszul}

The fact that a brane is infra-red empty in the region of $x$ 
where ${\bf T}(x)$ is non-vanishing suggests a way to 
represent a D-brane wrapped on a submanifold of $X$ as
a complex of vector bundles of the entire space $X$
\cite{Ashoke,WittenK}. 
For example, consider a holomorphic line bundle ${\mathcal L}$
with a section $f$. A D-brane wrapped on the zero locus of $f$, the divisor
$D_f$, may be represented by
the complex
$$
{\mathcal L}^{-1}\stackrel{f}{\longrightarrow}{\mathcal O},
$$
where ${\mathcal O}$ is at R-degree $0$.
In this paper, we do not attempt to construct and analyze
worldsheet boundary conditions
for B-type D-branes wrapped on submanifolds nor do we show that
the D-branes associated with complexes of vector bundles are really
D-isomorphic to such lower dimensional branes.
Rather, we take the following indirect route.
For each complex submanifold there is a coherent sheaf supported on it,
and we take a complex of vector bundles
that is quasi-isomorphic to that sheaf as the worldsheet
 {\it definition} of the D-brane wrapped on that submanifold. For example, 
the above complex is indeed quasi-isomorphic to the coherent sheaf
${\mathcal O}_{D_f}$ supported on the divisor $D_f$.
To be precise, a general D-brane can be represented as a {\it complex}
of coherent sheaves and it is known that there exists a complex
of vector bundles on $X$ that is quasi-isomorphic to it
(cf Section~\ref{sec:math}).
Thus, we still have a `definition' that applies to the most general D-brane.
Consistency of this proposal is provided by  our results of the
previous subsection: as long as complexes of vector bundles are concerned,
D-isomorphisms are nothing else but quasi-isomorphisms.

As another example, and as the example that plays a key r\^ole 
in this paper, we introduce a complex that represents a point on
$X$. For concretness we consider Euclidean space $X=\C^n$
with coordinates $x_1,...,x_n$ and a D0-brane at the origin
$\mathfrak{p}=\{x^1=\cdots =x^n=0\}$.
Let us recall the Clifford algebra (\ref{Cliffordalg}) for $\eta_i$
and $\bareta_i$ and its
representation $\cp_n$, the Clifford module, generated by the vector
$|0\rangle$ that is annihilated by all $\eta_i$'s.
We consider the brane with the Chan-Paton space $\cp_n$ and
the tachyon profile given by
\beq
Q(x)=\sum_{i=1}^nx_i\eta_i.
\eeq
This defines a sequence of linear maps
of the subspaces of $\cp_n$:
$$
\C\bareta_1\cdots\bareta_n|0\rangle
\stackrel{Q}{\longrightarrow}\cdots
\stackrel{Q}{\longrightarrow}
\bigoplus_{i<j}\C\bareta_i\bareta_j|0\rangle
\stackrel{Q}{\longrightarrow}
\bigoplus_{i=1}^n\C\bareta_i|0\rangle
\stackrel{Q}{\longrightarrow}
\C|0\rangle
$$
such that $Q^2=0$. Namely, we have a complex of trivial vector bundles,
\beq
{\mathscr K}\,:~~
{\mathcal O}\stackrel{Q}{\longrightarrow}
{\mathcal O}^{\oplus n}\stackrel{Q}{\longrightarrow}
{\mathcal O}^{\oplus {n\choose 2}}\to
\cdots\to
{\mathcal O}^{\oplus {n\choose 2}}\stackrel{Q}{\longrightarrow}
{\mathcal O}^{\oplus n}\stackrel{Q}{\longrightarrow}
{\mathcal O}.
\label{Koszul}
\eeq
A complex of the form (\ref{Koszul}) is called a {\it Koszul complex}.
Note that the boundary potential is
\beq
\{Q,Q^{\dag}\}=\left(\sum_{i=1}^n|x_i|^2\right)\cdot {\rm id}^{}_{\cp_n}.
\label{D0potential}
\eeq
It vanishes precisely at the origin $\mathfrak{p}$, and therefore
represents a D0-brane at the origin. Indeed, the tachyon profile
${\bf T}=iQ-iQ^{\dag}$ is nothing but the ``Atiyah-Bott-Shapiro
construction for the
D0-brane'' \cite{WittenK} (see for eaxmple \cite{blin}).
In particular, it has the proper Ramond-Ramond charge as well as
the correct space of chiral ring elements with other branes.

If we considered a theory in which the point $x=0$ is excised
in some way, then the boundary potential (\ref{D0potential}) would be nowhere
vanishing, and hence the Koszul complex ${\mathscr K}$
must represent a brane that is empty in the infra-red limit.
In particular, a brane can be modified by binding such ${\mathscr K}$'s
without changing its low energy behaviour.
This operation will be used frequently later in this paper.

A Koszul complex can also be used to construct 
the fractional branes in an
orbifold theory --- D0-branes stuck at the orbifold fixed points.
We consider the orbifold of the Euclidean space $\C^n$ by a
finite group $\Gamma$ of linear transformations,
$\Gamma\owns\gamma:x_i\mapsto \sum_{j=1}^n\gamma_{ij}x_j$. 
We would like to find a representation $\rho$ of $\Gamma$
on the Clifford module $\cp_n$ such that the brane
$(\cp_n,Q)$ obeys the invariance condition
$$
\rho(\gamma)^{-1}\left(\sum_{i,j=1}^n\gamma_{ij}x_j
\eta_i\right)\rho(\gamma)=\sum_{j=1}^nx_j\eta_j,
$$
that is,
$\rho(\gamma)\eta_j \rho(\gamma)^{-1}=\sum_{i=1}^n\eta_i\gamma_{ij}$. 
In order to preserve the Clifford algebra relations,
we find that the $\bareta_i$'s must transform as
$\rho(\gamma)\bareta_i\rho(\gamma)^{-1}=\sum_{j=1}^n(\gamma^{-1})_{ij}
\bareta_{j}$.
If we decide that $|0\rangle\in \cp_n$ is a $\Gamma$-invariant vector, we find
that the vectors $\{\bareta_i|0\rangle\}$ transform as
$$
\rho(\gamma):\bareta_i|0\rangle\longmapsto
\sum_{j=1}^n\bareta_j|0\rangle
(\gamma^{-1})_{ij}.
$$
Let ${\mathscr R}$ be the defining representation of
$\Gamma$, that is, the space $\C^n$ regarded as a representation
of $\Gamma$. The basis elements $e_i$ of ${\mathscr R}$
transforms under $\gamma\in \Gamma$
as $e_i\mapsto \sum_{j=1}^ne_j\gamma_{ji}$.
Then, we find that the elements $\bareta_i|0\rangle$ transform
in the same way as the dual basis to $\{e_i\}$.
Namely, the space spanned by $\{\bareta_i|0\rangle\}$ can be identified with
${\mathscr R}^*$ as a representation of $\Gamma$.
Similarly, the space spanned by $\{\bareta_{i_1}\cdots\bareta_{i_r}|0\rangle
\}$ can be identified with $\wedge^r{\mathscr R}^*$.
Thus, we find that the Koszul complex can be
regarded as the complex of $\Gamma$-modules
\beq
\wedge^n{\mathscr R}^*\stackrel{Q}{\longrightarrow}
\wedge^{n-1}{\mathscr R}^*\to
\cdots\to
\wedge^2{\mathscr R}^*\stackrel{Q}{\longrightarrow}
{\mathscr R}^*\stackrel{Q}{\longrightarrow}
\C.
\eeq
We may also consider tensoring this with any representation $\rho$
of $\Gamma$.
We shall denote the brane associated with that complex by
${\mathcal O}_{\mathfrak{p}}(\rho)$.
For an Abelian group, such as $\Gamma\cong \Z_d$, we use the additive
notation ${\mathcal O}_{\mathfrak{p}}(\rho_m)={\mathcal O}_{\mathfrak{p}}(m)$.
For irreducible representations, 
these are the fractional branes stuck at the orbifold point $\mathfrak{p}$.
As a side remark and for later convenience, we note that
${\mathcal O}_{\mathfrak{p}}(\wedge^n{\mathscr R})$ can also be written as
\beq
\C\stackrel{Q}{\longrightarrow}
{\mathscr R}\stackrel{Q}{\longrightarrow}
\wedge^2{\mathscr R}\to
\cdots\to
\wedge^{n-1}{\mathscr R}\stackrel{Q}{\longrightarrow}
\wedge^{n}{\mathscr R}.
\eeq
This is what we would directly find if we span the Clifford module $\cp_n$
by the even and odd multiples of the $\eta_i$'s on
the state $|0\rangle'$ that is annihilated by all $\bareta_i$'s.

\subsection{The Landau-Ginzburg Case}
\label{subsec:DisomLG}

We next discuss D-isomorphisms in Landau-Ginzburg models.
We recall that the boundary interaction for a B-brane
is given by the formula (\ref{SWLLG}) for a matrix factorization
$Q(x)$ of the superpotential $W(x)$.

D-term deformations include deformations of the fibre
metric of the Chan-Paton space.
In particular, a similarity transformation by
an even (or R-symmetry preserving) matrix
\beq
Q(x)\to U(x)^{-1}Q(x) U(x)
\label{simiLG}
\eeq
is of this type.
The D-brane is empty at low energies 
if and only if the potential $U(x)={1\over 2}\{Q(x),Q(x)^{\dag}\}$ 
is positive definite for all $x$. 
For example, a matrix factorization of the form
\beq
Q(x)=\left(\begin{array}{cc}
0&{\bf 1}_r\\
W(x){\bf 1}_r&0
\end{array}\right)\quad
\mbox{or}\quad
\left(\begin{array}{cc}
0&W(x){\bf 1}_r\\
{\bf 1}_r&0
\end{array}\right)
\label{trivmf}
\eeq
has a positive definite boundary potential,
$\{Q(x),Q(x)^{\dag}\}=(1+|W(x)|^2){\rm id}_V>0$,
and can be ignored at low energies.
In the case where $W(x)$ is a quasi-homegeneous polynomial
and each variable $x_i$ has a positive degree,
any quasi-homogeneous
matrix factorization with $\{Q(x),Q(x)^{\dag}\}>0$ is
holomorphically isomorphic to
the direct sum of those of the type (\ref{trivmf}).

Let us prove the last statement. We take a matrix factorization
of size $2r$ by $2r$,
$$
Q(x)=\left(\begin{array}{cc}
0&f(x)\\
g(x)&0\end{array}
\right).
$$
Suppose $f(x)$ has rank $s$ at $x=0$.
Then, with a change of basis, that is, with a
similarity transformation (\ref{simiLG}),
it can be written as
$$
f(0)=\left(\begin{array}{cc}
{\bf 1}_s&0\\
0&0
\end{array}\right).
$$
If we move away from $x=0$, polynomials appear in the entries of $f(x)$
except at the first $s$ diagonals where ``$1$'' are.
This is because each entry must be quasi-homogeneous
and $1$ is the only polynomial of degree zero.
By standard linear algebra operations, one can erase all the
entries in the same raws and colums as these $1$'s, and one may
assume the form
$$
f(x)=\left(\begin{array}{cc}
{\bf 1}_s&0\\
0&A(x)
\end{array}\right).
$$
Then, $f(x)g(x)=g(x)f(x)=W(x){\bf 1}_r$ shows that $g(x)$ can be written
as
$$
g(x)=\left(\begin{array}{cc}
W(x){\bf 1}_s&0\\
0&B(x)
\end{array}\right),
$$
where $A(x)B(x)=B(x)A(x)=W(x){\bf 1}_{r-s}$.
The condition $\{Q,Q^{\dag}\}>0$ at $x=0$ requires that
$B(0)B(0)^{\dag}>0$ and $B(0)^{\dag}B(0)>0$.
Thus, again by raw and column operations, one may assume
$B(x)={\bf 1}_{r-s}$ which in turn requires $A(x)=W(x){\bf 1}_{r-s}$.
In this way, we have seen that $Q(x)$ is equivalent up to
similarity transformations to the sum of $(1,W)$'s and
$(W,1)$'s;
\beq
Q(x)\cong \left(\begin{array}{cc}
0&{\bf 1}_s\\
W(x){\bf 1}_s&0
\end{array}\right)\oplus
\left(\begin{array}{cc}
0&W(x){\bf 1}_{r-s}\\
{\bf 1}_{r-s}&0
\end{array}\right).
\eeq
This proves our claim.

We have learned that ``brane-antibrane annihilation'' in a LG model
means annihilation of $(1,W)$'s and $(W,1)$'s.
Using the same argument as in the non-linear sigma models,
we can show that two D-branes are related by
a chain of D-term deformations and brane-antibrane annihilations
if there is a degree zero state whose cone has positive potential
$\{Q,Q^{\dag}\}>0$.
Two such D-branes flow to the same fixed point in the infra-red limit.

\subsection{Kn\"orrer Periodicity}
\label{subsec:Knorrer}

So far, our focus was renormalization group flows of boundary interactions.
However, bulk interactions generically flow as well, and the interplay of
bulk and boundary RG flows is expected to have some
important consequences. For recent works on this subject
in systems with  ${\mathcal N}=2$ supersymmetry, 
see \cite{bulkboundary} and references therein. 
Here, we would like to focus on
the most primitive among bulk RG flows --- integrating out massive fields.
As far as we know, this has not been considered in full detail
in the literature.

Let us consider a Landau-Ginzburg model of $n+2$ variables,
$X_1,\ldots,X_n,U,V$ with superpotential of the form
\beq
W=W_L(X_1,...,X_n)+UV.
\eeq
We see that the fields $U$ and $V$ are massive and 
must be integrated out at an appropriate energy scale.
(We assume that $X_1,...,X_n$ are massless or have lower masses.)
The question is how the D-branes 
in the high energy theory including the variables $U$ and $V$
are related to the ones in the low energy theory
where $U$ and $V$ are gone. 
That is, what is the relation of matrix factorizations
of $W_L(x)+uv$ and matrix factorizations of $W_L(x)$?
A similar question in a mathematical context is solved and 
is known as
{\it Kn\"orrer periodicity} \cite{Knorrer}, which states that
the category of matrix factorizations of $W_L(x)$ and that of
$W_L(x)+uv$ are equivalent. 
We show below that this is indeed relevant to our question.
Moreover, we construct an explicit map of branes which
was not given in \cite{Knorrer}. 
This construction will play a very important r\^ole later in this paper.

Let us first consider the opposite problem: Given a brane 
in the low energy theory,
does it come from a brane in the high energy theory?
Namely, if $Q_L(x)$ is a matrix factorization of $W_L(x)$, 
is there a matrix factorization $Q(x,u,v)$
of $W_L(x)+uv$ such that the brane $Q$ flows to the brane $Q_L$ as
$U$ and $V$ are integrated out?
The answer is yes.
An important r\^ole is played by 
a certain brane of the theory of variables $U$ and $V$ only,
and with superpotential $W=UV$. 
For our purpose it is convenient to use its realization in terms of
boundary fermions $\eta,\bareta$ with boundary action
\beq
S_1(\eta,U,V)
=\int_{\partial\surface}
\dd t\left\{
i\bareta\dot{\eta}-{1\over 2}|v|^2-{1\over 2}|u|^2
+\Rp\Bigl(\psi^u\eta+\psi^v\bareta\Bigr)
\right\},
\label{UVbdry}
\eeq
where $\psi^u$ and $\psi^v$ are the boundary values of 
the fermionic components $\psi=\psi_++\psi_-$ of
$U$ and $V$. Upon quantization, $\eta$ and $\bareta$ obey the anticommutation
relations $\{\eta,\bareta\}=1$, $\eta^2=\bareta^2=0$, which is
represented on the two dimensional vector space spanned by
$|0\rangle$ and $\bareta|0\rangle$, where $|0\rangle$ is annihilated by
$\eta$. With respect to that basis,
$\eta$ and $\bareta$ are represented by the matrices
$$
\eta=\left(\begin{array}{cc}0&1\\0&0\end{array}\right),\qquad
\bareta=\left(\begin{array}{cc}0&0\\1&0\end{array}\right),
$$ 
and we find that the boundary interaction (\ref{UVbdry}) is
of the from (\ref{SWLLG}), with
$$
Q=u\eta +v\bareta=\left(\begin{array}{cc}0&u\\v&0\end{array}\right).
$$
There is a unique supersymmetric ground state in the Ramond sector of
the open string stretched between two copies of it.
Since the system has a finite correlation length, this implies that
the system formulated on the half-space also has a unique
supersymmetric ground state. (See Section~\ref{sec:vacuum}
for a more detailed reasoning and an explicit contruction.)
Now, given a brane 
${\rm P}\exp\left(-i\int_{\partial\surface}{\mathcal A}^L_t\dd t\right)$
of the low energy theory,
we consider the following brane of the high energy theory
\beq
{\rm P}\exp\left(-i\int_{\partial\surface}{\mathcal A}^L_t\dd t\right)
\exp\Bigl(iS_1(\eta,U,V)\Bigr).
\label{UVbrane}
\eeq
If we integrate out the fields $U,V$ and $\eta$, 
the factor $\exp(iS_1(U,V,\eta))$ simply drops since
the $(U,V,\eta)$ system has a unique ground
state of zero energy.
As a result, we simply get
back the original brane of the low energy theory.
Thus, (\ref{UVbrane}) is the brane we were looking for.
One can write it in the form (\ref{SWLLG}), with
$$
Q=Q_L+u\eta+v\bareta.
$$
If $Q_L$ is represented by a matrix
$$
Q_L=\left(\begin{array}{cc}0&f(x)\\g(x)&0\end{array}\right)
$$
with respect to a basis ${\bf b}_0=({\bf e}_0,{\bf o}_0)$ of the original
Chan-Paton space,
then $Q$ is represented by a matrix
\beq
Q=\left(\begin{array}{cc|cc}
0&f(x)&0&u\\
g(x)&0&-u&0\\
\hline
0&-v&0&g(x)\\
v&0&f(x)&0
\end{array}\right)
\label{UVmf}
\eeq
with respect to the basis ${\bf b}=({\bf e}_0\otimes |0\rangle,
{\bf o}_0\otimes |0\rangle, {\bf o}_0\otimes \bareta|0\rangle,
{\bf e}_0\otimes \bareta |0\rangle)$ of the 
Chan-Paton space of the brane
(\ref{UVbrane}). This map $Q_L\longmapsto Q$ is indeed the functor
given in \cite{Knorrer} that makes the equivalence of the
two categories.

Now we come to our main problem: Given a brane in the high energy theory,
what happens when $U$ and $V$ are integrated out?
We already know the answer if the given brane is of
the form (\ref{UVbrane}): it is 
${\rm P}\exp\left(-i\int{\mathcal A}^L_t\dd t\right)$.
Note that this is {\it not} the same as just
setting $u=v=0$ in the matrix factorization $Q(x,u,v)$.
If we simply did that, we would obtain
$$
Q|_{u=v=0}=\left(\begin{array}{cc|cc}
0&f(x)&0&0\\
g(x)&0&0&0\\
\hline
0&0&0&g(x)\\
0&0&f(x)&0
\end{array}\right)
=\left(\begin{array}{cc}
0&f(x)\\ g(x)&0\end{array}\right)
\oplus
\left(\begin{array}{cc}
0&g(x)\\ f(x)&0\end{array}\right)
$$
which is twice as much in size as the correct answer.
Rather, we should extract the first half block of it.
What happens to more general branes?
The key is the fact, shown in Kn\"orrer's paper \cite{Knorrer},
that any matrix factorization of $W=W_L(x)+uv$ is 
holomorphically isomorphic to the one of the form
(\ref{UVmf}) up to a decoupled sum of empty branes
$(W,1)$, $(1,W)$. Thus, the procedure is first
to find such a presentation and then take out the relevant block of
$Q|_{u=v=0}$.
However, this is not systematic and requires a lot of work
in the indivisual case.

Here, we present a general procedure to find the low energy brane
without explicitly finding a special presentation.
Let $Q$ be the matrix factorization of $W=W_L(x)+uv$ represented on
a Chan-Paton vector space $\cp$. 
We set $v=0$ but keep $u$ in $Q$,
\beq
\widehat{Q}=Q|_{v=0},
\label{UVtoIR}
\eeq
and regard $\widehat{Q}$ as
a matrix factorization of $W_L(x)$ represented
on the infinite dimensional Chan-Paton space
\beq
\widehat{\cp}=\cp\oplus u\cp\oplus u^2\cp\oplus u^3\cp\oplus\cdots.
\label{UVtoIRcp}
\eeq
Then, $(\widehat{Q},\widehat{\cp})$ is the matrix factorization
determining the low energy brane. It is infinite in size, but
the boundary potential $\{\widehat{Q},\widehat{Q}^{\dag}\}$
is mostly positive and has zero only in a finite dimensional subspace
of $\widehat{\cp}$.

Let us see how it works.
First, consider the case where $Q$ is already of the form (\ref{UVmf}).
With respect to the basis $({\bf b}, u{\bf b}, u^2{\bf b},\ldots)$
of $\widehat{\cp}$,
$\widehat{Q}$ is represented by the matrix
\beq
\widehat{Q}=\left(\begin{array}{cc|cc|cc|cc|cc|cc}
0&f&&&&&&&&&&\\
g&0&&&&&&&&&&\\
\hline
&&0&g&&&&&&&&\\
&&f&0&&&&&&&&\\
\hline
&&0&1&0&f&&&&&&\\
&&-1&0&g&0&&&&&&\\
\hline
&&&&&&0&g&&&&\\
&&&&&&f&0&&&&\\
\hline
&&&&&&0&1&0&f&&\\
&&&&&&-1&0&g&0&&\\
\hline
&&&&&&&&&&\ddots&
\end{array}\right),
\label{KnorrerhatQ}
\eeq
where unwritten entries are all zero.
We find that $\widehat{Q}$ consists mostly of
the direct sum of infinite copies of
the block
$$
Q_1=\left(\begin{array}{cc|cc}
0&g&&\\
f&0&&\\
\hline
0&1&0&f\\
-1&0&g&0
\end{array}\right).
$$
The potential of this block is
$$
\{Q_1,Q_1^{\dag}\}=\left(\begin{array}{cc|cc}
1+gg^{\dag}+f^{\dag}f\!\!&0&&\\
0&\!\! 1+ff^{\dag}+g^{\dag}g &&\\
\hline
&& 1+ff^{\dag}+g^{\dag}g \!\!&0\\
&&0&\!\! 1+gg^{\dag}+f^{\dag}f
\end{array}\right),
$$
which is everywhere positive.  
(Equivalently, there is a similarity transformation that makes
$Q_1$ into a direct sum of $(1,W_L)$'s and $(W_L,1)$'s.)
Thus, each of such blocks is empty
in the infra-red limit.
Therefore, only the first block ${\,0\,\,f\choose g\,\, 0\,}$
of $\widehat{Q}$ remains, and this is indeed the right answer.
Next, consider the empty branes
$$
Q_2=\left(\begin{array}{cc}
0&W\\
1&0
\end{array}\right),\qquad
Q_3=\left(\begin{array}{cc}
0&1\\
W&0
\end{array}\right).
$$
$\widehat{Q}_2, \widehat{Q}_3$ are direct sums of
infinite copies of $Q_2|_{v=0}, Q_3|_{v=0}$ and in particular have 
everywhere positive potentials. Thus, $\widehat{Q}_2$ and $\widehat{Q}_3$
are also empty. Since any matrix factorization is holomorphically
isomorphic to the sum of matrices of the form
(\ref{UVmf}), $Q_2$ and $Q_3$ \cite{Knorrer}, 
we find that the above procedure gives
the right answer to the low energy brane in the general case.

Some remarks are in order:

\noindent
(i) One may notice the asymmetry in the r\^ole of $u$ and $v$. 
We could have applied the above procedure by swapping
the two variables. If we did so, the resulting low energy brane would have
the opposite $\Z_2$-grading. This
asymmetry or ambiguity came from the fact that
the $\Z_2$-grading of branes in the high energy theory is
not given separately for the $x$-part and for the $(u,v)$-part.
What we have done above is to make a choice of the $\Z_2$-grading
in the $(u,v)$-part --- we declared that $|0\rangle$ is even and
$\bareta|0\rangle$ is odd.
The opposite choice, or equivalently, the swap of
$u$ and $v$, results in the opposite $\Z_2$-grading
of the low energy D-brane. 
One should remember that there is this choice dependence
in the map of branes from the high energy theory
to low energy theory.
If we want to be systematic, we need to fix one choice and use it
for all branes.

\noindent
(ii) The above procedure will turn out to be extremely powerful
and have a wide range of applications,
despite the fact that we need to invoke infinite size Chan-Paton factors.
The particularly important case is where the Landau-Ginzburg superpotential
is fibred over some base manifold (the case of non-linear
sigma model with superpotential). In such a situation,
it is in general impossible to find a presentation of $Q$
as (\ref{UVmf}) globally, and it is also very difficult to
find such a presentation locally and to patch them together.
However, the above procedure $(\cp,Q)\mapsto (\widehat{\cp},\widehat{Q})$
can be applied without difficulty
to such fibred situation and provides
a one-shot answer.
We will use this fibre-wise
construction in Section~\ref{sec:compactI}.

\noindent
(iii)
Infinite size Chan-Paton factors
which are effectively finite for the same reason as above
 had been discussed earlier 
in \cite{DTI,EdK}.

\subsubsection*{\it Note on Short Orbit Branes}

It is a good point to digress for a moment to
explain the matrix factorization realization of the short orbit 
RS-branes given in Section~\ref{subsub:LGO}. 
The ${\mathcal N}=2$ minimal model at level $k$
is realized as the infra-red limit of the single variable
LG model with superpotential $W=X^d$ with $d=k+2$. 
Alternatively, it can also be identified with the IR fixed point of the model
with two variables $W=X^d-Y^2$.
The minimal model with even level 
has two classes of Cardy branes, ordinary branes
and short orbit branes \cite{MMS}. Ordinary branes are realized by
matrix factorizations of $W=x^d$ correponding to
$x^{L+1}\cdot x^{d-L-1}$ while the short orbit branes
correspond to the matrix factorization
$$
Q=\left(\begin{array}{cc}
0&x^{d\over 2}-y\\
x^{d\over 2}+y&0
\end{array}\right)
$$
of $W=x^d-y^2$ \cite{book} (see also \cite{KapLi2}).
It was found in \cite{BH2} that there are an odd number of
fermionic zero modes in the open string stretched between ordinary
and short orbit branes. For that reason, short
orbit branes cannot coexist with ordinary branes if we want to
define a $\Z_2$-grading operator $(-1)^F$. This problem disappears for
a product branes with even number of short orbit 
factors in a product of minimal models. 
This is why we allow such short orbit branes in Gepner models
\cite{BHHW}.

Consider the product of two minimal models
with even levels,
$d_1=2m_1$ and $d_2=2m_2$.
A product of two short-orbit branes
is realized by the matrix factorization
$$
Q=\Bigl(x_1^{m_1}-y_1\Bigr)\eta_1
+\Bigl(x_1^{m_1}+y_1\Bigr)\bareta_1
+\Bigl(x_2^{m_2}-y_2\Bigr)\eta_2
+\Bigl(x_2^{m_2}+y_2\Bigr)\bareta_2
$$
of $W=x_1^{2m_1}-y_1^2+x_2^{2m_2}-y_2^2$.
We write $\eta_1={1\over 2}(\xi_1+i\xi_1')$, 
$\eta_2={1\over 2}(\xi_2+i\xi_2')$ for real $\xi_i$, $\xi_i'$
and then introduce $\eta={1\over 2}(\xi_1'+i\xi_2')$.
Writing $u=-y_2-iy_1$, $v=y_2-iy_1$, the above matrix can be written
as
$$
Q=x_1^{m_1}\xi_1
+x_2^{m_2}\xi_2
+u\eta+v\bareta,
$$
which is a matrix factorization of $W=x_1^{2m_1}+x_2^{2m_2}+uv$.
In this form, one can readily integrate out the $U,V,\eta$ system.
The result is
the matrix factorization
$
Q_L=x_1^{m_1}\xi_1+x_2^{m_2}\xi_2
$
of $W_L=x_1^{2m_1}+x_2^{2m_2}$.
This is why (\ref{shortorbitbrane}) represents a short orbit brane
in a Gepner model.

\subsection{D-brane Transport On The K\"ahler Moduli Space}
\label{subsec:whatwedo}

Bulk $(2,2)$ supersymmetric quantum field theories 
have two kinds of distinguished deformation parameters
--- chiral and twisted chiral parameters. For non-linear sigma models,
chiral parameters correspond to the complex structure of the target space
while twisted chiral parameters determine the complexified K\"ahler class.
The moduli space of $(2,2)$ theories up to bulk D-term deformations
is a direct product $\moduli_C\times \moduli_K$, 
where $\moduli_C$ and $\moduli_K$ are
parametrized by chiral and twisted chiral parameters respectively.
We shall refer to $\moduli_K$ and $\moduli_C$
as the {\it K\"ahler moduli space} and the {\it complex structure
moduli space} respectively, although the geometric interpretation is present
only around special corners of $\moduli_K$, called 
the large volume limits. Moving away from such a corner,
the $\alpha'$ corrections grow and the sigma model description eventually
becomes totally inadequate. There can also be corners of different type
which are described in terms of Landau-Ginzburg models or orbifolds thereof.
In general, a single K\"ahler moduli space may have multiple regions with
quite different descriptions. For example, the non-linear sigma model
on the quintic hypersurface $X_G=\{G(x)=0\}$ in $\CP^4$ and 
the Landau-Ginzburg orbifold $W=G(x)/\Z_5$ are at two opposite corners
of the same one dimensional K\"ahler moduli space. There are also examples
of $\moduli_K$ with several large volume limits corresponding to K\"ahler 
manifolds of different topology.
As discussed in the introduction, the main purpose of the paper
is to construct a family of boundary interactions in a family of bulk theories
defined in a region of $\moduli_K$
that encompass various corners with different interpretations.
We show below that there is a natural notion of ``parallel families''
of B-branes.

We first describe the ``parallel transport'' of D-branes along a path
in $\moduli_K$.
Deformations of the $(2,2)$ bulk theory inside $\moduli_K$
are generated by bulk twisted F-terms which are classified as 
D-terms from the point of view of the ${\mathcal N}=2_B$ supersymmetry. 
Namely, they are
 of the form ${\bf Q}{\bf Q}^{\dag}(\cdots)$ where ${\bf Q}$ and 
${\bf Q}^{\dag}$ are the ${\mathcal N}=2_B$ generators.
Also, we are interested in properties of D-branes that do
not change under ${\mathcal N}=2_B$ boundary D-term deformations.
This motivates us to take the following rule of D-brane transport:
{\it The bulk and boundary interactions must vary by
${\mathcal N}=2_B$ D-terms only.} 
To be more explicit, if $S(\tau)=S_{\rm bulk}(\tau)+S_{\rm bdry}(\tau)$
is a one parameter family of actions that realizes the transport, its 
variation must be a D-term
\beq
{\dd\over \dd \tau}S(\tau)
={\bf Q}{\bf Q}^{\dag}(\cdots).
\label{deform}
\eeq
The chiral sector of B-type D-branes does not change under the
bulk and boundary ${\mathcal N}=2_B$ D-term deformations.
Therefore, the rule is defined so that the chiral sector remains constant
under the transport.
When B-twist is possible, this means that the associated open topological
field theory remains invariant under the transport.

We assert that the rule (\ref{deform}) defines a ``flat connection''
on the ``bundle of D-branes'' over the K\"ahler moduli space $\moduli_K$
in a certain sense.
Let us first show that the D-brane transport obeying our rule
is unique up to boundary ${\mathcal N}=2_B$ D-term and bulk $(2,2)$
D-term deformations.
Let us consider two admissible transports
of a given interaction $S(0)$ over the same path. 
Infinitesimally, the transports can be written as
$S_i(\epsilon)=S(0)+\epsilon{\bf Q}{\bf Q}^{\dag}A_i$, $i=1,2$.
Then, $S_2(\epsilon)$ can be regarded as the deformation of
$S_1(\epsilon)$ by the term $\epsilon {\bf Q}{\bf Q}^{\dag}(A_2-A_1)$.
Since $S_1(\epsilon)$ and $S_2(\epsilon)$ represents the same
point of $\moduli_K$, the deformation term must have no twisted F-term
component. Thus, it is a boundary D-term plus possibly a bulk $(2,2)$ D-term.
By composition of this elementary process, we find that
admissible transports over the same path are related by a chain of
boundary D-term deformations, possibly with bulk $(2,2)$ D-term deformations.
One important point which is implicitly assumed in
this argument is that we can use a set of field variables whose
supersymmetry transformations 
do not depend on the twisted chiral parameters, 
at least inside the region of $\moduli_K$ we are considering.
This is to ensure that $\epsilon {\bf Q}{\bf Q}^{\dag}(A_2-A_1)$,
which is a D-term in the initial theory $S(0)$,
is also a D-term with respect to the ${\mathcal N}=2_B$ supersymmetry
of the deformed theory $S_1(\epsilon)$.
In the non-linear sigma model, this is indeed the case
since the sets of fields as well as their
supersymmetry transformation are fixed when the
complex structure of the target space is fixed.
By a similar argument, we find that the ``connection'' is flat:
under a deformation of the path in $\moduli_K$ with fixed initial
and final points, the result of D-brane transport changes only by boundary 
${\mathcal N}=2_B$ D-terms and bulk $(2,2)$ D-terms.
Note that this holds only for {\it continuous} deformations of the paths.
There can be non-trivial monodromies for topologically
non-trivial loops in $\moduli_K$.
In fact such monodromies are known to exist even at the level of
D-brane charge.

As we have discussed earlier in this section, our main interest is
in the properties of D-branes that do not change not only
under boundary D-term deformations but also under brane-antibrane
annihilation.
A natural question is whether 
brane-antibrane annihilation at one point of
the K\"ahler moduli space is sent to brane-antibrane annihilation
at another point under the parallel transport.
Formally, that must be the case.
Otherwise the chiral sector would change,
but the transport is defined so that it remains constant.
And this indeed appears plausible  
as long as the path stays inside a large volume regime
--- positivity of the boundary potential $\{Q,Q^{\dag}\}$
is unaffected under deformations of the K\"ahler class.
However, it is not at all obvious whether this continues to be the case
if the path goes out of one large volume regime and the size of the
target space becomes vanishingly small.

That the chiral sector does not change under D-brane transport
does not mean that the full theory remains constant.
In particular, the infra-red limit must depend on where we are on $\moduli_K$,
since it defines a ${\mathcal N}=2_B$ superconformal
boundary condition in the $(2,2)$ superconformal field theory
that really depends both on $\moduli_K$ and $\moduli_C$.
We expect a rich pattern of renormalization group
flows that change along the transport.
For example, let us consider a complex of two vector bundles,
$$
{\mathcal E}_1\stackrel{Q}{\longrightarrow}{\mathcal E}_2,
$$
which defines a parallel family of boundary interactions over a region of
$\moduli_K$. The region may be separated into two
by a wall of marginal stability. On one side of the wall
the map $Q$ is tachyonic so that the brane flows to a single indecomposable
${\mathcal N}=2_B$ superconformal boundary condition.
On the other side of the wall, $Q$ is irrelevant and vanishes in the 
infra-red limit. Then the brane splits into two
${\mathcal N}=2_B$ superconformal boundary conditions, one corresponding to 
${\mathcal E}_1$ and another corresponding to ${\mathcal E}_2$.
Note that in the latter case the above brane and the brane 
${\mathcal E}_1[1]\oplus{\mathcal E}_2$ are not D-isomorphic to each other
but still
flow to the same superconformal boundary interaction.

Other quantities that show dependence on $\moduli_K$
are the overlap of the boundary state 
with the R-R ground states
$
\langle i|{\mathcal B}\rangle.
$
These are defined as the path integrals over an A-twisted semi-infinite
cigar $\surface$
with an operator corresponding to the ground state $\langle i|$
inserted at the tip, see\cite{HIV}. This quantity does not depend on 
boundary D-term deformations as they would
insert an ${\mathcal N}=2_B$ D-term at the boundary which
is annihilated by the supersymmetric ground state:
$$
\delta\langle i|{\mathcal B}\rangle=
\langle i|\int_{\partial\surface}{\bf Q}{\bf Q}^{\dag}(\cdots)
|{\mathcal B}\rangle=0.
$$
On the other hand, as argued in \cite{OOY,HIV}, they do depend on
the twisted chiral parameters and satisfy a certain system of
differential equations.
The overlaps are called the generalized central charges 
and play important roles in the study of D-brane stability.

When the infra-red limits of the $(2,2)$ quantum field theories 
are used as the backgrounds for string compactification, 
the family of boundary interactions
obeying the rule (\ref{deform}) plays an important r\^ole
in spacetime physics.
Open string states for such families of boundary interactions
define open string fields that can be used everywhere
on the region of $\moduli_K$ under consideration.
This provides us with a basis to study spacetime D-term potentials
as a function of those open string field variables as well as
of closed string fields associated to $\moduli_K$ parameters.

An ideal framework to study the above issues is provided by 
linear sigma models.
They are $(2,2)$ supersymmetric gauge theories in $1+1$ dimensions
defined over the moduli space $\moduli_K\times\moduli_C$,
which includes large volume limits, Landau-Ginzburg orbifold points
as well as regions in between where 
neither a geometrical nor a Landau-Ginzburg
interpretation is totally absent.
It uses a single set of field variables with a fixed
supersymmetry transformation rule
and the dependence on the $\moduli_K\times\moduli_C$ 
moduli parameters appears only 
in the (twisted) F-term interactions of the action.
In this paper, we study parallel families of boundary
interactions using linear sigma models.
We will in fact encounter a sharp problem associated with
brane-antibrane annihilation, 
and will find a rather surprizing solution.
Also, our construction provides a starting point for the study
of D-brane stability and spacetime D-term potentials in intermediate
regimes of $\moduli_K$ where no useful description of the low energy
theory is available.

\section{Linear Sigma Models}
\label{sec:LSM}

In this section, we review the basic aspects of
$(2,2)$ supersymmetric linear sigma models in $1+1$ dimensions
\cite{phases}.
The main purpose is to fix notations and to introduce a class of
examples that will be used in this paper.
We will also obtain a new result on general multiparameter models
(Section~\ref{subsec:LSMdel}) that will play an important r\^ole.

\subsection{The Lagrangian}
\label{subsec:LSMlagrangian}

Let us consider a $2d$ $(2,2)$ supersymmetric
gauge theory with a compact Abelian gauge group
$T\cong U(1)_1\times U(1)_2\times\cdots\times U(1)_k$ 
and matter chiral superfields $\Phi=(\Phi_1,...,\Phi_N)$,
where $\Phi_i$ has charge $Q_i^a$ with respect to the $a^{\rm th}$
gauge group $U(1)_a$ ($a=1,...,k$ is the gauge index and $i=1,...,N$
is the `flavor' index).
We denote  the vector superfield for $U(1)_a$ by $V_a$, and
its curvature by
$\Sigma_a=\bD_+D_-V_a$ (a twisted chiral superfield).
The Lagrangian takes the following form
\beqa
{\mathcal L}&=&\int\dd^4\theta
\left(-{1\over 2}\sum_{a,b=1}^k(e^{-2})^{ab}
\overline{\Sigma}_a\Sigma_b
+\sum_{i=1}^N
\bPhi_i\e^{Q_i\cdot V}\Phi_i\right)
\nn
\\
&&
+{\rm Re}\int\dd^2\widetilde{\theta}
\left(-\sum_{a=1}^kt^a\Sigma_a\right)
+{\rm Re}\int \dd^2\theta \,W(\Phi)
\label{L}
\eeqa
The first term on the right hand side is the gauge kinetic term
where $e$ is the gauge coupling constant.
The second term
 is the matter kinetic term with
the minimal coupling to the gauge fields, where 
$Q_i\cdot V$ is a short hand notation for $\sum_{a=1}^kQ_i^aV_a$.
The third is a twisted superpotential term,
 where 
\beq
t^a=r^a-i\theta^a
\label{FItheta}
\eeq
is a complex combination
of the Fayet-Iliopoulos (FI) parameter $r^a$ and the theta angle
$\theta^a$ for the $a^{\rm th}$ gauge group $U(1)_a$.
The last term exists if there is a gauge invariant
holomorphic polynomial $W(\Phi)$ of $\Phi_1,...,\Phi_N$,
the superpotential.

Let us write down the Lagrangian in terms of the component fields.
We recall that a vector multiplet consists of a gauge field
$v_{\mu}$, a complex scalar $\sigma$, a Dirac fermion $\lambda_{\pm}$
and a real auxilary field $D$. A chiral multiplet consists of
a complex scalar $\phi$, a Dirac fermion $\psi_{\pm}$
and a complex auxiliary field $F$.
The component expressions for
the gauge kinetic term, 
the matter kinetic term,
and the twisted superpotential term
are given below 
\beqa
{\mathcal L}_{\rm g}&=&\int\dd^4\theta\left(-{1\over 2e^2}
\overline{\Sigma}\Sigma\right)+\mbox{total derivative}\nn\\
&=&
{1\over 2e^2}\,\left[
|\partial_0\sigma|^2-|\partial_{1}\sigma|^2
+i\blambda_-(\lrd_{\!\! 0}+\lrd_{\!\! 1})\lambda_-
+i\blambda_+(\lrd_{\!\! 0}-\lrd_{\!\! 1})\lambda_+
+v_{01}^2+D^2
\right],\nn\\
\label{Lg}\\
{\mathcal L}_{\rm m}&=&
\int\dd^4\theta\,\,\bPhi\e^V\Phi
+\mbox{total derivative}\nn\\
&=&
|D_0\phi|^2-| D_1\phi|^2
+i\bpsi_-(\lrD_{\!\! 0}+\lrD_{\!\! 1})\psi_-
+i\bpsi_+(\lrD_{\!\! 0}-\lrD_{\!\! 1})\psi_+
+\bphi D\phi+|F|^2
\nn\\
&&\,
-|\sigma \phi|^2-\bpsi_-\sigma\psi_+-\bpsi_+\bsigma\psi_-
-i\bphi\lambda_-\psi_++i\bphi\lambda_+\psi_-+i\bpsi_+\blambda_-\phi
-i\bpsi_-\blambda_+\phi,
\nn\\
\label{Lm}\\
{\mathcal L}_{{\rm FI}\,\theta}&=&
{\rm Re}\int\dd^2\widetilde{\theta}\Bigl(-t\Sigma\Bigr)
=-rD+\theta v_{01}.
\label{LFI}
\eeqa
Only the special case of $T=U(1)$ and with
just one charge $1$ matter field is presented, 
since the generalization is obvious.
The superpotential term is
\beqa
{\mathcal L}_W&=&{\rm Re}\int \dd^2\theta \,W(\Phi)
={\rm Re}\left[\,\,\sum_{i=1}^NF_i{\partial W\over\partial\phi_i}(\phi)
-\sum_{i,j=1}^N{\partial^2W\over\partial\phi_i\partial\phi_j}(\phi)
\psi_+^i\psi_-^j\,\,\right].
\label{LW}
\eeqa
In the above expressions, $\lrd_{\!\!\mu}$ and $\lrD_{\!\!\mu}$ are defined
as
\beq
\psi_1\lrd_{\!\!\mu}\psi_2:={1\over 2}\psi_1(\partial_{\mu}\psi_2)
-{1\over 2}(\partial_{\mu}\psi_1)\psi_2.
\eeq
If the worldsheet has no boundary, which is the case
within this section,
a total derivative can be ignored
and hence there is no need to distinguish
$\psi_1(\partial_{\mu}\psi_2)$ and $-(\partial_{\mu}\psi_1)\psi_2$.
However, later in this paper, we will consider 
worldsheets {\it with} boundary. Then a total derivative
is non-zero in general and the distinction is important.
The above choice of Lagrangian
is the one that will be used throughout this paper.

\subsection{Phases}
\label{subsec:LSMphases}

The classical potential for the scalar fields $\phi_i$ and $\sigma_a$
is obtained after integrating
out the auxiliary fields $D_a$ and $F_i$:
\beq
U=\sum_{i=1}^N\left|\sum_{a=1}^kQ^a_i\sigma_a\phi_i\right|^2
+{e^2\over 2}\sum_{a=1}^k\left(\sum_{i=1}^NQ^a_i|\phi_i|^2-r^a\right)^2
+\sum_{i=1}^N\left|{\partial W\over\partial\phi_i}(\phi)\right|^2.
\label{U}
\eeq
Here we assume $e^2_{ab}=\delta_{ab}e^2$ for simplicity.
One obtains some idea of the low energy theory
by looking at the vacuum locus,
$U=0$. It depends very much on the value of the FI parameters $r^a$
which enter into the middle term, the D-term potential.
If $r=(r^1,...,r^k)$ is in a certain domain,
the D-term equations
\beq
\sum_{i=1}^NQ^a_i|\phi_i|^2-r^a=0\qquad \forall a=1,\ldots, k,
\label{D}
\eeq
may require that the $\phi_i$ are non-zero and the matrix
$M^{ab}=\sum_{i=1}^NQ_i^aQ_i^b|\phi_i|^2$ has maximal rank $k$.
This means that the gauge group $T$ is completely
broken, or broken to a finite subgroup.
In particular, $U=0$ requires all $\sigma_a$ to be zero
by the first term in (\ref{U}).
However, this may fail at special values of $r=(r^1,...,r^k)$.
Also, it is possible that the equation $U=0$ has no solution
in some domain.
Thus, the space $\R^k_{\rm FI}$ spanned by
FI parameters is divided into a finite number of chambers.
There are loci in which
the rank of $M^{ab}$ for generic solutions to
(\ref{D}) are less than $k$.
They are parts of linear hypersurfaces and form
walls that divide $\R^k_{\rm FI}$
into chambers.
The chambers that admit solutions to the vacuum equation $U=0$
are called the {\it phases} and the walls separating them
are called the {\it phase boundaries}.
At a point $r$ in the interior of each phase, the continuous part
of the gauge group $T$ is broken everywhere on the vacuum locus $U=0$.

A {\it geometric phase} is a phase in which $T$ is completely broken at
any solution of $U=0$
and all modes transverse to $U=0$ are massive. 
In that case, the low energy theory is a non-linear sigma model
whose target space is the quotient $(U=0)/T$.
If the model has zero superpotential $W=0$, the space
$(U=0)/T$ is the symplectic quotient of $\C^N$ by $T$ with the moment map
equation (\ref{D}). As a complex manifold, it is the quotient
by the complexified gauge group $T_{\C}$,
\beq
X_r=(\C^N-\Delta_r)/T_{\C},
\label{Xr}
\eeq
where $\Delta_r$ 
consists of points whose $T_{\C}$ orbits do not pass through
solutions to (\ref{D}).
$\Delta_r$ is a union of linear subspaces of $\C^N$
 and depends only on the phase that
$r$ belongs to. We shall call it
the {\it deleted set} of the respective phase.
$X_r$ is a so called {\it toric manifold}.
If the superpotential is non-trivival,
the vacuum locus is a submanifold $M_r$ of this toric manifold $X_r$,
which is defined by the
F-term equations
\beq
{\partial W\over \partial \phi_i}=0\qquad \forall i=1,\ldots, N.
\eeq
In the $e^{-2}\to 0$ limit, the gauge field $v_{\mu}$ and the scalar field
$\sigma$ are expressed in terms of the matter fields as
\beqa
  \label{gaugeconn}
v_{a0}
&=&\sum_{b,i}M^{-1}_{ab}Q^b_i\left(
i\bphi_i\lrd_{\!\!0}\phi_i
+{1\over 2}(\bpsi_{i+}\psi_{i+}+\bpsi_{i-}\psi_{i-})\right),
\\
v_{a1}
&=&\sum_{b,i}M^{-1}_{ab}Q^b_i\left(
i\bphi_i\lrd_{\!\!1}\phi_i
+{1\over 2}(\bpsi_{i+}\psi_{i+}-\bpsi_{i-}\psi_{i-})\right),
\label{gaugeconn1}
\\
  \label{sigmaconn}
\sigma_a&=&-\sum_{b,i}M^{-1}_{ab}Q^b_i\bpsi_{i+}\psi_{i-}.
\eeqa
The
gauge field is the pull-back of the connection of the complex line
bundle ${\mathcal O}(e_a)$ over $X_r$ (or its restriction to
$M_r$) associated with the charge 1 representation for
the $a^{\rm th}$ $U(1)$.

If $W=0$, all phases are almost geometric in the sense that
all modes transverse to $U=0$ are massive and
the quotient $X_r=(U=0)/T$ has at most orbifold singularities.
When it can be realized as a global orbifold by an unbroken gauge group, 
$X_r=X'_r/\Gamma$,
the low energy theory is the orbifold theory 
in the standard sense \cite{DHVW}:
the $\Gamma$-gauged sigma model on $X_r'$.
Otherwise, no convenient description of the low energy theory 
is available today.
See for example \cite{SharpeStack} and references therein.
A singular $X_r$ makes a perfect sense as an {\it algebraic variety},
but that has no useful suggestion to the description of the theory.
It makes sense also as something called a {\it quotient stack},
and that seems to carry convenient structures, especially
when we consider D-branes. (See Section~\ref{sec:math}.)
In this paper, somewhat loosely,
we simply refer to the low energy
theory as ``the non-linear sigma model on
the toric variety $X_r$'',
having this subtlety in mind.

If the superpotential $W$ is non-trivial,
there are various phases in which some of the transverse modes
to $U=0$ are massless. The extreme cases are the so called
{\it Landau-Ginzburg phases}.
A Landau-Ginzburg phase is a phase in which
the vacuum locus $(U=0)/T$ is one point and all modes
transverse to the $T_{\C}$ orbit do not acquire mass from the
D-term potential.
In the limit where $r$ is scaled up to infinity,
the modes tangent to the $T_{\C}$ orbit decouple
and the theory reduces to the Landau-Ginzburg model
for the transverse modes, possibly with a residual discrete gauge
symmetry (Landau-Ginzburg orbifold).

\subsection*{\it RG Flows And Calabi-Yau Conditions}

Under the renormalization group (RG)
the FI parameters $r^1,...,r^k$ flow as
$r^a(\mu)=r^a(\mu')+\sum_{i=1}^NQ^a_i\log(\mu/\mu')$, so that
$r(\mu)$ runs along a straight line in $\R^k_{\rm FI}$.
In general, this induces a flow between different phases
or domains without solution to the vacuum equation $U=0$.
However, under the condition
\beq
\sum_{i=1}^NQ^a_i=0\quad\forall a=1,\ldots, k,
\label{conformal}
\eeq
the FI parameters do not run and are genuine parameters of the theory.
A related effect is the axial anomaly.
The classical action (\ref{L}) has axial $U(1)$ R-symmetry but this is
anomalously broken if the condition (\ref{conformal}) is violated.
As a consequence, the shift of theta angles,
$\theta^a\to\theta^a+\sum_{i=1}^NQ^a_i\Delta$,
is physically irrelevant as that can be absorbed by
a field redefinition using an axial rotation.
If the condition (\ref{conformal}) is indeed met,
there is no such anomaly and
all the $k$ theta angles $\theta^a$ are genuine parameters of the theory.
If the superpotential is quasi-homogeneous, namely if
it obeys
\beq
W(\lambda^{R_1}\Phi_1,\ldots,\lambda^{R_N}\Phi_N)
=\lambda^2W(\Phi_1,\ldots,\Phi_N),
\label{qh}
\eeq
for certain $R_1,...,R_N$ (called the R-charges of the fields
$\Phi_1,...,\Phi_N$),  
there is also a vector $U(1)$ R-symmetry.

If the two conditions (\ref{conformal}) and (\ref{qh})
are met, we have both axial and vector $U(1)$ R-symmetries.
They are expected to become a part of the superconformal symmetry
of the non-trivial infra-red fixed point of the RG flow.
In such a case, we have a family of superconformal field theories
parametrized by the FI-theta parameters $t^a=r^a-i\theta^a$
as well as the parameters that enter into the superpotential
$W(\Phi)$. In many cases these are the entire set of exactly marginal
parameters, but in many other cases there are extra parameters.

\subsection{Singularity}
\label{subsec:LSMsing}

Let us assume that the Calabi-Yau conditions (\ref{conformal}) and (\ref{qh})
are met and hence all $t^a$'s are genuine parameters of the theory.

On a phase boundary, a non-compact Coulomb branch emerges ---
the vacuum locus $U=0$ includes a point at which
there is an unbroken continuous subgroup of $T$
and the corresponding $\sigma_a$ is unconstrained.
This implies a singularity of the theory.
The story must be modified in the quantum theory
since it depends also on the theta angles $\theta^a$.
Actual existence of the Coulomb branch can be examined
by computing the quantum ground state energy at large values of
$\sigma$'s.
The result is
\beq
U_{\rm eff}={1\over 2}\sum_{a,b=1}^ke^2_{ab}(\sigma)
{\partial \overline{\widetilde{W}}_{\rm eff}\over\partial\bsigma_a}
{\partial\widetilde{W}_{\rm eff}\over\partial\sigma_b}.
\label{Ueff}
\eeq
$e^2_{ab}(\sigma)$ are the effective gauge coupling constants. They
approach their classical values as $|\sigma|\to \infty$.
$\widetilde{W}_{\rm eff}(\sigma)$ is the effective
twisted superpotential which is obtained by integrating out
the charged chiral multiplet fields. Its first derivatives are given by
\beq
{\partial\widetilde{W}_{\rm eff}\over\partial\sigma_a}(\sigma)
=-t^a
-\sum_{i=1}^NQ^a_i\log\left(\sum_{b=1}^kQ^b_i\sigma_b\right).
\label{deri}
\eeq
The imaginary part of these are the effective background electric fields
and enter into (\ref{Ueff}) as the electrostatic energy
\cite{Coleman}. To be precise, the potential (\ref{Ueff})
changes if $\theta^a$ is shifted by $2\pi$, and the lowest among all
is the actual ground state energy. In particular, the potential is zero if
the derivatives (\ref{deri})
vanish modulo $2\pi i$ times integers \cite{phases,Coleman}.
Namely, the vacuum equation is
\beq
\prod_{i=1}^N\left(\sum_{b=1}^kQ^b_i\sigma_b\right)^{Q_i^a}=\e^{-t^a}.
\label{Coulomb}
\eeq
Under the condition (\ref{conformal}), this equation is invariant under
the uniform rescaling of $\sigma_a$'s, so that existence of one
solution means existence of a non-compact Coulomb branch.
Eliminating $\sigma_a$'s we obtain an equation for
$\e^{-t^a}$'s that defines the locus where there is a quantum Coulomb branch.
The theory is singular there since the wavefunctions spread
over the Coulomb branch and are not normalizable. 
In general, there are additional singular loci coming from
mixed Coulomb-Higgs branches \cite{MP}.

Let us denote the set of singular points by $\mathfrak{S}$.
It is a union of hypersurfaces in the set $(\C^{\times})^k$
of all FI-theta parameters $\{(\e^{t^1},...,\e^{t^k})\}$.
The K\"ahler moduli space is the complement
$$
\mathfrak{M}_K=(\C^{\times})^k\setminus \mathfrak{S}.
$$
In order to get some idea of how $\moduli_K$ looks like,
it is useful to introduce two projections of $(\C^{\times})^k$ ---
the projection to the FI parameters (Log)
and the projection to the theta parameters (Arg):
\beq
\mbox{
\setlength{\unitlength}{0.7mm}
\begin{picture}(70,35)(-25,1)
\put(0,29){$(\C^{\times})^k$}
\put(-18,17){Log}
\put(20,17){Arg}
\put(3,25){\vector(-1,-1){18}}
\put(10,25){\vector(1,-1){18}}
\put(-24,0){$\R^k_{\rm FI}$}
\put(26,0){$(S^1)^k_{\theta}$}
\end{picture}
}
\eeq
The image of the singular loci $\mathfrak{S}$ under the map Log 
is called the Amoeba 
of $\mathfrak{S}$ \cite{GKZ}.
It is a domain on $\R^k_{\rm FI}$ with tentacles which asymptote to the
classical phase boundaries at large $|r|$, possibly shifted by some
finite amount.  The image of $\mathfrak{S}$ under the
map Arg is called the co-Amoeba
or the Alga of $\mathfrak{S}$ \cite{Passare,Kris}.
We will see in some examples that the Alga of
$\mathfrak{S}$ has a non-empty complement,
which has an important consequence for
D-brane transport at the center of the
moduli space $\moduli_K$.

\bigskip

\subsection{Examples}
\label{subsec:LSMexamples}

In what follows, we introduce some examples
which will accompany us throughout this paper.
We place emphasis on the
phase structure and the deleted sets $\Delta_r$.
In Example (C), we introduce a method to find the deleted sets
which turns out to be useful for describing coherent sheaves
on toric varieties.
In Examples (A) and (C), one can consider a compact theory with
a non-trivial superpotential $W$,
but one may also consider a non-compact theory with $W=0$.

\subsection*{(A) Calabi-Yau Hypersurface In $\CP^{N-1}$}

\begin{figure}[htb]
\centerline{\includegraphics{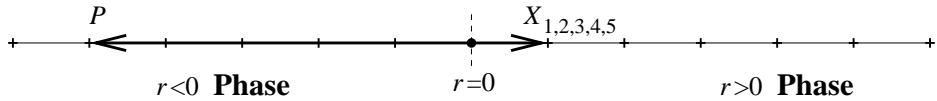}}
\caption{The Phases of the model (A) for $N=5$.}
\label{quinticA}
\end{figure}

The first example has gauge group $U(1)$ and $(N+1)$ fields
$P,X_1,\ldots,X_N$ with charge $-N,1,\ldots,1$.
We consider the superpotential
$W=PG(X_1,\ldots,X_N)$ where $G(X)$ is a homogeneous
polynomial of degree $N$.
There are two phases as shown in Fig.~\ref{quinticA}.
We plot the lattice of gauge charges (called
the Picard lattice) inside $\R^k_{\rm FI}$, 
for a reason that will become clear later.
The deleted sets are
\beqa
&&\Delta_+=\{x_1=x_2=\cdots=x_N=0\}\qquad\mbox{for $r>0$},
\nn\\
&&\Delta_-=\{p=0\}\qquad \mbox{for $r<0$}.
\nn
\eeqa
The exact location of the singularity is
$\e^{t}=(-N)^N$.
In the $r\gg 0$ phase, the theory 
reduces to the non-linear sigma model on
the Calabi-Yau hypersurface $X_G=\{G=0\}$ in the projective space
$\CP^{N-1}$.
In the $r\ll 0$ phase, the vacuum manifold $(U=0)/U(1)$
is a one point with $p\ne 0$ and $x=0$ which breaks the gauge group
to $\Gamma=\{\omega\in U(1)|\omega^N=1\}\cong\Z_N$.
All the $X_i$'s are massless there.
At $r\to -\infty$, the theory reduces to the Landau-Ginzburg
orbifold of the $N$ variables $X_1,\ldots, X_N$
with the superpotential $W=G(X)$ and the orbifold group $\Gamma$.
If there is no superpotential, $r\gg 0$ is the geometric phase on
the total space of ${\mathcal O}(-N)$ (the canonical bundle)
over $\CP^{N-1}$, while the $r\to -\infty$ limit yields the free orbifold
$\C^N/\Gamma$.

\subsection*{(B) The Resolved Conifold, 
	${\mathcal O}(-1)\oplus {\mathcal O}(-1)\to \CP^1$}

\begin{figure}[ht]
\centerline{\includegraphics{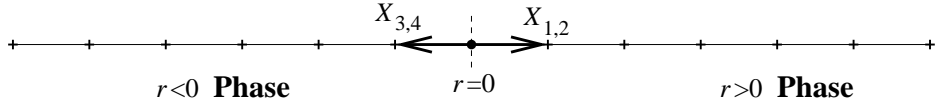}}
\caption{The Phases of the model (B).}
\label{conifoldA}
\end{figure}
The second example has $U(1)$ gauge group and four fields 
$X_1,X_2,X_3,X_4$ with charge
$1,1,-1,-1$. There are two phases as shown in Fig.~\ref{conifoldA}.
The deleted sets are
\beqa
&&\Delta_+=\{x_1=x_2=0\}\qquad\mbox{for $r>0$},
\nn\\
&&\Delta_-=\{x_3=x_4=0\}\qquad \mbox{for $r<0$}.
\nn
\eeqa
The exact location of the singularity is $\e^t=1$.
The $r\gg 0$ phase is the non-linear sigma model 
on the total space of ${\mathcal O}(-1)\oplus{\mathcal O}(-1)$
over $\CP^1$ (a resolved conifold), while the $r\ll 0$ phase 
is that of another resolved conifold.
The difference is that in the $r\gg 0$ pase,
$x_1,x_2$ span the base $\CP^1$ and $x_3,x_4$ span fibres,
while in the $r\ll 0$ phase $\CP^1$ is spanned by $x_3,x_4$
and the fibres are spanned by $x_1,x_2$.
They are both geometric phases and the transition between them
is called the {\it flop}.

\subsection*{(C) A two parameter model}

The third example has two K\"ahler parameters.
It has $U(1)^2$ gauge group and
seven charged fields as shown below:
$$
\begin{array}{cccccccc}
      & P & X_1 & X_2 & X_3 & X_4 & X_5 & X_6\\
U(1)_1&-4 &  0  &  0  &  1  &  1  &  1  &  1 \\
U(1)_2& 0 &  1  &  1  &  0  &  0  &  0  & -2
\end{array}
$$

\begin{figure}[htb]
\centerline{\includegraphics{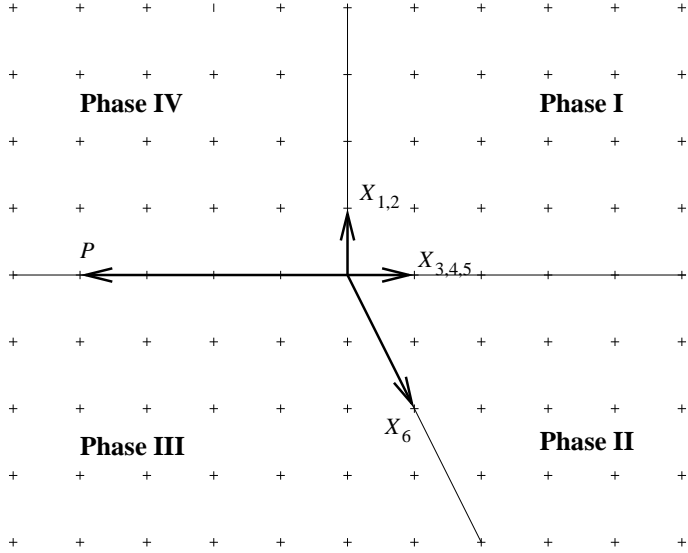}}
\caption{The Phases of the two parameter model (C).}
\label{twoparaA}
\end{figure}

Using this example, we introduce a method to find
the phases and the deleted sets,
without writing down the D-term equations, but by looking
at the charge lattice embedded into $\R^{k}_{\rm FI}$.
Not only facilitating the problem with a geometric picture,
this method turns out to be useful also when we describe coherent sheaves
on toric varieties (see Section~\ref{sec:math}).

First of all, the phase boundaries are domains of the hypersurfaces
which are in positive linear spans of
$(k-1)$ charge vectors of some of the fields.
In the present example there are four boundaries correspnding to four
charge vectors.
Accordingly, there are four phases as shown in Fig.~\ref{twoparaA}.

Next we describe how to find the deleted set, say in Phase I.
Let us take any hyperplane
through the origin such that Phase I is in one of the two halves
of $\R^k_{\rm FI}$ separated by the plane.
Then we forbid the common zero of
those $x_i$'s that are on the same side as Phase I.
This is required by the D-term equation
for the $U(1)$ subgroup with respect to which
the charge vectors in that hyperplane is neutral.
For the vertical hyperplane in Fig.~\ref{twoparaA},
the charge vectors on the same side as
 Phase I are those of $X_3,X_4,X_5,X_6$. Thus $\{x_3=x_4=x_5=x_6=0\}$
is deleted. (The relevant subgroup is $U(1)_1$ and the equation
is $|x_3|^2+|x_4|^2+|x_5|^2+|x_6|^2-4|p|^2=r^1$. 
Since $r^1$ is positive in Phase I,
$x_3=x_4=x_5=x_6=0$ is indeed forbidden.)
By other choices of the hyperplane, we find that we also
delete $\{x_1=\ldots=x_6=0\}$ and $\{x_1=\ldots=x_5=0\}$ as well as
$\{x_1=x_2=0\}$. But the latter subspace 
contains the previous two, so that the
deleted set for Phase I is the union of 
$\{x_1=x_2=0\}$ and $\{x_3=x_4=x_5=x_6=0\}$.
Repeating this procedure provides the deleted sets for the other
phases as well: 
\beqa
&&\Delta_{\rm I}=\{x_1=x_2=0\}\cup\{x_3=x_4=x_5=x_6=0\},
\nn\\
&&\Delta_{\rm II}=\{x_1=x_2=x_3=x_4=x_5=0\}\cup\{x_6=0\},
\nn\\
&&\Delta_{\rm III}=\{p=0\}\cup\{x_6=0\},
\nn\\
&&\Delta_{\rm IV}=\{p=0\}\cup\{x_1=x_2=0\}.
\nn
\eeqa

\begin{figure}[htb]
\centerline{\includegraphics{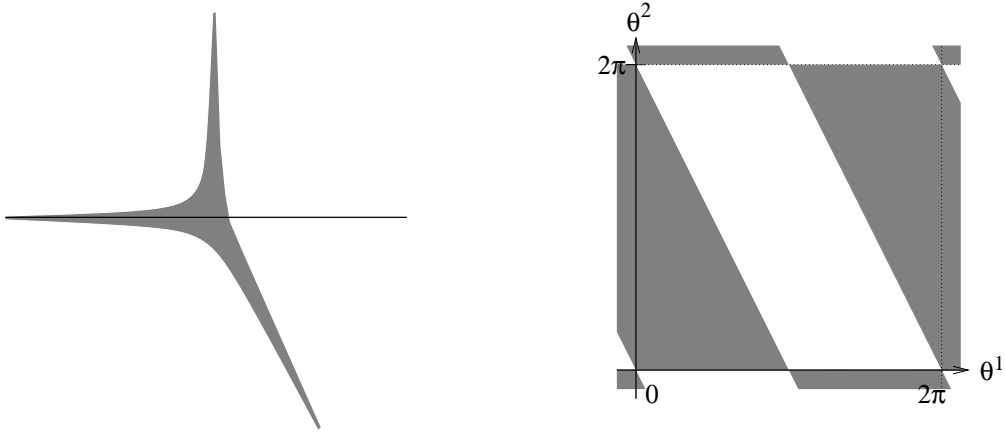}}
\caption{The Amoeba (Left) and Alga (Right) of $\mathfrak{S}$
of the two parameter model (C).}
\label{alamtwo}
\end{figure}

The exact location of the singularity $\mathfrak{S}$ is
the union of
$$
\e^{-t^1}=4^{-4}(1-2u),\qquad
\e^{-t^2}=\frac{u^2}{(1-2u)^2},
$$
where $u:=\sigma_2/\sigma_1$, and
$$
\e^{-t^2}=2^{-2}.
$$
The latter component comes from the mixed Higgs-Coulomb branch, where
$\sigma_2$ as well as fields neutral to
$U(1)_2$ are large.
The Amoeba and Alga of the singular locus are
depicted in Fig.~\ref{alamtwo}. 
The Alga is found by looking at the behaviour of
$\theta^1$ and $\theta^2$ near the special points,
$u=0, {1\over 2}$ and $\infty$, that correspond to the tentacles
of the Amoeba, i.e., to the phase boundaries. (We learned this technique
from M.~Passare.)
Note that the Alga has a non-trivial
complement. 

Without the superpotential, the low energy theory is the sigma model
on a non-compact Calabi-Yau variety $X_r$
with or without orbifold singularity.
In Phase I, the unborken subgroup is everywhere trivial $\{1\}$
and the space $X_r$ is the total space of 
a holomorphic line bundle on a smooth compact manifold $B_r$.
The base manifold $B_r$ is the compact toric manifold
corresponding to the model without
the field $P$, and the line bundle is ${\mathcal O}(-4,0)$.
In Phase II, the unbroken subgroup is $\{(1,\pm 1)\}\cong \Z_2$
at $x_1=x_2=0$, and the space $X_r$ is the total space of
the line bundle ${\mathcal O}(-8)$ over the weighted projective space
${\rm W}\CP^4_{[11222]}$.
In Phase III, the unbroken subgroup is $\{(\e^{2\pi i/4},\e^{2\pi i/8})\}
\cong \Z_8$ at $x_1=\cdots=x_5=0$,
and the space $X_r$ is the orbifold $\C^5/\Z_8$.
In Phase IV, the unbroken subgroup is $\{(\e^{2\pi i/4},1)\}\cong \Z_4$,
and the space $X_r$ is the orbifold of ${\mathcal O}_{\CP^1}(-2)\times\C^3$
by $\Z_4$.

We can also consider a theory with superpotential
$$
W=PG(X_1,...,X_6)
$$
where $G(X)$ is a homogeneous polynomial of bidegree $(4,0)$, such as
$G(X)=X_1^8X_6^4+X_2^8X_6^4+X_3^4+X_4^4+X_5^4$. 
In Phase I and Phase II the low energy theory is
the non-linear sigma model on the hypersurface
$\{G\!=\!0\}$ of $B_r$ and ${\rm W}\CP^4_{[11222]}$ respectively.
In Phase III it is the $\Z_8$ orbifold of the Landau-Ginzburg model
with superpotential $W=G(X_1,...,X_5,1)$.
In Phase IV it is a non-linear Landau-Ginburg orbifold.

\subsection*{(D) Resolution Of $A_{N-1}$ Singularity}

As the final example, we take the $U(1)^{N-1}$ gauge theory
with the following matter content:
\beq
\begin{array}{ccccccccccc}
         &X_1&X_2&X_3&X_4&\cdots&\!\!X_{N-2}\!\!&\!\!X_{N-1}\!\!&\!\!
X_N\!\!&\!\!X_{N+1}\\
U(1)_1   & 1 & -2& 1 & 0 &\cdots&   0   & 0     & 0 &  0\\
U(1)_2   & 0 & 1 & -2& 1 &\cdots&   0   &  0    & 0 &  0\\
\vdots   &   &   &   \ddots&\ddots&\ddots& &&  & &   \\
\vdots   &   &   &   &  &\ddots&\ddots&\ddots &  &   \\
U(1)_{N-2}&0 & 0 & 0 & 0 &\cdots&  1  &  -2   & 1 & 0 \\
U(1)_{N-1}&0 & 0 & 0 & 0 &\cdots&  0  &  1   &  -2& 1
\end{array}
\label{Acontent}
\eeq
This theory describes the $A_{N-1}$ singularity and its various
resolutions. The case $N=2$ is identical to the case $N=2$
in Example (A).

\begin{figure}[htb]
\psfrag{c1}{$C_1$}
\psfrag{c2}{$C_2$}
\psfrag{c3}{$C_3$}
\psfrag{c-2}{$C_{N-2}$}
\psfrag{c-1}{$C_{N-1}$}
\centerline{\includegraphics{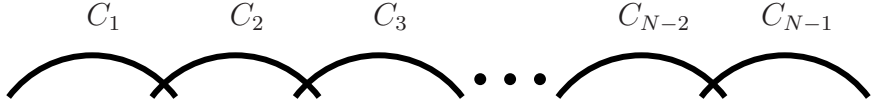}}
\caption{The exceptional divisor of the resolved $A_{N-1}$ singularity.}
\label{fig:Ediv}
\end{figure}

There is a phase, {\it the orbifold phase},
where the deleted set $\Delta_{\rm orb}$
is the union of $\{x_i=0\}$ for $i=2,...,N$.
The non-zero values of these $x_i$'s breaks the gauge group 
to a discrete subgroup $\{(\omega,\omega^2,...,\omega^{N-1}); \omega^N=1\}$
which is isomorphic to $\Z_N$. In the limit $r^a\to-\infty$,
the low energy theory is the free orbifold 
$X_{\rm orb}=\C^2/\Z_N$ 
where $\C^2$ is spanned by $x_1,x_{N+1}$ and the group $\Z_N$
acts on it by $(x_1,x_{N+1})\to (\omega x_1,\omega^{-1}x_{N+1})$.
If we introduce the orbifold invariants by
$x=x_1^N$, $y=x_{N+1}^N$ and $z=x_1x_{N+1}$, they satisfy
the equation
$$
xy=z^N.
$$
The singularity at the origin is called the $A_{N-1}$ singularity.
The opposite limit is {\it the large volume phase} where all $r^a$ are
positive. The deleted set $\Delta_{\rm res}$ is a union of
$\{x_i=x_j=0\}$ for all $i,j$ such that $|i-j|\geq 2$.
The low energy theory is the sigma model on the full resolution
$X_{\rm res}$ of the $A_{N-1}$ singularity.
The projection map $\pi:X_{\rm res}\to X_{\rm orb}$ is described by
$x=x_1^Nx_2^{N-1}x_3^{N-3}\cdots x_N$,
$y=x_2x_3^2\cdots x_N^{N-1}x_{N+1}^N$,
$z=x_1x_2\cdots x_{N+1}$.
The pre-image of the singular point $x=y=z=0$ is the exceptional divisor
which is a chain of 2-spheres $C_1,\ldots, C_{N-1}$, where
$C_i$ is defined by the equation $x_{N-i+1}=0$.
They intersect according to the $A_{N-1}$ Dynkin diagram as in 
Figure~\ref{fig:Ediv}.

In between, there are other phases corresponding to partial resolutions. 
A convenient way to label the phases is to asign $0$ or $1$
to each node of the $A_{N-1}$ Dynkin diagram,
depending on whether the node is resolved or not.
For example, the orbifold phase is $(11\cdots 1)$ and the fully 
resolved phase is $(00\cdots 0)$. The phase where only $C_1$ is resolved 
is
$(011\cdots 1)$
and the phase where only $C_2$ is unresolved is
$(010\cdots 0)$.
In total, there are $2^{N-1}$ phases. 
If there is a sequence of $m$ $1$'s in the label, that means
that there is an $A_{m}$ singularity in that phase.
For example, the phase $(1101)$ in the
$N=5$ model has one $A_2$ singularity and one $A_1$
singularity. 

A phase boundary corresponds to a blow down or blow up of one $C_i$.
In the above labelling system, that is the boundary between
the phases where the $i$-th node flips between $0$ and $1$.
Let us describe where it sits inside the space $\R^{N-1}_{\rm FI}$.
We know that it is a part of the hyperplane spanned by the charge vectors
of a subset of $(N-1)-1$ variables, but which subset?
It is the set of all $x_j$'s except 
$x_{N-i+1}$, $x_{N-i_++1}$ and $x_{N-i_-+1}$ where
$i_-<i$ and $i_+>i$ are the labels of the resolved nodes which are closest
to $i$ on the left and on the right. When there is no resolved node
on the left ({\it resp}. right) of $i$ we set $i_-=0$ ({\it resp}. $i_+=N$).
This shows that there is at least one phase boundary in the hyperplane
spanned by the charge vectors of any subset of $(N-2)$ 
charge vectors.

\begin{figure}[htb]
\centerline{\includegraphics{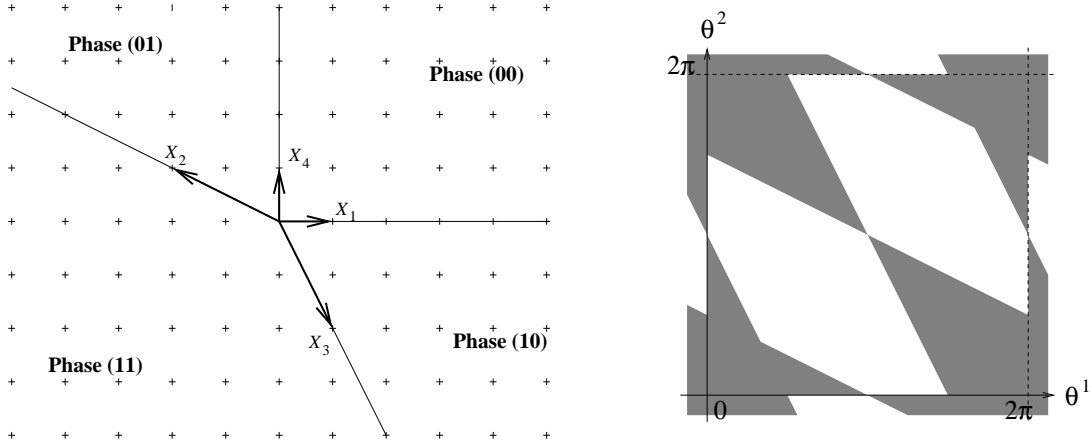}}
\caption{The phases (left) and the Alga of
$\mathfrak{S}$ (right) of the $A_2$ model.}
\label{fig:A2}
\end{figure}

For illustration, let us consider models with small $N$'s.
For the $A_2$ model, we can analyze the system very explicitly
as a two parameter model.
Figure~\ref{fig:A2} (left) shows the phases.
As we have learned in Example (C), we can read off the deleted set
at each phase:
\beqa
&&
\Delta_{(00)}=\{x_1=x_3=0\}\cup\{x_1=x_4=0\}\cup\{x_2=x_4\},
\nn\\
&&\Delta_{(01)}=\{x_2=0\}\cup\{x_1=x_4=0\},\nn\\
&&\Delta_{(10)}=\{x_3=0\}\cup\{x_1=x_4=0\},\nn\\
&&\Delta_{(11)}=\{x_2=0\}\cup\{x_3=0\}.\nn
\eeqa
That $x_2=0$ ({\it resp}. $x_3=0$) 
is not allowed means that $C_2$ ({\it resp}. $C_1$) is not resolved,
thus justifying the labelling of the phases. 
Note that each of the phase boundaries is spanned by the
charge vector of a particular variable, in accord with the general
description given above.
The singular locus is found in the by-now standard way.
Its Alga is shown in Figure~\ref{fig:A2} (right).

\begin{figure}[htb]
\centerline{\includegraphics{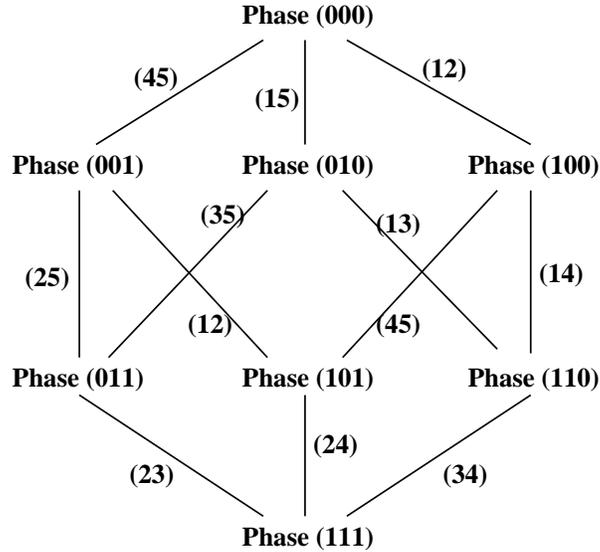}}
\caption{The phases and phase boundaries of the
$A_3$ model}
\label{fig:A3}
\end{figure}

For higher $N$ the analysis becomes increasingly complicated,
and it is not illuminating to draw the picture of phases 
on a two-dimensional sheet.
Figure~\ref{fig:A3} shows the labels of phases and their boundaries
in the $A_3$ model.
The label $(jk)$ of the phase boundary
means that it is a domain on the hyperplane
 spanned by the charge vectors of $x_j$ and $x_k$,
inside their positive span.
It is the positive span itself, unless it is on the hyperplane
spanned by $(12)$ and $(45)$.
The positive spans of $(12)$ and $(45)$ intersect on a half-line
which separates each of them into two phase boundaries.
That is why ``$(12)$'' and ``$(45)$'' appear twice
in Figure~\ref{fig:A3}.

\subsection{Phase Boundary And Deleted Sets}
\label{subsec:LSMdel}

To finish the review of
linear sigma models, we point out a relation between the deleted sets
of adjacent phases. This relation will play an important r\^ole
later in this paper.

Let us consider two phases, say, Phase I and Phase II, which are separated
by a phase boundary.
Let $T^u$ be the unbroken subgroup at that phase boundary, and we choose
basis elements $(e^1,...,e^k)$ of the Lie algebra of $T$ so that
the first element $e^1$ generates $T^u$. We choose its sign so that
$r^1$ is positive in Phase I and negative in Phase II.
We denote by $\Delta_+^{\rm I,II}$ the common zeroes of $\phi_i$'s
that are positively charged under the subgroup $T^u$, $Q^1_i>0$.
Similarly, we denote by $\Delta_-^{\rm I,II}$ the common zeroes
of $\phi_i$'s that are negatively charged, $Q_i^1<0$.
Obviously $\Delta_+^{\rm I,II}$ is in $\Delta_{\rm I}$,
and $\Delta_-^{\rm I,II}$ is in $\Delta_{\rm II}$.
We claim that
\beq
\begin{array}{l}
\Delta_{\rm I}=\Delta^{{\rm I,II}}_+\cup(\Delta_{\rm I}\cap \Delta_{\rm II}),
\\[0.2cm]
\Delta_{\rm II}=\Delta^{{\rm I,II}}_-\cup(\Delta_{\rm I}\cap \Delta_{\rm II}).
\end{array}
\label{importantre}
\eeq

It is straightforward to check such a relation in examples. Let us
take, say, Example (C), 
 and look at the boundary between Phase I and Phase IV.
The basis of the gauge group
is already chosen in the way just mentioned, and we find
$\Delta_+^{\rm I,IV}=\{x_3=x_4=x_5=x_6=0\}$ and 
$\Delta_-^{\rm I,IV}=\{p=0\}$.
On the other hand, we have
$\Delta_{\rm I}\cap\Delta_{\rm IV}=\{x_1=x_2=0\}$.
Indeed the relation
(\ref{importantre}) holds for Phase I and Phase IV.
It is very easy to check it for other phase boundaries.

In general, the relation (\ref{importantre}) can be proved as follows.
Let $\phi$ be an element of $\Delta_{\rm I}$ that is not in
$\Delta_{\rm II}$. 
This means that the $T_{\C}$ orbit of $\phi$ does not 
include a solution of the D-term equation
at any point in Phase I,
and also, at any point in Phase II there is some element $g\in T_{\C}$
such that $g\phi$ solves the D-term equation there.
Namely, the D-term image of the $T_{\C}$-orbit of $\phi$
includes the Phase II entirely but does not meet Phase I.
This D-term image is known to be convex in 
$\R^k_{\rm FI}$ \cite{Convexity}, and hence
is entirely on the same side as Phase II with respect to the
hyperplane spanned by the I-II boundary. 
In particular, we have 
$$
\sum_iQ_i^1|z^{Q^1_i}\phi_i|^2<0\qquad \forall z\in \C^{\times}.
$$
This is possible only if $\phi_i=0$ for each $i$ such that
$Q_i^1>0$. Namely, $\phi$ must be in the set
$\Delta_+^{\rm I,II}$. This shows $\Delta_{\rm I}\setminus
(\Delta_{\rm I}\cap\Delta_{\rm II})\subset \Delta_+^{\rm I,II}$,
which is equivalent to
$\Delta_I=\Delta_+^{\rm I,II}\cup(\Delta_{\rm I}\cap\Delta_{\rm II})$.
The proof of the other relation is similar.

\section{D-branes In Linear Sigma Models}
\label{sec:LSMbranes}

In this section, we write down boundary interactions in linear sigma models
that preserve ${\mathcal N}=2_B$ supersymmetry as well as
vector $U(1)$ R-symmetry.
The building block is the Wilson line brane that carries a definite
gauge charge. 
Interactions of Wilson lines can also be introduced just as in
the systems considered in Section~\ref{sec:branecat}.
In a system with vanishing superpotential, they are
given by complexes of Wilson line branes.
If the superpotential $W$ is non-vanishing, they are given 
by gauge invariant and homogeneous matrix factorizations of $W$.
We take a first look at the boundary conditions on the bulk fields, 
postponing
the full consideration to Section~\ref{sec:GRR}. 
We end the section by describing the chiral sector
in the theory with vanishing gauge coupling $e=0$.

\subsection{The Bulk Action And Boundary Counter Terms}
\label{subsec:LSMvariation}

${\mathcal N}=2_B$
supersymmetry variation of the bulk action
is a boundary term which is in general non-vanishing 
unless a boundary condition is used.
We would like to find a boundary counter term 
whose variation cancels it.
First, we focus on the gauge kinetic term,
matter kinetic term and the FI-theta term.
The counter term for the superpotential F-term
is written using boundary degrees of freedom and will be fully considered
in Section~\ref{subsec:compactbranes}.
The kinetic terms and FI-theta terms
are $(2,2)$ D-terms and twisted F-terms. 
As explained in Section~\ref{subsec:whatwedo},
they
are classified as D-terms with
respect to the ${\mathcal N}=2_B$ supersymmetry.
Thus, we should be able to
write the invariant action
in the form $\int\dd t\,{\bf Q}{\bf Q}^{\dag}(\cdots)$,
where
${\bf Q}=\overline{\bf Q}_++\overline{\bf Q}_-$,
${\bf Q}^{\dag}={\bf Q}_++{\bf Q}_-$ are the ${\mathcal N}=2_B$ generators.
Action of such a form would be manifestly
${\mathcal N}=2_B$ invariant and automatically lead us to find the
needed boundary counter term.

A hint to find such a form lies in the original manifestly $(2,2)$
invariant action written in terms of the superspace intergals (\ref{L}).
Up to a total derivative, bulk D-term
$\int\dd^4\theta K$ is equal to
${\bf Q}_+{\bf Q}_-\overline{\bf Q}_-\overline{\bf Q}_+k
\sim {1\over 2}{\bf Q}{\bf Q}^{\dag}[\overline{\bf Q}_+,{\bf Q}_+]k$,
where $k$ is the lowest component of the superfield $K$.
Similarly, for a twisted chiral superfield
$\widetilde{W}$,
the twisted F-term
$\int\dd^2\widetilde{\theta}\,\widetilde{W}$ can be written as
${\bf Q}{\bf Q}^{\dag}\left(\widetilde{W}|_{\rm lowest}\right)$ 
again up to a total derivative.
This motivates us to take the following as the manifestly
${\mathcal N}=2_B$ invariant action:
\beqa
S_{\rm g}+S_{\rm m}+S_{{\rm FI}\,\theta}
&=&{1\over 4\pi}\int_{\surface}\dd^2\s\,
{\bf Q}{\bf Q}^{\dag}[\overline{\bf Q}_+,{\bf Q}_+]
\left(-{1\over 2e^2}\sum_{a=1}^k|\sigma_a|^2
+\sum_{i=1}^N|\phi_i|^2\right)
\nn\\
&&+{1\over 2\pi}{\rm Re}\int_{\surface}\dd^2\s\,{\bf Q}{\bf Q}^{\dag}
\left(\,-\sum_{a=1}^kt^a\sigma_a\,\right).
\label{Smani}
\eeqa
Despite its appearance, it includes the boundary counter term.
Let us write it down more explicitly. 
For the case of the $U(1)$ gauge theory with a 
single charge $1$ matter, we find
\beqa
S_g
&=&{1\over 2\pi}\int_{\surface}\dd^2\s\,{\mathcal L}_g
+{1\over 4\pi e^2}
\int_{\partial\surface}\dd t\,\left[{1\over 2}\partial_1|\sigma|^2
+\Ip(\sigma)D+\Rp(\sigma)v_{01}\right],
\label{Sgggg}
\\
S_{\rm m}
&=&{1\over 2\pi}\int_{\surface}\dd^2\s\,{\mathcal L}_m
+{1\over 2\pi}\int_{\partial\surface}\dd t\,
\left[{i\over 2}(\bpsi_-\psi_+-\bpsi_+\psi_-)+\Ip(\sigma)|\phi|^2\right],
\label{Smmmm}
\\
S_{{\rm FI}\,\theta}
&=&{1\over 2\pi}\int_{\surface}\dd^2\s\,{\mathcal L}_{{\rm FI}\,\theta}
+{1\over 2\pi}\int_{\partial\surface}\dd t\,
\Ip\Bigl(\,-t\sigma\,\Bigr),
\label{SFIFI}
\eeqa
where ${\mathcal L}_{\rm g}$, ${\mathcal L}_{\rm m}$,
${\mathcal L}_{{\rm FI}\,\theta}$ are the expressions given in
(\ref{Lg}), (\ref{Lm}), (\ref{LFI}) respectively.
The counter term in (\ref{Smmmm}) is the gauge theory version of the
 ``standard
boundary term'' (\ref{L0bdry}) in non-linear sigma models and LG models.
For the general case, the total boundary counter term 
is expressed as
\beqa
S_{\rm tot}^{\rm c.t.}
&=&{1\over 2\pi}\int_{\partial\surface}\dd t\,\left\{
{1\over 2 e^2}
\sum_{a=1}^k\left[{1\over 2}\partial_1|\sigma_a|^2
+\Ip(\sigma_a)D_a+\Rp(\sigma_a)(v_a)_{01}\right]\right.
\nn\\
&&\left.
+{i\over 2}\sum_{i=1}^N\Bigl(\bpsi_{i-}\psi_{i+}-\bpsi_{i+}\psi_{i-}\Bigr)
+\Ip
\sum_{a=1}^k\left[\left(\sum_{i=1}^NQ_i^a|\phi_i|^2-t^a\right)\sigma_a\right]
\right\}.
\label{Lct}
\eeqa
This counter term had been found earlier in \cite{blin,HKLM}.

As always, there is a freedom to add boundary D-terms,
$$
\Delta S_{\rm bdry}=\int_{\partial\surface}\dd t\,
{\bf Q}{\bf Q}^{\dag}\lambda.
$$
This would have no effect to the low energy theory
as long as $\lambda$ is small enough at infinity in 
the field space.
But addition of such a term
will alter the theory if $\lambda$ is large at infinity.
In what follows, we will not consider such
``large D-terms''.

\subsection{The Wilson Line}
\label{subsec:LSMwilson}

\newcommand{\p}{\pi}
\newcommand{\Y}{\Psi}
\newcommand{\cA}{{\cal A}}
\newcommand{\cB}{{\cal B}}
\newcommand{\cI}{{\cal I}}
\newcommand{\tta}{\theta}
\newcommand{\bbH}{\mathbb{H}}
\newcommand{\dl}{\partial}
\newcommand{\Ga}{\Gamma}
\newcommand{\STr}{\mathrm{Str}}
\newcommand{\diag}{\mathrm{diag}}
\newcommand{\tach}{{\bf T}}
\newcommand{\vf}{\varphi}
\newcommand{\de}{\delta}
\newcommand{\la}{\lambda}
\newcommand{\brst}{{\de_{\textrm{\tiny BRST}}}}

Next, we consider boundary interactions that are by themselves
${\mathcal N}=2_B$ invariant but that are not D-terms. 
The simplest one is
\beq
{1\over 2}\int_B\dd\theta\dd\btheta\, V
=-\Bigl[\,v_0-{\rm Re}(\sigma)\,\Bigr].
\label{Wilson}
\eeq
This is manifestly ${\mathcal N}=2_B$ invariant, but is not
invariant under the $U(1)$ gauge transformation
$iv_0\to iv_0+g\partial_0g^{-1}$.
However, the exponentiated action, the Wilson line
$$
W_q(t_f,t_i)=\exp\left(-i
\int^{t_f}_{t_i}q\Bigl[\,v_0-{\rm Re}(\sigma)\,\Bigr]\dd t\right)
$$
transforms as
$$
W_q(t_f,t_i)\longrightarrow
g(t_f)^q\cdot W_q(t_f,t_i)\cdot g(t_i)^{-q}
$$
and hence is gauge covariant whenever the number $q$
is an integer. We denote the brane supporting this Wilson line by
$$
\wilson(q).
$$
The simple interpretation is that the Chan-Paton space 
$\cp$ carries charge $q$ under the gauge group.
Since the superfield $V$ is not gauge invariant,
the Wilson line is {\it not a D-term} despite its appearance
(\ref{Wilson}).

For a $U(1)^k$ theory, 
a choice of $k$-tuple of integers defines the Wilson line brane
$\wilson(q^1,...,q^k)$ in the same way.
Note that the bulk theta term can be converted into a boundary term
by Stokes theorem:
$$
\int_{\surface}\theta\,v_{01}\,\dd^2\s
=-\int_{\partial\surface}\theta\,v_{0}\,\dd t,
$$
Including this into the counter term (\ref{Lct}) and the Wilson line term,
we have the following boundary Lagrangian
\beqa
  \label{gaugeboundary}
S_{\rm bdry}&=&S^{\rm c.t.}_{\rm g}
+{1\over 2\pi}\int_{\partial\surface}\dd t\,
\left\{\,
{i\over 2}\sum_{i=1}^N\Bigl(\bpsi_{i-}\psi_{i+}-\bpsi_{i+}\psi_{i-}\Bigr)
\right.
\nn\\
&&\left.+\sum_{a=1}^k\left(\sum_{i=1}^NQ_i^a|\phi_i|^2-r^a\right)
{\rm Im}(\sigma_a)
-\sum_{a=1}^k(\theta^a+2\pi q^a)
\Bigl[\,(v_a)_0-{\rm Re}(\sigma_a)\,\Bigr]\,\right\}.
\nn\\
\eeqa
The expression (\ref{gaugeboundary}) makes it manifest that
the theory depends only on the combination $\theta^a+2\pi q^a$,
or equivalently, that the theory does not change under 
\beq
  \label{thetaqequiv}
  \theta^a \rightarrow \theta^a + 2\pi m^a  , \quad \mathrm{and} \quad
  q^a  \rightarrow q^a - m^a,
\eeq
for integers $m^1,...,m^k$.

The $2\pi$ periodicity in the theta parameters may be lost
if the worldsheet $\surface$ has a boundary. One way to see this is 
to note that the integral
$$
{1\over 2\pi}\int_{\surface}v_{12}\,\dd^2\s
$$ 
is not necessarily an integer unless a boundary condition like
$v|_{\partial\surface}=0$ is imposed. In the present system, we decide
not to impose such a boundary condition nor 
$(v_0-\Rp(\sigma))|_{\partial\surface}=0$, and hence the theta parameters
indeed do not have the $2\pi$ periodicity.
We will discuss more on the boundary conditions in later sections.
Thus, $\theta^a\to\theta^a+2\pi m^a$ with no change in
$q^a$ is a non-trivial operation. In particular, the Wilson line brane
$\wilson(q^1,...,q^k)$ makes an invariant sense
only when the theta parameters are specified as a real number (not
just modulo  $2\pi\Z$).

\subsection{Interaction Of Wilson Lines}
\label{subsec:noncompactbranes}

\newcommand{\LSM}{{\cal L}}

It turns out that the Wilson line branes serve as the building blocks of
more general supersymmetric boundary interactions.
The first step of the generalization is
to take the direct sum
\beq
\brane=\bigoplus_{i=1}^n \wilson(q_i).
\eeq
Namely, instead of just ${\mathcal A}_t=\sum_{a=1}^kq^a(v_{a0}-\Rp(\sigma_a))$
we consider the matrix valued boundary interaction
\beq
{\mathcal A}_t=\sum_{a=1}^k\left(\begin{array}{ccc}
q^a_1&&\\
&\ddots&\\
&&q^a_n
\end{array}\right)\Bigl[\, (v_a)_0-{\rm Re}(\sigma_a)\,\Bigr].
\eeq
Under the gauge transformation by a $U(1)^k$ valued function
$g=(g_1,...,g_k)$,
it transforms as
$i\cA_t\to i\cA_t+\rho(g)\partial_t\rho(g)^{-1}$
with
\beq
\rho(g)=\left(\begin{array}{ccc}
g^{q_1}&&\\
&\ddots&\\
&&g^{q_n}
\end{array}\right),
\eeq
where $g^q:=g_1^{q^1}\cdots g_k^{q^k}$.
Simply put, the Chan-Paton space $\cp$ of the brane
$\oplus_{i=1}^k\wilson(q_i)$
carries the representation $\rho$ of the gauge group $T\cong U(1)^k$.
Note that the $\cA_t$ can be written succinctly as
$$
\cA_t=\sum_{a=1}^k\rho_*(e^a)\Bigl[\,(v_a)_0-{\rm Re}(\sigma_a)\,\Bigr]
=\rho_*\Bigl(v_0-{\rm Re}(\sigma)\Bigr),
$$
where $\rho_*$ is the infinitesimal form of $\rho$, defined by
$\rho_*(X)=-i{\dd \over \dd t}\rho(\e^{itX})\Bigr|_{t=0}$ for an
element $iX$ of the Lie algebra of the gauge group.

Just as in various $(2,2)$ theories considered in Section~\ref{sec:branecat},
we may also introduce a $\Z_2$ graded sum of Wilson line branes
$\brane=\wilson^{\rm ev}\oplus\wilson^{\rm od}$
along with a tachyon profile $Q$ that represents an interaction between 
Wilson lines.
Namely we introduce a $\Z_2$ graded Chan-Paton space
$$
  \cp = \cp^{\rm ev} \oplus \cp^{\rm od},
$$
carrying a representation $\rho$ of the gauge group,
with an odd operator $Q$ on $\cp$ that depends holomorphically on
the fields $\phi_1,..,\phi_N$.
Then one can write the boundary interaction 
\beq
  \label{tachconn}
  \cA_t = \rho_*\Bigl(v_0-{\rm Re}(\sigma)\Bigr)
+\frac12 \{Q,Q^{\dag}\}
-\frac12 \sum_{i=1}^N\psi^i {\partial\over \partial\phi_i} Q 
+ \frac12 \sum_{i=1}^N \bpsi_i{\partial \over \partial\bphi_i}Q^\dag 
\eeq
First of all, we
would like $\cA_t$ to transform under the gauge transformation as
\beq
i\cA_t\longrightarrow \rho(g) i\cA_t\rho(g)^{-1}
+\rho(g) \partial_t\rho(g)^{-1}.
\eeq
This is the case if and only if $Q$ satisfies
\beq
\rho(g)^{-1}Q(g\cdot \phi)\rho(g)=Q(\phi),
\label{Qgaugeinv}
\eeq
where $g\cdot \phi$ is the gauge transform of 
$\phi=(\phi_1,...,\phi_N)$, given by $(g^{Q_1}\phi_1,...,g^{Q_N}\phi_N)$.
Under this condition,
$\cA_t$ is ${\mathcal N}=2_B$ supersymmetric if and only if
$Q$ squares to the identity times a constant, $Q^2 = c \cdot{\rm id}_{\cp}$.

\subsubsection*{\it R-symmetry}

The linear sigma model without superpotential has a vector
$U(1)$ R-symmetry. 
We consider the R-symmetry that acts trivially on $\phi_i$'s even though
the model usually has other symmetries with which 
the R-symmetry could be dressed.
This is motivated by the fact that the bulk theory generically
reduces to a large volume sigma model for which we decided to
respect the R-symmetry such that the target coordinates have R-charge zero.
(See Section~\ref{subsubsec:R-symmetry}.)
We restrict our attention to branes that preserve this vector
$U(1)$ R-symmetry.

Since $Q$ enters into the supercharge ${\bf Q}$,
we would like $Q$ to have R-charge 1.
Namely, we would like to have an action of $U(1)$ R-symmetry
on the Chan-Paton space $\cp$, given by a matrix
$R(\lambda)$ such that
\beq
R(\lambda)Q(\phi)R(\lambda)^{-1}=\lambda Q(\phi).
\label{Rsymm}
\eeq
This together with $Q^2=c\cdot {\rm id}_{\cp}$ requires
\beq
Q^2=0.
\label{LMnil}
\eeq
We would like the R-action to commute with the gauge group action,
\beq
R(\lambda)\rho(g)=\rho(g) R(\lambda).
\label{Gcommu}
\eeq
Under the condition (\ref{Rsymm}), the $Q$-dependent part of
$\cA_t$ is invariant under the R-symmetry when combined with 
the conjugation by $R(\lambda)$, and under the condition (\ref{Gcommu}),
the remanining part $\rho_*(v_0-{\rm Re}(\sigma))$
is also invariant.
As in the case of non-linear sigma models (see
Section~\ref{subsubsec:R-symmetry}),
we may assume that $R(\lambda)$ is a genuine representation of
$U(1)$ and is compatible with the $\Z_2$ grading,
so that the eigenvalues are 
$\lambda^j$ for some integer $j$ which is even ({\it resp.} odd)
for elements of $\cp^{\rm ev}$ ({\it resp.} $\cp^{\rm od}$).
If we denote by $\cp^j$ the $R(\lambda)=\lambda^j$ part of
$\cp$, we have
\beq
  \label{Zgrading}
  \cp =
  \bigoplus_{j=j_{min}}^{j_{max}} \cp^j.
\eeq
with
$$
  \cp^{\rm ev} = \bigoplus_{j:~even} \cp^j,\quad
  \mathrm{and} \quad
  \cp^{\rm od} = \bigoplus_{j:~odd} \cp^j.
$$
By the commutativity with gauge group (\ref{Gcommu}),
each $\cp^j$ corresponds to a direct sum $\brane^j$ of Wilson line branes.
If we order the subspaces $\brane^j$ by increasing R-charge the
interaction $Q$ has the block-off diagonal form:
\beq
  \label{Qoffdiag}
  Q = \left(\begin{array}{cccccc}
        0 & d^{j_{max}} & 0 & \ldots & 0 & 0 \\
        0 & 0 & d^{j_{max}-1} & \ldots & 0 & 0 \\
        \vdots & &\ddots & \ddots && \vdots \\
        0 & 0 & 0 & \ldots &  0 & d^{j_{min}+1} \\
        0 & 0 & 0 & \ldots &  0 & 0 
      \end{array}\right)  ,
\eeq
where $d^j$ denotes the interaction term between
$\brane^{j-1}$ and $\brane^j$. 

As usual, there is an ambiguity in the choice of $R(\lambda)$ ---
it can be replaced by $\lambda^2R(\lambda)$ without violating any of
the above condition. However, this of course does not change the
physical property of the brane. Nevertheless it is sometimes
useful to keep this information of the R-symmetry action.
An {\it R-graded D-brane} $\lsmB$, 
a brane with this additional information,
is determined by the triple
$(\brane,Q,R)$ or equivalently by the quartuple 
$(\cp,Q,\rho,R)$. Note that the information on
$\lsmB$ can nicely be encoded as a complex
\beq
  \label{LSMcomplex}
  {\mathcal C}(\lsmB):~~~~~
  \cdots \stackrel{d^{j-1}}{\longrightarrow}
  \brane^{j-1} \stackrel{d^{j}}{\longrightarrow}
  \brane^{j} \stackrel{d^{j+1}}{\longrightarrow}
  \brane^{j+1} \stackrel{d^{j+2}}{\longrightarrow}
  \cdots \quad.
\eeq
Each $\brane^j$ is a direct sum of Wilson line branes.

\subsection{Matrix Factorizations}
\label{subsec:compactbranes}

Let us now discuss D-branes in the linear sigma model 
with a non-zero superpotential $W$.
The supersymmetry transformation of
the bulk F-term is given by the Warner term:
\beq
\delta\int_{\surface}\dd^2\s\,{\mathcal L}_W
=-{\rm Re}\int_{\partial\surface}
\dd t\,\sum_{i=1}^N\bepsilon \psi_i{\partial W\over\partial\phi_i}
\label{LSMWarner}
\eeq
As in LG models, the cancellation is 
done by a matrix factorization of $W$.
Let us consider a $\Z_2$ graded sum of Wilson line branes
with a polynomial tachyon profile.
Namely, a $\Z_2$ graded
Chan-Paton space, $\cp=\cp^{\rm ev}\oplus\cp^{\rm od}$, 
an odd operator $Q$ on $\cp$ which is a polynomial in
$\phi=(\phi_1,...,\phi_N)$, and a representation $\rho$ of
the gauge group $T\cong U(1)^k$ on $\cp$, 
which obey the gauge invariance condition
\beq
\rho(g)^{-1}Q(g\cdot \phi)\rho(g)=Q(\phi).
\label{MFgaugeinv}
\eeq
Then the corresponding boundary interaction
$$
  \cA_t = \rho_*\Bigl(v_0-{\rm Re}(\sigma)\Bigr)
+\frac12 \{Q,Q^{\dag}\}
-\frac12 \sum_{i=1}^N\psi^i {\partial\over \partial\phi_i} Q 
+ \frac12 \sum_{i=1}^N \bpsi_i{\partial \over \partial\bphi_i}Q^\dag 
$$
transforms under the ${\mathcal N}=2_B$ supersymmetry as
\beq
\delta{\mathcal A}_t\,=\,
-{\rm Re}\left\{\,\sum_{i=1}^N
\left(\bepsilon \psi_i{\partial \over\partial \phi_i}Q^2\right)
-[\bepsilon Q^{\dag},Q^2]
\,\right\}
+i{\mathcal D}_t\Bigl(\bepsilon Q+\epsilon Q^{\dag}\Bigr)
-i\Bigl(\dot{\bepsilon}Q+\dot{\epsilon}Q^{\dag}\Bigr).
\label{delALSM}
\eeq
The first term cancels the Warner term if and only if
$Q$ is a matrix factorization of $W$,
\beq
Q^2= W\cdot {\rm id}_{\cp}.
\eeq
(The second and the third terms of (\ref{delALSM}) are as before.)

We focus our attention to bulk theories with a vector $U(1)$ R-symmetry
with an integrality that is compatible with the statistics of
operators.
Vector R-symmetry requires that the
superpotential $W$ is quasi-homogeneous.
Namely, there is a one-parameter group of
linear transformations of the variables
$\phi\mapsto R_{\lambda}(\phi)$ commuting with the gauge symmetry
\beq
R_{\lambda}(g\cdot \phi)=g\cdot R_{\lambda}(\phi), 
\eeq
such that
\beq
W(R_{\lambda}(\phi))=\lambda^2 W(\phi).
\label{quasiho}
\eeq
Note that there is an ambiguity in $R_{\lambda}$
 --- it can be modified by the gauge symmetry.
Integrality compatible with statistics means that
the R-charges of operators or of NS-NS states are integral
and are conguruent modulo $2$ to their spins.
This is the case when $R_{\lambda}$ for $\lambda=\e^{\pi i}$
is trivial, 
\beq
R_{\e^{\pi i}}(\phi)=\phi,
\label{congru}
\eeq
or can be made trivial with the help of modification by a 
gauge transformation, in the phase where the theory reduces to a sigma model
with large volume limit.
Thus, we assume both (\ref{quasiho}) and (\ref{congru}) in the rest of
this paper.
This is certainly the case when the superpotential is designed
to engineer a hypersurface $W=PG(X)$ or a complete intersection 
of hypersurfaces $W=\sum_{\beta=1}^lP_{\beta}G_{\beta}(X)$ in 
a toric variety: 
$R_{\lambda}(p,x)=(\lambda^2p,x)$ satisfies the condition.

We would like the branes to respect this R-symmetry and integrality.
The condition of R-symmetry is that $Q$ has R-charge $1$. Namely,
there is a one parameter group of
linear operators $R(\lambda)$ on the Chan-Paton space
$\cp$ commuting with the gauge symmetry
\beq
R(\lambda)\rho(g)=\rho(g) R(\lambda),
\label{GcommuMF}
\eeq
such that
\beq
R(\lambda)Q(R_{\lambda}(\phi))R(\lambda)^{-1}=\lambda Q(\phi).
\eeq
Integrality means that the R-grading of NS-states and operators 
is integral and reduces modulo 2 to the original $\Z_2$ grading.
Let us take two branes with R-symmetry, $(\cp_i,Q_i,\rho_i)$
with $R_i$, $i=1,2$.
The requirement in the polynomial sector is
\beq
R_2(\e^{\pi i})\vp(R_{\e^{\pi i}}(\phi))R_1(\e^{\pi i})^{-1}
=(-1)^{|\vp|}\vp(\phi)
\eeq
for any polynomial $\vp$ of $\phi_1,...,\phi_N$
with values in $\Hom(\cp_1,\cp_2)$ which is gauge invariant.
In the frame where (\ref{congru}) holds, this means that
$R_1(\e^{\pi i})$ and $R_2(\e^{\pi i})$ 
are equal to the $\Z_2$ grading operators
$\sigma_1$ and $\sigma_2$ of
$\cp_1$ and $\cp_2$
(up to a common constant which can be set equal to $1$).
Thus, we require that
\beq
\qquad R(\e^{\pi i})=\sigma_{\cp}\quad (\mbox{the $\Z_2$-grading of $\cp$}),
\label{bcongru}
\eeq
in the frame where (\ref{congru}) holds.
In particular, this means $R(\e^{2\pi i})=(\sigma_{\cp})^2={\rm id}_{\cp}$.
Namely, the eigenvalues of $R(\lambda)$ are $\lambda^j$ where $j$ 
is an integer which is even ({\it resp}. odd) on $\cp^{\rm ev}$
({\it resp}. $\cp^{\rm od}$).
As usual, the replacement 
$R(\lambda)\to \lambda^2R(\lambda)$ does not change the physical
property of the branes,
and we call a brane with the additional information $R(\lambda)$
an {\it R-graded D-brane}.

As an example, let us explicitly
write down the data of an R-graded D-brane
in the $U(1)$ gauge theory with the fields $P,X_1,...,X_N$
of charge $-N,1,\ldots,1$ having the superpotential
$W=PG(X_1,...,X_N)$ where $G(X)$ is a homogeneous polynomial of degree $N$.
Choosing a basis of the Chan-Paton space where
the first half are even and the latter half are odd,
it is given by
$(Q,\rho,R)$
\beqa
&&Q(p,x)=\left(\begin{array}{cc}
0&f(p,x)\\
g(p,x)&0
\end{array}\right),
\qquad Q(p,x)^2=pG(x){\bf 1}_{2\ell}
\\
&&\rho(g)=\left(\begin{array}{ccc}
g^{q_1}&&0\\
&\ddots&\\
0&&g^{q_{2\ell}}
\end{array}\right),
\\
&&R(\lambda)
=\left(\begin{array}{ccc}
\lambda^{j_1}&&0\\
&\ddots&\\
0&&\lambda^{j_{2\ell}}
\end{array}\right),
\qquad
R(\e^{\pi i})=\left(\begin{array}{cc}
{\bf 1}_{\ell}&0\\
0&-{\bf 1}_{\ell}
\end{array}\right),
\eeqa
such that
\beq
\rho(g)^{-1}Q(g^{-N}p,g x)\rho(g)
=Q(p,x),
\label{ginv}
\eeq
and 
\beq
R(\lambda)Q(\lambda^2p,x)R(\lambda)^{-1}
=\lambda Q(p,x).
\label{homog}
\eeq

\subsection{A First Look At The Boundary Condition}
\label{subsec:firstlookBC}

Let us take a first look at the boundary condition of the bulk fields.
We postpone the discussion of the vector multiplet fields to
Section~\ref{sec:GRR}, as that requires some understanding of
the quantum theory.
Here we focus on the condition on the matter chiral multiplet fields,
treating the vector multiplet fields as backgrounds.

It is straightforward to check that the following set of
boundary conditions are invariant under the
${\mathcal N}=2_B$ supersymmetry:
\beq
\begin{array}{l}
D_1\phi-{\rm Im}(\sigma)\phi=0,\\
\psi_+=\psi_-,\\
D_1(\psi_++\psi_-)-(\blambda_++\blambda_-)\phi-{\rm Im}(\sigma)(\psi_++\psi_-)
=0,\\
F=0.
\end{array}
\label{bcChi}
\eeq
Here we considered charge $1$ fields in a $U(1)$ gauge theory.
The generalization is obvious.
If there is no superpotential $W$ and no
boundary interaction $Q$, this is also compatible with the variational
equation of the action: The $\Ip(\sigma)|\phi|^2$ term in the boundary
counter term (\ref{Lct}) deforms the ordinary Neumann condition
$D_1\phi=0$ to $D_1\phi=\Ip(\sigma)\phi$.

If there is a superpotential $W$ or a boundary interaction $Q$,
the variational equation changes and the boundary conditions
will be modified. However, as long as we can treat the interactions
$W$ and $Q$ as perturbation, we can still use the conditions
(\ref{bcChi}) without modifications.
In particular, when we discuss the effective theory on the Coulomb branch
in Section~\ref{sec:vacuum},
the charged matter multiplets are heavy and hence $W$ and $Q$,
which are generically of high powers in the matter fields, can be treated
as perturbation.
In the literature, perturbative
treatment of boundary interaction is widely used. 
In studying the renormalization group flow of the spacetime 
electromagnetic potential, 
one can use either the ``correct'' (mixed Dirichlet-Neumann)
boundary condition or the ``incorrect'' Neumann condition \cite{abouelsaood}.
For a non-Abelian gauge group, the latter approach turns out to
be more efficient and leads us quickly to the Yang-Mills equation
at leading order in the $\alpha'$ expansion \cite{nARG}.

The treatment of the auxiliary field needs some care
in this approach. 
The boundary condition $\overline{F}=0$ from (\ref{bcChi})
may appear too strong in the presence of the superpotential $W$:
It appears to require $W'=0$ at the boundary since the bulk
equation of motion reads $W'=-\overline{F}$.
But that would be inconsistent with the free boundary condition for $\phi$
and would completely change our picture of boundary interaction
based on matrix factorization.
However, if we carefuly think about the meaning of the equation of motion
and the boundary condition,
we immediately find that there is no need to require $W'=0$. 
To illustrate it, we 
consider the following toy model:
$$
\int\dd F_1\cdots\dd F_n\,\exp\left(-{1\over 2}\sum_{i,j=1}^nA_{ij}F_iF_j
+\sum_{i=1}^nB_iF_i\right).
$$
The equations of motion for $F_i$'s are $F_i=\sum_{j=1}^n(A^{-1})_{ij}B_j$.
If we impose the condition $F_1=0$, we simply loose the $F_1$ integral
and obtain a different answer as a function of $B_j$'s.
But we never require $B_j$'s to satisfy
$\sum_{j=1}^n(A^{-1})_{1j}B_j=0$.
However, there is still some subtlety we may need to be aware of.
It concerns the supersymmetry variation in the version when the
auxiliary fields are eliminated. 
To show the essential point,
we describe it in a LG model without gauge interaction.
The auxiliary field $\overline{F}$ appears in
the $\bepsilon$-variation of the fermion $\bpsi_{\pm}$:
$$
\delta\bpsi_{\pm}=\pm\bepsilon \overline{F}.
$$
In the bulk theory after the auxiliary fields are integrated,
we simply set $\overline{F}=-W'$ in this variation.
However, in the presence of a boundary and with the above boundary condition,
we need to set
$$
\delta\bpsi_{\pm}=\left\{
\begin{array}{ll}
\mp\bepsilon W'&\mbox{in the interior}\\
0&\mbox{at the boundary}.
\end{array}\right.
$$
Therefore the variation is in general discontinuous at the boundary.
In particular, the variation of the bulk Lagrangian density
 will have a delta function supported at the boundary.
In fact, this delta function is crucial in obtaining the Warner term
$$
\delta S_{\rm bulk}=-\int_{\partial\surface}\dd t\left(
{1\over 2}\bepsilon\psi W'\right),
$$
with the correct normalization. Without the discontinuity 
in $\delta\bpsi_{\pm}$, that would be
off by a factor of ${1\over 2}$.
In the previous treatment where we did not specify the boundary condition
and we used the variation $\delta\bpsi_{\pm}$ without discontinuity,
the correct Warner term
results with additional contribution from the variation of
the standard boundary term
($S^{\rm c.t.}_{\rm matter}$ in the gauge theory version).
With the boundary condition, that term is absent.

\subsection{The Ultra-Violet Limit: $e=0$}
\label{subsec:UVlimit}

\newcommand{\cR}{{\cal R}}
\newcommand{\cH}{{\cal H}}
\newcommand{\grA}{{gr\!\!-\!\!A}}

In the strict limit of vanishing gauge coupling $e\to 0$,
which corresponds to the ultra-violet limit, the
linear sigma model becomes particularly simple. 
With the standard field
redefinition, $V \rightarrow e V$,
we see that 
the vector multiplet fields decouple from the rest of the system
in the limit $e\to 0$.
Alternatively, the infinite kinetic terms simply freeze 
the gauge multiplet fields, and we are left with the matter sector.
The boundary condition (\ref{bcChi}) then
becomes the standard Neumann condition
\beq
\begin{array}{l}
\partial_1\phi=0,\\
\psi_+=\psi_-,\\
\partial_1(\psi_++\psi_-)=0,\\
F=0.
\end{array}
\label{bcChi0}
\eeq
The D-term potential is turned off and
the matter sector is completely
independent of the K\"ahler moduli. 
The remnant of the gauge theory is that
we respect the gauge symmetry and require that physical observables must be
gauge invariant.
For example, the boundary
interaction $Q$ must respect the gauge invariance condition,
(\ref{Qgaugeinv}) or (\ref{MFgaugeinv}).
In the following we study the chiral ring of D-branes in the matter
sector in this limit.

We emphasize that the 
limit $e^2\rightarrow 0$ is not smooth even in the chiral sector. In
fact, we turned off the D-term potential and altered the theory at
infinity in field space, so that even quantities like the Witten index
may jump. 
Nevertheless, we study the chiral ring in this limit as it plays
an important r\^ole, as we will see later,
 in describing the theory 
with finite gauge coupling.

We first consider the theory with vanishing superpotential.
Let us take two D-branes $\lsmB_i=(\cp_i,Q_i,\rho_i,R_i)$,
$i=1,2$. To find the space of chiral ring elements
in the $\lsmB_1$-$\lsmB_2$ sector,
we use the zero mode approximation.
Namely, the operators in this sector are antiholomorphic
forms with values in $Hom(\cp_1,\cp_2)$ that are gauge invariant,
\beq
{\mathcal H}_{\rm zero}(\lsmB_1,\lsmB_2)
=\left[\Omega^{0,\bullet}(\C^N,Hom(\cp_1,\cp_2))\right]^T
\label{LSMzero}
\eeq
and the supercharge acts as the Dolbeault-like operator
$i{\bf Q}^{\rm zero}\vp=\bartial\vp
+Q_2\vp-(-1)^{|\vp|}\vp Q_1$. 
Here, the gauge group action is determined by
the action of $T$ on $\C^N$ as well as on $Hom(\cp_1,\cp_2)$ via
$\rho_1$ and $\rho_2$.
Note that $R_1$ and $R_2$ together with
the form-degree determines a
$\Z$-grading on the space under which ${\bf Q}^{\rm zero}$ has degree $1$.
The space of chiral ring elements of R-degree $p$ is the $p$-th
cohomology group
$\Ho^p(\lsmB_1,\lsmB_2)=H^p_{{\bf Q}^{\rm zero}}
({\mathcal H}^{\bullet}_{\rm zero}(\lsmB_1,\lsmB_2))$.

As before, using the property that any $\bartial$-closed form of
positive degree is $\bartial$-exact on $\C^N$,
we can use the holomorphic or polynomial truncation where the space
(\ref{LSMzero}) is replaced by the space of
$T$-invariant holomorphic or polynomial functions with values in 
$Hom(\cp_1,\cp_2)$ and the supercharge action is simply
$i{\bf Q}^{\it pol}\vp=Q_2\vp-(-1)^{|\vp|}\vp Q_1$. 
In this context, it is convenient to introduce the notion of graded
rings and graded modules.
The gauge group action introduces a grading by $k$-integers,
or $\Z^k$-grading, in the polynomial ring
$$
\ring=\C[\phi_1,...,\phi_N].
$$
An element of $\ring$ has degree $(n_1,...,n_k)$ if it has charge
$(n_1,...,n_k)$ under the gauge group $T$. For example, the variable
$\phi_i$, which has gauge charge $Q_i^a$ under the $a$-th
$U(1)$ factor of $T\cong U(1)^k$, has degree $(Q_i^1,...,Q_i^k)$.
The degree is additive with respect to the product, and in this sense
it is a graded ring. (More on mathematics will
be discussed in Section~\ref{sec:math}.)
For each representation $(\cp,\rho)$ of the gauge group $T$,
we introduce an $\ring$-module
$$
  M = \cR \otimes_{\C} \cp. 
$$
It is a graded $\ring$-module: an element of $M$ has degree
$(n_1,...,n_k)$ if it has gauge charge $(n_1,...,n_k)$.
For example, if $v\in\cp$ transforms as $\rho(g)v=g^qv
=g_1^{q^1}\cdots g_k^{q^k}\cdot v$, then
$1\otimes v\in M$ has degree $q=(q^1,...,q^k)$.
A homomorphism of graded $\ring$-modules
$f:M_1\to M_2$ is a homomorphism of $\ring$-modules
that preserves the $\Z^k$-grading information. Namely, it sends
an element of $M_1$ of a certain degree to an element of
$M_2$ of the same degree. For the $M_i$ that come from
representations $(\cp_i,\rho_i)$ of $T$, this is equivalent to
$f(b\otimes \rho(g)v_1)=(b\otimes \rho_2(g))f(1\otimes v_1)$
for $b\in \ring$ and $v_1\in \cp_1$.
In this language, the space of $T$-invariant polynomial functions with values
in $Hom(\cp_1,\cp_2)$ is linearly isomorphic to
the space of homomorphisms of graded $\ring$-modules,
\beq
\left[\Gamma_{\it pol}(\C^N,Hom(\cp_1,\cp_2))\right]^T\cong
\HomgrR(M_1,M_2).
\eeq
Since $\cp_i$ are $\Z$-graded by the R-symmetry $R_i$ commuting with $T$, 
$\cp_i=\oplus_j\cp_i^j$,
the $\ring$-modules
$M_i$ are also $\Z$-graded, $M_i=\oplus_jM_i^j$, where
$M_i^j=\ring\otimes \cp^j_i$.
Thus the above space is also $\Z$-graded
$\HomgrR^p(M_1,M_2)=\oplus_j\HomgrR(M_1^j,M_2^{j+p})$.
This of course corresponds to the
R-grading of ${\mathcal H}_{\rm zero}(\lsmB_1,\lsmB_2)$
restricted to the polynomial sector.
In particular, the space of chiral ring elements with R-degree $p$
is given by the $p$-th cohomology group
$$
\Ho^p(\lsmB_1,\lsmB_2)=H^p_{{\bf Q}^{\it pol}}
(\HomgrR^{\bullet}(M_1,M_2)).
$$
For a brane $(\cp,Q,\rho,R)$ or $(\brane,Q,R)$, the complex of Wilson line
branes (\ref{LSMcomplex}) can alternatively be represented as the complex of
graded $\ring$-modules
\beq
  \label{modulecomplex}
 {\mathcal C}:~~~ \ldots \stackrel{d^{j-1}}{\longrightarrow}
  M^{j-1} \stackrel{d^{j}}{\longrightarrow}
  M^{j} \stackrel{d^{j+1}}{\longrightarrow}
  M^{j+1} \stackrel{d^{j+2}}{\longrightarrow}
  \ldots \quad .
\eeq
Note that the information of the gauge charge (or Wilson line)
is encoded in the $\Z^k$-grading information of each $M^j$.
In this language, the space of chiral ring elements is given by the space
of chain maps of the complexes ${\mathcal C}_1$ and ${\mathcal C}_2$
of graded $\ring$-modules up to homotopy
\beq
\Ho^p(\lsmB_1,\lsmB_2)
\cong \HomHogrR({\mathcal C}_1,{\mathcal C}_2[p]).
\eeq
Here $[p]$ is the shift in the R-grading which involves the sign flip
of $Q_2$ if $p$ is odd.
In short, the D-brane category in the chiral sector for the
linear sigma model at $e^2=0$
is the homotopy category of the category of complexes of graded
$\ring$-modules. We will revisit the latter in Section~\ref{sec:math}.

Let us consider the theory with non-zero superpotential.
The data of a D-brane $(\cp,Q,\rho,R)$ can be encoded in
a two-periodic sequence of maps of graded $\ring$-modules
\beq
{\mathcal C}:\qquad\cdots\stackrel{Q}{\longrightarrow}M^{\rm ev}
\stackrel{Q}{\longrightarrow}M^{\rm od}
\stackrel{Q}{\longrightarrow}M^{\rm ev}
\stackrel{Q}{\longrightarrow}M^{\rm od}
\stackrel{Q}{\longrightarrow}\cdots\qquad
\eeq
The $\ring$-module $M^{\rm ev/od}:=\ring\otimes_{\C}\cp^{\rm ev/od}$
can also be regarded as a module over the graded ring
$\sring=\ring/(W)$ in which any multiple of the
 superpotential $W$ is equal to zero.
Then, it is a {\it complex} of
graded $\sring$-modules.
Let us take two such complexes
${\mathcal C}_i$ corresponding to two branes
$(\cp_i,Q_i,\rho_i)$, $i=1,2$.
The space of chiral ring elements for this sector
is the space of cochain maps of the complexes of
graded $\sring$-modules up to homotopies
\beqa
&&\Ho^{\rm ev}(\lsmB_1,\lsmB_2)
=\HomHogrS({\mathcal C}_1,{\mathcal C}_2),\nn\\
&&\Ho^{\rm od}(\lsmB_1,\lsmB_2)
=\HomHogrS({\mathcal C}_1,{\mathcal C}_2[1]),\nn
\eeqa
where ${\mathcal C}_2[1]$ is the shift of
${\mathcal C}_2$ by one (with the sign flip of $Q_2$).
When the two branes are R-graded, these spaces are also $\Z$-graded
$$
\Ho^{\rm ev}(\lsmB_1,\lsmB_2)=
\oplus_{p:{\rm even}}
\Ho^p(\lsmB_1,\lsmB_2),\qquad
\Ho^{\rm od}(\lsmB_1,\lsmB_2)=
\oplus_{p:{\rm odd}}
\Ho^p(\lsmB_1,\lsmB_2).
$$
The product structure is given by the composition of chain maps.

\newcommand{\simleq}{
\begin{array}{c}
<\\[-0.4cm]
\sim
\end{array}}
\newcommand{\simgeq}{
\begin{array}{c}
>\\[-0.4cm]
\sim
\end{array}}

\section{The Vacuum Energy/Charge On The Interval}
\label{sec:vacuum}

In this section, we study properties of the quantum
vacuum of certain massive field
theories formulated on an interval and on the half-space. 
The focus is the energy 
and charge densities of the vacuum state.
For the most part, in
 Sections~\ref{subsec:largesigma} 
through \ref{subsec:gaugedynamics},
we study
 theories that appear in the Coulomb branch with
large values of $\sigma$ in linear sigma models.
In Section~\ref{subsec:massfromW}, we consider
a different type of theories
 --- theories at $\sigma=0$ with superpotential mass terms.
This section has an independent flavor and can be read independently of
the rest of the paper. 
The reader can skip this section in the first reading
as the result will be quoted when it is used.

\noindent
{\bf Notation and convention:} In this section alone, we shall use $x$ for 
the spatial coordiate of the worldsheet (instead of $\s$ that is used
in other sections).
Also, we take the convention that the Lagrangian enters
into the path-integral weight as $\exp\left(i\int L\dd t\right)$,
unlike in other sections where it enters  as
$\exp\left({i\over 2\pi}\int L\dd t\right)$.

\subsection{The System At Large $\sigma$}
\label{subsec:largesigma}

In Section~\ref{sec:GRR}, we will study 
the boundary conditions on the Coulomb branch,
and for that we need to know the effective potential for $\sigma$ 
in the presence of a boundary.
For this purpose, we compute the energy and charge of the ground state in
the matter sector defined on the interval, $0\leq x\leq L$,
for a fixed large value of $\sigma$.
The bulk Lagrangian for the case of a single
charge 1 field is
\beqa
L_{\rm bulk}&=&\int^L_0\Biggl\{
|D_0\phi|^2-|D_1\phi|^2
+
i\bpsi_-\bigl(\lrD_0+\lrD_1\bigr)\psi_-
+i\bpsi_+\bigl(\lrD_0-\lrD_1\bigr)\psi_+
\nn\\
&&\qquad\quad
-|\sigma|^2|\phi|^2+D|\phi|^2
-\bpsi_-\sigma\psi_+-\bpsi_+\bsigma\psi_-\Biggr\}
\dd x.
\label{Lmatbulk}
\eeqa
We are interested in the boundary condition preserving an
${\mathcal N}=2_B$ supersymmetry. We consider both
Ramond and Neveu-Schwarz (NS) sectors. 
In Ramond sector the two boundaries preserve the same
supercharge (say $\oQ_++\oQ_-$ both at $x=0$ and $x=L$), while
in NS sector
the preserved supercharges are opposite (say $\oQ_++\oQ_-$ at $x=L$
and $\oQ_+-\oQ_-$ at $x=0$). The boundary interaction
that preserves the supersymmetry $\oQ_++\oQ_-$ at the right boundary
$x=L$ is
\beq
L^+_{\rm right\,\,bdry}=
\left[\, {\rm Im}(\sigma)|\phi|^2+{i\over 2}\bpsi_-\psi_+
-{i\over 2}\bpsi_+\psi_-\,\right]_{x=L}
\label{rightbt}
\eeq
with boundary condition
\beq
D_1\phi=\Ip(\sigma)\phi,\quad
\psi_+=\psi_-,\quad
D_1(\psi_-+\psi_+)=\Ip(\sigma)(\psi_-+\psi_+).
\label{rightbc}
\eeq
At the left boundary, $x=0$,
the interaction preserving $\oQ_+\pm\oQ_-$ is
\beq
L^{\mp}_{\rm left\,\,bdry}
=
\mp \left[\,{\rm Im}(\sigma)|\phi|^2+{i\over 2}\bpsi_-\psi_+
-{i\over 2}\bpsi_+\psi_-\,\right]_{x=0}
\label{leftbt}
\eeq
with boundary condition
\beq
D_1\phi=\pm \Ip(\sigma)\phi,\quad
\psi_+=\pm \psi_-,\quad
D_1(\psi_-\pm \psi_+)=\pm \Ip(\sigma)(\psi_-\pm \psi_+).
\label{leftbc}
\eeq

We first turn off the gauge field and auxiliary field,
 $v_{\mu}=D=0$, and assume that $\sigma$ is large and constant.
We often use the notaion
$$
M=|\sigma|,\quad
S={\rm Im}(\sigma),\quad
\sqrt{M^2-S^2}=|{\rm Re}(\sigma)|.
$$

\subsection{Mode Expansions}
\label{subsec:modeexpand}

We denote $b=(\psi_++\psi_-)/\sqrt{2}$ and $c=(\psi_--\psi_+)/\sqrt{2}$.
The boundary condition on the fields $\phi, b, c$
in Ramond and NS sectors are 
\beqa
&&(\mbox{\underline{R}}):\quad
\left.
\begin{array}{l}
\partial_1\phi=S\phi,
\\
\partial_1b=Sb,
\\
c=0
\end{array}\right\}\quad
\mbox{both at $x=0,L$},
\label{Rbc}
\\
&&(\mbox{\underline{NS}}):\quad
\left.\begin{array}{l}
\partial_1\phi=-S\phi
\\
b=0,
\\
\partial_1c=-Sc,
\end{array}\right\}\,\,
\mbox{at $x=0$},\qquad
\left.\begin{array}{l}
\partial_1\phi=S\phi
\\
\partial_1b=Sb,
\\
c=0
\end{array}\right\}\,\,
\mbox{at $x=L$},
\label{NSbc}
\eeqa
If we use these conditions, the total Lagrangian can be written as
$$
L_{\rm bulk}+L_{\rm bdry}=\int_0^L\Biggl[\,
|\dot{\phi}|^2-\bphi\Bigl(-\partial_1^2+|\sigma|^2\Bigr)\phi
+i\Psi^{\dag}\dot{\Psi}
-\Psi^{\dag}{\mathcal D}\Psi\,\Biggr]\dd x^1,
$$
where
$$
\Psi:=\left(\begin{array}{cc}
b\\
c
\end{array}\right),\quad
{\mathcal D}:=\left(\begin{array}{cc}
\Rp(\sigma)&\!\!-i\partial_1-i\Ip(\sigma)\\
-i\partial_1+i\Ip(\sigma)\!\!&-\Rp(\sigma)
\end{array}\right)
$$
In both Ramond and NS sectors, the kinetic operators,
$-\partial_1^2$ for $\phi$ and ${\mathcal D}$ for $\Psi$,
are hermitian with respect to the standard inner product.
Thus, we can expand the fields 
by eigenvectors of these operators.

\subsubsection{Ramond Sector}

We first consider the Ramond sector (\ref{Rbc}). 
For the scalar field $\phi$,
the boundary conditions 
on the plane wave $f(x)=a\sin(kx)+b\cos(kx)$
read
$$
ka=Sb,\qquad
ka\cos(kL)-kb\sin(kL)=Sa\sin(kL)+Sb\cos(kL).
$$
Using the first equation, the second equation simplifies to
$(k^2+S^2)\sin(kL)=0$ and we find $k=\pi n/L$, $n=1,2,3,...$.
Thus, the plane waves obeying the boundary condition are
\beq
f_n(x)={k_n\cos(k_n x)+S\sin(k_n x)\over\sqrt{k_n^2+S^2}},\qquad
k_n={\pi n\over L}.
\eeq
There is also a single non-oscillating mode:
\beq
f_0(x)=\sqrt{SL\over \e^{2SL}-1}\e^{Sx}.
\eeq
For $S\gg 1/L$ ({\it resp.} $S\ll -1/L$), this mode is 
localized near the right
({\it resp}. left) boundary. It is a constant mode for $S=0$.

The functions $f_n(x)$, $n=0,1,2,...$ are normalized as
$$
\int_0^Lf_n(x)f_m(x)\dd x ={L\over 2}\delta_{n,m}.
$$
Since $f_n(x)$ are eigenfunctions for the kinetic
operator $-\partial_1^2$, we can expand the field $\phi$ as
$$
\phi(x)=\sum_{n=0}^{\infty}\sqrt{2\over L}\phi_nf_n(x),
$$
so that the Lagrangian is written as
\beq
L_{\rm boson}=\sum_{n=0}^{\infty}\left(\,|\dot{\phi}_n|^2
-(M^2+k_n^2)|\phi_n|^2\,\right),
\label{RbL}
\eeq
where it is understood that $k_0^2=-S^2$.

Let us next study the mode expansion of fermions.
The boundary condition for $b$ is exactly the same as for the scalar $\phi$
and thus it can be expanded by $f_n(x)$, $n=0,1,2,...$.
The condition of $c$ is easier to solve:
\beq
g_n(x)=\sin(k_nx),\qquad k_n={\pi n\over L}\quad (n=1,2,3...).
\eeq
On the subspace $b\propto f_n(x)$ and $c\propto g_n(x)$, for $n=1,2,3,...$,
the kinetic operator is written as
$$
{\mathcal D}=\left(\begin{array}{cc}
\Rp(\sigma)&-i\sqrt{k_n^2+S^2}\\
i\sqrt{k_n^2+S^2}&-\Rp(\sigma)
\end{array}\right),
$$
and it is diagonalized by
\beq
\Psi_{n,\pm }(x)=\left(\begin{array}{cc}
\sqrt{\sqrt{M^2+k_n^2}\pm \Rp(\sigma)\over L\sqrt{M^2+k_n^2}}f_n(x)\\
\pm i\sqrt{\sqrt{M^2+k_n^2}\mp \Rp(\sigma)\over L\sqrt{M^2+k_n^2}}g_n(x)
\end{array}\right),\quad {\mathcal D}=\pm\sqrt{M^2+k_n^2}.~~
\eeq
The $n=0$ mode is non-oscillating,
\beq
\Psi_0(x)=\left(\begin{array}{c}
\sqrt{2\over L}f_0(x)\\
0
\end{array}\right),\quad
{\mathcal D}=\Rp(\sigma).
\eeq
If we expand the field $\Psi$ as
$$
\Psi(x)=b_0\Psi_0(x)+\sum_{n=1}^{\infty}
\left(b_{n,+}\Psi_{n,+}(x)
+b_{n,-}^{\dag}\Psi_{n,-}(x)\right)
$$
the Lagrangian is expressed as
\beqa
L_{\rm fermion}&=&
ib_0^{\dag}\dot{b}_0-\Rp(\sigma)b_0^{\dag}b_0\nn\\
&&\!\!\!\!\!\!\!\!\!\!\!\!\!\!\!\!
+\sum_{n=1}^{\infty}\left(\,
ib_{n,+}^{\dag}\dot{b}^{}_{n,+}
+ib_{n,-}^{\dag}\dot{b}^{}_{n,-}
-\sqrt{M^2+k_n^2}\,b_{n,+}^{\dag}b^{}_{n,+}
+\sqrt{M^2+k_n^2}\,b^{}_{n,-}b_{n,-}^{\dag}\,\right).
\qquad
\label{RfL}
\eeqa

\subsubsection{Neveu-Schwarz Sector}\label{subsub:NSmodes}

\begin{figure}[tb]
\centerline{\includegraphics{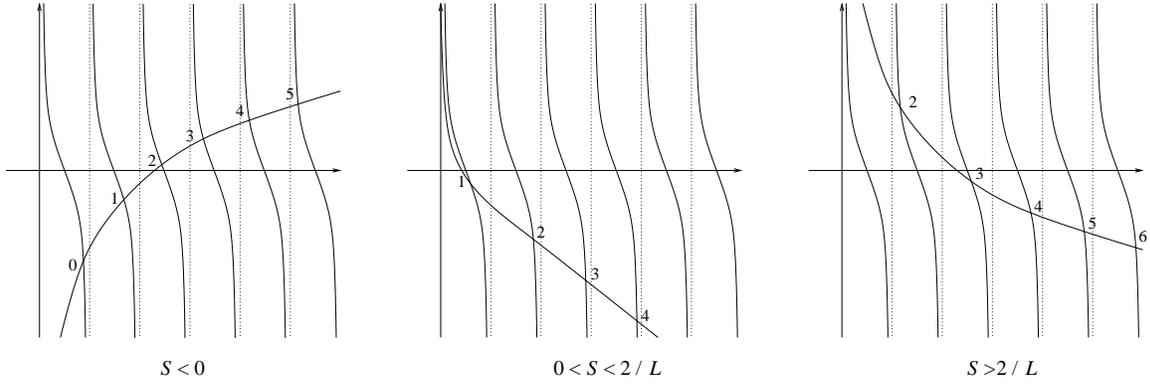}}
\caption{Flow of modes as $S$ is varied --- the scalar field (NS-sector).}
\label{sflow1}
\end{figure}

Let us now consider the NS sector (\ref{NSbc}).
For the plane-wave $h(x)=a\sin(kx)+b\cos(kx)$, the boundary 
condition for $\phi$ reads
$$
ka=-Sb, \qquad
ka\cos(kL)-kb\sin(kL)=Sa\sin(kL)+Sb\cos(kL).
$$
We find that the allowed wavenumbers $k$ are
\beq
\cot(kL)={S\over 2k}-{k\over 2S}.
\label{coteqn}
\eeq
There are solutions $k_n$ labelled by an integer $n$ such that
$$
k_n\to {\pi n\over L}\qquad\mbox{as $n\to\infty$},
$$
but the precise value of $k_n$ deviates from ${\pi n/L}$ for small
$n$. Also, the starting number $n$ depends on the value of $S$
(See Fig.~\ref{sflow1}):
\beqa
S < 0&&n=0,1,2,3,4,\ldots\nn\\
0\leq S< 2/L&&n=\quad 1,2,3,4,\ldots\nn\\
S\geq 2/L&&n=\qquad  2,3,4,\ldots\nn
\eeqa
For each such $k_n$ we have the plane wave
\beq
h_n(x)={k_n\cos(k_nx)-S\sin(k_nx)\over \sqrt{k_n^2+S^2}}
\left(\mbox{$1-{2S\over L(k_n^2+S^2)}$}\right)^{-1/2}.
\eeq
There may also be 
a non-oscillating mode spanned by exponential functions
$h(x)=a\e^{\kappa x}+b\e^{-\kappa x}$. The condition reads
$$
\kappa(a-b)=-S(a+b),\quad
\kappa (\e^{\kappa L}a-\e^{-\kappa L}b)
=S(\e^{\kappa L}a+\e^{-\kappa L}b),
$$
from which we find that the allowed value of $\kappa$ is
\beq
\coth(\kappa L)={S\over 2\kappa}+{\kappa\over 2S}.
\label{kappa1}
\eeq
\begin{figure}[tb]
\centerline{\includegraphics{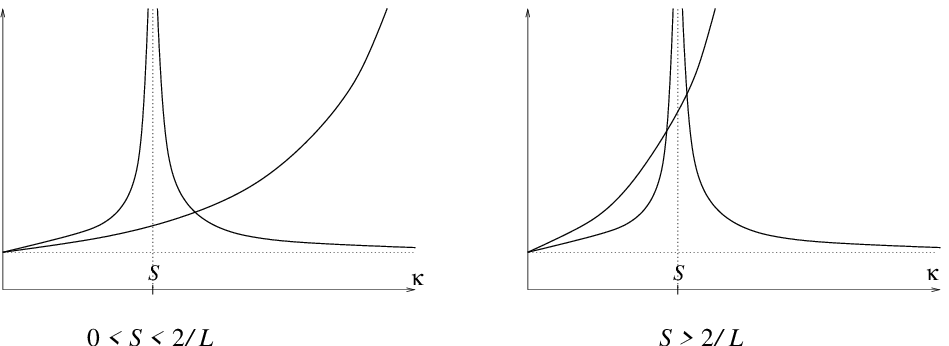}}
\caption{Solving (\ref{kappa1}), i.e., $\e^{2\kappa L}
=\left({\kappa+S\over \kappa-S}\right)^2$.}
\label{deceq1}
\end{figure}
See Fig.~\ref{deceq1} for the pattern of solutions
to this equation.
For $S<0$, there is no solution other than $\kappa=0$
for which the function $h(x)$ is zero.
At $S=0$, there is one solution $\kappa=\kappa_0=0$
for which $h(x)$ is a constant mode.
For $0< S<2/L$, the solution is at $\kappa=\kappa_0>S$.
For $S\geq 2/L$, there are two solutions, one
at $\kappa=\kappa_0>S$ and another at $\kappa=\kappa_1<S$.
These two values of $\kappa$ converge to $S$ rapidly as
$SL$ grows;
$$
\kappa_{0,1}\sim S\tanh^{\mp 1}(SL/2),\qquad SL\gg 1.
$$
These additional modes can be written as
\beqa
&&h_0(x)=\sqrt{\kappa_0L\over 
2(1-\e^{-2\kappa_0L}+2\kappa_0 L\e^{-\kappa_0 L})}
\Bigl(\e^{\kappa_0(x-L)}+\e^{-\kappa_0x}\Bigr)
\qquad S\geq 0,
\label{expoh0}\\
&&h_1(x)=\sqrt{\kappa_1 L\over 
2(1-\e^{-2\kappa_1L}-2\kappa_1 L\e^{-\kappa_1 L})}
\Bigl(-\e^{\kappa_1(x-L)}+\e^{-\kappa_1x}\Bigr)
\quad S\geq 2/L.
\label{expoh1}
\eeqa
Note that $h_0(x)=\sqrt{1\over 2}$ at $S=0$
and $h_1(x)=\sqrt{3\over 2}{L-2x\over L}$ at $S=2/L$.
For $S\gg 1/L$, these modes are localized near the two boundaries.

The modes are labeled by 
$n=0,1,2,...$ at any value of $S$
--- for $S<0$ they are all oscillating modes while
for $S>2/L$ the first two are exponential modes.
These functions are normalized as
$$
\int_0^Lh_n(x)h_m(x)=\delta_{n,m}{L\over 2},
\qquad n,m=0,1,2,3,...
$$
If we expand the field as 
$$
\phi(x)=\sum_{n=0}^{\infty}\sqrt{2\over L}\phi_nh_n(x),
$$
the Lagrangian reads as
\beq
L_{\rm boson}=\sum_{n=0}^{\infty}\left(\,|\dot{\phi}_n|^2
-(M^2+k_n^2)|\phi_n|^2\,\right),
\label{NSbL}
\eeq
where it is understood that
$k_0^2=-\kappa_0^2$ and/or $k_1^2=-\kappa_1^2$ whenever it applies.

\begin{figure}[tb]
\centerline{\includegraphics{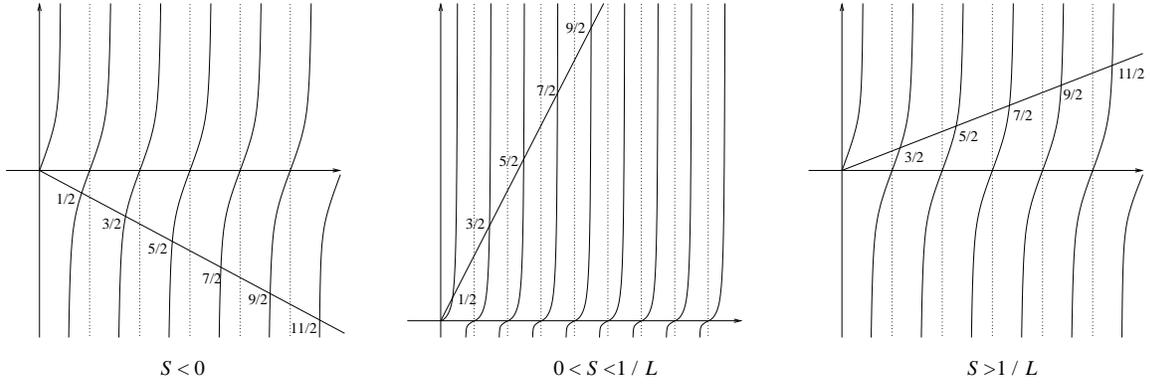}}
\caption{Flow of modes as $S$ is varied 
--- the Dirac fermion (NS sector): the scale for
the case $0<S< 1/L$ is different.}
\label{sflow2}
\end{figure}

The boundary conditions of the fermions $b$, $c$ are satisfied by
$\sin(kx)$ and $\sin(k(x-L))$ respectively, provided $k$ obeys the equation
\beq
\tan(kL)={k\over S}.
\label{taneqn}
\eeq
There are solutions $k_r$ labelled by a half integer $r$
such that
$$
k_r\to {\pi r\over L}\qquad \mbox{as $r\to \infty$}.
$$
The starting number $r$ depends on the value of $S$
(See Fig.~\ref{sflow2}):
\beqa
S< 1/L &&r={1\over2},{3\over 2},{5\over 2},{7\over 2},\ldots,\nn\\
S\geq 1/L &&r=\quad\, {3\over 2},{5\over 2},{7\over 2},\ldots,\nn
\eeqa
For each such $k_r$, we have the plane-waves
\beqa
&&g_r(x)=\sin(k_rx)\left(\mbox{$1-{S\over L(k_r^2+S^2)}$}\right)^{-1/2},\nn\\
&&h_r(x)=g_r(L-x)
={k_r\cos(k_rx)-S\sin(k_rx)\over \sqrt{k_r^2+S^2}}
\left(\mbox{$1-{S\over L(k_r^2+S^2)}$}\right)^{-1/2},
\nn
\eeqa
which are normalised as
$$
\int_0^Lg_r(x)g_s(x)\dd x
=\int_0^Lh_r(x)h_s(x)\dd x
=\delta_{r,s}{L\over 2}.
$$
On this subspace of wavenumber $k_r$, the kinetic operator
${\mathcal D}$ is expressed as
$$
{\mathcal D}=\left(\begin{array}{cc}
\Rp(\sigma)&i\sqrt{k_r^2+S^2}\\
-i\sqrt{k_r^2+S^2}&-\Rp(\sigma)
\end{array}\right),
$$
and it is diagonalized by
\beq
\Psi_{r,\pm}(x)=\left(\begin{array}{c}
\sqrt{\sqrt{M^2+k_r^2}\pm\Rp(\sigma)\over L\sqrt{M^2+k_r^2}}g_r(x)\\
\mp i\sqrt{\sqrt{M^2+k_r^2}\mp\Rp(\sigma)\over L\sqrt{M^2+k_r^2}}h_r(x)
\end{array}\right),\quad {\mathcal D}=\pm\sqrt{M^2+k_r^2}.~~
\eeq
There may also be eigenmodes spanned by exponential functions
$\e^{\pm \kappa x}$ --- there is indeed such a mode 
for $\kappa$ solving
\beq
\tanh(\kappa L)={S\over\kappa}.
\label{kappa2}
\eeq
\begin{figure}[tb]
\centerline{\includegraphics{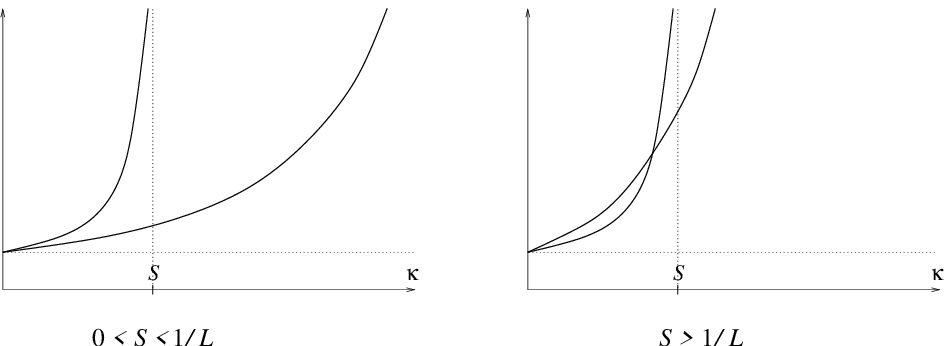}}
\caption{Solving (\ref{kappa2}), i.e., $\e^{2\kappa L}
={S+\kappa\over S-\kappa}$.}
\label{deceq2}
\end{figure}
There is no non-zero solution if $S<1/L$, and
there is a single non-zero solution when $S> 1/L$ (see Fig.~\ref{deceq2}).
It approaches $S$ rapidly as $SL$ grows,
$$
\kappa\sim S\tanh(SL),\qquad SL\gg 1.
$$
Thus, this exponential mode can be regarded as the replacement of
the $r={1\over 2}$ oscillating mode that is missing for
$S\geq 1/L$. The modes are obtained by simply replacing
$k_{1\over 2}$ by $i\kappa$ in $\Psi_{{1\over 2},\pm}(x)$, explicitly,
\beqa
\Psi_{{1\over 2},\pm}(x)
&=&{2\sqrt{\kappa}\e^{-\kappa L}\over
\sqrt{1-\e^{-4\kappa L}-4\kappa L\e^{-2\kappa L}}}
\left(\begin{array}{c}
\sqrt{\sqrt{M^2-\kappa^2}\pm\Rp(\sigma)\over \sqrt{M^2-\kappa^2}}
\sinh(\kappa x)\\
\pm i\sqrt{\sqrt{M^2-\kappa^2}\mp\Rp(\sigma)\over \sqrt{M^2-\kappa^2}}
\sinh(\kappa(x-L))
\end{array}\right),\nn\\
&&\quad {\mathcal D}=\pm\sqrt{M^2-\kappa^2}.~~
\label{expoPsi}
\eeqa
In the limit $S\searrow 1/L$, $\kappa$ approaches
zero but $\Psi_{{1\over 2},\pm}$ approaches a non-zero linear function.
Thus, for any value of
$S$ the modes are parametrized by positive half-integers
 $r={1\over 2},{3\over 2},....$.
If we expand the field $\Psi$ as
$$
\Psi(x)=\sum_{r={1\over 2}.{3\over 2},...}\left(b_{r,+}\Psi_{r,+}(x)
+b_{r,-}^{\dag}\Psi_{r,-}(x)\right),
$$
the Lagrangian is expressed as
\beq
L_{\rm fermion}=\sum_{r={1\over 2},{3\over 2},...}
\left(\,
ib_{r,+}^{\dag}\dot{b}^{}_{r,+}
+ib_{r,-}^{\dag}\dot{b}^{}_{r,-}
-\sqrt{M^2+k_r^2}\,b_{r,+}^{\dag}b^{}_{r,+}
+\sqrt{M^2+k_r^2}\,b^{}_{r,-}b_{r,-}^{\dag}\,\right),
\label{NSfL}
\eeq
where it is understood that $k_{1\over 2}^2=-\kappa^2$ 
if $S\geq 1/L$.

\subsection{The Ground State}
\label{subsec:groundstate}

Let us next quantize the system. Looking at the Lagrangian, 
(\ref{RbL}) and (\ref{RfL}) in the Ramond sector, (\ref{NSbL}) 
and (\ref{NSfL}) in the NS sector, it is clear what to do.

The bosonic system is simply the sum of harmonic oscillators.
For each mode $n$ we introduce creation and annihilation operators
\beq
[a_n^i,(a_m^j)^{\dag}]=\delta^{i,j}\delta_{n,m},\qquad
[a_n^i,a_m^j]=
[(a_n^i)^{\dag},(a_m^j)^{\dag}]=0,
\eeq
where we need two sets, $i,j=1,2$,
since we have {\it complex} (two real) oscillators. 
The variables $\phi_n=(\phi^1_n+i\phi^2_n)/\sqrt{2}$
and their conjugate momenta
can be written as
$$
\phi^i_n={1\over \sqrt{2\sqrt{M^2+k_n^2}}}(a^i_n+(a_n^i)^{\dag}),
\quad
\dot{\phi}^i_n=\sqrt{\sqrt{M^2+k_n^2}\over 2}(-ia^i_n+i(a_n^i)^{\dag}),
$$
or equivalently, for $\phi=(\phi^1+i\phi^2)/\sqrt{2}$,
\beqa
&&\phi^i(x)=\sum_n\sqrt{1\over L\sqrt{M^2+k_n^2}}(a_n^i+(a_n^i)^{\dag})
F_n(x),
\nn\\
&&\dot{\phi}^i(x)=\sum_n\sqrt{\sqrt{M^2+k_n^2}\over L}(-ia_n^i
+i(a_n^i)^{\dag})
F_n(x),
\nn
\eeqa
where $F_n(x)=f_n(x)$ ({\it resp}. $h_n(x)$) in the Ramond sector
({\it resp}. NS sector).
The Hamiltonian of the system is the standard one
\beq
H_{\rm boson}=\sum_{n=0}^{\infty}\sum_{i=1,2}
\sqrt{M^2+k_n^2}\left((a_n^i)^{\dag}a_n^i
+{1\over 2}\right).
\eeq

For the fermion system,
it is simply to require the anticommutation relations
to the mode coefficients
\beq
\{b^{}_{n,\alpha},b^{\dag}_{m,\beta}\}=\delta_{n,m}\delta_{\alpha,\beta},
\qquad
\{b^{}_{n,\alpha},b^{}_{m,\beta}\}
=\{b^{\dag}_{n,\alpha},b^{\dag}_{m,\beta}\}
=0,
\eeq
where the indices $\alpha,\beta$ are for $\pm$.
In the Ramond sector, we also have
the non-oscillating modes, $b_0$ and $b_0^{\dag}$, which obey
\beq
\{b_0,b_0^{\dag}\}=1,\quad b_0^2=(b_0^{\dag})^2=0,
\eeq
and anticommute with all others, $b_{n,\alpha}$ and $b_{n,\alpha}^{\dag}$.
The Hamiltonian is given by
\beq
H_{\rm fermion}
=\sum_{n,\alpha}\sqrt{M^2+k_n^2}
\left(b_{n,\alpha}^{\dag}b_{n,\alpha}-{1\over 2}\right),
\eeq
with the addition of 
\beq
H_{\rm fermion \,\,0}=\Rp(\sigma)\left(b_0^{\dag}b_0-{1\over 2}\right),
\eeq
in the Ramond sector.
Here, we have chosen the standard operator ordering
$b^{\dag}b\to {1\over 2}[b^{\dag},b]=b^{\dag}b-{1\over 2}$.

The ground state of the system is 
the state $|0\rangle$ annihilated by all 
annihilation operators
\beq
a_n^i|0\rangle=b_{n,\alpha}|0\rangle=0.
\eeq
In the Ramond sector,
we need an additional condition --- if $\Rp(\sigma)>0$,
the ground state must be annihilated by $b_0$ while 
it must be annihilated by $b_0^{\dag}$ if
$\Rp(\sigma)<0$:
\beq
\begin{array}{ll}
b_0|0\rangle=0,&
\Rp(\sigma)>0,\\
b_0^{\dag}|0\rangle=0,&
\Rp(\sigma)<0.
\end{array}
\label{fgst}
\eeq

The energy gap to the first excited state is
$\sqrt{M^2+k_0^2}$. In the Ramond sector $k_0^2$ is understood as $-S^2$
and thus the gap is $\sqrt{M^2-S^2}=|\Rp(\sigma)|$.
For $\Rp(\sigma)=0$, the gap vanishes.
In this case,
the $n=0$ exponential modes for the scalars have vanishing potential, 
see (\ref{RbL}), and the ground state wavefunction
is not normalizable in the $\phi_0$ directions.
In addition, there is a two-fold degeneracy from the exponential mode from
the fermion.
In the NS sector, $k_0^2$ is non-negative for $S\leq 0$ but is
negative, $k_0^2=-\kappa_0^2$, for $S>0$.
As $\Rp(\sigma)$ approaches zero while $S$ is positive, 
$M^2+k_0^2=M^2-\kappa_0^2$ turns negative (this occurs when
$|\Rp(\sigma)|\ll 2S\e^{-SL/2}$ provided $SL\gg 1$).
In such a case, the potential for the $n=0$ exponential mode
is unbounded below, see (\ref{NSbL}), 
and there is no ground state in the system.
The appearance of the zero mode or unstable mode at special values of
$\sigma$
is extremely important and plays a crucial r\^ole
later in this paper.
For now, we focus on the cases where there is a unique normalizable
ground state and study its energy and charge.

\subsection{The Energy}
\label{subsec:energy}

Let us compute the energy and its density of the ground state.

\subsubsection{Total Energy}

\subsubsection*{R sector}

The computation in the Ramond sector is extremely simple:
For each non-zero mode, the positive energy from
the boson and the negative energy from the fermion cancel out;
$$
\sqrt{M^2+k_n^2}\left({1\over 2}+{1\over 2}\right)
+\sqrt{M^2+k_n^2}\left(-{1\over 2}-{1\over 2}\right)
=0.
$$
The ground state energy from the bosonic zero mode is
$\sqrt{M^2+k_0^2}\left({1\over 2}+{1\over 2}\right)=\sqrt{M^2-S^2}
=|\Rp(\sigma)|$.
The ground state energy from the fermionic zero mode is
$\Rp(\sigma)\left(-{1\over 2}\right)$ if $\Rp(\sigma)>0$
and $\Rp(\sigma)\left({1\over 2}\right)$ if $\Rp(\sigma)<0$ (see
(\ref{fgst})), that is,
$-{1\over 2}|\Rp(\sigma)|$ for any case.
They fail to cancel against each other,
$
|\Rp(\sigma)|-{1\over 2}|\Rp(\sigma)|={1\over 2}|\Rp(\sigma)|.
$
Thus, the gound state energy in the Ramond sector is
\beq
E_{\rm vac}=
{1\over 2}|\Rp(\sigma)|
\label{EvacR}
\eeq

\subsubsection*{NS Sector}

The energy of the ground state in the NS sector
is less straightforward to compute, but it is possible to
find the answer
in the limit $L\to\infty$.

Let us first start with the $S=0$ case where the bosons have
integer modes $k_n={\pi n\over L}$, $n=0,1,2,...$
 and fermions have half-integer modes $k_r={\pi r\over L}$,
$r={1\over 2}, {3\over 2},{5\over 2},...$.
The ground state energy is
\beq
E_{\rm vac}=\sum_{n=0}^{\infty}\sqrt{M^2+k_n^2}
-\sum_{r={1\over 2},{3\over 2},...}\sqrt{M^2+k_r^2}
\label{EvacNS0}
\eeq
This is infinity minus infinity and we need an appropriate regularization
to define the sum.
We do this using the fact that the sum over
\beqa
I_n&:=&\sqrt{M^2+\left({\pi n\over L}\right)^2}
-{1\over 2}\sqrt{M^2+\left({\pi \left(n-{1\over 2}\right)\over L}\right)^2}
-{1\over 2}\sqrt{M^2+\left({\pi \left(n+{1\over 2}\right)\over L}\right)^2}
\nn\\
&&=-{M^2\over 2}\left({\pi \over 2L}\right)^2\left(
M^2+\left({\pi n\over L}\right)^2\right)^{-{3\over 2}}
+{1\over L^3}O\left({1\over \left({\pi n\over L}\right)^2}\right)
\nn
\eeqa
is finite.
Namely, we evaluate the sum (\ref{EvacNS0}) as
\beqa
E_{\rm vac}&=&
{1\over 2}\sum_{n=0}^{\infty}\left(
\sqrt{M^2+\left({\pi n\over L}\right)^2}
-\sqrt{M^2+\left({\pi \left(n+{1\over 2}\right)\over L}\right)^2}\right)
\nn\\
&&+{1\over 2}\left\{M+\sum_{n=1}^{\infty}\left(
\sqrt{M^2+\left({\pi n\over L}\right)^2}
-\sqrt{M^2+\left({\pi \left(n-{1\over 2}\right)\over L}\right)^2}
\right)
\right\}
\nn\\
&=&{M\over 2}+{1\over 2}\left(M-\sqrt{M^2+\left({\pi\over 2L}\right)^2}\right)
+\sum_{n=1}^{\infty}I_n
\nn\\
&=&{M\over 2}+O\left({1\over ML^2}\right)+O\left({1\over L}\right)
\nn
\eeqa
Thus, in the limit $L\to \infty$ we have
\beq
\lim_{L\to\infty}E_{\rm vac}={M\over 2}={1\over 2}|\Rp(\sigma)|.
\label{limEvacNS0}
\eeq
The validity of this regularization will be examined momentarily.

Let us move on to the case with non-zero $S$.
We are interested in the limit of large $L$ for a fixed $S$, so that
we may assume $|S|\gg 1/L$.
We divide the total energy into two parts,
the part $E'_{\rm vac}$ coming from the oscillating or constant
modes and the part coming from the exponential modes.
The latter part is present for $S>0$ but is absent
for $S\leq 0$.
To find $E'_{\rm vac}$,
we first compute the derivative
${\partial\over \partial S}E_{\rm vac}'$ and then integrate,
$$
{\partial E'_{\rm vac}\over\partial S}=
{\sum_n}'{k_n\over\sqrt{M^2+k_n^2}}{\partial k_n\over \partial S}
-{\sum_r}'{k_r\over\sqrt{M^2+k_r^2}}{\partial k_r\over \partial S}.
$$
The sum is over $n=0,1,2,...$, $r={1\over 2},{3\over 2},...$ if $S\leq 0$
while it omits $n=0,1$ and $r={1\over 2}$ if $S\gg 1/L$.
The equations determining $k_n$ and $k_r$ are in (\ref{coteqn}) 
and (\ref{taneqn}) respectively, from which it follows
$$
{\partial k_n\over \partial S}
=-{1\over L}{2k_n\over k_n^2+S^2}(1+\cdots),\qquad
{\partial k_r\over \partial S}
=-{1\over L}{k_r\over k_r^2+S^2}(1+\cdots),
$$
where $+\cdots$ are terms that vanish as $L\to \infty$ (in what follows such
terms will not be mentioned).
Thus we find
\beqa
{\partial E'_{\rm vac}\over \partial S}
&=&-2{\sum_n}'{1\over L}{k_n^2\over \sqrt{M^2+k_n^2}(k_n^2+S^2)}
+{\sum_r}'{1\over L}{k_r^2\over \sqrt{M^2+k_r^2}(k_r^2+S^2)}
\nn\\
&\stackrel{L\to\infty}{\longrightarrow}&
-\int_0^{\infty}{\dd k\over \pi}{k^2\over\sqrt{M^2+k^2}(k^2+S^2)}
\nn
\eeqa
This is logarithmically divergent
 and we introduce a cut-off $\Lambda\gg M$:
\beqa
{\partial E'_{\rm vac}\over \partial S}
&=&
-\int_0^{\Lambda}{\dd k\over \pi}\left({1\over \sqrt{M^2+k^2}}
-{S^2\over \sqrt{M^2+k^2}(k^2+S^2)}\right)
\nn\\
&=&-{1\over \pi}\log\left({2\Lambda\over M}\right)
+\int_0^{\Lambda}{\dd k\over \pi}{S^2\over \sqrt{M^2+k^2}(k^2+S^2)}
\nn\\
&=&-{1\over \pi}\log\left({2\Lambda\over M}\right)
+{1\over \pi}\tan\theta\left(\pm {\pi\over 2}-\theta\right),
\qquad
S\begin{array}{c}>\\[-0.4cm]
<
\end{array}
0.
\nn
\eeqa
where $S=M\sin\theta$, with $0< \theta <\pi$ or $-\pi<\theta<0$. 
Integrating, we find for $\pm S>0$
\beqa
\lefteqn{
E'_{\rm vac}(S)-E'_{\rm vac}(0_{\pm})}\nn\\
&=&-{S\over \pi}\left[\log\left({2\Lambda\over M}\right)+1\right]
\mp{\sqrt{M^2-S^2}\over 2}\pm {M\over 2}
+{\sqrt{M^2-S^2}\over \pi}\arctan\left({S\over\sqrt{M^2-S^2}}\right).
\nn
\eeqa
Here, arctangent is assumed to take values between $-{\pi\over 2}$
and ${\pi\over 2}$.
Let us now determine $E'_{\rm vac}(0_{\pm})$. We recall that the spectrum
of oscillating modes is continuous as $S$ approaches $0$ from below
(the lowest mode converges to the zero mode).
Thus we have $E'_{\rm vac}(0_-)=E'_{\rm vac}(0)=E_{\rm vac}(0)={M\over 2}$.
On the other hand, we loose {\it two} bosonic and
{\it one} fermionic oscillating modes as $S$ is increased beyon $2/L$.
Since each mode comes with a pair
(complex for boson and $\pm$ for fermion), we have
$$
E'_{\rm vac}(0_+)=-2M+M+E'_{\rm vac}(0)=-{M\over 2}.
$$
Using these we find
\beqa
E'_{\rm vac}(S)
\!\!&=&\!\!
-{S\over \pi}\left[\log\left({2\Lambda\over M}\right)+1\right]
\mp{\sqrt{M^2-S^2}\over 2}
+{\sqrt{M^2-S^2}\over \pi}\arctan\left({S\over\sqrt{M^2-S^2}}\right),
\nn\\
&&\qquad\qquad\qquad\,\,
\mbox{for}\,\,\,\begin{array}{l}
S>0\\[-0.2cm]
S\leq 0.
\end{array}
\nn
\eeqa
If $S>0$, which actually means $S\gg 2/L$,
we need to add the contribution from
the exponential modes,
$$
\sqrt{M^2-\kappa_0^2}+\sqrt{M^2-\kappa_1^2}-\sqrt{M^2-\kappa^2}
\simeq \sqrt{M^2-S^2}.
$$
After the addition, the term $\mp {\sqrt{M^2-S^2}\over 2}$ in $E'_{\rm vac}$
becomes just $+{\sqrt{M^2-S^2}\over 2}$ for any value of $S$.
In this way, we find that
the total energy is
\beq
E_{\rm vac}
=
-{\Ip(\sigma)\over \pi}\left[\log\left({2\Lambda\over |\sigma|}\right)+1\right]
+{|\Rp(\sigma)|\over 2}
+{|\Rp(\sigma)|\over \pi}\arctan\left({\Ip(\sigma)\over |\Rp(\sigma)|}\right).
\label{EvacNS}
\eeq

\subsubsection{Energy Density}
\label{subsub:Edensity}

Let us next compute the vacuum energy density, defined as
the vacuum expectation value
\beq
{\mathcal E}(x)=\langle 0|{\mathcal H}(x)|0\rangle
\eeq
of the Hamiltonian density
$$
{\mathcal H}(x)=|\dot{\phi}(x)|^2
+\bphi(x)(-\partial_x^2+M^2)\phi(x)
+{1\over 2}[\Psi(x)^{\dag},{\mathcal D}\Psi(x)].
$$

\subsubsection*{R Sector}

In the Ramond sector it is
\beqa
{\mathcal E}_{\rm R}(x)
&=&2\sum_{n=0}^{\infty}{\sqrt{M^2+k_n^2}\over L}f_n(x)^2
-\sum_{n=0}^{\infty}{\sqrt{M^2+k_n^2}\over L}(f_n(x)^2+g_n(x)^2)
\nn\\
&=&\sum_{n=0}^{\infty}{\sqrt{M^2+k_n^2}\over L}(f_n(x)^2-g_n(x)^2)
\nn\\
&=&\sum_{n=1}^{\infty}{\sqrt{M^2+k_n^2}\over L}{k_n^2\cos(2k_nx)+
k_nS\sin(2k_nx)\over k_n^2+S^2}
+{\sqrt{M^2-S^2}\over L}{SL\over \e^{2SL}-1}\e^{2Sx}
\nn
\eeqa
Let us focus on the region near the left boundary
$x=0$, and take the $L\to\infty$ limit.
In this limit, the last term becomes
$$
{\sqrt{M^2-S^2}\over L}{SL\over \e^{2SL}-1}\e^{2Sx}
\longrightarrow
\left\{\begin{array}{ll}
0&S> 0\\
{M\over 2L}\to 0&S=0\\
-\sqrt{M^2-S^2}S\e^{2Sx}&S<0.
\end{array}\right.
$$
Thus, the density can be written in the limit as
\beq
\lim_{L\to\infty}
{\mathcal E}_{\rm R}(x)=\int_0^{\infty}{\dd k\over \pi}\sqrt{M^2+k^2}
{k^2\cos(2kx)+kS\sin(2kx)\over k^2+S^2}-\theta_{S<0}|\Rp(\sigma)|S\e^{2Sx}.
\label{ERnear0}
\eeq
If $x$ is strictly away from the boundary, $x>0$, one may proceed
the computation as follows:
\begin{figure}[htb]
\centerline{\includegraphics{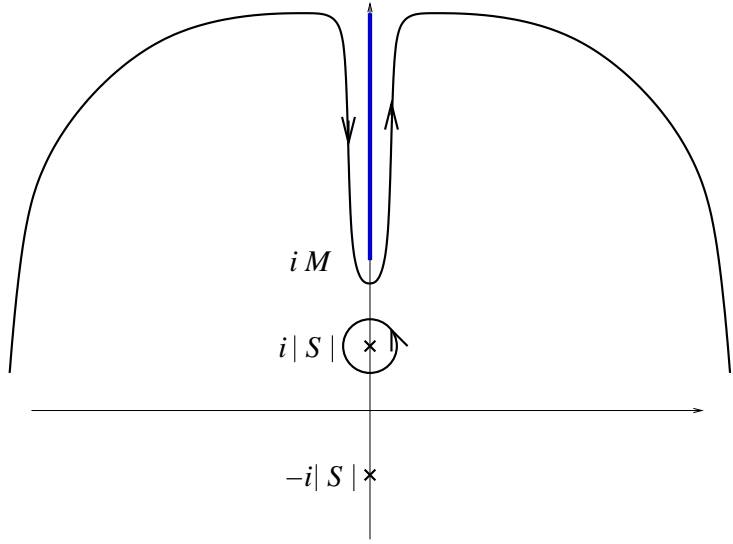}}
\caption{Deformed contour.}
\label{contour1}
\end{figure}
\beqa
\lim_{L\to\infty}{\mathcal E}_{\rm R}(x)&=&
\int_0^{\infty}
{\dd k\over \pi}\sqrt{M^2+k^2}{k\over 2}\left[{k-iS\over k^2+S^2}
\e^{2ikx}+{k+iS\over k^2+S^2}\e^{-2ikx}\right]
-\theta_{S<0}|\Rp(\sigma)|S\e^{2Sx}
\nn\\
&=&\int_{-\infty}^{\infty}{\dd k\over 2\pi}\sqrt{M^2+k^2}{k\over k+iS}\e^{2ikx}
-\theta_{S<0}|\Rp(\sigma)|S\e^{2Sx}
\nn\\
&=&\int_M^{\infty}{\dd iK\over 2\pi}\left(i\sqrt{K^2-M^2}-(-i\sqrt{K^2-M^2})
\right){iK\over iK+iS}\e^{-2Kx}
\nn\\
&&+\theta_{S<0}{2\pi i\over 2\pi}\sqrt{M^2-S^2}
(-iS)\e^{2Sx}-\theta_{S<0}|\Rp(\sigma)|S\e^{2Sx}
\nn\\
&=&-\int_M^{\infty}{\dd K\over \pi}\sqrt{K^2-M^2}{K\over K+S}\e^{-2Kx}.
\label{intprocess}
\eeqa
From the second to the third line, the contour is deformed
as in Fig.~\ref{contour1} using the fact that $x$ is strictly
positive. 
Note that there is a simple pole at $k=-iS$ only when $S$ is negative, 
which produces the additional term $\theta_{S<0}(\cdots)$,
which in turn is canceled against the contribution from
the exponential mode.

We see that the density is localized near the boundary $x=0$ with widths $1/M$.
Let us compute the total of this localized energy
by integrating the density {\it after} the limit $L\to \infty$.
(This is different from integration before the limit,
which simply reproduces the total energy given in (\ref{EvacR}).)
Since the integration domain includes $x=0$, 
we must go back to the expression (\ref{ERnear0})
before the contour deformation of Fig.~\ref{contour1}.
Also, we need to cut off the momentum integral at $k=\Lambda$:
\beqa
E_{x\sim 0}\!\!&=&\!
\lim_{R\to\infty}\int_0^R\dd x\,\lim_{L\to\infty}{\mathcal E}_R(x)
\Bigr|_{{\rm cut \,off}\,\Lambda}
\nn\\
&=&\!\!\lim_{R\to\infty}\int_0^R\dd x
\left[\int_0^{\Lambda}{\dd k\over \pi}\sqrt{M^2+k^2}{k^2\cos(2kx)+kS\sin(2kx)
\over k^2+S^2}-\theta_{S<0}|\Rp(\sigma)|S\e^{2Sx}\right]
\nn\\
&=&\lim_{R\to\infty}\int_0^{\Lambda}
{\dd k\over \pi}{\sqrt{M^2+k^2}\over k^2+S^2}
\left[k^2{\sin(2kR)\over 2k}+S{-\cos(2kR)+1\over 2}\right]
+\theta_{S<0}{|\Rp(\sigma)|\over 2}.
\nn
\eeqa
At this stage we use
$$
\lim_{R\to\infty}{\sin(2kR)\over 2k}={\pi\over 2}\delta(k),
$$
and also the fact that
$$
\int_0^{\infty}{\dd k\over \pi}{\sqrt{M^2+k^2}\over k^2+S^2}
\cos(2kR)
$$
decays exponentially as $RM, RS\to\infty$ --- as can be shown by
deforming the contour as in (\ref{intprocess}).
Then we find
\beq
E_{x\sim 0}
=S\int^{\Lambda}_0{\dd k\over 2\pi}{\sqrt{M^2+k^2}\over k^2+S^2}
+\delta_{S,0}{M\over 4}+\theta_{S<0}{|\Rp(\sigma)|\over 2}.
\eeq
The integral on the right hand side is defined to be zero
for $S=0$. For $S\ne 0$, it can be evaluated;
\beqa
\lefteqn{S\int_0^{\Lambda}{\dd k\over 2\pi}{\sqrt{M^2+k^2}\over k^2+S^2}}
\nn\\
&=&{S\over 2\pi}\left[\log\left({2\Lambda\over M}\right)+1\right]
-{\sqrt{M^2-S^2}\over 2\pi}\arctan\left({S\over\sqrt{M^2-S^2}}\right)
+{1\over 4}{\rm sgn}(S)\sqrt{M^2-S^2}.
\nn
\eeqa
Thus the localized energy at $x=0$ is
\beq
E_{x\sim 0}
={\Ip(\sigma)\over 2\pi}\left[\log\left({2\Lambda\over |\sigma|}\right)
+1\right]
-{|\Rp(\sigma)|\over 2\pi}
\arctan\left({\Ip(\sigma)\over |\Rp(\sigma)|}\right)
+{1\over 4}|\Rp(\sigma)|.
\label{ERloc0}
\eeq

Let us next compute the energy localized near the right boundary $x=L$,
where we use the coordinate $y=x-L\leq 0$.
Namely, we take the limit $L\to\infty$ of ${\mathcal E}_{\rm R}(x)$
 keeping $y$ finite.
Since the wavenumbers are $k_n={\pi n/L}$, we have equations
$\sin(2k_nx)=\sin(2k_ny)$ and $\cos(2k_nx)=\cos(2k_ny)$
that allow us to write the energy density in the $y$ coordinate:
$$
{\mathcal E}_{\rm R}(x)
=\sum_{n=1}^{\infty}{\sqrt{M^2+k_n^2}\over L}{k_n^2\cos(2k_ny)+
k_nS\sin(2k_ny)\over k_n^2+S^2}+
{\sqrt{M^2-S^2}\over L}{SL\over 1-\e^{-2SL}}\e^{2Sy}
$$
The limit can now be taken straightforwardly:
\beq
\lim_{L\to\infty}{\mathcal E}_R(x)
=\int_0^{\infty}{\dd k\over \pi}\sqrt{M^2+k^2}
{k^2\cos(2ky)+kS\sin(2ky)\over k^2+S^2}+\theta_{S>0}|\Rp(\sigma)|S\e^{2Sy}.
\eeq
Proceeding as before, we find
\beq
\lim_{L\to\infty}{\mathcal E}_{\rm R}(x)=
-\int_M^{\infty}{\dd K\over \pi}\sqrt{K^2-M^2}
{K\over K-S}\e^{2Ky}\qquad\mbox{for $y<0$ strictly},
\eeq
and
\beqa
E_{x\sim L}\!\!&=&\!\!\!\!\lim_{R\to\infty}\int_{-R}^0
\dd y\left[
\int_0^{\Lambda}{\dd k\over \pi}\sqrt{M^2+k^2}
{k^2\cos(2ky)+kS\sin(2ky)\over k^2+S^2}
+\theta_{S>0}|\Rp(\sigma)|S\e^{2Sy}\right]
\nn\\
&=&-S\int_0^{\Lambda}{\dd k\over 2\pi}{\sqrt{M^2+k^2}\over k^2+S^2}
+\delta_{S,0}{M\over 4}+\theta_{S>0}{|\Rp(\sigma)|\over 2}
\nn\\
&=&-{\Ip(\sigma)\over 2\pi}\left[\log\left({2\Lambda\over |\sigma|}\right)
+1\right]
+{|\Rp(\sigma)|\over 2\pi}
\arctan\left({\Ip(\sigma)\over |\Rp(\sigma)|}\right)
+{1\over 4}|\Rp(\sigma)|
\label{ERlocL}
\eeqa
Note that
$$
E_{x\sim 0}+E_{x\sim L}={|\Rp(\sigma)|\over 2}=E_{\rm vac},
$$
see (\ref{EvacR}). 
This means that the total energy
comes purely from the energies localized at the two boundaries. 
This is consistent with the fact
that the energy density vanishes in the bulk
of the interval
${1\over M}\ll x\ll L-{1\over M}$.

\subsubsection*{NS Sector}

Let us next discuss the energy density of the ground state
in the NS sector (\ref{NSbc}). Actually, we know what to expect: At 
the right boundary $x=L$, the boundary condition
 is exactly the same as the Ramond boundary condition
and thus we should obtain
the same answer as ${\mathcal E}_{\rm R}(x)$. 
The NS condition 
at the left boundary $x=0$ is obtained from the Ramond condition at
the right boundary $x=L$, by
$\psi_{\pm}(x)\to \pm \bpsi_{\mp}(L-x)$, $\phi(x)\to \phi(L-x)$.
Therfore we expect that ${\mathcal E}_{\rm NS}(x)$ for $x$ close to $0$ is 
obtained from ${\mathcal E}_{\rm R}(x)$ at $x$ close to $L$,
by the replacement $x\to L-x$. To summarize, the expectation is
\beqa
\lim_{L\to\infty}{\mathcal E}_{\rm NS}(x)\!\!\!&=&\!\!
\left\{\begin{array}{l}
\displaystyle \int_0^{\infty}{\dd k\over \pi}\sqrt{M^2+k^2}
{k^2\cos(2ky)+kS\sin(2ky)\over k^2+S^2}
+\theta_{S>0}|\Rp(\sigma)|S\e^{2Sy}\\
\qquad\qquad\qquad\qquad\qquad\qquad\qquad\qquad\qquad
 x \simleq L,
\\[0.3cm]
\displaystyle \int_0^{\infty}{\dd k\over \pi}\sqrt{M^2+k^2}
{k^2\cos(2kx)-kS\sin(2kx)\over k^2+S^2}
+\theta_{S>0}|\Rp(\sigma)|S\e^{-2Sx}
\\
\qquad\qquad\qquad\qquad\qquad\qquad\qquad\qquad\qquad
 x \simgeq 0,
\end{array}\right.
\label{expfrmR}
\\
&=&\left\{
\begin{array}{ll}
\displaystyle
-\int_M^{\infty}{\dd K\over \pi}\sqrt{K^2-M^2}
{K\over K-S}\e^{2Ky}
&\,\,\,\, x< L\,\,\,\,\mbox{strictly},
\\[0.3cm]
\displaystyle
-\int_M^{\infty}{\dd K\over \pi}\sqrt{K^2-M^2}
{K\over K-S}\e^{-2Kx}
&\,\,\,\, x> 0 \,\,\,\,\mbox{strictly},
\end{array}
\right.
\nn
\eeqa
and
\beqa
\lefteqn{E_{x\sim 0}=E_{x\sim L}=
-S\int_0^{\Lambda}{\dd k\over 2\pi}{\sqrt{M^2+k^2}\over k^2+S^2}
+\delta_{S,0}{M\over 4}+\theta_{S>0}{|\Rp(\sigma)|\over 2}}
\nn\\
&=&
-{\Ip(\sigma)\over 2\pi}\left[\log\left({2\Lambda\over |\sigma|}\right)
+1\right]
+{|\Rp(\sigma)|\over 2\pi}
\arctan\left({\Ip(\sigma)\over |\Rp(\sigma)|}\right)
+{1\over 4}|\Rp(\sigma)|
\eeqa
The sum of the two is
is nothing but the total energy (\ref{EvacNS}) of the
NS vacuum.

One can indeed verify the above expectations.
We split the density into two parts, the part ${\mathcal E}'_{\rm NS}(x)$
from the oscilating or constant modes, and the part 
${\mathcal E}^{\it exp}_{\rm NS}(x)$
from the exponential modes that are present only for $S>0$.
We first consider the contribution from the oscillating/constant modes.
By definition we have
$$
{\mathcal E}'_{\rm NS}(x)=2{\sum_n}'{\sqrt{M^2+k_n^2}\over L}h_n(x)^2
-{\sum_r}'{\sqrt{M^2+k_r^2}\over L}(g_r(x)^2+h_r(x)^2).
$$
See Section~\ref{subsub:NSmodes} for
the quantities that appears here. In particular,
$k_n$ and $k_r$ are defined by the equations
(\ref{coteqn}) and (\ref{taneqn}).
Let us look at the behaviour near the left boundary $x\simgeq 0$,
taking the limit $L\to\infty$ for a finite $x$.
In this limit, the difference between neighboring wavenumbers
disappears $|k_{n}-k_{r=n\pm {1\over 2}}|<{\pi\over L}\to 0$,
and we have $h_n(x)\sim h_r(x)$ for a finite $k_n\sim k_r$.
Thus we have
\beqa
{\mathcal E}'_{\rm NS}(x)&=&{\sum_n}' {\sqrt{M^2+k_n^2}\over L}
(h_n(x)^2-g_{n\pm {1\over 2}}(x)^2)+\cdots\nn\\
&=&{\sum_n}'{\sqrt{M^2+k_n^2}\over L}{k_n^2\cos(2k_nx)-
k_nS\sin(2k_nx)\over k_n^2+S^2}+\cdots
\nn\\
&\stackrel{L\to\infty}{\longrightarrow}&
\int_0^{\infty}{\dd k\over \pi}\sqrt{M^2+k^2}
{k^2\cos(2kx)-kS\sin(2kx)\over k^2+S^2},\qquad
x \simgeq 0,
\nn
\eeqa
where $+\cdots$ are terms that vanish in the $L\to\infty$ limit. 
Let us now look near the right boundary $x\simleq L$: take the limit
$L\to\infty$ with $y=x-L$ kept finite.
To do this, we write everything as a function of $y$.
Using the defining equation of $k_n$ and
$k_r$ we find
\beqa
&&
h_n(x)=\pm{k_n\cos(k_ny)+S\sin(k_ny)\over\sqrt{k_n^2+S^2}}(1+\cdots)
\nn\\
&& 
g_r(x)=\pm {k_r\cos(k_ry)+S\sin(k_ry)\over\sqrt{k_r^2+S^2}}(1+\cdots)
\nn\\
&&
h_r(x)=-\sin(k_ry)(1+\cdots).
\nn
\eeqa
In the limit $L\to \infty$ the sum over the modes becomes 
an integral over $k$ where the difference of
$k_n$ and $k_r$ disappears.
Noting that
$h_n(x)^2\sim g_r(x)^2\ne h_r(x)^2$ for $k_n\sim k_r$,
we find
\beqa
{\mathcal E}'_{\rm NS}(x)&=&{\sum_n}' {\sqrt{M^2+k_n^2}\over L}
(h_n(x)^2-h_{n\pm {1\over 2}}(x)^2)+\cdots\nn\\
&=&{\sum_n}'{\sqrt{M^2+k_n^2}\over L}{k_n^2\cos(2k_ny)+
k_nS\sin(2k_ny)\over k_n^2+S^2}+\cdots
\nn\\
&\stackrel{L\to\infty}{\longrightarrow}&
\int_0^{\infty}{\dd k\over \pi}\sqrt{M^2+k^2}
{k^2\cos(2ky)+kS\sin(2ky)\over k^2+S^2},\qquad
x \simleq L.
\nn
\eeqa
We next consider the contribution from the exponential modes, 
which are given by
(\ref{expoh0}), (\ref{expoh1}) and (\ref{expoPsi}).
These functions behave as follows, in the limit
$L\to \infty$ for finite
$x$ and finite $y=x-L$,
\beqa
\mbox{(finite $x$)}:
&&\!\!\!\left\{
\begin{array}{l}
h_0(x), h_1(x)\to\sqrt{SL\over 2}\e^{-Sx},\sqrt{SL\over 2}\e^{-Sx}\\
\Psi_{{1\over 2},+}(x),\Psi_{{1\over 2},-}(x)\to 
\left\{\begin{array}{ll}
\left(\begin{array}{c}
0\\0
\end{array}\right), \left(\begin{array}{c}
0\\i\sqrt{2S}\e^{-Sx}
\end{array}\right)&\Rp(\sigma)>0\\
\left(\begin{array}{c}
0\\-i\sqrt{2S}\e^{-Sx}
\end{array}\right),\left(\begin{array}{c}
0\\0
\end{array}\right)&\Rp(\sigma)<0
\end{array}\right.
\end{array}
\right.
\nn\\
\mbox{(finite $y$)}:
&&\!\!\!\left\{
\begin{array}{l}
h_0(x), h_1(x)\to\sqrt{SL\over 2}\e^{Sy},-\sqrt{SL\over 2}\e^{Sy}\\
\Psi_{{1\over 2},+}(x),\Psi_{{1\over 2},-}(x)\to 
\left\{\begin{array}{ll}
\left(\begin{array}{c}
\sqrt{2S}\e^{Sy}\\0
\end{array}\right), \left(\begin{array}{c}
0\\0
\end{array}\right)&\Rp(\sigma)>0\\
\left(\begin{array}{c}
0\\0
\end{array}\right),\left(\begin{array}{c}
\sqrt{2S}\e^{Sy}\\0
\end{array}\right)&\Rp(\sigma)<0
\end{array}\right.
\end{array}
\right.
\nn
\eeqa
The contribution to the energy density is
\beqa
{\mathcal E}^{\it exp}_{\rm NS}(x)
&=&\sqrt{M^2-\kappa_0^2}{2\over L}h_0(x)^2
+\sqrt{M^2-\kappa_1^2}{2\over L}h_1(x)^2
\nn\\
&&
-{\sqrt{M^2-\kappa^2}\over 2}\Bigl(\Psi_{{1\over 2},+}(x)^{\dag}
\Psi_{{1\over 2},+}(x)
+\Psi_{{1\over 2},-}(x)^{\dag}
\Psi_{{1\over 2},-}(x)\Bigr)
\nn\\
&\stackrel{x\, {\rm finite}}{\longrightarrow}&
|\Rp(\sigma)|S\e^{-2Sx}+|\Rp(\sigma)|S\e^{-2Sx}
-{|\Rp(\sigma)|\over 2}2S\e^{-2Sx}
=|\Rp(\sigma)|S\e^{-2Sx}
\nn\\
&\stackrel{y\, {\rm finite}}{\longrightarrow}&
|\Rp(\sigma)|S\e^{2Sy}+|\Rp(\sigma)|S\e^{2Sy}
-{|\Rp(\sigma)|\over 2}2S\e^{2Sy}
=|\Rp(\sigma)|S\e^{2Sy}.
\nn
\eeqa
The sum
${\mathcal E}'_{\rm NS}(x)+{\mathcal E}^{\it exp}_{\rm NS}(x)$ 
leads to the expected answer (\ref{expfrmR}) in the limit
$L\to\infty$.

\subsection{The Charge}
\label{subsec:charge}

Let us compute the charge of the ground state.
The charge density operator is defined by
\beq
j^0
={i\over 2}\{\bphi,\dot{\phi}\}
-{i\over 2}\{\phi,\dot{\bphi}\}
+{1\over 2}[\Psi^{\dag},\Psi]
\eeq
and we want to compute the eigenvalue of the total charge
$$
Q=\int_0^Lj^0(x)\dd x
$$
of the ground state $|0\rangle$ and the vacuum expectation value
of the density
$$
\rho(x)=\langle 0|j^0(x)|0\rangle.
$$

\subsubsection*{R Sector}

Using the mode expansion, we find that the density operator is
expressed as
\beqa
j^0(x)\!\! &=&\!\!\!\!
\sum_{n,m=0}^{\infty}
{i\over L}\left({M^2+k_m^2\over M^2+k_n^2}\right)^{1\over 4}\!
f_n(x)f_m(x)\left[
(a^1_n+a^{1\dag}_n)(a^2_m-a^{2\dag}_m)
-(a^2_n+a^{2\dag}_n)(a^1_m-a^{1\dag}_m)\right]
\nn\\
&&\!\!+
\sum_{n,m=0}^{\infty}
\Biggl(
{1\over 2}\left[b_{n,+}^{\dag},b_{m,+}\right]
\Psi_{n,+}(x)^{\dag}\Psi_{m,+}(x)
-{1\over 2}\left[b_{m,-}^{\dag},b_{n,-}\right]
\Psi_{n,-}(x)^{\dag}\Psi_{m,-}(x)
\nn\\
&&\qquad\qquad +b_{n,-}b_{m,+}
\Psi_{n,-}(x)^{\dag}\Psi_{m,+}(x)
+b_{n,+}^{\dag}b_{m,-}^{\dag}
\Psi_{n,+}(x)^{\dag}\Psi_{m,-}(x)
\Biggr),
\label{j0modes}
\eeqa
in which we set $\Psi_{0,+}(x)=\Psi_0(x)$ and
$\Psi_{0,-}(x)=0$, and
the total charge is
\beq
Q=\sum_{n=1}^{\infty}\left[\,
i\left(a_n^{1\dag}a_n^2-a_n^{2\dag}a_n^1\right)
+b_{n,+}^{\dag}b_{n,+}
-b_{n,-}^{\dag}b_{n,-}\,\right]
+\left[\,
i\left(a_0^{1\dag}a_0^2-a_0^{2\dag}a_0^1\right)
+b_0^{\dag}b_0-{1\over 2}\,\right].
\label{QtotR}
\eeq
The ground state $|0\rangle$ is annihilated by $b_0$ if $\Rp(\sigma)>0$
and by $b_0^{\dag}$ if $\Rp(\sigma)<0$. Thus, it has charge
\beq
Q_{\rm vac}=\left\{\begin{array}{cl}
-{1\over 2}&\quad \Rp(\sigma)>0\\
{1\over 2}&\quad \Rp(\sigma)<0
\end{array}\right.
\,\,=\,-{1\over 2}{\rm sgn}\Rp(\sigma).
\label{QvacR}
\eeq
The charge density is
\beqa
\rho_{\rm R}(x)&=&
\sum_{n=1}^{\infty}\left[\,\, 
-{1\over 2}\Psi_{n,+}(x)^{\dag}\Psi_{n,+}(x)
+{1\over 2}\Psi_{n,-}(x)^{\dag}\Psi_{n,-}(x)\,\,\right]
-{1\over 2}{\rm sgn}\Rp(\sigma)\Psi_0(x)^{\dag}\Psi_0(x)
\nn\\
&=&-\sum_{n=1}^{\infty}{\Rp(\sigma)\over L\sqrt{M^2+k_n^2}}(f_n(x)^2-g_n(x)^2)
-{1\over 2}{\rm sgn}\Rp(\sigma){2S\e^{2Sx}\over \e^{2SL}-1}
\nn\\
&=&-\sum_{n=1}^{\infty}{\Rp(\sigma)\over L\sqrt{M^2+k_n^2}}
{k_n^2\cos(2k_n x)+k_nS\sin(2k_nx)\over
k_n^2+S^2}
-{\rm sgn}\Rp(\sigma){S\e^{2Sx}\over \e^{2SL}-1}.
\label{rhoR}
\eeqa
Let us look at the neighborhood of the left boundary $x=0$ by taking
the limit $L\to\infty$ for a finite $x$. In this limit we have
\beq
\lim_{L\to\infty}\rho_{\rm R}(x)
=-\int_0^{\infty}{\dd k\over \pi}{\Rp(\sigma)\over\sqrt{M^2+k^2}}
{k^2\cos(2kx)+kS\sin(2kx)\over k^2+S^2}
+\theta_{S<0}{\rm sgn}\Rp(\sigma)S\e^{2Sx}.
\eeq
If $x$ is strictly away from the boundary $x>0$, it is
\beq
\lim_{L\to\infty}\rho_{\rm R}(x)=
-\int_M^{\infty}{\dd K\over\pi}{\Rp(\sigma)K\over\sqrt{K^2-M^2}(K+S)}
\e^{-2Kx}.
\eeq
We see that the charge is localized near the boundary with width $1/M$.
The total localized charge is
\beqa
Q_{x\sim 0}&=&\lim_{R\to\infty}\int_0^R\lim_{L\to\infty}\rho_{\rm R}(x)\dd x
\nn\\
&=&{\rm sgn}(\Rp(\sigma))\left[\,{1\over 2\pi}
\arctan\left({\Ip(\sigma)\over |\Rp(\sigma)|}
\right) -{1\over 4}\, \right].
\label{QloclR}
\eeqa
Let us next look at the neighborhood of the right boundary $x=L$.
In the limit $L\to\infty$ for a finite $y=x-L$, we have
\beq
\lim_{L\to\infty}\rho_{\rm R}(x)
=-\int_0^{\infty}{\dd k\over \pi}{\Rp(\sigma)\over\sqrt{M^2+k^2}}
{k^2\cos(2ky)+kS\sin(2ky)\over k^2+S^2}
-\theta_{S>0}{\rm sgn}\Rp(\sigma)S\e^{2Sy}.
\eeq
If $x$ is strictly away from the boundary $x<L$, i.e. $y<0$, it is
\beq
\lim_{L\to\infty}\rho_{\rm R}(x)=
-\int_M^{\infty}{\dd K\over\pi}{\Rp(\sigma)K\over\sqrt{K^2-M^2}(K-S)}
\e^{2Ky}.
\eeq
Again, the charge is localized near the boundary.
The total is
\beqa
Q_{x\sim L}&=&\lim_{R\to\infty}\int_{-R}^0\lim_{L\to\infty}\rho_{\rm R}(x)\dd y
\nn\\
&=&{\rm sgn}(\Rp(\sigma))\left[\,-{1\over 2\pi}
\arctan\left({\Ip(\sigma)\over |\Rp(\sigma)|}
\right)- {1\over 4}\, \right].
\label{QRlocL}
\eeqa
We note that the sum of the localized charges
$Q_{x\sim 0}+Q_{x\sim L}$ reproduces the total charge given by
(\ref{QvacR}).

\subsubsection*{NS Sector}

\begin{figure}[htb]
\centerline{\includegraphics{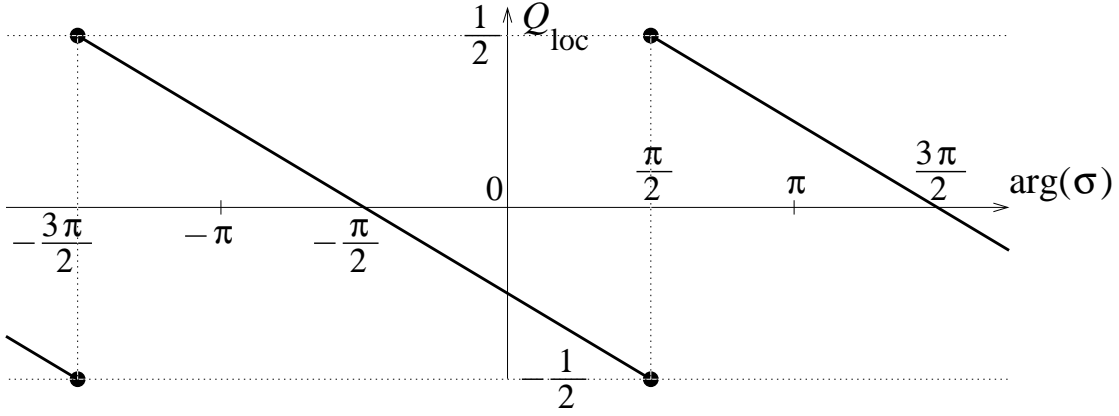}}
\caption{Charge localized near the right boundary of an interval}
\label{fig:Qloc1}
\end{figure}

The charge density operator $j^0(x)$ can be expanded in terms of
the modes as in (\ref{j0modes}), from which it follows that
\beq
Q=\sum_ni\left(\,a_n^{1\dag}a_n^2-a_n^{2\dag}a_n^1\,\right)
+\sum_r\left(\,b_{r,+}^{\dag}b_{r,+}
-b_{r,-}^{\dag}b_{r,-}\,\right).
\eeq
In particular, the vacuum has zero total charge,
\beq
Q_{\rm vac}=0.
\eeq
The charge density of the ground state
can be determined either directly or by employing the symmetry
argument as in the computation of the energy density.
In the latter method, we consider the transformation
$\psi_{\pm}(x)\to \pm\bpsi_{\mp}(L-x)$, $\phi(x)\to \bphi(L-x)$
that maps the Ramond boundary condition at $x=L$ to the
NS boundary condition at $x=0$. Note that it flips the sign of
the charge density, $j^0(x)\to -j^0(x-L)$.
In either way we find
\beqa
\rho_{\rm NS}(x)\!\!\!\!\!&=&\!\!\!\!\!
-{\sum_r}'{\Rp(\sigma)\over L\sqrt{M^2+k_r^2}}(g_r(x)^2-h_r(x)^2)
-{1\over 2}\theta_{S>0}\left(|\!|\Psi_{{1\over 2},+}(x)|\!|^2
-|\!|\Psi_{{1\over 2},-}(x)|\!|^2\right)
\nn\\
&\stackrel{L\to\infty}{\longrightarrow}&\!\!
\left\{\begin{array}{l}
\displaystyle -\int_0^{\infty}{\dd k\over \pi}
{\Rp(\sigma)\over\sqrt{M^2+k^2}}
{k^2\cos(2ky)+kS\sin(2ky)\over k^2+S^2}
-\theta_{S>0}{\rm sgn}\Rp(\sigma)S\e^{2Sy}
\\
\qquad\qquad\qquad\qquad\qquad\qquad\qquad\qquad\qquad\qquad\qquad\qquad
x \simleq L,
\\[0.3cm]
\displaystyle \int_0^{\infty}{\dd k\over \pi}
{\Rp(\sigma)\over \sqrt{M^2+k^2}}
{k^2\cos(2kx)-kS\sin(2kx)\over k^2+S^2}
+\theta_{S>0}{\rm sgn}\Rp(\sigma)S\e^{-2Sx}
\\
\qquad\qquad\qquad\qquad\qquad\qquad\qquad\qquad\qquad\qquad\qquad\qquad
x \simgeq 0,
\end{array}\right.
\nn\\
&=&\left\{
\begin{array}{ll}
\displaystyle
-\int_M^{\infty}{\dd K\over \pi}
{\Rp(\sigma)K\over\sqrt{K^2-M^2}(K-S)}\e^{2Ky}
&\qquad x< L\,\,\,\,\mbox{strictly},
\\[0.3cm]
\displaystyle
\int_M^{\infty}{\dd K\over \pi}
{\Rp(\sigma)\over \sqrt{K^2-M^2}(K-S)}\e^{-2Kx}
&\qquad x> 0\,\,\,\,\mbox{strictly},
\end{array}
\right.
\nn
\eeqa
and
\beq
Q_{x\sim L}=-Q_{x\sim 0}
=
{\rm sgn}(\Rp(\sigma))\left[\,-{1\over 2\pi}
\arctan\left({\Ip(\sigma)\over |\Rp(\sigma)|}
\right)- {1\over 4}\, \right]
\label{QlocNS}
\eeq
In Fig.~\ref{fig:Qloc1}, we plot the graph of the 
function (\ref{QRlocL})=(\ref{QlocNS})
that shows the charge localized at the right boundary, 
for both Ramond and NS sectors.

\subsection{Note On Supersymmetry}
\label{subsec:SUSYnote}

Let us comment on the supersymmetry of the states.
The system in the Ramond sector has an ${\mathcal N}=2_B$ supersymmetry
which transforms the fields as
\beqa
&&\delta\phi=\sqrt{2}\epsilon b,\nn\\
&&\delta b=-\sqrt{2}\bepsilon(i\partial_t\phi+\Rp(\sigma)\phi),\nn\\
&&\delta c=i\sqrt{2}\bepsilon(\partial_x\phi-\Ip(\sigma)\phi).
\nn
\eeqa
By Noether procedure we find the supercharges generating them;
\beq
{\bf Q}={\bf Q}_0+\sum_{n=1}^{\infty}{\bf Q}_n
\label{expaQ}
\eeq
where
\beqa
{\bf Q}_0&=&i\sqrt{2}\left(-i\dot{\bphi}_0+\Rp(\sigma)\bphi_0\right)b_0
\nn\\[0.2cm]
&=&\left\{\begin{array}{ll}
i\sqrt{2|\Rp(\sigma)|}
\Bigl(a_0^{1\dag}-ia_0^{2\dag}\Bigr)b_0&\,\,\,\Rp(\sigma)>0,\\[0.2cm]
-i\sqrt{2|\Rp(\sigma)|}\Bigl(a_0^1-ia_0^2\Bigr)b_0&\,\,\,\Rp(\sigma)<0,
\end{array}\right.
\\[0.2cm]
{\bf Q}_n&=&i\sqrt{\sqrt{M^2+k_n^2}+\Rp(\sigma)\over\sqrt{M^2+k_n^2}}
\left(-i\dot{\bphi}_n+\sqrt{M^2+k_n^2}\phi_n\right)b_{n,\uparrow}
\nn\\
&&+i\sqrt{\sqrt{M^2+k_n^2}-\Rp(\sigma)\over\sqrt{M^2+k_n^2}}
\left(-i\dot{\bphi}_n-\sqrt{M^2+k_n^2}\phi_n\right)b^{\dag}_{n,\downarrow}
\nn\\
&=&
i\sqrt{\sqrt{M^2+k_n^2}+\Rp(\sigma)}b_{n,+}
\left(a_n^{1\dag}-ia_n^{2\dag}\right)
-i\sqrt{\sqrt{M^2+k_n^2}-\Rp(\sigma)}b^{\dag}_{n,-}
\left(a_n^1-ia_n^2\right)
\nn\\
\eeqa
The supercharges obey the supersymmetry relations
with equivariant parameter $\Rp(\sigma)$;
\beqa
&\displaystyle {1\over 2}\{{\bf Q},{\bf Q}^{\dag}\}=H+\Rp(\sigma)Q,
\label{twistedSUSY}\\
&{\bf Q}^2={\bf Q}^{\dag 2}=0.
\eeqa
The operator $Q$ on the right hand side of the first equation is
the total charge operator (\ref{QtotR}) (not a supercharge).

We note that each term of ${\bf Q}$ and ${\bf Q}^{\dag}$ 
has an annihilation operator.
This is obvious for the $n\ne 0$ modes
but is true also for the $n=0$ mode; see the property (\ref{fgst}) of
the ground state.
Therefore, the ground state is supersymmetric, i.e., annihilated by 
both of the supercharges
$$
{\bf Q}|0\rangle={\bf Q}^{\dag}|0\rangle=0.
$$
The energy and the charge of the Ramond ground state is computed in
(\ref{EvacR}) and (\ref{QvacR});
The energy is
$E_{\rm vac}=|\Rp(\sigma)|/2$ while the charge is 
$Q_{\rm vac}=-{\rm sgn}(\Rp(\sigma))/2$. These are perfectly consistent
with the supersymmetry algebra (\ref{twistedSUSY}) and
the fact that the state is supersymmetric.
The ground state has positive energy
but it does not mean that supersymmetry is broken in this matter sector.

\subsection{The Case Of The Half-Space}
\label{subsec:half}

Let us comment on the theory formulated on the half-space, say,
\beq
x\leq 0.
\label{halfsp}
\eeq
To be specific, we consider the boundary conditions
which are the same as in the ``right boundary $x=L$'' of the finite interval. 
Namely
we add the term (\ref{rightbt}) at the boundary $x=0$ and impose
the associated boundary conditions (\ref{rightbc}) there:
$$
\partial_1\phi=S\phi,\quad
\psi_+=\psi_-,\quad
\partial_1(\psi_++\psi_-)=S(\psi_++\psi_-).
$$

As in the finite interval there are oscillating modes and
exponential modes.
Oscillating modes for the fields $\phi$ and $b=(\psi_++\psi_-)/\sqrt{2}$,
or $c=(\psi_--\psi_+)/\sqrt{2}$ are
\beq
f_k(x)={k\cos(kx)+S\sin(kx)\over k^2+S^2},\quad \mbox{or}\quad
g_k(x)=\sin(kx),
\label{osci}
\eeq
for any positive $k$.
In addition, $\phi$ and $b$ may have exponential modes, proportional to
\beq
\e^{Sx}.
\label{decay}
\eeq
They are normalizable when $S$ is positive. Since they decay rapidly
as $x\to -\infty$, we shall call them {\it decaying modes}.
They have a frequency $\sqrt{M^2-S^2}=|\Rp(\sigma)|$
and become zero modes when $S=M$, that is,
when $\sigma=i|\sigma|$.
If $S$ is negative, the function $\e^{Sx}$ 
grows exponentially and will not be considered.

Mode expansion and quantization is straightforward.
It is simply to consider the $L\to\infty$ limit of the 
finite interval theory --- we only have to discard the modes that vanishes
in that limit (such as the exponential mode in the Ramond sector of
the $S<0$ theory, and a linear combination of the
two exponential modes of the NS sector in the $S>1/L$ and $S>2/L$ theory).
The fields are expanded as
\beqa
\phi^i(x)&=&
\int_0^{\infty}{\dd k\over \sqrt{\pi}}
{1\over \sqrt[4]{k^2+M^2}}\Bigl(a^i(k)+a^i(k)^{\dag}\Bigr)
f_k(x)
\nn\\
&&\quad
+\theta_S{1\over \sqrt[4]{M^2-S^2}}\Bigl(a^i_D+a_D^{i\dag}\Bigr)
\sqrt{S}\e^{Sx},
\\
\lpi_{\phi^i}(x)&=&
\int_0^{\infty}{\dd k\over \sqrt{\pi}}
\sqrt[4]{k^2+M^2}\Bigl(-ia^i(k)+ia^i(k)^{\dag}\Bigr)
f_k(x)
\nn\\
&&\quad
+\theta_S\sqrt[4]{M^2-S^2}\Bigl(-ia^i_D+ia_D^{i\dag}\Bigr)
\sqrt{S}\e^{Sx},
\\
\Psi(x)&=&\int_0^{\infty}{\dd k\over\sqrt{\pi\sqrt{k^2+M^2}}}
\left[\Psi_{\uparrow,k}(x)b_{\uparrow}(k)+
\Psi_{\downarrow,k}(x)b_{\downarrow}(k)^{\dag}\right]
+\theta_S\sqrt{2S}b_D\Psi_D(x),
\nn\\
\eeqa
where $\Psi_{\alpha,k}(x)$ and $\Psi_D(x)$ are the eigen modes of
the kinetic operator ${\mathcal D}$ defined by
\beqa
&&
\Psi_{\uparrow,k}=\left(\begin{array}{cc}
\sqrt{{\sqrt{k^2+M^2}+\Rp(\sigma)}}f_k(x)\\
i\sqrt{\sqrt{k^2+M^2}-\Rp(\sigma)}g_k(x)
\end{array}\right),\quad {\mathcal D}=\sqrt{k^2+M^2}~~\\
&&
\Psi_{\downarrow,k}=\left(\begin{array}{cc}
\sqrt{{\sqrt{k^2+M^2}-\Rp(\sigma)}}f_k(x)\\
-i\sqrt{\sqrt{k^2+M^2}+\Rp(\sigma)}g_k(x)
\end{array}\right),\quad {\mathcal D}=-\sqrt{k^2+M^2}.~~~\\
&&
\Psi_D=\left(\begin{array}{cc}
\e^{Sx}\\
0\end{array}\right),\quad {\mathcal D}=\Rp(\sigma).
\eeqa
The mode operators obey the (anti-)commutation relations
\beqa
&&[a^i(k_1),a^j(k_2)^{\dag}]=\delta^{i,j}\delta(k_1-k_2),\nn\\
&&([a^i_D,a_D^{j\dag}]=\delta^{i,j}, \quad 
[a^i_D,a^j(k)^{\dag}]=0\,\ {\rm etc}),
\nn\\
&&\{b_{\alpha}(k_1),b_{\beta}(k_2)^{\dag}\}=\delta_{\alpha\beta}
\delta(k_1-k_2),\quad
\{b_{\alpha}(k_1),b_{\beta}(k_2)\}=0,\nn\\
&&(\, \{b_D,b_D^{\dag}\}=1,\quad \{b_D,b_{\alpha}(k)^{\dag}\}=0\,\,\,
{\rm etc}),\nn
\eeqa
so that the fields obey the canonical
(anti-)commutation relations
\beqa
&&[\phi^i(x_1),\lpi_{\phi^j}(x_2)]=i\delta^i_j\delta(x_1-x_2),
\nn\\
&&[\phi^i(x_1),\phi^j(x_2)]=[\lpi_{\phi^i}(x_1),\lpi_{\phi^j}(x_2)]=0.
\nn\\
&&\{b(x_1),b(x_2)^{\dag}\}=\delta(x_1-x_2),
\quad\{b(x_1), b(x_2)\}=0,
\nn\\
&&\{c(x_1),c(x_2)^{\dag}\}=\delta(x_1-x_2),
\quad\{c(x_1), c(x_2)\}=0,
\nn\\
&&\{b(x_1),c(x_2)^{\dag}\}=0,...
\nn
\eeqa
Note that we need the decaying mode (in the case $S>0$)
for this to work.

The ground state is the state annihilated by all the annihilation operators
$a^i(k),b_{\alpha}(k)$ and, in the case $S>0$,
\beq
\begin{array}{ll}
b_D|0\rangle=0&\,\,\, \mbox{if $\Rp(\sigma)>0$},\\
b_D^{\dag}|0\rangle=0&\,\,\,\mbox{if $\Rp(\sigma)<0$}.
\end{array}
\eeq
This is because
the decaying mode contributes to the total Hamiltonian by
$$
H_D={1\over 2}\Rp(\sigma)[b_D^{\dag},b_D].
$$
which has value $-{1\over 2}\Rp(\sigma)$
({\it resp.} $+{1\over 2}\Rp(\sigma)$)
for the state annihilated by $b_D$ ({\it resp.} $b_D^{\dag}$). 
The energy gap to the first excited state is $M$ for $S\leq 0$
and is $|\Rp(\sigma)|$ for $S>0$ from the decaying modes
of $\phi$ and $b$.
It vanishes at $\sigma=i|\sigma|$ where the decaying modes become zero modes.
In particular, the ground states are two-fold degenerate
and non-normalizable.

We can easily obtain the energy (density)
and the charge (density) of the vacuum --- we simply use the result
of the $L\to\infty$ limit of the finite interval theory:
\beqa
E&=&-{\Ip(\sigma)\over 2\pi}\left[\log\left({2\Lambda\over |\sigma|}\right)
+1\right]
+|\Rp(\sigma)|\left[\,{1\over 4}+
{1\over 2\pi}\arctan\left({\Ip(\sigma)\over |\Rp(\sigma)|}\right)
\,\right]
\label{Ehalfspace}
\\
&&\Biggl(\,{\mathcal E}(x)=
-{1\over \pi}\int_M^{\infty}\dd K\sqrt{K^2-M^2}{K\over K-S}\e^{2Kx}
\qquad x<0\,\,\,\,\mbox{strictly}\,\Biggr),
\nn\\
Q&=&-{\rm sgn}(\Rp(\sigma))\left[\,{1\over 4}+{1\over 2\pi}
\arctan\left({\Ip(\sigma)\over |\Rp(\sigma)|}\right)\,\right]
\\
&&\Biggl(\,\rho(x)=-{\Rp(\sigma)\over \pi}
\int_M^{\infty}{\dd K\over \sqrt{K^2-M^2}}{K\over K-S}\e^{2Kx}\qquad x<0
\,\,\,\,\mbox{strictly}\,\Biggr).
\nn
\eeqa
The formula (\ref{Ehalfspace}) agrees with
the result obtained in \cite{BFSV}.
Alternatively, one can find these results independently of
the finite interval computation, directly from the above mode expansions.
Then, we can learn about the finite interval
theory. In particular, we can reproduce all of the results on
the energy (density) and charge (density) at least for the case
$L$ is very large.

\subsection{Cases With Several Matters}
\label{subsec:SeveralMatter}

Let us record the results for the case
where we have $n$ matter fields with charges $Q_1,...,Q_n$.
We just have to sum the results for the single field cases
with
$M=|Q_i\sigma|$, $S=\Ip(Q_i\sigma)$.

If the theory is formulated on the half space (\ref{halfsp}),
the energy and the charge are localized near the boundary, 
and the totals are
\beqa
E&=&
{1\over 2\pi}\sum_{i=1}^n\left\{
Q_i\Ip(\sigma)\left[\log\left({|Q_i\sigma|\over 2\Lambda}\right)-1\right]
+Q_i|\Rp(\sigma)|\arctan\!\left({\Ip(\sigma)\over |\Rp(\sigma)|}\right)
+{|Q_i|\over 4}|\Rp(\sigma)|\right\}
\nn\\
&=&{1\over 2\pi}\Ip\left\{\sum_{i=1}^nQ_i\sigma
\left[\log\left({Q_i\sigma\over 2\Lambda}\right)-1\right]\right\}
+\sum_{i=1}^n{|Q_i|\over 4}|\Rp(\sigma)|,
\label{Ehalfm}\\
Q&=&
-{{\rm sgn}\Rp(\sigma)\over 2\pi}\sum_{i=1}^n
Q_i\arctan\left({\Ip(\sigma)\over |\Rp(\sigma)|}\right)
-\sum_{i=1}^n{|Q_i|\over 4}{\rm sgn}\Rp(\sigma),
\label{Qhalfm}
\eeqa
where ``arctan'' is assumed to take values between
$-{\pi\over 2}$ and ${\pi\over 2}$.

If formulated on the interval,
the energy and the charge are localized near the two boundaries.
The ones at the right boundary, $x=L$, are the same as on the
half space, (\ref{Ehalfm}) and (\ref{Qhalfm}), for both NS and 
Ramond sectors.
The ones at the left boundary, $x=0$, depend on the sector.
In the NS sector, the energy is the same while the charge is opposite,
in comparison to (\ref{Ehalfm}) and (\ref{Qhalfm}).
In the Ramond sector, the energy and the charge are
opposite to (\ref{Ehalfm}) and (\ref{Qhalfm}) except the last terms
involving the $|Q_i|$'s.
The total energy and charge are the sum of those localized
ones:
\beqa
&&E_{\rm R\, vac}=
\sum_{i=1}^n{|Q_i|\over 2}|\Rp(\sigma)|,\qquad
Q_{\rm R\, vac}=-\sum_{i=1}^n{|Q_i|\over 2}{\rm sgn}(\Rp(\sigma)),
\label{EQvacRmulti}\\[0.2cm]
&&
E_{\rm NS\, vac}=2E,\qquad
Q_{\rm NS\, vac}=0.\label{EQvacNSmulti}
\eeqa

Assuming the Calabi-Yau condition
$$
\sum_{Q_i>0}Q_i=-\sum_{Q_i<0}Q_i=:{\mathscr S},
$$
the above formulae simplify, in particular
\beqa
&&E={1\over 2\pi}\sum_{i=1}^nQ_i\log|Q_i|\Ip(\sigma)
+{{\mathscr S}\over 2}|\Rp(\sigma)|,\\
&&Q=
-{\rm sgn}\Rp(\sigma){{\mathscr S}\over 2}.
\eeqa

\subsection{Gauge Dynamics}
\label{subsec:gaugedynamics}

So far, we have set all the fields in the gauge multiplet zero,
except the constant mode of the scalar component $\sigma$.
If we want to obtain the effective potential for
$\sigma$ we need to integrate out the gauge field and the auxiliary field.

Let us consider the system on the interval $0\leq x\leq L$
with charges $q_l$ and $q_r$ on the left and the right boundaries
respectively. 
The Lagrangian is given by
\beq
L=\int_0^L\dd x\,
\left\{{1\over 2e^2}(v_{01}^2+D^2)+{\theta\over 2\pi} v_{01}\right\}
-q_r v_0\Bigr|_{x=L}+q_l v_0\Bigr|_{x=0}
+L_{\rm matter}
\label{startL}
\eeq
where $L_{\rm matter}$ is the matter sector Lagrangian
--- in the case of a single chiral multiplet with charge $1$, it is
the sum of the bulk part (\ref{Lmatbulk}) and the boundary parts
(\ref{rightbt}), (\ref{leftbt}).
Integrating out the auxiliary field $D$ simply produces
the D-term potential
$$
U_D={e^2\over 2}\left(|\phi|^2-r\right)^2.
$$

When we quantize a gauge theory
in operator formalism, it is best to switch
to the canonical formulation. The action in the canonical formulation
is given by (see e.g. \cite{Faddeev})
\beqa
S&=&\int\dd t\int_0^L\dd x\left\{
{1\over e^2}(\partial_0v_1-\partial_1v_0)E_1-{1\over 2e^2}(E_1)^2
-v_0 j^0+{\theta\over 2\pi}(\partial_0v_1-\partial_1v_0)\right\}
\nn\\
&&-\int \Bigl( q_r v_0\Bigr|_{x=L}-q_l v_0\Bigr|_{x=0}\Bigr)\, \dd t
+S_{\rm matter},
\eeqa
where $j^0$ is the charge density
\beq
j^0=i\bphi\lpi_{\bphi}-i\phi\lpi_{\phi}+\bpsi_-\psi_-+\bpsi_+\psi_+,
\eeq
and $S_{\rm matter}$ is the matter action
\beqa
S_{\rm matter}&=&\int\dd t\int_0^L\dd x
\Biggl\{\, \lpi_{\phi}\dot{\phi}+\lpi_{\bphi}\dot{\bphi}
-|\lpi_{\phi}|^2-|D_1\phi|^2-|\sigma|^2|\phi|^2-U_D
\nn\\
&&\qquad\qquad\qquad
+i\bpsi_-(\partial_0-\!\lrD_{\!\! 1})\psi_-
+i\bpsi_+(\partial_0-\!\lrD_{\!\! 1})\psi_+
-\bpsi_-\sigma\psi_+-\bpsi_+\bsigma\psi_-\Biggr\}
\nn\\
&&\,+\,\,
\mbox{boundary terms, (\ref{rightbt}) + (\ref{leftbt}).}
\label{matterS}
\eeqa
It is easy to see that we get back the system with the Lagrangian
(\ref{startL}) after integrating out
$E_1$, $\lpi_{\phi}$, $\lpi_{\bphi}$.
On the other hand, in the Hamiltonian formulation, they play the
role of conjugate momenta for $v_1$, $\phi$ and $\bphi$.
($i\bpsi_{\pm}$ are conjugate momenta for $\psi_{\pm}$).
$v_0$ is a non-dynamical variable and imposes the Gauss law constraints.

We choose the boundary condition such that 
$v_0$ is allowed to vary at the boundary.
The Gauss law constraints are then
\beqa
&&{1\over e^2}\partial_1E_1=j^0 \qquad (\mbox{in the bulk}),\\
&&{1\over e^2}E_1+{\theta\over 2\pi}
+q_r=0\qquad (\mbox{on the right boundary})\quad\mbox{and},\\
&&{1\over e^2}E_1+{\theta\over 2\pi}
+q_l=0\qquad (\mbox{on the left boundary}).
\eeqa
They can be solved by
\beq
{1\over e^2}E_1(x)=-\left({\theta\over 2\pi}+q_l\right)
+\int_0^x j^0(x')\dd x'\nn\\
\,\stackrel{\rm or}{=}\,
-\left({\theta\over 2\pi}+q_r\right)+\int^x_Lj^0(x')\dd x'
\label{formulaE1}
\eeq
which is consistent if and only if
the following condition is satisfied:
\beq
Q=\int_0^Lj^0(x')\dd x'=q_l-q_r.
\label{GaussQ}
\eeq
For a given pair of Chan-Paton charges,
this is the condition on the state of the matter system:
the charge of that state must agree with the difference of the 
Chan-Paton charges at the left and the right boundaries.

The Hamiltonian density of the total system, with the gauge condition
$v_1=0$, is given by
\beqa
{\mathcal H}(x)&=&{1\over 2e^2}E_1(x)^2+U_D(\phi)
+{\mathcal H}_{\rm matter}(x)\nn\\
&=&{e^2\over 2}\left({\theta\over 2\pi}+q_l-\int_0^xj^0(x')\dd x'\right)^2
+{e^2\over 2}\Bigl(|\phi(x)|^2-r\Bigr)^2
+{\mathcal H}_{\rm matter}(x)
\label{Hdensity}
\eeqa
where 
${\mathcal H}_{\rm matter}$ is the Hamiltonian density of
the matter sector.
The task is to find the ground state of
the Hamiltonian 
$\int_0^L{\mathcal H}(x)\dd x$
and its energy.
This is a hard problem.
An exception is the case of a Dirac fermion with
$\sigma=0$ (massless Schwinger model)
where the diagonalization can be done exactly.
For the massive model, 
there are various approximation methods to treat this
problem depending on the range of parameters, such as the relation
of the gauge coupling constant $e$ and the mass $|\sigma|$
of the matter fields.
See for example \cite{Coleman}.
Although this is a very interesting problem in its own right,
we do not develop a thorough study in the present paper.

Instead, we take just one approximation method, which leads to the 
following answer for the energy density of the ground state:
\beq
{\mathcal E}_{\rm vac}(x)
={e^2\over 2}\left({\theta\over 2\pi}+q_l
-\int_0^x\langle j^0(x')\rangle\dd x'\right)^2
+{e^2\over 2}\Bigl(\langle |\phi(x)|^2\rangle-r\Bigr)^2
+{\mathcal E}_{\rm vac}^{\rm matter}(x).
\label{Edensitytot}
\eeq
In the above expression,
$\langle {\mathcal O}(x) \rangle$ stands for the expectation value
with respect to the ground state $|0\rangle$ of the matter sector,
$\langle 0|{\mathcal O}(x)|0\rangle$.
Note that $\langle j^0(x)\rangle$ is the charge density
$\rho(x)$ which we have computed in Section~\ref{subsec:charge}.
Also, ${\mathcal E}_{\rm vac}^{\rm matter}(x)$ in (\ref{Edensitytot})
is the energy density of the ground state in the matter sector, 
which was obtained in Sections~\ref{subsub:Edensity} and \ref{subsec:half}.

Let us describe the origin of (\ref{Edensitytot})
and estimate the error. The 
exact answer would be obtained by integrating out the matter fields as well as 
the fields $v_{\mu}$ and $D$, in a fixed background of
$\sigma$. One way to perform this is to first integrate out
the matter fields for a fixed general profile for $v_{\mu}(x)$ and $D(x)$,
and then integrate out the latter.
At the first step, we may
treat the coupling $-v_0(x)j^0(x)+D(x)|\phi(x)|^2$
perturbatively and keep only the first order terms, that is,
repace it with 
$-v_0(x)\langle j^0(x)\rangle +D(x)\langle |\phi(x)|^2\rangle$.
This is the approximation that leads to (\ref{Edensitytot}).
The discarded part starts with terms of the
form
$$
L{v_{01}^2\over \Delta E},\qquad L{D^2\over \Delta E},
$$
where $\Delta E$ is the energy gap between
 the ground state and the first excited
state. The rest are of higher order in $v_{01}$ and $D$.
These errrors shift the gauge coupling as
$$
{1\over e^2}\,\longrightarrow \,{1\over e^2}\,
+\,{\rm constant}\cdot{L\over \Delta E}.
$$
As long as the shift is small compared to $1/e^2$,
we may say that (\ref{Edensitytot}) is a valid approximation,
with an error given by power series in $e^2L\over \Delta E$.
Recall that the energy gap is of the order of the real part of 
$\sigma$, $\Delta E\sim |\Rp(\sigma)|$,
provided the imaginary part is positive, $\Ip(\sigma)> 0$,
in the NS sector, and  for any case in the Ramond sector.
In such a case, the condition for the validity of (\ref{Edensitytot}) is
\beq
|\Rp(\sigma)|\gg e^2L.
\label{pertcond}
\eeq
This is in addition to the condition
$|\sigma|\gg e$ which is already assumed in the bulk theory.
In Section~\ref{subsec:nature}, we will obtain the same condition from
a different perspective.

The gap $\Delta E\sim |\Rp(\sigma)|$ in the matter sector
comes from the exponential modes (the deacying modes) 
localized near the boundary.
Recall that they become the zero modes or unstable modes
when $\Rp(\sigma)$ vanishes.
In such a case, the whole idea of the effective action 
for the vector multiplet breaks down.
We must treat those modes on the same footing as the vector multiplet
fields.
For example, we may integrate out the oscillating modes only, leaving
the decaying modes dynamical.
Then (\ref{Edensitytot}) is a Hamiltonian density operator
that involves the decaying modes and the $\sigma$ fields.
Note that we no longer have an unstable potential 
for the decaying modes in such a treatment.
Although this problem is important and
interesting on its own right, we do not attempt to solve it here.

Thus, let us assume (\ref{pertcond}).
Even in such a case, one may still wonder if the expression 
(\ref{Edensitytot}) makes sense, since 
$\rho(x)=\langle j^0(x)\rangle$ as well as
$\langle |\phi(x)|^2\rangle$ diverges at the boundary and their squares
may be dangerous. To examine this
and also to find how the answer bahaves as a function of $\sigma$, 
let us compute the total
of the electrostatic and D-term energies.
This is enough to see whether there is a problem locally, since
these two terms are positive semi-definite.

Let us first compute it in the Ramond sector.
The mode expansion leads to the expression
$$
\langle |\phi(x)|^2\rangle=\sum_n{1\over L}{f_n(x)^2\over \sqrt{M^2+k_n}^2}.
$$
Using this we find that the energy from the D-term potential
is
\beqa
E_D&=&
{e^2\over 2}L\left(\sum_{n=0}^{\infty}{1\over 2L\sqrt{M^2+k_n^2}}-r(\Lambda)
\right)^2
\nn\\
&&+{e^2\over 4}\sum_{n=1}^{\infty}\left({1\over 4L(M^2+k_n^2)}
-{S^2\over L|\Rp(\sigma)|\sqrt{M^2+k_n^2}(k_n^2+S^2)}\right)
\nn\\
&&
+{e^2S\over 8\Rp(\sigma)^2}\left({1\over\tanh(SL)}-{1\over SL}\right)
\label{EDR},
\eeqa
where it is understood that $k_0^2=-S^2$ so that 
$\sqrt{M^2+k_0^2}=|\Rp(\sigma)|$.
Also, using the expression (\ref{rhoR}) for
$\langle j^0(x')\rangle=\rho_R(x')$, we find that
the electrostatic energy is
\beqa
E_{\it es}\!\!&=&\!\!
{e^2\over 2}L\left({\theta\over 2\pi}+q_l+
\sum_{n=1}^{\infty}{\Rp(\sigma)S\over 2L\sqrt{M^2+k_n^2}(k_n^2+S^2)}
+{{\rm sgn}\Rp(\sigma)\over 4}
\left({2\over 1-\e^{2SL}}+{1\over SL}\right)
\right)^2\nn\\
&&{e^2\over 8}\sum_{n=1}^{\infty}\left(
{\Rp(\sigma)^2\over 2L(M^2+k_n^2)(k_n^2+S^2)}
-{|\Rp(\sigma)|\over L\sqrt{M^2+k_n^2}(k_n^2+S^2)}\right)
\nn\\
&&+{e^2\over 32S}\left({1\over\tanh(SL)}-{1\over SL}\right).
\label{EesR}
\eeqa
Despite its appearance, this is continuous at $S=0$, with
\beqa
E_D&\stackrel{S=0}{=}&
{e^2\over 2}L\left(\sum_{n=0}^{\infty}{1\over 2L\sqrt{M^2+k_n^2}}-r(\Lambda)
\right)^2
+{e^2\over 16}\sum_{n=1}^{\infty}{1\over L(M^2+k_n^2)},
\nn\\
E_{\it es}&\stackrel{S=0}{=}&
{e^2\over 2}L\left({\theta\over 2\pi}+q_l+{1\over 4}{\rm sgn}\Rp(\sigma)
\right)^2
+{e^2\over 16}\sum_{n=1}^{\infty}{k_n^2\over L(k_n^2+M^2)(\sqrt{M^2+k_n^2}
+M)^2}.
\nn
\eeqa
To obtain the last expression, we have used the formula
$$
\sum_{n=1}^{\infty}{1\over n^2}={\pi^2\over 6}.
$$
The behaviour at large $L$ is
\beqa
E_D&=&{e^2\over 2}L\left(
{1\over 2\pi}\log\left({2\Lambda\over |\sigma|}\right)-r(\Lambda)\right)^2
+{e^2\over 32|\sigma|}
+{e^2\Ip(\sigma)\over 4\pi \Rp(\sigma)^2}\arctan\left({\Ip(\sigma)\over
|\Rp(\sigma)|}\right),
\nn\\
E_{\it es}
&=&{e^2\over 2}L\left({\theta\over 2\pi}+q_l-Q_{x\sim 0}
\right)^2
-{e^2\over 32|\sigma|}
+{e^2\over 8\pi \Ip(\sigma)}\arctan\left({\Ip(\sigma)\over
|\Rp(\sigma)|}\right).
\nn
\eeqa
Here $Q_{x\sim 0}$ is the charge localized near the left boundary
(\ref{QloclR}).
The result is regular everywhere away from the locus $\Rp(\sigma)=0$
which we excluded by (\ref{pertcond}).
The quantity inside the parenthesis
in the extensive part ${e^2\over 2}L(\cdots)^2$ should be defined as
the series that appears in the first expressions
(\ref{EDR}), (\ref{EesR}) ---
an error of order $1/L$ there would affect the non-extensive part. 
Note that it is finite as the cut-off is removed,
since the FI parameter depends on it as
$r(\Lambda)=r(\mu)+{1\over 2\pi}\log\left({\Lambda\over \mu}\right)$.
The sum of the D-term and electrostatic energies,
$E_{\it D,es}=E_{\it es}+E_D$, is thus
\beq
E_{\it D,es}
={e^2\over 2}L|t_{\it eff}|^2
+{e^2\over 8\pi}\left(
{2\Ip(\sigma)\over \Rp(\sigma)^2}+{1\over \Ip(\sigma)}\right)
\arctan\left({\Ip(\sigma)\over
|\Rp(\sigma)|}\right),
\eeq
where
$$t_{\it eff}
=r(\Lambda)-{1\over 2\pi}\log\left({2\Lambda\over |\sigma|}\right)
-i\left[{\theta\over 2\pi}+q_l-Q_{x\sim 0}\right]+O(1/L).
$$

Let us next write down the result for the NS sector.
We have to treat
the three cases, $S>0$, $S=0$ and $S<0$, separately.
The main point is that it is continuous at $S=0$, with
\beqa
E_D\!\!&\stackrel{S=0}{=}&\!\!
{e^2\over 2}L\left(\sum_{n=0}^{\infty}{1\over 2L\sqrt{M^2+k_n^2}}-r(\Lambda)
\right)^2
+{e^2\over 16}\sum_{n=0}^{\infty}{1\over L(M^2+k_n^2)}
\nn
\\
E_{\it es}\!\!&\stackrel{S=0}{=}&\!\!
{e^2\over 2}L\left({\theta\over 2\pi}+q_l-{\rm sgn}\Rp(\sigma)
\left[{1\over 4} -{1\over L}I
\right]
\right)^2
+{e^2\over 16}\sum_{r}^{\infty}{k_r^2\over L(k_r^2+M^2)(\sqrt{M^2+k_r^2}
+M)^2}
\nn
\eeqa
with
$$
I:=\sum_r{1\over 2L\sqrt{M^2+k_r^2}(\sqrt{M^2+k_r^2}+M)}.
$$
To obtain this, we used the formula
$$
\sum_{r={1\over 2},{3\over 2},...}{1\over r^2}={\pi^2\over 2}.
$$
The behaviour at large $L$ for the general value of $S=\Ip(\sigma)$ is
\beqa
E_D&=&{e^2\over 2}L\left(
{1\over 2\pi}\log\left({2\Lambda\over |\sigma|}\right)-r(\Lambda)\right)^2
+{e^2\over 32|\sigma|}
+{e^2\Ip(\sigma)\over \Rp(\sigma)^2}Q_{x\sim 0}^2,
\nn\\
E_{\it es}
&=&{e^2\over 2}L\left({\theta\over 2\pi}+q_l-Q_{x\sim 0}
\right)^2
-{e^2\over 32|\sigma|}
+{e^2\over 2\Ip(\sigma)}\left[Q_{x\sim 0}^2-{1\over 16}\right],
\nn
\eeqa
where $Q_{\sim 0}$ is the localized charge given now by
(\ref{QlocNS}). The result is regular everywhere away from
the excluded locus $\Rp(\sigma)=0$.
The sum of them is
\beq
E_{\it D,es}={e^2\over 2}L|t_{\it eff}|^2
+{e^2\over 2}\left[\left(
{2\Ip(\sigma)\over \Rp(\sigma)^2}+{1\over \Ip(\sigma)}\right)
Q_{x\sim 0}^2-{1\over 16\Ip(\sigma)}\right],
\eeq
where 
$t_{\it eff}$ is as in the Ramond sector, with $Q_{x\sim 0}$ 
from (\ref{QlocNS}).

The main conclusion of these computations is that,
in both R and NS sectors, the D-term and electrostatic energy
of the ground state behaves as
\beq
E_{\it D,es}={e^2\over 2}L|t_{\it eff}|^2
+{e^2\over |\sigma|}A.
\label{anssummary}
\eeq
$A$ is a function of $\Ip(\sigma)/\Rp(\sigma)$, independent of $L$,
which is regular everywhere except $\Rp(\sigma)\to0$.
This also holds in the theory with multiple fields of various charges.

This concludes that there is no danger from taking the square of a quantity
that diverges at the boundary.
Furthermore, 
we see that the dominant part of the vacuum energy, for 
large $\sigma$ with a fixed $L$,
is the localized energy from the matter sector
which grows linearly
with $\sigma$. Compare (\ref{anssummary}) 
with the expressions obtained in earlier subsections, 
for example (\ref{Ehalfm}).

\subsubsection*{An Anomaly}

Although it is not directly relevant for the main discussion of this paper,
we would like to comment on an anomaly in the Ramond sector
that comes from the Gauss law constraint. 
We consider a $U(1)$ gauge theory with $n$ matter fields 
of charges $Q_1,\ldots, Q_n$, which we assume integers
with q.c.d.$(Q_i)=1$.

The Gauss law implies that the total charge $Q$
of the state in the matter sector must match the difference of the charges
$q_l$ and $q_r$ which we place on the two boundaries, (\ref{GaussQ}).
We usually take both $q_l$ and $q_r$ to be integers, and hecne $Q$ must
be an integer.
The total charge of a state in the matter sector
is an integer plus the charge of the ground state. 
In the NS sector, we found that the charge of the ground state vanishes,
see (\ref{EQvacNSmulti}). Thus, the charges of all states are integers.
In the Ramond sector, on the other hand,
the charge of the ground state is obtained in
(\ref{EQvacNSmulti}), from which we conclude
that the charges of all states are
$\sum_{i=1}^nQ_i/2$ modulo integers.
Thus, we find a conflict with the Gauss law when
\beq
\sum_{i=1}^nQ_i\quad\mbox{is odd.}
\eeq
In the context of linear sigma models, this is precisely
the case when the corresponding toric variety is not a spin
manifold. In an open string Ramond sector of
the non-linear sigma model, we need to have 
a spin structure of the target space in order 
to quantize the fermionic zero modes.
The above anomaly is understood as the Coulomb branch
counterpart of this.
There is of course no problem when the Calabi-Yau condition is assumed,
$\sum_{i=1}^nQ_i=0$.

\subsection{Mass From Superpotential}
\label{subsec:massfromW}

We now turn to a different system:
the LG model of two variables $U,V$ with superpotential
$$
W=2MUV
$$
where $M$ is taken to be real positive.
We also consider a D-brane correponding to the matrix factorization
$$
Q=\left(\begin{array}{cc}
0&\sqrt{2M}u\\
\sqrt{2M}v&0
\end{array}\right).
$$
We may regard this as a part of the matter sector of a
$U(1)$ gauge theory in which $U$ and $V$ have charges $-1$ and $1$. 
If we write the matrix as 
$Q=\sqrt{2M}(u\eta+v\bareta)$, then $\eta$ and $\bareta$ have gauge
charges $1$ and $-1$ respectively.
In this section we study the property of the ground state of the open string
whose both ends have the brane $Q$.
In particular, we are interested in the energy and charge density of the
ground state, especially in the limit where the length $L$
of the string
is taken to be very large.

Before doing any computation, we already know quite a lot
about the ground state. First of all, there is just one chiral ring element
proportional to the identity
\beq
\left(\begin{array}{cc}
1&0\\
0&1
\end{array}\right).
\label{chrelt}
\eeq
Thus, by the spectral flow, we know that there is a unique supersymmetric 
ground state. In particular the ground state energy is zero
$$
E_{\rm vac}=0.
$$
At this point we recall that any quantum field theory formulated
on a compact space, such as the interval $[0,L]$ we are considering,
behaves like quantum mechanics. In particular, any symmetry present in the
system cannot be spontaneously broken by the ground state $|0\rangle$.
In the present open string system, there are two symmeties in sight.
One is the $U(1)$ charge symmetry (under which $U,V$ has charge $-1,1$).
The other is the parity symmetry $x\to L-x$ that swaps 
the left and the right boundaries.
That the $U(1)$ symmetry is unbroken means that
the ground state has charge zero:
$$
Q_{\rm vac}=0.
$$
That the parity symmetry is unbroken means that the energy and charge density
of the ground state is symmetric under $x\to L-x$:
$$
{\mathcal E}(x)={\mathcal E}(L-x),\qquad
\rho(x)=\rho(L-x).
$$
Finally, one more important fact is that the bulk theory has a mass gap.
This in particular means that any local observable ${\mathcal O}(x)$
approaches the value of the bulk vacuum when it 
is {\it far enough} from the boundary, 
that is, when $1/M\ll x\ll L-1/M$. In particular we expect
$$
{\mathcal E}(x)\longrightarrow 0,\quad {\rho}(x)\longrightarrow 0
$$
in that regime. Namely ${\mathcal E}(x)$ and 
$\rho(x)$ might have a non-trivial profile but that is confined in 
a region of width $1/M$ near the two boundaries. Moreover, the parity 
symmetry as well as $E_{\rm vac}=Q_{\rm vac}=0$ tell us that
the total energy and the charge accumulated 
near each of the boundaries vanish.
Thus, for a long distance observer,
it looks as if the energy and the charge density identically vanishes
without any delta function support at the boundaries.
In what follows, we confirm this expectation by an explicit computation.

 The system (\ref{Lmatbulk}) we considered
in the bulk part of this section does not have the parity symmetry
$x\to L-x$. To be precise, there would be a parity symmetry only when
it is combined with
a sign flip of of $\sigma$ and a complex conjugation of
$\phi$ and $\psi$.
Thus, the energy and charge densities are not symmetric under 
the parity but only so when the appropriate operation is applied.

Explcit quantization of a related system had been done in
\cite{book} and later also in \cite{KapLi1}. 
These works studied the Dirichlet boundary condition $U=0$
in the same bulk theory. That brane can be obtained from
 the above matrix factorization by replacing
$\sqrt{2M}u$, $\sqrt{2M}v$ by $\zeta\sqrt{2M}u$, $\zeta^{-1}\sqrt{2M}v$
and taking the limit $\zeta\to\infty$.

\subsubsection{The Ground State}

The Lagrangian of the open string system reads as
\beqa
L&=&\int_0^L\Biggl[\,\,|\dot{u}|^2-|u'|^2+|\dot{v}|^2-|v'|^2-M^2|u|^2-M^2|v|^2
\nn\\
&&\qquad\,\,
+i\opsi^u_-(\lrd_{\!\!\! t}+\lrd_{\!\!\! x})\psi^u_-
+i\opsi^u_+(\lrd_{\!\!\! t}-\lrd_{\!\!\! x})\psi^u_+
\nn\\
&&\qquad\,\,
+i\opsi^v_-(\lrd_{\!\!\! t}+\lrd_{\!\!\! x})\psi^v_-
+i\opsi^v_+(\lrd_{\!\!\! t}-\lrd_{\!\!\! x})\psi^v_+
\nn\\
&&\qquad\,\,
-iM(\psi^u_+\psi^v_-+\psi^v_+\psi^u_-)
+iM(\opsi^v_-\opsi^u_++\opsi^u_-\opsi^v_+)\,\,\Biggr]
\dd x
\nn\\
&&+\Biggl[-{i\over 2}(\psi^u_+\opsi^u_-+\opsi^u_+\psi^u_-)
-{i\over 2}(\psi^v_+\opsi^v_-+\opsi^v_+\psi^v_-)\,\Biggr]^L_0
\nn\\
&&+\Biggl[\,\,i\bareta_L\dot{\eta}_L+\sqrt{2M}{\rm Re}
(\psi^u\eta_L+\psi^v\bareta_L)-M|u|^2-M|v|^2\,\,\Biggr]_{x=L}
\nn\\
&&-\Biggl[\,\,i\bareta_0\dot{\eta}_0+\sqrt{2M}{\rm Re}
(\psi^u\eta_0+\psi^v\bareta_0)+M|u|^2+M|v|^2\,\,\Biggr]_{x=0}
\eeqa
Note that the boundary fermions
$\eta_0,\bareta_0$ have the opposite orientation of time.
In particular, they have the ``wrong'' sign kinetic term and
obey the non-standard hermiticity relation
$$
\eta_0^{\dag}=-\bareta_0,
$$
with respect to the standard orientation of time.

We take the standard supersymmetric Neumann condition
for the bulk fields:
\beq
\left.
\begin{array}{r}
\partial_1u=\partial_1v=0,\\
\psi^u_+-\psi^u_-=\psi^v_+-\psi^v_-=0,\\
\partial_1(\psi^u_++\psi^u_-)=\partial_1(\psi^v_++\psi^v_-)=0,
\end{array}\right\}
\quad\mbox{at both $x=0,L$}.
\label{bcs}
\eeq
Note that this does not agree with the one coming from
the variational principle. For example, the condition on $u$
from the variational 
principle would be
$$
\partial_1 u+Mu=0~~\mbox{at $x=L$;}~~~
\partial_1u-Mu=0~~\mbox{at $x=0$},
$$
but we take the Neumann condition $\partial_1u=0$ at both boundaries.
However, we will find no problem in quantization.
In fact, the boundary conditions from the variational principle
will show up in an interesting way.

This system has $U(1)$ symmetry: $u,\psi^u_{\pm},
\bareta_L,\bareta_0$
has charge $-1$ and $v,\psi^v_{\pm},\eta_L,\eta_0$ has charge $1$.
There is also a parity symmetry
\beqa
&&u(x)\to i u(L-x),\quad v(x)\to i v(L-x),\nn\\
&&\psi^u_{\pm}(x)\to i\psi^u_{\mp}(L-x),\quad
\psi^v_{\pm}(x)\to i\psi^v_{\mp}(L-x),
\nn\\
&&\eta_L\to i\eta_0,\quad\bareta_L\to i\bareta_0,\quad
\eta_0\to i\eta_L,\quad\bareta_0\to i\bareta_L,
\nn
\eeqa
that commutes with the $U(1)$ symmetry.

The action is quadratic in all fields and the quantization is straightforward
provided we find a clever choice of variables.
The bosonic part is a decoupled sum of four copies (two from $u$
and two from $v$) of the real scalar field $\phi$ with Lagrangian
$$
L_B={1\over 2}(\dot{\phi},\dot{\phi})-{1\over 2}(\phi,{\mathcal D}_B\phi)
$$
where $(\phi_1,\phi_2):=\int_0^L\dd x\phi_1(x)^*\phi_2(x)$
and
\beq
{\mathcal D}_B=-\partial_1^2+M^2+M\delta(x-L)+M\delta(x).
\eeq
The fermionic system decomposes into two sectors. Let us take the
linear combination $\psi^{\pm}_{\alpha}
=(\psi^v_{\alpha}\pm \opsi^u_{\alpha})/\sqrt{2}$ where $\alpha=\pm$
 is the spinor index and then introduce 
$$
\Psi^{+}=\left(\begin{array}{c}
b^+\\
c^+\\
\eta_0
\end{array}\right),\qquad
\Psi^{-}=\left(\begin{array}{c}
b^-\\
c^-\\
\eta_L
\end{array}\right),
$$
for
$b^{\pm}:=(\psi^{\pm}_-+\psi^{\pm}_+)/\sqrt{2}$,
$c^{\pm}:=(\psi^{\pm}_--\psi^{\pm}_+)/\sqrt{2}$.
The Lagrangian can be written as
$$
L_F=i(\Psi^+,\dot{\Psi}^+)+i(\Psi^-,\dot{\Psi}^-)
-(\Psi^+,{\mathcal D}^+\Psi^+)-(\Psi^-,{\mathcal D}^-\Psi^-)
$$
where $(\Psi_1,\Psi_2):=
\int^L_0\dd x (b_1(x)^{\dag}b_2(x)+c_1(x)^{\dag}c_2(x))+\eta_1^{\dag}\eta_2$
and
\beqa
{\mathcal D}^+&=&
\left(\begin{array}{ccc}
0&-i\partial_1+iM&\sqrt{2M}\delta(x)\\
-i\partial_1-iM&0&0\\
\sqrt{2M}{\rm ev}_0&0&0
\end{array}\right),\\
{\mathcal D}^-&=&
\left(\begin{array}{ccc}
0&-i\partial_1-iM&\sqrt{2M}\delta(x-L)\\
-i\partial_1+iM&0&0\\
\sqrt{2M}{\rm ev}_L&0&0
\end{array}\right).
\eeqa
Here ${\rm ev}_*$ is the evaluation map, ${\rm ev}_Lb=b(L)$ and 
${\rm ev}_0b=b(0)$.

The kinetic operators ${\mathcal D}_B$, ${\mathcal D}^{\pm}$
are hermitian operators in the space of 
functions defined by (\ref{bcs}): Neumann for $u,v,b^{\pm}$ and Dirichlet
for $c^{\pm}$. 
Thus, they have eigenvectors with real eingenvalues.
Let us diagonalize these operators. To this end, we introduce
projection operators onto Neumann and Dirichlet functions
$$
(P_cf)(x)=\sum_{n=0}^{\infty}c_n(x)(c_n,f),\qquad
(P_sf)(x)=\sum_{n=1}^{\infty}s_n(x)(s_n,f),
$$
where $c_0(x)={1\over \sqrt{L}}$, $c_n(x)=\sqrt{2\over L}\cos({\pi nx\over L})$
and $s_n(x)=\sqrt{2\over L}\sin({\pi nx\over L})$.
They obey the following relations
\beqa
&&(P_cf)'(x)=(P_sf')(x),\label{Pcfp}\\
&&(P_sf)'(x)=(P_cf')(x)-\delta(x-L)f(L)+\delta(x)f(0),\label{Psfp}\\
&&P_cP_sP_c=P_c.\label{PcPs}
\eeqa

Using the first two, we find
$$
{\mathcal D}_BP_cf
=P_c(-f''+M^2f)+\delta(x-L)(f'+Mf)-\delta(x)(f'-Mf).
$$
Thus, if $f$ obeys the boundary condition 
\beq
\begin{array}{ll}
f'(x)+Mf(x)=0&\mbox{at $x=L$}\\
f'(x)-Mf(x)=0&\mbox{at $x=0$},
\end{array}
\label{bcf}
\eeq
${\mathcal D}_B$ is simply represented by  $-f''+M^2f$.
At this point, we notice that this boundary condition for
$f$ is precisely the same as the boundary condition for
the boson in the system considered earlier: the NS condition
 (\ref{NSbc}) for $\phi$ with $S=-M$. In particular, for a plane wave
$f(x)=a\sin(kx)+b\cos(kx)$ 
the allowed wavenumbers $k$ are determined by the equation
(\ref{coteqn}) with $S=-M$.
As before we denote the solutions by $k_n$, $n=0,1,2,...$,
which approach ${\pi n\over L}$ as $n\to\infty$.

For the fermion $\Psi=(P_cf,P_sg,\beta)$, 
the equations
${\mathcal D}^{\pm}\Psi=\lambda\Psi$ both lead to
$$
P_c(-f''+M^2f)+\delta(x-L)(f'+Mf)
-\delta(x)(f'-Mf)=\lambda^2P_cf.
$$
This can be obtained by eliminating $P_sg$ and $\eta$ from the three
equations and using (\ref{Pcfp}) (\ref{Psfp}) and (\ref{PcPs}).
Again we find the standard eingenvalue problem
$-f''+M^2f=\lambda^2f$ provided $f$ obeys the
boundary condition (\ref{bcf}).
In particular, we have plane wave solutions
with eigenvalue $\lambda=\pm\sqrt{k_n^2+M^2}$, where $k_n$ solves 
(\ref{coteqn}) with $S=-M$.

Normalized modes are given as follows:
For bosons, we have
\beq
h_n(x)=\alpha_n(P_cf_n)(x),\qquad {\mathcal D}_B=k_n^2+M^2
\eeq
and for fermions
\beqa
&&\Psi^+_{n,\pm}(x)={\alpha_n\over \sqrt{L}}
\left(\begin{array}{c}
(P_cf_n)(x)\\
\pm i(-1)^n (P_sg_n)(x-L)\\
\pm {\sqrt{2M}k_n\over k_n^2+M^2}
\end{array}\right),\qquad
{\mathcal D}^+=\pm\sqrt{k_n^2+M^2},\\
&&\Psi^-_{n,\pm}(x)={\alpha_n\over \sqrt{L}}
\left(\begin{array}{c}
(P_cf_n)(x)\\
\pm i(P_sg_n)(x)\\
\pm (-1)^n{\sqrt{2M}k_n\over k_n^2+M^2}
\end{array}\right),\qquad
{\mathcal D}^-=\pm\sqrt{k_n^2+M^2},
\eeqa
where
$$
\alpha_n=\mbox{$\left(1+{2M\over L(k_n^2+M^2)}\right)^{-{1\over 2}}$},\quad
f_n(x)={k_n\cos(k_nx)+M\sin(k_nx)\over \sqrt{k_n^2+M^2}},\quad
g_n(x)=\sin(k_n x).
$$
Expanding the fields as
\beqa
&&\phi^i(x)=\sum_{n=0}^{\infty}\phi^i_{n}\sqrt{2\over L}h_n(x),\quad
i=1,2,3,4,
\nn\\
&&\Psi^{\pm}(x)=\sum_{n=0}^{\infty}
\left(b_{n,+}^{\pm}\Psi^{\pm}_{n,+}(x)
+b_{n,-}^{\pm\,\,\dag}\Psi^{\pm}_{n,-}(x)\right),
\nn
\eeqa
the Lagrangian can be written as
\beqa
L&=&\sum_{n=1}^{\infty}\left[\,\,
\sum_{i=1}^4\left({1\over 2}(\dot{\phi}^i_{n})^2
-{k_n^2\!+\!M^2\over 2}(\phi^i_{n})^2\right)\right.
\nn\\
&&\quad\left.
+\sum_{\epsilon=\pm}\left(ib_{n,+}^{\epsilon\,\,\dag}\dot{b}_{n,+}^{\epsilon}
+ib_{n,-}^{\epsilon\,\,\dag}\dot{b}_{n,-}^{\epsilon}
-\sqrt{k_n^2\!+\!M^2}b_{n,+}^{\epsilon\,\,\dag}b_{n,+}^{\epsilon}
+\sqrt{k_n^2\!+\!M^2}b_{n,-}^{\epsilon}b_{n,-}^{\epsilon\,\,\dag}
\right)\,\,\right].\nn
\eeqa
The Hamiltonian is
\beq
H=\sum_{n=0}^{\infty}\left[\,\,
\sum_{i=1}^4\sqrt{k_n^2+M^2}\left(a_n^{i\,\dag}a_n^i+{1\over 2}\right)
+\sum_{\epsilon,\alpha}\sqrt{k_n^2+M^2}
\left(b_{n,\alpha}^{\epsilon\,\,\dag}b_{n,\alpha}^{\epsilon}-{1\over 2}\right)
\,\,\right],
\eeq
where $a^{i\,\dag}_n$, $a^i_n$, $b^{\epsilon\,\dag}_{n,\alpha}$,
$b^{\epsilon}_{n,\alpha}$ are the creation and annihilation operators 
obeying the standard (anti-)commutation relations.
There is a unique ground state $|0\rangle$ characterized by
$$
a^i_n|0\rangle=b^{\epsilon}_{n,\alpha}|0\rangle=0.
$$

Let us discuss the energy and its density of the ground state $|0\rangle$.
The bosonic and fermionic contributions to the energy
cancel at each level: 
$4{\sqrt{k_n^2+M^2}\over 2}-4{\sqrt{k_n^2+M^2}\over 2}=0$.
Therefore the total energy of the ground state vanishes: 
\beq
E_{\rm vac}=0
\label{Evacvanish}
\eeq
This is of course a consequence of supersymmetry.
To find the energy density, let us introduce the notation
$$
\left[\Psi_1(x)^{\dag}\stackrel{x_*}{,}\Psi_2(x)\right]
:=[b_1(x)^{\dag},b_2(x)]+[c_1(x)^{\dag},c_2(x)]
+\delta(x-x_*)[\eta_1^{\dag},\eta_2].
$$
Then, the energy density operator can be written as
\beqa
{\mathcal H}(x)&=&\sum_{i=1}^4\left({1\over 2}\dot{\phi}^i(x)
+{1\over 2}\phi^i(x){\mathcal D}_B\phi^i(x)\right)\nn\\
&&+{1\over 2}\left[\Psi^-(x)^{\dag}
\stackrel{L}{,}
{\mathcal D}^-\Psi^-(x)\right]
+{1\over 2}\left[\Psi^+(x)^{\dag}
\stackrel{0}{,}
{\mathcal D}^+\Psi^+(x)\right]
\eeqa
The energy density of the ground state is 
\beqa
\lefteqn{{\mathcal E}(x):=\langle 0|{\mathcal H}(x)|0\rangle=}
\label{EdensitySup}
\\
&&\sum_{n=0}^{\infty}{\sqrt{k_n^2+M^2}\over L}\alpha_n^2\Biggl(
2f_n(x)^2-g_n(x)^2-g_n(x\!-\!L)^2
-{2Mk_n^2\over (k_n^2+M^2)^2}(\delta(x\!-\!L)
+\delta(x))\Biggr).
\nn
\eeqa
It is indeed symmetric under $x\to L-x$, as one can see by using
$f_n(L-x)=(-1)^nf_n(x)$.

Let us next discuss the $U(1)$ charge of the ground state.
The charge density operator is defined by
\beqa
j^0(x)&=&{i\over 2}\{\overline{v},\dot{v}\}-{i\over 2}\{v,\dot{\overline{v}}\}
-{i\over 2}\{\overline{u},\dot{u}\}+{i\over 2}\{u,\dot{\overline{u}}\}
\nn\\
&&+{1\over 2}\left[\Psi^-(x)^{\dag}\stackrel{L}{,}\Psi^-(x)\right]
+{1\over 2}\left[\Psi^+(x)^{\dag}\stackrel{0}{,}\Psi^+(x)\right].
\nn
\eeqa
The total charge as well as its density of the ground state vanish
\beqa
&&Q_{\rm vac}=0,\\
&&\rho(x):=\langle 0|j^0(x)|0\rangle=0.
\eeqa
This matches with the expectation.

\subsubsection{$L\to \infty$ Limit}

Let us look at the energy density in the limit where $L$ is very large
compared to $1/M$.
We first look at the region close to the
left boundary $x=0$, that is, $0\leq x\ll L$.
The formula (\ref{EdensitySup}) can be rewritten as
\beqa
{\mathcal E}(x)&=&\sum_{n=0}^{\infty}{\sqrt{k_n^2+M^2}\over L}
{2k_n^2\alpha_n^2\over (k_n^2+M^2)^2}
\Biggl(\,(k_n^2-M^2)\cos(2k_nx)+2k_nM\sin(2k_nx)
\nn\\
&&\qquad\qquad\qquad\qquad\qquad\qquad
-M(\delta(x\!-\!L)
+\delta(x))\Biggr).
\nn
\eeqa
In the limit $L\to\infty$ the sum over $n$ turns into
integral over the momentum $k$. 
If $x$ is close to but strictly away from $x=0$,
the contour of the integral can be deformed and we have
\beq
{\mathcal E}(x)=-2\int^{\infty}_M{\dd K\over \pi}{\sqrt{K^2-M^2} K^2\over
(K+M)^2}\e^{-2Kx}
\eeq
We see that it decays exponentially as $\e^{-2Mx}$ or faster.
This is of course a consequence of the mass gap of the bulk theory.
By the symmetry $x\to L-x$, we find the same behaviour near the right boundary
$x=L$. Namely, the energy density is non-trivial only in the region of width
$1/M$ near the two boundaries.
Since the total energy vanishes
(\ref{Evacvanish}), we learn that the energies localized
near the left and the right boundaries vanish indivisually.
Indeed that can be confirmed by a direct computation:
\beqa
E_{x\sim 0}\!\!&:=&\!
\lim_{R\to \infty}\int^R_0{\mathcal E}(x)\dd x\nn\\
\!\!&=&\!\lim_{R\to\infty}\int^{\Lambda}_0{\dd k\over \pi}
\sqrt{k^2+M^2}{k\over (k^2+M^2)^2}\Bigl((k^2\!-\!M^2)\sin(2kR)
+2kM(1-\cos(2kR)\Bigr)
\nn\\
&&\qquad
-\int^{\Lambda}_0{\dd k\over \pi}\sqrt{k^2+M^2}{2Mk^2\over (k^2+M^2)^2}.
\nn
\eeqa
The last term comes from the delta function at $x=0$  --- the one at
$x=L$ of course does not contribute.
The two lines cancel out and we have $E_{x\sim 0}=0$.

\subsubsection{$L\to 0$ Limit}

For completeness,
let us consider the opposite regime, $ML\ll 1$.
In the limit $ML\to 0$, the momenta $k_n$ approach
the standard value
$$
k_nL\to \pi n,\qquad n=0,1,2,3,\ldots.
$$
In particular, the $n=0$ mode approaches the constant mode
while the $n\geq 1$ modes approach the standard plane wave modes
$\cos({\pi n\over L}x),\sin({\pi n\over L}x)$, for $0\leq x\leq L$.
To be more precise, using the defining equation (\ref{coteqn}), we find that
$k_0$ diverges as
$$
k_0^2\to {2M\over L},
$$
but it is still true that the mode approches a constant,
$\cos(k_0x)\to 1$ and $\sin(k_0x)\to 0$,
as long as $x$ is in the interval $[0,L]$.
Let us look at the ground state $|0\rangle$ 
in this zero mode sector.

In the limit $ML\ll 1$, the fermionic zero mode becomes
$$
\Psi^+_{0,\pm}(x),\Psi^-_{0,\pm}(x)\longrightarrow
{1\over\sqrt{2L}}\left(\begin{array}{c}
1\\0\\\pm\sqrt{L}
\end{array}\right).
$$
Thus, we can write $b^{\epsilon}={1\over \sqrt{2L}}
(b_{0,+}^{\epsilon}+b_{0,-}^{\epsilon\,\,\dag})$,
$\eta_L={1\over \sqrt{2}}(b_{0,+}^--b_{0,-}^{-\,\dag})$,
$\eta_0={1\over \sqrt{2}}(b_{0,+}^+-b_{0,-}^{+\,\dag})$,
Since the ground state $|0\rangle$
is annihilated by $b_{n,\alpha}^{\epsilon}$, 
it is annihilated by
$$
b^-+{1\over\sqrt{L}}\eta_L,\quad
b^{-\dag}-{1\over\sqrt{L}}\eta_L^{\dag},\quad
b^++{1\over\sqrt{L}}\eta_0,\quad
b^{+\dag}-{1\over\sqrt{L}}\eta_0^{\dag}.
$$
It follows that the zero mode ground state is of the form
$$
|0\rangle^{\rm zero}\propto
\left(b^{+\dag}-{1\over \sqrt{L}}\eta_0^{\dag}\right)
\left(b^{-\dag}-{1\over\sqrt{L}}\eta_L^{\dag}\right)|0\rangle_1^{},
$$
where $|0\rangle^{}_1$ is the state annihilated by 
$\eta_0,\eta_L,b^+,b^-$ as well as the bosonic 
annihilation operators. (Note the normalization
$\{b^{\epsilon_1},b^{\epsilon_2\dag}\}
=\delta^{\epsilon_1,\epsilon_2}/\sqrt{L}$.)
Recalling the definition of $b^{\pm}$, and using the matrix
representation of
$\eta_L$, $\eta_0$, etc,
\beqa
&&\eta_L\left(\begin{array}{cc}
a&b\\c&d
\end{array}\right)=\left(\begin{array}{cc}
0&1\\0&0
\end{array}\right)
\left(\begin{array}{cc}
a&b\\c&d
\end{array}\right),
\nn\\
&&
\eta_0\left(\begin{array}{cc}
a&b\\c&d
\end{array}\right)=\left(\begin{array}{cc}
(-1)^aa&-(-1)^bb\\-(-1)^cc&(-1)^dd
\end{array}\right)\left(\begin{array}{cc}
0&1\\0&0
\end{array}\right),
\nn
\eeqa
we find that the ground state is
given by
\beq
|0\rangle^{\rm zero}
=\left(\begin{array}{cc}
|0\rangle'-{L\over 2}\opsi^v\opsi^u|0\rangle'&-\sqrt{L}\opsi^v|0\rangle'
\\[0.2cm]
\sqrt{L}\opsi^u|0\rangle'&|0\rangle'+{L\over 2}\opsi^v\opsi^u|0\rangle'
\end{array}\right).
\label{zeromodevac}
\eeq
Here, $|0\rangle'$ is the state of the bulk zero mode sector that is 
annihilated by $\psi^u,\psi^v$ as well as the bosonic annihilation operators.
As a wave function of $u,v$, it is proportional to
$\e^{-\sqrt{2ML}(|u|^2+|v|^2)}$.
The state (\ref{zeromodevac}) is indeed annihilated by
the zero mode supercharge
\beqa
i{\bf Q}^{\rm zero}\!\!&=&\!\!\opsi^u\partial_{\overline{u}}
+\opsi^v\partial_{\overline{v}}+\sqrt{2M}(\eta_Lu+\bareta_Lv)
-\sqrt{2M}(\eta_0u+\bareta_0v),\nn\\
i{\bf Q}^{{\rm zero}\dag}\!\!&=&\!\!\psi^u\partial_{u}
+\psi^v\partial_{v}-\sqrt{2M}(\bareta_L\overline{u}+\eta_L\overline{v})
-\sqrt{2M}(\bareta_0\overline{u}+\eta_0\overline{v}).\nn
\eeqa
Note the normalization $\{\psi^u,\opsi^u\}=\{\psi^v,\opsi^v\}=2/L$.
It indeed corresponds to (\ref{chrelt}):
$$
|0\rangle^{\rm hol}=
\left(\begin{array}{cc}
1&0\\0&1
\end{array}\right)
$$
in the holomorphic truncation.

\section{The Grade Restriction Rule}
\label{sec:GRR}

We now discuss the first 
main quantum effect of the linear sigma model with boundary.
The goal of the paper is to construct parallel families of boundary 
interactions over the bulk of the K\"ahler moduli space,
including in particular the boundaries between different phases.
At each phase boundary, at least one $U(1)$ subgroup of the gauge group
is unbroken,
and the corresponding Coulomb branch has a bounded bulk potential
which vanishes exactly at the singular point.
The focus of study will be the behaviour of boundary interactions 
and boundary conditions on that Coulomb branch.
This leads us to {\it the grade restriction rule} (or 
{\it the band restriction rule}), which classifies
the Chan-Paton representation of the gauge group for the D-brane
that can be transported across the phase boundary.
This is the main result of this paper.

\subsection{A-branes In LG-Models}
\label{subsec:ALG}

The theory on the Coulomb branch is
described in terms of the twisted chiral superfield $\Sigma=\bD_+D_-V$
and has a superpotential which is classically 
$\widetilde{W}(\Sigma)=-t\Sigma$.
Note that B-type boundary conditions on twisted chiral superfields
are like A-type boundary conditions on chiral superfields.
To pave the way to discuss the boundary condition of bulk fields
on the Coulomb branch, we briefly digress 
to reexamine the general requirement for A-branes in Landau-Ginzburg models.
For simplicity, we consider the LG model of $n$ chiral superfields
spanning a flat Euclidean space $\C^n$, with some polynomial superpotential
$W$.

It was argued in \cite{GJS,HIV}
that an A-brane in a LG model is a Lagrangian submanifold
whose image in the $W$-plane must be
parallel to the real line, 
or equivalently
\beq
\Ip \,W={\rm constant}
\label{ImWcond}
\eeq
on the brane.
It is definitely true that (\ref{ImWcond}) must be
satisfied as long as we use the standard bulk action without a boundary term
and impose the standard D-brane boundary condition. By
{\it the standard D-brane boundary condition} we mean Dirichlet 
on normal and Neumann on tangent coordinates along with the condition
on fermions that follows from supersymmetry.
However, the requirement (\ref{ImWcond}) is relaxed if 
we modify the action by a suitable boundary interaction which leads to a
non-standard boundary condition.

Let us first examine the ${\mathcal N}=2_A$ supersymmetry of the 
boundary conditions themselves, which 
does not depend on the detail of the action.
It is convenient to use the real components of the chiral superfields,
$x^I,\psi^I_{\pm},f^I$
($I=1,...,2n$), which are related to the complex ones as
 $\phi^i=x^{2i-1}+ix^{2i}$,
$\psi_{\pm}^i=\psi_{\pm}^{2i-1}+i\psi_{\pm}^{2i}$, 
$F^i=-i(f^{2i-1}+if^{2i})$. We also introduce
$$
N^I:=\partial_1x^I-f^I\qquad I=1,\ldots,2n,
$$
in additon to $\psi^I=\psi^I_++\psi^I_-,\tpsi^I=\psi^I_+-\psi^I_-$.
The variation of these component fields reads
\beq
\begin{array}{c}
\delta x^I=i\epsilon_1\psi^I+i\epsilon_2{\mathcal J}^I_{\,\,J}\tpsi^J,
\\[0.2cm]
\delta\psi^I=-2\epsilon_1\dot{x}^I+2\epsilon_2{\mathcal J}^I_{\,\,J}N^J,
\\[0.2cm]
\delta\tpsi^I=-2\epsilon_1N^I+2\epsilon_2{\mathcal J}^I_{\,\,J}\dot{x}^J,
\\[0.1cm]
\delta N^I=i\epsilon_1\dot{\tpsi^I}+i\epsilon_2{\mathcal J}^I_{\,\,J}
\dot{\psi}^J,
\end{array}
\eeq
where ${\mathcal J}^I_{\,\,J}$ is the complex structure
(multiplication by $i=\sqrt{-1}$ written in real coordinate):
${\mathcal J}^{2i}_{\,\,2i-1}=-{\mathcal J}^{2i-1}_{\,\,\,2i}=1$ and
all other entries are zero.
$\epsilon_1$ is the parameter of the
${\mathcal N}=1$ supersymmetry 
while $\epsilon_2$ is the extension to ${\mathcal N}=2_A$.
We find an invariant set of boundary conditions of the form\footnote{
There are more general boundary conditions
which eventually lead to coisotropic branes \cite{KapOr}. However, they 
require at least complex dimension two, $n\geq 2$.
Since we are primarily interested in the Coulomb branch, 
especially for $U(1)$ gauge theories,
 we shall focus on the conditions of 
the type (\ref{Bcof}).}
\beq
x\in L, \qquad \psi\in {\rm T}_xL,\quad \tpsi\in {\mathcal J}_x{\rm T}_xL,
\quad N\in{\mathcal J}_x{\rm T}_xL,
\label{Bcof}
\eeq
for some submanifold $L$ of $\R^{2n}$.
This is consistent and complete when ${\rm T}_xL$ and
${\mathcal J}_x{\rm T}_xL$ have no overlap (except the origin $x$)
and span the whole of ${\rm T}_x\R^{2n}\cong\R^{2n}$,
$$
{\rm T}_xL\cap{\mathcal J}_x{\rm T}_xL=\{x\},\qquad {\rm T}_xL
+{\mathcal J}_x{\rm T}_xL=\R^{2n}.
$$
Such an $L$ is called a {\it totally real} submanifold of 
$(\R^{2n}, {\mathcal J})=\C^n$.
It must be middle dimensional, $\dim_{\R}L=n$.
To summarize, we have
an invariant set of boundary conditions (\ref{Bcof}) for each
totally real submanifold of $\C^n$.

Let us next examine the invariance of the action.
We first take the standard bulk action $S_{\rm bulk}$ without a boundary term. 
Its variation is
\beqa
\delta S_{\rm bulk}&=&\int_{\partial\surface}\dd t\,\Biggl[\,
{i\over 2}\epsilon_1
\left(g(\dot{x},\tpsi)-g(N,\psi)+2\psi^I\partial_I\Ip(W)
\right)
\nn\\
&&\qquad \quad +{i\over 2}\epsilon_2\left(g(\dot{x},
{\mathcal J}\psi)-g(N,{\mathcal J}\tpsi)
+2({\mathcal J}\tpsi)^I\partial_I\Ip(W)\right)\,\Biggr].
\eeqa
Vanishing of the term $\epsilon_1g(\dot{x},\tpsi)$ 
requires that ${\rm T}_xL$ and ${\mathcal J}_x{\rm T}_xL$ must be
orthogonal to each other,
$$
{\rm T}_xL\perp {\mathcal J}_x{\rm T}_xL.
$$
For a totally real submanifold $L$, this is nothing but the condition
that it is a {\it Lagrangian} submanifold with respect to the symplectic
structure $\omega(v,w)=g({\mathcal J}v,w)$.
Once that is satisfied, the only remaining condition is that
$\psi^I\partial_I\Ip W=({\mathcal J}\tpsi)^I\partial_I\Ip W=0$.
Since the vectors $\psi$ and ${\mathcal J}\tpsi$
are tangent to $L$, this means that
$\Ip W$ is locally constant on $L$. This is how the requirement
(\ref{ImWcond}) arises.
It is easy to see that the boundary condition (\ref{Bcof})
is consistent with the variational equation from the standard action:
Since the action has no boundary term,
the variational equation requires that $\partial_1 x$ must be normal
to the brane. Since $N=\partial_1 x-f\in {\mathcal J}{\rm T}L$
is normal to $L$ we find that $\partial_1x$ and $f$ must be independently
normal to $L$. If we use the bulk equation for the auxiliary field
\beq
f^I=-g^{IJ}\partial_J\Ip W,
\label{eomF}
\eeq
the condition that $f$ is normal to $L$ is automatically satisfied
provided $\Ip W$ is constant on $L$.
This is the basic story of \cite{HIV}.

The standard bulk action $S_{\rm bulk}$ can be modified by a boundary
term so that it is  ${\mathcal N}=2_A$ invariant without use of
any boundary condition.
In fact one can write down a manifestly ${\mathcal N}=2_A$
invariant action as follows:
\beqa
S_{\rm tot}&=&{1\over 2}\int_{\surface}\dd^2\s\,{\bf Q}_A{\bf Q}^{\dag}_A
[{\bf Q}_+,\overline{\bf Q}_+]K(\phi,\bphi)
-\Rp\int_{\surface}\dd^2\s\,{\bf Q}_A{\bf Q}^{\dag}_A W(\phi)
\nn\\
&=&S_{\rm bulk}
+\int_{\partial\surface}\dd t \,\left(\,
{1\over 2}\sum_{I,J=1}^{2n}g_{IJ}x^IN^I
-\Ip\, W\,\right)
\label{mani}
\eeqa
Here we used the K\"ahler potential 
$K={1\over 2}\sum_{I,J=1}^{2n}g_{IJ}x^Ix^J$.
If we use this action, obviously, no requirement should arise from the 
${\mathcal N}_A=2$ invariance. This time, 
however, a condition comes from the consistency of
the boundary condition (\ref{Bcof}) and the variational equation
from (\ref{mani}), which reads
$$
{1\over 2}g_{IJ}(\delta x^I N^J+x^I\delta N^J)
-\delta x^I(g_{IJ}\partial_1 x^J+\partial_I\Ip W)
+{i\over 4}g_{IJ}(\delta\psi^I\tpsi^J+\delta\tpsi^I\psi^J)=0.
$$
From the fermion terms, we find again the condition that
${\mathcal J}_x{\rm T}_xL$ must be orthogonal to ${\rm T}_xL$, that is, 
$L$ must be a Lagrangian submanifold.
If we use the bulk equation for the auxiliary field (\ref{eomF}),
the $\delta x^I$-terms combine to give
$g_{IJ}\delta x^IN^J=0$,
which is again satisfied under ${\mathcal J}_x{\rm T}_xL\perp {\rm T}_xL$.
This leaves us with the equation $\sum_{I,J=1}^{2n}g_{IJ}x^I\delta N^J=0$.
This is not satisfied for an arbitrary Lagrangian submanifold $L$
--- it has to be a Lagrangian plane that goes through the origin $x=0$,
that is,
a linear Lagrangian subspace of $\R^{2n}$.
Note that we have the modified Neumann condition
\beq
\partial_1x^{I_t}+g^{I_tJ}\partial_J\Ip W=0,
\label{modNe}
\eeq
in the direction tangent to the brane.

We have found that the manifestly invariant action (\ref{mani})
admits only a very spacial class of D-branes --- linear Lagrangian subspaces. 
In fact, one can modify the action by adding a boundary term that is
by itself ${\mathcal N}_A=2$ invariant. For example, we may add
a boundary D-term
\beqa
\Delta S_{\rm bdry}&=&
\int_{\partial\surface}\dd t\, {\bf Q}_A{\bf Q}_A^{\dag}h
\label{bDh}\\
&=&\int_{\partial\surface}\dd t\,
\left({\mathcal J}^I_{\,\,J}N^J\partial_Ih
+{i\over 2}({\mathcal J}\tpsi)^I\psi^J\partial_I\partial_Jh\right)
\nn
\eeqa
for some function $h$ of $\R^{2n}$. We see that this changes the
boundary term of (\ref{mani}) as
$$
g_{IJ}x^IN^J\to g_{IJ}\left(x^I-2\omega^{IK}\partial_Kh\right)N^J
$$
which corresponds to a Hamiltonian deformation of the brane $L$.
Thus, {\it D-term deformations on the boundary 
generate Hamiltonian deformations of the brane.}
If $h$ is linear in $x^I$'s then this simply corrsponds to
a parallel displacement of $L$. If $h$ is quadratic it corresponds to
a symplectic rotation of $L$. For a more general function, 
we obtain a more general Lagrangian submanifold. This is one way to obtain
more general D-branes than just linear Lagrangian subspaces.
Alternatively, we may simply take
\beq
S_{\rm tot}=S_{\rm bulk}+\int_{\partial\surface}\dd t\,\Bigl(-\Ip\,W
\,\Bigr).
\label{notmani}
\eeq
It is not automatically  ${\mathcal N}_A=2$ invariant but
the invariance requires only the Lagrangian condition
${\rm T}_xL\perp {\mathcal J}_x{\rm T}_xL$.
The variational equation is then solved again by (\ref{Bcof})
that includes the modified Neumann condition (\ref{modNe}).
Thus, we find an A-brane for {\it any} Langrangian submanifold $L$.
The imaginary part of $W$ does not have to be a constatnt on $L$.

In the new formulation, 
with the action (\ref{mani}) or (\ref{notmani}),
the system has a potential energy at the boundary
\beq
V_{\rm bdry}=\Ip\, W.
\label{bpot}
\eeq
The imaginary part of a holomorphic function
is unbounded below and above on $\R^{2n}$.
Thus, depending on the asymptotic direction of the brane $L$, 
the boundary potential can be unbounded below.
To avoid any problem associated with it, 
we propose to require that $\Ip\, W$ must be bounded below
on the brane $L$. 
Necessity of such a constraint
is not so obvious since there is also a bulk potential
$
U_{\rm bulk}={1\over 4}g^{i\bj}\partial_iW\partial_{\bj}\overline{W}.
$
An exception is the case where $W$ is linear, which is  
the main focus of the present section: the bulk potential is constant whereas
the boundary potential is linear. In this case,
it is absolutely necessary to 
require that $\Ip W$ is bounded below at every infinity of the brane $L$.
When $W$ is quadratic or higher, it is less clear whether we need
the constraint. However, as we shall see below,
it is preferable to keep this requirement also in the general case.

If we loose the requirement (\ref{ImWcond}),
the set of allowed D-branes is considerably expanded.
However, we may identify the branes that
are related by boundary D-term deformations.
We have seen that a Hamiltonian deformation of the brane corresponds to
the boundary D-term of the form (\ref{bDh}).
Here we must be careful ---
(\ref{bDh}) can be regarded as a deformation only when its effect is
small at infinity in the field space. 
In the absence of a superpotential,
we need to assume that the function $h$ approaches a constant at infinity
 or a linear function at most. 
For example, a quadratic function $h$ corresponds to a rotation
of the brane and cannot be regarded as a ``deformation'' since 
it results in an indefinitely
large move at infinity.
The parallel displacement, corresponding to a linear $h$, is the marginal case
where one may or may not regard it as a deformation.
In the presence of bulk and boundary potentials, 
a higher power of $h$ is allowed as long as the effect is small relative 
to the effect of the potentials. 
In fact, as long as the potentials grow fast enough along the brane $L$,
the wavefunction in that infinity direction is damped exponentially, so that
the Hamiltonian deformation for almost any function $h$
can be safely regarded as a boundary D-term deformation.

An important invariant under the boundary D-term deformation
is the overlap of the boundary state $|B_L\rangle$
with a R-R ground state $|i\rangle$, known as the 
{\it generalized central charge}.
(Here we assume that
the polynomial
$W$ is quadratic or higher so that there are supersymmetric ground states
as many as ${\rm deg} W-1$.)
It is realized as the path integral on a semi-infinite
cigar and is represented as the integral:
\beq
\langle i|B_L\rangle=\int_L\e^{-\beta\, \Ip\,W}\omega_i(\beta).
\label{gZ}
\eeq
Here $\beta$ is the circumferemce of the boundary circle of the cigar 
and $\omega_i(\beta)$ is the differential form on $\R^{2n}$ of middle degree
that corresponds to the ground state $|i\rangle$. To be precise,
the form $\omega_i(\beta)$ realizes the ground state of the supersymmetric
quantum mechanics obtained by compactification on
the circle of circumference $\beta$.
See \cite{book} for details. 
It is normalizable and has the asymptotic behaviour
$|\!|\omega_i(\beta)|\!|\sim \e^{-\beta|W|}$. 
Although it indeed decays exponentially at infinity,
the factor $\e^{-\beta\, \Ip\,W}$ coming from the boundary potential
(\ref{bpot}) can be dangerous if $L$ extends to the direction
with $\Ip\, W=-|W|$.
Thus, Hamiltonian deformations 
of $L$ across that dangerous hypersurface cannot be regarded as boundary
D-term deformations. Therefore, we should better avoid that hypersurface. 
Given the freedom to use boundary D-term deformation, we may even
require that $\Ip W$ grows at every asymptotic direction of $L$.
In that case the integral (\ref{gZ})
can be recast into an integral of a holomorphic differential
$\int_L\e^{i\beta W}\Omega_i(\beta)$  \cite{book}.

\begin{figure}[tb]
\centerline{\includegraphics{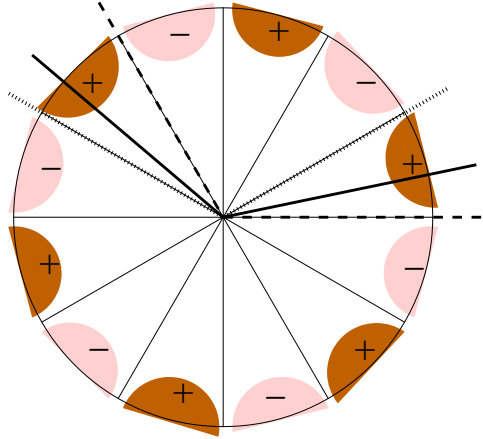}}
\caption{Some A-branes in LG model with $W=\Phi^6$.}
\label{fig:ALG1}
\end{figure}

This picture of A-branes in LG models is consistent with
all known results based on (\ref{ImWcond}) and also on other methods. 
It actually explains some of the puzzles in the old picture.
For example, 
in \cite{HIV} only branes with
$\Ip W=$ constant and $\Rp W\to +\infty$ were considered, and 
those with $\Rp W\to -\infty$
were completely ignored.
In the new picture,
those branes with $\Rp W\to +\infty$
and those with $\Rp W\to -\infty$ are connected 
via D-term deformations through the region with large positive $\Ip W$.
There is a one to one correspondence between the 
D-isomorphism classes in such two sets of branes.
For illustration, let us consider the LG
model for the minimal model $W=\Phi^{k+2}$.
If we require $\Ip W=0$, a brane is a union of two rays in the directions
$\e^{\pi mi/(k+2)}$, $m=0,...,2k+3$. However, only rays in the directions
$\e^{2\pi mi/(k+2)}$, $m=0,...,k+1$ were considered in \cite{HIV}.
In the new picture, each ray can be in any direction bewteen
$\e^{2\pi mi/(k+2)}$ and $\e^{\pi (2m+1)i/(k+2)}$ so that $\Ip W$ 
is bounded below at infinity. 
Fig.~\ref{fig:ALG1} shows the case $k=4$. 
The dashed line is 
an example of a brane taken 
in \cite{HIV} while the centipede line satisfies $\Ip W=0$ but was ignored.
In the new picture any brane is allowed as long as it has asymptotic
region with positive $\Ip W$.
The solid line is such an example and it indeed connects
the dashed line and the centipede line.
For the branes preserving the opposite supercharge
${\bf Q}_A'=\overline{\bf Q}_+-{\bf Q}_-$ (or for the branes
preserving the same supercharge but on the left boundary), 
the boundary potential has the opposite sign, $V^{\rm bdry}=-\Ip\, W$,
and thus the requirement is that $\Ip\, W$ is bounded above
on $L$. Then, the rays must be between 
$\e^{\pi (2m-1)i/(k+2)}$ and
$\e^{2\pi mi/(k+2)}$. This alternation pattern has been observed
in the geometrical picture of D-branes in
the $U(1)$ gauged $SU(2)$ WZW model (Kazama-Suzuki model)
that realizes the minimal model \cite{MMS}.

\subsection{Asymptotic Condition In The Coulomb Branch}
\label{subsec:AsympC}

We now come back to the study of boundary linear sigma models.
Our focus will be the boundary condition on the Coulomb branch.
As discussed in the previous section,
we need to specify a Lagrangian submanifold $L$
on which the boundary potential is bounded below.
We first consider a model with $U(1)$ gauge group and assume the 
Calabi-Yau condition
$$
\sum_{i=1}^nQ_i=0.
$$

In the region with large $\sigma$, it is appropriate to
integrate out the charged matter fields.
This yields corrections to
the potential and kinetic terms and also produces
higher derivative terms. 
Corrections to the kinetic term and higher derivative terms
are written as power series in $e/|\sigma|$ and
$k_{\mu}/|\sigma|$ and are negligible in the low energy limit
in the region $|\sigma|\gg e$.
Therefore, we can use the classical kinetic term
(\ref{Sgggg}) in that regime.
Since it is written in the manifestly supersymmetric form
with quadratic K\"ahler potential,
as discussed in the previous section,
the Lagrangian $L$ must be a linear Lagrangian subspace.
Namely, it must be asymptotically a straight line in the Coulomb branch.
Of course the correction will be large 
in the region $|\sigma|<e$, and there is no reason for
$L$ to be straight there.
Thus, the Lagrangian $L$ is a bent line, coming in toward the origin
from one asymptotic direction and going out to
infinity in another direction.
In order for the brane to be coupled to the Higgs
branch theory, $L$ should better go through the region $|\sigma|\ll e$.

The correction to the potential can be non-zero even in the 
asymptotic region.
In the bulk, the twisted superpotential
is corrected simply by the shift of the FI-theta parameter,
$\widetilde{W}=-t\Sigma\to
\widetilde{W}_{\it eff}=-t_{\it eff}\Sigma$
$$
t_{\it eff}
=t+\sum_{i=1}^nQ_i\log (Q_i)
\quad\mbox{mod $2\pi i\Z$}.
$$
The $2\pi \Z$ ambiguity of the imaginary part is fixed by the 
boundary charge. 
Quantum corrections to the boundary potential and the boundary charge 
were analyzed in  Section~\ref{sec:vacuum}.
For the Wilson line brane with charge $q$,
the effective boundary potential is
\beqa
V^{\rm bdry}_{\it eff}&=&
{1\over 2\pi}\left(r+\sum_{i=1}^nQ_i\log|Q_i|\right)\Ip(\sigma)
-\left({\theta\over 2\pi}+q\right)\Rp(\sigma)
+{1\over 4}\sum_{i=1}^n|Q_i\Rp(\sigma)|,
\nn\\
\label{bpotLSM}
\eeqa
and the effective charge is
\beq
q_{\it eff}=q-{\rm sgn}\Rp(\sigma)\sum_{i=1}^n{|Q_i|\over 4}.
\label{bchLSM}
\eeq
The potential can indeed be written as
$V^{\rm bdry}_{\it eff}=-{1\over 2\pi}\Ip\,\widetilde{W}_{\it eff}$
with $t_{\it eff}=r_{\it eff}-i(\theta+2\pi q_{\it eff})$
where 
$r_{\it eff}=r+\sum_{i=1}^nQ_i\log |Q_i|$.
(In comparison with (\ref{bpot}),
the factor of ${1\over 2\pi}$ is just a convention and the sign
is the difference between chiral and twisted chiral superpotentials.)

Note that there is a discontinuity in the effective charge
and singularity in the potential at $\Rp(\sigma)=0$, that is,
on the imaginary line. 
This is due to the appearance of zero modes 
from the charged matter sector:
The decaying modes
localized at the boundary
become zero modes exactly at $\sigma=i|\sigma|$ for positively charged
fields and at $\sigma=-i|\sigma|$ for negatively charged
fields.
In those directions,  no matter how large $|\sigma|$ is,
the matter fields are not really decoupled from
the low energy dynamics.
The effective description
purely in terms of $\Sigma$ breaks down at $\sigma=\pm i|\sigma|$.
In this sense we shall sometimes call the imaginary $\sigma$ line
a {\it singular line}.

The asymptotic lines for $L$ must be such that the boundary potential
(\ref{bpotLSM}) is bounded below. 
Let us depict the region of positive boundary potential,
for various values of $r$ and a fixed ${\theta\over 2\pi}+q$.
The behaviour depends very much on the relation of
${\theta\over 2\pi}+q$ and the charge shift $\pm{1\over 2}{\mathscr S}$,
\beq
{\mathscr S}:={1\over 2}\sum_{i=1}|Q_i|=\sum_{Q_i>0}Q_i.
\eeq
We consider three cases separately.

\begin{figure}[h]
\psfrag{lpos}{\footnotesize $r_{\it eff}\gg 0$}
\psfrag{pos}{\footnotesize $r_{\it eff}>0$}
\psfrag{zero}{\footnotesize $r_{\it eff}= 0$}
\psfrag{neg}{\footnotesize $r_{\it eff}< 0$}
\psfrag{lneg}{\footnotesize $r_{\it eff}\ll 0$}
\centerline{\includegraphics{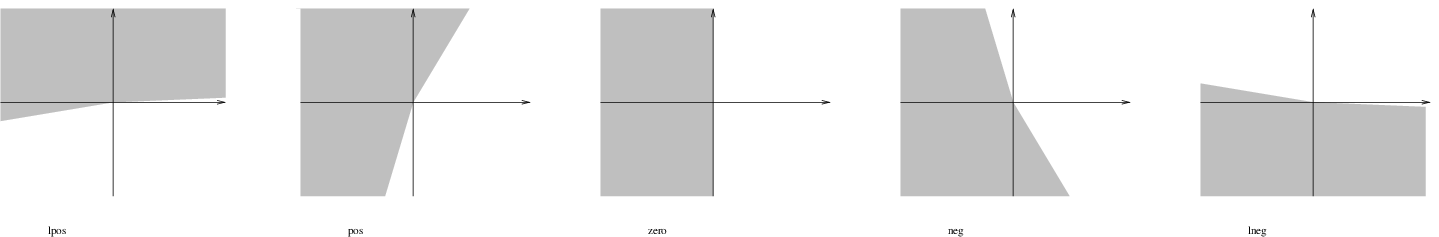}}
\caption{${\theta\over 2\pi}+q>{1\over 2}{\mathscr S}$}
\label{fig:Vpos}
\end{figure}

\medskip

\begin{figure}[h]
\psfrag{lpos}{\footnotesize $r_{\it eff}\gg 0$}
\psfrag{pos}{\footnotesize $r_{\it eff}>0$}
\psfrag{zero}{\footnotesize $r_{\it eff}= 0$}
\psfrag{neg}{\footnotesize $r_{\it eff}< 0$}
\psfrag{lneg}{\footnotesize $r_{\it eff}\ll 0$}
\centerline{\includegraphics{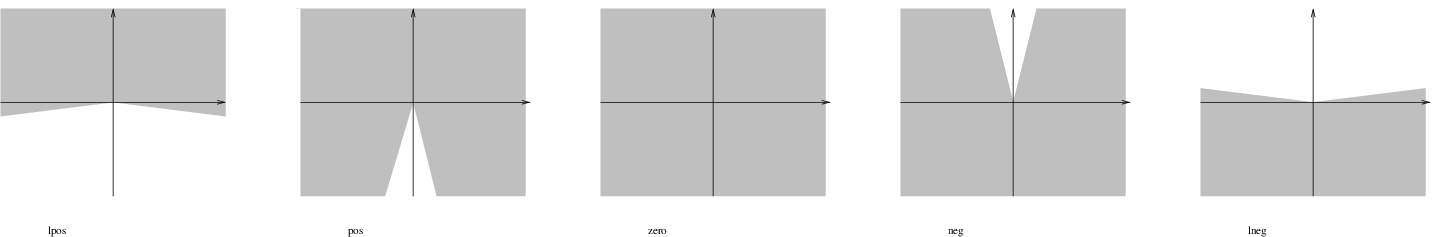}}
\caption{$-{1\over 2}{\mathscr S}<
{\theta\over 2\pi}+q<{1\over 2}{\mathscr S}$}
\label{fig:Vgrr}
\end{figure}

\medskip

\begin{figure}[h]
\psfrag{lpos}{\footnotesize $r_{\it eff}\gg 0$}
\psfrag{pos}{\footnotesize $r_{\it eff}>0$}
\psfrag{zero}{\footnotesize $r_{\it eff}= 0$}
\psfrag{neg}{\footnotesize $r_{\it eff}< 0$}
\psfrag{lneg}{\footnotesize $r_{\it eff}\ll 0$}
\centerline{\includegraphics{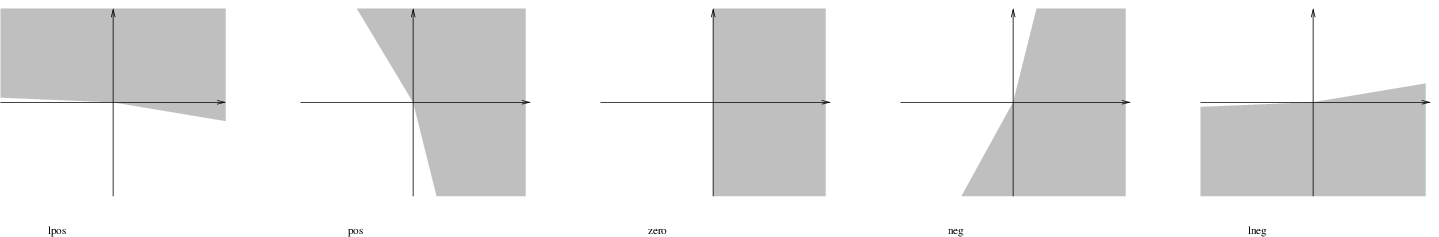}}
\caption{${\theta\over 2\pi}+q<-{1\over 2}{\mathscr S}$}
\label{fig:Vneg}
\end{figure}

\noindent
The boundary potential is positive in the shaded
regions.
If ${\theta\over 2\pi}+q$ is larger than ${{\mathscr S}\over 2}$
or smaller than $-{{\mathscr S}\over 2}$,
the positive potential region rotates roughly
by $180$ degrees as $r$ is changed from
$+\infty$ to $-\infty$.
Not a single Lagrangian
can stay inside the region for all values of $r$. 
On the other hand, if $-{1\over 2}{\mathscr S}<
{\theta\over 2\pi}+q<{1\over 2}{\mathscr S}$, 
the region does not rotate. In particular, the real line
is always inside.

\begin{figure}[h]
\centerline{\includegraphics{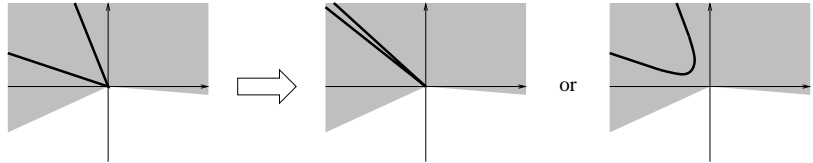}}
\caption{Annihilation of the brane.}
\label{fig:annihilation}
\end{figure}

As discussed in the previous section, we can deform the brane $L$
as long as they stay inside the admissible region --- that would be 
a boundary D-term deformation. If we use this freedom we notice that
some of the configurations should be regarded as trivial. For example, 
we can consider deformations as shown in
Figure~\ref{fig:annihilation}.
When the two rays are deformed to coincide,
brane-antibrane annihilation takes place.
Alternatively, 
the brane may be detached from the center of the Coulomb branch.
Then, it is decoupled from the Higgs branch theory
and will eventually disappear to infinity.
The existence of such deformations 
means that the starting configuration (left) should be regarded
as trivial. 
This does not apply if the two rays of $L$ are on the opposite
side of the singular line at $\sigma=\pm i|\sigma|$.
In order to move the two rays until they coincide,
either one or both of them must hit the singular line.
Also, even if the brane is detached from the origin, it will still 
intersect with the singular line and continue to be coupled to 
the matter sector.

\subsection{Rules Of D-Brane Transport}
\label{subsec:rule}

We now describe rules of D-brane transport. We first consider models with 
$U(1)$ gauge group and next the cases of higher rank gauge groups.

\subsubsection{$U(1)$ Gauge Group ---The Grade Restriction}

We are interested in transporting branes
along a path that goes from the large volume phase
$r\gg 0$ to the small volume phase $r\ll 0$, or in the opposite direction.
The path should better avoid the singular points
at $r_{\it eff}=0$ and $\theta\in 2\pi\Z+\pi{\mathscr S}$.
\begin{figure}[h]
\centerline{\includegraphics{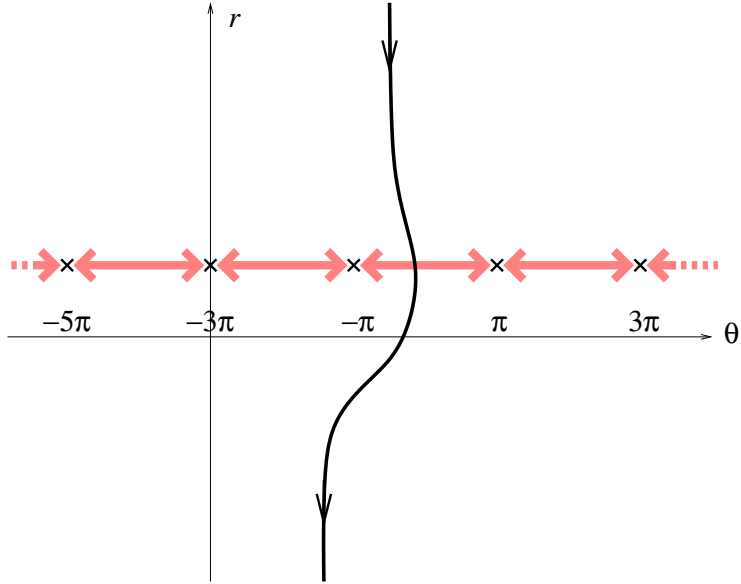}}
\caption{A path through one of the windows.}
\label{fig:apath}
\end{figure}
Thus, it must go through the windows between the singular points.
For simplicity, we consider paths that goes just once through one 
of the windows, as shown in Figure~\ref{fig:apath}. (The Figure is
for the case where
${\mathscr S}$ is odd. For even ${\mathscr S}$, the windows
are at $2\pi n<\theta<2\pi(n+1)$, $n\in\Z$).
More complicated paths are just some combinations of such simple paths.
We claim 

\medskip
\noindent
{\bf The Grade Restriction Rule}:\\[0.2cm]
{\it The Wilson line brane $\wilson(q)$ can be transported smoothly between 
the two phases if and only if the inequality
\beq
-{{\mathscr S}\over 2}<{\theta\over 2\pi}+q<{{\mathscr S}\over 2}
\label{GRR}
\eeq
holds in the window.
The same can be said on a complex of Wilson line branes $\wilson(q_i)$
or a matrix factorization of the superpotential:
each $q_i$ must obey the inequality (\ref{GRR}) in the window
in order for the brane to be transported smoothly.}

\medskip
\noindent
For a given choice of window, this rule selects ${\mathscr S}$
consequtive integers as the allowed set of boundary charges $q$. For example, 
in a theory with ${\mathscr S}=3$, the allowed set is
$\{-1,0,1\}$ if the window is at $-\pi<\theta<\pi$.
If we change the window to $-3\pi<\theta<-\pi$, the set changes
to $\{0,1,2\}$. Below, we provide a derivation of this rule.

\begin{figure}[h]
\centerline{\includegraphics{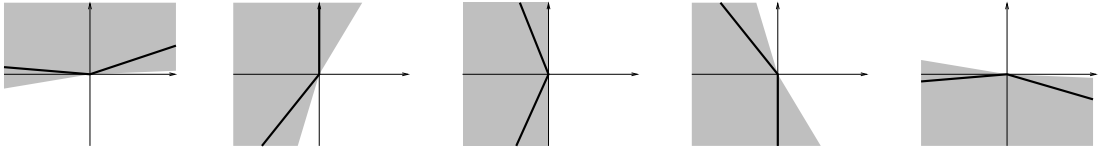}}
\caption{Rotation of $L$ in the case 
where $q$ is too large for (\ref{GRR}).}
\label{fig:Pos1}
\end{figure}

Suppose the inequality (\ref{GRR}) is not satisfied on the window.
To be concerete, suppose that $q$ is too large for that.
As we move along the path, say from $r\gg 0$ to $r\ll 0$,
the region of positive boundary potential rotates
counter-clockwise as in Figure~\ref{fig:Vpos}.
The rays of $L$ must be rotated
so that they are always inside that region.
We depict an example of such a rotation
in Figure~\ref{fig:Pos1}.
The starting configuration (left) is chosen so that the two rays
are on the opposite side of the singular line $\sigma=\pm i|\sigma|$.
As we have discussed above, this is required for the brane to be non-trivial.
At some moment before we go through the window,
the ray that was on the right of the singular line must overlap with 
the singular line on $\sigma=i|\sigma|$ (second from left).
At that moment, something non-trivial must happen to the brane.
If nothing had happened, then after that we would have a configuration
where both of the rays are on the left of the singular line (center).
But that configuration is trivial as the brane would annihilate by 
admissible deformations.
This is in contradiction to the fact that 
the starting configuration was non-trivial.
Somthing non-trivial must also happen when the other ray overlaps
with the singular line on $\sigma=-i|\sigma|$ (second from right).
At this stage, we cannot exactly tell {\it what} really happens
except that it has to do with the zero modes from the matter sector. 
Later in this paper, we will discuss this point
from a different perspective.

Let us now consider the case where
 the inequality (\ref{GRR}) is satisfied.
If the path is a straight line at a constant $\theta$
obeying (\ref{GRR}), then
there is no need to move $L$ as the real line
\beq
\Ip(\sigma)=0
\label{Imsigma}
\eeq
is always inside the admissible region.
If the path is curved as in Figure~\ref{fig:apath},
then it may happen that the inequality is not satisfied
along a part of the path.
Then, we may need to deform $L$ away from the real line.
However, as long as the inequality is satisfied on the window at
$r_{\it eff}=0$,
the admissible region is always on both sides
of the singular line $\sigma=\pm i|\sigma|$,
and the two rays do not have to overlap
with the line, never
at any point of the path.
Thus, the brane can be continuously deformed from one phase to another.

To summarize, we can consider branes
for arbitrary boundary charges $q_i$. They are non-trivial as long as 
the two rays of $L$ are on opposite sides of
the singular line $\sigma=\pm i|\sigma|$.
However, only those obeying the grade restriction rule
can be continuously transported between the two phases through the window.
For a component of a brane which does not satisfy the grade restriction rule,
the asymptotic condition on the Coulomb branch
$L$ must overlap with the singular line
$\sigma=\pm i|\sigma|$ along the way,
and {\it something non-trivial} is expected to happen.
This is the meaning of the grade restriction rule.

\subsubsection{Higher Rank Gauge Group --- The Band Restriction}

We now consider the model with higher rank gauge group 
$T\cong U(1)^k$ ($k\geq 2$).
We are interested in transporting branes along a path that goes
from one phase to another
in the multi-dimensional K\"ahler moduli space $\moduli_K$.
Here we focus on paths that go from one phase to an adjacent phase
through the phase boundary in the asymptotic region, that is, 
the region with large values of FI parameters.
Any two phases can be connected by a combination of such paths.

In the asymptotic region, there is exactly one $U(1)$ subgroup
of the gauge group $T$ that can be unbroken at the phase boundary. 
The remaining gauge group is completely Higgsed and can be ignored.
Repeating what we have done before for the unbroken $U(1)$ subgroup,
we find a rule of D-brane transport.
Let us arrange the 
basis of the Lie algebra of $T$ so that the first factor is the unbroken
$U(1)$. 
Then, the coordinate $r^1$ is transverse to the phase boundary.
The singular locus is asymptotically at
$r_{\it eff}^1=r^1+\sum_{i=1}^nQ_i^1\log|Q^1_i|=0$
and $\theta^1\in 2\pi \Z+{\pi}{\mathscr S}^1$ where
\beq
{\mathscr S}^1:=\sum_{Q^1_i>0}Q^1_i.
\eeq
We fix one of the windows
at $r^1_{\it eff}=0$ and
consider a path that goes though it.
We have

\medskip
\noindent
{\bf The Band Restriction Rule}:\\[0.2cm]
{\it The Wilson line brane $\wilson(q^1,...,q^k)$ 
can be transported smoothly across the phase boundary 
if and only if the inequality
\beq
-{{\mathscr S}^1\over 2}<{\theta^1\over 2\pi}+q^1<{{\mathscr S}^1\over 2}
\label{BRR}
\eeq
holds in the window. 
The same can be said for a complex of Wilson line branes $\wilson(\vec{q_i})$
or a matrix factorization of the superpotential:
each $q^1_i$ must obey the inequality (\ref{BRR}) in the window.}

\begin{figure}[h]
\psfrag{1}{\footnotesize $q^1$}
\psfrag{2}{\footnotesize $q^2$}
\centerline{\includegraphics{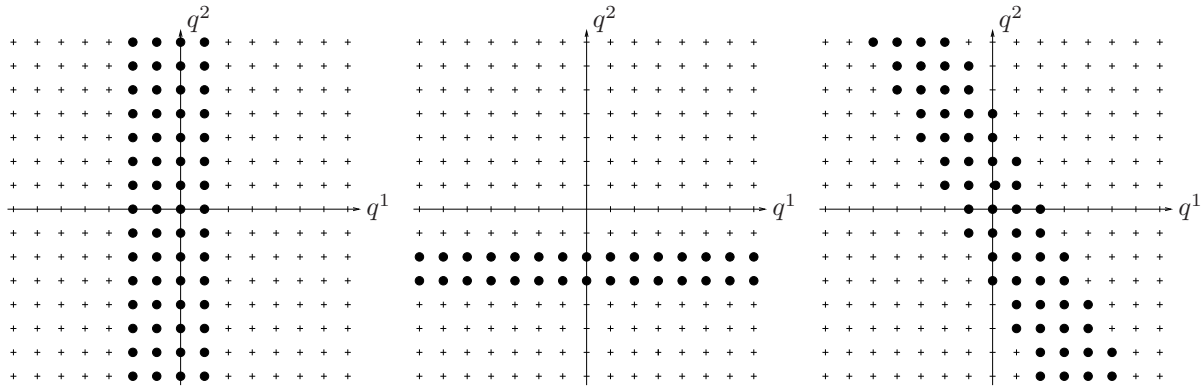}}
\caption{Examples of bands in the two parameter model.}
\label{fig:band}
\end{figure}

\medskip
\noindent
There is no condition on $q^2,...,q^n$.
For a choice of window, this selects a band 
of width ${\mathscr S}^1$ in the lattice of charge vectors.
For example, let us consider the two parameter model (Example (C)
in Section~\ref{subsec:LSMexamples}), and look at the
boundary between Phase I and Phase IV. The unbroken subgroup
is the original $U(1)_1$, with the number ${\mathscr S}^1=4$.
If we choose the window $0<\theta^1<2\pi$ at
$r_{\it eff}^1=r^1-4\log 4=0$, $r^2\gg 0$, then the band is
$\{(q^1,q^2)\,|\,q^1=-2,-1,0,1\}$ as shown in 
Figure~\ref{fig:band} (left).
The Figure also shows a sample of bands at other phase boundaries;
the I-II or III-IV boundary (middle) and
the II-III boundary (right).

\subsection{The Nature Of The Rule}
\label{subsec:nature}

The above derivation of the grade/band restriction rule
is certainly concerned with the transport of D-branes
in a linear sigma model with {\it finite gauge coupling} $e$
which is formulated on a worldsheet of {\it finite (spatial) volume}.

In fact, the effective boundary potential (\ref{bpotLSM})
as well as the effective boundary charge (\ref{bchLSM}) 
ceases to be valid if we take the infinite coupling or infinite
volume limit.
The validity of a perturbative
treatment of the gauge interaction that leads to these expressions 
had been analyzed in Section~\ref{subsec:gaugedynamics}.
Here we would like to discuss it from a different
perspective. As before, we consider
the strip of width $L$.
There is a bulk contribution to the total energy
\beq
E_{\rm bulk}={e^2L\over 2}r_{\it eff}^2
+{e^2L\over 2}\left({\theta\over 2\pi}+q_b\right)^2.
\eeq
The second term is the electrostatic energy:
$q_b$ is the charge at the (right) boundary that determines
the value of the electric field via the Gauss law.
The boundary charge $q_b$ can be changed if we choose a different state
from the matter sector. Such a change may reduce the electrostatic energy
$E_{\it es}$,
but it may increase the energy of the matter sector at the same time.
The effective potential and charge are obtained by finding 
the configuration that minimizes the {\it total} energy.
The results (\ref{bpotLSM}) and (\ref{bchLSM}) are obtained
by ignoring the contribution of $E_{\it es}$ in the minimization
problem.
That is valid when the excitation energy in the matter sector
is much larger than 
the possible change in $E_{\it es}$ by the change of boundary charge.
The separation of the energy levels is of order $|\sigma|$ for 
oscillating modes and $|\Rp(\sigma)|$ for decaying modes.
Thus, the region of validity is
\beq
|\Rp(\sigma)|\gg e^2L.
\eeq
This is indeed the same as the condition (\ref{pertcond}) we obtained earlier.
In the opposite regime, $|\Rp(\sigma)|\ll e^2L$, while $|\sigma|$
is still much larger than the gauge coupling $e$,
the electrostatic bulk energy $E_{\it es}$ is much more important
than the boundary potential. Then, the state in the matter sector 
must be such that the boundary charge $q_b$ minimizes the
electrostatic energy.
Thus, ``the singular {\it line}''  of purely imaginary $\sigma$, which played
the crucial r\^ole for the grade/band restriction rule,
is actually a singular {\it band} of width $\sim e^2L$. This band spreads over
the entire $\sigma$-plane if we take 
the infinite volume limit $L\to \infty$ or 
infinite gauge coupling limit $e\to \infty$, or any limit 
that sends
$e^2L$ to infinity.

The key question is what is the rule of D-brane transport
in a theory formulated on
a worldsheet of infinite volume, or in the conformal field
theory to which the linear sigma model flows to
in the infra-red limit.
Does the grade restriction rule still hold?
Is there some phase transition that is not visible
at finite volume and finite gauge coupling?

A similar issue was discussed in the bulk theory \cite{phases}
concerning the meaning of the Coulomb branch analysis
to the phase transition in the infinite volume system.
It was first noted that the existence of a phase transition
depends on the particular path
in the multi-parameter space of theories. 
Then the focus was concentrated on the paths inside conformal field theories
for which we know the answer from mirror symmetry.
It was argued that the finite volume and finite coupling
theory is enough to find all possible singularities of the conformal
field theories:
if the singularity is purely due to the vanishing of
the Coulomb branch potential ($E_{\rm bulk}$ above),
then the potential only seems to grow in the infra-red limit $e\to\infty$
if that was non-zero for finite $e$ and finite $L$.
The only possible problem in this argument was
the conflict of the condition $|\sigma|\gg e$ for the Coulomb branch
analysis and the infra-red limit $e\to \infty$.

The situation is much more subtle in the
boundary theory.
It is certainly the case that the existence of a singularity
depends on the particular path
in the multi-parameter space of theories. 
The grade/band restriction rule is derived for
the path in the $(r,\theta)$ space at fixed, {\it finite}
values of $e$ and $L$: there is no singularity in the transport of the brane
if and only if any of the boundary charges
of the brane is inside the grade/band restriction range 
(\ref{GRR}) or (\ref{BRR}).
Does it tell something about transport of the corresponding 
conformally invariant brane
in the corresponding family of conformal field theories?
For branes with charges outside the grade/band restriction range, 
the answer is simple.
We do not even have a family of boundary conformal field theories
from such a family of linear sigma model branes.
This is so as long as they are defined as the infra-red limit of
boundary QFTs with finite $e$ and finite $L$.
The grade restriction rule is valid
as long as $e^2L$ is finite, no matter how large it is.
There is a problem already at finite volume and finite coupling,
and hence there is nothing to study.

Things are non-trivial for branes with charges 
inside the grade/band restriction range.
First of all, we do have a family of boundary conformal field
theories from such a linear sigma model brane.
The question is then whether there is a singularity along the way.
There is no compelling argument like the one for the bulk theory that
shows the non-existence of a singularity.
On the contrary, it is {\it expected} that there are
singularities or more appropriately transition points.
We know many examples of D-brane decays when the closed string background
is dialed through walls of marginal stability:
on one side of the wall the brane is stable and on the other side
it decays to, say, two stable D-branes. 
From the worldsheet point of view, the point of marginal stability
can be regarded as a transition point in the
family of boundary conformal field theories.
Thus, it is not that there is no singularity if the grade/band
restriction rule is obeyed.
But it is that a grade restricted brane
provides a family of boundary conformal field theories.
It provides us with a starting point to study stability and D-brane decay.

\subsection{The Full Boundary Conditions}
\label{subsec:fullBC}

Let us explicitly write down the boundary conditions on the bulk fields.
For the Wilson line brane $\wilson(q)$ in a
$U(1)$ gauge theory, the Lagrangian submanifold
$$
\Ip(\sigma)=0
$$
is in the admissible region as long as $(r,\theta)$ is
in the strip domain $-{{\mathscr S}\over 2}<
{\theta\over 2\pi}+q<{{\mathscr S}\over 2}$, $-\infty<r<\infty$.
Thus, we take it as a part of the boundary condition of the vector 
multiplet fields.
The boundary conditions on other component fields, (\ref{Bcof}), read
$\lambda_++\lambda_-=0$ and $\partial_1\Rp(\sigma)+v_{01}=0$.
Actually, we may take $v_1=0$ as an extra condition
since its ${\mathcal N}=2_B$ variation is proportional 
to that of $\Ip(\sigma)$.
In this way, we find the following set of boundary conditions
\beq
\begin{array}{l}
\Ip(\sigma)=0,\\
\lambda_++\lambda_-=0,\\
\partial_1(v_0-\Rp(\sigma))=0,\\
v_1=0.
\end{array}
\label{bcVec}
\eeq
Another set of conditions follows from the Gauss law constraint:
\beq
\begin{array}{l}
\displaystyle \!\!\!\!
-{1\over e^2}v_{01}=\theta+2\pi q,\\[0.2cm]
\partial_1(\lambda_+-\lambda_-)=0,\\
\partial_1(D+\partial_1\Ip(\sigma))=0.
\end{array}
\label{BCGauss}
\eeq
If we use (\ref{bcVec}), the conditions on the matter chiral multiplet fields
(\ref{bcChi}) simplify to
\beq
\begin{array}{l}
\partial_1\phi=0,\\
\psi_+=\psi_-,\\
\partial_1(\psi_++\psi_-)=0,\\
F=0.
\end{array}
\label{bcChisimp}
\eeq
These are the full set of boundary conditions on the bulk fields.
These are invariant under the ${\mathcal N}=2_B$ supersymmetry and 
are compatible with the variational equation.

At a point $(r,\theta)$ away from the strip domain,
only one half of the line $\Ip(\sigma)=0$ is in the admissible region.
As we have discussed, the two asymptotic lines of 
$L$ must be on the opposite sides of the singular line $\sigma=\pm i|\sigma|$.
Thus, we take $L$ to be the one obtained from $\Ip(\sigma)=0$ by rotating
one of the asymptotic lines so that it is in the admissible region. 
This is possible as long as $r_{\it eff}\ne 0$.
In particular, for $r\gg 0$ or $r\ll 0$, we can take $L$ to be
very close to the real line $\Ip(\sigma)=0$. 
The other set of conditions (\ref{BCGauss}) can still be imposed.
For the matter sector, we need to use the original condition
(\ref{bcChi}) for the part of $L$ away from $\Ip(\sigma)=0$.

\subsubsection*{Remarks}

\noindent
{\bf (i)}~~
The condition on the matter fields (\ref{bcChisimp})
is identical to the ultra-violet boundary condition (\ref{bcChi0}).
At this moment we do not know the significance of this observation.
One possibility is that the $e=0$ ultra-violet theory
can be used to study some of the properties of the grade restricted branes. 
Indeed, as we will see in later sections,
the $e=0$ theory provides correct results for
the space of chiral ring elements for any pair of grade restricted branes.

\noindent
{\bf (ii)}~~ 
When there are bulk and boundary interactions from the superpotential
$W$ and the tachyon profile $Q$, the variational equation is changed and
the boundary conditions should be modified. 
However, we can still use the above 
boundary conditions as long as we treat these interactions as perturbations.
Recall the remark on this approach given in 
Section~\ref{subsec:firstlookBC}. Here, we would like to
comment on the treatment of the Gauss law.
Suppose a brane consists of more than one Wilson lines with 
various different charges $q_i$. The Gauss law constraint
shows that the bulk fields obey different boundary conditions
for different Chan-Paton components of the same brane.
This may look strange at first sight
but a moment of reflection reveals that
there is nothing wrong, at least in perturbation theory.
In the path-intergal formalism, we expand the path ordered
exponential into a Dyson series.
Each term of the series is a product of matrices of the form
$\rho_*(v_0-\Rp(\sigma))$,
$\psi^i\partial_iQ$, $\{Q,Q^{\dag}\}$, etc, which are
inserted at points of the worldsheet boundary.
These points separate the boundary into
segments --- each segment corresponds to one of the Chan-Paton
components.
Then, it is simply that we impose different
conditions on different parts of the worldsheet boundary.
In the operator formalism, the situation is even clearer.
The Chan-Paton space is already 
a part of the quantum Hilbert space ---
an open string wavefunction is a product of
the bulk part and the Chan-Paton factor.
It is simply that the bulk fields obey boundary conditions
that depend on the Chan-Paton factor.

\noindent
{\bf (iii)}~~ 
That one can take an extra condition $v_1=0$
is a special feature of the Lagrangian submanifold $\Ip(\sigma)=0$.
In fact a similar property is possessed by the singular line
$$
\Rp(\sigma)=0
$$ 
on which the decaying modes from the matter sector become zero modes.
One can take an extra condition $v_0=0$, since its 
${\mathcal N}=2_B$ variation is identical to that of
$\Rp(\sigma)$ (that is why the combination $v_0-\Rp(\sigma)$
is supersymmetric). 
The ${\mathcal N}=2_B$ invariant set of conditions including these
is
\beq
\begin{array}{l}
{\rm Re}(\sigma)=0,\\
\lambda_+-\lambda_-=0,\\
D+\partial_1{\rm Im}(\sigma)=0,\\
v_0=0.
\end{array}
\label{bcVec2}
\eeq
This may be completed by another set of conditions:
\beq
\begin{array}{l}
\partial_1(\lambda_++\lambda_-)=0,\\
\partial_1(v_{01}+\partial_1{\rm Re}(\sigma))=0,\\
\partial_1{\rm Im}(\sigma)=0.
\end{array}
\label{bcVec2p}
\eeq
In fact, (\ref{bcVec}) and (\ref{bcVec2}) are the only
possible ${\mathcal N}=2_B$ invariant sets that include
two independent equations of the form $f=0$ and
$gv_0+hv_1=0$ where $f,g,h$ are real valued functions
of $\sigma$ and $\bsigma$.
We do not yet understand the significance of this observation.

\section{Non-Compact Models}
\label{sec:noncompact}

\newcommand{\lsm}{\mathrm{LSM}}

In this section, we achieve our main goal of the paper:
using linear sigma models, we construct
parallel families of D-branes over regions of the 
K\"ahler moduli space $\moduli_K$ that encompass various different phases.
The key result of the previous section,
the grade (or band) restriction rule, plays
the most important role in the construction.
This section focuses on models without superpotential, whose low energy
theories are non-linear sigma models on toric Calabi-Yau varieties,
possibly with orbifold singularities.

In Section~\ref{subsec:LEBC} we study 
how D-branes of the linear sigma
model descend to the low energy theories deep inside the phases
of the K\"ahler moduli space. 
We proceed in two steps:
First, we determine the low
energy boundary interaction of a given complex of Wilson
line branes (\ref{LSMcomplex}). 
Second, we take into account D-term
deformations and brane-antibrane annihiliation,
named {\it D-isomorphisms} in Section~\ref{sec:D-isom},
which do not change
the low energy behaviour of boundary interactions.

In Section~\ref{subsec:grrrevisited}, we apply
 the grade (or band) restriction rule
and transport D-branes along paths that connect adjacent phases,
achieving our main goal of the paper.
As a consequence,
this leads to one to one correspondences between D-isomorphism classes
of D-branes at different phases. Mathematically, this results in
 equivalences of
the derived categories of different toric varieties.

As an application, we obtain an honest
understanding of D-brane monodromies associated to closed
loops in K\"ahler moduli space, which we shall consider in
Section~\ref{subsec:NCmonodromy}. 
In Section~\ref{subsec:moreex}, we demonstrate the power
of construction in typical examples: the (local) flop
transition and the McKay correspondence.
In Section~\ref{subsec:center}, we discuss families of
D-branes over the central region of the 
moduli space where multiple phase boundaries meet.
We find a simple construction which works under a certain condition.

\newcommand{\cO}{{\cal O}}
\newcommand{\cE}{{\cal E}}
\newcommand{\cF}{{\cal F}}
\newcommand{\DDbar}{{D\bar D}}

\subsection{Low Energy Boundary Conditions}
\label{subsec:LEBC}

We first study the low energy behaviour of D-branes of the linear sigma model,
assuming that we are deep in some phase in the K\"ahler moduli
space where the gauge group is broken either completely or
to its discrete subgroup.
At energies well below the gauge coupling $e$,
the theory reduces to the bottom of the scalar potential,
the vacuum manifold $X_r= (\C^N-\Delta_r)/T_\C$, possibly with orbifold
points.
Thus, we are interested in how the linear sigma model branes
descend to D-branes in the non-linear sigma model on $X_r$.

The low energy limit is achieved by the limit $e\to \infty$, in which 
the vector multiplet loses the kinetic term in the bulk.
The auxiliary field $D$ becomes a Lagrange multiplier field that
imposes the D-term equation strictly.
The components $v_{\mu}$ and $\sigma$
become auxiliary fields and the effect of elimination is
to set them equal to their classical values,
see (\ref{gaugeconn})-(\ref{sigmaconn}).
We claim that, under appropriate boundary condition on the
vector multiplet, this picture remains the same also 
in the presence of a boundary with the boundary interaction
(\ref{Lct}) or (\ref{gaugeboundary})
that includes $D$, $v_{\mu}$ and $\sigma$.
For concreteness, we consider a $U(1)$ gauge theory
formulated on the left-half plane,
$\surface=\{(t,s)|s\leq 0\}$, and put the Wilson line brane
$\wilson(q)$ at the boundary.

We first note that $D$ and $v_1$ appear only in the boundary counter term
$S_{\rm g}^{\rm c.t.}$
for the gauge kinetic terms, see (\ref{Sgggg}) and (\ref{gaugeboundary}).
As we have discussed in Section~\ref{subsec:ALG},
the boundary condition for the gauge multiplet fields are chosen
so that this counter term vanishes. Thus, $D$ and $v_1$ do not appear
in the boundary interaction.
In particular, the effect of integrating them out
remains the same as in the bulk: impose the D-term equation
as a constraint and set $v_1$ equal to the classical value 
(\ref{gaugeconn1}).
The part of the action that depends on $v_0$ is
\beqa
\lefteqn{{1\over 2\pi}\int_{\surface}\left(v_0^2|\phi|^2
-2iv_0\bpsi\lrd_{\!\!\! t}\phi-v_0(\bpsi_+\psi_++\bpsi_-\psi_-)\right)
\dd t\dd\s
-{1\over 2\pi}\int_{\partial\surface}(\theta+2\pi q)v_0\dd t}
\nn\\
&&={1\over 2\pi}\int_{\surface}\left(v_0^2|\phi|^2
-2iv_0\bpsi\lrd_{\!\!\! t}\phi-v_0(\bpsi_+\psi_++\bpsi_-\psi_-)
-(\theta+2\pi q)v_0\delta(\s)\right)
\dd t\dd\s.
\nn
\eeqa
For simplicity of notation, we consider only a single charge $1$ matter field.
Completing the square and integrating out $v_0$, we are left with
\beqa
\lefteqn{-{1\over 2\pi}\int_{\surface}{1\over |\phi|^2}\left(i\bphi
\lrd_{\!\!\! t}\phi+{1\over 2}(\bpsi_+\psi_++\bpsi_-\psi_-)+
{1\over 2}(\theta+2\pi q)\delta(\s)\right)^2\dd t\dd\s}
\nn\\
&&=-{1\over 2\pi}\int_{\surface}
{1\over |\phi|^2}\left(i\bphi
\lrd_{\!\!\! t}\phi+{1\over 2}(\bpsi_+\psi_++\bpsi_-\psi_-)\right)^2
\dd t\dd\s
\nn\\
&&\qquad-{1\over 2\pi}\int_{\partial\surface}\left[
(\theta+2\pi q)\left(i\bphi
\lrd_{\!\!\! t}\phi+{1\over 2}(\bpsi_+\psi_++\bpsi_-\psi_-)\right)
+{1\over 4}(\theta+2\pi q)^2\delta(0)\right]\dd t.
\nn
\eeqa
The effect is simply to set $v_0$ equal to the classical
value of the bulk theory (\ref{gaugeconn}),
except that we acquire the boundary term that involves $\delta(0)$.
This looks problematic.
For $\sigma$ integration,
we find
\beqa
\lefteqn{
{1\over 2\pi}\int_{\surface}\Bigl(-|\sigma|^2|\phi|^2
-\bpsi_-\sigma\psi_+-\bpsi_+\bsigma\psi_-+(\theta+2\pi q)\Rp(\sigma)\delta(\s)
\Bigr)\dd t\dd\s}
\nn\\
\longrightarrow&&
{1\over 2\pi}\int_{\surface}{1\over |\phi|^2}|\bpsi_+\psi_-|^2\dd t\dd\s
\nn\\&&
-{1\over 2\pi}\int_{\partial\surface}
\left[(\theta+2\pi q){1\over 2}(\bpsi_-\psi_++\bpsi_-\psi_+)
-{1\over 4}(\theta+2\pi q)^2\delta(0)
\right]\dd t
\nn
\eeqa
Again, we find the unwanted $\delta(0)$ at the boundary. However, we notice
that it has the opposite sign and exactly the same absolute value
compared to the $\delta(0)$ term from the $v_0$-integration. Thus, they
completely cancel out.  
This cancellation occurs precisely when both
$v_0$ and $\Rp(\sigma)$ are unconstrained at the boundary.
And this is indeed the type of boundary conditions we impose,
as we have discussed in Sections~\ref{subsec:AsympC} and \ref{subsec:fullBC}.
Otherwise, we may need to keep an uncancelled $\delta(0)$
in the low energy theory. 
Our choice of boundary conditions saves us from such
a complication.
The appearance of $\delta(0)$ at the boundary from elimination of
 auxiliary fields was found earlier in \cite{Peskin}
where the significance of such a term to supersymmetry is discussed.

To summarize, the effect of intergrating out $v_0$ and $\sigma$
also remains the same as in the bulk:
set them equal to their classical values (\ref{gaugeconn}), (\ref{sigmaconn}).
Accordingly, the boundary interaction for the
Wilson line brane $\brane(q)=\brane(q^1\ldots q^k)$ 
is expressed as
\[
  \cA_t = \sum_{a=1}^k  q^a \bigl[(v_a)_0 -
  \textrm{Re}(\sigma_a)\bigr] =
  \dot{x}^IA^{(q)}_I(x)-{i\over 4}F^{(q)}_{IJ}(x)\psi^I\psi^J
   ,
\]
where the term $\dot x^I A^{(q)}_I(x)$
is the pull-back of the connection on the
holomorphic line bundle $\cO(q)=\cO(q^1,\ldots, q^k)$ over $X_r$.
This is nothing but the boundary interaction for the D-brane
in the non-linear sigma model supporting the line bundle ${\mathcal O}(q)$,
see (\ref{gauge}). Thus, we find that the Wilson line brane descends to the
brane supporting the line bundle
$$
\wilson(q_1,\ldots, q_k)\longrightarrow {\mathcal O}(q_1,\ldots, q_k).
$$ 
Things are as simple as this also for more general branes,
direct sums of
Wilson line branes $\brane = \oplus_i \brane(q_i)$
with interaction $Q$.
The corresponding boundary interaction (\ref{tachconn}) directly descends to
the boundary interaction (\ref{newA}) of a D-brane
in the non-linear sigma model, associated with the vector bundle
$E=\oplus_i \cO(q_i)$ with connection $A={\rm diag}(A^{(q_i)})$
and tachyon ${\bf T}=iQ-iQ^{\dag}$.
The R-symmetry also descends straightforwardly.
Thus, the graded D-brane 
$\lsmB=({\mathcal V},Q,\rho,R)$ directly descends to the graded D-brane
$\nlsmB=(E,A,Q,R)$. In other words,
a complex of Wilson line branes, 
\[
  {\mathcal C}(\lsmB) :~~
  \cdots \stackrel{d^{j-1}}{\longrightarrow}
  \brane^{j-1} \stackrel{d^{j}}{\longrightarrow}
  \brane^{j} \stackrel{d^{j+1}}{\longrightarrow}
  \brane^{j+1} \stackrel{d^{j+2}}{\longrightarrow}
  \cdots
\]
with $\brane^j = \oplus_{i=1}^{n_j} \brane(q_{ij})$, becomes
a complex of holomorphic vector
bundles over the target space $X_r$, 
\beq
  \label{lowenergy}
{\mathcal C}(\nlsmB) :~~ \cdots\stackrel{d^{j-1}}{\longrightarrow}
{\mathcal E}^{j-1}\stackrel{d^{j}}{\longrightarrow}
{\mathcal E}^{j}\stackrel{d^{j+1}}{\longrightarrow}
{\mathcal E}^{j+1}\stackrel{d^{j+2}}{\longrightarrow}\cdots  ,
\eeq
where the component of R-degree $j$ is a direct sum of line bundles,
$\cE^j = \oplus_{i=1}^{n_j} \cO(q_{ij})$.

In the above discussion, we have assumed that $X_r$ is smooth, but
everything goes through in the presence of orbifold loci as well, provided
we consider a part of the gauge charges $(q^1,...,q^k)$
as the data of the orbifold group action
on the Chan-Paton factors rather than the data determining the vector bundle
with connection. 
When $X_r$ can be realized as a global orbifold of a space $X'_r$,
we may regard ${\mathcal O}(q^1,...,q^k)$ as 
an equivariant line bundle over $X'_r$. 
In general, the correct mathematical name for it
is a line bundle over the quotient stack $[(\C^N-\Delta_r)/T_{\C}]$,
but not a sheaf over the algebraic varity $X_r$
(see \cite{AKO} and Section~\ref{sec:math}).
In what follows, somewhat loosely we call such objects
simply ``line bundles or vector bundles over the 
toric variety $X_r$''.

Let us introduce some notation.
We denote by $\mathfrak{D}(\C^N,T)$ the set of
graded D-branes in the linear sigma model. We denote by $D(X_r)$ 
the set of
graded D-branes in the low energy theory with target space $X_r$.
The map of D-branes obtained above is denoted by
\beq
  \pi_r : \mathfrak{D}(\C^N,T) \longrightarrow D(X_r) .
\label{LSMtoLE}
\eeq
In the following,
we shall indicate the
degree $0$ component of a complex
by underlining it, $\cE^0 = \underline{\cE}$, 
when there is a room of confusion. Also, a complex consisting of 
a single vector bundle ${\mathcal E}$ at R-degree $j$ will be denoted by
${\mathcal E}[-j]$. For the one at degree 0, we may simply denote it by
${\mathcal E}$ when there is no room of confusion.

\subsection*{\it D-isomorphisms}

Let us now apply the concepts that are introduced in 
Section~\ref{sec:D-isom} to D-branes in the low energy theory.
Namely, we regard D-branes to be isomorphic in $D(X_r)$
(or simply {\it D-isomorphic})
if they are related by a combination of
D-term deformations and brane-antibrane annihilation.
Isomorphic D-branes in $D(X_r)$ flow to
the same infra-red fixed point, 
although the converse is not true in general.
The map (\ref{LSMtoLE}) is therefore the operation of 
modding out D-branes of the linear sigma model by D-isomorphism relations.
In what follows, we study such isomorphism
relations and see how they depend on the phases
of the K\"ahler moduli space.

Recall from Section~\ref{sec:D-isom} that
D-isomorphisms can be characterized as follows.
For two D-branes in the linear sigma model, $\lsmB_1$ and $\lsmB_2$,
a D-isomorphism of the image D-brane in
$D(X_r)$ may be represented by a degree $0$ map
$\varphi$ in the linear sigma model
whose cone 
$$
  Q_{C(\varphi)} = \left(\begin{array}{cc}
      -Q_1    &  0 \\
      \varphi & Q_2
    \end{array} \right)
$$
has positive definite boundary potential,
$\{Q_{C(\varphi)},Q_{C(\varphi)}^{\dag}\}>0$,
everywhere on $X_r$.
Notice that everywhere-positivity of the boundary potential
depends heavily on the deleted set $\Delta_r$, which
is determined by the bulk D-term equation.
This is how the D-isomorphism relations of linear sigma model branes
depend on the phase that we are in.

Let us illustrate the dependence in Example 
(A) with $N=3$, namely, $U(1)$ gauge theory with fields $P,X_1,X_2,X_3$
of charge $-3,1,1,1$ respectively.
The model has two phases,
$r\gg 0$ and $r\ll 0$.
The deleted sets are
$$
\begin{array}{ll}
\Delta_+=\{x_1=x_2=x_3=0\},\,\,& \mbox{in the ~$r\gg 0$ phase},\\
\Delta_-=\{p=0\},&\mbox{in the ~$r\ll 0$ phase}.
\end{array}
$$
Let us consider the
D-brane, $\lsmB_+$, given by the complex%
\footnote{In the following we use the short-hand notation 
$\brane(-1)\stackrel{X}{\longrightarrow}
 \brane(0)^{\oplus 3}\stackrel{X}{\longrightarrow}
 \brane(1)^{\oplus 3}\stackrel{X}{\longrightarrow}
 \brane(2)
$
for Koszul-like complexes (\ref{emptybrane}).}
\beq
  \label{emptybrane}
  {\mathcal C}(\lsmB_+) :~~ 
  \brane(-1) \stackrel{\left(\mbox{\tiny $\!\!\!
                \begin{array}{c}
		  x_1\\ x_2\\ x_3
		\end{array}$}\!\!\!\right)}
	{-\!\!\!-\!\!\!-\!\!\!-\!\!\!\longrightarrow}
  \brane(0)^{\oplus 3}  \stackrel{\left(\mbox{\tiny $\!\!\!
                \begin{array}{ccc}
		 \!   0  \!\!\!&\!\!\!\!\!\!\!  -x_3 \!\!&\!\! x_2 \\ 
          \! x_3 \!\!\!\!&\!\!\!\!\!\!   0  \!\!\!\!&\!\!\!\!\!  -x_1 \\ 
		 \!  -x_2 \!\!&\!\!\!\! x_1 \!\!\!\!&\!\!\!   0 \! 
		\end{array}$}\!\!\!\right)}
	{-\!\!\!-\!\!\!-\!\!\!-\!\!\!\longrightarrow}
    \brane(1)^{\oplus 3}   \stackrel{\left(x_1, x_2, x_3 \right)}
	{-\!\!\!-\!\!\!-\!\!\!-\!\!\!\longrightarrow}
  \underline{\brane(2)} .
\eeq
The boundary potential is
\beq
 \label{invtach}
 \{ Q_+, Q_+^\dagger \} ~=~ \left(|x_1|^2 +|x_2|^2 +|x_3|^2\right)
 \cdot \id_{{\mathcal V}_+}.
\eeq
In the low energy theory of the positive volume phase $r\gg 0$,
the potential is strictly positive since the point
$x_1=x_2=x_3=0$ is deleted. As a
consequence complete brane-antibrane annihilation takes place:
$\lsmB_+$ is D-isomorphic to the empty brane,
$$
\pi_+(\lsmB_+)\cong 0.
$$
On the other hand, we may view $\lsmB_+$ as a result of binding two D-branes,
$$
  {\mathcal C}(\lsmB_1) :~~
  \brane(-1) \stackrel{X}
	{\longrightarrow}
  \brane(0)^{\oplus 3}  \stackrel{X}
	{\longrightarrow}
  \underline{\brane(1)}^{\oplus 3}\quad\mbox{and}\quad
  {\mathcal C}(\lsmB_2)=\brane(2) , 
$$
by the right-most map $\varphi=(x_1,x_2,x_3)$ in (\ref{emptybrane}), via 
the cone construction.
Then,
the positivity of $\{Q_+,Q_+^{\dag}\}$ tells us that $\lsmB_1$ and 
$\lsmB_2$ determine isomorphic low energy D-branes in $D(X_r)$:
$$
  \pi_+(\lsmB_1) ~~\cong~~ \pi_+(\lsmB_2) .
$$
Let us next study these branes in the negative volume phase
$r\ll 0$, where the deleted set is $\Delta_-=\{p=0\}$
and the low energy theory is the free orbifold $X_-=\C^3/\Z_3$.
Now the $x_i$'s are allowed to vanish at the same time,
and hence the boundary potential for $\lsmB_+$ 
is no longer positive everywhere. 
In fact, the D-brane $\lsmB_+$ descends
directly to the fractional D-brane
(see Section~\ref{subsec:Koszul}), 
$\cO_{\mathfrak{p}}(\overline{2})$,
which is localized at the orbifold point 
$\mathfrak{p}=\{x_1=x_2=x_3=0\}$,
$$
  \pi_-(\lsmB_+) ~~\cong~~ \cO_{\mathfrak{p}}(\overline{2}) .
$$
Moreover, we find that
$\pi_-(\lsmB_1)$ and $\pi_-(\lsmB_2)$ are not isomorphic in $D(X_-)$.

A complementary example is provided by the complex
$$
  {\mathcal C}(\lsmB_-) :~~
  \brane(q\!+\!3) \stackrel{p}{\longrightarrow} \underline{\brane(q)} ,
$$
for some $q\in\ZZ$. The associated boundary potential is
$\{Q_-,Q_-^{\dag}\} = |p|^2 \cdot \id_{{\mathcal V}_-}$. 
In the orbifold phase, $r\ll 0$, it is positive everywhere,
and hence the image $\pi_-(\lsmB_-)$ is empty in the infra-red limit.
This can also be interpreted as the D-isomorphism
\beq
  \label{z3qism}
  \pi_-(\brane(q\!+\!3))
  ~~\cong~~ 
  \pi_-(\brane(q))  ,
\eeq
which reflects the breaking of the
gauge group $U(1)$ to the discrete subgroup $\ZZ_3$.
On the other hand, at large volume the boundary potential
$\{Q_-,Q_-^{\dag}\}$ vanishes at the $p=0$ locus --- 
the exceptional divisor $E\cong \CP^2$,
so that we find
$$
\pi_+(\lsmB_-)\cong \cO_E(q),
$$
where
$\cO_E(q)$ is the line bundle supported on $E$
equipped with the restriction of the gauge connection of the
line bundle $\cO(q)$ over $X_+$.
Of course, the D-branes $\pi_+(\wilson(q\!+\! 3))={\mathcal O}(q\!+\! 3)$ and
$\pi_+(\wilson(q))={\mathcal O}(q)$ are not isomorphic in $D(X_+)$.
The ``difference'' is represented by
$\cO_E(q)$.

Let us summarize our findings on the low energy behaviour of
D-branes deep inside the phases of the linear sigma model.
As we have seen explicitly in our examples,
there are {\it phase dependent} low energy relations among D-branes.
In a general model with gauge group $T=U(1)^k$, 
there are a multitude of phases, and we have a pyramid 
as depicted below.
The maps $\pi_{\rm I},\pi_{\rm II},\ldots$ 
are projections that mod out the linear sigma model branes
by the low energy D-isomorphism relations.
And the D-isomorphism relations are governed by
the deleted sets $\Delta_{\rm I},\Delta_{\rm II},\ldots\,$,
which determine the vacuum manifolds
$X_{\rm I}, X_{\rm II},\ldots$ and hence
the tachyon condensation patterns in the respective phases. 
\begin{center}
\begin{picture}(150,100) \thicklines
  \put(40,89){$\mathfrak{D}(\C^N,T)$}
  \put(60,85){\vector(-2,-3){35}}
  \put(67,82){\vector(-1,-3){22}}
  \put(73,82){\vector(1,-3){22}}
  \put(80,85){\vector(2,-3){35}}
  \put(24,53){$\pi_{\rm I}$}
  \put(38,42){$\pi_{\rm II}$}
  \put(67,40){$\pi_{\rm III}$}
  \put(104,53){$\pi_{\rm IV}$}
  \put(3,21){$D(X_{\rm I})$}
  \put(25,5){$D(X_{\rm II})$}
  \put(80,5){$D(X_{\rm III})$}
  \put(100,23){$D(X_{\rm IV})$}
  \textcolor{blue}{
    \put(65,25){\line(-3,1){55}}
    \put(65,25){\line(-4,-1){60}}
    \put(65,25){\line(1,-5){5}}
    \put(65,25){\line(6,-1){65}}
    \put(65,25){\line(4,1){60}}
    \put(60,50){\circle*{0.5}}
    \put(65,51){\circle*{0.5}}
    \put(70,51){\circle*{0.5}}
    \put(75,50){\circle*{0.5}}
    \put(80,49){\circle*{0.5}}
  }
\end{picture}
\end{center}

\subsection{Crossing Phase Boundaries By Grade (Band) Restriction Rule}
\label{subsec:grrrevisited}

Let us now consider the problem of transporting D-branes
back and forth between different phases,
along paths that cross the phase boundaries.
It would be a hard task if we
tried to do this directly within the infra-red conformal field theory.
Going away from the large volume limits
there are huge perturbative as well as non-perturbative corrections
that will completely blur the geometric picture of D-branes, and
we do not really have a good or convenient description of
the low energy theories near the phase boundaries.
Linear sigma models allow us to circumvent this problem:
they provide us with a simple and explicit 
UV-description of the bulk theory over regions of the moduli space
including the phase boundaries.
The grade restriction rule allows us to extend this advantage
to description of D-branes.
Together with the tachyon condensation pattern
that we just found, it will provide a beautiful solution 
to the problem of D-brane
transport across phase boundaries.

The first step is to lift a given D-brane in the low energy theory to
a D-brane in the linear sigma model. Let us denote such a lift of the
boundary interactions by $\omega_r: D(X_r) \to \mathfrak{D}(\C^N,T)$. By
definition the composition $\pi_r \circ \omega_r$ has to map a brane
to an isomorphic brane,
symbolically,  $\pi_r \circ \omega_r \cong\id_{D(X_r)}$.
For an arbitrary D-brane in $D(X_r)$,
the existence of a lift to $\mathfrak{D}(\C^N,T)$
is guaranteed, because any complex of vector bundles is
  D-isomorphic to a complex of the form (\ref{lowenergy}).
(See Section \ref{sec:math} for this point and extension to complexes of
coherent sheaves.) 
In view of the isomorphism relation of D-branes in $D(X_r)$ through D-term
deformations and brane-antibrane annihilation, we notice that
$\omega_r$ is highly ambiguous, i.e., for any $\nlsmB \in D(X_r)$
there are infinitely many D-branes $\lsmB \in \mathfrak{D}(\C^N,T)$ 
with the property that
$\pi_r(\lsmB)\cong \nlsmB$. However, as long as we transport
the D-brane along a path in K\"ahler moduli space that stays within a
given phase, the tachyon condensation pattern 
does not change, and any D-brane $\lsmB$ such that $\pi_r(\lsmB)\cong\nlsmB$
will do as a lift.

However, when we move to another phase, the tachyon condensation pattern
does change. Then, a different lift could result in a different transport.
And that is indeed the case!
In fact, we already know plenty of examples 
that clearly exhibits the dependence on the choice of lift.
Let us consider Example (A) with $N=3$.
In the positive volume phase $r\gg 0$, we may lift an empty brane 
in $D(X_+)$ to an empty brane in the linear sigma model
or to the brane 
$\lsmB_+$ that was defined in (\ref{emptybrane}).
In the negative volume phase $r\ll 0$, the empty brane of course descends to
an empty brane in $D(X_-)$
but the brane $\lsmB_+$ descends to the fractional brane
${\mathcal O}_{\mathfrak{p}}(\overline{2})$ which is
not empty in the infra-red limit.
As another example, the brane ${\mathcal O}(\overline{q})\in D(X_-)$
may be lifted to $\wilson(q)$ or $\wilson(q+3)$ but their
images in the positive volume phase are completely different,
${\mathcal O}(q)\not\cong {\mathcal O}(q+3)\in D(X_+)$.

We now encounter the problem discussed in
Section~\ref{subsec:whatwedo}: the parallel transport
of D-branes does not seem to preserve the D-isomorphism classes.
What does this mean? Do the D-isomorphism relations break down
somewhere along the way?
This is the point where
the grade (band) restriction rule comes to the rescue: 
It is simply that some of the linear sigma model branes cannot be transported
to the other phase,
in the sense described in Section~\ref{subsec:rule}.
It is {\it not} the D-isomorphism relation but the transport itself
that breaks down.

Suppose that Phase I and Phase II share a phase boundary, and
let us fix a window $w$ in the space of FI-theta parameters
for paths that connect the two phases.
The window $w$ defines the subset
${\cal T}^w_{\rm I,II} \subset \mathfrak{D}(\C^N,T)$
consisting of grade (or band) restricted D-branes,
and in fact only the D-branes therein can smoothly get 
through.
Hence, in order to be able to transport a low energy
D-brane $\nlsmB$ along a path
that passes through the window $w$, 
we have to make sure that it
is lifted to a grade (or band) restricted D-brane
 $\lsmB$,
i.e., we need a lift 
$\omega^w_{\rm I,II}:D(X_{\rm I})\rightarrow {\cal T}^w_{\rm I,II}$
such that
$\pi_{\rm I} \circ \omega^w_{\rm I,II} \cong \id_{D(X_{\rm I})}$.
For the transport in the opposite direction,
we need the corresponding lift with 
I and II exchanged.
Diagramatically, we can associate the following hat diagram
to the phase boundary with window $w$:
\beq
\begin{picture}(150,99) \thicklines
  \put(43,85){$\mathfrak{D}(\C^N,T)$}
  \put(66,70){$\cup$}
  \put(60,55){${\cal T}^w_{\rm I,II}$}
  \put(60,80){\vector(-2,-3){45}}
  \put(80,80){\vector(2,-3){45}}
  \put(18,45){$\pi_{\rm I}$}
  \put(108,45){$\pi_{\rm II}$}
  \put(32,13){\vector(2,3){25}}
  \put(108,13){\vector(-2,3){25}}
  \put(42,20){$\omega^w_{\rm I,II}$}
  \put(78,20){$\omega^w_{\rm II,I}$}
  \put(0,-1){$D(X_{\rm I})$}
  \put(105,-1){$D(X_{\rm II})$}
\end{picture}
\label{hatdiagram1}
\eeq
It is not \emph{a priori} clear whether the map
$\omega^w_{\rm I,II}$ to the subset ${\cal T}^w_{\rm I,II}$
exists for every D-brane in $D(X_{\rm I})$. And even if it exist,
it may not be unique.
We will mathematically prove that it
indeed exists in the next section, but for now we
illustrate the main point by examples. 
As we will see, for the case $T=U(1)$,
where the grade restriction rule applies,
the lift is also unique up to decoupled
additon of trivial brane-antibrane 
pairs with complete tachyon condensation.
For the higher rank case, $T=U(1)^k$ with $k>1$,
where the band restriction applies, 
the uniqueness of the lift is lost. 
However, this is not harmful in that
the ambiguity does not matter once we compose the lift with
the projections $\pi_{\rm II}$ 
or $\pi_{\rm I}$.
Namely, the composite maps
\beqa
&F^w_{\rm I,II}:D(X_{\rm I})\stackrel{\omega^w_{\rm I,II}}{\longrightarrow}
{\mathcal T}^w_{\rm I,II}\stackrel{\pi_{\rm II}}{\longrightarrow} 
D(X_{\rm II}),
\nn\\
&F^w_{\rm II,I}:D(X_{\rm II})\stackrel{\omega^w_{\rm II,I}}{\longrightarrow}
{\mathcal T}^w_{\rm I,II}\stackrel{\pi_{\rm I}}{\longrightarrow} 
D(X_{\rm I}),
\nn
\eeqa
induce maps of D-isomorphism classes of branes that do not depend on the
choice of lifts.
In particular, they are inverses of each other,
$$
F^w_{\rm II,I}\circ F^w_{\rm I,II}\cong {\rm id}_{D(X_{\rm I})},
\qquad
F^w_{\rm I,II}\circ F^w_{\rm II,I}\cong {\rm id}_{D(X_{\rm II})}.
$$

Let us illustrate the main points using
Example (A) with $N=3$. The singular points on the FI-theta parameter space 
are $(r,\theta)=(3\log 3, \pi+2\pi n)$ ($n\in \Z$). 
Let us choose the window $w=\{-\pi<\theta<\pi\}$ at the phase boundary
$r=3\log 3$.
The corresponding grade restriction rule on the Picard lattice $\Z$
is $\mathfrak{C}^w=\{-1,0,1\}$ 
and hence the subset ${\mathcal T}^w$ is generated by
$$
  \wilson(-1),~\wilson(0),~\wilson(1).
$$
We start at positive volume, $r \gg 0$, with the
holomorphic line bundle $\cO(2)$ over $X_+$. The most
na\"\i ve lift to the linear sigma model is the D-brane
${\mathcal C}(\lsmB_2)=\brane(2)$; but, as we have
seen previously, the D-brane 
$$
  {\mathcal C}(\lsmB_1) :~~ \brane(-1) \stackrel{X}
	{\longrightarrow}
  \brane(0)^{\oplus 3}  \stackrel{X}
	{\longrightarrow}
  \underline{\brane(1)}^{\oplus 3} 
$$ 
also satisfies $\pi_+(\lsmB_1) \cong\cO(2)$. 
In fact, there are infinitely many D-branes $\lsmB$ with
$\pi_+(\lsmB) \cong \cO(2)$. However, among
those the D-brane $\lsmB_1$ is special in that it is an object in
the grade restricted subset ${\cal T}^w$, whereas, for
instance, $\lsmB_2$ is not. We conclude that
 $\lsmB_1$ is the right representative
to cross the phase boundary through the window $w$. After arriving at
the orbifold phase, $r \ll 0$, we apply the projection
$\pi_-$ to $\lsmB_1$ and obtain the low energy D-brane in $D(X_-)$:
$$
  \cO(\overline{-1}) \stackrel{X}{\longrightarrow}
  \cO(\overline{0})^{\oplus 3}  \stackrel{X}{\longrightarrow}
  \underline{\cO(\overline{1})}^{\oplus 3}  .
$$
This is the result of transporting
the D-brane ${\mathcal O}(2)\in D(X_+)$ through the window $w$.

We next start at $r\ll 0$, with the equivariant line bundle
${\mathcal O}(\overline{2})$ over $\C^3$. 
Again this can be lifted to infinitely many branes
$\wilson(2+3n)$, $n\in \Z$, but only one of them, $\wilson(-1)$,
is in the grade restriction range. Thus, the transport of
${\mathcal O}(\overline{2})\in D(X_-)$ through
the window $w$ results in ${\mathcal O}(-1)\in D(X_+)$.

As another example, let us consider the fractional brane
${\mathcal O}_{\mathfrak{p}}(\overline{2})\in D(X_-)$. 
We have found in (\ref{liftfrac}) that
its na\"\i ve lift is
the brane $\lsmB_+$ given in (\ref{emptybrane}).
However, the rightmost entry, $\wilson(2)$, is not
in the grade restriction range.
But we can replace $\lsmB_+$ by
$$
\wilson(-1)\stackrel{X}{\longrightarrow}
\wilson(0)^{\oplus 3}\stackrel{X}{\longrightarrow}
\wilson(1)^{\oplus 3}\stackrel{pX}{\longrightarrow}
\underline{\wilson(-1)}
$$
using the D-isomorphism relation 
$p:\wilson(2)\stackrel{\cong}{\longrightarrow}\wilson(-1)$
in the $r\ll 0$ phase.
This new D-brane can be transported safely through the window
$w$, and we obtain the D-brane
$
{\mathcal O}(-1)\stackrel{X}{\longrightarrow}
{\mathcal O}(0)^{\oplus 3}\stackrel{X}{\longrightarrow}
{\mathcal O}(1)^{\oplus 3}\stackrel{pX}{\longrightarrow}
\underline{{\mathcal O}(-1)}
$
as the large volume image of ${\mathcal O}_{\mathfrak{p}}(\overline{2})$.

The key step
is to find a lift of a low energy D-brane
to the grade restricted subset
in the phase of the starting point.
How can we find such a lift in general? 
It is always possible to find {\it some} lift to a complex of Wilson line 
branes in $\mathfrak{D}(\C^N,T)$, but that may not be grade restricted.
The point is that this complex can always be changed into a grade restricted 
one by binding infra-red empty D-branes to it, so that the D-isomorphism 
class is preserved. In the $r\gg 0$ phase, one can do so using the D-branes
\beq
\label{EXempty}
\wilson(n)\stackrel{X}{\longrightarrow}
\wilson(n+1)^{\oplus 3}\stackrel{X}{\longrightarrow}
\wilson(n+2)^{\oplus 3}\stackrel{X}{\longrightarrow}
\wilson(n+3)\quad \textrm{for~all} ~  n\in\Z.
\eeq
By binding these D-branes to the original complex,
we can eliminate the Wilson line branes $\wilson(q)$ 
whose charges are too large or too small, 
and we can repeat this procedure until
the resulting complex fits into the grade restriction range.
The complex (\ref{EXempty}) has the right length so 
that one can make sure that the process of decreasing 
or increasing the charges does not overshoot.
In the $r\ll 0$ phase, the same r\^ole is played by
$$
\wilson(n)\stackrel{p}{\longrightarrow}\wilson(n-3).
$$

Let us describe the corresponding empty branes 
in the general one-parameter model
with the fields $X_1,...,X_l$, carrying positive charges 
$Q_1,...,Q_l$, and the fields $Y_1,...,Y_{l'}$, carrying 
negative charges $-Q'_1,...,-Q'_{l'}$. In view of the 
Calabi-Yau condition (\ref{conformal}) we have
$\sum_{i=1}^kQ_i={\mathscr S}=\sum_{j=1}^{l'}Q_j'$. 
In the $r\gg 0$ phase, 
any D-brane $\lsmB$ can be brought into the grade restriction range
by using the Koszul complex
\beq
{\mathscr K}_+\,:~~\wilson(0)\stackrel{X}{\longrightarrow}
\wilson_+\stackrel{X}{\longrightarrow}
\wedge^2\wilson_+\stackrel{X}{\to}\cdots
\stackrel{X}{\to}
\wedge^{l-1}\wilson_+\stackrel{X}{\longrightarrow}
\wedge^{l}\wilson_+=\wilson({\mathscr S})
\label{X-Koszul}
\eeq
and its shifts ${\mathscr K}_+(n)$, where 
$\wilson_+:=\oplus_{i=1}^l\wilson(Q_i)$.
In the $r\ll 0$ phase, this can be done using
the Koszul complex
\beq
{\mathscr K}_-\,:~~\wilson(0)\stackrel{Y}{\longrightarrow}
\wilson_-\stackrel{Y}{\longrightarrow}
\wedge^2\wilson_-\stackrel{Y}{\to}\cdots
\stackrel{Y}{\to}
\wedge^{l'-1}\wilson_-\stackrel{Y}{\longrightarrow}
\wedge^{l'}\wilson_-=\wilson(-{\mathscr S})
\label{Y-Koszul}
\eeq
and its shifts ${\mathscr K}_-(n)$, where 
$\wilson_-:=\oplus_{j=1}^{l'}\wilson(-Q'_j)$.

In models with higher rank gauge groups,
we mentioned that the band restricted lift is not unique but that
the non-uniqueness does not matter in the end.
Let us illustrate this subtle point
using the two-parameter model (C).
There are four phases as depicted in Fig.~\ref{twoparaA}.
Let us focus on Phases III and IV. We recall the deleted sets there:
\beqa
&&\Delta_{\rm III}=\{x_6=0\}\cup\{p=0\},\nn\\
&&\Delta_{\rm IV}=\{x_1=x_2=0\}\cup\{p=0\}.\nn
\eeqa
The unbroken gauge group at the III-IV phase boundary is the subgroup 
$U(1)_2$ which has width ${\mathscr S}=2$.
The asymptotic singular locus in this 
direction is at $\theta^2 \in 2\pi\Z$. Let us choose the 
window $-2\pi <\theta^2 <0$ for which the band restriction rule
is
$$
q^2=0,1.
$$
Every brane in Phase IV can be lifted to a complex of Wilson line branes
obeying this band restriction rule. This can be done by reducing or
increasing the charge $q^2$ using the complex
$$
{\mathscr K}_+^{{\rm III},{\rm IV}}\,:~~
\wilson(0,0)\stackrel{x_1\choose x_2}{-\!\!\!-\!\!\!-\!\!\!\longrightarrow}
\!\!\!
\begin{array}{c}
\wilson(0,1)\\[-0.15cm]
\oplus\\[-0.15cm]
\wilson(0,1)
\end{array}\!\!\!
\stackrel{(-x_2,x_1)}{-\!\!\!-\!\!\!-\!\!\!\longrightarrow}
\wilson(0,2)
$$
or its shifts ${\mathscr K}_+^{{\rm III},{\rm IV}}(n,m)$. They are 
D-isomorphic to the empty brane in the low energy theory, since
$\{x_1=x_2=0\}$ is a part of the deleted set $\Delta_{\rm IV}$.
Similarly, every brane in Phase III
can be lifted to a complex of band restricted Wilson line branes
using the complex
$$
{\mathscr K}_-^{{\rm III},{\rm IV}}\,:~~
\wilson(0,0)\stackrel{X_6}{\longrightarrow}\wilson(1,-2),
$$
or its shifts ${\mathscr K}_-^{{\rm III},{\rm IV}}(n,m)$,
which are empty in the low energy theory since
$\{x_6=0\}$ is a part of the deleted set $\Delta_{\rm III}$.
However, in both phases the lift is not unique. The reason
is that there are additional branes that are empty in the low energy
theory, i.e.,
$$
{\mathscr K}\,:~~
\wilson(0,0)\stackrel{p}{\longrightarrow}\wilson(-4,0)
$$
and its shifts ${\mathscr K}(n,m)$. 
One can modify the lift using the latter branes without
changing the charge $q^2$. 
From the structure of the deleted sets,
this is obviously the only non-trivial ambiguity of the lifts
in both phases.
Now, the point is that this ambiguity is common to 
the two phases.
In particular, it
does not matter when the D-brane is projected down to
the low energy theory even after coming
to the other side of the phase boundary.
(Of course, when reduced to the low energy theory
in a different phase, say Phase I, in which $\{p=0\}$
is not a part of the deleted set, modification by ${\mathscr K}$
results in a totally different brane.)

This is the general situation.
The key point is the relation proved in Section~\ref{subsec:LSMdel}:
\beq
\begin{array}{l}
\Delta_{\rm I}=\Delta^{{\rm I,II}}_+\cup(\Delta_{\rm I}\cap \Delta_{\rm II})
\\[0.2cm]
\Delta_{\rm II}=\Delta^{{\rm I,II}}_-\cup(\Delta_{\rm I}\cap \Delta_{\rm II})
\end{array}
\eeq
One can find a lift in the band restriction range using the Koszul
complexes associated with $\Delta^{{\rm I,II}}_+$ and
$\Delta^{{\rm I,II}}_-$ in Phases I and II respectively.
There are genuine ambiguities in the lifts but 
they are from $\Delta_{\rm I}\cap \Delta_{\rm II}$ and are common
to both phases. Thus, one can go back and forth between Phases I and II
without worrying about the ambiguity.

\subsection{Monodromies}
\label{subsec:NCmonodromy}

\newcommand{\mono}{\mathrm{M}}
\newcommand{\vanishing}{\mathfrak{V}}

Now that we learned a way to transport
D-branes across phase boundaries, we next
study transport of D-branes 
along non-trivial closed loops
in the K\"ahler moduli space $\mathfrak{M}_K$.
This yields an operation known as {\it monodromy}.

\subsection*{\it Models With Gauge Group $T=U(1)$}

The K\"ahler moduli space of the linear sigma model with
a single $U(1)$ gauge group
 is complex one-dimensional and has three special
points: the `positive volume limit'
$r \rightarrow \infty$, the `negative volume limit'  
$r \rightarrow -\infty$, and the singular point $t_{\it eff}=0$.
We describe the monodromies around each of these points.

Monodromies around the positive and negative volume
limits $r\to\pm \infty$ are rather
straightforward. They are simple shifts of the theta parameter by $\pm 2\pi$.
Since the theta parameter enters into the boundary interaction
of the Wilson line brane $\wilson(q)$
in the combination $\theta+2\pi q$, shift of $\theta$ by $\pm 2\pi$
is equivalent to the shift of $q$ by $\pm 1$ while $\theta$ is kept intact.
This shows that the monodromy is a shift in the gauge charge;
\beq
\mono_{\theta\to\theta\pm 2\pi}({\mathcal E},Q,R)~=~
({\mathcal E}(\pm 1),Q,R).
\eeq

\begin{figure}[htb]
\centerline{\includegraphics{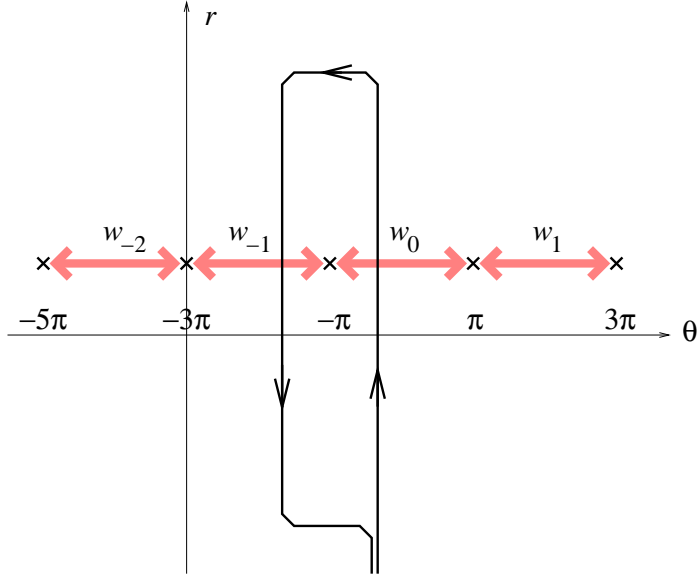}}
\caption{A loop around a singular point in $\mathfrak{M}_K$}
\label{fig:monodromy1}
\end{figure}

The monodromy around the singular point is less straightforward 
and hence is more interesting.
We recall that the simgular point is at
$\e^t=\prod_iQ_i^{Q_i}$, i.e., at
$r=\sum_iQ_i\log|Q_i|$ and $\theta\equiv \pi{\mathscr S}$ mod $2\pi\Z$,
where ${\mathscr S}=\sum_{Q_i>0}Q_i$.
Let us consider a loop that starts from a point deep inside
the negative volume phase $r\ll 0$,
goes once around the singular point {\it counter clockwise},
and comes back to the starting point.
This can be represented by a path in the $(r,\theta)$ space 
as depicted in Fig~\ref{fig:monodromy1}
(the figure
is for the case where ${\mathscr S}$ is odd so that the singularity is at
$\theta\in \pi(2\Z+1)$).
 The path starts at a point in $r\ll 0$
and goes to $r\gg 0$
 through a window $w_0$, makes a turn
and comes back to $r\ll 0$ through the next window
$w_{-1}$.
To transport a brane in $D(X_-)$ along this path,
we first lift it to a linear sigma model brane that respects the
grade restriction rule of the window $w_0$ and then move it to
$r\gg 0$ along the path that goes through $w_0$.
Once inside the positive volume phase,
using the D-isomorphism relation in $D(X_+)$, we switch the linear sigma model
brane to another one that obeys the grade restriction rule
of the window $w_{-1}$. And then we move it back to the negative volume
phase along the path through the window $w_{-1}$.
This gives the result of the monodromy along the loop around 
the singular point.

Let us illustrate this operation in Example (A) with $N=3$.
In this example, the width ${\mathscr S}$ is $3$ and 
the grade restriction rules for some
of the windows in Fig~\ref{fig:monodromy1} are
\beqa
&&{\mathcal T}^{w_{-1}}=\langle \wilson(0),\wilson(1),\wilson(2)\rangle,
\nn\\
&&{\mathcal T}^{w_0}=\langle \wilson(-1),\wilson(0),\wilson(1)\rangle,
\nn\\
&&{\mathcal T}^{w_1}=\langle \wilson(-2),\wilson(-1),\wilson(0)\rangle.
\nn
\eeqa
Let us take a D-brane at the orbifold point in the K\"ahler moduli
space, say the $\ZZ_3$-equivariant line bundle $\cO(\overline{2})$
over $\C^3$. The lift to the linear sigma 
model can be given by $\brane(q)$ for any $q \in 3\ZZ+2$. 
Since we first go to the positive volume phase through the window $w_0$,
we must choose the one in $\langle \wilson(-1),\wilson(0),\wilson(1)\rangle$, 
namely, $\wilson(-1)$.
Inside the $r\gg 0$ phase, we wish to find a different linear sigma
model brane
that is D-isomorphic to $\wilson(-1)$ at low energies
and lies in the subset $\langle \wilson(0),\wilson(1),\wilson(2)\rangle$
associated with the window $w_{-1}$.
In view of the fact that the brane $\lsmB_+$ given by
(\ref{emptybrane})
is D-isomorphic to the empty brane in this phase,
we find that the complex
\beq
\underline{\wilson(0)^{\oplus 3}}\stackrel{X}{\longrightarrow}
\wilson(1)^{\oplus 3}\stackrel{X}{\longrightarrow}
\wilson(2)
\label{temppp}
\eeq
is the one we wanted.
That is, this is the representative that can go through the window $w_{-1}$.
Once we are back in the $r\ll 0$ phase, we can project it to
the low energy theory.
In this way, we find  the monodromy image of ${\mathcal O}(\overline{2})$
along this loop (let us call it $L$)
\beq
\mono_L({\mathcal O}(\overline{2}))~=~
  \underline{\cO(\overline{0})^{\oplus 3}} \stackrel{X}{\longrightarrow}
  \cO(\overline{1})^{\oplus 3} \stackrel{X}{\longrightarrow} 
\cO(\overline{2})  .
\eeq
It is as easy as this in any example.

When presented in a slightly different way, the above example
leads to a general recipe to determine the monodromy action.
The replacement of
$\wilson(-1)$ by the complex (\ref{temppp})
in the $r\gg 0$ phase can be understood as
binding to $\wilson(-1)$ the empty brane $\lsmB_+$:
$$
\begin{array}{c}
\wilson(-1)\\[0.28cm] 
\\
\end{array}\!\!\!
\begin{array}{c}
\stackrel{X}{\longrightarrow}\\[-0.15cm]
\!\!\mbox{\footnotesize id}\!\!\searrow\,\,\\[0.4cm]
\end{array}\!\!\!
\begin{array}{c}
\wilson(0)^{\oplus 3}\\[-0.15cm]
\oplus\\[-0.15cm]
\underline{\wilson(-1)}\end{array}\!\!\!\!
\begin{array}{c}
\stackrel{X}{\longrightarrow}\\[0.25cm] 
\\\end{array}\!\!\!\!
\begin{array}{c}
\wilson(1)^{\oplus 3}\\[0.28cm] 
\\\end{array}\!\!\!\!
\begin{array}{c}
\stackrel{X}{\longrightarrow}\\[0.25cm] 
\\\end{array}\!\!\!\!
\begin{array}{c}
\wilson(2)\\[0.28cm]
\\\end{array}
$$
Once we are back in the $r\ll 0$ phase,
we may bind to this the empty brane
$\lsmB_-$, given by the complex
$\wilson(2)\stackrel{p}{\longrightarrow} \wilson(-1)$:
\vspace{-0.5cm}
$$
\begin{array}{c}
\\
\\
\wilson(-1)\\[0.28cm] 
\\
\end{array}\!\!\!
\begin{array}{c}
\\
\\
\stackrel{X}{\longrightarrow}\\[-0.15cm]
\!\!\mbox{\footnotesize id}\!\!\searrow\,\,\\[0.4cm]
\end{array}\!\!\!
\begin{array}{c}
\\
\\
\wilson(0)^{\oplus 3}\\[-0.15cm]
\oplus\\[-0.15cm]
\underline{\wilson(-1)}\end{array}\!\!\!\!
\begin{array}{c}
\\
\\
\stackrel{X}{\longrightarrow}\\[0.25cm] 
\\\end{array}\!\!\!\!
\begin{array}{c}
\wilson(2)\\[-0.15cm]
\oplus\\[-0.15cm]
\wilson(1)^{\oplus 3}\\[0.02cm] 
\\\end{array}\!\!\!\!\!\!
\begin{array}{c}
\stackrel{p}{\longrightarrow}\\[-0.15cm]
\,\,\,\,\searrow\!\mbox{\footnotesize id}\\[-0.15cm]
\stackrel{X}{\longrightarrow}\\[0.05cm] 
\\\end{array}\!\!\!\!
\begin{array}{c}
\wilson(-1)\\[-0.15cm]
\oplus\\[-0.15cm]
\wilson(2)\\[0.02cm]
\\\end{array}
$$
The net result is binding to the original brane $\wilson(-1)$
the brane $\vanishing$ given by the complex
\beq
\begin{array}{c}
\\[0.28cm]
\underline{\wilson(-1)}
\end{array}\!\!\!
\begin{array}{c}
\\[0.28cm]
\stackrel{X}{\longrightarrow}
\end{array}\!\!\!
\begin{array}{c}
\\[0.28cm]
\wilson(0)^{\oplus 3}
\end{array}\!\!\!\!
\begin{array}{c}
\\[0.28cm]
\stackrel{X}{\longrightarrow}
\end{array}\!\!\!\!
\begin{array}{c}
\wilson(2)\\[-0.15cm]
\oplus\\[-0.15cm]
\wilson(1)^{\oplus 3}
\end{array}\!\!\!\!\!\!
\begin{array}{c}
\stackrel{p}{\longrightarrow}\\[-0.15cm]
\,\,\,\,\searrow\!\mbox{\footnotesize id}\\[-0.15cm]
\stackrel{X}{\longrightarrow}
\end{array}\!\!\!\!
\begin{array}{c}
\wilson(-1)\\[-0.15cm]
\oplus\\[-0.15cm]
\wilson(2)
\end{array}
\label{BV}
\eeq
\\
by the cochain map $\vp:\vanishing\to \wilson(-1)$ that maps the left most
$\wilson(-1)$ identically to $\wilson(-1)$.
Collapsing the trivial brane-antibrane pair
$\wilson(2)\stackrel{\rm id}{\longrightarrow}\wilson(2)$,
according to the procedure from
(\ref{complexD}) to (\ref{complexD'}),
the brane $\vanishing$ could also be presented as
$$
\vanishing\cong~~\underline{\wilson(-1)}\stackrel{X}{\longrightarrow}
\wilson(0)^{\oplus 3}\stackrel{X}{\longrightarrow}
\wilson(1)^{\oplus 3}\stackrel{pX}{\longrightarrow}
\wilson(-1).
$$
Note that the map $\vp$ is the only cochain map from
$\vanishing$ to
$\wilson(-1)$, and this 
is true also when projected to $D(X_-)$.
Thus, we find that the monodromy image of 
${\mathcal O}(\overline{2})\in D(X_-)$
is simply the bound state with the brane $\pi_-(\vanishing)$ via the
unique map $\pi_-(\vanishing)\to {\mathcal O}(\overline{2})$:
$$
\mono_L({\mathcal O}(\overline{2}))~=~
{\rm Cone}\Bigl(\pi_-(\vanishing)\to {\mathcal O}(\overline{2})\Bigr).
$$
It is now clear what to do for a general brane ${\mathcal B}\in D(X_-)$.
We first lift it to a brane $\lsmB$ in the grade restricted subset
$\langle \wilson(-1),\wilson(0),\wilson(1)\rangle$ and bind a copy of
$\vanishing$ at each appearance of the factor $\wilson(-1)$ 
in $\lsmB$ that fails to 
obey the new grade restriction rule 
$\langle \wilson(0),\wilson(1),\wilson(2)\rangle$.
The end result is binding
as many copies of $\pi_-(\vanishing)$
as the number of $\wilson(-1)$'s in $\lsmB$. 
In fact, for each $\wilson(-1)$ at R-degree $j$ there
is a chiral ring element
$\Ho^j(\pi_-(\vanishing),{\mathcal B})$, and vice versa.
(This will become clear from our consideration
in Section~\ref{sec:math}.)
Thus, we find that the monodromy action is given by
\beq
\mono_L({\mathcal B})~=~
{\rm Cone}\left(\bigoplus_{j\in\Z}
\Ho^j(\pi_-(\vanishing),{\mathcal B})\otimes \pi_-(\vanishing)[-j]
\longrightarrow {\mathcal B}\right).
\label{monoL}
\eeq

Let us next consider the monodromy along the same loop but with the 
opposite orientation.
The loop goes around the singular point, now {\it clockwise}.
Let us see what happens to the brane $\wilson(-1)$ again.
Since we first go through the window $w_{-1}$, binding with
$\lsmB_-$ must be done in advance:
$$
~~~~
~~~~
~~~~~~~~
~~~~
~~~~
\begin{array}{c}
\underline{\wilson(-1)}\\[-0.075cm] 
\oplus\\[-0.075cm]
\wilson(2)
\end{array}\!\!\!\!\!
\begin{array}{c}
\\[-0.1cm]
\,\,\,\searrow\!\mbox{\footnotesize id}\\
\stackrel{\!p\,}{\longrightarrow}
\end{array}\!\!\!\!
\begin{array}{c}
\\[-0.15cm]
\\
\wilson(-1).\end{array}
$$
Once inside $r\gg 0$, we must bind the empty brane
$\lsmB_+$ before coming back to $r\ll 0$:
$$
\begin{array}{c}
\\
\\
\\[0.28cm]
\wilson(-1)
\end{array}\!\!\!
\begin{array}{c}
\\
\\
\\[0.28cm]
\stackrel{X}{\longrightarrow}
\end{array}\!\!\!
\begin{array}{c}
\\
\\
\\[0.28cm]
\wilson(0)^{\oplus 3}
\end{array}\!\!\!\!
\begin{array}{c}
\\
\\
\\[0.28cm]
\stackrel{X}{\longrightarrow}
\end{array}\!\!\!\!
\begin{array}{c}
\underline{\wilson(-1)}\\
\oplus\\
\wilson(2)\\[-0.15cm]
\oplus\\[-0.15cm]
\wilson(1)^{\oplus 3}
\end{array}\!\!\!\!\!\!
\begin{array}{c}
\\
\,\,\,\searrow\!\mbox{\footnotesize id}\\
\stackrel{p}{\longrightarrow}\\[-0.15cm]
\,\,\,\,\searrow\!\mbox{\footnotesize id}\\[-0.15cm]
\stackrel{X}{\longrightarrow}
\end{array}\!\!\!\!
\begin{array}{c}
\\
\\
\wilson(-1)\\[-0.15cm]
\oplus\\[-0.15cm]
\wilson(2).
\end{array}~~~~
~~~
$$
The net result is binding the brane $\vanishing$ in (\ref{BV})
again, 
but now with an arrow in the opposite direction
--- from the given brane $\wilson(-1)$ to $\vanishing$.
The monodromy action on a general brane ${\mathcal B}\in D(X_-)$
is
$$
\mono_{-L}({\mathcal B})~=~
{\rm Cone}\left({\mathcal B}\longrightarrow
\bigoplus_{j\in \Z}\Ho^j({\mathcal B},\pi_-(\vanishing))\otimes
\pi_-(\vanishing[j]))\right)[-1].
$$
It is obviously the inverse of $\mono_L$: the compositions
$\mono_{-L}\circ\mono_L$ and
$\mono_L\circ\mono_{-L}$
simply bind an empty brane to a given brane.

\begin{figure}[htb]
\centerline{\includegraphics{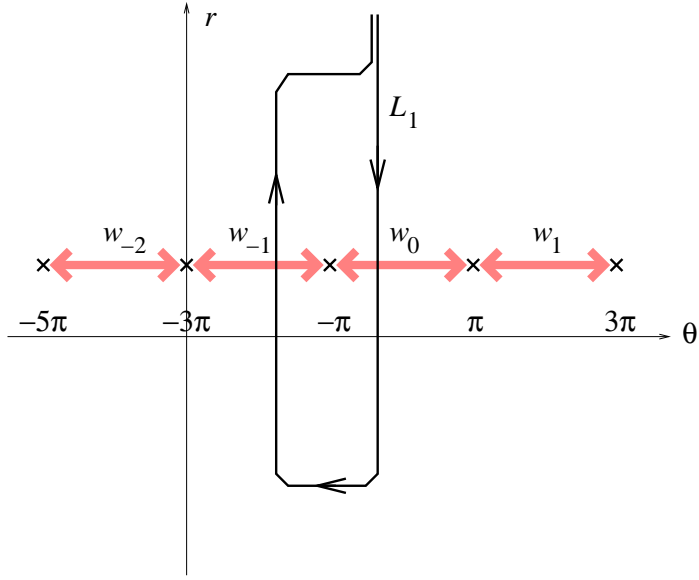}}
\caption{Another loop around the same point}
\label{fig:monodromy2}
\end{figure}

In the same way, we can study monodromies
along loops with a base point in the $r\gg 0$ phase, such as
$L_1$ as despicted in Fig.~\ref{fig:monodromy2}.
The result is again binding
the brane $\vanishing$:
\beqa
&&\mono_{L_1}({\mathcal B}')~=~
{\rm Cone}\left({\mathcal B}'\longrightarrow
\bigoplus_{j\in \Z}\Ho^j({\mathcal B}',\pi_+(\vanishing))\otimes
\pi_+(\vanishing[j]))\right)[-1],
\nn\\
&&\mono_{-L_1}({\mathcal B}')~=~
{\rm Cone}\left(\bigoplus_{j\in\Z}
\Ho^j(\pi_+(\vanishing),{\mathcal B}')\otimes 
\pi_+(\vanishing)[-j]
\longrightarrow {\mathcal B}'\right).
\nn
\eeqa

We have seen that the monodromy around the singular point 
$(r,\theta)=(3\log 3,-\pi)$
is
to bind a certain number of copies of the brane $\vanishing$.
In the $r\gg 0$ phase, the brane $\vanishing$
(with a shift by $3$) reduces at low energies to
$$
{\mathcal O}(2)\stackrel{p}{\longrightarrow}
{\mathcal O}(-1)~~\cong~{\mathcal O}_E(-1),
$$
the D-brane wrapped on the exceptional divisor $E\cong \CP^2$
and supporting a non-trivial $U(1)$ gauge connection.
In the negative volume phase $r\ll 0$,
it reduces to
$$
{\mathcal O}(\overline{-1})\stackrel{X}{\longrightarrow}
{\mathcal O}(\overline{0})^{\oplus 3}\stackrel{X}{\longrightarrow}
{\mathcal O}(\overline{1})^{\oplus 3}\stackrel{X}{\longrightarrow}
{\mathcal O}(\overline{-1})~~\cong~
{\mathcal O}_{\mathfrak{p}}(\overline{2}),
$$
which is one of the three fractional branes.
Note also that the monodromies around the same singular point
are described more naturally
in terms of another brane $\widetilde{\vanishing}$
\beqa
&&\left[\begin{array}{c}
\wilson(2)\\[-0.15cm]
\oplus\\[-0.15cm]
\underline{\wilson(-1)}
\end{array}\!\!\!\!\!\!
\begin{array}{c}
\stackrel{p}{\longrightarrow}\\[-0.15cm]
\mbox{\footnotesize id}\!\!\!\nearrow\,\,\,\\[-0.15cm]
\stackrel{X}{\longrightarrow}
\end{array}\!\!\!\!
\begin{array}{c}
\wilson(-1)\\[-0.15cm]
\oplus\\[-0.15cm]
\wilson(0)^{\oplus 3}
\end{array}\!\!\!
\begin{array}{c}
\\[0.28cm]
\stackrel{X}{\longrightarrow}
\end{array}\!\!\!\!
\begin{array}{c}
\\[0.28cm]
\wilson(1)^{\oplus 3}
\end{array}\!\!\!
\begin{array}{c}
\\[0.28cm]
\stackrel{X}{\longrightarrow}
\end{array}\!\!\!
\begin{array}{c}
\\[0.28cm]
\wilson(2)
\end{array}\right]
\nn\\[0.2cm]
&\cong&~~~~~
\underline{\wilson(2)}\stackrel{pX}{\longrightarrow}
\wilson(0)^{\oplus 3}\stackrel{X}{\longrightarrow}
\wilson(1)^{\oplus 3}\stackrel{X}{\longrightarrow}
\wilson(2),
\nn
\eeqa
if we would like to present the branes 
using the grade restricted set
$\langle \wilson(0),\wilson(1),\wilson(2)\rangle$
rather than $\langle \wilson(-1),\wilson(0),\wilson(1)\rangle$.
This brane is essentially the same as
$\vanishing$, since
$\pi_+(\vanishing)\cong\pi_+(\widetilde{\vanishing}[-2])$
and $\pi_-(\vanishing)\cong\pi_-(\widetilde{\vanishing})$.

It is easy to find the analogue of $\vanishing$ for any model
with $U(1)$ gauge group.
Consider the model with fields $X_1,...,X_l$ and $Y_1,...,Y_{l'}$
of positive charges $Q_1,...,Q_l$ and negative charges $-Q'_1,...,-Q'_{l'}$.
The brane $\vanishing$ relevant for the monodromy
around the singular point
is obtained by concatenation of 
the two Koszul-type complexes (\ref{X-Koszul})-(\ref{Y-Koszul}):
$$
\vanishing~=~{\rm Cone}\Bigl({\mathscr K}_-({\mathscr S})[-l']\longrightarrow
 {\mathscr K}_+\Bigr)[-l]
$$
by the identity map of the left-most $\wilson({\mathscr S})$ 
of ${\mathscr K}_-({\mathscr S})$
to the right-most $\wilson({\mathscr S})$ 
of ${\mathscr K}_+$. We may need to make an
appropriate shift in the gauge charge, $\vanishing(n)$,
depending on the precise value of the theta parameter.
In the positive ({\it resp.} negative) volume phase,
 it reduces to a brane wrapped on the
locus $Y_1=\cdots=Y_{l'}=0$ ({\it resp.} $X_1=\cdots=X_l=0$).

In \cite{Strominger} it was argued from the spacetime
point of view that the monodromy around a singular point of the moduli
space is governed by binding copies of the 
D-brane that becomes massless at that point. 
This approach was recast into the language of derived category
in \cite{SeTh,Horja} following Kontsevich's suggestion
and studied further in \cite{Douglas,AspDouglas,AKH}.
Our result directly confirms a part of this picture: the monodromy around a
singular point is the binding of a brane $\vanishing$.
A comparison with Strominger's spacetime picture then tells us that
our brane $\vanishing$ is the one that becomes massless at the singular point.
In the past, the brane that becomes massless was identified only
in the mirror description --- it is the A-brane wrapped on the vanishing cycle.
There was not even an attempt to do this 
for B-branes, except for identification of
the Ramond-Ramond charges.
This is because the usual methods were based on
the non-linear sigma model description, which certainly breaks down
near the singularity of the K\"ahler moduli space $\mathfrak{M}_K$.
In the present work, with the input from the spacetime picture,
we have directly identified in the linear sigma model description
the B-brane that becomes massless at the singular point.
In particular, the D-brane $\vanishing$ in our Example (A) 
confirms the results from the mirror computation of the central charge 
in \cite{Aspinwall}.

Our approach to describe monodromies has 
a technical advantage over the approach that was previously used 
in the literature such as \cite{SeTh,Horja,AKH}, 
which starts from a formula like
(\ref{monoL}). From our point of view this formula is a consequence
of a more general construction, i.e., a loop around a singular point
must go through two different windows, and hence
we must change the linear sigma model 
representatives according to the grade restriction rule.
In this procedure,
we do not have to compute the chiral ring spectrum
$\Ho^j(\pi_-(\vanishing),{\mathcal B})$ nor to construct the cone.
That is particularly advantageous in situations where
$\Ho^j(\pi_-(\vanishing),{\mathcal B})$ is infinite dimensional and 
the formula like (\ref{monoL}) does not make sense. We will find such examples 
in models with higher rank gauge group.

\subsection*{\it Multi-parameter Models}

For models with higher rank gauge groups, $T=U(1)^k$ with $k>1$,
the story is essentially the same, and our approach again provides 
an efficient way to find monodromies of loops
in the K\"aher moduli space $\mathfrak{M}_K$.

A shift of theta parameters $\theta^a\to \theta^a+2\pi n^a$ 
at a point deep inside a phase
does, as in the $U(1)$ case, shifts the gauge charges, ${\mathcal E}\to
{\mathcal E}(n^1,...,n^k)$.
The monodromy of a loop around 
the singular locus $\mathfrak{S}$ in an asymptotic direction can be
found as follows. An asymptotic direction corresponds to a phase boundary
where all but one $U(1)$ subgroup of the gauge group is completely Higgsed.
The loop can be regarded as a loop in the FI-theta parameter space
of that single $U(1)$ subgroup. Then, we simply apply what we have done
previously in the model with a single $U(1)$ gauge group.

\begin{figure}[htb]
\centerline{\includegraphics{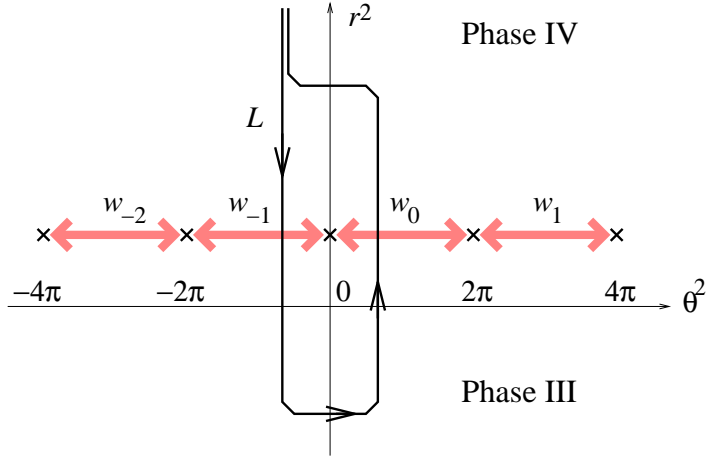}}
\caption{A loop around the singular locus in the two-parameter model (C)}
\label{fig:monodromy3}
\end{figure}

Let us illustrate the monodromy around the singular locus $\mathfrak{S}$
in Example (C).
We look at the asymptotic region
corresponding to the III-IV phase boundary, where
the unbroken gauge group is the subgroup
$U(1)_2$ and the singular locus is
at $\theta^2 \in 2\pi\Z$ and $r^2 = 2\log 2$. 
We consider the loop $L$ as depicted in Fig.~\ref{fig:monodromy3}.
The band restriction rules at the two relevant windows are:
\beqa
w_{-1}:&& q^2=0,1,\nn\\
w_0:&& q^2=-1,0.
\nn
\eeqa
Let us take the brane ${\mathcal O}(m,1)$ in Phase IV. 
Its lift that obeys
the band restriction rule of the window $w_{-1}$ is $\wilson(m,1)$.
Here $m$ could be replaced by $m\pm 4, m\pm 8,$ etc, since
$\{p=0\}$ is a part of the deleted set (this applies also to Phase III).
This can be moved safely to Phase III. Once inside that phase we switch to a
representative that obeys the band restriction rule of the next window
$w_0$. 
This is done by binding the complex ${\mathscr K}^{\rm III,IV}_-(m,1)$
introduced in Section~\ref{subsec:grrrevisited}
to $\wilson(m,1)$:
\beq
\label{bindKminus}
\begin{array}{c}
\wilson(m,1)\\[0.3cm]
\\
\end{array}\!\!\!\!\!\!
\begin{array}{c}
\stackrel{x_6}{\longrightarrow}\\[-0.15cm]
\mbox{\footnotesize id}\!\searrow\,\,\,\\[0.35cm]
\end{array}\!\!\!\!\!\!
\begin{array}{c}
\wilson(m+1,-1)\\[-0.15cm]
\!\!\oplus\\[-0.15cm]
\!\!\!\!\underline{\wilson(m,1)}
\end{array}  .
\eeq
Then, one can move $\wilson(m+1,-1)$ back to Phase IV through $w_0$.
After doing that, we again bring it into the form consisting only of the 
band restricted Wilson lines $\wilson(*,0)$ and $\wilson(*,1)$. 
This is done by binding
$$
{\mathcal K}^{\rm III,IV}_+(m+1,-1)\,:~~
\wilson(m+1,-1)\stackrel{x_1\choose x_2}{-\!\!\!-\!\!\!-\!\!\!\longrightarrow}
\!\!\!
\begin{array}{c}
\wilson(m+1,0)\\[-0.15cm]
\oplus\\[-0.15cm]
\wilson(m+1,0)
\end{array}\!\!\!
\stackrel{(-x_2,x_1)}{-\!\!\!-\!\!\!-\!\!\!\longrightarrow}
\wilson(m+1,1)
$$
via the identity map to $\wilson(m+1,-1)$.
The outcome is the total binding of
$$
\vanishing_m\,:~~
\underline{\wilson(m,1)}
\stackrel{x_6{x_1\choose x_2}}{-\!\!\!-\!\!\!-\!\!\!\longrightarrow}
\!\!\!
\begin{array}{c}
\wilson(m+1,0)\\[-0.15cm]
\oplus\\[-0.15cm]
\wilson(m+1,0)
\end{array}\!\!\!
\stackrel{(-x_2,x_1)}{-\!\!\!-\!\!\!-\!\!\!\longrightarrow}
\wilson(m+1,1)
$$
to the original brane $\wilson(m,1)$ via the map
$\vanishing_m\to \wilson(m,1)$ that sends
$\wilson(m,1)$ in $\vanishing_m$
to $\wilson(m,1)$ by the identity.
Thus, we found that the monodromy is given by
$$
\mono_L({\mathcal O}(m,1))~=~
{\rm Cone}\Bigl(\pi_{\rm IV}(\vanishing_m)\to {\mathcal O}(m,1)\Bigr).
$$
Note that the isomorphism class of $\pi_{\rm IV}(\vanishing_m)$
depends on $m$; in fact
$\pi_{\rm IV}(\vanishing_m)\cong \pi_{\rm IV}(\vanishing_{m'})$
if and only if $m \equiv m'$ modulo $4$.
This holds in Phase III as well.
Thus, for a general brane ${\mathcal B}$ in Phase IV,
the monodromy along the loop $L$ 
is binding a copy of $\pi_{\rm IV}(\vanishing_m)$ 
with $m=0,1,2$ or $3$ for each
$\wilson(n,1)$ in the lift $\lsmB$ to 
${\mathcal T}^{w_{-1}}_{\rm III,IV}$, where $n = m$ mod $4$.

When we try to write this monodromy action in the form
\beq
\label{monohigherrank}
\mono_L({\mathcal B})~=~
{\rm Cone}\left(\bigoplus_{m=0,1,2,3}\bigoplus_{j\in \Z}
\Ho^j(\pi_{\rm IV}(\vanishing_m),{\mathcal B})
\otimes\pi_{\rm IV}(\vanishing_m)[-j]\longrightarrow {\mathcal B}\right),
\eeq
we run into troubles. 
This is in view of the fact that in the present situation the 
D-branes $\vanishing_m$ are non-compact, 
so that infinite dimensional spaces of chiral ring elements 
appear in (\ref{monohigherrank}). For instance,  
$\Ho^0(\pi_{\rm IV}(\vanishing_m),{\mathcal O}(m,1))$ 
is spanned by the gauge invariant monomials 
$\{1,px_3^4,px_3^3x_4,\ldots\}$ that map 
${\mathcal O}(m,1)$ in 
$\pi_{\rm IV}(\vanishing_m)$ to ${\mathcal O}(m,1)$, 
whereas only the first element $1$ played a r\^ole in (\ref{bindKminus}). 

This is a quite general feature of multi-parameter models, 
and the monodromy formula (\ref{monohigherrank}) 
is only applicable if the D-branes that induce the 
monodromy action are compactly supported. 
As an example the D-branes inducing the monodromy 
at the I-IV phase boundary of the two-parameter model (C) are of this kind.
We want to stress that our method does not rely on computing 
the spectrum, and thus applies to \emph{any} phase boundaries 
in multi-parameter models.

\subsection{More Examples}
\label{subsec:moreex}

\subsubsection{Flop Transition Of The Resolved Conifold}

Let us consider Example (B), the $U(1)$ gauge theory with 
four fields 
$X,Y,U,V$ which carry charges $1,1,-1,-1$. It provides a local model
for a flop transition. The deleted sets of the two phases $r\gg 0$ and 
$r\ll 0$ are
$$
\Delta_+=\{x=y=0\},\qquad
\Delta_-=\{u=v=0\}.
$$
For $r\gg 0$ the fields $x,y$ form the base $\CP^1$
and $u,v$ span the fibre, whereas for $r\ll 0$ it is the other way
around:
$$
X_+=\left[ \begin{array}{c} 
    \cO(-1)^{\oplus 2}_{u,v}\\ 
    \downarrow \\ 
    \CP^1_{[x:y]}
  \end{array} \right]
  \quad \stackrel{\textrm{flop}}{\longleftarrow\!\!\longrightarrow}\quad
X_-=\left[  \begin{array}{c} 
    \CP^1_{[u:v]}\\ 
    \uparrow \\ 
     \cO(-1)^{\oplus 2}_{x,y}
  \end{array}   \right]
$$
In view of the deleted sets, we find that
the branes
\beqa
&&{\mathscr K}_+\,:~~
 \wilson(0)
  \stackrel{\left(\mbox{\tiny $\!\!\!\begin{array}{r}y\\ -x
      \end{array}$}\!\!\!\right)}{-\!\!\!-\!\!\!-\!\!\!\longrightarrow} 
  \wilson(1)^{\oplus 2}
  \stackrel{\left(x,y\right)}{-\!\!\!-\!\!\!-\!\!\!\longrightarrow} 
  \underline{\wilson(2)}
\nn\\
&&{\mathscr K}_-\,:~~
 \wilson(2)
  \stackrel{\left(\mbox{\tiny $\!\!\!\begin{array}{r}v\\ -u
      \end{array}$}\!\!\!\right)}{-\!\!\!-\!\!\!-\!\!\!\longrightarrow} 
  \wilson(1)^{\oplus 2}
  \stackrel{\left(u,v\right)}{-\!\!\!-\!\!\!-\!\!\!\longrightarrow} 
  \underline{\wilson(0)}
\nn
\eeqa
are empty at low energies in the $r\gg 0$ and 
$r\ll 0$ phases, respectively.
The model has a singular point at $\e^{r-i\theta}=1$.

We would like to transport branes back and forth between the two phases.
Let us choose the window $w=\{-2\pi<\theta<0\}$
at the phase boundary,
for which the grade restriction rule is given by
$$
  {\cal T}^w = \langle\, \brane(0), \brane(1) \,\rangle 
$$
In what follows, we consider compactly supported D-branes on one
side of the flop and find what they turn into on the other side. 

\subsubsection*{\it D$\,2$-Branes}

\paragraph{Phase $r\gg 0$:} Let us consider
$\cO_{E_+}(0) \in D(X_+)$,
the D2-brane wrapped on the zero section
$E_+=\{u=v=0\} \cong \CP^1_{[x:y]} \subset X_+$ and supporting
a trivial line bundle.
It can be represented by the complex of vector bundles
\beq
  \label{conifold1}
  {\mathcal O}(2)
  \stackrel{\left(\mbox{\tiny $\!\!\!\begin{array}{r}v\\ -u
      \end{array}$}\!\!\!\right)}{-\!\!\!-\!\!\!-\!\!\!\longrightarrow} 
  {\mathcal O}(1)^{\oplus 2}
  \stackrel{\left(u,v\right)}{-\!\!\!-\!\!\!-\!\!\!\longrightarrow} 
  \underline{{\mathcal O}(0)}.
\eeq
A lift to the linear sigma model is obtained by replacing
each line bundle ${\mathcal O}(q)$ in this complex
by the Wilson line brane $\wilson(q)$ --- this gives ${\mathscr K}_-$.
In order to transport it to the other phase $r\ll 0$ through the window
$w$, we have to rewrite it as a grade restricted complex,
i.e., we have to replace the Wilson line brane
$\brane(2)$ by something made of $\wilson(0)$ and $\wilson(1)$. 
This can be done using the brane ${\mathscr K}_+$ which is
empty at low energies in this phase.
Namely, we bind the empty brane ${\mathscr K}_+$,
written in a different basis, to our brane
${\mathscr K}_-$ by the cochain map ${\mathscr K}_-\to {\mathscr K}_+$
that sends $\wilson(2)$ to 
$\wilson(2)$ by the identity:
\beq
  \label{conifold2}
  \begin{array}{c}
    \\[-2pt] \\[-2pt] \brane(0)
  \end{array}
  \begin{array}{c} 
    \\[-2pt] \\[-9pt]
    \stackrel{x\choose y}{\longrightarrow} 
  \end{array}\!\!
  \begin{array}{c}
    \brane(2)\\[-2pt]
    \oplus\\[-2pt]
    \brane(1)^{\oplus 2}
  \end{array}\!\!
  \begin{array}{r}
  \stackrel{v\choose -u}{\longrightarrow}\\[-0.15cm]
  \searrow\!\!\!\!{}^\textrm{\tiny id} \\
  \stackrel{\mbox{\tiny $(-y,x)$}}{\longrightarrow} \\[0.1cm]
  \end{array}\!\!
  \begin{array}{c}
  \brane(1)^{\oplus 2}\\[-2pt]
  \oplus\\[-2pt]
  \brane(2)
  \end{array}\!\!
  \begin{array}{c}
  \stackrel{\mbox{\tiny $(u,v)$}}{\longrightarrow} 
  \\[-2pt] \\[-2pt] \
  \end{array}\!\!
  \begin{array}{c}
  \underline{\brane(0)}\\[-2pt] \\[-2pt] \
  \end{array} \ .
\eeq
Cancelling the trivial 
pair $\wilson(2)\stackrel{\rm id}{\longrightarrow}\wilson(2)$
 as in (\ref{complexD}) and
(\ref{complexD'}), we obtain the following
grade restricted representative of ${\mathcal O}_{E_+}(0)$:
$$
  {\mathcal C}(\lsmB_1)\,:~~
  \brane(0)
  \stackrel{\left(\mbox{\tiny $\!\!\!\begin{array}{c}x\\ y
      \end{array}$}\!\!\!\right)}{-\!\!\!-\!\!\!-\!\!\!\longrightarrow} 
    \brane(1)^{\oplus 2}
  \stackrel{\left(\!\!\!\!\mbox{\tiny $\begin{array}{ccc}
          yv &  -xv\\ 
         -yu &   xu 
      \end{array}$}\!\!\!\!\right)}{-\!\!\!-\!\!\!-\!\!\!\longrightarrow} 
  \brane(1)^{\oplus 2}
  \stackrel{\left(u,v\right)}{-\!\!\!-\!\!\!-\!\!\!\longrightarrow} 
  \underline{\brane(0)} \ .
$$
Another example is provided by the D-brane 
$\cO_{E_+}(-1) \in D(X_+)$, again wrapped on the zero section $E_+$,
but now supporting a non-tivial line bundle. 
The na\"\i ve lift is ${\mathscr K}_-(-1)$, but this includes
the Wilson line brane $\wilson(-1)$,
which is outside the grade restriction rule.
This time we glue the empty brane ${\mathscr K}_+(-1)$ by the map
${\mathscr K}_+(-1)\to {\mathscr K}_-(-1)$ and then cancel
$\wilson(-1)\stackrel{\rm id}{\longrightarrow}\wilson(-1)$.
This leads to the following grade restricted representative of
${\mathcal O}_{E_+}(-1)$:
$$
  {\mathcal C}(\lsmB_2)\,:~~
  \brane(1)
  \stackrel{\left(\mbox{\tiny $\!\!\!\begin{array}{c}u\\ v
      \end{array}$}\!\!\!\right)}{-\!\!\!-\!\!\!-\!\!\!\longrightarrow} 
    \brane(0)^{\oplus 2}
  \stackrel{\left(\!\!\!\!\mbox{\tiny $\begin{array}{ccc}
          yv &  -uy\\ 
         -vx &   xu 
      \end{array}$}\!\!\!\!\right)}{-\!\!\!-\!\!\!-\!\!\!\longrightarrow} 
  \underline{\brane(0)}^{\oplus 2}
  \stackrel{\left(x,y\right)}{-\!\!\!-\!\!\!-\!\!\!\longrightarrow} 
  \brane(1) \ .
$$

\paragraph{Phase $r\ll 0$:} After crossing the phase boundary 
the tachyon condensation pattern changed. Now the complexes 
${\mathscr K}_-(n)$ are trivial branes-antibrane systems,
and they can be eliminated from $\lsmB_1$ and $\lsmB_2$. For instance,
if we restore the identity map in $\lsmB_1$ to write it in the form
(\ref{conifold2}) and then eliminate the trivial upper line, we see
that one ends up with $\pi_-({\mathscr K}_+[1])$. This is isomorphic to
the D2-brane $\cO_{E_-}(2)[1] \in D(X_-)$ wrapped on the zero section
$E_- = \{x=y=0\} \cong \CP^1_{[u:v]}$.
Repeating these arguments
for $\lsmB_2$ we find the following maps:
$$
  \cO_{E_+}(0)~~
  \begin{array}{c}
    \textrm{\tiny $\pi_+$} \\[-8pt]
    \longleftarrow  \\ [-13pt]
    \longrightarrow \\[-9pt]
    \textrm{\tiny $\omega^w_{+,-}$}
  \end{array}~~
  \lsmB_1 ~~
  \begin{array}{c}
    \textrm{\tiny $\pi_-$} \\[-8pt]
    \longrightarrow  \\ [-13pt]
    \longleftarrow \\[-9pt]
    \textrm{\tiny $\omega^w_{-,+}$}
  \end{array}~~
  \cO_{E_-}(2)[1] \ ,
$$
and
$$
  \cO_{E_+}(-1)~~
  \begin{array}{c}
    \textrm{\tiny $\pi_+$} \\[-8pt]
    \longleftarrow  \\ [-13pt]
    \longrightarrow \\[-9pt]
    \textrm{\tiny $\omega^w_{+,-}$}
  \end{array}~~
  \lsmB_2 ~~
  \begin{array}{c}
    \textrm{\tiny $\pi_-$} \\[-8pt]
    \longrightarrow  \\ [-13pt]
    \longleftarrow \\[-9pt]
    \textrm{\tiny $\omega^w_{-,+}$}
  \end{array}~~
  \cO_{E_-}(1)[-1] \ .
$$
We find that  D2-branes wrapped on the zero section $E_+$ in one phase
are mapped to  D2-branes wrapped on the zero section $E_-$
in the other phase.
Recall that the R-degree for D-branes reduces modulo two to
 the $\ZZ_2$-degree that distinguishes branes and antibranes.
 The shifts, by $[1]$ resp. $[-1]$, of the R-degrees on the
right-hand side, therefore, indicate that the branes on $X_+$ turned into
antibranes on $X_-$.

\subsubsection*{\it D$\,0$-Branes}

\paragraph{Phase $r\gg 0$:} Next, we study a D0-brane at a point on
$E_+ \subset X_+$, say at $\mathfrak{p}=\{u=v=P= 0\}$, where
$P = \alpha x + \beta y$ and $[\alpha\!:\!\beta] \in \CP^1$
parametrizes the location of the D$0$-brane. 
A linear sigma model lift of this D$0$-brane $\cO_{\mathfrak{p}}$ can
be realized as the Koszul complex
\beq
  \label{conifold3}
  \brane(1)
  \stackrel{\left(\mbox{\tiny $\!\!\!\begin{array}{c}P \\ u \\ v
      \end{array}$}\!\!\!\right)}{-\!\!\!-\!\!\!-\!\!\!\longrightarrow} 
  \begin{array}{c}
    \brane(2)\\[-5pt]
    \oplus\\[-5pt]
    \brane(0)^{\oplus 2}
  \end{array}
  \stackrel{\left(\!\!\!\!\mbox{\tiny $\begin{array}{ccc}
    -v \!\!&\!\!   0 \!\!&\!\!  P\\ 
     u \!\!&\!\!  -P \!\!&\!\!  0\\ 
     0 \!\!&\!\!   v \!\!&\!\! -u
      \end{array}$}\!\!\!\!\right)}{-\!\!\!-\!\!\!-\!\!\!-\!\!\!-\!\!\!-\!\!\!\longrightarrow} 
  \begin{array}{c}
    \brane(1)^{\oplus 2}\\[-5pt]
    \oplus\\[-5pt]
    \brane(-1)
  \end{array}
  \stackrel{\left(u,v,P\right)}{-\!\!\!-\!\!\!-\!\!\!\longrightarrow} 
  \underline{\brane(0)} \ .
\eeq
This can be viewed as the D2-brane $\cO_{E_+}(0)$ and the 
anti-D2-brane $\cO_{E_+}(-1)[1]$ coupled by a tachyon proportional to
$P = \alpha x + \beta y$:
\begin{center}
\begin{picture}(300,50) 
  \put(0,0){$\wilson(1)$}
   \put(52,9){\footnotesize ${\,u\choose \,v}$}
  \put(34,4){\vector(1,0){45}}
  \put(86,0){$\wilson(0)^{\oplus 2}$}
    \put(145,7){\tiny $(v,-u)$}
  \put(132,4){\vector(1,0){45}}
  \put(185,0){$\wilson(-1)$}
     \put(42,22){\tiny $P$}
     \put(34,12){\vector(2,1){45}}
     \put(119,25){\tiny $\left(\!\!\begin{array}{cc}
                          \!\!0\!\!&\!\!P\!\!\\
                         \!\!-P\!\!&\!\!0\!\!
                         \end{array}\!\right)$}
     \put(142,13){\vector(2,1){35}}
     \put(238,22){\tiny $P$}
     \put(225,11){\vector(2,1){45}}
         \put(97,18){$\oplus$}
         \put(194,18){$\oplus$}
  \put(90,37){$\wilson(2)$}
    \put(146,46){\footnotesize ${-v\choose u}$}
  \put(131,41){\vector(1,0){45}}
  \put(184,37){$\wilson(1)^{\oplus 2}$}
    \put(238,44){\tiny $(u,v)$}
  \put(225,41){\vector(1,0){45}}
  \put(277,37){$\wilson(0)$}
\end{picture}
\end{center}
We can glue in the trivial Koszul complexes ${\mathscr K}_+$ and
${\mathscr K}_+(-1)$ in order to cancel the Wilson lines
$\wilson(2)$ and $\wilson(-1)$ outside the grade restriction range.
For this it is best to write the Koszul complexes using a
new coordinate system $(P,Q)$, where 
$P$ is as above and $Q$ is a new linear coordinate,
so that $P$ and $Q$ cannot simultaneously vanish on $X_+$.
After the gluing, we have
\begin{center}
\begin{picture}(300,130)(0,-37)
  \put(86,-37){$\wilson(-1)$}
    \put(155,-28){\footnotesize ${-Q\choose P}$}
  \put(131,-33){\vector(1,0){45}}
  \put(184,-37){$\wilson(0)^{\oplus 2}$}
    \put(238,-30){\tiny $(P,Q)$}
  \put(225,-33){\vector(1,0){45}}
  \put(277,-37){$\wilson(1)$.}

\put(225,-22){\vector(1,1){45}}
   \put(250,-5){\tiny $(0,-{\rm id})$}

      \put(142,-17){\footnotesize id}
      \put(140,-23){\vector(2,1){35}}
         \put(97,-19){$\oplus$}
         \put(194,-19){$\oplus$}

   \put(40,50){\footnotesize ${0\choose -{\rm id}}$}
\put(34,20){\vector(1,1){45}}

  \put(0,0){$\wilson(1)$}
   \put(52,9){\footnotesize ${\,u\choose \,v}$}
  \put(34,4){\vector(1,0){45}}
  \put(86,0){$\wilson(0)^{\oplus 2}$}
    \put(145,7){\tiny $(v,-u)$}
  \put(132,4){\vector(1,0){45}}
  \put(185,0){$\wilson(-1)$}
     \put(53,25){\tiny $P$}
     \put(34,11){\vector(2,1){45}}
     \put(119,25){\tiny $\left(\!\!\begin{array}{cc}
                          \!\!0\!\!&\!\!P\!\!\\
                         \!\!-P\!\!&\!\!0\!\!
                         \end{array}\!\right)$}
     \put(142,13){\vector(2,1){35}}
     \put(238,22){\tiny $P$}
     \put(225,11){\vector(2,1){45}}
         \put(97,18){$\oplus$}
         \put(194,18){$\oplus$}
  \put(90,37){$\wilson(2)$}
    \put(155,46){\footnotesize ${-v\choose u}$}
  \put(131,41){\vector(1,0){45}}
  \put(184,37){$\wilson(1)^{\oplus 2}$}
    \put(238,44){\tiny $(u,v)$}
  \put(225,41){\vector(1,0){45}}
  \put(277,37){$\wilson(0)$}

        \put(142,58){\footnotesize id}
        \put(140,53){\vector(2,1){35}}
           \put(97,55){$\oplus$}
           \put(194,55){$\oplus$}

  \put(0,74){$\wilson(0)$}
   \put(52,83){\footnotesize ${\,P\choose \,Q}$}
  \put(34,78){\vector(1,0){45}}
  \put(86,74){$\wilson(1)^{\oplus 2}$}
    \put(145,81){\tiny $(-Q,P)$}
  \put(132,78){\vector(1,0){45}}
  \put(185,74){$\wilson(2)$}

\end{picture}
\end{center}
Note that we need to include non-zero maps
$\wilson(1)\to\wilson(1)^{\oplus 2}$ and $\wilson(0)^{\oplus 2}\to\wilson(0)$
in order to have a complex.
Eliminating the trivial brane-antibrane pairs
$\wilson(q)\stackrel{\rm id}{\longrightarrow}\wilson(q)$
(there are four of them), we obtain the grade restricted representative
$$
  {\mathcal C}(\lsmB_{\mathfrak{p}})\,:~~
  \brane(0)
  \stackrel{\left(\mbox{\tiny $\!\!\!\begin{array}{c}P \\ uQ \\ vQ
      \end{array}$}\!\!\!\right)}{-\!\!\!-\!\!\!-\!\!\!\longrightarrow} 
  \begin{array}{c}
    \brane(1)\\[-5pt]
    \oplus\\[-5pt]
    \brane(0)^{\oplus 2}
  \end{array}
  \stackrel{\left(\!\!\!\!\mbox{\tiny $\begin{array}{ccc}
    -vQ \!\!&\!\!   0 \!\!&\!\!  P\\ 
     uQ \!\!&\!\!  -P \!\!&\!\!  0\\ 
     0 \!\!&\!\!   Qv \!\!&\!\! -Qu
      \end{array}$}\!\!\!\!\right)}{-
\!\!\!-\!\!\!-\!\!\!-\!\!\!-\!\!\!-\!\!\!\longrightarrow} 
  \begin{array}{c}
    \brane(1)^{\oplus 2}\\[-5pt]
    \oplus\\[-5pt]
    \brane(0)
  \end{array}
  \stackrel{\left(Qu,Qv,P\right)}{-\!\!\!-\!\!\!-\!\!\!\longrightarrow} 
  \underline{\brane(1)} \ .
$$
\paragraph{Phase $r\ll 0$:} The transport to the other phase
through the window $w$
yields a quite exotic D-brane in $D(X_-)$. There is no way to eliminate
trivial brane-antibrane pairs, like (\ref{conifold1}), from
$\lsmB_{\mathfrak{p}}$ by D-term deformations. Moreover, it is not
possible to rewrite 
$\pi_-(\lsmB_{\mathfrak{p}})$ as a complex of compactly supported
D-branes on $E_-$, as one might suspect. 
The best we can tell about $\pi_-(\lsmB_{\mathfrak{p}})$ is that as a
complex of coherent sheaves its non-trivial cohomology is given by 
$H^{-1}(\pi_-(\lsmB_{\mathfrak{p}})) \cong \cO_{E_-}(2)[1]$ as well as 
$H^{0}(\pi_-(\lsmB_{\mathfrak{p}})) \cong \cO_{E_-}(1)$, which is in
line with the result for the D$2$-branes that we found previously.
Actually, we have at hand an explicit example of a perverse (point)
sheaf. This class of objects was studied
in the context of flop transitions and derived categories by 
Bridgeland in \cite{Flops}. From the latter point of view our
example D-brane $\lsmB_{\mathfrak{p}}$ was investigated in 
\cite{Aspinwall}.

\subsubsection{McKay Correspondence}
\label{subsub:McKay}

Suppose we consider a linear sigma model which, in a particular
phase, reduces at low energies
to a free orbifold $X_{\rm orb} \cong \C^n/\Gamma$
with a finite subgroup $\Gamma\subset SL(n,\C)$. 
In the present context of Abelian
gauge groups, $T=U(1)^k$, the group $\Gamma$ arises 
as the unbroken subgroup of $T$ and as such
 must be Abelian. In view of the fact that the 
representation $\rho$ of $T$ on the Chan-Paton space descends to a
representation of the subgroup $\Gamma$,
the low energy D-branes on the orbifold are
actually given in terms of complexes of $\Gamma$-equivariant vector
bundles on $\C^n$:
$$
  D(X_{\rm orb}) ~\cong~ D_{\Gamma}(\C^n)  .
$$
The other phases correspond to partial or complete crepant%
\footnote{The property `crepant' (opposite to `{\it dis}crepant'),
 which says that the canonical line bundle remains trivial, 
  is ensured by the Calabi-Yau condition (\ref{conformal}).}
resolutions, $X_{\rm res}$, of the orbifold singularity.
Transport of branes along a path from $X_{\rm orb}$
to $X_{\rm res}$ in the K\"ahler moduli
space $\moduli_K$
leads to a map
of low energy boundary conditions up to
D-term deformations and brane-antibrane annihilation:
\beq
 \label{McKay1}
F: D_{\Gamma}(\C^n) ~~
 \longrightarrow
 ~~D(X_{\rm res})  .
\eeq
By construction, the chiral sector is preserved under our transports.
As remarked earlier,
the chiral sector of low energy boundary conditions $D(X_r)$
gives rise to the derived category of
coherent sheaves ${\bf D}(X_r)$
--- the objects
are elements in $D(X_r)$ and as the
morphisms we only take chiral ring elements. 
(Note that ${\bf D}(X_r)$ has smaller information than $D(X_r)$.
For example, it does not depend on
$\moduli_K$ at all, while $D(X_r)$ does.)
As a consequence of our map
(\ref{McKay1}), we find
 a modern version of McKay correspondence:\\[5pt]
\emph{Given a finite Abelian group $\Gamma \subset SL(n,\C)$ and a crepant
  resolution $X_{res}$ of the quotient $\C^n/\Gamma$, there exists
  an equivalence of derived categories:}%
\beq
 \label{McKay2}
\Phi:  {\bf D}_{\Gamma}(\C^n) \stackrel{\cong}{\longrightarrow}
  {\bf D}(X_{res})  .
\eeq
For arbitrary $n$ the equivalence
(\ref{McKay2}) was shown (as a special case) in \cite{Kawamata}. 
For $n \leq 3$, but $\Gamma$ also non-Abelian, it was proven in \cite{BKR}.
McKay correspondence was also discussed by physicists
in \cite{EmMk,PhysMcKay}.

We illustrate these maps (\ref{McKay1}) in three examples.

\subsubsection*{(A) Resolution of the orbifold $\C^N/\ZZ_N$}

The simplest example of an orbifold resolution is provided by
Example (A),
the $U(1)$ gauge theory with the fields $P,X_1,\ldots,X_N$ of charge
$-N,1,\ldots,1$.
The low energy theory is the orbifold $X_- \cong \C^N/\ZZ_N$
in the $r\ll 0$ phase (the orbifold phase),
whereas it is the non-linear sigma model on
the total space of the line bundle $\cO(-N)$ over $\CP^{N-1}$
in the $r\gg 0$ phase (the large volume phase).

We choose the window $w:-N\pi<\theta<-N\pi+2\pi$ at the phase boundary
so that the grade
restriction rule is
$$
{\mathcal T}^w=\langle\,\wilson(0),\wilson(1),\ldots,\wilson(N-1)\,\rangle.
$$
We transport branes from the orbifold phase to
the large volume phase along a path that goes through this window.
We first consider the equivariant line bundles ${\mathcal O}(\bi)\in D(X_-)$
parametrized by a mod-$N$ integer $\bi\in \Z_N$.
 As its lift to the linear sigma model,
we take $\wilson(i)$ where $i$ is the
integer in the grade restriction range $\{0,1,...,N-1\}$
that reduces modulo $N$ to $\bi$. 
It can be tranported safely to the large volume phase where it descend to
the line bundle ${\mathcal O}(i)\in D(X_+)$.
Thus, we find the following transportation rule:
\beq
\quad F^w_{-,+}:{\mathcal O}(\bi)\,\longmapsto\,
{\mathcal O}(i),\qquad i=0,1,\ldots, N-1.
\eeq
Next, let us consider the
fractional branes, ${\mathcal O}_{\mathfrak{p}}(\bi)\in D(X_-)$, 
parametrized again by 
a mod-$N$ integer $\bi\in \Z_N$. 
These are D-branes stuck at the $\Z_N$-fixed point 
$\mathfrak{p} = \{x_1=\cdots=x_N=0\}$. 
A lift to the linear sigma model
may be realized as the Koszul complex made of Wilson line branes
\beq
  \label{z3koszul}
\lsmB_{\bi}:~~  \wilson(i\!-\!N) \stackrel{X}{\longrightarrow}
\cdots\stackrel{X}{\longrightarrow}
   \wilson(i\!-\!2)^{\oplus {N\choose 2}} 
  \stackrel{X}{\longrightarrow}
    \brane(i\!-\!1)^{\oplus N}   
\stackrel{X}{\longrightarrow}
  \underline{\wilson(i)}.
\eeq
However, for any choice of $i$,
at least one of the Wilson line branes in this complex is outside the grade
restriction range. Thus, we must find appropriate replacements
from the subset ${\mathcal T}^w$.
This is straightforward. In the order of
${\mathcal O}_{\mathfrak{p}}(\bar 0)$,${\mathcal O}_{\mathfrak{p}}(\bar 1)$,
${\mathcal O}_{\mathfrak{p}}(\bar 2)$,..., 
${\mathcal O}_{\mathfrak{p}}(\overline{N-1})$,
the grade restricted lifts are given as follows:
{\small
\beqa
\lsmB'_{\bar 0}:&&\!\!
\wilson(0)\stackrel{X}{\longrightarrow}\cdots
\stackrel{X}{\longrightarrow}
\wilson(N\!-\!3)^{\oplus {N\choose 3}}\stackrel{X}{\longrightarrow}
\wilson(N\!-\!2)^{\oplus {N\choose 2}}\stackrel{X}{\longrightarrow}
\wilson(N\!-\!1)^{\oplus N}\stackrel{pX}{\longrightarrow}
\underline{\wilson(0)}
\nn\\
\lsmB'_{\bar 1}:&&\!\!
\wilson(1)\stackrel{X}{\longrightarrow}\cdots
\stackrel{X}{\longrightarrow}
\wilson(N\!-\!2)^{\oplus {N\choose 3}}\stackrel{X}{\longrightarrow}
\wilson(N\!-\!1)^{\oplus {N\choose 2}}\stackrel{pX}{\longrightarrow}
\wilson(0)^{\oplus N}\stackrel{X}{\longrightarrow}
\underline{\wilson(1)}
\nn\\
\lsmB'_{\bar 2}:&&\!\!
\wilson(2)\stackrel{X}{\longrightarrow}\cdots
\stackrel{X}{\longrightarrow}
\wilson(N\!-\!1)^{\oplus {N\choose 3}}\stackrel{pX}{\longrightarrow}
\wilson(0)^{\oplus {N\choose 2}}\stackrel{X}{\longrightarrow}
\wilson(1)^{\oplus N}\stackrel{X}{\longrightarrow}
\underline{\wilson(2)}
\nn\\
\cdots&&\qquad\cdots\qquad\cdots
\nn\\
\lsmB'_{\overline{N-1}}:&&\!\!
\wilson(N\!-\!1)\stackrel{pX}{\longrightarrow}
\wilson(0)^{\oplus N}\stackrel{X}{\longrightarrow}
\cdots
\stackrel{X}{\longrightarrow}
\wilson(N\!-\!2)^{\oplus N}\stackrel{X}{\longrightarrow}
\underline{\wilson(N\!-\!1)}
\nn\\[-0.4cm]
\label{liftfrac}
\eeqa
}

\vspace{-0.7cm}
\noindent
These branes can be transported safely to the large volume.
Once we are in that phase, we can just reduce them to the 
low energy theory ---
it is simply to replace $\wilson(q)$ by ${\mathcal O}(q)$.
This is essentially the end of the story.
We can, however, simplify the result.
The point is that these lifts $\lsmB'_{\bi}$
can be presented as the bound state of the brane $\lsmB_{\bi}$ in
(\ref{z3koszul}) with another, simpler, brane $\lsmB''_{\bi}$
that is
empty in the low enery theory in the phase $r\ll 0$. 
After being transported to the $r\gg 0$ phase, 
the part $\lsmB_{\bi}$ in such a presentation
is empty in the infra-red limit since
$\{x_1=\cdots=x_N=0\}$ is deleted.
Thus, we are left with the other part $\lsmB''_{\bi}$:
$\pi_+(\lsmB'_{\bi})\cong \pi_+(\lsmB''_{\bi})$.
For example, the last few terms of
$\lsmB'_{\bar 2}$ can be presented as:
{\small
\begin{center}
\begin{picture}(350,75)(-50,0)

\put(106,70){$\wilson(0)^{\oplus {N\choose 2}}$}
\put(168,76){\tiny $X$}
\put(160,72){\vector(1,0){30}}
\put(203,70){$\wilson(1)^{\oplus N}$}
\put(258,76){\tiny $X$}
\put(250,72){\vector(1,0){30}}
\put(287,70){$\wilson(2)$}

\put(122,52){$\oplus$}
\put(215,52){$\oplus$}
\put(77,55){\tiny $p$}
\put(70,47){\vector(2,1){30}}
\put(167,55){\tiny $p$}
\put(160,47){\vector(2,1){30}}
\put(257,55){\tiny $p$}
\put(250,47){\vector(2,1){30}}

\put(15,35){$\wilson(N)^{\oplus {N\choose 2}}$}
\put(78,41){\tiny $-X$}
\put(70,37){\vector(1,0){30}}
\put(100,35){$\wilson(N\!+\!1)^{\oplus N}$}
\put(168,41){\tiny $-X$}
\put(160,37){\vector(1,0){30}}
\put(197,35){$\wilson(N\!+\!2)$}

\put(295,35){$\oplus$}

\put(35,18){$\oplus$}
\put(122,18){$\oplus$}
\put(215,18){$\oplus$}
\put(86,20){\footnotesize id}
\put(70,28){\vector(2,-1){30}}
\put(176,20){\footnotesize id}
\put(160,28){\vector(2,-1){30}}
\put(266,20){\footnotesize id}
\put(250,28){\vector(2,-1){30}}

\put(-42,0){$\cdots$}
\put(-15,6){\tiny $X$}
\put(-25,2){\vector(1,0){30}}
\put(8,0){$\wilson(N\!-\!1)^{\oplus {N\choose 3}}$}
\put(80,6){\tiny $X$}
\put(70,2){\vector(1,0){30}}
\put(105,0){$\wilson(N)^{\oplus {N\choose 2}}$}
\put(170,6){\tiny $X$}
\put(160,2){\vector(1,0){30}}
\put(195,0){$\wilson(N\!+\!1)^{\oplus N}$}
\put(262,6){\tiny $X$}
\put(250,2){\vector(1,0){30}}
\put(285,0){$\underline{\wilson(N\!+\!2)}$}

\end{picture}
\end{center}
}
\noindent
The bottom line is the brane $\lsmB_{\bar 2}$ which is infra-red
empty in the phase $r\gg 0$. The above two lines 
form the brane $\lsmB''_{\bar 2}$, which is infra-red empty in
the phase $r\ll 0$. To see that this is indeed isomorphic to
$\lsmB'_{\bar 2}$ (even upstairs in $\mathfrak{D}(\C^{N+1},U(1))$),
eliminate first the leftmost brane-antibrane pair
 using the standard procedure (from (\ref{complexD}) to
(\ref{complexD'})). Then the remaining two brane-antibrane pairs
simply decouple, thus yielding $\lsmB'_{\bar 2}$.
It is clear from this example that
the attatched brane $\lsmB_{\bi}''$ for general $\bi\in\Z_N$
is the cone of
$\mathfrak{A}_i(N)\stackrel{p}{\to}\mathfrak{A}_i$, where
\beq
\mathfrak{A}_i\,:~~\wilson(0)^{\oplus {N\choose i}}
\stackrel{X}{\longrightarrow}
\wilson(1)^{\oplus {N\choose i-1}}
\stackrel{X}{\longrightarrow}\cdots
\stackrel{X}{\longrightarrow}
\wilson(i-1)^{\oplus N}\stackrel{X}{\longrightarrow}
\underline{\wilson(i)},
\label{defAAA}
\eeq
with $i\in \{0,1,...,N-1\}$.
In the low energy theory at $r\gg 0$,
this brane $\lsmB_{\bi}''$ becomes
the following complex supported at 
the exceptional divisor $E=\{P=0\}\cong \CP^{N-1}$:
$$
\pi_+(\lsmB''_{\bi})\cong~~
{\mathcal O}_E(0)^{\oplus {N\choose i}}
\stackrel{X}{\longrightarrow}
{\mathcal O}_E(1)^{\oplus {N\choose i-1}}
\stackrel{X}{\longrightarrow}\cdots
\stackrel{X}{\longrightarrow}
{\mathcal O}_E(i-1)^{\oplus N}\stackrel{X}{\longrightarrow}
\underline{{\mathcal O}_E(i)}
$$
This is the simplified version of the large volume image
of the fractional brane ${\mathcal O}_{\mathfrak{p}}(\bi)$.
Actually, one can further simplify it using the Euler sequence of 
$\CP^{N-1}$:
$$
0\longrightarrow{\mathcal O}\longrightarrow
{\mathcal O}(1)^{\oplus N}
\longrightarrow T_{\CP^{N-1}}\longrightarrow 0
$$
and its various dual versions, which show that
$$
\underline{{\mathcal O}^{\oplus {N\choose i}}}
\stackrel{X}{\longrightarrow}
{\mathcal O}(1)^{\oplus {N\choose i-1}}
\stackrel{X}{\longrightarrow}\cdots
\stackrel{X}{\longrightarrow}
{\mathcal O}(i-1)^{\oplus N}\stackrel{X}{\longrightarrow}
{\mathcal O}(i)
$$
is quasi-isomorphic to $\Omega^i(i)$
where $\Omega^i$ is the sheaf of holomorphic $i$-forms on $\CP^{N-1}$.
Therefore,
we find that the image brane $\pi_+(\lsmB''_{\bi})$
can also be written as
$\Omega^i_{E}(i)[i]$. 
To summarize, we find
\beq
\quad 
F^w_{-,+}:{\mathcal O}_{\mathfrak{p}}(\bi)\,
\longmapsto\, \Omega^i_{E}(i)[i],\qquad i=0,1,\ldots, N-1.
\label{McKay}
\eeq
This is the form that was conjectured 
to be the large volume image of the fractional brane
${\mathcal O}_p(\bi)$
in the literature 
\cite{EmJau,EmMk},
 based
on R-R charge, mirror symmetry and mathematical construction of the
equivalence
\cite{BKR,ItoNakajima}. We have finally 
proved that conjecture from the purely worldsheet
point of view. Actually, we have proved a precise version:
we now know that the above correspondence is with respect 
to the path through the particular window $w$. If we had chosen a
different homotopy
class of paths,
we would have a different correspondence.

\subsubsection*{(C) A two parameter model}

\begin{figure}[htb]
\centerline{\includegraphics{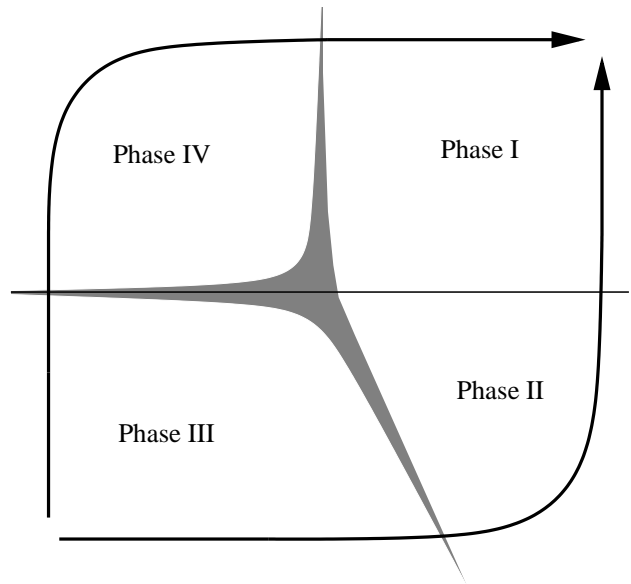}}
\caption{Two classes of route from Phase III to Phase I} 
\label{fig:routes1}
\end{figure}

As the next example of orbifold resolution,
we consider the two parameter model (C) which has four phases,
including an orbifold phase (Phase III)
and a large volume phase of a smooth Calabi-Yau manifold (Phase I).
We transport the branes from the orbifold phase to the large volume phase.
The two phases do not share a phase boundary, so that we consider
routes through Phase II or Phase IV, crossing
two phase boundaries
--- as depicted in Fig.~\ref{fig:routes1}.
In either case,
there is an infinite number of homotopy classes
coming from the choice of a window at each phase boundary.
The unbroken subgroups $T_u$ and the widths ${\mathscr S}$ 
at the phase boundaries are
$$
\begin{array}{ccccc}
&\mbox{I-II}&\mbox{II-III}&\mbox{III-IV}&\mbox{IV-I}
\\
T_u&\{(1,g)\}&\{(g^2,g)\}&\{(1,g)\}&\{(g,1)\}
\\
{\mathscr S}&2&8&2&4
\end{array}
$$
At the I-II boundary and the III-IV boundary,
we choose the window
$
w':-2\pi<\theta^2<0.
$
At the IV-I boundary we choose
$
w'':-4\pi<\theta^1<-2\pi.
$
At the II-III boundary, we consider two windows
\beqa
&&w_0\,:~-10\pi<2\theta^1+\theta^2<-8\pi,
\nn\\
&&
w_1\,:~-8\pi<2\theta^1+\theta^2<-6\pi.
\nn
\eeqa
For these windows we find the following band restriction rules
\beqa
\mathfrak{C}_{\rm I,II}^{w'}=
\mathfrak{C}_{\rm III,IV}^{w'}\!\!&=&\!\!\{q^2=0,1\},\nn\\
\mathfrak{C}_{\rm IV,I}^{w''}\!\!&=&\!\!\{q^1=0,1,2,3\},\nn\\
\mathfrak{C}_{\rm II,III}^{w_0}\!\!&=&\!\!\{2q^1+q^2=1,2,3,4,5,6,7,8\}\nn\\
\mathfrak{C}_{\rm II,III}^{w_1}\!\!&=&\!\!\{2q^1+q^2=0,1,2,3,4,5,6,7\}.\nn
\eeqa

As a first exercise, 
we take the $\Z_8$-equivariant line bundle
${\mathcal O}(\bar 0)$ in the orbifold phase.
It can be lifted to the Wilson-line brane $\wilson(a,b)$ 
where $a$ and $b$ are any pair of integers such that
$2a+b=0$ modulo 8. 
Let us consider transporting it to Phase III through the window
$w_0$. The band restriction rule dictates us to choose
$a,b$ such that $2a+b=8$. Other than that it is arbitrary.
Before transporting it to Phase I through $w'$, we should represent
such $\wilson(a,b)$ as a complex of Wilson line branes 
$\wilson(q^1,q^2)$ with $q^2=0,1$, using the
D-isomorphism relations in Phase II.
This can be done simply by requiring $b=0$ or $1$, without using any
non-trivial D-isomorphism. In view of the relation
$2a+b=8$ it must be that $b=0$. This uniquely fixes $(a,b)=(4,0)$.
In this way we obtain the transportation rule
$$
F^{w'}_{\rm II,I}\circ F^{w_0}_{\rm III,II}:{\mathcal O}(\bar 0)
\longmapsto {\mathcal O}(4,0).
$$
If instead we choose the window $w_1$ at the III-II boundary, we must
have $2a+b=0$ in the first step and therefore the transportation is
$$
F^{w'}_{\rm II,I}\circ F^{w_1}_{\rm III,II}:{\mathcal O}(\bar 0)
\longmapsto {\mathcal O}(0,0).
$$
Let us next consider the route through Phase IV.
To cross the III-IV boundary through window $w'$, we need to band restrict
to $b=0$ or $1$. In the present case where $2a+b=0$ mod $8$, $b$ must be $0$.
Before transporting the brane to Phase I through $w''$, 
we should represent it as a complex of Wilson line branes 
$\wilson(q^1,q^2)$ with $q^1=0,1,2,3$, using the
D-isomorphism relations in Phase IV. Again this is done
without any effort by simply setting $a=0$ in $\wilson(a,0)$. This 
gives
$$
F^{w''}_{\rm IV,I}\circ F^{w'}_{\rm III,IV}:{\mathcal O}(\bar 0)
\longmapsto {\mathcal O}(0,0).
$$
Notice that $F^{w'}_{\rm II,I}\circ F^{w_1}_{\rm III,II}$
and $F^{w''}_{\rm IV,I}\circ F^{w'}_{\rm III,IV}$
give the same result but
$F^{w'}_{\rm II,I}\circ F^{w_0}_{\rm III,II}$ 
gives a different one.
In fact, this holds for any brane, as can be seen as follows.

In the present model, we actually have a {\it grade} restriction rule.
We notice that the intersections of the relevant bands
are as follows:
\beqa
\mathfrak{C}^{w_0}_{\rm III,II}\cap \mathfrak{C}^{w'}_{\rm II,I}
\!\!&=&\!\!\{(0,1),(1,0),(1,1),(2,0),(2,1),(3,0),(3,1),(4,0)\},\nn\\
\mathfrak{C}^{w_1}_{\rm III,II}\cap \mathfrak{C}^{w'}_{\rm II,I}
=
\mathfrak{C}^{w'}_{\rm III,IV}\cap \mathfrak{C}^{w''}_{\rm IV,I}
\!\!&=&\!\!\{(0,0),(0,1),(1,0),(1,1),(2,0),(2,1),(3,0),(3,1)\}.\nn
\eeqa
\begin{figure}[htb]
\centerline{\includegraphics{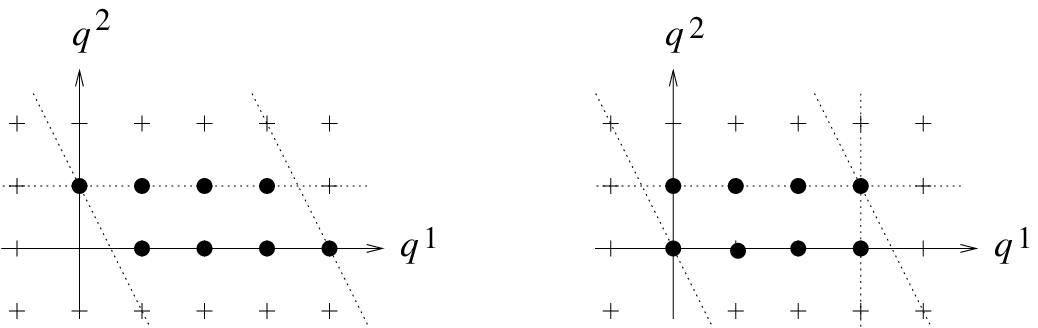}}
\caption{The grade restriction rule for the route
III$\stackrel{w_0}{\to}$II$\stackrel{w'}{\to}$I (Left),
III$\stackrel{w_1}{\to}$II$\stackrel{w'}{\to}$I (Right),
III$\stackrel{w'}{\to}$IV$\stackrel{w''}{\to}$I (Right)} 
\label{fig:intersection}
\end{figure}\\
The key point is that, in any phase, 
any brane can be written as a complex of Wilson line branes with
$(q^1,q^2)$ from either one of the two sets,
by using D-isomorphisms in that phase.
For example, in Phase III where the deleted set is
$
\Delta_{\rm III}=\{p=0\}\cup\{x_6=0\},
$
one can use the D-isomorphisms
$$
\wilson(q^1,q^2)\stackrel{p}{\longrightarrow}\wilson(q^1-4,q^2),\qquad
\wilson(q^1,q^2)\stackrel{x_6}{\longrightarrow}\wilson(q^1+1,q^2-2),
$$
to bring any $(q^1,q^2)$ into the finite set.
Once that is done,
the brane can be transported safely along
the respective route.
Since the finite sets of charges $\{(q^1,q^2)\}$ are equal for the routes
III$\stackrel{w_1}{\to}$II$\stackrel{w'}{\to}$I and 
III$\stackrel{w'}{\to}$IV$\stackrel{w''}{\to}$I, the transports
result in the same brane. The set is different for
the route III$\stackrel{w_0}{\to}$II$\stackrel{w'}{\to}$I 
and hence the map of branes is different.

\begin{figure}[htb]
\centerline{\includegraphics{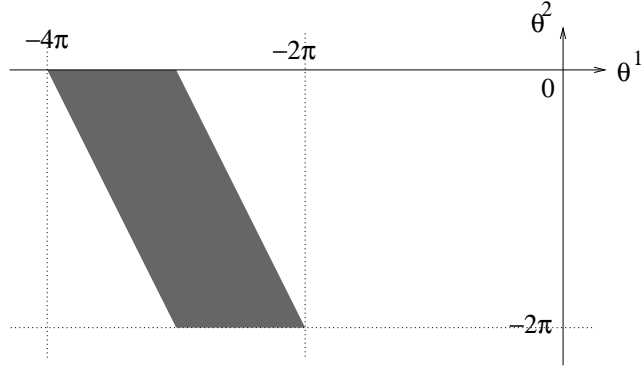}}
\caption{Overlap of the windows $w_1$, $w'$ and $w''$}
\label{fig:overlap}
\end{figure}

The fact that the two routes,
 III$\stackrel{w_1}{\to}$II$\stackrel{w'}{\to}$I and 
III$\stackrel{w'}{\to}$IV$\stackrel{w''}{\to}$I,
give rise to the same map of branes may imply that these routes are homopotic
to each other. Indeed they are!
To see this let us look at the windows $w_1$, $w'$ and $w''$
for these routes.
Fig.~\ref{fig:overlap} shows the overlap of these windows.
Comparing with Fig.~\ref{alamtwo}, we find that the overlap
precisely matches with the complement of the Alga of
the singular locus $\mathfrak{S}$.
Note that any point of the alga complement 
determines a global section of
the entire plane $\R^2_{\rm FI}$ of FI parameters
$(r^1,r^2)$ to the K\"ahler moduli space
$\moduli_K=(\C^{\times})^2\setminus\mathfrak{S}$.
Such a section defines a homotopy of the two routes under consideration.
On the other hand, the overlap of
the windows $w_0$, $w'$ and $w''$ is contained inside the Alga of
$\mathfrak{S}$. This means that the attempted homotopy of the routes
III$\stackrel{w_0}{\to}$II$\stackrel{w'}{\to}$I and
III$\stackrel{w'}{\to}$IV$\stackrel{w''}{\to}$I 
will intersect with the singular locus $\mathfrak{S}$
and cannot really define a homotopy.

These observations have some significance on D-brane transport along paths 
through the central region of the moduli space $\moduli_K$,
as will be discussed in Section~\ref{subsec:center}.

\subsubsection*{\it Transport Of The Fractional Branes}

Let us transport the fractional branes ${\mathcal O}_{\mathfrak{p}}(\bi)$
of the orbifold $\C^5/\Z_8$ along the route 
III$\stackrel{w_1}{\to}$II$\stackrel{w'}{\to}$I.
As remarked above, we can use the grade restriction rule
$q^1=0,1,2,3$, $q^2=0,1$ for this path.
Recall that
 the fractional brane ${\mathcal O}_{\mathfrak{p}}(\bar 0)$ is given by
$$
{\mathcal O}(\bar 0)\stackrel{X}{\longrightarrow}
{\mathscr R}\stackrel{X}{\longrightarrow}
\wedge^2{\mathscr R}\stackrel{X}{\longrightarrow}
\wedge^3{\mathscr R}\stackrel{X}{\longrightarrow}
\wedge^4{\mathscr R}\stackrel{X}{\longrightarrow}
{\mathcal O}(\bar 0)
$$
where ${\mathscr R}={\mathcal O}(\bar 1)^{\oplus 2}\oplus 
{\mathcal O}(\bar 2)^{\oplus 3}$ is the defining representation.
The other ones ${\mathcal O}_{\mathfrak{p}}(\bi)$ are the tensor product
of this complex with ${\mathcal O}(\bi)$.
As a lift of ${\mathcal O}_{\mathfrak{p}}(\bar 0)$ 
to the linear sigma model, we may take
$\lsmB_{\bar 0}$ given by
{\small
\begin{center}
\begin{picture}(300,65)(17,-5)

\put(-35,30){$\wilson(0,0)$}

\put(9,45){\tiny $X$}
\put(14,27){\tiny $Y$}
\put(0,37){\vector(3,1){29}}
\put(0,29){\vector(3,-1){29}}

\put(35,45){$\wilson(0,1)^{\oplus 2}$}
\put(55,30){$\oplus$}
\put(35,15){$\wilson(1,0)^{\oplus 3}$}

\put(88,57){\tiny $X$}
\put(96,43){\tiny $Y$}
\put(88,27){\tiny $X$}
\put(96,10){\tiny $Y$}
\put(82,50){\vector(3,1){25}}
\put(82,45){\vector(3,-1){25}}
\put(82,20){\vector(3,1){25}}
\put(82,12){\vector(3,-1){25}}

\put(110,60){$\wilson(0,2)$}
\put(130,45){$\oplus$}
\put(110,30){$\wilson(1,1)^{\oplus 6}$}
\put(130,15){$\oplus$}
\put(110,0){$\wilson(2,0)^{\oplus 3}$}

\put(171,58){\tiny $Y$}
\put(164,42){\tiny $X$}
\put(171,28){\tiny $Y$}
\put(164,12){\tiny $X$}
\put(171,-4){\tiny $Y$}
\put(157,60){\vector(3,-1){25}}
\put(157,35){\vector(3,1){25}}
\put(157,30){\vector(3,-1){25}}
\put(157,5){\vector(3,1){25}}
\put(157,-2){\vector(3,-1){25}}

\put(185,45){$\wilson(1,2)^{\oplus 3}$}
\put(205,30){$\oplus$}
\put(185,15){$\wilson(2,1)^{\oplus 6}$}
\put(205,0){$\oplus$}
\put(185,-15){$\wilson(3,0)$}

\put(246,43){\tiny $Y$}
\put(239,27){\tiny $X$}
\put(246,13){\tiny $Y$}
\put(239,-3){\tiny $X$}
\put(232,45){\vector(3,-1){25}}
\put(232,20){\vector(3,1){25}}
\put(232,15){\vector(3,-1){25}}
\put(232,-10){\vector(3,1){25}}

\put(260,30){$\wilson(2,2)^{\oplus 3}$}
\put(280,15){$\oplus$}
\put(260,0){$\wilson(3,1)^{\oplus 2}$}

\put(321,28){\tiny $Y$}
\put(314,12){\tiny $X$}
\put(307,30){\vector(3,-1){25}}
\put(307,5){\vector(3,1){25}}

\put(336,15){$\wilson(3,2)$}

\end{picture}
\end{center}
}
\noindent
where we use ``$X$'' for $x_1, x_2$ and ``$Y$'' for $x_3,x_4,x_5$.
The lifts of
other branes ${\mathcal O}_{\mathfrak{p}}(\bi)$ 
can be obtained by tensoring with
$\wilson(n,0)$ if $i=2n$ and with $\wilson(n,1)$ if $i=2n+1$,
with $n=0,1,2,3$.
We denote such lifts by $\lsmB_{\bi}$.

Before transporting it to Phase I, we have to grade restrict the branes.
That can be done simply by multiplying
appropriate powers of $p$ and $x_6$ to the arrows. 
We denote this grade restricted version of $\lsmB_{\bi}$ by $\lsmB_{\bi}'$.
For example, for the brane $\lsmB_{\bar 0}$ we replace the upper-right entries
$\wilson(0,2)$, $\wilson(1,2)$, $\wilson(2,2)$ by
$\wilson(1,0)$, $\wilson(2,0)$, $\wilson(3,0)$ and replace the $X$'s
that go into them by $x_6X$. Also, we replace the rightmost entry
$\wilson(3,2)$ by $\wilson(0,0)$ and substitute the arrows $Y$, $X$ 
going into it
by $pY$, $px_6X$.
Thus, 
$\lsmB_{\bar 0}'$ is given by
{\small
\begin{center}
\begin{picture}(300,65)(17,-5)

\put(-35,30){$\wilson(0,0)$}

\put(9,45){\tiny $X$}
\put(14,27){\tiny $Y$}
\put(0,37){\vector(3,1){29}}
\put(0,29){\vector(3,-1){29}}

\put(35,45){$\wilson(0,1)^{\oplus 2}$}
\put(55,30){$\oplus$}
\put(35,15){$\wilson(1,0)^{\oplus 3}$}

\put(86,57){\tiny $x_6X$}
\put(96,43){\tiny $Y$}
\put(88,27){\tiny $X$}
\put(96,10){\tiny $Y$}
\put(82,50){\vector(3,1){25}}
\put(82,45){\vector(3,-1){25}}
\put(82,20){\vector(3,1){25}}
\put(82,12){\vector(3,-1){25}}

\put(110,60){$\wilson(1,0)$}
\put(130,45){$\oplus$}
\put(110,30){$\wilson(1,1)^{\oplus 6}$}
\put(130,15){$\oplus$}
\put(110,0){$\wilson(2,0)^{\oplus 3}$}

\put(171,58){\tiny $Y$}
\put(162,42){\tiny $x_6X$}
\put(171,28){\tiny $Y$}
\put(164,12){\tiny $X$}
\put(171,-4){\tiny $Y$}
\put(157,60){\vector(3,-1){25}}
\put(157,35){\vector(3,1){25}}
\put(157,30){\vector(3,-1){25}}
\put(157,5){\vector(3,1){25}}
\put(157,-2){\vector(3,-1){25}}

\put(185,45){$\wilson(2,0)^{\oplus 3}$}
\put(205,30){$\oplus$}
\put(185,15){$\wilson(2,1)^{\oplus 6}$}
\put(205,0){$\oplus$}
\put(185,-15){$\wilson(3,0)$}

\put(246,43){\tiny $Y$}
\put(237,27){\tiny $x_6X$}
\put(246,13){\tiny $Y$}
\put(239,-3){\tiny $X$}
\put(232,45){\vector(3,-1){25}}
\put(232,20){\vector(3,1){25}}
\put(232,15){\vector(3,-1){25}}
\put(232,-10){\vector(3,1){25}}

\put(260,30){$\wilson(3,0)^{\oplus 3}$}
\put(280,15){$\oplus$}
\put(260,0){$\wilson(3,1)^{\oplus 2}$}

\put(321,28){\tiny $pY$}
\put(308,12){\tiny $px_6X$}
\put(307,30){\vector(3,-1){25}}
\put(307,5){\vector(3,1){25}}

\put(336,15){$\wilson(0,0)$}

\end{picture}
\label{page:B0}
\end{center}
}
\noindent
Once that is done and the D-brane is 
transported into Phase I, we simply go to the
low energy theory, replacing $\wilson(q^1,q^2)$
by ${\mathcal O}(q^1,q^2)$. That's all.

However, just as in Example (A), we may simplify the image
D-branes in the large volume phase. 
The strategy is the same:
We present
the grade restricted lift $\lsmB_{\bi}'$
as a bound state of $\lsmB_{\bi}$ and another brane
$\lsmB_{\bi}''$ that is infra-red empty in Phase III. Once transported
into Phase I, then, this time the part $\lsmB_{\bi}$ is infra-red empty
and we are left with $\pi_{\rm I}(\lsmB_{\bi}'')$.
For example, $\lsmB_{\bar 0}'$ described in the previous paragraph
can be presented as the bound state of $\lsmB_{\bar 0}$ and
{\small
\begin{center}
\begin{picture}(300,60)(17,5)

\put(-30,30){$\lsmB_{\bar 0}''\,:$}

\put(39,45){$\wilson(0,2)$}

\put(88,57){\tiny $X_6$}
\put(96,43){\tiny $Y$}
\put(82,50){\vector(3,1){25}}
\put(82,45){\vector(3,-1){25}}

\put(110,60){$\wilson(1,0)$}
\put(130,45){$\oplus$}
\put(110,30){$\wilson(1,2)^{\oplus 3}$}

\put(171,58){\tiny $Y$}
\put(164,42){\tiny $X_6$}
\put(171,28){\tiny $Y$}
\put(157,60){\vector(3,-1){25}}
\put(157,35){\vector(3,1){25}}
\put(157,30){\vector(3,-1){25}}

\put(185,45){$\wilson(2,0)^{\oplus 3}$}
\put(205,30){$\oplus$}
\put(185,15){$\wilson(2,2)^{\oplus 3}$}

\put(246,43){\tiny $Y$}
\put(239,27){\tiny $X_6$}
\put(246,13){\tiny $Y$}
\put(232,45){\vector(3,-1){25}}
\put(232,20){\vector(3,1){25}}
\put(232,15){\vector(3,-1){25}}

\put(265,60){$\wilson(4,0)$}
\put(280,45){$\oplus$}
\put(260,30){$\wilson(3,0)^{\oplus 3}$}
\put(280,15){$\oplus$}
\put(264,0){$\wilson(3,2)$}

\put(317,64){\tiny $p$}
\put(320,45){\footnotesize id}
\put(307,62){\vector(1,0){25}}
\put(307,57){\vector(1,-1){30}}

\put(315,30){\tiny $Y$}
\put(314,12){\tiny $X_6$}
\put(307,30){\vector(3,-1){25}}
\put(307,5){\vector(3,1){25}}

\put(336,60){$\wilson(0,0)$}
\put(350,38){$\oplus$}
\put(336,15){$\wilson(4,0)$}

\end{picture}
\end{center}
}
\noindent
by the map $\lsmB_{\bar 0}''\to\lsmB_{\bar 0}$ that sends the bottom line
of $\lsmB_{\bar 0}''$ with entries
$\wilson(*,2)$ to the upper-right line of $\lsmB_{\bar 0}$
by the identity maps.
Note that $\lsmB_{\bar 0}''$ is infra-red empty in Phase III since
it consists of $x_6$-Koszul and $p$-Koszul complexes.
This is why we can replace $\lsmB_{\bar 0}'$ with the above bound state
(which we denote by $\widetilde{\lsmB}_{\bar 0}'$).
On the other hand,
$\lsmB_{\bar 0}$ is infra-red empty in Phase I with
$\Delta_I=\{X=0\}\cup\{Y=x_6=0\}$ since it consists of $X$-Koszul complexes.
Thus we find
$\pi_I(\widetilde{\lsmB}_{\bar 0}')\cong
\pi_I(\lsmB_{\bar 0}'').$ Note that
$\lsmB_{\bar 0}''$ also includes a $(Y,x_6)$-Koszul complex as a part.
Dropping them, we are left with
$$
{\mathcal O}(4,0)\stackrel{p}{\longrightarrow}{\mathcal O}(0,0).
$$
This is quasi-isomorphic to the sheaf ${\mathcal O}_E(0,0)$ supported
at the exceptional divisor $E=\{p=0\}$.
In this way we find that the large volume image of
the fractional brane ${\mathcal O}_{\mathfrak{p}}(\bar 0)$
is the D-brane wrapped on the divisor
$E$ with a trivial line bundle.
Repeating the same procedure we find simple representatives
of the large volume images of all the 
fractional branes ${\mathcal O}_{\mathfrak{p}}(\bi)$:
$$
{\rm Cone}\left(
{\mathscr A}_i(4,0)\stackrel{p}{\longrightarrow}{\mathscr A}_i(0,0)\right)
$$
where ${\mathscr A}_i$ are the complexes of vector bundles 
given below. 
{\small
\beqa
{\mathscr A}_0\,=\,
{\mathcal O}(0,0),&&
\nn\\[0.1cm]
{\mathscr A}_1\,=\,
{\mathcal O}(0,0)^{\oplus 2}\stackrel{\mbox{\tiny $X$}}{\longrightarrow}
\underline{{\mathcal O}(0,1)},&&
\nn\\[0.1cm]
{\mathscr A}_2\,=\,
{\mathcal O}(0,0)\stackrel{{0\choose X}}{\longrightarrow}
\begin{array}{c}
{\mathcal O}(0,0)^{\oplus 3}\\[-0.2cm]
\oplus\\[-0.2cm]
{\mathcal O}(0,1)^{\oplus 2}
\end{array}
\stackrel{\mbox{\tiny $(Y,x_6X)$}}{\longrightarrow}
\underline{{\mathcal O}(1,0)},&&
\nn\\[0.1cm]
{\mathscr A}_3\,=\,
\begin{array}{c}
{\mathcal O}(0,0)^{\oplus 6}\\[-0.2cm]
\oplus\\[-0.2cm]
{\mathcal O}(0,1)
\end{array}\stackrel{\left(\mbox{\tiny $\begin{array}{cc}
\!\!\! X\!\!\! &\!\!\! 0\!\!\!\\
\!\!\! Y\!\!\! &\!\!\! x_6X\!\!\!
\end{array}
$}\right)}{\longrightarrow}
\begin{array}{c}
{\mathcal O}(0,1)^{\oplus 3}\\[-0.2cm]
\oplus\\[-0.2cm]
{\mathcal O}(1,0)^{\oplus 2}
\end{array}\stackrel{\mbox{\tiny $(Y,X)$}}{\longrightarrow}
\underline{{\mathcal O}(1,1)},&&
\nn\\[0.1cm]
{\mathscr A}_4\,=\,
{\mathcal O}(0,0)^{\oplus 3}
\stackrel{\left(\mbox{\tiny $
\begin{array}{c}
\!\!\! 0\!\!\!\\
\!\!\! X\!\!\!\\
\!\!\! Y\!\!\!
\end{array}
$}\right)}{\longrightarrow}
\begin{array}{c}
{\mathcal O}(0,0)^{\oplus 3}\\[-0.2cm]
\oplus\\[-0.2cm]
{\mathcal O}(0,1)^{\oplus 6}\\[-0.2cm]
\oplus\\[-0.2cm]
{\mathcal O}(1,0)
\end{array}
\stackrel{\left(\mbox{\tiny $
\begin{array}{ccc}
\!\!\! Y\!\!\!&\!\!\! x_6X\!\!\!&\!\!\! 0\!\!\!\\
\!\!\! 0\!\!\!&\!\!\! Y\!\!\!&\!\!\! X\!\!\!
\end{array}
$}\right)}{\longrightarrow}
\begin{array}{c}
{\mathcal O}(1,0)^{\oplus 3}\\[-0.2cm]
\oplus\\[-0.2cm]
{\mathcal O}(1,1)^{\oplus 2}
\end{array}
\stackrel{\mbox{\tiny $(Y,x_6X)$}}{\longrightarrow}
\underline{{\mathcal O}(2,0)},&&
\nn\\[-0.2cm]
& \\[-0.2cm]
{\mathscr A}_5\,=\,
\begin{array}{c}
{\mathcal O}(0,0)^{\oplus 6}\\[-0.2cm]
\oplus\\[-0.2cm]
{\mathcal O}(0,1)^{\oplus 3}
\end{array}
\stackrel{\left(\mbox{\tiny $
\begin{array}{cc}
\!\!\! X\!\!\!&\!\!\! 0\!\!\!\\
\!\!\! Y\!\!\!&\!\!\! x_6X\!\!\!\\
\!\!\! 0\!\!\!&\!\!\! Y\!\!\!
\end{array}
$}\right)}{\longrightarrow}
\begin{array}{c}
{\mathcal O}(0,1)^{\oplus 3}\\[-0.2cm]
\oplus\\[-0.2cm]
{\mathcal O}(1,0)^{\oplus 6}\\[-0.2cm]
\oplus\\[-0.2cm]
{\mathcal O}(1,1)
\end{array}
\stackrel{\left(\mbox{\tiny $
\begin{array}{ccc}
\!\!\! Y\!\!\!&\!\!\! X\!\!\!&\!\!\! 0\!\!\!\\
\!\!\! 0\!\!\!&\!\!\! Y\!\!\!&\!\!\! x_6X\!\!\!
\end{array}
$}\right)}{\longrightarrow}
\begin{array}{c}
{\mathcal O}(1,1)^{\oplus 3}\\[-0.2cm]
\oplus\\[-0.2cm]
{\mathcal O}(2,0)^{\oplus 2}
\end{array}
\stackrel{\mbox{\tiny $(Y,X)$}}{\longrightarrow}
\underline{{\mathcal O}(2,1)},&&
\nn\\[0.1cm]
{\mathscr A}_6\,=\,
{\mathcal O}(0,0)^{\oplus 3}
\stackrel{\left(\mbox{\tiny $
\begin{array}{c}
\!\!\! 0\!\!\!\\
\!\!\! X\!\!\!\\
\!\!\! Y\!\!\!
\end{array}
$}\right)}{\longrightarrow}\!\!\!
\begin{array}{c}
{\mathcal O}(0,0)\\[-0.2cm]
\oplus\\[-0.2cm]
{\mathcal O}(0,1)^{\oplus 6}\\[-0.2cm]
\oplus\\[-0.2cm]
{\mathcal O}(1,0)^{\oplus 3}
\end{array}\!\!\!
\stackrel{\left(\mbox{\tiny $
\begin{array}{ccc}
\!\!\! Y\!\!\!&\!\!\! x_6X\!\!\!&\!\!\! 0\!\!\!\\
\!\!\! 0\!\!\!&\!\!\! Y\!\!\!&\!\!\! X\!\!\!\\
\!\!\! 0\!\!\!&\!\!\! 0\!\!\!&\!\!\! Y\!\!\!
\end{array}
$}\right)}{\longrightarrow}\!\!\!
\begin{array}{c}
{\mathcal O}(1,0)^{\oplus 3}\\[-0.2cm]
\oplus\\[-0.2cm]
{\mathcal O}(1,1)^{\oplus 6}\\[-0.2cm]
\oplus\\[-0.2cm]
{\mathcal O}(2,0)
\end{array}\!\!\!
\stackrel{\left(\mbox{\tiny $
\begin{array}{ccc}
\!\!\! Y\!\!\!&\!\!\! x_6X\!\!\!&\!\!\! 0\!\!\!\\
\!\!\! 0\!\!\!&\!\!\! Y\!\!\!&\!\!\! X\!\!\!
\end{array}
$}\right)}{\longrightarrow}\!\!\!
\begin{array}{c}
{\mathcal O}(2,0)^{\oplus 3}\\[-0.2cm]
\oplus\\[-0.2cm]
{\mathcal O}(2,1)^{\oplus 2}
\end{array}\!\!\!\!
\stackrel{\mbox{\tiny $(Y,x_6X)$}}{\longrightarrow}\!\!
\underline{{\mathcal O}(3,0)},&&
\nn\\[0.1cm]
{\mathscr A}_7\,=\,
\begin{array}{c}
{\mathcal O}(0,0)^{\oplus 2}\\[-0.2cm]
\oplus\\[-0.2cm]
{\mathcal O}(0,1)^{\oplus 3}
\end{array}\!\!\!
\stackrel{\left(\mbox{\tiny $
\begin{array}{cc}
\!\!\! X\!\!\!&\!\!\! 0\!\!\!\\
\!\!\! Y\!\!\!&\!\!\! x_6X\!\!\!\\
\!\!\! 0\!\!\!&\!\!\! Y\!\!\!
\end{array}
$}\right)}{\longrightarrow}\!\!\!
\begin{array}{c}
{\mathcal O}(0,1)\\[-0.2cm]
\oplus\\[-0.2cm]
{\mathcal O}(1,0)^{\oplus 6}\\[-0.2cm]
\oplus\\[-0.2cm]
{\mathcal O}(1,1)^{\oplus 3}
\end{array}\!\!\!
\stackrel{\left(\mbox{\tiny $
\begin{array}{ccc}
\!\!\! Y\!\!\!&\!\!\! X\!\!\!&\!\!\! 0\!\!\!\\
\!\!\! 0\!\!\!&\!\!\! Y\!\!\!&\!\!\! X\!\!\!\\
\!\!\! 0\!\!\!&\!\!\! 0\!\!\!&\!\!\! Y\!\!\!
\end{array}
$}\right)}{\longrightarrow}\!\!\!
\begin{array}{c}
{\mathcal O}(1,1)^{\oplus 3}\\[-0.2cm]
\oplus\\[-0.2cm]
{\mathcal O}(2,0)^{\oplus 6}\\[-0.2cm]
\oplus\\[-0.2cm]
{\mathcal O}(2,1)
\end{array}\!\!\!
\stackrel{\left(\mbox{\tiny $
\begin{array}{ccc}
\!\!\! Y\!\!\!&\!\!\! X\!\!\!&\!\!\! 0\!\!\!\\
\!\!\! 0\!\!\!&\!\!\! Y\!\!\!&\!\!\! x_6X\!\!\!
\end{array}
$}\right)}{\longrightarrow}\!\!\!
\begin{array}{c}
{\mathcal O}(2,1)^{\oplus 3}\\[-0.2cm]
\oplus\\[-0.2cm]
{\mathcal O}(3,0)^{\oplus 2}
\end{array}\!\!\!
\stackrel{\mbox{\tiny $(Y,X)$}}{\longrightarrow}\!\!
\underline{{\mathcal O}(3,1)}.&&
\nn
\label{LVfrac}
\eeqa
}
In particular, the images are complexes of sheaves
$({\mathscr A}_i)_E$ supported at the divisor $E$
(the pushforward of the complex ${\mathscr A}_i|_E$ over
$E$ by the embedding map $E\hookrightarrow X$).

The complexes ${\mathscr A}_i$ for high values of $i$ can be 
simplified using the $X$- and the $(Y,x_6)$-Koszul complexes.
In particular, there is a duality relation
$$
{\mathscr A}_i\cong {\mathscr A}_{7-i}^T(-1,1)[4],
$$
where ${\mathscr A}^T$ is the transpose of the complex ${\mathscr A}$
in which the dual of the R-degree zero component is defined to have 
R-degree zero.
The Chern characters of ${\mathscr A}_i|_E$ are
\beqa
&&{\rm ch}({\mathscr A}_0)|_E=1,\nn\\
&&{\rm ch}({\mathscr A}_1)|_E=-1+L,\nn\\
&&{\rm ch}({\mathscr A}_2)|_E=-3+H-2L+{1\over 2}H^2+{1\over 6}H^3
+{1\over 12}v_E,\nn\\
&&{\rm ch}({\mathscr A}_3)|_E=3-H-2L-{1\over 2}H^2+HL-{1\over 6}H^3
+{1\over 2}H^2L+{1\over 12}v_E,\nn\\
&&{\rm ch}({\mathscr A}_4)|_E
=\left(3+H+2L-{1\over 2}H^2+HL+{1\over 6}H^3
-{1\over 2}H^2L+{1\over 12}v_E\right)\e^{-H+L},\nn\\
&&{\rm ch}({\mathscr A}_5)|_E
=\left(-3-H+2L+{1\over 2}H^2-{1\over 6}H^3
+{1\over 12}v_E\right)\e^{-H+L},\nn\\
&&{\rm ch}({\mathscr A}_6)|_E
=\left(-1-L\right)\e^{-H+L},\nn\\
&&{\rm ch}({\mathscr A}_7)|_E=\e^{-H+L}.\nn
\\\label{ChfracB}
\eeqa
Here $H$ and $L$ are the first Chern classes of ${\mathcal O}(1,0)$
and ${\mathcal O}(0,1)$ and $v_E$ is a volume form of the toric variety
$E$. They obey homology relations
$L^2=0$, $H^4=2H^3L=2v_E$, and $\int_Ev_E=1$. 

\subsubsection*{(D) Resolutions Of $A_{N-1}$ Singularity}

As the final example, we consider resolutions of the $A_{N-1}$
singularity, which provide historically the first
example of McKay correspondence.
We are particularly interested in connecting the orbifold phase
$(11\cdots 1)$
and the fully resolved phase 
$(00\cdots 0)$ via some other phases corresponding to
partial resolutions.
See Section~\ref{subsec:LSMexamples} for the labelling of phases
and remarks on the phase boundaries.
We consider a route that goes through the particular
sequence of phases: 
$$
(00\cdots 000)\longleftrightarrow
(00\cdots 001)\longleftrightarrow
(00\cdots 011)\longleftrightarrow
\cdots
\longleftrightarrow
(01\cdots 11)\longleftrightarrow
(11\cdots 11)
$$
We see that there are $N$ phases and $(N-1)$ phase boundaries.
The first phase boundary corresponds to blowing down (up)
the right most divisor $C_{N-1}=\{x_2=0\}$. Applying the general rule, we see
that the phase boundary in $\R^{N-1}_{\rm FI}$
is a domain in the hyperplane spanned by the charge
vectors of all $x_j$'s but $x_1,x_2,x_3$. The unbroken
subgroup is therefore $T^u_1=U(1)_1$.
At the $i$-th phase boundary,
change occurs at $C_{N-i}=\{x_{i+1}=0\}$ and the hyperplane is spanned
by all $x_j$'s but $x_1,x_{i+1}, x_{i+2}$. The unbroken 
subgroup is 
$$
T^u_i=\{(g,g^2,g^3,\ldots,g^i,1,1,\ldots,1)\,|\,g\in U(1)\,\},
$$
which has width ${\mathscr S}_i=(i+1)$.
Thus, the band restriction rule at the $i$-th phase boudary is
$$
-{i+1\over 2}<{\theta^1+2\theta^2+\cdots +i\theta^i\over 2\pi}
+q^1+2q^2+\cdots +iq^i<{i+1\over 2}.
$$
There is a non-empty region of the space of theta parameters such
that the following set of charges
are band restricted at all the
$(N-1)$ phase boundaries.
\beq
\mathfrak{C}=\{\,0,\,e_1,\,e_2,\,\ldots,\,e_{N-1}\,\}
\label{Atypeset}
\eeq
($e_i$ is the charge vector $q$ where all $q^j=0$ but $q^i=1$.)
For example check the value
$\theta^i=-\pi/i$ ($i=1,...,(N-1)$).
At each of the $N$ phases on the route, any brane 
is D-isomorphic to a complex of Wilson line branes with charges 
in this finite set $\mathfrak{C}$.
For example, at the orbifold phase, all the equivariant line bundles
${\mathcal O}(\bi)$ can be realized as the low energy images
of $\wilson(e_i)$, where it is understood that $e_0=0$.
Thus, we can use this set to define the grade restriction rule
for the transport of D-branes along this route, say
from the orbifold phase to the fully resolved phase.

Let us see the large volume images of the fractional branes
${\mathcal O}_{\mathfrak p}(\bi)$:
$$
{\mathcal O}(\bi)
\stackrel{{x_1\choose x_{N+1}}}{-\!\!\!-\!\!\!-\!\!\!\longrightarrow}
\begin{array}{c}
{\mathcal O}(\overline{i+1})\\
\oplus\\
{\mathcal O}(\overline{i-1})
\end{array}
\stackrel{(-x_{N+1},x_1)}{-\!\!\!-\!\!\!-\!\!\!\longrightarrow} 
{\mathcal O}(\bi)
$$
Its lift such that all the charges are from
the set (\ref{Atypeset}) is
$$
\wilson(e_i)
\stackrel{{\,a\,\choose b}}{-\!\!\!-\!\!\!-\!\!\!\longrightarrow}
\begin{array}{c}
\wilson(e_{i+1})\\
\oplus\\
\wilson(e_{i-1})
\end{array}
\stackrel{(-c,d)}{-\!\!\!-\!\!\!-\!\!\!\longrightarrow} 
\wilson(e_i),
$$
where
\beqa
&
a=x_1x_2\cdots x_{i+1},& c=x_{i+2}\cdots x_{N+1},\nn\\
&
b=x_{i+1}\cdots x_{N+1},& d=x_1x_2\cdots x_i.
\nn
\eeqa
For the $i=0$ case, we set
$e_{-1}:=e_{N-1}$ so that
$b=x_{N+1}$ and $d=x_1x_2\cdots x_N$
while $a=x_1$ and $c=x_2\cdots x_{N+1}$ remains valid.
The large volume image is obtained simply by replacing
$\wilson(q)$ by ${\mathcal O}(q)$.
As before, we can simplify this image by taking out and eliminating
infra-red empty complexes. 
For the $i=0$ case, we take out and eliminate a Koszul complex
for a pair of variables $(x_1,x_{N+1})$ which cannot vanish
in the large volume regime.
This shows that the large volume image of
the fractional brane ${\mathcal O}_{\mathfrak{p}}$ 
for the trivial representation is
\beq
{\mathcal O}(e_1+e_{N-1})
\stackrel{x_2x_3\cdots x_N}{-\!\!\!-\!\!\!-\!\!\!\longrightarrow} 
\underline{{\mathcal O}(0)}.
\eeq
This is the structure sheaf of the entire exceptional divisor
$E=C_1+\cdots+C_{N-1}$.
For $i>0$, we take out three Koszul complexes associated with
three pairs of variables,
$(x_1\cdots x_i, x_{i+3}\cdots x_{N+1})$,
$(x_1\cdots x_i, x_{i+2})$ and
$(x_{i+1},x_{i+3}\cdots x_{N+1})$, which cannot vanish
in the large volume phase 
since the sets $\{x_j=x_l=0\}$ with $|j-l|>2$ are deleted.
In this way, we find the following large volume image of the 
fractional brane ${\mathcal O}_{\mathfrak{p}}(\bi)$
for the $i$-th representation 
\beq
{\mathcal O}(e_i+e_{i+1}-e_{i+2})
\stackrel{x_{i+1}}{-\!\!\!\longrightarrow} 
{\mathcal O}(e_{i-1}-e_i+2e_{i+1}-e_{i+2})
\to \underline{0}.
\eeq
This is a sheaf supported at the component $C_{N-i}=\{x_{i+1}=0\}$
of the exceptional divisor, shifted by $1$ to the left.
Thus, we very explicitly recontructed 
the original McKay correspondence --- one-to-one correspondence between
the non-trivial irreducible representations of $\Z_N$ and
the irreducible components of the exceptional divisor
of the resolution of the $A_{N-1}$ singularity.

\subsection{Center Of Multiparameter Moduli Space}
\label{subsec:center}

In models with higher rank gauge groups, $k>1$,
we have so far discussed D-brane transport across phase boundaries 
only in the asymptotic regions where all but one $U(1)$ subgroup 
is completely broken. However,
it is of course an important problem to construct a parallel family 
of boundary interactions over the central region of the moduli space 
where multiple phase boundaries meet.
Although we do not attempt to find a general solution
in this paper, we have something to say about this problem.

We made an interesting observation in the two parameter model (C):
There is a finite set of Wilson line branes which obeys the band 
restriction rule at all the phase boundaries, with the following
properties;
\begin{itemize}
\item[(i)]
In each phase
the set is maximal generating, that is,
there is no low energy D-isomorphism relation 
among the Wilson line branes in the set
and any brane is D-isomorphic to a complex of sums of them.
\item[(ii)]
The overlap $\mathfrak{Z}$ of windows
for which the set obeys the band restriction rule
precisely matches with a copy of the complement of the 
 Alga of the singular locus $\mathfrak{S}$, that is, the
values of the theta parameters that are missed by $\mathfrak{S}$.
\end{itemize}
The property (ii) means that
the region $\R^k_{\rm FI}\times \mathfrak{Z}$ in the FI-theta parameter space
does not meet the singular locus.
Due to the flatness of our connection, 
the outcome of D-brane transport from one phase to another,
along any path inside this region, is the same as
the result of transport along paths that stay in the asymptotic region.
This tempts us to ask: do the Wilson line branes in this set
define smooth families of boundary interactions over
the entire region $\R^k_{\rm FI}\times \mathfrak{Z}$?
Can we use it to construct the parallel family of an arbitrary brane
over this central region?
In fact, existence of such a finite set
of Wilson lines is not limited to the example (C) but holds in
many multiparameter models.
Thus, let us discuss this question in a general context
of $U(1)^k$ gauge theory with matter fields $\Phi_i$
with charge $Q^a_i$ ($a=1,...,k$, $i=1,...,N$).

Let us look at the effective boundary potential on the Coulomb branch.
Introducing
$M_i(\sigma):=\sum_{a=1}^kQ^a_i\sigma_a$, it is written as follows:
\beqa
V_{\it eff}^{\rm bdry}&=&
{1\over 2\pi}\sum_{a=1}^kr^a\Ip(\sigma_a)
-\sum_{a=1}^k\left({\theta^a\over 2\pi}+q^a\right)\Rp(\sigma_a)
\nn\\
&&
+\sum_{i=1}^N{1\over 2\pi}\Ip M_i(\sigma)\Bigl(\log|M_i(\sigma)|-1\Bigr)
\nn\\
&&+\sum_{i=1}^N\left\{
{1\over 4}|\Rp M_i(\sigma)|+{1\over 2\pi}
|\Rp M_i(\sigma)|\arctan\left({\Ip M_i(\sigma)\over |\Rp M_i(\sigma)|}\right)
\right\}.
\nn
\eeqa
This is not valid on the complex hyperplanes $M_i(\sigma)=0$
where the $i$-th field becomes massless,
and there is also a singularity at the real hyperplanes $\Rp M_i(\sigma)=0$,
$\Ip M_i(\sigma)>0$ where the $i$-th field has a normalizable zero mode
localized at the boundary.
We shall call the latter the singular hyperplanes.
We would like to find a Lagrangian submanifold that asymptotes to
Lagrangian planes on which the boundary 
potential is bounded below. Also, we would like the Lagrangian 
planes to avoid meeting with the singular hyperplanes as we 
vary the FI and theta parameters. Can we find such a family of Lagrangians?
This problem is technically complicated.
So let us simplify the problem 
by choosing a particular Lagrangian submanifold
\beq
\Ip(\sigma_a)=0,\quad\forall a.
\label{allimsigma}
\eeq
{\it Is the boundary potential bounded below on this submanifold ?}
Note that it has no danger of meeting the singular hyperplanes.
On this Lagrangian plane, the effective boundary potential is
given by
\beq
V_{\it eff}^{\rm bdry}
=-\sum_{a=1}^k\left({\theta^a\over 2\pi}+q^a\right)\sigma_a
+\sum_{i=1}^N
{1\over 4}|M_i(\sigma)|.
\label{Vbdryred}
\eeq
This is a piecewise linear function which has corners at
the hyperplanes $M_i(\sigma)=0$. 
It is bounded below if and only if it is bounded below
at each one dimensional intersection of $(k-1)$ hyperplanes
$M_{i_1}(\sigma)=\cdots =M_{i_{k-1}}(\sigma)=0$.
Note that such a one-dimensional intersection
is the direction of the $U(1)$ subgroup 
which is unbroken by the values of 
$\Phi_{i_1},\ldots,\Phi_{i_{k-1}}$,
namely the unbroken $U(1)$ at the phase boundary spanned by
the charge vectors of these fields.
And the boundary potential (\ref{Vbdryred}) restricted
on that line is the same as the boundary potential 
for the theory of that $U(1)$ gauge group only.
Therefore, the boundary potential (\ref{Vbdryred})
is bounded below for some values of the theta parameters
if and only if the band restriction rule
is satisfied at all the phase boundaries for those values of the theta
parameters.
Under such a circumstance, we have a smooth family of
Wilson line branes $\wilson (q^1,\ldots,q^k)$ over the entire space
of FI-parameters $\R^k_{\rm FI}$.

This is exactly the situation we had in Example (C). Moreover, we have
a {\it set} of such Wilson line branes with the property (i).
Thus, indeed 
 we can construct the parallel family of
 an arbitrary brane over the central region
of the moduli space, using this set just like 
the grade restricted
set in one-parameter models.

\begin{figure}[htb]
\psfrag{NA}{$\mathfrak{C}_A$}
\psfrag{NB}{$\mathfrak{C}_B$}
\centerline{\includegraphics{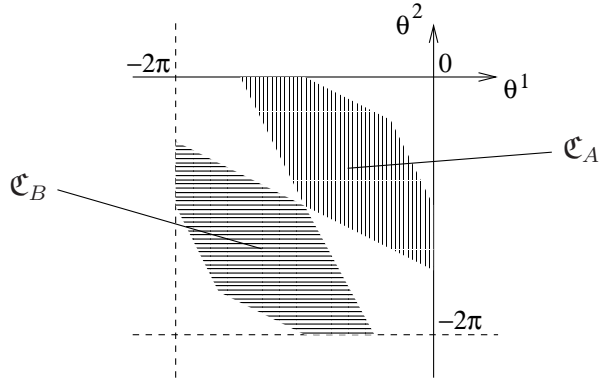}}
\caption{Overlap of the windows for the two sets}
\label{fig:A2overlap}
\end{figure}

As mentioned above, there are many other models
with such a finite set of Wilson line branes.
To see just one, 
let us take the two parameter model in Example (D) 
--- the $A_2$ singularity and its resolutions.
In this model, there are two sets of different kinds.
One is $\mathfrak{C}_A=\{(0,0),(1,0),(0,1)\}$
and the other is $\mathfrak{C}_B=\{(1,0),(0,1),(1,1)\}$.
In Figure~\ref{fig:A2overlap}
we show the overlaps of the windows at all the phase boundaries
for which the sets obey the band restriction rule.
We see that they are non-empty, and moreover coincides with
a copy of the complement of the Alga of the singular locus 
$\mathfrak{S}$ which is shown in Figure~\ref{fig:A2} (right).
In fact, within two parameter models,
it is difficult to come up with a model which does
{\it not} have such a finite set.
This makes us wonder if this is a general phenomenon.
Is there always a finite set with the property (i) and (ii)?
Can we always define D-brane transport across the center of 
the moduli space, with the simple boundary condition (\ref{allimsigma})?

The first counter example was shown to us by Mikael Passare.
It is a $U(1)^2$ theory with the following matter content:
\beq
\begin{array}{ccccc}
      &X_1&X_2&X_3&X_4\\
U(1)_1& 0 & 1 & 2 &-3\\
U(1)_2& 1 & 0 &-4 & 3
\end{array}
\label{Passare}
\eeq
It is straightforward to check that there is no
finite set of Wilson line branes 
which satisfies both of the properties (i) and (ii), and
that the Alga indeed has no complement.
This model is actually a truncated version of a model with more parameters
--- the five parameter model of the $A_5$ singularity: the 
$U(1)^5$ gauge theory
with seven fields $X_1,\ldots,X_7$ as shown in (\ref{Acontent}). 
If we are in a phase
where $\{x_2=0\}$, $\{x_5=0\}$ and $\{x_6= 0\}$
are among the deleted set,
then we obtain the same theory as (\ref{Passare}) by eliminating
the three variables $X_2, X_5$ and $X_6$.
By a direct computation, one can show that a
set with the properties (i) and (ii) does not exist
in the full $U(1)^5$ gauge theory either. 
This example
shows that we cannot always define D-brane transport
across the center of the moduli space with a finite set of Wilson line branes,
at least with the simple boundary condition (\ref{allimsigma}).
In such a model, a more general boundary condition is needed.

However, as an interesting observation, we would like to point out that
there is a finite set of Wilson line branes with just the property (i)
in all the examples we considered so far.
For example, in the model of the $A_{N-1}$ singularity,
the set $\mathfrak{C}$ given in (\ref{Atypeset})
is maximal generating in all the phases.

We emphasize that the problem we have discussed
in this subsection has nothing to do with finding a correspondence between 
D-branes at different phases --- 
We had established a completely general contruction
of D-brane transport between arbitrary pair of phases,
using paths which stay in the asymptotic region.
The main point of the present discussion was how to
cover the central region of the moduli space.
We found that there is a simple way to do so, provided that
there is a finite set of Wilson line branes with the properties (i) and (ii).
Such a set exists in many examples, but not always.

\section{Some Mathematical Background}
\label{sec:math}

\newcommand{\cM}{\mathcal{M}}

Before taking the next step into models with superpotentials, we would
now like to pause to explain some mathematical background that will
shed light on the previous discussions and  facilitate the
subsequent ones. We shall be exploring the relation
of holomorphic line bundles, or more generally coherent sheaves,
to modules and a simple construction of D-brane 
categories which follows. We use the linear sigma model as our guide
throughout. The discussion in this section is a mixture of review
material, some new approaches to existing results and also some
original material.

Consider the manifold $Y = \C^n$ with coordinates $x_1, \ldots, x_n$. 
The main object of our discussion will be the ring of polynomial
functions on $Y$, written as  
\beq
\cR = \C[ x_1, \ldots x_n ] \, .
\eeq 
Many geometric properties of the space $Y$ have a natural algebraic
description in terms of the ring $\cR$. 

Points of $Y$ are in one-to-one correspondence with {\it maximal,
  prime ideals} of $\cR$. For example, the point 
$\mathfrak{p}: \, \, \{ x_1  = a_1, \ldots, x_n = a_n \}$ 
corresponds to the set
of polynomials $\cI_{\mathfrak{p}} \subset R$ which vanish at $\mathfrak{p}$. 
This set of
polynomials is an {\it ideal} of $\cR$ since it is closed under
addition and also under multiplication by arbitrary polynomials in
$\cR$. (It is {\it prime} since multiplying two polynomials from
outside the set stays outside and {\it maximal} since it is not a
subset of any bigger prime ideal --- just as a point is indecomposable
and minimal among complex (algebraic) subspaces.) 

We are interested in D-branes or sheaves on $Y$. The simplest example
is the space filling brane given by the structure sheaf $\cO$. On an
open set of $Y$, the local sections of $\cO$ are given by holomorphic
functions with no poles inside the set. (More precisely, in the
current algebraic context we should restrict to rational functions
(ratios of polynomials) with no poles in the open set.)  

The global sections of this sheaf (or any other sheaf on $Y$) form a
module over $\cR$. This simply says that we can add global sections
and multiply them by polynomials. In the special case of the structure
sheaf, $\cO$, the global sections are rational functions on the whole
of $Y$, i.e., they are just polynomials. Thus the structure sheaf $\cO$
corresponds to the module $\cR$ itself, which is the simplest example
of a module over $\cR$.    

Ideals of $\cR$ provide a richer class of $\cR$-modules. In particular
the maximal, prime ideals $\cI_{\mathfrak{p}}$, discussed above, give rise to
$\cR$-modules $\cI_{\mathfrak{p}}$ for each point of $Y$. 
We can also consider the
cokernel of the map which embeds $\cI_{\mathfrak{p}}$ 
in $\cR$ to form a further
class of modules $\cR/\cI_{\mathfrak{p}}$.   
These correspond to the structure sheaves of points (D0-branes) of $Y$
as can be confirmed by comparing the action of $\cR$ on global
sections of the sheaves. (Polynomials in the ideal $\cI_{\mathfrak{p}}$ 
annhilitate
the module $\cR/\cI_{\mathfrak{p}}$ just as polynomials vanishing at $p$ act as
multiplication by zero on the D0-brane at $\mathfrak{p}$.) 

There is more to say here --- such as how the local sections of a sheaf
can be recovered from the corresponding module (by {\it localizing}
the module on the ring of polynomial functions which have no zeros on
the relevant open set) --- but we refer elsewhere for details \cite{AlgGeom}.  

We have touched on an important point above which is that given a map
between modules, we can always form its cokernel and  kernel
which are themselves modules. This property means that the set of
$\cR$-modules forms an {\it abelian category.} The same is true (by
definition) for coherent sheaves on $Y$. 

The key result for our purposes is the following: 
i) the set of coherent sheaves on $Y$ is in one-to-one correspondence 
with the set of $\cR$-modules and 
ii) the set of maps between a pair of sheaves 
is identical to 
the set of maps between the corresponding modules. More precisely 
{\it the category of coherent sheaves on $Y$ and the category of 
$\cR$-modules are equivalent as abelian categories.} 

We shall elaborate more on this below but roughly the statement says 
that on a simple enough space such as $Y$, we can learn everything 
about sheaves by studying global sections. This is an enormous simplification! 

The lesson of the linear sigma model approach, in this context, is that we 
can go a long way by working with simple spaces. The remaining ingredient 
which we need in order to go further is gauge symmetry.

\subsection{$\C^\times$ Action On $\C^n$ --- Grading Of The Coordinate Ring}
\label{subsec:Ccrossaction}

In this section we will be describing a construction, 
due to Serre \cite{Serre}, 
of sheaves on $\CP^{n-1}$ in terms of graded $\cR$-modules.
Consider the $\C^\times$ action on $Y$
\beq 
x_i \rightarrow \lambda x_i, \,\, \, \, i = 1, \ldots , n .
\eeq
This induces a $\Z$-grading on the ring $\cR$ such that the grading 
(degree) of a ring element is given by its representation under $\C^\times$. 
Thus the degree zero elements are complex numbers, the degree one elements 
are homogeneous linear polynomials in the $x_i$ and so on. 

To mod out by $\C^\times$, we simply restrict to graded $\cR$-modules 
and consider only maps of degree zero between modules. (A graded module 
of a graded ring is simply a module with a grading such that 
multiplication by $\cR$ respects grading in the obvious way.) 

To find the graded $\cR$-module $M$ corresponding to a particular sheaf 
$\cM$ we start as before: the degree zero elements of $M$ are given by 
the global sections of $\cM$. However, this information alone will no 
longer be enough to reconstruct $\cM$ as a sheaf. 
To see this, consider the structure sheaf $\cO$ on $\CP^{n-1}$, whose 
global sections are $\C$-valued constants. On a local patch, there are 
further gauge invariant sections which are rational functions of degree 
zero, realised as a ratio of homogenous polynomials of degree $q$. 
In order to retrieve these extra local sections it is clear that we will 
need to keep also `global sections of degree $q$' when we construct the 
module $M$. More precisely, these are the global sections of $\cM$ 
tensored with $\cO(q)$. These provide the degree $q$ elements of the 
associated module $M$. 

In this way it is clear that the structure sheaf $\cO$ on $\CP^{n-1}$ 
is once again associated with the module $\cR$, considered as a graded 
module over the graded ring $\cR$. Other sheaves can be identified with 
graded modules in the same way.

We have given a simple description of coherent sheaves on $Y/\C^\times$, 
as graded $\cR$-modules. The remaining detail to take care of is the 
deletion of the origin, $\Delta = \{x_1\!=\!\dots\!=\!x_n\!=\!0\}$, 
so that we move to $\CP^{n-1} \cong (Y-\Delta)/\C^\times$. We would like 
to think of two sheaves as being the same on $\CP^{n-1}$ if they disagree 
only at the origin of $Y$. Thus, we would like to mod out by sheaves with 
support at the origin of $Y$.

Let us examine the structure sheaf of the origin in a little more detail. 
This corresponds to the graded module given by just the degree zero part 
of $\cR$. As a vector space this module is isomorphic to $\C$ and this is 
how we shall refer to it. The $\cR$-module $\C$ is annihilated by $\cR^+$, 
the set of all positive degree elements of $\cR$ (the set of all polynomials 
that vanish at the origin.) 

There are various other modules which are annihilated by $\cR^+$, or more 
subtlely, by higher powers of $\cR^+$. In the former class, we can consider 
the module $\C$ shifted in degree so that it lives in degree $q$. This module 
is usually labelled $\C(-q)$ since it corresponds to the structure sheaf 
of the origin tensored with $\cO(-q)$. In the latter class, we could consider 
for example the module given by all the degree zero and degree one elements 
of $\cR$. We might label this module as $\cR_{\leq 1}$, consisting of 
polynomials of degree $\leq 1$. Clearly this module is not annihilated 
directly by $\cR^+$ but is annihilated by $(\cR^+)^2$ and thus should 
also correspond to a sheaf which vanishes away from the origin (otherwise 
we should be able to construct arbitrary degree polynomials in $\cR^+$ 
which do not annihilate the sheaf.)

In general, it should be clear that graded $\cR$-modules which 
are localized at the origin correspond to modules whose grading is bounded 
above. These modules (and only these) will be annihilated by sufficiently 
high powers of $\cR^+$. These form the subcategory of (graded) {\it torsion} 
modules and it is these that we should mod out by in order to recover 
sheaves on $\CP^{n-1}$. Modding out by torsion modules simply means that 
we consider two graded modules as being the same if they agree for 
sufficiently high degree.

We are now ready to state Serre's classic result: 
{\it the category of coherent sheaves on $\CP^{n-1}$ 
is isomorphic as an abelian category to the category 
of graded $\cR$-modules modulo torsion modules.}  
        
\subsection{Generalization To Toric Manifolds} 
\label{subsec:generaltoric}

The generalization to toric manifolds is straightfoward. 
Following the route suggested by the linear sigma model, 
we construct toric manifolds as quotients of $Y - \Delta_r$ 
by $(\C^\times)^k$, where $\Delta_r$ is some deleted set. 
The $(\C^\times)^k$ action on $Y$ gives rise to $k$ different 
gradings on $\cR$ and we consider modules which are graded with 
respect to all $k$ gradings. (In particular, maps between modules 
should be degree zero with respect to each grading.)

Dealing with the deleted sets in the different phases appears more 
complicated but in fact turns out to be just as simple as for $\CP^{n-1}$. 
First, note that in general we want to consider $\Z$-gradings on $\cR$ 
rather than just $\N$-gradings as before --- in other words, some of our 
coordinate fields, $x_1, \ldots, x_n$, may have negative charges with
respect to any subgroup $\C^\times \subset (\C^\times)^k$. Associated
with a single $\C^\times$ gauge group, we may consider both
$(+)$-torsion and $(-)$-torsion modules, i.e., modules with bounded
above or bounded below grading, respectively.

In each low energy phase 
of the linear sigma model we shall be deleting the set of points
$\Delta_r$ where 
either all positively or all negatively charged fields under
particular $\C^\times$'s are set to zero. 
(Recall the description of the deleted set $\Delta_r$ given 
in Section~\ref{subsec:LSMexamples}.)
In terms of modules this amounts 
to moding out by either
$(+)$-torsion or $(-)$-torsion modules under the corresponding gradings.
These are the modules whose supports lie in the deleted set $\Delta_r$. 

At this point we have reached a very simple unified description of 
sheaves in the various phases of a linear sigma model. To summarise:
{\it the category of coherent sheaves in each phase is equivalent, as an 
abelian category, to the category of multi-graded $\cR$-modules 
modulo a particular set of torsion subcategories associated to
$\Delta_r$.}

This generalization of Serre's construction was originally
given by Cox \cite{Cox} 
in the case where the quotient $X_r=(Y-\Delta_r)/(\C^{\times})^k$
is smooth.
When it has (orbifold) singularities,
the above description is simply wrong
if we regard $X_r$ as an algebraic variety,
as shown in \cite{Cox} with an explicit example.
As discussed in Section~\ref{subsec:LSMphases},
we do not really have a convenient description of the
low energy theory unless $X_r$ is a smooth manifold
or a global orbifold, and
there is no physical reason to believe that we should take 
coherent sheaves
of $X_r$ as an algebraic variety as the data for D-branes.
Rather, we understand the theory as the low energy limit of
our super-renormalizable gauge theory.
As such, we should take $(\C^{\times})^k$-equivariant 
coherent sheaves over $Y-\Delta_r$ as the D-brane data,
and we are indeed doing so in the above discussion.
Mathematically, the category of such objects
is known to be equivalent to the category of coherent sheaves
on the so-called quotient stack
$[(Y-\Delta_r)/(\C^{\times})^k]$ \cite{AKO}.
(This last reference also gives a description of sheaves as
graded modules mod torsions.)
Thus, in the above statement,
``coherent sheaves'' should be understood as those 
on the quotient stack.

In the next subsection we will introduce complexes and a description 
of the derived category which turns out to be very simple in this 
setting. This will lead us to a straightforward argument that the 
derived categories of the different phases are in fact equivalent 
in the Calabi-Yau case.  

\subsection{Complexes}
\label{subsec:complexes}

So far we have been discussing branes corresponding to individual 
sheaves and their description in terms of modules. We now wish to discuss
more general brane configurations consisting of complexes of sheaves. 
In fact, since we already have an equivalence between sheaves and 
modules we might as well carry out the discussion in terms of 
complexes of modules. 

We start once again with the case of $Y = \C^n$ where things 
are particularly simple. We will see that by representing 
arbitrary complexes in terms of `free resolutions' there is 
no need to introduce any of the complications of 
the derived category and in particular no need for nontrivial 
quasi-isomorphisms. All branes 
on $Y$ can be represented as complexes of free modules and in 
this representation, the morphisms (chiral ring elements)
between branes are just ordinary 
${\bf Q}$-cohomology classes where ${\bf Q}$ 
represents the operator induced from differentials of the complexes.

A {\it free} $\cR$-module, $\cR^{\oplus a}$  is a direct sum of $a$ 
copies of $\cR$. Every $\cR$-module $M$ has a surjection from
 a free module $\cR^{\oplus a}$ onto itself for some $a$. To produce such a 
surjection we simply choose a generating set for the module 
and map the unit of a different copy of $\cR$ onto each generator, 
much as we would do for a vector space. The difference from the case 
of vector spaces is that there will in general be relations between 
the generators. We can express these relations by writing the kernel 
$K$ of the surjection to form a short exact sequence:
\beq
0 \rightarrow K \rightarrow \cR^{\oplus a} \rightarrow M \rightarrow 0.
\eeq  
(Note that to avoid clutter we do not write explicit names of maps 
on arrows unless needed for clarity.) 
We can now start again and take a surjection from a free module 
$\cR^{a_1}$ onto $K$, with kernel $K_1$
\beq
0 \rightarrow K_1 \rightarrow \cR^{\oplus a_1} \rightarrow K \rightarrow 0.
\eeq
If we combine the surjection from $\cR^{a_1}$ onto $K$ and the 
injection from $K$ into $\cR^{\oplus a}$ into a single step then we form 
a single exact sequence as the reader should verify
\beq
0 \rightarrow K_1 \rightarrow \cR^{\oplus a_1} \rightarrow 
\cR^{\oplus a} \rightarrow 
M \rightarrow 0.
\eeq   
Continuing in this way, taking surjections from free modules onto 
the successive kernels we finally achieve a free resolution of $M$  
(the fact that this process terminates in a finite number of steps 
is a result of Hilbert's syzygy theorem)
\beq
0 \rightarrow \cR^{\oplus a_j} \rightarrow \cR^{\oplus a_{j-1}} 
\rightarrow \ldots 
\rightarrow \cR^{\oplus a_1} \rightarrow \cR^{\oplus a}
 \rightarrow M \rightarrow 0.
\eeq 
Stepping ahead a little, we should think of this as giving a 
representation of $M$ as a complex of free modules in the 
derived category of $\cR$-modules. We would now like to do 
the same for an arbitrary complex of $\cR$-modules 
\beq  
0 \rightarrow M_0 \rightarrow M_1 \rightarrow \ldots \rightarrow 
M_k \rightarrow 0.
\eeq 
We present the argument for a complex of lengh two 
$ 0 \rightarrow M \rightarrow N \rightarrow 0$ but the generalization 
should be clear. We know that for $M$ and $N$ individually we can 
construct free resolutions 
\beq 
\label{M}
0 \rightarrow P^{-j} \rightarrow P^{-j+1} \rightarrow \ldots 
\rightarrow P^{-1} \rightarrow P^{0} \rightarrow M \rightarrow 0,
\eeq
\beq 
\label{N}
0 \rightarrow Q^{-j} \rightarrow Q^{-j+1} \rightarrow \ldots 
\rightarrow Q^{-1} \rightarrow Q^{0} \rightarrow N \rightarrow 0.
\eeq
We have changed notation so that the $P$'s and $Q$'s are all free 
modules of the form $\cR^{\oplus s}$ for various values of $s$. We have also 
assumed that the two resolutions are of the same length (otherwise 
we can always pad the shorter one with zeroes at the beginning).

So to begin with, let us `turn off' the map between $M$ and $N$, 
allowing us (trivially) to build the following exact sequence 
in which all the maps are taken from (\ref{M}) and (\ref{N}) above
\beq
\label{MplusN}
0 \rightarrow 
\begin{matrix}
0 & \rightarrow & Q^{-j} & \rightarrow &  & Q^{-1} & \rightarrow & Q^0 & 
\rightarrow & N \\ 
\oplus &  & \oplus &  & \ldots & \oplus &  & \oplus &  & \oplus  \\ 
P^{-j} & \rightarrow & P^{-j+1} & \rightarrow &  & P^0 & \rightarrow & M & 
\rightarrow & 0 
\end{matrix} 
\rightarrow 0.
\eeq
Next we would like to turn the map from $M$ to $N$ back on. We will
need to simultaneously turn on some maps from the $P$'s to the $Q$'s
in (\ref{MplusN}) so that we still get a complex (i.e., so that the
differential still squares to zero.)
\beq
\label{freeres}
0 \rightarrow \begin{matrix}
0 & \rightarrow & Q^{-j} & \rightarrow &  & Q^{-1} & \rightarrow & Q^0 & 
\rightarrow & N \\ 
\oplus & \nearrow & \oplus & \nearrow & \ldots & \oplus & \nearrow & \oplus & 
\nearrow & \oplus  \\ 
P^{-j} & \rightarrow & P^{-j+1} & \rightarrow &  & P^0 & \rightarrow & M & 
\rightarrow & 0 
\end{matrix} 
\rightarrow 0.
\eeq
To see that extra maps from $P$ to $Q$ will be necessary just inspect 
the rightmost part of the diagram
and notice that the two-step map from $P^0$ to $N$ through $M$ will 
not be zero in general. 
To correct this we introduce a compensating map from $P^0$ 
to $N$ through $Q^0$. 

As we now explain, the property of $P^0$ which allows 
us to construct this compensating map, is that it is a free module. 
Suppose for simplicity that $P^0$ is equal to $\cR$. 
To completely specify a map from $\cR$ to another module 
it is sufficient to specify the image of the generating 
element $1_{\cR}$ of  $\cR$. 

Consider the map $P^0 = \cR \rightarrow N$ and 
suppose $1_{\cR} \rightarrow n  \in N $. 
Since $Q^0$ maps surjectively onto $N$, we must have some 
$q \in Q^0$ such that $q \rightarrow n$. Then if we map 
$1_{\cR} \rightarrow -q$ as our map from $P^0$ to $Q^0$, 
we can cancel the unwanted contribution. The generalization 
to construct (\ref{freeres}) is straightforward using the fact 
that all $P^i$ are free.

Once again, there is an interpretation of the free resolution 
(\ref{freeres}) in the framework of the derived category. As previously 
noted, the resolutions of $M$ and $N$ individually correspond to 
representations of these modules in the derived category by complexes 
of free modules $P^{\cdot}$ and $Q^{\cdot}$.  
Similarly, we should think of the 
combined complex with $P^{\cdot}$ `bound' to $Q^{\cdot}$ 
by the additional maps 
introduced in (\ref{freeres}) as being an equivalent representation 
of the object $0 \rightarrow M \rightarrow N \rightarrow 0$ in the 
derived category.

We have now described something quite interesting: {\it every object 
in the derived category of $\cR$-modules can be represented as a 
complex of free modules.}
This is a nice result on its own since it means that to describe 
arbitrary D-branes on $\C^n$ it is sufficient to use complexes 
built only out of copies of the structure sheaf  $\cO$. 

To go further, we would like to describe the space of chiral ring elements. 
In fact, this is where we really benefit from writing things in terms 
of free modules. Working with complexes of free modules, there are no 
non-trivial quasi-isomorphisms to worry about and all morphisms between 
complexes are just ordinary ${\bf Q}$-cohomology classes. We shall now sketch 
a proof of this statement.

First we show that there are no non-trivial exact sequences built out 
of free modules only, or more precisely that any such exact sequence 
is a direct sum of trivial exact sequences of the form
\beq
\label{identity}
0 \rightarrow \cR \rightarrow \cR \rightarrow 0
\eeq 
where the map in the diagram above is the identity. To see this it is 
once again sufficient to look at the generators $1_{\cR}$. 
Suppose we have an exact sequence of free modules and, for 
simplicity, suppose that the `rightmost' non-zero module is 
a single copy of $\cR$. Since this is the rightmost module in 
an exact sequence, the map onto it must be a surjection and in particular 
the generator $1_{\cR}$ must have a preimage. This preimage can only be 
the generator $1_{\cR'}$ of the corresponding preimage module and this 
assures that the map is in fact (a $\C$-valued multiple of) the identity 
as in (\ref{identity}). The generalization when the rightmost module 
is $\cR^a$ is straighforward.

Given this property, it then follows by standard arguments\footnote{See 
for example Lemma 1.6 in \cite{Kapranov}.} that all morphisms in the derived 
category between complexes of free modules are represented by ordinary 
cochain maps modulo homotopies (i.e., ${\bf Q}$-cohomology classes.)  

This result concludes our rather lengthy review of the derived category 
of $\C^n$! We have seen that this derived category is in fact a very 
simple beast, being equivalent to the category of complexes of free 
$\cR$-modules with morphisms given by ordinary ${\bf Q}$-cohomology 
classes. We would 
now like to extend some of these results to the graded case and use them 
to describe branes in more interesting spaces than $\C^n$. We will find 
that some of the nice properties of free modules are lost when we mod out 
by torsion. The resolution to these problems is intimately related to 
the grade restriction phenomenon which we have already encountered 
in the physical setting.

\subsection{Graded Case}
\label{subsec:gradedcase}

Some parts of the analysis of the previous section generalize immediately 
to the graded case, in which we treat $\cR$ as a graded ring and allow 
only degree zero maps between graded modules. It remains true that given 
a graded module $M$, there exists a surjection from a free module onto $M$. 
However, in this case the set of free modules is richer since we need to 
take  into account the grading. We label by $\cR(q)$ the free module which 
is isomorphic to $\cR$ except that the unit $1_{\cR}$ is in degree $-q$ . 
In the case of $\CP^{n-1}$ this corresponds to the sheaf $\cO(q)$. 
A general free module is a direct sum of one-dimensional modules of this type. 
(In the multi-graded case we have one-dimensional modules 
$R(q_1,q_2,...,q_k)$ labelled according to the degree of $1_{\cR}$ for 
all $k$ gradings.)

Whilst switching to graded modules presents no new difficulties, modding 
out by torsion modules does introduce complications. The prime difficulty 
arises because there now exist non-trivial exact sequences of free modules. 
This will invalidate our argument that morphisms 
between complexes are ordinary 
${\bf Q}$-cohomology classes. 
Furthermore, the non-trivial exact sequences which 
appear will depend on which phase we are in, so that an equivalence 
between phases might appear unlikely. As we shall see, grade restriction 
provides a route around the problem. Before we demonstrate the solution, 
we should first take a closer look at the problem.

As an example, we set $n=4$ and consider the ring $\cR = \C[x_1,x_2,x_3,p]$ 
with degrees $(1,1,1,-3)$.  (We have chosen an example in which the 
charges sum to zero for later convenience, but for the current discussion 
this is not important.)
We consider the phase in which we mod out by modules with bounded above 
grading. Thus we are deleting the set $\{x_1=x_2=x_3=0\}$ and considering 
branes on $\cO_{\CP^2}(-3)$.
 
In order to find a non-trivial exact sequence of free modules after 
modding out by torsion modules, we can proceed as follows. First, take a 
sheaf which is supported on the deleted set (a $(+)$-torsion module) 
which will be identified with zero after modding out. 
The simplest example is the structure sheaf of the deleted set,  
which corresponds to the module $\C[p]$ (i.e., the module consisting 
of polynomials of $p$ only). This module is annihilated by ring elements 
containing any of the positively charged fields $x_i$. We may also 
consider a version of this module shifted in degree, $\C[p](q)$. 
For the moment let us choose $q=3$. 

The idea is to take a free resolution of this module as we have
described earlier. When we then mod out by torsion, the original
module gets deleted and we are left with an exact sequence of free
modules. Here is the (minimal) free resolution: 
\beq
 0 \longrightarrow  \cR(0) \stackrel{\left(\mbox{\tiny $\!\!\!
                \begin{array}{c}
		  x_1\\ x_2\\ x_3
		\end{array}$}\!\!\!\right)}
	{-\!\!\!-\!\!\!\longrightarrow}
  \cR(1)^{\oplus 3}  \stackrel{\left(\mbox{\tiny $\!\!\!\!
                \begin{array}{ccc}
		   \! 0  \!\!\!&\!\!\!\!\!  -x_3 \!\!\!&\!\!\! x_2\! \\ 
		  x_3 \!\!\!\!\!&\!\!\!\!   0  \!\!\!\!&\!\!\!\!\!  -x_1 \\ 
		 \!  -x_2 \!\!\!&\!\!\! x_1 \!\!\!\!\!&\!\!\!   0 \!
		\end{array}$}\!\!\!\!\right)}
	{-\!\!\!-\!\!\!-\!\!\!-\!\!\!-\!\!\!-\!\!\!\longrightarrow}
    \cR(2)^{\oplus 3}   \stackrel{\left(x_1, x_2, x_3 \right)}
	{-\!\!\!-\!\!\!-\!\!\!\longrightarrow}
  \cR(3) \longrightarrow \C[p](3) \longrightarrow 0 ,
\eeq   
which turns into an exact sequence of free modules after modding out by torsion
\beq
\label{relation1}
0 \rightarrow \cR(0) \rightarrow \cR(1)^{\oplus 3} \rightarrow 
\cR(2)^{\oplus 3} \rightarrow \cR(3) \rightarrow 0
\eeq

In the other phase of the model we mod out by $(-)$-torsion modules,
i.e., those with bounded below grading. This corresponds to deleting
the set $\{p=0\}$ and studying branes on $\C^3/\Z_3$. In this case,
the structure sheaf of the deleted set is represented by the module 
$\C[x_1,x_2,x_3]$, which has free resolution
\beq
0 \rightarrow \cR(3) \stackrel{p}{\rightarrow} \cR(0) \rightarrow
\C[x_1,x_2,x_3] \rightarrow 0 
\eeq
leading to the exact sequence of free modules (after modding out by 
$(-)$-torsion):
\beq
\label{relation2}
0 \rightarrow \cR(3) \rightarrow \cR(0) \rightarrow 0  .
\eeq

We have found different `relations' between free modules in the two phases 
which would be expected to lead to different derived categories. We shall 
return to this problem in the next section but first conclude this 
discussion with a couple of positive remarks.

Objects in the derived category modulo torsion modules can be represented 
as ordinary complexes of graded $\cR$-modules. In other words we can pick 
a representative of each torsion equivalence class, which is an ordinary 
complex of modules (by `cutting off' the modules in the sequence below 
a sufficiently high degree.)

Furthermore, the isomorphisms which we previously used to represent 
complexes of arbitrary modules as complexes of free modules, are still 
available to us. We should think of the process of modding out as just 
adding some new isomorphisms, in which case we have not changed the set 
of underlying objects or lost any of the original morphisms. Thus, 
the conclusion that objects of the derived category can be represented 
as complexes of free modules remains true. 
This is a remarkable statement and so we repeat it: 
{\it any object in any of the low energy phases of the linear sigma model 
can be represented as a complex of free $\cR$-modules.}

Thus the boundary conditions we constructed earlier are the most general 
that we need to consider in order to describe a general brane configuration. 
All of the discussion so far applies equally well to the multi-graded 
$U(1)^k$ case. 

Next we would like to learn more about the morphisms in the different phases 
in order to study equivalences between phases. In the $U(1)$ case we will 
surprisingly find a complete description of the set of morphisms 
in terms of an ordinary ${\bf Q}$-cohomology of maps between complexes of 
(a restricted set of) free modules, whilst in the $U(1)^k$ case we will 
not be able to be so explicit. In either case, the analysis will be 
sufficient to prove equivalence of the derived category of the different 
phases under the Calabi-Yau condition. The new ingredient that 
we shall need is grade restriction.

\subsection{Grade Restriction}
\label{subsec:mathGRR}

We start by examining the origins of grade restriction in the model of 
the previous subsection.
The essential cause of our problems is that we have kept too many free 
modules of the form $\cR(q)$. In order to describe general elements of 
the derived category in a `nice' way we need to employ a reduced set of 
free modules. 

The basic problem shows up in terms of relations (exact sequences) 
between free modules. We have seen examples of this, in (\ref{relation1}), 
relevant to the $\cO_{\CP^2}(-3)$ phase and in (\ref{relation2}), 
relevant to the $\C^3/\Z_3$ phase. 

The existence of these relations is not only an inconvenience, but is
also an obstacle to proving equivalence between phases. This is
because, as illustrated in the two examples above, the available
relations depend on the choice of phase.  

This is where the grade restriction comes to the rescue. In both of the cases 
descibed above we have a relation between $\cR(3)$ and a complex of 
modules from the set $\{ \cR(0), \cR(1), \cR(2) \}$. If we restrict 
to the set of free modules $\{ \cR(0), \cR(1), \cR(2) \}$ then we claim 
i) to have no notrivial relations (in the shape of exact sequences) 
between this restricted set of free modules and ii) to have just enough 
modules to generate the most general object in the derived category. 
These statements are supposed to hold in either phase.

It is easy to convince oneself of statement i) - essentially that 
(\ref{relation1}) and (\ref{relation2}) are the `minimal' relations 
possible in either phase.  A proof can be provided, for example, 
by the explicit computation of Ext groups on projective space in 
arbitrary dimension. 

So we focus on statement ii). 
The claim is that an arbitrary element of the derived category in
either phase can be represented as a complex of free modules from the
set $\{ \cR(0), \cR(1), \cR(2) \}.$ (In general if the sum of positive
charges of the coordinate fields $x_i$ is $\,{\mathscr S}_+$ 
we restrict to the set 
$\{ \cR(0), \cR(1), \ldots \cR({\mathscr S}_+-1) \}$.) 

We have already seen that any object in the derived category can 
be represented as a complex of free modules $\cR(q)$ if $q$ is 
allowed to take arbitrary values. So roughly, we need to show 
that any free module can be expressed in terms of the restricted set 
$\{ \cR(0), \cR(1), \cR(2) \}$. 

The idea behind this is quite simple. Suppose we are in the 
$\cO_{\CP^2}(-3)$ phase. We may use relation (\ref{relation1}) 
to write $\cR(3)$ in terms of the restricted set. 
Recall that in the derived category this exact sequence should 
be thought of as an equivalence between $\cR(3)$ and the complex
\beq
\label{R3}
0 \rightarrow \cR(0) \rightarrow \cR(1)^{\oplus 3} \rightarrow
\cR(2)^{\oplus 3} \rightarrow 0 
\eeq    
Next we consider $\cR(4)$. We may use (\ref{relation1}) shifted 
in degree in order to write $\cR(4)$ in terms of 
$\{ \cR(1), \cR(2), \cR(3) \}$
\beq
\label{R4}
0 \rightarrow \cR(1) \rightarrow \cR(2)^{\oplus 3} \rightarrow 
\cR(3)^{\oplus 3} \rightarrow 0
\eeq    
To write $\cR(4)$ in terms of the grade restricted set, we may then 
replace each of the $\cR(3)$'s by modules from the restricted set, as above. 

To do this explicitly we want to `bind' (\ref{R4}) to three copies 
of (\ref{relation1}) and then `annihilate' pairs of $\cR(3)$'s. 
We show how to bind to the first copy of (\ref{relation1}). 

Consider the identity map from one of the $\cR(3)$'s in (\ref{R4}) 
to $\cR(3)$ in (\ref{relation1}). We claim that this map can be 
extended to a map between complexes. (Note that we are talking here 
about an honest cochain map between complexes of modules rather than 
some more general morphism in the derived category in which case the statement 
would be trivial.)
\beq
\label{bind}
0 \rightarrow \begin{matrix}
0 & \rightarrow & \cR(0) & \rightarrow  & \cR(1)^{\oplus 3} & \rightarrow & 
\cR(2)^{\oplus 3} & \rightarrow & \cR(3) \\ 
\oplus & \nearrow & \oplus & \nearrow  & \oplus & \nearrow & \oplus &
\nearrow  {\tiny \left( \begin{matrix}
1 \\ 
0 \\ 
0
\end{matrix} \right)} & \oplus  \\ 
0 & \rightarrow & \cR(1) & \rightarrow  & \cR(2)^{\oplus 3} & \rightarrow & 
\cR(3)^{\oplus 3} & \rightarrow & 0 
\end{matrix} 
\rightarrow 0 
\eeq
Every map in this diagram is an honest map between modules and 
the whole argument can be carried out even prior to modding out by torsion. 
However, in that case, the complex (\ref{relation1}) represents 
a non-zero module and so the object (\ref{bind}) produced after 
binding would not be equivalent to (\ref{R4}). Once we mod out, 
the operation becomes an equivalence since we are binding to 
an object with support on the deleted set.  

To construct the cochain map used above, we use a small extension of 
the argument we gave earlier in the construction of equation (\ref{freeres}). 
That construction allowed us to build a map from a complex of free modules 
to an exact sequence of modules. In the current situation, we want to build 
a map from a complex of free modules (\ref{R4}), but in this case the complex 
we are mapping to fails to be exact in the rightmost position (as a complex 
of modules prior to modding out.) However, the failure of exactness is only 
in the lowest degree and does not cause a problem since
the maps to $\cR(3)$ in (\ref{R4}) are all at least linear in the positively 
charged fields. We leave it to the reader to check the details.

 The `annihilation' step is to remove the trivial pair 
$ 0 \rightarrow \cR(3) \rightarrow \cR(3) \rightarrow 0$ from (\ref{bind}), 
to leave an equivalent complex in which we have reduced the number of 
$\cR(3)$'s
\beq
0 \rightarrow \begin{matrix}
0 & \rightarrow & \cR(0) & \rightarrow  & \cR(1)^{\oplus 3} & \rightarrow & 
\cR(2)^{\oplus 3} & \rightarrow & 0 \\ 
\oplus & \nearrow & \oplus & \nearrow  & \oplus & \nearrow & \oplus & \nearrow
  & \oplus  \\ 
0 & \rightarrow & \cR(1) & \rightarrow  & \cR(2)^{\oplus 3} & \rightarrow & 
\cR(3)^{\oplus 2} & \rightarrow & 0 
\end{matrix} 
\rightarrow 0 
\eeq 
By repeating twice more for the remaining $\cR(3)$'s we finally write 
$\cR(4)$ as a complex of modules from the grade restricted set.

It should be clear how to extend this argument for $\cR(q)$, $q>4$ by
induction. More generally, given any object in the derived category,
represented as a complex of free modules, we can remove step-by-step
all of the modules $\cR(q)$ with $q \geq 3$ by repeating this
construction many times. (At each step we remove the remaining free
module of highest degree which is positioned furthest to the right in
the complex so that all maps to this module are at least linear in the
positively charged fields and we can argue as above.) 

For $\cR(q)$ with $q<0$ there is a similar argument using the sequence 
(\ref{relation1}) shifted in degree by $-1$ and read from left to right 
as a relation between $\cR(-1)$ and the restricted set. Once again, 
by shifting degrees and iterating, we may extend this argument to 
the general case of $q<0$. The binding process in this case involves maps 
{\it from} shifted versions of (\ref{relation1}) {\it to} the complex 
that we are grade restricting. To construct the relevant maps we can 
use a trick of `dualizing' to put ourselves in the situation above, 
applying the binding procedure above and then dualizing back. 
Dualizing in this context means replacing all free modules $\cR(q)$ 
by $\cR(-q)$ and reversing the direction of all maps. 
 
This completes the proof that all objects in the derived category 
in the $\cO_{\CP^2}(-3)$ phase can be written as complexes of 
grade restricted modules. The same argument holds in the $\C^3/\Z_3$ phase 
although in this case it is rather more straightforward to see 
since according to (\ref{relation2}), every pair $R(q)$ and $R(q+3)$ are 
isomorphic and so we can always replace $q$ with $q$ mod $3$. 
Also, although we treated a specific example, the same arguments work 
just as well for the general $U(1)$ linear sigma model. We are now ready 
to state our general result: 

{\it Suppose we have a $U(1)$ linear sigma model with positively charged 
fields $x_i$ and negatively charged fields $y_j$, such that the sum of 
charges of the $x_i$ is $\,{\mathscr S}_+$ and the sum of charges of 
the $y_j$ is $\,{\mathscr S}_-$. 
Let $\cR = \C[x_i;y_j]$ be the associated graded polynomial ring. 

The model has two low energy phases. In the first phase, which
corresponds to deleting the locus $\{x_i=0, \, \forall \, i \}$ the
branes are given by the derived category of graded $\cR$-modules
modulo those with bounded above grading. Every object of this
category can be represented as a complex of free modules from the
grade restricted set $\{ \cR(0), \cR(1), \ldots, \cR({\mathscr S}_+ -1)\}$. 
Furthermore, morphisms in this representation are ordinary cochain 
maps modulo homotopies (${\bf Q}$-cohomology classes) and there are no
non-trivial quasi-isomorphisms. 

Similar statements hold for the second phase and the general brane in
this phase can be written as a complex of free modules from the grade
restricted set $\{ \cR(0), \cR(1), \ldots, \cR({\mathscr S}_- - 1) \}$. In the
Calabi-Yau case, $\,{\mathscr S}_- = \, {\mathscr S}_+$,
 this gives an explicit equivalence
between the derived categories of the two phases.}

Parts of this result have appeared previously in the literature.
For the case of $\CP^{n-1}$, the description of an arbitrary brane as a
complex of $\{ \cR(0), \cR(1), \ldots, \cR(n - 1) \}$ follows from
Beilinson's work \cite{Beilinson}. 
The equivalence for $\cO_{\CP^2}(-3)$ and $\C^3/\Z_3$
is a special case of the McKay
correspondence by Bridgeland-King-Reid \cite{BKR},
as already mentioned in Section~\ref{subsec:moreex}. 
Bondal-Orlov \cite{BO} and Bridgeland
\cite{Flops} studied cases where all
fields have degree $\pm 1$ and proved equivalence between phases under
the Calabi-Yau condition.

A.~King pointed out to the authors that
the result as stated above appeared previously in
a work by Van den Bergh, see \cite{vdBergh} Section 8.
Thus, we cannot claim it as our own original result,
even though we were not aware of that work at the point we
obtained it. 
Nonetheless, our approach to the problem, based on the physics of 
linear sigma model, naturally leads to a number of important
generalizations.
One of them is generalization to multi-graded cases, which we discuss next.
Another is extension to systems with superpotential, which we will
discuss in Section~\ref{sec:compactI}, where we obtain
equivalences of the derived category of a Calabi-Yau manifold
and different type of triangulated categories, such
as the category of matrix factorizations.
All these equivalences are unified under the same principle
--- the grade restriction rule.

\subsection{Multi-Graded Case}
\label{subsec:multigradedcase}

We would like to describe the effect of crossing phase boundaries in
the general $U(1)^k$ linear sigma model. 
Since only a single $U(1)$ is unbroken at
each boundary, we already have all the necessary tools at our disposal.

We write the general one-dimensional free module as 
$\cR(q^1, \ldots, q^k)$ to display the grading under each
$U(1)$. We focus on
the first $U(1)$ and suppose that the sum of positive 
({\it resp}. negative)
charges of the variables under this $U(1)$ is $\,{\mathscr S}_{+}$ 
({\it resp}. $\,{\mathscr S}_{-}$) 
as before.

Suppose that we are in a phase in which we need to mod out by the
$(+)$-torsion modules (those whose grading is bounded above under this
$U(1)$). According to our previous discussion, we can describe an
arbitrary brane in this phase as a complex of free modules 
$\cR(q^1, \ldots, q^k)$, where $q^1$ can be restricted to lie in the range
$\{ 0,1,\ldots, \,{\mathscr S}_+-1 \}$. We treat 
phases in which the D-term for this $U(1)$ takes the opposite sign 
in a similar way --- restricting
$q^1$ to lie between $0$ and $({\mathscr S}_- \!- \! 1)$.

Since we still have to mod out by torsion modules for the remaining
$U(1)$'s, we expect to have relations (in the form of exact sequences)
between the set of free modules for which only $q^1$ has been grade
restricted, i.e., relations within the band restricted set. Thus, we
can no longer claim to have a simple description
of the derived category in which morphisms are 
ordinary ${\bf Q}$-cohomology classes (homotopy classes of
cochain maps). However, for the purposes of describing equivalences between
neighbouring phases this is not necessary. This is because 
when we move between adjacent phases, we only switch the sign of a single 
D-term and so switch from $(+)$-torsion to $(-)$-torsion for
a single $U(1)$.
The remaining $U(1)$'s are not affected, 
and hence 
modding out by remaining torsion modules is the {\it same} procedure 
for both phases.
This is manifest in the relation (\ref{importantre}) between
the deleted sets.

Thus it becomes clear how we are to compare branes in different
phases. Each time we wish to cross a phase boundary 
we grade restrict with respect to
the associated $U(1)$. We may then pass freely into the next
phase. If the sum
of positive and negative charges of the variables
are equal for the $U(1)$ in question,
there will be no new relations between branes after crossing the
boundary and we will have an equivalence of categories. {\it If the
  Calabi-Yau condition (\ref{conformal}) holds so that the sum of charges 
  is zero, then it follows, by crossing a boundary at a time,
  that the derived categories in all low energy phases are
  equivalent.}

We emphasize again that, unlike in the single graded case,
this proof of equivalence does not imply nor rely on
 a simple description of derived
categories in terms of a finite set of rank one free modules.
In fact, there is a toric variety which does not admit  
such a set (the counter example \cite{HillePerling} 
to King's conjecture \cite{King}).
However, as mentioned in Section~\ref{subsec:center},
there {\it is} such a finite set
in all the Calabi-Yau examples we considered so far.

A different construction of derived equivalences 
in the general multi-graded case 
had been given by Kawamata in \cite{Kawamata}.

\subsection{Generalizations}
\label{subsec:generalizations}

Although we shall not develop these here, there are various directions 
in which one could generalize the preceding analysis. One possibility 
is to allow a non-Abelian group action $G$ on $Y = \C^n$ and thus 
decompose $\cR$ into representations of $G$. This line of development 
has been initiated by Kapranov \cite{Kapranov,Kapranov2} who generalizes 
the Belinson result for sheaves on $\CP^n$ to Grassmanians. Clearly, 
these developments will be relevant for describing branes 
in non-Abelian linear sigma models.

Another direction is to consider more general choices of the ring $\cR$. 
Much of the basic theory can be developed even in the case of 
non-commutative rings \cite{ArtinZhang} and it would be fascinating 
to develop a physical interpretation of this work. However, we leave 
these directions to future research and return now to our main focus 
which is the study of D-branes in Abelian linear sigma models.

\section{Compact Models} 
\label{sec:compactI}

\newcommand{\hyper}{M}

In this section, we consider the problem of D-brane transport
in systems that arise from linear sigma models with superpotential. 
Applying the grade restriction rule, 
we find a rule of D-brane transport along paths on the K\"ahler moduli space.
This leads, for example, to one-to-one correspondences between
 D-isomorphism classes of D-branes in Landau-Ginzburg orbifolds and those
in geometric regimes. 
The basic idea of transport itself is identical to 
the one in the non-compact models of
Section~\ref{sec:noncompact}.
Besides having matrix factorizations instead of complexes,
the main new feature is that, depending on the phase,
some of the bulk fields may acquire masses from superpotential F-terms
and therefore must be integrated out.

\subsection{The D-Brane Transport}
\label{subsec:compactLEBC}

\newcommand{\MFarrow}[2]{\begin{picture}(50,20)(0,20)
  \put(22,33){{\small $#1$}}
    \put(5,27){\vector(1,0){40}}    
  \put(45,23){\vector(-1,0){40}}    
  \put(22,13){{\small $#2$}}
  \end{picture}}
\newcommand{\MFarrowshort}[2]{\begin{picture}(32,20)(0,20)
  \put(12,31){{\small $#1$}}
    \put(5,27){\vector(1,0){22}}    
  \put(27,23){\vector(-1,0){22}}    
  \put(14,14){{\small $#2$}}
  \end{picture}}
\newcommand{\sMFarrow}[2]{\begin{picture}(50,20)(0,20)
  \put(22,29){{\small $#1$}}
  \put(5,27){\vector(1,0){40}}    
  \put(45,24){\vector(-1,0){40}}    
  \put(22,17){{\small $#2$}}
  \end{picture}}
\newcommand{\ssMFarrow}[2]{\begin{picture}(50,20)(0,20)
  \put(22,31){{\small $#1$}}
  \put(5,27){\vector(1,0){40}}    
  \put(45,23){\vector(-1,0){40}}    
  \put(22,12){{\small $#2$}}
  \end{picture}}
\newcommand{\dMFarrow}[2]{\begin{picture}(50,10)(0,20)
  \put(11,22){{\small $#1$}}
  \put(45,32){\vector(-2,-1){40}}
  \put(5,8){\vector(2,1){40}}
  \put(34,15){{\small $#2$}}
  \end{picture}}
\newcommand{\MFmorph}[2]{\begin{picture}(50,30)(0,20)
  \put(10,10){{\small $#1$}}
  \put(5,15){\vector(2,1){40}}    
  \put(45,15){\vector(-2,1){40}}    
  \put(10,35){{\small $#2$}}
  \end{picture}}
\newcommand{\MFmorphA}[2]{\begin{picture}(50,25)(0,15)
  \put(0,14){{\small $#1$}}
  \put(45,35){\vector(-3,-2){40}}    
  \put(5,35){\vector(3,-2){40}}    
  \put(42,14){{\small $#2$}}
  \end{picture}}
\newcommand{\MFlsm}{\mathfrak{MF}_W(\C^N,T)}
\newcommand{\MFlsmT}{\mathfrak{MF}_W({\cal T}^w_{{\rm I},{\rm II}})}
\newcommand{\MFT}{\mathfrak{MF}_W({\cal T}^w)}

To start with, let us briefly recall the r\^ole of the superpotential
in the bulk theory.
The classical vacuum configuration in the Higgs branch is govened by
the D-term and F-term contributions to the potential (\ref{U}).
Away from phase boundaries, the D-term potential confines
dynamics to a toric, non-compact Calabi-Yau variety $X_r$.
The F-term
potential determined by the superpotential $W$ restricts the vacuum
further to a complex subvariety $\hyper_r \subset X_r$.
Depending on the phase, a part of the transverse modes
of $\hyper_r$ in $X_r$ may acquire mass from the superpotential F-terms.
In the case where all transverse modes are massless, 
the low energy theory is a Landau-Ginzburg model
with superpotential $W$ over the non-compact toric variety $X_r$.
In the other extreme where all transverse modes are massive,
we obtain a non-linear sigma model on the vacuum manifold $M_r$.
The nature of low energy theory depends very much on the phase
and the pattern is in general very complex.

However, one could always choose to use the description as
a LG model on the non-compact toric variety $X_r$.
This description is most natural if
the energy scale, $e\sqrt{r}$,
set by the D-terms is much higher than the one, $m_W$, 
set by the F-terms, and we consider 
an intermediate scale, $e\sqrt{r} \gg \mu \gg m_W$.

Like in the non-compact situation we are taking the large gauge
coupling limit, so that we can integrate out the gauge multiplet
algebraically (cf. (\ref{gaugeconn}) and (\ref{sigmaconn})). The
matrix factorization $(\cp,Q,\rho,R)$ in the
linear sigma model then becomes a matrix factorization of $W$
over the toric variety $X_r$. Let us denote the set of matrix
factorizations of $W$ over $X_r$ by $MF_W(X_r)$, and refer to the
set of matrix factorizations in the linear sigma model as
$\MFlsm$. Two D-branes in $\MFlsm$ that 
are related by D-isomorphisms will flow to the same D-brane in the
infra-red limit, so that we define $MF_W(X_r)$ as the set of matrix
factorizations of the linear sigma model up to D-isomorphisms.
Let us introduce the corresponding projection:
$$
  \pi_r:\MFlsm\rightarrow MF_W(X_r) .
$$
Similarly to Section~\ref{subsec:LEBC} we have a pyramid of maps:
\begin{equation}
  \label{pic:pyramid}
\begin{picture}(150,100) \thicklines
  \put(35,90){$\MFlsm$}
  \put(60,85){\vector(-2,-3){36}}
  \put(67,82){\vector(-1,-3){24}}
  \put(73,82){\vector(1,-3){24}}
  \put(80,85){\vector(2,-3){36}}
  \put(24,53){$\pi_{\rm I}$}
  \put(38,40){$\pi_{\rm II}$}
  \put(68,39){$\pi_{\rm III}$}
  \put(104,53){$\pi_{\rm IV}$}
  \put(-20,22){$MF_W(X_{\rm I})$}
  \put(5,-1){$MF_W(X_{\rm II})$}
  \put(80,-1){$MF_W(X_{\rm III})$}
  \put(100,22){$MF_W(X_{\rm IV})$}
  \textcolor{blue}{
    \put(65,25){\line(-3,1){55}}
    \put(65,25){\line(-4,-1){60}}
    \put(65,25){\line(1,-5){7}}
    \put(65,25){\line(6,-1){65}}
    \put(65,25){\line(4,1){60}}
    \put(60,50){\circle*{0.5}}
    \put(65,51){\circle*{0.5}}
    \put(70,51){\circle*{0.5}}
    \put(75,50){\circle*{0.5}}
    \put(80,49){\circle*{0.5}}
  }
\end{picture}
\end{equation}

Let us illustrate these projections in Example (A) with
$N=3$. We pick the superpotential $W = p \cdot G(x)$ with the cubic fermat
polynomial $G(x) = x_1^3+x_2^3+x_3^3$. 
In the $r\to-\infty$ limit,
this gives the LG-model with superpotential $G(x)$ over the orbifold 
$X_- \cong \C^3/\ZZ_3$, namely, a LG orbifold.
At $r \gg 0$ and at the intermediate energy scale
$\mu$ we have a LG-model with potential $W=pG(x)$ over 
$X_+$, the total space of
the line bundle $\cO(-3)\rightarrow \CP^2$. 
This superpotential gives masses to the fibre coordinate $p$ and
to the transverse mode of $x$'s to the hypersuface $G(x)=0$, 
but we choose to keep both of them. 
At lower energies $\mu \ll m_F$,
it is more appropriate to integrate them out,
and we have the sigma model on the 
elliptic curve, $\{p\!=\!G\!=\!0\}\subset X_+$.)

In order to describe D-isomorphism realtions in the two phases,
we first consider matrix factorizations that are
infra-red empty, i.e., the ones for which the boundary potential 
$\{Q,Q^\dagger\}$ is strictly positive over $X_r$.
They can be constructed from the Koszul complexes for the deleted sets,
and will play the analogous r\^ole as the latter played
in the non-compact models.
Recall that our R-charge assignment
is $R_\lambda(p,x_1,x_2,x_3) = (\lambda^2 p,x_1,x_2,x_3)$.

In the $r \ll 0$ phase,
the deleted set is $\Delta_- = \{p=0\}$. Let us consider 
the brane $\lsmB_-$ given by
$$
Q=\left(\begin{array}{cc}
0&G(x)\\
p&0
\end{array}\right),
\quad
\rho(g)=\left(\begin{array}{cc}
g^2&0\\
0&g^{-1}
\end{array}\right),
\quad
R(\lambda)=\left(\begin{array}{cc}
1&0\\
0&\lambda^{-1}
\end{array}\right).
$$
For simplicity, we use the following notation to encode this data:
\beq
  \label{LGtrivialMF}
  \lsmB_-:~~
  \Biggl(\wilson(2)^{}_{{}_0} \MFarrow{p}{G} 
\wilson(-1)^{}_{{}_{-1}}\Biggr)
\eeq
The subscript of each Wilson line brane labels the R-charge.
This brane has strictly positive potential 
$\{Q,Q^\dagger\} = (|p|^2+|G|^2)\cdot\id_2$ 
over $X_-$ where $p$ is non-zero, and therefore is infra-red empty. 
Note that this matrix factorization is obtained from the complex
$\wilson(2)\stackrel{p}{\longrightarrow}\wilson(-1)$,
which defined an empty brane in the non-compact model with $W=0$,
by adding the arrow in the opposite direction,
$\wilson(2)\displaystyle \mathop{\longleftarrow}_{G}\wilson(-1)$.

Let us next consider the $r\gg 0$ phase where the deleted set is
$\Delta_+=\{x_1\!=\!x_2\!=\!x_3\!=\!0\}$.
Just as above, we consider adding some arrows to a complex corresponding to
an empty brane in the non-compact model, such as
$
\wilson(-1)\stackrel{X}{\rightarrow}
\wilson(0)^{\oplus 3}\stackrel{X}{\rightarrow}
\wilson(1)^{\oplus 3}\stackrel{X}{\rightarrow}
\wilson(2),
$
which can be realized as $X=\sum_{i=1}^3x_i\bareta_i$
on the Clifford module.
If we add the arrows $pX^2=\sum_{i=1}^3px_i^2\eta_i$ in the opposite 
direction, we find a matrix factorization
$Q=X+pX^2$ of $W=\sum_{i=1}^3px_i^3$.
With an appropriate choice of R-charge, say
$-2$ on the Clifford vacuum $|0\rangle$, we have a graded matrix factorization.
Presented in the similar way as (\ref{LGtrivialMF}), it is
\beq
  \label{LVtrivialMF}
  \lsmB_+:~~\Biggl(
  \begin{array}{c}
    \wilson(-1)^{}_{{}_{-2}} \\[-5pt]
    \oplus    \\[-5pt]
    \wilson(1)^{\oplus 3}_{{}_0}
  \end{array}
  \MFarrow{g_+}{f_+}
  \begin{array}{c}
    \wilson(0)^{\oplus 3}_{{}_{-1}} \\[-5pt]
    \oplus    \\[-5pt]
    \wilson(2)^{}_{{}_1}
  \end{array}\Biggr)  ,
\eeq
with
$$
  g_+ =  \left(\!\!\begin{array}{cccc}
    x_1 & 0 & px_3^2 & -px_2^2 \\[-2pt]
    x_2 & -px_3^2 & 0 & px_1^2 \\[-2pt]
    x_3 & px_2^2 & -px_1^2 & 0 \\[-2pt]
    0 & x_1 & x_2 & x_3
  \end{array}\!\!\right) ,~~
  f_+=\left(\!\!\begin{array}{cccc}
    px_1^2 & px_2^2 & px_3^2 & 0 \\[-2pt]
    0 & -x_3 & x_2 & px_1^2\\[-2pt]
    x_3 & 0 & -x_1 & px_2^2 \\[-2pt]
    -x_2 & x_1 & 0 & px_3^2 
  \end{array}\!\!\right) .
$$
The boundary potential is strictly positive on $X_+$,
$\{Q,Q^{\dag}\}=\sum_{i=1}^3(|x_i|^2+|px_i^2|^2)>0$,
and the brane is indeed infra-red empty
in the $r\gg 0$ phase.

Let us bind these two branes using a map
from $\lsmB_-$ to $\lsmB_+$:
\beq
~~
\begin{array}{ccccccc}
&&&&
\wilson(2)_{{}_0}
&
\!\!\MFarrow{\!\!\!-p\,}{\!\!\!-G\,}\!\!
&
\wilson(-1)_{{}_{-1}}\\
&&&&\oplus&
\!\!\MFmorphA{X^2}{\id}\!\!
&\oplus\\
\wilson(-1)_{{}_{-2}}
&
\!\!\ssMFarrow{X}{pX^2}\!\!
&
\wilson(0)_{{}_{-1}}^{\oplus 3}
&
\!\!\ssMFarrow{X}{pX^2}\!\!
&
\wilson(1)_{{}_0}^{\oplus 3}
&
\!\!\ssMFarrow{X}{pX^2}\!\!
&
\wilson(2)_{{}_1}
\end{array}
\label{bindingMF}
\eeq
Following the procedure from
(\ref{complexD}) to (\ref{complexD'}), which applies also to
matrix factorizations, we can erase the trivial pair
$\wilson(2)\stackrel{\id}{\to}\wilson(2)$.
The result is
$$
\wilson(-1)_{{}_{-2}}~\,\ssMFarrow{X}{pX^2}~\,
\wilson(0)^{\oplus 3}_{{}_{-1}}~\,\ssMFarrow{X}{pX^2}~\,
\wilson(1)^{\oplus 3}_{{}_0}~\,\ssMFarrow{pX}{X^2}~\,
\wilson(-1)_{{}_{-1}},\qquad
$$
or more explicitly
$$
  \lsmB_0:~~\Biggl(
  \begin{array}{c}
    \wilson(-1)^{}_{{}_{-2}} \\[-5pt]
    \oplus    \\[-5pt]
    \wilson(1)^{\oplus 3}_{{}_0}
  \end{array}
  \MFarrow{g_0}{f_0}
  \begin{array}{c}
    \wilson(0)^{\oplus 3}_{{}_-1} \\[-5pt]
    \oplus    \\[-5pt]
    \wilson(-1)^{}_{{}_{-1}}
  \end{array}\Biggr)  ,
$$
with
\beq
  \label{masslessMF}
  g_0 = \left(\!\!\begin{array}{cccc}
    x_1 & 0 & px_3^2 & -px_2^2 \\[-2pt]
    x_2 & -px_3^2 & 0 & px_1^2 \\[-2pt]
    x_3 & px_2^2 & -px_1^2 & 0 \\[-2pt]
    0 & px_1 & px_2 & px_3
  \end{array}\!\!\right) ,~~
  f_0 = \left(\!\!\begin{array}{cccc}
    px_1^2 & px_2^2 & px_3^2 & 0 \\[-2pt]
    0 & -x_3 & x_2 & x_1^2\\[-2pt]
    x_3 & 0 & -x_1 & x_2^2 \\[-2pt]
    -x_2 & x_1 & 0 & x_3^2 
  \end{array}\!\!\right) .
\eeq

In the $r\ll 0$ phase where the brane $\lsmB_-$ is empty, this process
results in the low energy D-isomorphism
\beq
  \label{QQplus}
  \pi_-\bigl(\lsmB_+\bigr) ~\cong~ \pi_-\bigl(\lsmB_0 \bigr),
\eeq
which trades the Wilson line component $\wilson(2)_{{}_1}$
for $\wilson(-1)_{{}_{-1}}$.
The shift in the R-charge is due to the fact
that the varibale $p$ carries R-charge $2$ and that $(f_0,g_0)$ 
is obtained from $(f_+,g_+)$ by relocating $p$.
In general, given a matrix factorization in the Landau-Ginzburg
phase, we can always eliminate the Wilson line component with
largest charge $\wilson(q_{max})$ in favour of
$\wilson(q_{max}\!-\!3)$, as well as the 
smallest charge component $\wilson(q_{min})$ by
$\wilson(q_{min}\!+\!3)$. This reflects the broken 
gauge symmetry in the Landau-Ginzburg orbifold model, and tells us
that we can simply set $p=1$ in order to obtain
$\ZZ_3$-graded matrix factorizations. In the next subsection.
we will see how this works in a
general model.

In the $r\gg 0$ phase where the brane $\lsmB_+$ is infra-red empty,
the above process results in the D-isomorphism
\beq
  \label{QQminus}
  \pi_+\bigl(\lsmB_-\bigr) ~\cong~ \pi_+\bigl(\lsmB_0\bigr)  .
\eeq
The Wilson line component
$\wilson(2)$ in $\lsmB_-$ is traded for others $\wilson(q)$ with
$q=-1,0,1$.
Just as in this case, we can always lower (and increase)
the gauge charges $q$, by binding with 
$\lsmB_+$ or with its shifts in gauge and R-charges.
Each binding process changes the charge by at most $3$.

\subsubsection*{\it Grade Restriciton Rule}

We now transport D-branes across phase boundaries.
Let us consider two adjacent phases, say Phase I and Phase II,
and fix a window $w$ at the phase boundary.
The grade restriciton rule (or more generally
the band restriction rule) of Section~\ref{sec:GRR} tells us that we
cannot transport arbitrary matrix factorizations along a path that goes
through $w$. Only those branes whose Wilson line components
$\wilson(q^1,\ldots,q^k)$ are contained in the band restriction rule
can be transported,
i.e., $(q^1,\ldots,q^k)$ must be in the band
$\mathfrak{C}^w_{{\rm I},{\rm II}} \subset \ZZ^k$. We denote the set of
matrix factorizations in the band by $MF_W({\cal T}^w_{{\rm I},{\rm II}})$. 
Just like (\ref{hatdiagram1}) in the non-compact case,
 we can associate to a given window $w$, a hat diagram:
\beq
  \label{hatdiagram2}
\begin{picture}(150,100) \thicklines
  \put(52,87){$\MFlsm$}
  \put(80,72){$\cup$}
  \put(56,57){$\MFlsmT$}
  \put(63,80){\vector(-2,-3){45}}
  \put(105,80){\vector(2,-3){45}}
  \put(21,45){$\pi_{\rm I}$}
  \put(133,45){$\pi_{\rm II}$}
  \put(40,13){\vector(2,3){25}}
  \put(128,13){\vector(-2,3){25}}
  \put(50,20){$\omega^w_{{\rm I},{\rm II}}$}
  \put(98,20){$\omega^w_{{\rm II},{\rm I}}$}
  \put(-6,-1){$MF_W(X_{\rm I})$}
  \put(123,-1){$MF_W(X_{\rm II})$}
\end{picture}
\eeq
The lifts and projections compose to the maps
\beqa
&&  F^w_{{\rm I},{\rm II}}= \pi_{\rm II} \circ \omega^w_{{\rm I},{\rm II}}:
 MF_W(X_{\rm I}) \longrightarrow MF_W(X_{\rm II}),
\nn\\
&&  F^w_{{\rm II},{\rm I}}= \pi_{\rm I} \circ \omega^w_{{\rm II},{\rm I}}:
 MF_W(X_{\rm II}) \longrightarrow MF_W(X_{\rm I}),
\nn
\eeqa
which are inverse to each other,
$F^w_{{\rm II},{\rm I}}\circ F^w_{{\rm I},{\rm II}}\cong
{\rm id}_{MF_W(X_{\rm I})}$ and
$F^w_{{\rm I},{\rm II}}\circ F^w_{{\rm II},{\rm I}}\cong
{\rm id}_{MF_W(X_{\rm II})}$.
They induce a one-to-one correspondence between 
D-isomorphisms classes in $MF_W(X_{\rm I})$ and $MF_W(X_{\rm II})$.
The key step in the hat diagram (\ref{hatdiagram2}) 
is the lift from
the low energy matrix factorizations in a given phase, say $MF_W(X_{\rm I})$, 
to the grade restricted
subset $\MFlsmT$. A random lift $\lsmB=({\cal V},Q,\rho,R)$ of a matrix
factorization in $MF_W(X_{\rm I})$ to the linear sigma model does not comply
with the band restriction rule. We can, however, bind trivial D-branes to
$\lsmB$ in orde to eliminate Wilson line components $\wilson(q)$ outside the
band, without changing the D-isomorphism class in $MF_W(X_{\rm I})$.

In the example at hand, the trivial branes at small volume,
$r\ll 0$, are the matrix factorizations
$$
  \Biggl(\wilson(q+3)^{}_{{}_{j+1}} 
\MFarrow{p}{G} \wilson(q)^{}_{{}_j}\Biggr)
$$
for any $q \in \ZZ$ and $j\in \ZZ$.
In the large volume phase, $r\gg 0$, the trivial branes are
$$
  \Biggl(
  \begin{array}{c}
    \wilson(q)^{}_{{}_{j-1}} \\[-5pt]
    \oplus    \\[-5pt]
    \wilson(q+2)^{\oplus 3}_{{}_{j+1}}
  \end{array}
  \MFarrow{g_+}{f_+}
  \begin{array}{c}
    \wilson(q+1)^{\oplus 3}_{{}_{j}} \\[-5pt]
    \oplus    \\[-5pt]
    \wilson(q+3)^{}_{{}_{j+2}}
  \end{array}\Biggr)
$$
for any $q \in \ZZ$ and $j\in \ZZ$.
Note that both D-branes have just the right width in the
range of gauge charges in order to enable the restriction to $\mathfrak{C}^w$
for some window $w$ at the phase boundary. 
The binding process (\ref{bindingMF}) is an example where the Wilson line
component $\wilson(2)$ is eliminated in order to fit the matrix
factorization into the grade restriction rule $\mathfrak{C}^w = \{-1,0,1\}$
associated to the window $w = \{-\pi < \theta < \pi\}$. Indeed, the matrix
factorization $\lsmB_0$ is an object in $\MFT$, and as a result of
relations (\ref{QQplus}) and (\ref{QQminus}) we find:  
\beq
  \label{GRRexample}
F^w_{+,-}:
\pi_+\Bigl(\wilson(2)^{}_{{}_0} \MFarrow{p}{G}\wilson(-1)^{}_{{}_{-1}}\Bigr)
\longmapsto
  \pi_-\Biggl(
  \begin{array}{c}
    \wilson(-1)^{}_{{}_{-2}} \\[-5pt]
    \oplus    \\[-5pt]
    \wilson(1)^{\oplus 3}_{{}_0}
  \end{array}
  \MFarrow{g_+}{f_+}
  \begin{array}{c}
    \wilson(0)^{\oplus 3}_{{}_{-1}} \\[-5pt]
    \oplus    \\[-5pt]
    \wilson(2)^{}_{{}_1}
  \end{array}\Biggr) .
\eeq

Just as in this example, in any model,
the trivial matrix factorizations to be used
for grade or band restriction processes
are obtained from the trivial complexes in the non-compact version.
For the general model with $U(1)$ gauge group (using the notation
in Section~\ref{subsec:grrrevisited}),
they are obtained from the $X$-Koszul and $Y$-Koszul complexes
by adding arrows in the oppostite directions
\beqa
{\mathscr K}_+^{\rm mf}\!\!\!\!&=&\!\!\!\!
\left(\,
\wilson(0)\MFarrowshort{X}{a}\wilson_+
\MFarrowshort{X}{a}
\wedge^2\wilson_+
\MFarrowshort{X}{a}
\cdots
\MFarrowshort{X}{a}
\wedge^{l}\wilson_+\right),\qquad
\label{trivplus}\\
{\mathscr K}_-^{\rm mf}\!\!\!\!&=&\!\!\!\!
\left(\wilson(0)
\MFarrowshort{Y}{b}\wilson_-
\MFarrowshort{Y}{b}
\wedge^2\wilson_-
\MFarrowshort{Y}{b}
\cdots
\MFarrowshort{Y}{b}
\wedge^{l'}\wilson_-\right).\qquad
\label{trivminus}
\eeqa
Here $a=(a_1,...,a_l)$ and $b=(b_1,...,b_{l'})$ are such that
$$
W=\sum_{i=1}^lx_ia_i(x,y)=\sum_{j=1}^{l'}y_jb_j(x,y).
$$
Since $W$ is gauge invariant,
such $a_i$'s and $b_j$'s always exist if
$W$ is assumed to be reduced with respect to the
non-trivially charged fields (i.e. $W=0$ for $x=y=0$).
We have neglected to put
the R-charge information in (\ref{trivplus}) and (\ref{trivminus}).

For a model with higher rank gauge group, $T=U(1)^k$ with $k>1$,
at each phase boundary
we can construct similar matrix factorizations of the part of
$W$ which is reduced with respect to
the relevant fields at that phase boundary.
Then, we can take the tensor product with any matrix factorization for
the remaining part of $W$ to make a trivial D-brane.
As in the non-compact version, there are amibuities in
the grade restricted lifts $\omega^w_{\rm I,II}$
and $\omega^w_{\rm II,I}$, but they are due to common
deleted sets of the two pahses and hence do not matter in the end.

\subsubsection*{\it Integrating Out Heavy Fields}

We have achieved our goal also in models with superpotential --- we have
a complete rule of D-brane transport across phase boundaries.
However, the ``low energy'' theories in the above discussion
are non-linear LG models on toric verieties $X_r$, which are
in general different from the typical description,
such as nonilinear sigma models without potential or 
LG models on linear space.
In order to have a useful transportation rule,
we need to fill this gap.

The main gap exists in large volume phases where all the modes transverse to 
$\hyper_r$ in $X_r$ acquire masses from the
F-term superpotential $W$. For scales below the F-term mass,
we can integrate out these massive modes and we obtain
the non-linear sigma model on $\hyper_r$.
Matrix factorizations must
turn into complexes of holomorphic vector bundles over $\hyper_r$.
The pyramid (\ref{pic:pyramid}) is then enhanced by maps 
$MF_W(X_r) \rightarrow D(\hyper_r)$ in such phases, 
which associate to each matrix factorization at the intermediate scale 
a geometric D-brane on the subvariety $\hyper_r$: 
\begin{center}
\begin{picture}(150,130) \thicklines
  \put(35,120){$\MFlsm$}
  \put(60,115){\vector(-2,-3){36}}
  \put(67,112){\vector(-1,-3){24}}
  \put(73,112){\vector(1,-3){24}}
  \put(80,115){\vector(2,-3){36}}
  \put(24,83){$\pi_{\rm I}$}
  \put(38,70){$\pi_{\rm II}$}
  \put(68,69){$\pi_{\rm III}$}
  \put(104,83){$\pi_{\rm IV}$}
  \put(-20,52){$MF_W(X_{\rm I})$}
  \put(6,29){$MF_W(X_{\rm II})$}
  \put(35,25){\vector(0,-1){20}}
  \put(20,-7){$D(\hyper_{\rm II})$}
  \put(78,29){$MF_W(X_{\rm III})$}
  \put(102,25){\vector(0,-1){20}}
  \put(87,-7){$D(\hyper_{\rm III})$}
  \put(100,52){$MF_W(X_{\rm IV})$}
  \textcolor{blue}{
    \put(65,55){\line(-3,1){55}}
    \put(65,55){\line(-4,-1){60}}
    \put(65,55){\line(1,-5){7}}
    \put(65,55){\line(6,-1){65}}
    \put(65,55){\line(4,1){60}}
    \put(60,80){\circle*{0.5}}
    \put(65,81){\circle*{0.5}}
    \put(70,81){\circle*{0.5}}
    \put(75,80){\circle*{0.5}}
    \put(80,79){\circle*{0.5}}
  }
\end{picture}
\end{center}
We will find what the maps $MF_W(X_r)\longrightarrow D(M_r)$ are, and see that
they induce a one-to-one correspondence
between geometric D-branes in $D(M_r)$ and matrix factorizations in
$MF_W(X_r)$.
Unlike in the bulk theory, where integrating out
the massive fields essentially corresponds to setting them equal to zero,
the presence of a boundary enriches the story significantly.

A gap also exists at Landau-Ginzburg orbifold points where the usual low
energy description is in terms of LG models on linear spaces.
This has nothing to do with F-terms but has to do with D-terms.
At a LG orbifold point, some of the linear sigma model
fields simply decouple by acquiring infinite mass from 
the Higgs mechanism, and we are left with the LG model for the rest of
the fields. Thus, it is more convenient to use the description
that includes only those massless linear fields, rather than to work with
the set of all fields and implement the D-isomorphism relation,
such as (\ref{QQplus}), all the time.

Before attacking the main problem to find the maps 
$MF_W(X)\longrightarrow D(M)$
in large volume phases,
we first fill this gap at LG orbifold points.

\subsection{Landau-Ginzburg Orbifold Point}
\label{subsec:LGOpt}

We need to establish the equivalence of the two descriptions
at a Landau-Ginzburg point,
one with the full set of fields 
subject to the D-term relation and the other with the reduced set of fields
without any relation.
This problem
also exists in the model without superpotential $W=0$,
at a free orbifold point,
 and could have been
discussed in Section~\ref{sec:noncompact}. The treatment we will
provide below can be easily adapted for that case as well, and
we will leave that to the reader.  

We first discuss the problem in a specific class of examples --- Example (A).
Once that is done, the generalization is straightforward
and will be described only briefly.

\subsubsection{The $U(1)$ Theory}

Eaxmple (A) is a $U(1)$ gauge theory with 
$(N+1)$ fields $P,X_1,\ldots, X_N$ of charge $-N,1,\ldots,1$.
The superpotential is $W=PG(X)$ where $G(X)$ is a degree $N$
polynomial of $X_1,\ldots,X_N$.
The limit $r\to-\infty$ is the Landau-Ginzburg orbifold
point where $p$ has a vacuum expectation value that breaks the gauge group
$U(1)$ to the subgroup $\Z_N$ of $N$-th roots of unity.
The $P$ and the gauge multiplets decouple together by
acquiring infinite Higgs mass.

\subsubsection*{\it Elimination Of $P$}
\label{subsec:LSMtoLG}

\newcommand{\QLG}{\overline{Q}}
\newcommand{\rhoLG}{\overline{\rho}}
\newcommand{\RLG}{\overline{R}}

Elimination of $P$ is straightforward --- simply set it equal to its
expectation value which can be chosen to be 1.
Of course, since the field $p$ carries R-charge $2$, we must
change the gauge for the R-symmetry transformation. It is done as follows.

Let $(\cp,Q,\rho,R)$ be a D-brane in the linear sigma model.
We set
\beqa
&&\QLG(x):=Q(1,x),
\nn\\
&&\rhoLG(\omega):=\rho(\omega), \qquad \mathrm{with}\quad \omega^N = 1,
\label{LSMtoLG}
\\
&&\RLG(\lambda):=R(\lambda)\rho(\lambda^{-2/N}).
\nn
\eeqa
$\QLG(x)$ is of course a matrix factorization of $W=G(x)$.
It is invariant under the orbifold group action,
$\rhoLG(\omega)\QLG(\omega\cdot x)\rhoLG(\omega)^{-1}
=\QLG(x)$, and has R-charge $1$,
\beqa
\RLG(\lambda)\QLG(\lambda^{2/N}x)
\RLG(\lambda)^{-1}
\!&=&\!R(\lambda)\rho(\lambda^{2/N})^{-1}
Q(1,\lambda^{2/N}x)\rho(\lambda^{2/N})
R(\lambda)^{-1}\nn\\
&=&\!R(\lambda) Q(\lambda^2,x)
R(\lambda)^{-1}
\nn\\
&=&\!\lambda Q(1,x)
\,=\,\lambda\QLG(x).
\nn
\eeqa
The operators
$\rhoLG(\omega)$ and $R(\e^{i\alpha})$ of course commute
and satisfy
$$
\RLG(\e^{\pi i})\rhoLG(\e^{2\pi i/N})
=R(\e^{\pi i})\rho(\e^{-2\pi i/N})\rho(\e^{2\pi i/N})
=\sigma_{\cp},
$$
Thus, we obtained a D-brane $(\cp,\QLG,\rhoLG,\RLG)$ 
in the LG orbifold.

\subsubsection*{\it Inclusion Of $P$}

Let us next find the map in the opposite direction ---
from a D-brane in the LG orbifold to a D-brane in
$MF_W(X_{\rm orb})$. Actually, we will be able to 
find a map directly into the grade restricted
subset $\MFT$ in the linear sigma model, for any window $w$
at the phase boundary.
This will be particularly useful for D-brane transport
from the LG point to the large volume regime.
We recall that the set $\mathfrak{C}^{w}$
of admissible charges for a window $w$ is a set of $N$ consequetive
integers. In particular, it has the property
\beq
q,q'\in \mathfrak{C}^w
\Longrightarrow |q-q'|\leq N-1.
\label{Nw}
\eeq

Let $(\cp,\QLG,\rhoLG,\RLG)$ be a D-brane in the LG orbifold.
One can choose a basis in which $\RLG(\lambda)$ and $\rhoLG(\omega)$
are simultaneously diagionalized, with eigenvalues $\lambda^{\RLG_{\bf i}}$
and $\omega^{\overline{q}_{\bf i}}$ respectively.
Here we use the indices ${\bf i}$, ${\bf j},\ldots$
to label the Chan-Paton basis vectors.
$\RLG_{\bf i}$ is a rational number and
$\overline{q}_{\bf i}$ is a mod $N$ integer.
Since $\RLG(\e^{\pi i})\rhoLG(\e^{2\pi i/N})=\sigma_{\cp}$,
we find that $\e^{\pi i \RLG_i}\e^{2\pi i \overline{q}_i/N}$ is
$1$ or $-1$ depending on whether the basis element is even
($\sigma_{\bf i}=+1$) or odd ($\sigma_{\bf i}=-1$).
Namely, $\RLG_{\bf i}+2\overline{q}_{\bf i}/N$ is an even or odd
integer depending on $\sigma_{\bf i}=+1$ or $\sigma_{\bf i}=-1$.
In fact, for such an $\RLG_{\bf i}$
the equation
\beq
\RLG_{\bf i}=R_{\bf i}-{2q_{\bf i}\over N},
\label{qeq}
\eeq
is uniquely solved by integers $(R_{\bf i},q_{\bf i})$ from the range
\beqa
&&q_{\bf i}\in \mathfrak{C}^w,
\nn\\
&& R_{\bf i}\in \left\{\begin{array}{ll}
2\Z& \sigma_{\bf i}=+1\\
2\Z+1& \sigma_{\bf i}=-1.
\end{array}\right.
\nn
\eeqa
Let us now put
\beq
\rho(g):=\left(\begin{array}{ccc}
g^{q_1}&&0\\
&\ddots&\\
0&&g^{q_{2\ell}}
\end{array}\right),
\qquad
R(\lambda)
:=\left(\begin{array}{ccc}
\lambda^{R_1}&&0\\
&\ddots&\\
0&&\lambda^{R_{2\ell}}
\end{array}\right).
\label{defrhoRlsm}
\eeq
Then we have $\RLG(\lambda)=R(\lambda)
\rho(\lambda^{-2/N})$ and $q_{\bf i}=\overline{q}_{\bf i}$ mod $N$, that is,
$\rho(\omega)=\rhoLG(\omega)$ for an $N$-th root of unity $\omega$.
Hence, by the orbifold invariance of
$\QLG(x)$ we have 
$$
\rho(\omega)^{-1}\QLG(\omega x)\rho(\omega)=\QLG(x),\qquad
\mbox{if $\omega^N=1$}.
$$
This means that 
$
\rho(z)^{-1}\QLG(z x)\rho(z)
$
depends on $z$ only though $z^N$. 
Since $\QLG(z x)$ is a polynomial of
$z x_i$'s and since conjugation by $\rho(z)$ can change the
power of $z$ only at most by $z^{\pm (N-1)}$
(see (\ref{Nw})), we find that 
$
\rho(z)^{-1}\QLG(z x)\rho(z)
$
has no negative powers
of $z^N$, that is,
it is a polynomial in $z^N$,
$$
\rho(z)^{-1}\QLG(z x)\rho(z)
=\QLG_0(x)+z^N \QLG_1(x)+z^{2N} \QLG_2(x)+\cdots.
$$
$\QLG_i(x)$ is the component of
$\QLG(x)$ such that $\rho(z)^{-1}\QLG_i(z x)\rho(z)=
z^{i\, N}\QLG_i(x)$.
Now we define $Q(p,x)$ by replacing $z^N$ by $p$:
\beq
Q(p,x):=
\QLG_0(x)+p\QLG_1(x)+p^2\QLG_2(x)+\cdots.
\label{defQlsm}
\eeq
Let us perform some checks on this $Q(p,x)$.
Using the gauge transformation property of $\QLG_i(x)$,
we see that the equation 
$\QLG(x)^2=G(x)\cdot {\rm id}_{\cp}$ decomposes into
the following set of equations
\beqa
&&
\QLG_0(x)^2=0,\nn\\
&&
\QLG_0(x)\QLG_1(x)+\QLG_1(x)\QLG_0(x)=G(x){\rm id}_{\cp},\nn\\
&&
\QLG_0(x)\QLG_2(x)+\QLG_1(x)^2+\QLG_2(x)\QLG_0(x)=0,\nn\\
&&
\cdots\cdots
\nn
\eeqa 
These are compiled into a single equation
$$
Q(p,x)^2=pG(x)\cdot {\rm id}_{\cp},
$$
which says that $Q(p,x)$ is indeed a matrix factorization of
$pG(x)$.
Gauge invariance
and R-symmetry simply follow from the definition,
\beqa
\lefteqn{\rho(g)^{-1}Q(g^{-N}p,g x)\rho(g)}\nn\\
&=&\QLG_0(x)+(g^{-N}p)(g^N\QLG_1(x))
+(g^{-N}p)^2(g^{2N}\QLG_2(x))+\cdots\nn\\
&=&Q(p,x),\nn\\[0.1cm]
\lefteqn{R(\lambda)Q(\lambda^2p,x)
R(\lambda)^{-1}}\nn\\
&=&\RLG(\lambda)\rho(\lambda^{2/N})
Q(\lambda^2p,x)\rho(\lambda^{2/N})^{-1}
\RLG(\lambda)^{-1}
\nn\\
&=&
\RLG(\lambda) Q(p,\lambda^{2/N}x)
\RLG(\lambda)^{-1}
=\lambda Q(p,x).
\nn
\eeqa

To summarize, $(\cp,Q,\rho,R)$ given by
(\ref{defrhoRlsm}) and (\ref{defQlsm})
has all the properties to define
 a D-brane in the linear sigma model.
We have constructed a map that sends D-branes in the LG orbifold
to grade restricted D-branes in the linear sigma model.
It is clear that this
is the inverse of the map 
$(\cp,Q,\rho,R)\mapsto (\cp, \QLG,\rhoLG,\RLG)$
which was obtained by setting $p=1$ and modifiying the R-symmetry by
an appropriate gauge transformation.

\subsubsection*{\it Example}

If we apply the map $(\cp,Q,\rho,R)\mapsto (\cp, \QLG,\rhoLG,\RLG)$
to the D-branes $\lsmB_-$, $\lsmB_+$ and $\lsmB_0$ 
that were introduced in (\ref{LGtrivialMF}),
(\ref{LVtrivialMF}) and (\ref{masslessMF}), we find
\beqa
&&  \lsmB_-\longmapsto
  \Biggl({\mathcal O}(\overline{2})^{}_{{}_{-{4\over 3}}} 
\MFarrow{{\rm id}}{G} 
{\mathcal O}(\overline{2})^{}_{{}_{-{1\over 3}}}\Biggr),
\nn\\
&&
  \lsmB_+,\lsmB_0\longmapsto
\Biggl(
  \begin{array}{c}
    {\mathcal O}(\overline{2})^{}_{{}_{-{4\over 3}}} \\[-5pt]
    \oplus    \\[-5pt]
    {\mathcal O}(\overline{1})^{\oplus 3}_{{}_{-{2\over 3}}}
  \end{array}
  \MFarrow{g}{f}
  \begin{array}{c}
    {\mathcal O}(\overline{0})^{\oplus 3}_{{}_{-1}} \\[-5pt]
    \oplus    \\[-5pt]
    {\mathcal O}(\overline{2})^{}_{{}_{-{1\over 3}}}
  \end{array}\Biggr),
\nn
\eeqa
where $(f,g)$ are obtained from $(f_+,g_+)$ or $(f_0,g_0)$
by setting $p=1$. 
The subscripts are the Chan-Paton R-charges $\overline{R}_{\bf i}$ 
in the low energy theory.
We indeed find the empty brane $(1,G)$ as the image of $\lsmB_-$.
Also, $\lsmB_+$ and $\lsmB_0$ are mapped to the same D-brane.
(The image is equal to the Recknagel-Schomerus brane
${\mathcal B}_{(0,0,0),2,1}$ in the notation of Section~\ref{subsub:LGO}.)
If we apply the inverse map to the two LG branes obtained above,
with the grade restriction rule $\mathfrak{C}^w=\{-1,0,1\}$, 
we find respectively
$$
\Biggl(\wilson(-1)^{}_{{}_{-2}}\MFarrow{{\rm id}}{pG}
\wilson(-1)^{}_{{}_{-1}}\Biggr),
\quad
\Biggl(
  \begin{array}{c}
    \wilson(-1)^{}_{{}_{-2}} \\[-5pt]
    \oplus    \\[-5pt]
    \wilson(1)^{\oplus 3}_{{}_0}
  \end{array}
  \MFarrow{g_0}{f_0}
  \begin{array}{c}
    \wilson(0)^{\oplus 3}_{{}_-1} \\[-5pt]
    \oplus    \\[-5pt]
    \wilson(-1)^{}_{{}_{-1}}
  \end{array}\Biggr).
$$
The latter is $\lsmB_0$.

\subsubsection{General Case}

In a general linear sigma model,
a Landau-Ginzburg orbifold phase occurs when
the gauge group $U(1)^k$ is broken to a finite subgroup $\Gamma$
by non-zero values of $k$ fields, say $Y_1,\ldots, Y_k$,
and the D-term equation can be solved for arbitrary values of
the remaining fields, $X_1,\ldots, X_{N-k}$.
The deleted set is then
\beq
\Delta=\bigcup_{i=1}^k\, \{y_i=0\}.
\label{orbdelset}
\eeq
There is a fractional
change of basis of the gauge group that brings
the charges into the following form
$$
\widetilde{Q}_{Y_i}^a=-\delta^a_i,\qquad \widetilde{Q}^a_{X_j}\geq 0.
$$
In such a basis, the orbifold phase is $r^a<0$.
Indeed, the D-term equation can be solved at $r^a<0$ for arbitrary $x_j$'s
by setting $|y_i|^2=\sum_{j}\widetilde{Q}^i_{X_j}|x_j|^2-r^i>0$.
In what follows, we use the original integral basis, with $Q_j^a\in\Z$.
The only thing we need to know is that the $k\times k$ matrix
$Q^a_{Y_i}$ is invertible over rational numbers
and that $\widetilde{Q}^b_{X_j}:=-[(Q_Y)^{-1}Q_X]^b_j$ are non-negative.

The low energy theory is the LG orbifold with variables
$X=(X_1,...,X_{N-k})$ and superpotential
$$
W(X)=W(X_1,\ldots,X_{N-k},1,...,1).
$$
The orbifold group $\Gamma$ consists of elements $g=(g_1,...,g_k)$
of $U(1)^k$ obeying 
$$
\prod_{a=1}^kg_a\!\!{}^{Q^a_{Y_i}}=1,\quad\forall i=1,\ldots, k.
$$ 
Recall from (\ref{congru})
that the R-charges $R_{X_j}, R_{Y_i}$ of the fields $X_j,Y_i$
are even integers.
The R-symmetry action on $Y_j$'s can be trivialized by dressing with the
gauge transformation
$g_s(\lambda)=(\lambda^{s_1},\ldots,\lambda^{s_k})$
where $(s_1,\ldots, s_k)$ is the unique rational solution to the system of
$\Q$-linear equations
$$
\sum_{a=1}^ks_aQ^a_{Y_i}+R_{Y_i}=0,\quad\forall i=1,\ldots, k.
$$
The R-charges of $X_i$ in the low energy LG model are thus
$$
\overline{R}_{X_j}=R_{X_j}+\sum_{a=1}^ks_aQ^a_{X_j}.
$$
The element
$g_s(\e^{\pi i})=(\e^{\pi is_1},\ldots,\e^{\pi i s_k})$
acts trivially on $y_i$'s,
$\prod_a\e^{\pi i s_aQ^q_{Y_j}}=\e^{-\pi i R_{Y_j}}=1$,
and acts on $x_i$'s by the phase
$
\prod_a\e^{\pi i s_aQ^a_{X_j}}
=\e^{\pi i R_{X_j}}\e^{\pi i \sum_as_aQ^a_{X_j}}
=\e^{\pi i \overline{R}_{X_j}},
$
where we used the fact that the $R_{Y_i}$'s and the $R_{X_j}$'s are
even integers.
Namely, it belongs to the orbifold group $\Gamma$
and acts on the Landau-Ginzburg fields $x_j$
in the same way as the R-symmetry at $\lambda=\e^{\pi i}$.
Therefore, we can identify $g_s(\e^{\pi i})
=(\e^{\pi is_1},\ldots,\e^{\pi i s_k})$
as the element $\omega_{-1}\in \Gamma$ that was
defined in Section~\ref{subsub:LGO}.

\subsubsection*{\it Elimination Of $Y_i$'s}

Given a brane
$(\cp,Q,\rho,R)$ in the theory with full set of variables,
we set
\beqa
&&
\overline{Q}(x):=Q(x_1,...,x_{n-k},1,...,1)
\nn\\
&&
\overline{\rho}(\omega):=\rho(\omega),\qquad\mbox{with $\omega\in \Gamma$},
\label{LSMtoLG2}\\
&&
\overline{R}(\lambda):=R(\lambda)\rho(g_s(\lambda))^{-1}.
\nn
\eeqa
This $(\cp,\overline{Q},\overline{\rho},\overline{R})$
defines a D-brane in the LG orbifold.

\subsubsection*{\it Inclusion Of $Y_i$'s}

Let $(\cp,\overline{Q},\overline{\rho},\overline{R})$ be a D-brane
in the LG orbifold.
We denote the eigenvalues of
 $\overline{R}(\lambda)$ and
$\overline{\rho}(\omega)$
as $\lambda^{\overline{R}_{\bf i}}$ and 
$\prod_{a}\omega_a^{\overline{q}^a_{\bf i}}$.
It follows from 
$\overline{R}(\e^{\pi i})\overline{\rho}(\omega_{-1})=\sigma_{\cp}$ 
that $\overline{R}_{\bf i}+\sum_{a=1}^ks_a\overline{q}^a_{\bf i}$
is an even (odd) integer if
${\bf i}$ labels an even (odd) Chan-Paton vector.
Analogously to the $T=U(1)$ case,
we would like to find an even (odd) integer $R_{\bf i}$ and a set of integers
$q_{\bf i}^a\equiv \overline{q}^a_{\bf i}$
satisfying the equation
$$
\overline{R}_{\bf i}=R_{\bf i}-\sum_{a=1}^ks_aq^a_{\bf i}.
$$
There is a unique solution if we require $q^a_{\bf i}$
to be in a bounded range
such that 
$\widetilde{q}^a_{\bf i}:=-[Q_Y^{-1}q]^a_{\bf i}$ 
is in an open interval of width $1$.
Since the deleted set is
the union of the hypersurfaces $y_j=0$,
(\ref{orbdelset}),
 this bound precisely matches
with a fundamental domain of the Chan-Paton charges with respect to
the D-isomorphism relations
of the Wilson line branes.

Once the solution $(q^a_{\bf i},R_{\bf i})$
is found for each ${\bf i}$, we define $\rho(g)$ and $R(\lambda)$ as in the
$T=U(1)$ case. 
It follows from the orbifold invariance of $\overline{Q}(x)$ that
$\rho(g)^{-1}\overline{Q}(gx)\rho(g)$ depends on
$g=(g_1,\ldots,g_k)$ only through
$$
\qquad h_j:=\prod_{a=1}^kg_a\!\!{}^{-Q^a_{Y_j}},\quad i=1,\ldots, k.
$$
It is in fact a polynomial of $h_1,\ldots, h_k$,
due to the bound on $q^a_{\bf i}$, as one can see 
in the fractional basis of $U(1)^k$ mentioned at the beginning.
In particular, one can write $\overline{Q}(x)
=\sum_{\vec{n}}\overline{Q}_{\vec{n}}(x)$ such that
$\rho(g)^{-1}\overline{Q}_{\vec{n}}(gx)\rho(g)=h^{\vec{n}}\cdot
\overline{Q}_{\vec{n}}(x)$ where $\vec{n}$ is a $k$-tuple of
non-negative integers $n_1,...,n_k$,
and $h^{\vec{n}}$ is a short-hand notation for
the product $h_1^{n_1}\cdots h_k^{n_k}$.
Setting
$$
Q(x,y)=\sum_{\vec{n}}y_1^{n_1}\cdots y_k^{n_k}\overline{Q}_{\vec{n}}(x),
$$
we obtain a D-brane data $(\cp, Q,\rho,R)$ in the linear sigma model.
We have constructed the inverse to the map 
$(\cp, Q,\rho,R)\mapsto (\cp, \overline{Q},\overline{\rho},\overline{R})$
in (\ref{LSMtoLG2}).

\subsection{Geometric Phase}
\label{subsec:LVphase}

We now attack the problem of integrating out the bulk fields
that acquire mass from the superpotential F-term.
In fact, we had already considered the same problem in the context of
a simple Landau-Ginzburg model in Section~\ref{subsec:Knorrer},
where we integrated out two variables, $U$ and $V$, which enter
into the term $UV$ of superpotential.
We shall refer to our result of Section~\ref{subsec:Knorrer},
$(\cp,Q)\mapsto (\widehat{\cp},\widehat{Q})$ given in (\ref{UVtoIR})
and (\ref{UVtoIRcp}), as the {\it Kn\"orrer map}.
Our current situation can be regarded as
a massive LG model fibred over the subvariety
$M\subset X$, with a non-degenerate quadratic superpotential as a function of
the transverse coordinates. 
Thus, we may simply apply the Kn\"orrer map fibrewise in this setup.

In the first part of this subsection, 
we directly apply the Kn\"orrer map
and obtain the map of matrix factorizations
in $MF_W(X)$ to geometric D-branes in $D(M)$.
In the second part, we reformulate the Kn\"orrer map
and find the inverse map, from geometric D-branes
in $D(M)$ to matrix factorizations in $MF_W(X)$.

\subsubsection{$MF_W(X)\to D(M)$}
\label{subsub:LVred}

We first consider Example (A) with the superpotential
\beq
W=PG(X_1,\ldots, X_N).
\label{ExAW}
\eeq
The geometric phase is $r\gg 0$, and the description at the intermediate scale
is the LG model with this superpotential over the total space
$X=X_+$ of the holomorphic line numdle ${\mathcal O}(-N)\to \CP^{N-1}$.
The superpotential is critical at the submanifold
$\{P=G=0\}\subset X$, which is a degree $N$ hypersurface $M$ of 
the projective space $\CP^{N-1}$. 
We may say that we have a LG model with superpotential $W=PG$
fibred over the hypersurface $M$.
Thus, we can simply apply the Kn\"orrer map fibrewise.
We obviously have two options, one with the identification $U=P$ and $V=G$
and the other with $U=G$ and $V=P$. The two must give rise to the
same answer provided we do everything correctly.
In this first part, we take the former option.
The latter option turns out to be useful to find the inverse map
and will be considered in the second part.

Let $(\cp,Q,\rho,R)$ be a linear sigma model brane representing an element of
$MF_W(X)$. Following the Kn\"orrer map we put
\beqa
&&
\widehat{\cp}=\cp\oplus p\cp\oplus p^2\cp\oplus p^3\cp\oplus\cdots,
\label{hatcp}
\\
&&\widehat{Q}=Q|_{G(x)=0}.
\label{hatQ}
\eeqa
$\widehat{Q}$ is regarded as acting on the infinite dimensional space
$\widehat{\cp}$. The gauge and the R-symmetry actions
$\widehat{\rho}$, $\widehat{R}$ on the new Chan-Paton space $\widehat{\cp}$
are naturally determined by the respective charges of $p$,
up to an overall shift ambiguity. 
This ambiguity is fixed by setting
\beqa
&&
\widehat{\rho}=\rho\quad\mbox{on $\cp\subset \widehat{\cp}$},
\label{hatrho}
\\
&&
\widehat{R}=R\quad\mbox{on $\cp\subset \widehat{\cp}$}.
\label{hatR}
\eeqa
As we have discussed in Section~\ref{subsec:Knorrer}, we need
to make a choice of the overall R-charge assignment, and 
(\ref{hatR}) is just one choice.
As for the gauge group action, on the other hand,
there is no such ambiguity. We will see that
(\ref{hatrho}) is the correct identification, as long as
the B-field on $M$
is given in terms of the theta parameter by
\beq
B=\theta+N\pi.
\label{B-shift}
\eeq

$(\widehat{\cp},\widehat{Q},\widehat{\rho},\widehat{R})$ is the data of
the D-brane in the low energy sigma model.
The Chan-Paton space $\widehat{\cp}$ with
representations $(\widehat{\rho},\widehat{R})$ corresponds to
a graded vector bundle $\widehat{\mathcal E}$ over $M$;
a one-dimensional subspace of $\widehat{\cp}$ of gauge charge
$q$ and R-charge $j$ corresponds to a rank one subbundle
of $\widehat{\mathcal E}$ at degree $j$
which is isomorphic to the line bundle ${\mathcal O}(q)$
on $M$. 
$\widehat{Q}$ acts on $\widehat{\mathcal E}$ as a degree 1 bundle map.
Since we are on the hypersurface $\{G(x)=0\}$ it squares to zero
$$
\widehat{Q}^2=0.
$$
Thus, we have a complex of vector bundles.
Since $p$ has R-charge $2>0$, the subspace
$p^m\cp\subset \widehat{\cp}$ for larger $m$
corresponds to a subbundle with larger R-degree.
Therefore, if the original Chan-Paton space $\cp$
was finite dimensional, 
the degree $j$ subbundle ${\mathcal E}^j$ has finite rank for each $j$,
and 
the complex $(\widehat{\mathcal E},\widehat{Q})$
is bounded from the left, but unbounded to the right;
\beq
0\to{\mathcal E}^{j_{\rm m}}
\stackrel{d_{j_{\rm m}}}{\longrightarrow}
{\mathcal E}^{j_{\rm m}+1}\to\cdots\to
{\mathcal E}^j
\stackrel{d_j}{\longrightarrow}
{\mathcal E}^{j+1}\to\cdots\,\, .
\label{semiinfinite}
\eeq
Here $j_{\rm m}$ is the minimum of the R-charges of $\cp$.
Although it is an unbounded complex, it is exact at large enough
R-degrees. 
This follows from the corresponding property of
the Kn\"orrer map, see Section~\ref{subsec:Knorrer}.
Indeed, the infra-red empty blocks of the matrix $\widehat{Q}$ 
shown in (\ref{KnorrerhatQ})
are on subspaces of large R-charges if we assign R-charge 2 to 
the variable $u$.
There is also an alternative, algebraic proof of the exactness at
large degrees, which will be given momentarily.
In either way, we know that it is quasi-isomorphic to a bounded complex of
coherent sheaves, such as
$$
0\to{\mathcal E}^{j_{\rm m}}
\stackrel{d_{j_{\rm m}}}{\longrightarrow}
{\mathcal E}^{j_{\rm m}+1}\to\cdots\to
{\mathcal E}^{l}
\stackrel{d_{l}}{\longrightarrow}
{\rm Ker}\, d_{l+1}\to 0,
$$
for some large $l$.

Let us illustrate the procedure using the brane
$\lsmB_-$ in the $N=3$ case, which was introduced in (\ref{LGtrivialMF}).
The original Chan-Paton space $\cp$ is two-dimensional, and let us denote
the basis vectors by ${\bf e}_{2,0}$, ${\bf e}_{-1,-1}$.
The basis elements $p^m{\bf e}_{2,0}$, $p^m{\bf e}_{-1,-1}$
of the new Chan-Paton space $\widehat{\cp}$
have (gauge,R)-charges $(2+3m,2m)$, $(-1+3m,-1+2m)$ respectively
and thus correspond to the line bundles
${\mathcal O}(2+3m)^{}_{{}_{2m}}$, ${\mathcal O}(-1+3m)^{}_{{}_{-1+2m}}$
over the elliptic curve $M$.
$\widehat{Q}$ maps the basis elements as follows
$$
p^m{\bf e}_{2,0}\mapsto p^{m+1}{\bf e}_{-1,-1},\quad
p^m{\bf e}_{-1,-1}\mapsto 0.
$$
Thus, we find the complex
$$
0\to {\mathcal O}(-1)_{-1}\stackrel{0}{\longrightarrow}
{\mathcal O}(2)_0\stackrel{\rm id}{\longrightarrow}
{\mathcal O}(2)_1\stackrel{0}{\longrightarrow}
{\mathcal O}(5)_2\stackrel{\rm id}{\longrightarrow}
{\mathcal O}(5)_3\stackrel{0}{\longrightarrow}\cdots
$$
We find infinite copies of trivial brane-antibrane pairs.
Eliminating them all, we are left with the finite complex
$0\to{\mathcal O}(-1)_{{}_{-1}}\to 0$.
Hence, the Kn\"orrer map yields
$$
\pi_+(\lsmB_-)\longmapsto
{\mathcal O}(-1)[1].
$$
Let us consider another example, the brane
$\lsmB_+$ given in (\ref{LVtrivialMF}).
As above, for each Wilson line brane $\wilson(q)_i$ we have
an infinite series of line bundles ${\mathcal O}(q+3m)_{i+2m}$
on the elliptic curve. We first write the entries in (\ref{LVtrivialMF}),
from left to right with respect to the R-charge,
draw the arrows from the $p$-independent part of 
$(f_+,g_+)$,
and replace
the notation ``$\wilson$'' by ``${\mathcal O}$'':
$$
\begin{array}{ccccccc}
{\mathcal O}(-1)^{}_{{}_{-2}}\!\!\!\!&\stackrel{\!{}_{{}^{}_X}}{\to}&
\!\!{\mathcal O}(0)^{\oplus 3}_{{}_{-1}}\!\!&&&&\\
&&&\searrow\!\!\!\!\!{}^{{}^{{}_X}}\!\!\!&&&\\
&&&&\!\!{\mathcal O}(1)^{\oplus 3}_{{}_0}\!\!&\stackrel{\!{}_{{}_X}}{\to}
&\!\!{\mathcal O}(2)^{}_{{}_1} \\
\end{array}
$$ 
We next write its copies, shifted to the right by two
and tensored by ${\mathcal O}(3)$, 
and draw the arrows including the $p$-dependent part of
$(f_+,g_+)$:
$$
\qquad\qquad\begin{array}{cccccccccccc}
{\mathcal O}(-1)^{}_{{}_{-2}}\!\!\!\!&\stackrel{\!{}^{}_X}{\to}
&\!\!{\mathcal O}(0)^{\oplus 3}_{{}_{-1}}\!\!&
\,\stackrel{\!{}_{{}_{X^2}}}{\to}\!&\!\!{\mathcal O}(2)^{}_{{}_0}\!\!
&\stackrel{\!{}_{{}_X}}{\to}&\!\!{\mathcal O}(3)^{\oplus 3}_{{}_1}\!\!
&\,\stackrel{\!{}_{{}_{X^2}}}{\to}\!&\!\!{\mathcal O}(5)^{}_{{}_2}\!\!
&\stackrel{\!{}_{{}_X}}{\to}&\!\!{\mathcal O}(6)^{\oplus 3}_{{}_3}\!\!
&\,\stackrel{\!{}_{{}_{X^2}}}{\to}\!\\
&&&\searrow\!\!\!\!\!{}^{{}^{{}_X}}\!\!\!
&\oplus&\!\!\!{}^{{}_{{}^{X^2}}}\!\!\!\!\!\nearrow 
&\oplus&\searrow\!\!\!\!\!{}^{{}^{{}_X}}\!\!\!
&\oplus&\!\!\!{}^{{}_{{}^{X^2}}}\!\!\!\!\!\nearrow 
&\oplus&\searrow\!\!\!\!\!{}^{{}^{{}_X}}\!\!\!
\\
&&&&\!\!{\mathcal O}(1)^{\oplus 3}_{{}_0}\!\!&\stackrel{\!{}_{{}_X}}{\to}
&\!\!{\mathcal O}(2)^{}_{{}_1}\!\!&\,\stackrel{\!{}_{{}_{X^2}}}{\to}\!
&\!\!{\mathcal O}(4)^{\oplus 3}_{{}_2}\!\!&\stackrel{\!{}_{{}_X}}{\to}
&\!\!{\mathcal O}(5)^{}_{{}_3}\!\!&\,\stackrel{\!{}_{{}_{X^2}}}{\to}\! 
\end{array}\cdots
$$
This is the complex $(\widehat{\mathcal E}_+,\widehat{Q}_+)$.
It is exact. The exactness can be proven algebraically, but one may also 
prove it by computing the potential $\{\widehat{Q}_+,\widehat{Q}_+^{\dag}\}$
and showing that it is positive everywhere: 
The potential is $|x|^2+|x|^4=\sum_i(|x_i|^2+|x_i|^4)$
except at the first few terms, ${\mathcal O}(-1)_{{}_{-2}}$,
${\mathcal O}(0)^{\oplus 3}_{{}_{-1}}$, and
${\mathcal O}(1)^{\oplus 3}_{{}_0}$,
where it is $|x|^2$, $|x|^2\delta_{i,j}+\bar x_i^2x_j^2$, and
$(|x|^2+|x|^4)\delta_{i,j}-x_i^2\bar x_j^2$ respectively.
All the eigenvalues are bounded from below by $|x|^2$
and are positive as long as $(x_1,x_2,x_3)\ne (0,0,0)$.
Therefore, the brane determined by 
$(\widehat{\mathcal E}_+,\widehat{Q}_+)$
is infra-red empty, which is indeed expected from $\pi_+(\lsmB_+)\cong 0$.

If we apply this procedure to the brane $\lsmB_0$ given in 
(\ref{masslessMF}), we immediately find the complex 
$(\widehat{\mathcal E}_0,\widehat{Q}_0)$:
$$
\qquad\qquad\begin{array}{cccccccccccc}
{\mathcal O}(-1)^{}_{{}_{-2}}\!\!\!\!&\stackrel{\!{}_{{}_X}}{\to}
&\!\!{\mathcal O}(0)^{\oplus 3}_{{}_{-1}}\!\!&
\,\stackrel{\!{}_{{}_{X^2}}}{\to}\!&\!\!{\mathcal O}(2)^{}_{{}_0}\!\!
&\stackrel{\!{}_{{}_X}}{\to}&\!\!{\mathcal O}(3)^{\oplus 3}_{{}_1}\!\!
&\,\stackrel{\!{}_{{}_{X^2}}}{\to}\!&\!\!{\mathcal O}(5)^{}_{{}_2}\!\!
&\stackrel{\!{}_{{}_X}}{\to}&\!\!{\mathcal O}(6)^{\oplus 3}_{{}_3}\!\!
&\,\stackrel{\!{}_{{}_{X^2}}}{\to}\!\\
&&\oplus&\searrow\!\!\!\!\!{}^{{}^{{}_X}}\!\!\!
&\oplus&\!\!\!{}^{{}_{{}^{X^2}}}\!\!\!\!\!\nearrow 
&\oplus&\searrow\!\!\!\!\!{}^{{}^{{}_X}}\!\!\!
&\oplus&\!\!\!{}^{{}_{{}^{X^2}}}\!\!\!\!\!\nearrow 
&\oplus&\searrow\!\!\!\!\!{}^{{}^{{}_X}}\!\!\!
\\
&&\!\!{\mathcal O}(-1)^{}_{{}_{-1}}\!\!&\,\stackrel{\!{}_{{}_{X^2}}}{\to}\!
&\!\!{\mathcal O}(1)^{\oplus 3}_{{}_0}\!\!&\stackrel{\!{}_{{}_X}}{\to}
&\!\!{\mathcal O}(2)^{}_{{}_1}\!\!&\,\stackrel{\!{}_{{}_{X^2}}}{\to}\!
&\!\!{\mathcal O}(4)^{\oplus 3}_{{}_2}\!\!&\stackrel{\!{}_{{}_X}}{\to}
&\!\!{\mathcal O}(5)^{}_{{}_3}\!\!&\,\stackrel{\!{}_{{}_{X^2}}}{\to}\! 
\end{array}\cdots
$$
Notice that it is almost the same as 
$(\widehat{\mathcal E}_+,\widehat{Q}_+)$
except for the term ${\mathcal O}(-1)^{}_{{}_{-1}}$ with a map to
${\mathcal O}(1)^{\oplus 3}_{{}_0}$. Namely, it is the cone of this map:
$$
(\widehat{\mathcal E}_0,\widehat{Q}_0)
={\rm Cone}\Bigl({\mathcal O}(-1)\stackrel{{}_{X^2}}{\longrightarrow}
(\widehat{\mathcal E}_+,\widehat{Q}_+)\Bigr).
$$
Since $(\widehat{\mathcal E}_+,\widehat{Q}_+)$ is empty, we see that 
our brane $(\widehat{\mathcal E}_0,\widehat{Q}_0)$
is D-isomorphic to just ${\mathcal O}(-1)[1]$ which is
the same as the result for $\lsmB_-$.
This is exactly what is expected from (\ref{QQminus}).

\subsubsection*{\it Theta Shift}

We now show that the Chan-Paton gauge charges 
of the low energy D-brane are given by (\ref{hatrho}) 
with theta parameter shift (\ref{B-shift}).
To this end, we put the derivation of the Kn\"orrer map
in Section~\ref{subsec:Knorrer}
into the context of a $U(1)$ gauge theory.
Thus, we assume that the variables $X_1,...,X_n,U,V$ carry some
$U(1)$ gauge charges so that the superpotential $W=W_L(X)+UV$
is gauge invariant.
We suppose that $U$ and $V$ have charges $Q_u$ and $-Q_u$ respectively.
Then, the boundary variables $\eta$ and $\bareta$ of the brane
(\ref{UVbdry}) must carry gauge charges $-Q_u$ and $+Q_u$ respectively.
Accordingly, the Chan-Paton vectors from the ($U,V$) sector have the
following gauge charges:
\beq
\begin{array}{c|cc}
{\rm vector}&|0\rangle&\bareta |0\rangle\\\hline
\\[-0.5cm]
q&\displaystyle {Q_u\over 2}&\displaystyle -{Q_u\over 2}
\end{array}
\label{chassign}
\eeq
Now suppose we have some brane ${\mathcal A}^L$ in the low energy theory
and consider the brane (\ref{UVbrane}) in the high energy theory.
If we integrate out the $(U,V,\eta)$-system, we get back exactly the
original brane ${\mathcal A}^L$ since the ground state of
the $(U,V,\eta)$-system has zero effective energy and zero effective
charge, see Section~\ref{subsec:massfromW}. 
Let us look into the form of Chan-Paton vectors in this context.
If ${\bf e}_i$ are Chan-Paton vectors of the original low energy brane
${\mathcal A}^L$, then the Chan-Paton vectors of the high energy brane
(\ref{UVbrane}) are of the form 
${\bf e}_i\otimes |0\rangle,{\bf e}_i\otimes \bareta |0\rangle$.
If ${\bf e}_i$'s have gauge charge $q_i$,
then the latter have gauge charges $q_i+{Q_u\over 2}, q_i-{Q_u\over 2}$.
These are the Chan-Paton charges that we see before integrating out
$U$ and $V$.
The Kn\"orrer map is the procedure that takes out 
${\bf e}_i\otimes |0\rangle$.
However, what we want as the Chan-Paton vectors
in the low energy theory are
just ${\bf e}_i$, but not ${\bf e}_i\otimes |0\rangle$.
Thus, to find the correct charge of
the low energy Chan-Paton vectors
we need to subtract the charge of $|0\rangle$ from
the high energy Chan-Paton vectors. This yields the rule
$$
q^L_i=q^H_i-{Q_u\over 2}.
$$
Since the theta parameter contributes to the boundary charge in the form
$q+{\theta\over 2\pi}$, one can relocate this shift to a shift of
the theta parameter
$$
\theta^L=\theta^H-Q_u\pi,
$$ 
now without shifting the $q_i$'s.
In the current set up where $U=P$ and hence $Q_u=-N$, 
this rule leads to (\ref{hatrho}) 
with (\ref{B-shift}).

A shift of the theta angle due to integration of
massive fields was first found in the context of closed string
topological A-model by Morrison and Plesser \cite{MP}.
There, only the shift modulo $2\pi$ matters and indeed
their result is consistent with ours: $\e^{i\Delta \theta}=(-1)^N$.
In the present context of
(physical or topological) open string theoy with B-type boundary conditions,
a shift by an integer multiple of $2\pi$ also matters.
The above result shows the precise shift, including the integral part.

\subsubsection*{\it Left Semi-Infinite Complexes}

That the semi-infinite complex (\ref{semiinfinite})
is exact at large enough degrees can also be proven purely algebraically. 
The essential point is a well known fact in 
the theory of matrix factorizations (Eisenbud \cite{Eisenbud}, 
Proposition 5.1):
{\it Let
$$
Q=\left(\begin{array}{cc}
0&f(x)\\
g(x)&0
\end{array}\right)
$$
be a matrix factorization of some polynomial $W(x)=W(x_1,...,x_n)$,
of size $2\ell\times 2\ell$.
This induces an infinite, 2-periodic complex of modules over the ring
$B=\C[x_1,...,x_n]/W(x)$ 
\beq
\cdots \stackrel{\bar g}{\longrightarrow}
B^{\oplus \ell}\stackrel{\bar f}{\longrightarrow}
B^{\oplus \ell}\stackrel{\bar g}{\longrightarrow}
B^{\oplus \ell}\stackrel{\bar f}{\longrightarrow}\cdots\,\, .
\label{tac}
\eeq
Then this complex is exact.} This is proven as follows.
Suppose some homomorphism 
$\bar\kappa:B^{\oplus k}\to B^{\oplus \ell}$,
induced by an $\ell\times k$ polynomial matrix $\kappa$,
obeys the equation $\bar f\circ \bar\kappa=0$.
This means that the matrix $f\circ \kappa$
is divisible by $W$, that is, there is a $\ell\times k$
matrix $\gamma$ such that $f\circ \kappa=W\gamma$.
Then we have $W\kappa=g\circ f\circ \kappa=Wg\circ\gamma$,
which means $\kappa=g\circ\gamma$. In particular
$\bar\kappa=\bar g\circ\bar\gamma$. This shows the exactness at
$B^{\oplus \ell}\stackrel{\bar g}{\longrightarrow}
B^{\oplus \ell}\stackrel{\bar f}{\longrightarrow}B^{\oplus \ell}$.
The exactness at the other position can be shown in the same way.

In fact, this infinite complex is the totally acyclic complex (\ref{TAC1})
that we met before in Section~\ref{subsec:LG},
where we studied the chiral ring for matrix factorizations. 
We recall that we denoted it by ${\mathcal C}_Q$.

If we replace $\widehat{\cp}$ by
\beq
\bigoplus_{m\in \Z}p^m\cp=\cdots\oplus p^{-2}\cp\oplus p^{-1}\cp\oplus
\cp\oplus p\cp\oplus p^2\cp\oplus\cdots
\label{bigV}
\eeq
in the definition of
$(\widehat{\cp}, \widehat{Q},\widehat{\rho},\widehat{R})$
for $\lsmB=(\cp,Q,\rho,R)$,
then we obtain an infinite complex of vector bundles
unbounded both in the left and right. 
If we regard it as a complex of modules over the ring
$B=\C[x_1,\ldots, x_N]/G(x)$, and if
we ignore the grading, it is 2-periodic
and coincides with the totally acyclic complex
${\mathcal C}_{\QLG}$
associated with the matrix factorization $\QLG=Q|_{p=1}$
of the polynomial $G(x)$. In particular, it is exact
and can be regarded as a graded version of ${\mathcal C}_{\QLG}$.
We denote it by ${\mathcal C}_{\lsmB}$.
Our semi-infinite complex (\ref{semiinfinite})
is obtained from this exact complex
by chopping off the semi-infinite part corresponding to the negative power
components $p^{-m}\cp$.
This proves that (\ref{semiinfinite}) is exact at large enough degrees.

The fact that
the complex $(\widehat{\mathcal E},\widehat{Q})$ is a part of the
exact complex ${\mathcal C}_{\lsmB}$ also
shows that there is an alternative description:
Take the remaining part (i.e. the negative power part 
$\oplus_{m>0}p^{-m}\cp$) and shift it 
by $1$ to the right.
This produces a semi-infinite complex
which is unbounded to the left, 
but bounded from the right. 
Of course, this left semi-infinite complex is quasi-isomorphic to
the right semi-infinite complex $(\widehat{\mathcal E},\widehat{Q})$.
For example, the left semi-infinite version of
the complex $(\widehat{\mathcal E}_0,\widehat{Q}_0)$ that represents
$\pi_+(\lsmB_0)$ is
$$
\cdots\begin{array}{ccccccccc}
{\mathcal O}(-7)^{}_{{}_{-5}}\!\!\!\!&\stackrel{\!{}_{{}_X}}{\to}
&\!\!{\mathcal O}(-6)^{\oplus 3}_{{}_{-4}}\!\!&
\,\stackrel{\!{}_{{}_{X^2}}}{\to}\!&\!\!{\mathcal O}(-4)^{}_{{}_{-3}}\!\!
&\stackrel{\!{}_{{}_X}}{\to}&\!\!{\mathcal O}(-3)^{\oplus 3}_{{}_{-2}}\!\!
&&
\\
\oplus &\!\!\!{}^{{}_{{}^{X^2}}}\!\!\!\!\!\nearrow
&\oplus&\searrow\!\!\!\!\!{}^{{}^{{}_X}}\!\!\!
&\oplus&\!\!\!{}^{{}_{{}^{X^2}}}\!\!\!\!\!\nearrow
&\oplus&\searrow\!\!\!\!\!{}^{{}^{{}_X}}\!\!\!
&\\
{\mathcal O}(-8)^{\oplus 3}_{{}_{-5}}
&\stackrel{\!{}_{{}_X}}{\to}&\!\!{\mathcal O}(-7)^{}_{{}_{-4}}\!\!
&\,\stackrel{\!{}_{{}_{X^2}}}{\to}\!
&\!\!{\mathcal O}(-5)^{\oplus 3}_{{}_{-3}}\!\!&\stackrel{\!{}_{{}_X}}{\to}
&\!\!{\mathcal O}(-4)^{}_{{}_{-2}}\!\!&\,\stackrel{\!{}_{{}_{X^2}}}{\to}\!
&\!\!{\mathcal O}(-2)^{\oplus 3}_{{}_{-1}}\!\!
\end{array}\qquad\qquad\qquad
$$
If we attatch ${\mathcal O}(-1)^{}_{{}_0}$ to this complex 
with the map $(x_1,x_2,x_3)$ from the right-most entry,
${\mathcal O}(-2)^{\oplus 3}_{{}_{-1}}$, the result is 
an exact complex. (This can be seen by showing that the bounday potential
$\{Q,Q^{\dag}\}$ is positive, or
by noting that it is the complement of the exact complex 
$(\widehat{E}_+,\widehat{Q}_+)$ in the exact complex
${\mathcal C}_{\lsmB_+}$.)
Namely, there is a quasi-isomorphism from this complex to
${\mathcal O}(-1)[1]$. Thus, we obtain again the same result for
$\pi_+(\lsmB_0)$.

\subsubsection*{\it Complete Intersection Of Hypersurfaces In Toric Variety}

It is straighforward to generalize the above construction to
the case where $M$ is a complete intersection of hypersurfaces
in a toric variety. Let us consider a $U(1)^k$ gauge theory
which has
$(N+l)$ fields $X_1,\ldots,X_N$, $P_1,\ldots,P_l$ with charge
$Q_1^a,\ldots,Q_N^a$, $-d_1^a,\ldots,-d_l^a$, and the superpotential
\beq
W=\sum_{\beta=1}^lP_{\beta}G_{\beta}(X_1,...,X_N)
\eeq
where $G(X)_{\beta}$ are homogeneous polynomials of degree
$d_{\beta}^a$ with respect to the $a$-th gauge group.
We assign R-charges $0$ to $X_i$'s and $2$ to $P_{\beta}$'s.
We suppose that there is a phase in which the gauge group is completely
broken by the values of the $X_i$'s
and the non-compact variety $X$ is the
total space of the rank $l$ vector bundle 
$\oplus_{\beta=1}^l{\mathcal O}(-\vec{d}_{\beta})$
over a compact toric manifold $X_B$.
Then, the low energy theory is the non-linear sigma model on
the critical locus $P_{\beta}=G_{\beta}=0$ ($\beta=1,\ldots,l$), 
which is the complete intersection $M$ of hypersurfaces
$G_{\beta}(x)=0$ in the toric variety $X_B$.
If we have a D-brane $(\cp,Q,\rho,R)$ in the linear sigma model,
then the brane in the low energy sigma model is given by
\beqa
&&\widehat{\cp}=\bigoplus_{m_1,...,m_l\geq 0}
p_1^{m_1}\cdots p_l^{m_l}\cp,
\label{Vhatci}\\
&&\widehat{Q}=Q|^{}_{G_1=\cdots=G_l=0}.
\eeqa
The gauge and R-symmetry representations
$\widehat{\rho}$ and $\widehat{R}$ are determined by the 
charges of $p_{\beta}$'s and by the initial condition:
$\widehat{\rho}=\rho$, $\widehat{R}=R$ in the subspace
$\cp\subset \widehat{\cp}$, provided the B-field is given by
\beq
B^a=\theta^a+\left(\sum_{\beta=1}^ld_{\beta}^a\right)\pi.
\label{shiftthetacompl}
\eeq
The brane $(\widehat{\cp},\widehat{Q},\widehat{\rho},\widehat{R})$
defines a semi-infinite complex 
$(\widehat{\mathcal E},\widehat{Q})$
of vector bundles over $M$ 
which is bounded from the left but unbounded to the right.
It is exact at large enough degrees and hence is quasi-isomorphic to
a finite complex.
This follows from the property of the Kn\"orrer map, but 
also can be shown algebraically.
Indeed, if we relax the condition on the range of the
sum in (\ref{Vhatci}), for example, if we 
include the sum over negative $m_1$ as well, then we have an exact
complex of modules over the complete intersection ring
$B=\C[x_1,\ldots,x_N]/(G_1,\ldots,G_l)$.
To see that, just regard $B$ as the hypersurface ring of
$A=\C[x_1,\ldots,x_N]/(G_2,\ldots,G_l)$ and apply Eisenbud's proof
to the matrix factorization of $G_1(x)$ over the ring $A$.
In particular, we have various other versions of
$(\widehat{\mathcal E},\widehat{Q})$
corresponding to various different ranges of the sum (\ref{Vhatci}).
As an example, consider the sum where all $m_i$'s run over negative integers.
From this, with shift by $l$ to the right,
we obtain a left semi-infinite complex that is quasi-isomorphic to
$(\widehat{\mathcal E},\widehat{Q})$.

\subsubsection{$D(M)\longrightarrow MF_W(X)$}
\label{subsub:LVlift}

We now construct the inverse map.
The key is to consider the opposite identification
in the Kn\"orrer map, $U=G$, $V=P$.
Setting $p=0$ is straightforward, but extracting power
series of $G$ from a given matrix factorization is hardly practical.
This motivates us to reformulate the Kn\"orrer map.

\subsubsection*{\it Reformulation Of The Kn\"orrer Map}

The reformulation is best described using the language of 
rings and modules that we introduced in Section~\ref{sec:math}. 
Let us first describe the original formulation
of the Kn\"orrer map in that language.

A matrix factorization $(\cp, Q)$ of the superpotential
$W=W_L(x)+uv$ can be regarded as the pair $(M,Q)$ where $M$ is a $\Z_2$-graded
free module over the polynomial ring $\ring=\C[x_1,...,x_n,u,v]$
and $Q$ is an odd endomorphism of $M$ that squares to the
multiplication map by $W$. 
Setting $v=0$ corresponds to replacing $M$ by the module
$M/v=M\otimes_{\ring}(\ring/(v))$ over the ring 
$\ring/(v)$ which is isomorphic to
$$
A=\C[x_1,\ldots,x_n,u].
$$
The Kn\"orrer map image
$(\widehat{M},\widehat{Q})$ is then obtained by
regarding $M/v$ as a module over the ring 
$B=\C[x_1,...,x_n]$ 
which does not include the variable $u$. 
If $M$ has rank $r$ over $\ring$, then
$\widehat{M}$ has infinite rank,
$
\widehat{M}=B^{\oplus r}\oplus uB^{\oplus r}\oplus
u^2B^{\oplus r}\oplus\cdots.
$
As we have seen, this infinite-size matrix factorization
is isomorphic to a finite one, which we denote by 
$(M_L,Q_L)$.

Now let us describe the new formulation.
The key step is to view $B$ as the quotient of $A$ by the ideal $(u)$,
$$
B=A/(u),
$$
and to {\it regard the $B$-module $M_L$ as an $A$-module} by the rule
$a\cdot m:=[a]m$, for $a\in A$ and $[a]\in B$. We denote the result by
$i_*(M_L,Q_L)$. It is a matrix factorization of
$W_L(x)$ over $A$. By itself, $i_*M_L$ is not a free $A$-module
but one can find its free resolution, using the canonical resolution
of $B=A/(u)$,
$$
0\longrightarrow A\stackrel{u\times}{\longrightarrow}A\longrightarrow B
\longrightarrow 0.
$$
Namely, we replace each $B$ in $M_L\cong B^{\oplus s}$ by
$A\oplus A$ and add the map $A\stackrel{u\times}{\longrightarrow}A$
to $Q_L$. In other words, we take the graded tensor product
$\cp_L\otimes_{\C} (\C|0\rangle \oplus \C\bareta|0\rangle)$ and 
consider the sum $Q_L+u\eta$. This free resolution
is nothing but $(M/(v),Q|_{v=0})$, if $Q$ is obtained from $Q_L$
in the way described in Section~\ref{subsec:Knorrer}.
In general, $Q$ is of such a form up to the trivial matrix factorizations
$(1,W)$, $(W,1)$, and the trivial pieces are certainly
trivial even after setting $v=0$.
Thus, we find 
\beq
(M/(v),Q|_{v=0})\cong \mbox{free resolution of $i_*(M_L,Q_L)$}.
\label{Knorrerref}
\eeq
This is the property that the low energy brane $(M_L,Q_L)$ must have.
This can be used to find the inverse map: given a matrix factorization
$(M_L,Q_L)$ of $W_L(x)$ over the ring $B$, 
we push it forward by $i_*$ to a matrix factorization over the ring $A$, 
and then take its free resolution. The final step is
to recover the variable $v$.

\subsubsection*{\it Hypersurface In Projective Space}

Let us apply the above reformulation to Example (A). With the identification
$u=G(x)$ and $v=p$, the rings that appear are
$\ring=\C[x_1,\ldots,x_N,p]$,
\beqa
&&A=\C[x_1,\ldots,x_N],\quad\mbox{and}\nn\\
&&B=\C[x_1,\ldots,x_N]/G(x).
\nn
\eeqa
Of course, everything is graded with respect to the gauge charge,
and we also work in the large volume phase $r\gg 0$
where the locus $\{x_1=\cdots =x_N=0\}$ is deleted.
Namely instead of graded $A$-modules
({\it resp}. $B$-modules)
we consider coherent sheaves over the projective space $\CP^{N-1}$
({\it resp.} the hypersurface $\{G(x)=0\}$).
In this context, the map $i_*$ sending $B$-modules to
$A$-modules becomes the pushforward of sheaves by the embedding
$i: \{G(x)=0\}\hookrightarrow \CP^{N-1}$.
Let $({\mathcal E}_L,Q_L)$ be a bounded complex of coherent sheaves
over the hypersurface $M=\{G=0\}$. The first step is
to push it forward to $\CP^{N-1}$ and take its free resolution.
At this stage, we have a bounded complex of vector bundles over
$\CP^{N-1}$ of the form
$$
{\mathcal C}:\qquad
0\to\cdots
\longrightarrow
\bigoplus_{i=1}^{b_{j-1}}{\mathcal O}(q_{i,{j-1}})
\longrightarrow
\bigoplus_{i=1}^{b_j}{\mathcal O}(q_{i,j})
\longrightarrow
\bigoplus_{i=1}^{b_{j+1}}{\mathcal O}(q_{i,{j+1}})
\longrightarrow
\cdots\to 0
$$
This determines the data $(\cp,Q_0,\rho,R)$ where 
$Q_0$ is a matrix polynomial of $x_1,...,x_N$ obeying
$$
Q_0(x)^2=0.
$$
Of course, $Q_0$ is gauge invariant and has R-charge 1 
with respect to the representation $(\rho,R)$
of the gauge and R-symmetry group.
The final step is to recover the variable $p$, that is, to find
the extension of $Q_0(x)$ to a matrix factorization 
of $pG(x)$,
$$
Q(p,x)=Q_0(x)+pQ_1(x)+p^2Q_2(x)+\cdots,
$$ 
which is gauge invariant and has R-charge $1$.
The condition $Q(p,x)^2=pG(x)\cdot {\rm id}_{\cp}$ can be 
decomposed into a set of equations
\beqa
&& \{Q_0,Q_1\}=G\cdot{\rm id}_{\cp},
\label{eq111}\\
&& \{Q_0,Q_n\}=-{1\over 2}\sum_{l+k=n\atop l,k\geq 1}
Q_lQ_k,\qquad n\geq 2.
\label{eq222}
\eeqa
The (gauge,R)-charge of $Q_m(x)$ must be $(Nm,1-2m)$. 
We shall find such $Q_m$'s recursively, starting with $Q_1$.
Multiplication by $G(x)$ defines a cochain map 
${\mathcal C}\to {\mathcal C}(N)$. 
Since ${\mathcal C}$ is a resolution of
 the complex $i_*({\mathcal E}_L,Q_L)$ supported at $G(x)=0$,
this map has to be null homotopic. This shows that there is a map
$Q_1:{\mathcal C}\to {\mathcal C}(N)$ of degree $-1$
such that $\{Q_0,Q_1\}$ equals the $G(x)$-multiplication.
Thus, we found $Q_1(x)$ of charge $(N,-1)$ obeying (\ref{eq111}).
The rest of $Q_l(x)$ are found by induction.
Suppose that we found $Q_2,...,Q_{m-1}$ that solves (\ref{eq222})
for all $n$ below $m$.
Then
$$
  [Q_0,\sum_{l+k=m\atop l,k\geq 1} 
                         \{Q_l, Q_k\}] =
  \sum_{l+k=m\atop l,k\geq 1} 
\Bigl(
    [\{Q_0,Q_l\}, Q_k] -[Q_l,\{Q_0,Q_k\}]
    \Bigr) = 0
$$
This means that $\sum\{Q_l, Q_k\}:{\mathcal C}\to {\mathcal C}(Nm)$ 
is a cochain map.
The degree of this map is $2-2m$ which is negative.
Since the Ext group ${\rm Ext}^j({\mathcal C},{\mathcal C})$ is zero for
any negative degree $j$, the cochain map $\sum\{Q_l, Q_k\}$
has to be null-homotopic.
This shows the existence of a degree $1-2m$ map 
$Q_m:{\mathcal C}\to {\mathcal C}(Nm)$ 
that satisfies (\ref{eq222}) for $n=m$.

In this way we obtain the data $(\cp,Q,\rho,R)$ of a brane in the 
linear sigma model. Obviously, this is the one that must descend to the
complex $({\mathcal E}_L,Q_L)$ that we started with.
There is one subtlety though, concerning the value of the theta parameter and
the overall shift of R-charge.
The theta parameter must be related to the B-field
of the low energy sigma model
by
\beq
\theta=B+N\pi.
\label{shiftthetaininverse}
\eeq
The choice of R-symmetry is a matter of convention. However, if we
want to be consistent with the one in (\ref{hatR}) that was used in the map
$MF_W(X)\to D(M)$, we need to replace $R(\lambda)$ by
\beq
R(\lambda)\to R(\lambda)\lambda.
\label{shiftRininverse}
\eeq
To see this, let us come back to the derivation of
the Kn\"orrer map and its reformulation.
The lift of a low energy brane $(\cp_L,Q_L)$ to
the high energy theory is given by
$\cp=\cp_L\otimes(\C|0\rangle\oplus \C\bareta|0\rangle)$ and 
$Q=Q_L+u\eta+v\bareta$. In the discussion of the map
$MF_W(X)\to D(M)$ we used the identification $u=p,v=G$,
while we used $u=G,v=p$ for the inverse map $D(M)\to MF_W(X)$. 
However, to compare the two stories, it is better to use a common
framework. We can actually use $u=p,v=G$ also in the discussion of
the inverse map --- we just have to 
change the notation as $\eta=\bareta',\bareta=\eta',
|0\rangle=\bareta'|0\rangle'$.
The charge assignment for $|0\rangle, \bareta|0\rangle$
was $(q,R)=(-{N\over 2},0),({N\over 2},1)$. 
The gauge charge is canonical but the R-charge is a choice.
This choice is the one corresponding to (\ref{hatR}).
Thus, 
in the inverse map, we must use
$$
\begin{array}{c|cc}
{\rm vector}&|0\rangle'&\bareta' |0\rangle'\\\hline
\\[-0.5cm]
(q,R)&({N\over 2},1)& (-{N\over 2},0)
\end{array}
$$
In view of this charge assignment,
the free resolution of $B=A/(v)$
must be
$$
v\eta':A(-\mbox{${N\over 2}$})^{}_{{}_0}{\longrightarrow}
A(\mbox{${N\over 2}$})^{}_{{}_1},
$$
instead of the standard one 
$A(-N)^{}_{{}_{-1}}{\longrightarrow}A(0)^{}_{{}_0}$
which was used above in obtaining $(\rho,R)$.
Thus, we must shift the charges as
\beq
(q_i,R_i)\to (q_i+\mbox{${N\over 2}$},R_i+1).
\label{gradeshift2}
\eeq
The shift of the R-charge is nothing but 
(\ref{shiftRininverse}). The shift of the gauge charge can be traded for
the shift of the B-field $B\to B+N\pi$ which gives
 (\ref{shiftthetaininverse}).

As an example we consider D-branes on the Fermat type
elliptic curve, 
$\hyper = \{G(x) = x_1^3+x_2^3+x_3^3=0\} \subset \CP^2$.
We first consider
the D2-brane with trivial gauge field, the structure
sheaf $\cO$ of the curve, with some value $B_*\in\R$ 
of the B-field.
Its pushforward to $\CP^2$ is the sheaf ${\mathcal O}_M$ supported
at $M\subset \CP^2$, and its
free resolution is given by
${\cal C}: \cO(-3) 
\stackrel{G}{\longrightarrow}\underline{\cO(0)}$.
It is easy to find its lift to the linear sigma model:
$$
  \qquad \Biggl({\wilson(-3)}^{}_{{}_0} \MFarrow{G}{p}
    \wilson(0)^{}_{{}_{1}}\Biggr),\qquad\theta=B_*+3\pi.
$$
If we apply the map $MF_W(X)\to D(M)$ to this brane, 
we obtain the complex on the curve $M$,
$$
{\mathcal O}(-3)^{}_{{}_0}\stackrel{0}{\longrightarrow}
{\mathcal O}(0)^{}_{{}_1}\stackrel{\rm id}{\longrightarrow}
{\mathcal O}(0)^{}_{{}_2}\stackrel{0}{\longrightarrow}
{\mathcal O}(3)^{}_{{}_3}\stackrel{\rm id}{\longrightarrow}
{\mathcal O}(3)^{}_{{}_4}\stackrel{0}{\longrightarrow}
\cdots,\qquad
B=B_*+6\pi.
$$
This is quasi-isomorphic to
${\mathcal O}(-3)$ with $B=B_*+6\pi$ and that is indeed
equivalent to ${\mathcal O}(0)$ with the original value $B=B_*$
of the B-field, the brane we started with.
A more elaborate example is the D0-brane at the point
 $\mathfrak{p} = \{x_1+x_2=x_3=0\}$, the
skyscraper sheaf $\cO_{\mathfrak{p}}$ on $\hyper$. 
A free resolution of its pushforward $i_*{\mathcal O}_{\mathfrak{p}}$ 
to $\CP^2$
is given by
$$
  \cO(q-1) 
  \stackrel{\tiny \left(\!\!\!\begin{array}{c}-x_3\\x_1\!+\!x_2 
      \end{array}\!\!\!\right)}{\longrightarrow}
  \cO(q)^2
  \stackrel{(x_1+x_2,x_3)}{\longrightarrow}
  \underline{\cO(q+1)}
$$
for any $q \in \ZZ$. This readily lifts to the matrix
factorization
$$
 \lsmB_p:~~\Biggl(\,\wilson(q)^{\oplus 2}_{{}_{0}}\,\,
    \MFarrow{g}{f} 
\begin{array}{c}
\wilson(q+1)^{}_{{}_1}\\[-5pt] \oplus\\[-5pt] 
\wilson(q-1)^{}_{{}_{-1}}\end{array}
      \Biggr),\qquad \theta\equiv B_*+3\pi,
$$
with
\beq
    g = \left(
\begin{array}{cc} x_1\!+\!x_2& x_3 \\ 
-px_3^2& p(x_1^2-x_1x_2+x_2^2)\end{array}\right),~~
    f = \left(
\begin{array}{cc} p(x_1^2-x_1x_2+x_2^2)& -x_3 \\ 
px_3^2& x_1\!+\!x_2\end{array}\right).
\label{D0MF}
\eeq

\subsubsection*{\it Complete Intersection Of Hypersurfaces In A Toric Variety}

The above construction extends straightforwardly to the case
of a complete intersection $M$
 of hypersurfaces 
in a toric variety: Given a complex of sheaves on $M$, 
push it forward to the ambient toric variety, and then
take its free resolution. This defines a data $(\cp,Q_0,\rho,R)$ such that
$Q_0^2=0$. The step to find its extension to a matrix factorization
$$
Q=\sum_{n_1,...,n_l\geq 0}p_1^{n_1}\cdots p_l^{n_l}Q_{\vec{n}}(x)
$$
of $\sum_{\beta=1}^lp_{\beta}G_{\beta}(x)$ is a line by line
generalization of the case of projective hypersyrface. The shift of 
the theta parameter is opposite to (\ref{shiftthetacompl}):
$$
\theta^a=B^a+\left(\sum_{\beta=1}^ld_{\beta}^a\right)\pi.
$$
In order to be consistent with the map $MF_W(X)\to D(M)$
given earlier, we also need to shift the R-charge  by $l$.

At this point, we would like to acknowledge the work by Avramov and
Buchweitz \cite{AB} on the relation between matrix factorizations and
modules on hypersurface rings, which follows earlier works
by Shamash \cite{Shamash} and Eisenbud \cite{Eisenbud}. Later, in
\cite{AG} their results were turned into computer algorithms. In the
following we compare our formulations of the Kn\"orrer map to the
latter reference.

In the terminology adapted to
the current context, (i) they constructed a matrix factorization
$(\cp,Q,\rho,R)$ from a given module over the
complete intersection ring $B=\C[x_1,...,x_N]/(G_1,...,G_l)$, (ii)
constructed the left semi-infinite complex of graded free $B$-modules 
corresponding to that matrix factorization, and (iii) proved that
that semi-infinite complex is quasi-isomorphic to the original $B$-module.
The construction of the inverse map $D(M)\to MF_W(X)$ presented here
is inspired from their work. Especially the reconstruction
of $Q(p,x)$ from $Q_0(x)$ is a copy of their proof for the part (i).
On the other hand, we constructed the map $MF_W(X)\to D(M)$, corresponding to
their part (ii), using a completely different method, and that was done
when we were not aware of their work.
Finally, we hope that it is clear that we have given an independent
derivation of their isomorphism (iii)
which is physically transparent --- the original
$B$-module and the semi-infinite complex are simply two different
ways to describe the low energy behaviour of the same
brane in the linear sigma model.

\subsection{CY/LG Correspondence --- More Examples}
\label{subsec:Excompact}

Combining the brane transportation rule described in 
Section~\ref{subsec:compactLEBC} with the reduction and lift
maps found in Sections \ref{subsec:LGOpt} and
\ref{subsec:LVphase}, we obtain a very explicit map between D-branes
of a LG orbifold point and those in a large volume phase.
Given a brane in the initial phase, we lift it
to the linear sigma model and find its 
 grade or band restricted representative with respect to the chosen path
in the K\"ahler moduli space. After transportation 
through phase boundaries, we reduce it to 
the low energy theory in the final phase.

For example, let us consider D-brane transport in Example (A)
with Fermat-cubic polynomial $G=x_1^3+x_2^3+x_3^3$, along a path
in the K\"ahler moduli space through the window $-\pi<\theta<\pi$ which 
corresponds to the
grade restriction rule $\{-1,0,1\}$. Take the Recknagel-Schomerus brane
${\mathcal B}_{(0,0,0),2,1}$ at the orbifold theory. We have seen 
in Section~\ref{subsec:LGOpt} that it lifts
to the brane $\lsmB_0$ given in (\ref{masslessMF}) which is
already grade restricted. In the large volume phase, we have seen that
this reduces to ${\mathcal O}(-1)[1]$. Thus the transportation gives
$$
{\mathcal B}_{(0,0,0),2,1}\in MF_{\Z_3}(G)
\longmapsto {\mathcal O}(-1)[1]\in D(M).
$$
As another example, let us consider the D0-brane
on the elliptic curve $M$ at the point $\mathfrak{p}$ 
given by $x_1+x_2=0, x_3=0$.
It is given by the skyscraper sheaf ${\mathcal O}_{\mathfrak{p}}$ of $M$ and
 lifts to the brane $\lsmB_{\mathfrak{p}}$ 
given in (\ref{D0MF}) with any value of $q$.
The one with $q=0$ is grade restricted and can be transported to 
the LG phase. By reduction, we find the brane at the LG orbifold point.
This process gives
$$
{\mathcal O}_{\mathfrak{p}}\in D(M)\longmapsto
\left({\mathcal O}(\overline{0})^{\oplus 2}_{{}_0}\,\,
\MFarrow{g}{f}
\begin{array}{c}
{\mathcal O}(\overline{1})^{}_{{}_{1\over 3}}\\[-0.3cm]
\oplus\\[-0.1cm]
{\mathcal O}(\overline{2})^{}_{{}_{-{1\over 3}}}
\end{array}\right)\in MF_{\Z_3}(G)
$$
where $g$ and $f$ are obtained by setting $p=1$ in (\ref{D0MF}).

In what follows, we consider more examples of D-brane transport.

\subsubsection{Fermat Quintic}

We first consider the Fermat quintic 
$$
G(x_1,...,x_5)=x_1^5+x_2^5+x_3^5+x_4^5+x_5^5.
$$
We will find the large volume image of the RS-branes with 
${\bf L}=(0,0,0,0,0)$
and of a class of permutation branes at the LG orbifold point, as well as
the LG image of the D4-brane wrapped on a divisor
of the quintic.
Throughout, we consider the paths through the window $w:-5\pi<\theta<-3\pi$
with the grade restriction rule $\mathfrak{C}^w=\{0,1,2,3,4\}$.

\subsubsection*{\it ${\bf L}=0^5$ RS-Branes}

We first consider the Recknagel-Schomerus branes 
${\mathcal B}_{{\bf L},q,r}$ with ${\bf L}=(0,0,0,0,0)$: 
$$
\QLG_{0^5}=\sum_{i=1}^5\Bigl(x_i\eta_i+x_i^4\bareta_i\Bigr)
$$
represented on the Clifford module $\cp_5$.
Recall from Section~\ref{subsub:LGO}
that the labels $q$ and $r$ specify the representations of
the orbifold group and the R-symmetry group.
In particular the R-charge of the vector $|0\rangle$ is
$$
\RLG_{|0\rangle}=-{q\over 5}+r.
$$
We will only look at those with $r=0$, 
as the others can be recovered by overall shifts of the R-degree.
The vector $|0\rangle$ is even for this choice.

We first consider the brane
${\mathcal B}_{0^5,0,0}$ for which $\RLG_{|0\rangle}=0$.
The R-charges of the other basis elements of $\cp_5$
can be found by noting that the $\bareta_i$'s have R-charge $-{3\over 5}$:
\begin{center}
\begin{tabular}{|c||c|c|c|c|c|c|}
\hline
vector&$|0\rangle$&$\bareta_i|0\rangle$&$\bareta_i\bareta_j|0\rangle$
&$\bareta_i\bareta_j\bareta_k|0\rangle$
&$\bareta_i\bareta_j\bareta_k\bareta_l|0\rangle$
&$\bareta_1\cdots\bareta_5|0\rangle$\\
\hline
$\RLG_{\bf i}$
&0&$-{3\over 5}$&$-{6\over 5}$&$-{9\over 5}$&$-{12\over 5}$&$-3$\\
\hline
\end{tabular}
\end{center}
The first step is to lift it to a grade restricted brane
in the linear sigma model. 
This is done by solving the equation (\ref{qeq}), that is,
$$
\RLG_{\bf i}=R_{\bf i}-2q_{\bf i}/5
$$
where $R_{\bf i}$ is an even (odd) integer for
an even (odd) Chan-Paton vector 
and $q_{\bf i}$ must be taken from the grade restriction
range $\mathfrak{C}^w=\{0,1,2,3,4\}$.
For the even vector 
$|0\rangle$ with $\RLG=0$ the equation is solved by $R=0$ and $q=0$.
This corresponds to the Wilson line brane $\wilson(0)^{}_{{}_0}$.
For the odd vectors $\bareta_i|0\rangle$ with
$\RLG=-3/5$, the solution must have odd $R$ and is given by $R=1$, $q=4$,
which yields $\wilson(4)^{}_{{}_1}$.
Since we have five elements $\bareta_1|0\rangle$, ..., $\bareta_5|0\rangle$,
we have the sum of five copies
 $\wilson(4)^{\oplus 5}_{{}_1}$.
The solutions for all the basis vectors of $\cp_5$
are listed in the following table:
\begin{center}
\begin{tabular}{|c||c|c|c|c|c|c|}
\hline
vector&$|0\rangle$&$\bareta_i|0\rangle$&$\bareta_i\bareta_j|0\rangle$
&$\bareta_i\bareta_j\bareta_k|0\rangle$
&$\bareta_i\bareta_j\bareta_k\bareta_l|0\rangle$
&$\bareta_1\cdots\bareta_5|0\rangle$\\
\hline
$R_{\bf i}$&0&$1$&$0$&$-1$&$-2$&$-3$\\
\hline
$q_{\bf i}$&0&4&3&2&1&0\\
\hline
&$\wilson(0)^{}_{{}_0}$&$\wilson(4)^{\oplus 5}_{{}_1}$
&$\wilson(3)^{\oplus 10}_{{}_0}$&$\wilson(2)^{\oplus 10}_{{}_{-1}}$
&$\wilson(1)^{\oplus 5}_{{}_{-2}}$&$\wilson(0)^{}_{{}_{-3}}$\\
\hline
\end{tabular}
\end{center}
The variable $p$ can be included to $\QLG_{0^5}$ simply by multiplying 
each entry by the right power of $p$ so that the degrees 
$R_{\bf i},q_{\bf i}$ match.
For example, the map $x_i^4\bareta_i$ sending $\wilson(0)^{}_{{}_0}$ 
to $\wilson(4)^{\oplus 5}_{{}_1}$ needs no power of $p$ while
the map $x_i^4\bareta_i$ sending $\wilson(4)^{\oplus 5}_{{}_1}$ to 
$\wilson(3)^{\oplus 10}_{{}_0}$ needs to be multiplied by
a single power of $p$.
In this way, we find the matrix factorization $Q(p,x)$ of $pG(x)$.
This completes the construction of
grade restricted lift $(\cp_5,Q,\rho,R)$ of the
brane ${\mathcal B}_{0^5,0,0}$.

The next step is to reduce $(\cp_5,Q,\rho,R)$
down to the low energy theory in the 
large volume phase, applying the map $MF_W(X)\to D(M)$ in 
Section~\ref{subsub:LVred}:
We first write the Wilson line branes in the above
table in the order of the R-charges $R_{\bf i}$ and then write its copies
successively,
shifting the position by $2$ and the gauge degree $q$ by $5$ at each step.
The arrows are determined by
 $Q(p,x)$ but they can also
be read directly from the original matrix factroization $\QLG_{0^5}$
with the aid of the gauge degree information.
This yields the following semi-infinite complex over the quintic
 hypersurface $M$\\[-0.2cm]
\begin{figure}[h]
\psfrag{o0}{${\mathcal O}(0)$}
\psfrag{o1-5}{${\mathcal O}(1)^{\oplus 5}$}
\psfrag{o5}{${\mathcal O}(5)$}
\psfrag{o2-10}{${\mathcal O}(2)^{\oplus 10}$}
\psfrag{o6-5}{$\underline{{\mathcal O}(6)^{\oplus 5}}$}
\psfrag{o3-10}{${\mathcal O}(3)^{\oplus 10}$}
\psfrag{o10}{${\mathcal O}(10)$}
\psfrag{o7-10}{${\mathcal O}(7)^{\oplus 10}$}
\psfrag{o4-5}{${\mathcal O}(4)^{\oplus 5}$}
\psfrag{o11-5}{${\mathcal O}(11)^{\oplus 5}$}
\psfrag{o8-10}{${\mathcal O}(8)^{\oplus 10}$}
\psfrag{op}{$\oplus$}
\psfrag{x}{\footnotesize ${}^{X}$}
\psfrag{x4}{\footnotesize ${}^{X^4}$}
\centerline{\includegraphics{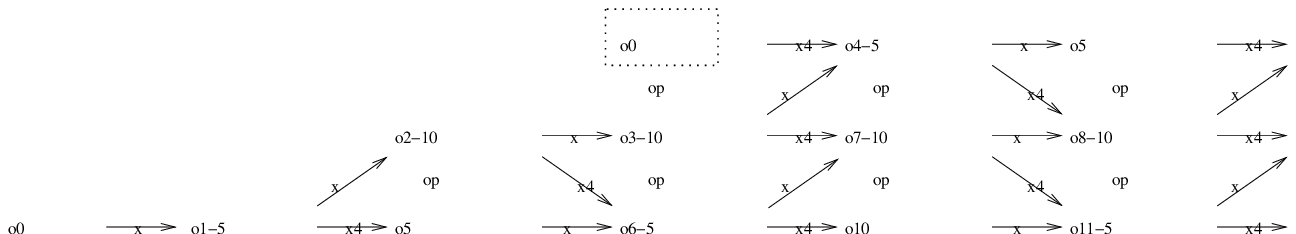}}
\end{figure}\\
This complex ${\mathcal C}=(\widehat{\mathcal E},\widehat{Q})$ 
is the large volume image of the
RS-brane ${\mathcal B}_{0^5,0,0}$.
As always, the 2-periodic part is exact and we can take out a finite length
complex of coherent sheaves. One way is to replace the 
infinite part of non-negative R-degrees
by the kernel sheaf of the map 
${\mathcal E}^0\to{\mathcal E}^1$.
However, in the present case there is a simpler complex that is 
quasi-isomorphic to ${\mathcal C}$.

To see this, let us digress for a moment to study the following brane
$\lsmB_1$ in the linear sigma model:
$$
Q_1=\sum_{i=1}^5\Bigl(x_i\eta_i+px_i^4\bareta_i\Bigr)
$$
We assign, say, charge $(q,R)=(5,5)$ to the Clifford vacuum $|0\rangle$,
so that the other vectors $\bareta_i|0\rangle$, $\bareta_i\bareta_j|0\rangle$, 
..., $\bareta_1\cdots\bareta_5|0\rangle$ have charges
$(4,4)$, $(3,3)$, ..., $(0,0)$ respectively.
Note that the boundary potential
 $\{Q_1,Q_1^{\dag}\}=\sum_{i=1}^5(|x_i|^2+|px_i^4|^2)$ is positive everywhere
in the large volume phase where $\{x_1=\cdots=x_5=0\}$ is deleted. 
Thus, we know that this brane is 
empty at low energies. It is the analog of the brane $\lsmB_+$
in the model for the Fermat cubic curve.
Applying the reduction map $MF_W(X)\to D(M)$,
we obtain the following complex\\[-0.2cm]
\begin{figure}[h]
\psfrag{o0}{\underline{${\mathcal O}(0)$}}
\psfrag{o1-5}{${\mathcal O}(1)^{\oplus 5}$}
\psfrag{o5}{${\mathcal O}(5)$}
\psfrag{o2-10}{${\mathcal O}(2)^{\oplus 10}$}
\psfrag{o6-5}{${\mathcal O}(6)^{\oplus 5}$}
\psfrag{o3-10}{${\mathcal O}(3)^{\oplus 10}$}
\psfrag{o10}{${\mathcal O}(10)$}
\psfrag{o7-10}{${\mathcal O}(7)^{\oplus 10}$}
\psfrag{o4-5}{${\mathcal O}(4)^{\oplus 5}$}
\psfrag{o11-5}{${\mathcal O}(11)^{\oplus 5}$}
\psfrag{o8-10}{${\mathcal O}(8)^{\oplus 10}$}
\psfrag{op}{$\oplus$}
\psfrag{x}{\footnotesize ${}^{X}$}
\psfrag{x4}{\footnotesize ${}^{X^4}$}
\centerline{\includegraphics{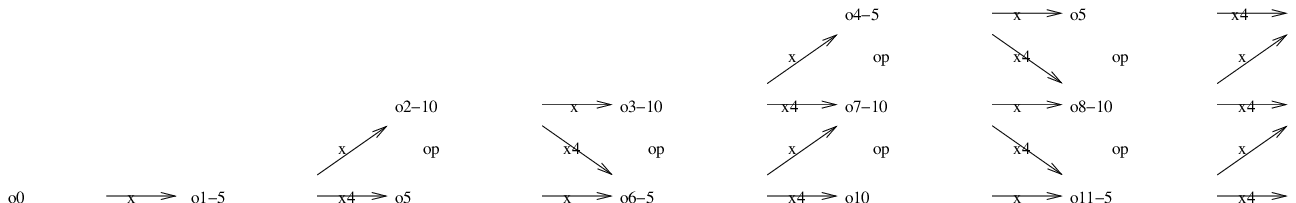}}
\end{figure}\\
One can check either by computing the potential 
$\{\widehat{Q}_1,\widehat{Q}_1^{\dag}\}$
or algebraically that it is exact and hence trivial at low energies,
as expected.
We denote this complex by ${\mathcal K}$.

Let us come back to the complex ${\mathcal C}$.
If we omit ${\mathcal O}(0)$ inside the
box of dashed lines, the remaning part is nothing but
the complex ${\mathcal K}$ shifted to the left by 3.
Since the latter is empty, we are simply left with
${\mathcal O}(0)$ in that box.
In other words,
we have the quasi-isomorphism relation
$$
{\mathcal C}={\rm Cone}({\mathcal O}(0)[-1]
\longrightarrow {\mathcal K}[3])~\cong ~{\mathcal O}(0)[0].
$$
Thus, the large volume image of the brane ${\mathcal B}_{0^5,0,0}$
is the structure sheaf ${\mathcal O}$ of the quintic hypersurface,
that is,
the D6-brane with trivial gauge bundle on it.

The large volume images of the other ${\bf L}=0^5$
 RS-branes ${\mathcal B}_{0^5,q,0}$
can be found in a similar way.
The semi-infinite complexes that we obtain for $q=1,2,3,4$ 
are respectively\\
\begin{figure}[h]
\psfrag{o1}{${\mathcal O}(1)$}
\psfrag{o2-5}{${\mathcal O}(2)^{\oplus 5}$}
\psfrag{o6}{${\mathcal O}(6)$}
\psfrag{o3-10}{${\mathcal O}(3)^{\oplus 10}$}
\psfrag{o0-5}{${\mathcal O}(0)^{\oplus 5}$}
\psfrag{o7-5}{${\mathcal O}(7)^{\oplus 5}$}
\psfrag{o4-10}{${\mathcal O}(4)^{\oplus 10}$}
\psfrag{o11}{${\mathcal O}(11)$}
\psfrag{o8-10}{${\mathcal O}(8)^{\oplus 10}$}
\psfrag{o5-5}{${\mathcal O}(5)^{\oplus 5}$}
\psfrag{o12-5}{${\mathcal O}(12)^{\oplus 5}$}
\psfrag{o9-10}{${\mathcal O}(9)^{\oplus 10}$}
\psfrag{op}{$\oplus$}
\centerline{\includegraphics{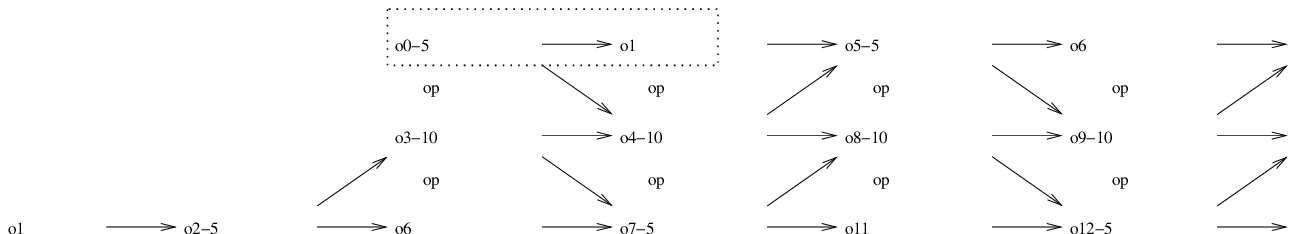}}
\vspace{0.5cm}
\psfrag{o2}{${\mathcal O}(2)$}
\psfrag{o3-5}{${\mathcal O}(3)^{\oplus 5}$}
\psfrag{o0-10}{${\mathcal O}(0)^{\oplus 10}$}
\psfrag{o7}{${\mathcal O}(7)$}
\psfrag{o4-10}{${\mathcal O}(4)^{\oplus 10}$}
\psfrag{o1-5}{${\mathcal O}(1)^{\oplus 5}$}
\psfrag{o8-5}{${\mathcal O}(8)^{\oplus 5}$}
\psfrag{o5-10}{${\mathcal O}(5)^{\oplus 10}$}
\psfrag{o12}{${\mathcal O}(12)$}
\psfrag{o9-10}{${\mathcal O}(9)^{\oplus 10}$}
\psfrag{o6-5}{${\mathcal O}(6)^{\oplus 5}$}
\psfrag{o13-5}{${\mathcal O}(13)^{\oplus 5}$}
\psfrag{o10-10}{${\mathcal O}(10)^{\oplus 10}$}
\psfrag{op}{$\oplus$}
\centerline{\includegraphics{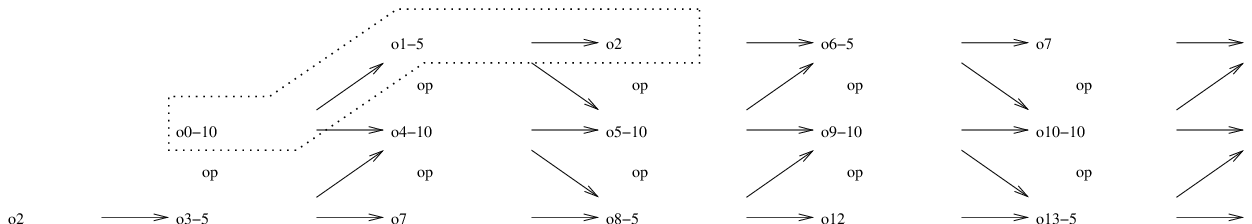}}
\vspace{0.5cm}
\psfrag{o3}{${\mathcal O}(3)$}
\psfrag{o0-10}{${\mathcal O}(0)^{\oplus 10}$}
\psfrag{o4-5}{${\mathcal O}(4)^{\oplus 5}$}
\psfrag{o1-10}{${\mathcal O}(1)^{\oplus 10}$}
\psfrag{o8}{${\mathcal O}(8)$}
\psfrag{o5-10}{${\mathcal O}(5)^{\oplus 10}$}
\psfrag{o2-5}{${\mathcal O}(2)^{\oplus 5}$}
\psfrag{o9-5}{${\mathcal O}(9)^{\oplus 5}$}
\psfrag{o6-10}{${\mathcal O}(6)^{\oplus 10}$}
\psfrag{o13}{${\mathcal O}(13)$}
\psfrag{o10-10}{${\mathcal O}(10)^{\oplus 10}$}
\psfrag{o7-5}{${\mathcal O}(7)^{\oplus 5}$}
\psfrag{o14-5}{${\mathcal O}(14)^{\oplus 5}$}
\psfrag{o11-10}{${\mathcal O}(11)^{\oplus 10}$}
\psfrag{op}{$\oplus$}
\centerline{\includegraphics{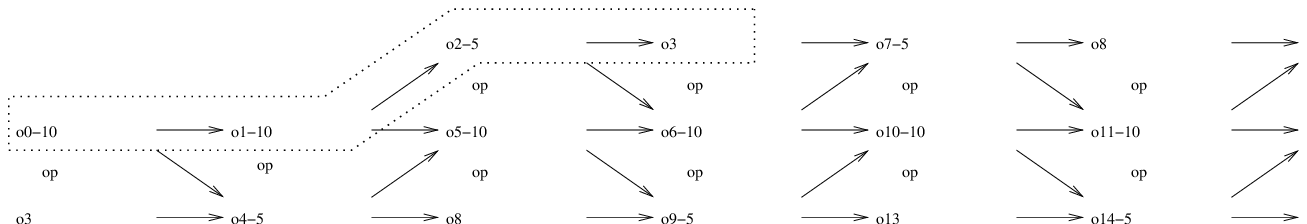}}
\vspace{0.5cm}
\psfrag{o0-5}{${\mathcal O}(0)^{\oplus 5}$}
\psfrag{o4}{${\mathcal O}(4)$}
\psfrag{o1-10}{${\mathcal O}(1)^{\oplus 10}$}
\psfrag{o5-5}{${\mathcal O}(5)^{\oplus 5}$}
\psfrag{o2-10}{${\mathcal O}(2)^{\oplus 10}$}
\psfrag{o9}{${\mathcal O}(9)$}
\psfrag{o6-10}{${\mathcal O}(6)^{\oplus 10}$}
\psfrag{o3-5}{${\mathcal O}(3)^{\oplus 5}$}
\psfrag{o10-5}{${\mathcal O}(10)^{\oplus 5}$}
\psfrag{o7-10}{${\mathcal O}(7)^{\oplus 10}$}
\psfrag{o14}{${\mathcal O}(14)$}
\psfrag{o11-10}{${\mathcal O}(11)^{\oplus 10}$}
\psfrag{o8-5}{${\mathcal O}(8)^{\oplus 5}$}
\psfrag{o15-5}{${\mathcal O}(15)^{\oplus 5}$}
\psfrag{o12-10}{${\mathcal O}(12)^{\oplus 10}$}
\psfrag{op}{$\oplus$}
\centerline{\includegraphics{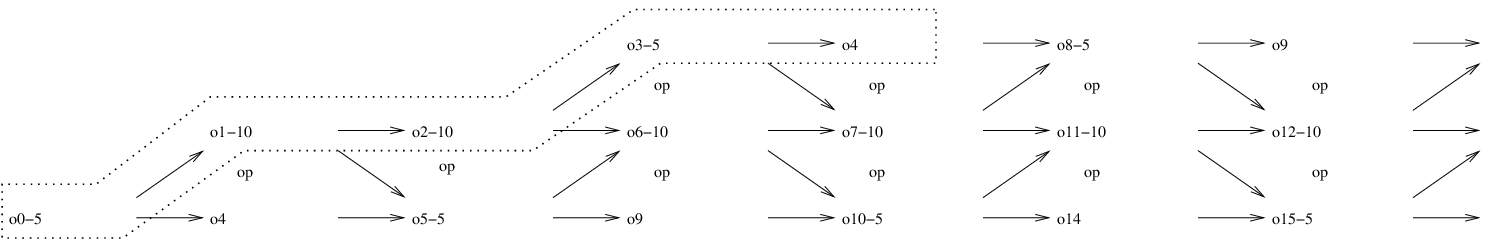}}
\end{figure}\\
We see that they are the same as the exact complexes ${\mathcal K}(q)[3]$
for $q=1,2,3,4$,
if we ignore the finite parts in the dashed box.
Thus, only the latter
\beqa
{\mathcal O}(0)^{\oplus 5}\stackrel{X}{\longrightarrow}
\underline{{\mathcal O}(1)},
&&
\nn\\
{\mathcal O}(0)^{\oplus 10}\stackrel{X}{\longrightarrow}
{\mathcal O}(1)^{\oplus 5}\stackrel{X}{\longrightarrow}
\underline{{\mathcal O}(2)},
&&
\nn\\
{\mathcal O}(0)^{\oplus 10}\stackrel{X}{\longrightarrow}
{\mathcal O}(1)^{\oplus 10}\stackrel{X}{\longrightarrow}
{\mathcal O}(2)^{\oplus 5}\stackrel{X}{\longrightarrow}
\underline{{\mathcal O}(3)},
&&
\nn\\
{\mathcal O}(0)^{\oplus 5}\stackrel{X}{\longrightarrow}
{\mathcal O}(1)^{\oplus 10}\stackrel{X}{\longrightarrow}
{\mathcal O}(2)^{\oplus 10}\stackrel{X}{\longrightarrow}
{\mathcal O}(3)^{\oplus 5}\stackrel{X}{\longrightarrow}
\underline{{\mathcal O}(4)},
&&
\nn
\eeqa
remain.
We further notice that they are quasi-isomorphic to the sheaves of sections
of the vector bundles $\wedge^q(T^*_{\PP^4}(1))$ of the ambient space
$\CP^4$, at R-degree $-q$, which are restricted to the hypersurface $M$.
Thus, the large volume images of the ${\bf L}=0^5$
Recknagel-Schomerus branes are:
\beq
{\mathcal B}_{0^5,q,0}\longmapsto \Omega^q_{\PP^4}(q)[q]\Bigl|_M,\qquad
\mbox{for ~$q=0,1,2,3,4$.}
\label{L=0maps}
\eeq
The value of the B-field is related to the theta parameter by
$$
B=\theta+5\pi.
$$
In particular, it is in the domain $0<B<2\pi$
for the chosen window.

We notice that the images (\ref{L=0maps}) are very similar to
the results (\ref{McKay}) of the transport of the fractional branes
${\mathcal O}_{\mathfrak{p}}(\bar q)$
in the non-compact version of the theory.
This observation leads us to an alternative way to
derive the same result. 
We first notice that the grade restricted lift
of the RS-brane ${\mathcal B}_{0^5,q,0}$ is the same as the
grade restricted lift $\lsmB_{\bar q}'$ of the fractional brane
${\mathcal O}_{\mathfrak{p}}(\bar q)$ shown in (\ref{liftfrac}),
except that we need to add the oppositely oriented
arrows, either $X^4$ or $pX^4$.
We next recall that $\lsmB_{\bar q}'$ is presented as the bound state
of the $X$-Koszul complex $\lsmB_{\bar q}$ 
and another complex $\lsmB_{\bar q}''$.
It is straightforward to find the matrix factorization version
of this bound state. It is a bound state
of the matrix factorization $\lsmB_1$,
obtained from the $X$-Koszul complex $\lsmB_{\bar q}$ by adding
the opposite arrows $pX^4$, and the 
matrix factorization $\lsmB''_{0^5,q,0}$ given by
$$
\mathfrak{A}_q(5)[-1]\,\MFarrow{p}{G}\,
\mathfrak{A}_q.
$$
Here $\mathfrak{A}_q$ is the complex defined in (\ref{defAAA}).
Since the brane $\lsmB_1$ is trivial in the large volume regime,
we can simply take the brane $\lsmB''_{0^5,q,0}$
as a lift of
the large volume image of the RS-brane.
At this stage, we apply the second formulation of the Kn\"orrer map:
Set $p=0$ and ask if that is presented as the pushforward of some complex
over the quintic hypersurface $G=0$. And yes it is!
It is the pushforward of the complex
associated with $\mathfrak{A}_q(5)[-1]$:
$$
{\mathcal O}(5)^{\oplus {5\choose q}}_{{}_{-q+1}}
\stackrel{X}{\longrightarrow}\cdots
\stackrel{X}{\longrightarrow}
{\mathcal O}(q+4)^{\oplus 5}_{{}_0}\stackrel{X}{\longrightarrow}
{\mathcal O}(q+5)^{}_{{}_1}
$$
Applying the R-charge shift (\ref{gradeshift2})
(backwards of course) and trading the difference in the
B-field shifts, (\ref{B-shift}) versus (\ref{shiftthetaininverse}), for
a charge shift,
we have
$$
{\mathcal O}(0)^{\oplus {5\choose q}}_{{}_{-q}}
\stackrel{X}{\longrightarrow}\cdots
\stackrel{X}{\longrightarrow}
{\mathcal O}(q-1)^{\oplus 5}_{{}_{-1}}\stackrel{X}{\longrightarrow}
{\mathcal O}(q)^{}_{{}_0}.
$$
This is precisely the result obtained above.

The pattern (\ref{L=0maps}) was conjected in \cite{BDLR} based on the analysis
of R-R charge and using mirror symmetry.
We have given a proof of the conjecture.

\subsubsection*{\it Other RS-Branes}

We do not explicitly write down the large volume
images of RS-branes for other ${\bf L}$'s, since it is very straightforward. 
However, we do indicate a way to extract a finite piece from
the semi-infinite complex.
Analogously to $\lsmB_+$ for the elliptic curve and to
$\lsmB_1$ for the ${\bf L}=0^5$ RS-branes, we consider
the brane $\lsmB_{{\bf L}}^{\rm triv}$ given by
$$
Q_{\bf L}^{\rm triv}=\sum_{i=1}^5\Bigl(x_i^{L_i+1}\eta_i
+px_i^{4-L_i}\bareta_i\Bigr)
$$
with any consistent choice of gauge and R-charge of Chan-Paton vectors.
The potential 
$$
\{Q_{\bf L}^{\rm triv},Q_{\bf L}^{{\rm triv}\dag}\}
=\sum_{i=1}^5\Bigl(|x_i|^{2(L_i+1)}+|p|^2|x_i|^{2(4-L_i)}\Bigr)
$$
is everywhere positive in the large volume phase.
Therefore, the semi-infinite complex ${\mathcal K}_{\bf L}$ 
obtained from this brane must be exact. 
The large volume image of a general RS-brane ${\mathcal B}_{{\bf L},q,r}$
is first given by a semi-infinite complex, and this is
presented as a bound state of a finite length complex
${\mathcal F}_{{\bf L},q,r}$ and the exact complex
${\mathcal K}_{\bf L}$ possibly with some shifts in gauge 
and R-degrees. 
Thus, one can extract the finite piece ${\mathcal F}_{{\bf L},q,r}$
as the large volume image of the RS-brane. Since
it is a complex of sums of line bundles ${\mathcal O}(q)$, we have proved
that the large volume images of the RS-branes are always restriction 
of finite complexes of vector bundles of the ambient space $\CP^4$.

\subsubsection*{\it Permutation Branes}

Next we consider some examples of permutation branes
of the Gepner model \cite{Permu} which were recently studied in terms of
matrix factorizations in \cite{ADDF,BG}. 
Take the following matrix factorization of $W=G(x)$;
$$
 \QLG = \sum_{i=1}^3 \Bigl(\,a_i \eta_i + b_i \bar\eta_i\,\Bigr) ,
$$
which is represented on the Clifford module $\cp_3$, where
\beqa
&&a_1=x_1+x_2,
\quad b_1=x_1^4-x_1x_2^3+x_1^2x_2^2-x_1x_2^3+x_2^4,\nn\\
&&a_2=x_3+x_4,\quad b_2=x_3^4-x_3x_4^3+x_3^2x_4^2-x_3x_4^3+x_4^4,\nn\\
&&a_3=x_5,\quad b_3=x_5^4.\nn
\eeqa
The brane at the Landau-Ginzburg orbifold point is specified by
the orbifold action and the R-charge of the Clifford vacuum $|0\rangle$.
Let ${\mathcal P}_{q,r}$ 
be the brane such that 
$\rho(\omega)|0\rangle=\omega^q|0\rangle$ and
$\RLG_{|0\rangle}=-2q/5+r$.
Let us start with the brane ${\mathcal P}_{0,0}$.
The R-charges $\RLG_{\bf i}$
of all the Chan-Paton vectors in the Landau-Ginzburg model 
as well as their grade restricted lifts in the linear sigma model
are listed
in the following table:
\begin{center}
\begin{tabular}{|c||c|c|c|c|}
\hline
vector&$|0\rangle$&$\bareta_i|0\rangle$&$\bareta_i\bareta_j|0\rangle$
&$\bareta_1\bareta_2\bareta_3|0\rangle$\\
\hline
$\RLG_{\bf i}$&$0$ &$-{3\over 5}$&$-{6\over 5}$&$-{9\over 5}$\\
\hline
$R_{\bf i}$&0&$1$&$0$&$-1$\\
\hline
$q_{\bf i}$&0&4&3&2\\
\hline
&$\wilson(0)^{}_{{}_0}$&$\wilson(4)^{\oplus 3}_{{}_1}$
&$\wilson(3)^{\oplus 3}_{{}_0}$&$\wilson(2)^{}_{{}_{-1}}$
\\
\hline
\end{tabular}
\end{center}
More explicitly, the grade restricted lift is given by
$$
\left(\begin{array}{ccc}
\wilson(0)^{}_{{}_0}&\sMFarrow{b}{pa}&\wilson(4)^{\oplus 3}_{{}_1}\\
\oplus&\dMFarrow{pb}{a}&\oplus\\
\wilson(3)^{\oplus 3}_{{}_0}&\sMFarrow{pb}{a}&\wilson(2)^{}_{{}_{-1}}
\end{array}\right),
$$
where $a$ and $b$ correspond to $\sum_ia_i\eta_i$ and
$\sum_ib_i\bareta_i$. Let us apply the reduction map to this brane
in the large volume phase. This time, we take the left semi-infinite
version:
\beq
\cdots
\begin{array}{cccccc}
\stackrel{b}{\to}&
\!\!{\mathcal O}(-6)^{\oplus 3}\!\!\!\!&\stackrel{a}{\to}&
\!\!{\mathcal O}(-5)\!\!&\stackrel{b}{\to}&
\!\!\underline{{\mathcal O}(-1)^{\oplus 3}}\,{}^{}_.\!\!\!\!
\\
\nearrow\!\!\!\!{}_a&
\oplus&\searrow\!\!\!\!{}^b&\oplus&\nearrow\!\!\!\!{}_a&
\\
\stackrel{b}{\to}&
\!\!{\mathcal O}(-3)\!\!&\stackrel{a}{\to}&
\!\!{\mathcal O}(-2)^{\oplus 3}\!\!\!\!&&
\end{array}
\label{permub}
\eeq
This is the large volume image of the permutation brane
${\mathcal P}_{0,0}$. 
The two-periodic part is exact and thus we can cut off
the degree $\leq -2$ parts and replace the degree $-1$ component
by the cokernel of the map from the degree $-2$ component.
This is one way to find a finite complex of coherent sheaves.
Alternatively, we may try to proceed as in the RS-branes: Take the 
matrix factorization $Q=\sum_i(a_i\eta_i+pb_i\bareta_i)$ and compare
its geometric image ${\mathcal K}$ with (\ref{permub}). 
This time, however, 
the potential for $Q$ is $\{Q,Q^{\dag}\}=\sum_i|a_i|^2$ for $p=0$
and fails to be positive everywhere. As a consequence the complex 
${\mathcal K}$ fails to be exact. But the potential is positive except
at $a_1=a_2=a_3=0$, which is a rational curve $D$ on the quintic $M$.
Accordingly, the complex ${\mathcal K}$ can be made exact by
adjoining the sheaf
${\mathcal O}_D$ supported at $D$. In this way we obtain the 
following exact complex
$$
\cdots
\begin{array}{cccccccccc}
\stackrel{b}{\to}&
\!\!{\mathcal O}(-6)^{\oplus 3}\!\!\!\!&\stackrel{a}{\to}&
\!\!{\mathcal O}(-5)\!\!&\stackrel{b}{\to}&
\!\!{\mathcal O}(-1)^{\oplus 3}\!\!&\stackrel{a}{\to}&
\!\!{\mathcal O}\!\!&\stackrel{q}{\to}&
\!\!{\mathcal O}_D\!\!
\\
\nearrow\!\!\!\!{}_a&
\oplus&\searrow\!\!\!\!{}^b&\oplus&\nearrow\!\!\!\!{}_a&&&&&
\\
\stackrel{b}{\to}&
\!\!{\mathcal O}(-3)\!\!&\stackrel{a}{\to}&
\!\!{\mathcal O}(-2)^{\oplus 3}\!\!\!\!&&&&&&
\end{array}
$$
Here $q$ is the restriction map to $D$.
We see that the complex (\ref{permub}) sits inside it.
In particular,
there is a quasi-isomorphism of (\ref{permub}) to
the complex $\underline{{\mathcal O}}\stackrel{q}{\to}{\mathcal O}_D$,
mapping the right-most ${\mathcal O}(-1)^{\oplus 3}$ to
${\mathcal O}$ by $a$. Thus, we can take 
$\underline{{\mathcal O}}\stackrel{q}{\to}{\mathcal O}_D$ as the large 
volume image of ${\mathcal P}_{0,0}$.

Repeating this procedure, we find the following simple images of
the permutation branes ${\mathcal P}_{q,0}$ for $q=0,1,2,3,4$:
\beqa
{\mathcal P}_{0,0}&\longmapsto&
\Bigl( \underline{{\mathcal O}}\stackrel{q}{\longrightarrow}
{\mathcal O}_D 
\Bigr)\nn\\
{\mathcal P}_{1,0}&\longmapsto&
\Bigl( {\mathcal O}^{\oplus 3}\stackrel{a}{\longrightarrow}
\underline{{\mathcal O}(1)}\stackrel{q}{\longrightarrow}
{\mathcal O}_D(1)
\Bigr)\nn\\
{\mathcal P}_{2,0}&\longmapsto&
\Bigl({\mathcal O}^{\oplus 3}\stackrel{a}{\longrightarrow}
{\mathcal O}(1)^{\oplus 3}\stackrel{a}{\longrightarrow}
\underline{{\mathcal O}(2)}\stackrel{q}{\longrightarrow}
{\mathcal O}_D(2)
\Bigr)\nn\\
{\mathcal P}_{3,0}&\longmapsto&
{\mathcal O}_D(-2)[1]\nn\\
{\mathcal P}_{4,0}&\longmapsto&
{\mathcal O}_D(-1)[1]\nn
\eeqa
Note that the images of ${\mathcal P}_{3,0}$ and
${\mathcal P}_{4,0}$ are the (anti-)D2-branes wrapped
on the curve $D$.
The Chern characters of these complexes $F{\mathcal P}_{q,0}$ are
$$
  \begin{array}{ccl}
    \mathrm{ch}(F{\mathcal P}_{0,0}) &=&
    1 - \frac{1}{5} H^2 - \frac{1}{5} H^3 ,\\[5pt]
    \mathrm{ch}(F{\mathcal P}_{1,0}) &=& 
    -2 + H + \frac{3}{10} H^2 - \frac{7}{30} H^3 ,\\[5pt]
    \mathrm{ch}(F{\mathcal P}_{2,0}) &=&  
    1 - H + \frac{3}{10} H^2 + \frac{7}{30} H^3 ,\\[5pt]
    \mathrm{ch}(F{\mathcal P}_{3,0}) &=& 
    -\frac 15 H^2 + \frac 15 H^3 ,\\[5pt]
    \mathrm{ch}(F{\mathcal P}_{4,0}) &=& 
    -\frac 15 H^2 .
  \end{array}  
$$
where we used that ${\rm ch}({\mathcal O}_D)={1\over 5}H^2+{1\over 5}H^3$.
This reproduces the result of \cite{BG}
on the K-theory charges of the large volume images
of the permutation branes
(up to an overall sign which can be traced to the shift
${\mathcal P}_{q,0}\to {\mathcal P}_{q,1}$).

\subsubsection*{\it D4-Brane}

\newcommand{\darrow}[1]{\begin{picture}(50,10)(0,20)
  \put(5,8){\vector(2,1){40}}
  \put(34,15){{\small $#1$}}
  \end{picture}}
\newcommand{\barrow}[1]{\begin{picture}(50,10)(0,20)
  \put(26,22){{\small $#1$}}
  \put(45,12){\vector(-2,1){40}}
  \end{picture}}
\newcommand{\MFarrowBig}[2]{\begin{picture}(50,40)(0,20)
  \put(17,40){{\small $#1$}}
  \put(5,27){\vector(1,0){40}}    
  \put(45,23){\vector(-1,0){40}}    
  \put(17,7){{\small $#2$}}
  \end{picture}}
\newcommand{\MFarrowShort}[2]{\begin{picture}(30,40)(0,20)
  \put(7,32){{\small $#1$}}
  \put(0,27){\vector(1,0){30}}    
  \put(30,23){\vector(-1,0){30}}    
  \put(7,15){{\small $#2$}}
  \end{picture}}
\newcommand{\MFarrowShorter}[2]{\begin{picture}(20,40)(0,20)
  \put(7,32){{\tiny $#1$}}
  \put(0,27){\vector(1,0){20}}    
  \put(20,23){\vector(-1,0){20}}    
  \put(7,15){{\tiny $#2$}}
  \end{picture}}
\newcommand{\MFmorphBig}[2]{\begin{picture}(50,30)(0,20)
  \put(10,10){{\small $#1$}}
  \put(5,15){\vector(2,1){40}}    
  \put(45,15){\vector(-2,1){40}}    
  \put(45,15){{\small $#2$}}
  \end{picture}}

As a final example for a D-brane on the quintic hypersurface 
(not necessarily of Fermat type), let us
consider a D$4$-brane ${\mathcal O}_H(0)$ on a hyperplane $H$, which is
determined by a linear form
$h(x) = \alpha \cdot X = \sum_{i=1}^5\alpha_i x_i$ with
parameters $\alpha \in \CP^4$. We take the B-field in the interval
$B\in (0,2\pi)$. 

We would like to find its image at the LG orbifold point. In order to
apply the map  
$D(M) \longrightarrow MF_W(X)$, we notice that, as an object on the
ambient space $\CP^4$, this D-brane is supported on the complete intersetion
$\{h(x)=0\} \cap \{G(x)=0\}$ and thus can be realized as Koszul
complex of these two polynomials.
For the lift to the linear sigma model, we have to
fill in the field $p$ to obtain a matrix
factorization. Taking into account the shifts
(\ref{shiftthetaininverse}) and (\ref{shiftRininverse}) of the B-field
and the R-charge, respectively, the lift is simply
$$
  \wilson(-1)^{}_{{}_{-1}}
  \MFarrowBig{\vspace*{-20pt}\footnotesize 
    \Big(\!\!\!\begin{array}{c}h\\[-3pt]-G\end{array}\!\!\!\Big)}{\!\!(0,-p)} 
  \begin{array}{c}\wilson(0)^{}_{{}_0}\\[-5pt] \oplus \\[-5pt]
    \wilson(4)^{}_{{}_0} 
  \end{array}
  \MFarrowBig{\!\!(G,h)}{\footnotesize
    \Big(\!\!\begin{array}{c}p\\[-3pt]0\end{array}\!\!\Big)} 
  \wilson(5)^{}_{{}_1} . 
$$
Here, we shifted the charges so that the theta parameter lies in the window 
$w:-5\pi <\theta< -3\pi$, with the grade restriction range
$\mathfrak{C}^w=\{0,1,2,3,4\}$. We see that the
Wilson line branes $\wilson(-1)_{-1}$ and $\wilson(5)_1$ must be
eliminated when we want to transport the D$4$-brane through
$w$. This can be done by binding the infra-red empty branes 
${\mathscr K}_+^{\rm mf}(-1)$ and ${\mathscr K}_+^{\rm mf}$, defined in
(\ref{trivplus}), to the matrix factorization. 
Let us consider the former process first:
$$
  \begin{array}{*{8}{c@{\hspace{2pt}}}c}
    &&&&\wilson(0)_{{}_0}^{}\\[-13pt]
    &&\wilson(-1)_{{}_{-1}}^{}&\MFarrow{}{}&
    \oplus&\MFarrow{}{}&
    \wilson(5)_{{}_1}^{}
    \\[-3pt]
    &&&&\wilson(4)_{{}_0}^{}\\[-20pt]
    &\darrow{\id}&&\darrow{\varphi_1}&&\darrow{\varphi_2}\\[5pt]
    \wilson(-1)_{{}_{-2}}^{}&\MFarrow{X}{pG'}&
    \wilson(0)^{\oplus 5}_{{}_{-1}}&
    \MFarrow{X}{pG'}&\wilson(1)^{\oplus 10}_{{}_0}&
    \MFarrow{X}{pG'}&\wilson(2)^{\oplus 10}_{{}_1}&
    \MFarrow{X}{pG'}&\wilson(3)^{\oplus 5}_{{}_2}\cdots. 
  \end{array}
$$
Here, $X$ and $G'$ are the short-hand notation for $\sum_ix_i\bareta_i$
and ${1\over 5}\sum_i\eta_i\partial_i G$ respectively, 
where $\eta_i$ and $\bareta_i$ are the Clifford generators
that are used to construct ${\mathscr K}_+^{\rm mf}$.
The binding map is given by
\beq
  \varphi_1 = \bigg(\!\!\begin{array}{c}-\alpha \\
    G' \end{array}\!\!\bigg), \qquad
  \varphi_2 = \alpha G',
  \nn 
\eeq
in which $\alpha=\sum_i\eta_i\alpha_i$.
Following the procedure from (\ref{complexD}) to (\ref{complexD'}),
we eliminate the trivial brane-antibrane pair
$\wilson(-1)_{{}_{-2}}\stackrel{\id}{\longrightarrow}
\wilson(-1)_{{}_{-1}}$:
$$
  \begin{array}{*{8}{c@{\hspace{2pt}}}c}
    &&\wilson(0)^{}_{{}_0}\\[-13pt]
    &&
    \oplus&\MFarrow{}{}&
    \wilson(5)_{{}_1}^{}
    \\[-3pt]
    &&\wilson(4)_{{}_0}^{}\\[-20pt]
    &\dMFarrow{\hspace*{-20pt}(0,pX)}{\varphi_1}&&\darrow{\varphi_2}\\[5pt]
    \wilson(0)^{\oplus 5}_{{}_{-1}}&
    \MFarrow{X}{pG'}&\wilson(1)^{\oplus 10}_{{}_0}&
    \MFarrow{X}{pG'}&\wilson(2)^{\oplus 10}_{{}_1}&
    \MFarrow{X}{pG'}&\wilson(3)^{\oplus 5}_{{}_2}&
    \MFarrow{X}{pG'}&\wilson(4)^{}_{{}_3}.
  \end{array}
$$
Let us next adjoin the matrix factorization ${\mathscr K}_+^{\rm mf}$
in order to eliminiate the $\wilson(5)_{{}_1}$. 
Starting with the identity to
$\wilson(5)_{{}_2}$ in ${\mathscr K}_+^{\rm mf}$,
we first find the arrows in the ordinary direction, from left to right,
so that we obtain a complex if we ignore the arrows linear in 
$p$ that goes in the opposite way.
Unlike in the first binding shown above,
this does not yet make a matrix factorization of $pG(x)$
--- $Q^2$ has entries other than $pG(x)\cdot \id$.
This problem itself can be fixed by introducing arrows linear in $p$,
but the new arrows procude a problem at different entries.
This process terminates by finding arrows which are cubic in $p$.
The final form of the grade restricted matrix factorization, obtained after
elimination of $\wilson(5)_{{}_1}\stackrel{\id}{\longrightarrow}
\wilson(5)_{{}_2}$, is

{\small
\begin{center}
\begin{picture}(450,100)(-10,0)

\put(0,90){$\wilson(0)_{{}_{-3}}$}
\put(30,95){\vector(1,0){20}}
\put(50,92){\vector(-1,0){20}}
\put(55,90){$\wilson(1)^{\oplus 5}_{{}_{-2}}$}
\put(95,95){\vector(1,0){20}}
\put(115,92){\vector(-1,0){20}}
\put(120,90){$\wilson(2)^{\oplus 10}_{{}_{-1}}$}
\put(165,95){\vector(1,0){20}}
\put(185,92){\vector(-1,0){20}}
\put(190,90){$\wilson(3)^{\oplus 10}_{{}_0}$}
\put(235,95){\vector(1,0){20}}
\put(255,92){\vector(-1,0){20}}
\put(260,90){$\wilson(4)^{\oplus 5}_{{}_1}$}

\put(220,69){\vector(2,1){42}}
\put(263,86){\vector(-2,-1){42}}
\put(195,60){$\wilson(0)_{{}_0}$}
\put(220,35){\vector(1,1){50}}
\put(195,30){$\wilson(4)_{{}_0}$}

\put(135,10){\vector(2,3){52}}
\put(140,10){\vector(1,1){50}}
\put(149,12){\vector(2,1){42}}
\put(194,30){\vector(-2,-1){42}}
\put(193,10){\vector(-2,3){52}}
\put(225,10){\vector(2,3){50}}
\put(270,10){\vector(-2,3){52}}
\put(254,7){\vector(-2,1){160}}
\put(323,8){\vector(-4,1){309}}
\put(326,9){\vector(-2,1){160}}
\put(405,11){\vector(-4,1){309}}

\put(120,0){$\wilson(0)^{\oplus 5}_{{}_{-1}}$}
\put(165,5){\vector(1,0){20}}
\put(185,2){\vector(-1,0){20}}
\put(190,0){$\wilson(1)^{\oplus 10}_{{}_0}$}
\put(235,5){\vector(1,0){20}}
\put(255,2){\vector(-1,0){20}}
\put(260,0){$\wilson(2)^{\oplus 10}_{{}_1}$}
\put(305,5){\vector(1,0){20}}
\put(325,2){\vector(-1,0){20}}
\put(330,0){$\wilson(3)^{\oplus 5}_{{}_2}$}
\put(370,5){\vector(1,0){20}}
\put(390,2){\vector(-1,0){20}}
\put(395,0){$\wilson(4)_{{}_3}$}

\end{picture}
\end{center}
}

\noindent
To avoid cluttering the diagram,
we do not include the detail of the arrows. It is
left as an exercise to the reader.

We can now perform the transport to the LG
orbifold point, where we set $p=1$ and shift the R-symmetry as
$\RLG_{\bf i} = R_{\bf i} - 2 q_{\bf i}/5$. The final result is the
matrix factorization of $G(x)$
on R-graded $\Z_5$-equivariant vector bundles on $\C^5$,
$$
\left({\mathcal E}_0\MFarrow{g}{f}{\mathcal E}_1\right),
$$
\beqa
{\mathcal E}_0
&=&{\mathcal O}(\bar 0)_{{}_0}
\oplus{\mathcal O}(\bar 4)_{{}_{-{8\over 5}}}
\oplus{\mathcal O}(\bar 1)^{\oplus 10}_{{}_{-{5\over 2}}}
\oplus{\mathcal O}(\bar 3)^{\oplus 5}_{{}_{4\over 5}}
\oplus{\mathcal O}(\bar 1)^{\oplus 5}_{{}_{-{12\over 5}}}
\oplus{\mathcal O}(\bar 3)^{\oplus 10}_{{}_{-{6\over 5}}},
\nn\\
{\mathcal E}_1
&=&{\mathcal O}(\bar 0)^{\oplus 5}_{{}_{-1}}
\oplus{\mathcal O}(\bar 2)^{\oplus 10}_{{}_{1\over 5}}
\oplus{\mathcal O}(\bar 4)_{{}_{7\over 5}}
\oplus{\mathcal O}(\bar 0)_{{}_{-3}}
\oplus{\mathcal O}(\bar 2)^{\oplus 10}_{{}_{-{9\over 5}}}
\oplus{\mathcal O}(\bar 4)^{\oplus 5}_{{}_{-{3\over 5}}},
\nn
\eeqa
where the matrices
are
$$
g\,=\,\left(\begin{array}{cccccc}
0  &  X  &  G' &  0  &  0  &  0\\
0  &  0  &  X  &  G' &  0  &  0\\
0  &  0  &  0  &  X  &  0  &  0\\
0  &  0  &  0  &\!\!-{1\over 4!}(G'')^4\alpha\! & G'& 0\\
0  &  0  & \!\!-{1\over 2}\alpha(G'')^2\! 
&\!{1\over 3}(G'')^3\alpha\! & X & G'\\
-G'&-\alpha& G''\alpha &  0  &  0  &X
\end{array}\right),
$$
$$
f\,=\,\left(\begin{array}{cccccc}
-\alpha  &  0  &  0 &  0  &  0  &  -X\\
G'  &  0  &  0  &  0 &  0  &  0\\
X  &  G'  &  0  &  0  &  0  &  0\\
0  &  X  &  G'  &  0  &  0  & 0\\
0  & \!\!-{1\over 3!}\alpha(G'')^3\!  & \!{1\over 4!}(G'')^4\alpha\! 
&  X& G' & 0\\
-\alpha G''&\!{1\over 2}(G'')^2\alpha\!& 0 &  0  &  X  &G'
\end{array}\right).
$$
Here $G''={1\over 5\cdot 4}\sum_{ij}\barxi_i\eta_j\partial_i\partial_jG$
in which $\xi_i$ and $\barxi_i$ are the Clifford generators to
realize the second ${\mathscr K}_+^{\rm mf}$
as $Q=\sum_i\left(\xi_ix_i+{1\over 5}\barxi_ip\partial_iG\right)$.
The $X$, $G'$, $\alpha$ in parts of the entries of the matrices are
$X=\sum_i\xi_ix_i$, $G'={1\over 5}\sum_i\barxi_i\partial_iG$
and $\alpha=\sum_i\barxi_i\alpha_i$, and the remaining parts are 
those using the $\bareta_i$'s and the $\eta_i$'s.
Which is which should be obvious.

This is a completely new result. Unlike the previous examples, 
there was no attempt in the literature to make an
educated guess for the LG image of the D$4$-brane ${\mathcal O}_H(0)$
from R-R charge and mirror symmetry considerations.

\subsubsection{Two-Parameter Model}

Let us consider the two parameter model 
--- Example (C) ---
 with superpotential 
$W=PG(X)$ where
$$
G(x_1,...,x_6)=x_1^8x_6^4+x_2^8x_6^4+x_3^4+x_4^4+x_5^4.
$$
This model has four phases as depicted in Fig.~\ref{twoparaA}.
Phase I is the large volume regime where the low energy theory is
the non-linear sigma model on the hypersurface $M=\{G=0\}$
in a toric variety.
At the point $\e^{-t_1}=\e^{-t_2}=0$ in Phase III, the theory reduces to
the LG model of five variables $X_1,..,X_5$
with Fermat type superpotential
$$
W=X_1^8+X_2^8+X_3^4+X_4^4+X_5^4,
$$
modulo the $\Z_8$ orbifold group $X_a\mapsto \omega X_a$,
$X_i\mapsto \omega^2 X_i$ ($\omega^8=1$) for $a=1,2,$ $i=3,4,5$.
The R-charges of the variables are
$\RLG[x_a]={1\over 4}$, $\RLG[x_i]={1\over 2}$.
We consider paths in the K\"ahler moduli space
between the LG orbifold point and the large volume phase via Phase II.
For a suitable choice of windows at the two phase boundaries
(i.e. window $w_1$ at the III-II boundary and window $w'$ at the II-I boundary;
see Section~\ref{subsub:McKay}),
we have the grade restriction rule:
\beq
\mathfrak{C}^w=\{(0,0),(0,1),(1,0),(1,1),(2,0),(2,1),(3,0),(3,1)\}.
\label{grrtwopara}
\eeq

\subsubsection*{\it ${\bf L}=0^5$ RS-Branes}

We consider the ${\bf L}=(0,0,0,0,0)$ RS-branes ${\mathcal B}_{0^5,q,r}$
at the LG orbifold point given by the matrix factorization
\beq
\QLG_{0^5}=\sum_{a=1,2}\left(x_a\eta_a+x_a^7\bareta_a\right)
+\sum_{i=3,4,5}\left(x_i\eta_i+x_i^3\bareta_i\right).
\label{2paraL=0}
\eeq
The R-charges of the Clifford generators
are $\RLG[\bareta_a]=-{3\over 4}$, $\RLG[\bareta_i]=-{1\over 2}$.
Using these, we find the R-charges of the Chan-Paton vectors 
in terms of the one $\RLG_{|0\rangle}$ for the vacuum vector 
$|0\rangle$.
For example, $\RLG[\bareta_a|0\rangle]=\RLG_{|0\rangle}-{3\over 4}$,
$\RLG[\bareta_i|0\rangle]=\RLG_{|0\rangle}-{1\over 2}$.
We focus on the branes ${\mathcal B}_{0^5,q,0}$ where
$|0\rangle$ is even.

Now let us consider the lift to the linear sigma model.
The first step is to solve the equation
\beq
\RLG_{\bf i}=R_{\bf i}-{2(2q_{\bf i}^1+q_{\bf i}^2)\over 8}
\label{Req2p}
\eeq
for each Chan-Paton vector.
There is a unique solution if we require $(q_{\bf i}^1,q_{\bf i}^2)$ to be
in the grade restriction range (\ref{grrtwopara}).
Once $(R_{\bf i}, q^1_{\bf i},q^2_{\bf i})$ 
are determined, we multiply each entry of the matrix $\overline{Q}_{0^5}$
by an appropriate power of $p$ and
$x_6$
so that the gauge invariance holds.
This leads to the matrix factorization $Q(p,x_1,..,x_6)$ 
of $W=pG(x)$
for the linear sigma model brane.

Let us first consider the brane 
${\mathcal B}_{0^5,0,0}$ with $\RLG_{|0\rangle}=0$.
The solution to the equation (\ref{Req2p})
is listed in the table below.
\begin{center}
\begin{tabular}{|c||c|c|c|c|c|c|}
\hline
vector
&$|0\rangle$
&$\bareta_a|0\rangle$
&$\bareta_i|0\rangle$
&$\bareta_1\bareta_2|0\rangle$
&$\bareta_a\bareta_i|0\rangle$
&$\bareta_i\bareta_j|0\rangle$\\
\hline
$\RLG$&0&$-{3\over 4}$&$-{1\over 2}$&$-{3\over 2}$&$-{5\over 4}$
&$-1$\\
\hline
$R$&0&1&1&0&0&0\\
\hline
$(q^1,q^2)$&(0,0)&(3,1)&(3,0)&(3,0)&(2,1)&(2,0)\\
\hline
image&$\wilson(0,0)$&$\wilson(3,1)^{\oplus 2}$&
$\wilson(3,0)^{\oplus 3}$&$\wilson(3,0)$&
$\wilson(2,1)^{\oplus 6}$&$\wilson(2,0)^{\oplus 3}$
\\
\hline
\end{tabular}
\end{center}
\begin{center}
\begin{tabular}{|c|c|c|c|c|c|}
\hline
$\bareta_1\bareta_2\bareta_i|0\rangle$
&$\bareta_a\bareta_i\bareta_j|0\rangle$
&$\bareta_3\bareta_4\bareta_5|0\rangle$
&$\bareta_1\bareta_2\bareta_i\bareta_j|0\rangle$
&$\bareta_a\bareta_3\bareta_4\bareta_5|0\rangle$
&$\bareta_1\cdots\bareta_5|0\rangle$\\
\hline
$-2$&$-{7\over 4}$&$-{3\over 2}$&$-{5\over 2}$&$-{9\over 4}$&$-3$\\
\hline
$-1$&$-1$&$-1$&$-2$&$-2$&$-3$\\
\hline
(2,0)&(1,1)&(1,0)&(1,0)&(0,1)&(0,0)
\\
\hline
$\wilson(2,0)^{\oplus 3}$&$\wilson(1,1)^{\oplus 6}$&$\wilson(1,1)$
&$\wilson(1,0)^{\oplus 3}$&$\wilson(0,1)^{\oplus 2}$&$\wilson(0,0)$\\
\hline
\end{tabular}
\end{center}
Applying the reduction map in the large volume regime, we find the 
following semi-infinite complex of vector bundles on the hypersurface
$M$:
\\[-0.5cm]
\begin{figure}[h]
\psfrag{o}{\footnotesize ${\mathcal O}$}
\psfrag{l-2}{\!\!\!\!\footnotesize ${\mathcal O}(0,1)^{\oplus 2}$}
\psfrag{h-3}{\!\!\!\!\footnotesize ${\mathcal O}(1,0)^{\oplus 3}$}
\psfrag{h4}{\!\!\!\!\footnotesize ${\mathcal O}(4,0)$}
\psfrag{h}{\!\!\!\!\footnotesize ${\mathcal O}(1,0)$}
\psfrag{hl-6}{\! \footnotesize ${\mathcal O}(1,1)^{\oplus 6}$}
\psfrag{h2-3}{~~~ \footnotesize ${\mathcal O}(2,0)^{\oplus 3}$}
\psfrag{h4l-2}{ \footnotesize  $\underline{{\mathcal O}(4,1)^{\oplus 2}}$}
\psfrag{h5-3}{\footnotesize ${\mathcal O}(5,0)^{\oplus 3}$}
\psfrag{h2-3}{\footnotesize ${\mathcal O}(2,0)^{\oplus 3}$}
\psfrag{h2l-6}{ \footnotesize  ${\mathcal O}(2,1)^{\oplus 6}$}
\psfrag{h3}{\!\!\!\!\footnotesize ${\mathcal O}(3,0)$}
\psfrag{h8}{\!\!\footnotesize ${\mathcal O}(8,0)$}
\psfrag{h5}{\!\!\footnotesize ${\mathcal O}(5,0)$}
\psfrag{h5l-6}{\ \footnotesize ${\mathcal O}(5,1)^{\oplus 6}$}
\psfrag{h6-3}{\footnotesize ${\mathcal O}(6,0)^{\oplus 3}$}
\psfrag{h3-3}{\footnotesize ${\mathcal O}(3,0)^{\oplus 3}$}
\psfrag{h3l-2}{ \footnotesize ${\mathcal O}(3,1)^{\oplus 2}$}
\centerline{\includegraphics[scale=1.20]{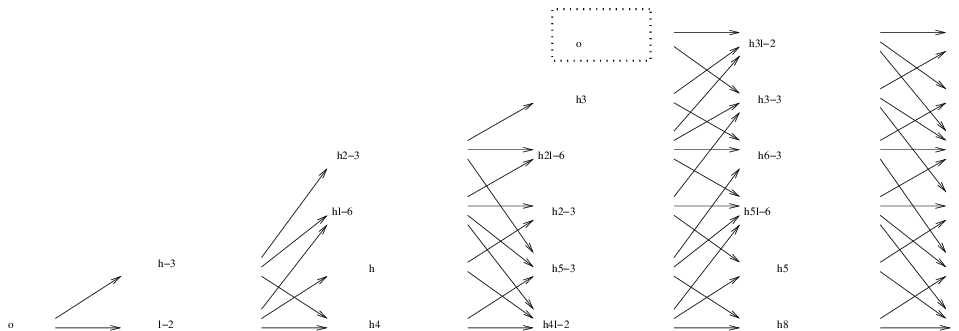}}
\end{figure}
\\[-0.2cm]
Each arrow corresponds to
$$
X=\sum_{a=1,2}x_a\eta_a,\quad
X^7=\sum_{a=1,2}x_a^7\bareta_a,\qquad
Y=\sum_{i=3,4,5}x_i\eta_i,\quad
Y^3=\sum_{i=3,4,5}x_i^3\bareta_i
$$
possibly with multiplication by a power of $x_6$.
The power of $x_6$ can be found by matching the gauge
charges of the source and the target.
For example, the arrows for the first few parts of the complex is
given as follows:\\
\begin{figure}[h]
\psfrag{o}{~~~\footnotesize ${\mathcal O}$}
\psfrag{l-2}{\!\!\!\!\footnotesize ${\mathcal O}(0,1)^{\oplus 2}$}
\psfrag{h-3}{\!\!\!\!\footnotesize ${\mathcal O}(1,0)^{\oplus 3}$}
\psfrag{h4}{\!\!\!\footnotesize ${\mathcal O}(4,0)$}
\psfrag{h}{\!\!\!\footnotesize ${\mathcal O}(1,0)$}
\psfrag{hl-6}{~~~~\footnotesize ${\mathcal O}(1,1)^{\oplus 6}$}
\psfrag{h2-3}{~~~\footnotesize ${\mathcal O}(2,0)^{\oplus 3}$}
\psfrag{x}{\footnotesize $X$}
\psfrag{y}{\footnotesize $Y$}
\psfrag{y3z}{\footnotesize $X^7x_6^4$}
\psfrag{xz}{\footnotesize $Xx_6$}
\psfrag{y3}{\footnotesize $Y^3$}
\centerline{\includegraphics[width=8cm,height=3.5cm]{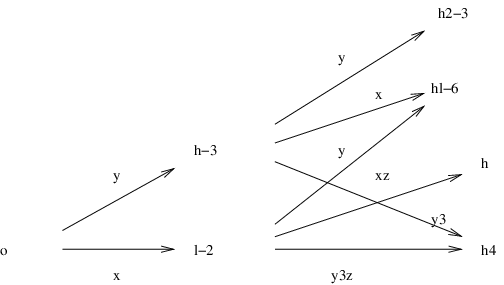}}
\end{figure}
\\
The part of the complex other than ${\mathcal O}$ in the dashed box
is exact. Thus, the whole complex is quasi-isomorphic to ${\mathcal O}$
at R-degree $0$.

One may repeat this procedure to find the large volume images of 
the other ${\bf L}=0^5$ RS-branes. However, as we have done in 
the quintic case, we may proceed more economically, using
what we have done for McKay correspondence.
The grade restricted lift
of the RS-brane ${\mathcal B}_{0^5,q,0}$, obtained by solving (\ref{Req2p}),
is the same as 
the grade restricted lift $\lsmB_{\bar q}'$ of the fractional brane
${\mathcal O}_{\mathfrak{p}}(\bar q)$, when we supplement it by
extra arrows going in the opposite direction.
For example, compare the above table with
the complex $\lsmB_{\bar 0}'$ written in page \pageref{page:B0}.
As we have seen in Section~\ref{subsub:McKay},
the complex $\lsmB_{\bar q}'$ is presented as the bound state
of a complex that is trivial in the large volume regime
and another complex of the form
$$
\mathfrak{A}_q(4,0)[1]\,\stackrel{p}{\longrightarrow}\,
\mathfrak{A}_q.
$$
Here $\mathfrak{A}_q$ is the Wilson line counterpart
of the complex ${\mathscr A}_q$ of 
vector bundles shown in (\ref{LVfrac}).
There is a complete parallel in the matrix factorization version:
the grade restricted lift
of the RS-brane can be presented as a bound state of
a matrix factorization that is trivial at large volume
and another matrix factorization given by
$$
\mathfrak{A}_q(4,0)[-1]\,\MFarrow{p}{G}\,
\mathfrak{A}_q.
$$
Applying the second Kn\"orrer map to this complex,
we find the large volume image.
The result is simply the restriction of the
large volume image of the fractional brane to the hypersurface $M=\{G=0\}$.
To summarize, the transport of the ${\bf L}=0^5$ RS-branes from the 
LG orbifold to the large volume regime results in the maps
$$
\qquad\qquad\qquad
{\mathcal B}_{0^5,q,0}\,\longmapsto\,
{\mathscr A}_q\Bigl|_M,\qquad \mbox{for ~$q=0,1,...,7$.}
$$
This can also be obtained
from the semi-infinite complex by eliminating
the exact pieces.
The Chern character of these branes are
obtained from those of ${\mathscr A}_q|_E$ listed in (\ref{ChfracB})
by restriction to $M$. 
It matches with the one in \cite{EmMk}, up to a 
translation of labelling.

\subsubsection*{\it Short Orbit Branes}

We next consider the short orbit branes
in the LG orbifold with $L_i+1={d_i\over 2}$ for all $i$, that is,
${\bf L}=(3,3,1,1,1)=:{\bf k\over 2}$.
Recall from Section~\ref{subsub:LGO}
that they are denoted by $\widehat{\mathcal B}_{{\bf k\over 2},q,r}$.
The associated matrix factorization is
$$
\QLG=x_1^4\, \xi_1+x_2^4\, \xi_2+x_3^2\, \xi_3+x_4^2\, \xi_4+x_5^2\, \xi_5
$$
represented on a complex module $\cp_3$ over the real Clifford algebra
$$
\qquad\qquad
\{\xi_i,\xi_j\}=2\delta_{i,j}\qquad \mbox{for~ $0\leq i,j\leq 5$.}
$$
The $\xi_i$'s are invariant under the R-symmetry and transform as
$\xi_i\to-\xi_i$ under the orbifold group generator.
The module $\cp_3$ is generated by a vector $|0\rangle$
which is annihilated by $\xi_1+i\xi_2$, $\xi_3+i\xi_4$,
$\xi_5+i\xi_0$.
It has dimension $2^3=8$.
The $\Z_2$-grading is such that even multiples of the $\xi_i$'s 
on $|0\rangle$ are even.

Let us first consider $\widehat{\mathcal B}_{{\bf k\over 2},0,0}$
where all the vectors of $\cp_3$ have $\RLG=0$.
The grade restricted 
lift to the linear sigma model is found by solving
$\RLG_{\bf i}=R_{\bf i}-2(2q_{\bf i}^1+q_{\bf i}^2)/8$:
\begin{center}
\begin{tabular}{|c||c|c|}
\hline
vector&even&odd\\
\hline
$\RLG_{\bf i}$&0&0\\
\hline
$R_{\bf i}$&0&1\\
\hline
$q_{\bf i}$&(0,0)&(2,0)\\
\hline
&$\wilson(0,0)^{\oplus 4}$&$\wilson(2,0)^{\oplus 4}$\\
\hline
\end{tabular}
\end{center}
The explicit factorization is
$$
\left(\,\wilson(0,0)^{\oplus 4}_{{}_0}\,\MFarrow{g}{pg}\,
\wilson(2,0)^{\oplus 4}_{{}_1}\,\right)
$$
with
$$
g=x_1^4x_6^2\, \xi_1+x_2^4x_6^2\, 
\xi_2+x_3^2\, \xi_3+x_4^2\, \xi_4+x_5^2\, \xi_5.
$$
The large volume image is
$$
0\to
\underline{{\mathcal O}(0,0)^{\oplus 4}}\stackrel{g}{\longrightarrow}
{\mathcal O}(2,0)^{\oplus 4}\stackrel{g}{\longrightarrow}
{\mathcal O}(4,0)^{\oplus 4}\stackrel{g}{\longrightarrow}
\cdots.
$$
Or, we may simply take the kernel of the first map.
Repeating the same procedure, we find the following transportarion rule
for the ${\bf L}={\bf {k\over 2}}$ short orbit branes
\beqa
&&\widehat{\mathcal B}_{{\bf {k\over 2}},0,0}
\,\longmapsto\,
{\rm Ker}\Bigl(g:{\mathcal O}(0,0)^{\oplus 4}\longrightarrow
{\mathcal O}(2,0)^{\oplus 4}\Bigr),\nn\\
&&\widehat{\mathcal B}_{{\bf {k\over 2}},1,0}
\,\longmapsto\,
{\rm Ker}\Bigl(g:{\mathcal O}(0,1)^{\oplus 4}\longrightarrow
{\mathcal O}(2,1)^{\oplus 4}\Bigr),\nn\\
&&\widehat{\mathcal B}_{{\bf {k\over 2}},2,0}
\,\longmapsto\,
{\rm Ker}\Bigl(g:{\mathcal O}(1,0)^{\oplus 4}\longrightarrow
{\mathcal O}(3,0)^{\oplus 4}\Bigr),\nn\\
&&\widehat{\mathcal B}_{{\bf {k\over 2}},3,0}
\,\longmapsto\,
{\rm Ker}\Bigl(g:{\mathcal O}(1,1)^{\oplus 4}\longrightarrow
{\mathcal O}(3,1)^{\oplus 4}\Bigr).\nn
\eeqa
The images of the remaining branes can be found
using $\widehat{\mathcal B}_{{\bf {k\over 2}},q+4,r}
\cong \widehat{\mathcal B}_{{\bf {k\over 2}},q,r-1}$.

\subsection{Monodromy}
\label{subsec:CompactMonodromy}

Let us now study monodromies in the compact models, 
that is, D-brane transport along 
non-trivial loops in the moduli space $\moduli_K$.
For this purpose it is best to stay at the intermediate energy
scale $\mu \gg m_W$ and simply work with the description of D-branes
through matrix factorizations, i.e., study the hat diagram
(\ref{hatdiagram2}). Then, 
the essential ideas carry over from the non-compact models.
When necessary, we can always translate the result
in terms of the low energy description.

\subsubsection*{\it Example (A)}

Let us first consider Example (A) with superpotential
$W=PG(X_1,...,X_N)$ for a degree $N$ polynomial $G(x_1,...,x_N)$.
The K\"ahler moduli space is the same as for the theory without
superpotential. It is complex one dimsnional and has three special points:
the large volume limit $r\to +\infty$, the LG orbifold point
$r\to -\infty$ and a singular point $\e^t=(-N)^N$.

The monodromies around the large volume limit and LG orbifold point 
come from shifting the theta parameter, $\theta \rightarrow \theta + 2\pi$. 
This is equivalent to keeping the theta parameter fixed and shifting the
representation of the gauge group:
\beq
  \label{LVmono}
  \mono_{\theta\rightarrow\theta+ 2\pi}
  \bigl(\pi_r({\cal V},Q,\rho(g),R)\bigr) = 
  \pi_r({\cal V},Q,g\rho(g),R)  ,
\eeq
i.e., the matrix factorization is tensored by $\wilson(1)$.
In the large volume regime we can express the monodromy in
terms of D-branes in $D(M)$, so that  (\ref{LVmono}) simply becomes tensoring
by $\cO(1)$.
In the Landau-Ginzburg phase, as
encountered in Section~\ref{subsec:LGOpt}, 
the vacuum expectation value, $p=1$, enforces a
shift of the R-symmetry action, i.e., 
$\RLG_{\bf i} = R_{\bf i} - 2q_{\bf i}/N$. 
Therefore, for a loop around the LG orbifold
point, the monodromy action (\ref{LVmono}) on $\ZZ_N$-graded matrix
factorizations becomes
$$
  \mono_{GP}
  ({\cal V},\bar Q,\bar\rho(\omega),\bar R(\lambda)) =
  ({\cal V},\bar Q,\omega\bar\rho(\omega) ,\lambda^{- 2/N}\bar
R(\lambda)).
$$
This yields the relation
$$
  \left(\mono_{GP}\right)^{N} \cong~ [2],
$$ 
where $[2]$ denotes the shift of R-charge by $-2$.
From the linear sigma model point
of view this is a consequence of the D-isomorphism
$\pi_-({\cal V},Q,g^N \rho(g),R)\cong\pi_-({\cal V},Q,\rho(g),\lambda^{-2}R)$,
which is a consequence of the fact that the cone of the map  
$$
  \begin{array}{ccc}
    {\cal V}^{od}(N) &
    \MFarrow{f}{g} &
    {\cal V}^{ev}(N) \\
  \begin{picture}(40,20)
    \put(20,15){\vector(0,-1){30}}
    \put(-15,0){\small $p \cdot \id_{od}$}
  \end{picture} &&
  \begin{picture}(40,20)
    \put(20,15){\vector(0,-1){30}}
    \put(25,0){\small $p \cdot \id_{ev}$}
  \end{picture}\\
    {\cal V}^{od}[2] &
    \MFarrow{f}{g} &
    {\cal V}^{ev}[2]
  \end{array}  
  \vspace*{1mm}
$$
is an empty D-brane.

\begin{figure}[tb]
\centerline{\includegraphics{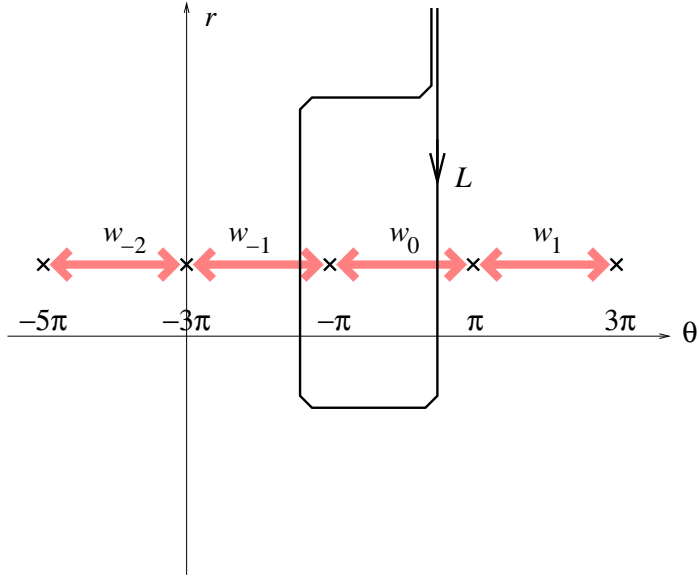}}
\caption{A loop around the singular point in $\mathfrak{M}_K$}
\label{fig:monodromy4}
\end{figure}
Next, we consider the monodromy around the singular point.
To be specific, we take the example with $N=3$ and the Fermat polynomial
$G=x_1^3+x_2^3+x_3^3$. 
Consider the loop $L$ as depicted in Fig~\ref{fig:monodromy4} which is
based at large volume and goes around 
the singular point. As we
follow $L$ we have to grade restrict first according to the window $w_0$
and then with respect to the window $w_{-1}$. 
The grade restriction rules for these windows are:
\beq
  \nonumber
  \mathfrak{C}^{w_0} = \{-1,0,1\} \quad \mathrm{and}\quad
  \mathfrak{C}^{w_{-1}} = \{0,1,2\}.
\eeq
Let us illustrate the monodromy around the singular point by 
looking at its action on the D0-brane, 
$\cO_{\mathfrak{p}} \in D(\hyper)$, at the point 
$\mathfrak{p}=\{x_1+x_2=x_3=0\}$ of
the elliptic curve. 
In Section \ref{subsub:LVlift} we found that it
is lifted to the matrix factorization
$$
  \lsmB_{\mathfrak{p}}:~~\Biggl(\begin{array}{c}
    \wilson(0)^{}_{{}_0}\\[-5pt] \oplus\\[-5pt] \wilson(0)^{}_{{}_0}
  \end{array}
    \MFarrow{g}{f} 
  \begin{array}{c}
    \wilson(1)^{}_{{}_1}\\[-5pt] \oplus\\[-5pt] \wilson(-1)^{}_{{}_{-1}}
  \end{array}\Biggr),
$$
where $f$ and $g$ are given in (\ref{D0MF}).
Following the loop $L$ we note that $\lsmB_{\mathfrak{p}}$ indeed fits
into the grade restricted set $\frak{MF}_W({\cal T}^{w_0})$, and thus
can be transported to the Landau-Ginzburg phase. When going back to
the large volume phase through the window $w_{-1}$ we note that the Wilson
line component $\wilson(-1)$ is in conflict with $\mathfrak{C}^{w_{-1}}$. 
So, we have to eliminate it first by binding 
$\lsmB_{\mathfrak{p}}$ with the brane $\lsmB_-$ given in (\ref{LGtrivialMF})
that
is trivial in the Landau-Ginzburg phase. The resulting matrix
factorization is
$$
  \lsmB'_{\mathfrak{p}}:~~\Biggl(\begin{array}{c}
    \wilson(0)^{}_{{}_0}\\[-5pt] \oplus\\[-5pt] 
    \wilson(0)^{}_{{}_0}
  \end{array}
    \MFarrow{g'}{f'} 
    \begin{array}{c}
      \wilson(1)^{}_{{}_1}\\[-5pt] \oplus\\[-5pt] 
      \wilson(2)^{}_{{}_1}
    \end{array}
  \Biggr),
$$
with
\beq
  \nonumber
  g' = \left(\begin{array}{cc} 
    x_1\!+\!x_2& x_3 \\ 
    -x_3^2& x_1^2-x_1x_2+x_2^2
  \end{array}\right),~~
  f' = p\left(\begin{array}{cc}
    x_1^2-x_1x_2+x_2^2& -x_3 \\ 
    x_3^2& x_1\!+\!x_2
  \end{array}\right).
\eeq
Back at the large volume regime, 
as a geometric D-brane in $D(M)$ this matrix factorization
becomes 
$$
    \underline{\cO(1)^{\oplus 2}}
    \stackrel{(x_1\!+x_2,x_3)}{-\!\!\!-\!\!\!-\!\!\!\longrightarrow} 
    \cO(2)
    ~ \longrightarrow ~
    \cO_{\mathfrak{p}}
$$
with the B-field in the range
$B\in (0,2\pi)$.
The monodromy along the loop $L$ is therefore:
$$
  \mono_{L}(\cO_{\mathfrak{p}}) ~\cong~
  \left(
    \underline{\cO(1)^{\oplus 2}}
    \stackrel{(x_1\!+x_2,x_3)}{-\!\!\!-\!\!\!-\!\!\!\longrightarrow} 
    \cO(2)
    ~ \longrightarrow ~
    \cO_{\mathfrak{p}}
  \right).
$$

In the spirit of \cite{Strominger} let us understand the monodromy
action in terms of binding massless D-branes
to the original brane. For that let us start with a D-brane in
$MF_W(M)$ represented via its lift in the grade
restricted set $\frak{MF}_W({\cal T}^{w_0})$. When we follow the loop
we have to apply grade restriction with respect to the window
$w_{-1}$, which means that we have to eliminate all components
$\wilson(-1)$ by binding of the matrix factorization $\lsmB_-$ given
in (\ref{LGtrivialMF}). When
we are back to the large volume phase we try to find again a
representative in $\frak{MF}_W({\cal T}^{w_0})$. So, we have to
eliminate the component
$\wilson(2)$ in $\lsmB_-$ in terms of binding
$\lsmB_+$ given in (\ref{LVtrivialMF}). In total, to every Wilson
line component $\wilson(-1)$ at R-degree $j$ in the original D-brane
$\lsmB$ we have to bind the matrix factorization, 
$$
  \lsmB_0:~~\Biggl(
  \begin{array}{c}
    \wilson(-1)^{}_{{}_{-2}} \\[-5pt]
    \oplus    \\[-5pt]
    \wilson(1)^{\oplus 3}_{{}_0}
  \end{array}
  \MFarrow{g_0}{f_0}
  \begin{array}{c}
    \wilson(0)^{\oplus 3}_{{}_{-1}} \\[-5pt]
    \oplus    \\[-5pt]
    \wilson(-1)^{}_{{}_{-1}}
  \end{array}\Biggr)  ,  
$$
with appropriate shifts in R-degree. The matrices $f_0$ and $g_0$ 
are defined in (\ref{masslessMF}). Chiral ring elements in
$\Ho^{j}(\pi_+(\lsmB),\pi_+(\lsmB_0)[-1])$ are in one-to-one correspondence
with the number of Wilson line component
$\wilson(-1)$ at R-degree $-j$.
This lets us write the monodromy action
around the singular point as
\beq
  \label{monoB0plus}
  \mono_{L}(\pi_+(\lsmB)) =    
  {\rm Cone}\left(\pi_+(\lsmB)\longrightarrow\bigoplus_{j\in\Z}
    \Ho^{j}(\pi_+(\lsmB),\pi_+(\lsmB_0))\otimes \pi_+(\lsmB_0)[j]
    \right)[-1].
\eeq
If we want to express this monodromy action in terms of objects in
$D(M)$, we need to map $\lsmB_0$ to its image as geometric D-brane. 
We know that the image is the line bundle $\cO$ with the
B-field in the interval $(0,2\pi)$. On a general D-brane
$\cB \in D(M)$ the monodromy action is therefore
$$
  \mono_{L}(\cB) =    
  {\rm Cone}\left(\cB\longrightarrow\bigoplus_{j\in\Z}
    \Ho^{j}(\cB,\cO)\otimes \cO[j]
    \right)[-1].
$$
In the context of derived categories this monodromy action was first
suggested by Kontsevich and further studied in 
\cite{SeTh,Horja,Douglas,AspDouglas,AKH}. 
Indeed, from considerations in the mirror
dual theory the D-brane $\cO$ was found to become massless at the
singular point, see for instance \cite{Aspinwall}.

\begin{figure}[htb]
\centerline{\includegraphics{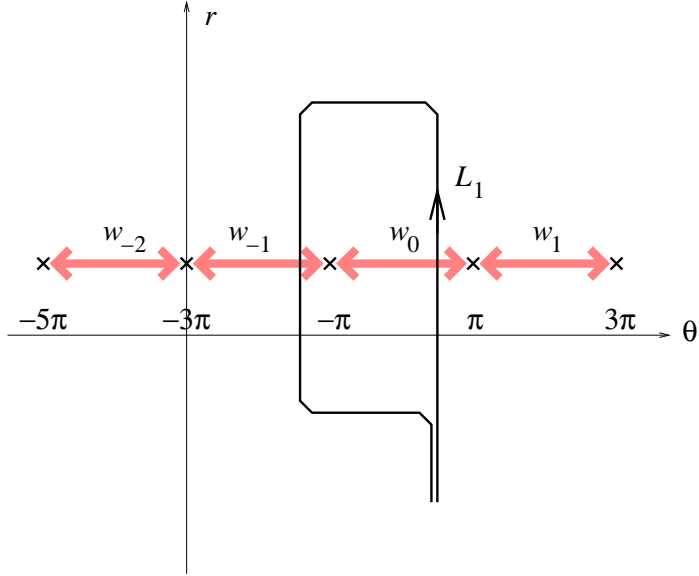}}
\caption{Another loop around the singular locus in  $\mathfrak{M}_K$}
\label{fig:monodromy5}
\end{figure}
Similar arguments show that the monodromy around the singular point
following the loop $L_1$ in Fig.~\ref{fig:monodromy5}, which has its
base point at small volume, is given by:
\beq
  \label{monoB0minus}
  \mono_{L_1}(\pi_-(\lsmB)) =
  {\rm Cone}\left(\bigoplus_{j\in\Z}
    \Ho^j(\pi_-(\lsmB_0),\pi_-(\lsmB))\otimes \pi_-(\lsmB_0)[-j]
    \longrightarrow\pi_-(\lsmB)\right).
\eeq
This monodromy action on matrix factorizations in Landau-Ginzburg
orbifold models was recently suggested in \cite{Jockers}. In fact,
$\pi_-(\lsmB_0)$ is one of the ${\bf L}=0^3$ RS-branes of the Gepner model
associated with the LG orbifold.

For general $N$, the brane that plays the r\^ole of $\lsmB_0$
is
\[{\small
\setlength{\unitlength}{0.7mm}
\begin{picture}(170,50)
\put(125,40){$\wilson(q\!+\!N)^{}_{{}_{-1}}$}
\put(160,46){\footnotesize $-p$}
\put(160,36){\footnotesize $-G$}
\put(155,43){\vector(1,0){15}}
\put(170,41){\vector(-1,0){15}}
\put(175,40){$\wilson(q)^{}_{{}_{-2}}$}
\put(153,29){\footnotesize $\id$}
\put(155,35){\vector(1,-1){15}}
\put(167,29){\footnotesize $G'$}
\put(170,35){\vector(-1,-1){15}}
\put(-26,10){$\wilson(q)^{}_{{}_{-\!N}}$}
\put(-5,16){\footnotesize $X$}
\put(-6,4){\footnotesize $pG'$}
\put(-7,13){\vector(1,0){12}}
\put(5,11){\vector(-1,0){12}}
\put(8,10){$\wilson(q\!+\!1)^{\oplus N}_{{}_{-\!N\!+\!1}}$}
\put(39,16){\footnotesize $X$}
\put(38,4){\footnotesize $pG'$}
\put(37,13){\vector(1,0){12}}
\put(49,11){\vector(-1,0){12}}
\put(52,10){$\wilson(q\!+\!2)^{\oplus {N\choose 2}}_{{}_{-N\!+\!2}}$}
\put(85,16){\footnotesize $X$}
\put(84,4){\footnotesize $pG'$}
\put(84,13){\vector(1,0){10}}
\put(93,11){\vector(-1,0){10}}
\put(98,10){$\cdots$}
\put(111,16){\footnotesize $X$}
\put(110,4){\footnotesize $pG'$}
\put(108,13){\vector(1,0){10}}
\put(118,11){\vector(-1,0){10}}
\put(121,10){$\wilson(q\!+\!\!N\!\!-\!1)^{\oplus N}_{{}_{-1}}$}
\put(160,16){\footnotesize $X$}
\put(159,4){\footnotesize $pG'$}
\put(158,13){\vector(1,0){12}}
\put(170,11){\vector(-1,0){12}}
\put(173,10){$\wilson(q\!+\!N)^{}_{{}_0}.$}
\end{picture}
}\]
It satisfies the grade restriction for 
$\mathfrak{C}^w =\{q,q+1,\ldots,q+N-1\}$ 
after eliminating the trivial pair and,
therefore, $\lsmB_0$ induces the monodromy around the singular point 
$r=N \log N$, $\theta/2\pi = -q-N/2$. In the large volume phase 
it becomes the
D-brane $\cO(q)$ with $B=\theta + N\pi$, and at the LG orbifold point it
becomes one of the Recknagel-Schomerus ${\bf L}=0^N$ RS-branes
if $G$ is the Fermat polynomial.

\subsubsection*{\it The General Case}

It is straightforward now to find 
``the vanishing cycle'',
the D-brane that 
induces the monodromy around the singular point,
in a general theory with one-dimensional 
moduli space
$\moduli_K$.
It is obtained by binding
two D-branes that are trivial at $r\ll 0$ and $r\gg 0$.
Such trivial branes had been written down
in (\ref{trivminus}) and (\ref{trivplus}) respectively.
The map which binds them together is
$$
\varphi:{\mathscr K}_-^{\rm mf}({\mathscr S})\to {\mathscr K}_+^{\rm mf}
$$
that sends the leftmost $\wilson({\mathscr S})$ 
to the rightmost $\wedge^k\wilson_+
\cong \wilson({\mathscr S})$
by the identity and sends $\wilson_-({\mathscr S})$
to $\wedge^{k-1}\wilson_+$ by the map $c$ associated with $c_{ij}(x,y)$
satisfying the identities
$$
a_i=\sum_{j=1}^lc_{ij}(x,y)y_j,\qquad
b_j=\sum_{i=1}^kx_ic_{ij}(x,y).
$$
Such $c_{ij}$ exists since $a_i$ have negative and $b_j$ have positive
charges. (In particular, we have $W=\sum_{i,j}x_ic_{ij}(x,y)y_j$.)
Here we have defined the R-charge of the 
trivial branes in such a way that $\wilson({\mathscr S})$ has R-charge zero 
in both ${\mathcal K}_-^{\rm mf}$ and ${\mathcal K}_+^{\rm mf}$.
It is easy to see that $\varphi$ is then a degree zero map.
The vanishing cycle is
thus
\beq
\vanishing^{\rm mf}={\rm Cone}\Bigl(\,\varphi:
{\mathcal K}_-^{\rm mf}({\mathscr S})
\longrightarrow {\mathcal K}_+^{\rm mf}\,\Bigr)
\eeq

Monodromies in theories with higher dimensional moduli space
$\moduli_K$ are described in the same way as in
one-parameter models. 
Loops around special points deep inside a 
K\"ahler cone correspond to shifts of the gauge charges:
the loop $\theta^a \rightarrow \theta^a + 2\pi$ induces
$\wilson(q^1,\ldots,q^a,\ldots,q^k)\rightarrow
\wilson(q^1,\ldots,q^a\!+\!1,\ldots,q^k)$ 
on every Wilson line component of the matrix factorization.
The monodromy around the singular locus $\mathfrak{S}$ between two
phases is controlled by the associated band restriction rule, say
$\mathfrak{C}^w_{\rm I,II}$. We start with a D-brane in Phase I and take
its representative in the band restricted set $\MFlsmT$. When we
circle around the singular locus the band will be shifted and some
Wilson line components will fall out of the new band. These components must
be eliminated through binding matrix factorizations 
${\mathscr K}_{-,A}^{\rm mf}$
associated with $\Delta^{\rm I,II}_-$,
which appears in the relation
(\ref{importantre}) between the deleted sets.
The subscript ``$A$'' stems from the fact that there are a
multitude of D-branes ${\mathscr K}_{-,A}^{\rm mf}$, 
which are related by shifts of
the gauge charges along the band. Then, back in Phase I we may 
band restrict again with respect to $\mathfrak{C}^w_{\rm I,II}$ 
using the matrix
factorization ${\mathscr K}_{+,A}^{\rm mf}$ that comes with
$\Delta^{\rm I,II}_+$. Altogether we bind D-branes 
$\vanishing^{\rm mf}_{A}
= {\rm Cone}({\mathscr K}_{-,A}^{\rm mf} \rightarrow 
{\mathscr K}_{+,A}^{\rm mf})$
to the original D-brane. 
For the loop in the opposite direction, the branes 
that are bound are
$\vanishing_A^{\prime {\rm mf}} 
= {\rm Cone}({\mathscr K}_{+,A}^{\rm mf} \rightarrow
{\mathscr K}_{-,A}^{\rm mf} )$.
One may find formulas analogous to (\ref{monoB0plus}) and
(\ref{monoB0minus}). However, we point out once again that our approach 
does not rely on determining chiral ring spectra between
D-branes and, therefore, facilidates  computations considerably.

\subsection{Relation To Orlov's Functors}
\label{subsec:relOrlov}

Suppose a linear sigma model reduces in one phase
to the non-linear sigma model on a Calabi-Yau manifold
$M_{\rm LV}$ and in another phase to the Landau-Ginzburg orbifold
with a superpotential $W_{\rm LG}$ and an orbifold group $\Gamma$.
We know how to transport D-branes
along any path $\tau$ in the moduli space $\moduli_K$ that passes
through phase boundaries and connects these two phases.
This leads to a map of low energy D-branes
\beq
F_{\tau}: MF_{\Gamma}^{}(W_{\rm LG})\longrightarrow D(M_{\rm LV}),
\eeq
and its inverse. Since the parallel transport preserves
the chiral sector, as the truncated version of the above map,
we have a functor between the categories of D-branes
\beq
\Phi_{\tau}:{\bf MF}_{\Gamma}^{}(W_{\rm LG})\longrightarrow
{\bf D}(M_{\rm LV}),
\label{ours}
\eeq
and its inverse, that is, an equivalence of the two categories.
Of course, this is just an example ---
we have an equivalence of D-brane categories 
for any pair of phases. For example, if the model has two large volume
phases
corresponding to different Calabi-Yau manifolds, $M_1$ and $M_2$,
we have an equivalence of the derived categories,
${\bf D}(M_1)\stackrel{\cong}{\longrightarrow} {\bf D}(M_2)$.
If it has two LG phases, we have an equivalence
${\bf MF}_{\Gamma_1}(W_1)\stackrel{\cong}{\longrightarrow}
{\bf MF}_{\Gamma_2}(W_2)$.

Equivalences of the type (\ref{ours}) were 
constructed by Dmitri Orlov \cite{Orlov} in the case of 
projective hypersurfaces.
Here we would like to comment on the relation of that work to ours.
Thus, we consider Example (A) with superpotential
$W=PG(X_1,\ldots, X_N)$ for which $\Gamma=\Z_N$ and
$W_{\rm LG}=G(X_1,\ldots,X_N)$. We shall denote the correspinding 
Calabi-Yau hypersurface $\{G=0\}\subset \CP^{N-1}$
by $M_G$.

\subsubsection*{\it Early Constructions}

As backgrounds, we first
list various different ways to describe the category
${\bf MF}(G)$ of matrix factorizations of $G(x)$, that is,
the category of B-type D-branes in LG model with superpotential $G(x)$
(without orbfiold for now).
The basic references for this material are Eisenbud \cite{Eisenbud}
and Buchweitz \cite{Ragnar}.
See also the book \cite{Yoshino}.
The main players are 
finitely generated modules over the ring
\beq
B=\C[x_1,\ldots,x_N]/G(x).
\label{ringB}
\eeq
The grading that exists in this ring,
associated with the $U(1)$ gauge symmetry, plays no r\^ole
in the present discussion.

Let us consider a matrix factorization of $G(x)$
$$
Q=\left(\begin{array}{cc}
0&f(x)\\
g(x)&0
\end{array}\right),\qquad
Q(x)^2=G(x)\cdot {\bf 1}_{2\ell}.
$$
To this, we shall associated a $B$-module
$$
M_Q={\rm coker} \left(\bar f:B^{\oplus \ell}\to B^{\oplus\ell}\right).
$$
This is an example of a {\it maximal Cohen-Macaulay (MCM) module}
over $B$. There are several ways to define the MCM condition, 
but for the ring $B$ as above,
it is enough to define it as a module which admits
such a presentation.
To the trivial matrix factorizations,
$(f,g)=(1,G)$ and $(G,1)$, we have the trivial module
$M_{(1,G)}=0$ and the free module $M_{(G,1)}=B$ respectively.
The category ${\bf MF}(G)$
is equivalent to the category of MCM $B$-modules
modulo the subcategory of projective modules.

As we have discussed in Section~\ref{subsec:LG} and \ref{subsub:LVred},
we can also associate to a matrix factorization $Q$
a 2-periodic exact sequence of free modules
\beq
{\mathcal C}_Q:~~\cdots \stackrel{\bar f}{\longrightarrow}
B^{\oplus \ell}\stackrel{\bar g}{\longrightarrow}
B^{\oplus \ell}\stackrel{\bar f}{\longrightarrow}
\underline{B^{\oplus \ell}}\stackrel{\bar g}{\longrightarrow}
B^{\oplus \ell}\stackrel{\bar f}{\longrightarrow}
B^{\oplus \ell}\stackrel{\bar g}{\longrightarrow}
B^{\oplus \ell}\stackrel{\bar f}{\longrightarrow}
\cdots\,\, .
~~~
\label{ugtac}
\eeq
As a matter of convention we place the target of one of
$\bar f$'s at degree 0 (the underlined $B^{\oplus\ell}$).
This is an example of a totally acyclic complex.
In general, a {\it totally acyclic complex} of $B$-modules
is an exact complex of projective modules of finite ranks, whose 
$B$-dual complex is also exact. 
We have seen in Section~\ref{subsec:LG} that
chiral ring elements
between the D-branes corresponding to two matrix factorizations,
$Q_1$ and $Q_2$,
are in one-to-one correspondence with
homotopy classes of cochain maps of the corresponding
complexes, ${\mathcal C}_{Q_1}$ and ${\mathcal C}_{Q_2}$, see (\ref{Htac}).
In fact, the category ${\bf MF}(G)$
is equivalent to the homotopy category 
of totally acyclic complexes
of $B$-modules, denoted by ${\bf TAC}(B)$.

The MCM module $M_Q$ ``fits in the middle'' of
the complex ${\mathcal C}_Q$,
$$
\begin{array}{cccccccccccccccc}
\cdots \!\!\!&\stackrel{\bar f}{\longrightarrow}\! &\!B^{\oplus \ell}\! &
\!\stackrel{\bar g}{\longrightarrow}\! &\!B^{\oplus \ell} \!&
\!\stackrel{\bar f}{\longrightarrow}\! &\!\underline{B^{\oplus \ell}} &
\stackrel{\bar g}{\longrightarrow} &B^{\oplus \ell}\! & 
\!\stackrel{\bar f}{\longrightarrow}\! &\!B^{\oplus \ell} \!&
\!\stackrel{\bar g}{\longrightarrow}\! &\!B^{\oplus \ell} \!&
\!\stackrel{\bar f}{\longrightarrow} &\!\!\!\cdots \\[-0.2cm]
 & & & & & & &\!\!\!\!\searrow\,\,\,\,\,\,\nearrow \!\!\!\!& 
& & & & & & &\\[-0.2cm]
 & & & & & & &\, M_Q \!& & & & & & & &\\[-0.1cm]
 & & & & & &\nearrow \!\!\!\!\!\!\!\!\!\!\!\!\!\!\!\!& 
&\!\!\!\!\!\!\!\!\!\!\!\!\!\!\!\!\searrow & & & & & & &\\[-0.3cm]
 & & & & & &0 \!\!\!\!\!& &\!\!\!\!\!0 & & & & & & &
\end{array}
$$
Namely, the two sequences, one ending at $M_Q$ and the other
starting with $M_Q$, are both exact,
and the triangle in the middle commutes.
In general, any MCM module fits
in the middle of a totally acyclic complex.
Such a complex is called the complete resolution of the MCM module.

As the final ingredient, we discuss the r\^ole of 
{\it perfect complexes}, that is,
bounded complexes of projective modules.
Let ${\bf D}(B)$ be the derived category of $B$-modules
consisting of complexes with bounded cohomologies.
The main claim is that the category ${\bf MF}(G)$
is equivalent to the derived category 
${\bf D}(B)$ modulo the subcategory ${\bf P}(B)$
consisting of perfect complexes.
The functor ${\bf MF}(G)\to {\bf D}(B)/{\bf P}(B)$
is straighforward:
To each matrix factorization $Q$ we associate
the one-term complex $M_Q[0]$.
Since $M_Q$ fits in the middle of
the totally acyclic complex ${\mathcal C}_B$, 
this object is quasi-isomorphic to 
the left-half of it, ${\mathcal C}_{\mathcal B}^{\leq 0}$,
 and also to a shift of the right-half,
${\mathcal C}_{\mathcal B}^{\geq 1}[1]$.
Since we are modding out by perfect complexes,
we can take any position to truncate ${\mathcal C}_Q$ into a half.
Namely, as the image of $Q$, we can take
the left semi-infinite complex ${\mathcal C}_Q^{\leq j}$ or 
the right semi-infinite complex ${\mathcal C}_Q^{\geq j}[1]$ for any $j\in \Z$
--- they are all isomorphic in ${\bf D}(B)/{\bf P}(B)$.
The other direction ${\bf D}(B)/{\bf P}(B)\to {\bf MF}(G)$
is less straighforward, but the idea is simple.
To each complex ${\mathcal C}\in{\bf D}(B)$, we first find a complex 
$P^{\cdot}$ of projective modules, bounded from the right,
that is quasi-isomorphic to it.
It is unbounded to the left unless ${\mathcal C}$ was perfect.
Since it has bounded cohomologies, it is exact at low enough degrees, 
say, at degrees $j\leq -j_*$.
Discarding a perfect part, it is isomorphic in 
${\bf D}(B)/{\bf P}(B)$ to the truncated complex $P^{\cdot\leq -j_*-1}$
which is exact except at the right-most term $P^{-j_*-1}$.
In fact, it is a truncated version of a totally acyclic complex.
This gives an object of ${\bf TAC}(B)$ and hence of ${\bf MF}(G)$.

To summarize, we have equivalences of categories
\beq
{\bf MF}(G)~\cong~
{\bf TAC}(B)~\cong~
{{\bf MCM}(B)\over {\rm projectives}}~\cong~
{\bf D}(B)/{\bf P}(B).
\label{MCMequiv}
\eeq
Some of these categories are naturally triangulated, and these
are equivalences of triangulated categories.
The part other than ${\bf MF}(G)$ 
holds also for a more general ring than (\ref{ringB}),
see  \cite{Ragnar}.
The category
${\bf D}(B)/{\bf P}(B)$ is also discussed more recently in \cite{Orlov0}
where it is called the category of singulariries.

\subsubsection*{\it Orlov's Construction}

\newcommand{\grB}{{\rm gr}\mbox{-}B}

Let us now describe the construction of \cite{Orlov}.
The main players are finitely generated {\it graded} modules over the 
{\it graded} ring
\beq
B=\C[x_1,\ldots,x_N]/G(x),
\label{grringB}
\eeq
where the $x_i$'s have degree $1$.
They form an Abelian category and we denote by
${\bf D}(\grB)$ the derived category consisting of complexes
with bounded cohomologies.
If we mod out these categories
by the subcategories consisting of (complexes of) torsion modules,
we obtain the category of coherent sheaves on the hypersurface
$M_G$ and its derived category ${\bf D}(M_G)$, just as in the case of toric
varieties as described in Section~\ref{sec:math}.
If we mod out the derived category ${\bf D}(\grB)$
by the subcategory ${\bf P}(\grB)$ consisting of perfect complexes,
then we obtain a category which is equivalent to ${\bf MF}_{\Z_N}^{}(G)$.
The last connection
is the graded version of (\ref{MCMequiv}) (see below for the precise
mapping). Thus, we have projections to the categories of our interest:
$$
\begin{picture}(125,67) 
  \put(38,62){${\bf D}(\grB)$}
  \put(53,53){\vector(-2,-3){25}}
  \put(69,53){\vector(2,-3){25}}
  \put(-3,30){perfect}
  \put(89,30){torsion}
  \put(-3,0){${\bf MF}_{\Z_N}^{}(G)$}
  \put(90,0){${\bf D}(M_G)$}
\end{picture}
$$

The next step is to consider lifts of
the downstairs categories, ${\bf MF}_{\Z_N}^{}(G)$ and ${\bf D}(M_G)$,
into the derived category of modules.
To obtain a finitely generated modules for the
latter, we need to choose a cut-off $i\in\Z$, that is, we 
restrict our attention to modules which are trivial
at (gauge) degree less than $i$. 
There is a natural functor from ${\bf D}(M_G)$ to 
the subcategory ${\bf D}(\grB_{\geq i})$ of complexes of such modules.
The image of this functor is denoted by
${\bf D}_i$. 
On the other hand, ${\bf MF}_{\Z_N}^{}(G)$ is still obtained
from ${\bf D}(\grB_{\geq i})$ by modding out by a certain subcategory
${\bf P}_{\geq i}$.
There is a complement of ${\bf P}_{\geq i}$
in ${\bf D}(\grB_{\geq i})$, denoted by ${\bf T}_i$, which is equivalent to
${\bf MF}_{\Z_N}^{}(G)$ under the projection functor.
The precise ${\bf P}_{\geq i}$ and ${\bf T}_i$ will be described momentarily.
The fact is that
the two subcategroies, ${\bf D}_i$ and ${\bf T}_i$, are equal under the 
Calabi-Yau condition, i.e., when the degree of $G$ is equal to the number
$N$ of variables. 
Thus, we have a ``hat-diagram''
\beq
\begin{picture}(165,90) 
  \put(56,81){${\bf D}(\grB_{\geq i})$}
  \put(80,65){$\cup$}
  \put(64,53){${\bf T}_i={\bf D}_i$}
  \put(68,74){\vector(-2,-3){40}}
  \put(100,74){\vector(2,-3){40}}
  \put(4,36){perfect}
  \put(130,36){torsion}
  \put(68,48){\vector(-2,-3){22}}
  \put(123,14){\vector(-2,3){22}}
  \put(56,21){$\cong$}
  \put(104,21){$\cong$}
  \put(0,0){${\bf MF}_{\Z_N}^{}(G)$}
  \put(118,0){${\bf D}(M_G)$}
\end{picture}
\label{orlovhat}
\eeq
which proves the equivalence.
(There is also a simple relation between ${\bf T}_i$ and ${\bf D}_i$
in the non-Calabi-Yau cases, which results
in a simple relation between ${\bf MF}_{\Z_N}^{}(G)$ and ${\bf D}(M_G)$.)

The subcategories
${\bf P}_{\geq i}$ and ${\bf T}_i$ of ${\bf D}(\grB_{\geq i})$
are defined as follows.
${\bf P}_{\geq i}$ is the category generated by free modules 
$B(\nu)$ for $\nu\leq -i$.
(Note that the lowest degree element, $1$, of
$B(\nu)$ has degree $-\nu$, and therefore $B(\nu)$
belongs to ${\bf D}(\grB_{\geq i})$ if $\nu\leq -i$.)
${\bf T}_i$
is the ``left-orthogonal'' of ${\bf P}_{\geq i}$ in ${\bf D}(\grB_{\geq i})$,
that is, the subcategory consisting of all objects
$L^{\cdot}$ such that 
$\Hom^{}_{{\bf D}(\grB_{\geq i})}(L^{\cdot},P^{\cdot})=0$
for all $P^{\cdot}\in {\bf P}_{\geq i}$.

In order to use this construction, we would better have some 
understanding of the subcategory ${\bf T}_i$.
If we were working in the full ${\bf D}(\grB)$ without a cut-off,
an example of an object in ${\bf T}_i$ would be
$B(\nu)$ for $\nu>-i$ since there are no morphisms from
$B(a)$ to $B(b)$ if $a>b$.
Similarly any complex built out of the $B(\nu)$ for $\nu>-i$
would be in ${\bf T}_i$. The latter remark is useful
when the resulting complex happens to be quasi-isomorphic
to an object in ${\bf D}(\grB_{\geq i})$
since this provides a way to construct objects in the real
${\bf T}_i\subset {\bf D}(\grB_{\geq i})$.

\subsubsection*{\it The Functor ${\bf MF}_{\Z_N}(G)\to {\bf D}(M_G)$}

Let us try to explicitly evaluate the functor  
$\Phi_i^{\rm Orlov}:{\bf MF}_{\Z_N}(G)\to {\bf D}(M_G)$ induced by
the contruction (\ref{orlovhat}).
The key step is of course to find a lift
of a given object in ${\bf MF}_{\Z_N}(G)$
to the subcategory ${\bf T}_i$.
 But first we need a lift to ${\bf D}(\grB)$, or more precisely, to
${\bf D}(\grB_{\geq i})$.

Let us pick a $\Z_N$-equivariant matrix factorization
${\mathcal B}=(\cp,\QLG,\rhoLG,\RLG)\in{\bf M}_{\Z_N}(G)$.
To this we can associate a totally acyclic complex ${\mathcal C}_{\mathcal B}$
of {\it graded} $B$-modules:
\beq
{\mathcal C}_{\mathcal B}:~~\cdots \stackrel{\bar f}{\longrightarrow}
{\mathcal C}_{\mathcal B}^{-2}\stackrel{\bar g}{\longrightarrow}
{\mathcal C}_{\mathcal B}^{-1}\stackrel{\bar f}{\longrightarrow}
{\mathcal C}_{\mathcal B}^0\stackrel{\bar g}{\longrightarrow}
{\mathcal C}_{\mathcal B}^1\stackrel{\bar f}{\longrightarrow}
{\mathcal C}_{\mathcal B}^2\stackrel{\bar g}{\longrightarrow}
{\mathcal C}_{\mathcal B}^3\stackrel{\bar f}{\longrightarrow}
\cdots\,\, .
~~~
\label{gtac}
\eeq
Namely, we first construct the complex
${\mathcal C}_{\QLG}$ of ungraded $B$-modules as before (\ref{ugtac}), 
and then provide it with a
grading using the information of R-symmetry
\beq
{\mathcal C}_{\mathcal B}^{j}=
\bigoplus_{\sigma_{\bf i}=(-1)^j}\mbox{$B({N\over 2}(j-\RLG_{\bf i}))$},\qquad
\mbox{for ~$j\in \Z$}.
\label{defCj}
\eeq
The numbers ${N\over 2}(j-\RLG_{\bf i})$ that appear here are all integers, 
since
$\e^{2\pi i(j-\RLG_{\bf i})/2}=\sigma_{\bf i}\e^{-\pi i \RLG_{\bf i}}$
are eigenvalues of
$\sigma_{\cp}\RLG(\e^{-\pi i})=\rhoLG(\e^{2\pi i/N})$ and hence must have
 order $N$.
The complex ${\mathcal C}_{\mathcal B}$ is 2-periodic up to a shift,
${\mathcal C}_{\mathcal B}^{j+2}={\mathcal C}_{\mathcal B}^j(N)$.
As the object in ${\bf D}(\grB)/{\bf P}(\grB)$ corresponding to
${\mathcal B}\in {\bf MF}_{\Z_N}(G)$ we can take the truncation of
${\mathcal C}_{\mathcal B}$ at any position, that is,
the left semi-infinite ${\mathcal C}_{\mathcal B}^{\leq j}$ 
or the right semi-infinite ${\mathcal C}_{\mathcal B}^{\geq j}[1]$ 
for any $j\in \Z$.
A representative that lands in ${\bf D}(\grB_{\geq i})$ is obtained 
by truncating far enough to the left.
For example, take the left semi-infinite 
complex ${\mathcal C}_{\mathcal B}^{\leq j_0}$
with sufficiently small $j_0$ so that it includes only
$B(\nu)$'s with $\nu<-i$.

Next, let us find an object $L$ which lies in ${\bf T}_i$
and fits into a distinguished triangle
\beq
P ~\longrightarrow ~ L
~ \longrightarrow ~ {\mathcal C}_{\mathcal B}^{\leq j_0}
\label{triPLC}
\eeq
for some $P\in {\bf P}_{\geq i}$.
As such an $L$, we propose to take the left semi-infinite complex 
obtained from ${\mathcal C}_{\mathcal B}$ by keeping {\it all}
$B(\nu)$'s with $\nu\leq -i$ and discarding all higher
$B(\nu)$'s.
It is obviously an object of ${\bf D}(\grB_{\geq i})$.
Also, ${\mathcal C}_{\mathcal B}^{\leq j_0}$ is obtained from
$L$ by discarding its subcomplex $P$ 
consisting of finitely many $B(\nu)$'s (with $\nu\leq -i$ of course).
Thus, $L$ indeed fits into a triangle (\ref{triPLC}) with
$P\in {\bf P}_{\geq i}$.
What is less obvious is that $L$ lies in ${\bf T}_i$.
To see this, let us consider the complement $L^c$ of 
$L$ in ${\mathcal C}_{\mathcal B}$, which is a right semi-infinite complex.
Since ${\mathcal C}_B$ is exact, $L^c[1]$ is quasi-isomorphic to 
$L\in {\bf D}(\grB_{\geq i})$.
Obviously, it consists of $B(\nu)$'s with $\nu>-i$.
By the remark at the end of the description of Orlov's construction,
the object $L\cong L^c[1]$ belongs to ${\bf T}_i$.
Thus, $L$ or equivalently $L^c[1]$ is the lift of 
${\mathcal B}\in {\bf MF}_{\Z_N}(G)$ that lands in ${\bf T}_i$.

To find its image in ${\bf D}(M_G)$, we just regard the complex 
of $B$-modules as 
a complex of vector bundles on the hypersurface $M_G$, by the replacement
$B(\nu)\to {\mathcal O}(\nu)$.

To summarize, we have a simple way to find
the large volume image of a given brane in the LG orbifold,
under the functor $\Phi_i^{\rm Orlov}$
induced by (\ref{orlovhat}):
{\it Take the corresponding totally acyclic complex and keep
only $B(\nu)$'s with $\nu\leq -i$. (Alternatively, keep only
$B(\nu)$'s with $\nu\geq -i+1$ and shift by one to the left.)
Then, regard the semi-infinite complex
as a complex of vector bundles on $M_G$.}

\subsubsection*{\it Comparison}

This is essentially the same as the result of our transport 
for a certain choice of window.
The key point is that for any linear sigma model lift 
$\lsmB=(\cp,Q,\rho,R)$ of a brane
${\mathcal B}=(\cp,\QLG,\rhoLG,\RLG)$
of the LG orbifold,
the totally acyclic complex ${\mathcal C}_{\lsmB}$ that
was introduced around Eqn~(\ref{bigV})
precisely matches with the 
totally acyclic complex ${\mathcal C}_{\mathcal B}$
defined by (\ref{gtac}) and (\ref{defCj}).
To see this, it is enough to rewrite the equation (\ref{qeq})
that relates $\lsmB$ and ${\mathcal B}$ as
$$
q_{\bf i}={N\over 2}(R_{\bf i}-\RLG_{\bf i}),
$$
and compare it with (\ref{defCj}).
The actual low energy image $(\widehat{\mathcal E},\widehat{Q})\in 
{\bf D}(M_G)$
 of $\lsmB$ is obtained from the trunction of 
${\mathcal C}_{\lsmB}$ that corresponds to
the truncation 
$$
\bigoplus_{m\in\Z}p^m\cp
~\, \longmapsto ~\,
\widehat{\cp}=\bigoplus_{m\geq 0}p^m\cp, 
$$
see (\ref{bigV}) and (\ref{hatcp}).
If the lift $\lsmB$ obeys the grade restriction rule $\{a,\ldots, a+N-1\}$,
the truncation is to keep all $B(\nu)$'s with $\nu\geq a$ and to discard
all lower $B(\nu)$'s.
The result is precisely the image of ${\mathcal B}$
in the subcategory ${\bf T}_i$ with $a=-i+1$, shifted by one to the right.
Note that the above grade restriction rule corresponds to
to the window, 
$-{N\over 2}-a<{\theta\over 2\pi}<-{N\over 2}-a+1$.
To summarize, the functor induced by the construction
(\ref{orlovhat}) is related to ours (\ref{ours}) by
\beq
\Phi_i^{\rm Orlov}=[1]\circ \Phi_{\tau}
\eeq
where $\tau$ is a path through the window
$i-{N\over 2}-1<{\theta\over 2\pi}<i-{N\over 2}$.

\section*{Acknowledgement}

We would like to thank
Denis Auroux,
Alexey Bondal, 
Ilka Brunner,
Ragnar Buchweitz, 
Alastair Crow,
David Eisenbud, 
Tohru Eguchi,
Matthias Gaberdiel,
Dongfeng Gao,
Ezra Getzler,
Paul Horja,
Kazuo Hosomichi,
Amer Iqbal,
Misha Kapranov,
Anton Kapustin,
Ludmil Katzarkov,
Kris Kennaway,
Alastair King,
Yi Lin,
Emil Martinec,
Grisha Mikhalkin,
Greg Moore,
Dave Morrison, 
Dmitri Orlov,
Mikael Passare,
Hubert Saleur,
David Tong,
Cumrun Vafa and 
Johannes Walcher
for useful discussions, correspondence, and earlier collaborations
that are relevant to the present work.
We especially want to thank Ragnar Buchweitz for patient and 
clear explanations on homological algebra and other mathematics,
from the very basic to advanced.

We are grateful to the Fields Institute, Toronto
and the Perimeter Institute, Waterloo, as well as the coorganizers of
a program there, ``The Geometry of String Theory'' (2004-2005), 
which brought the authors together. We are happy to thank the participants
of the program for fruitful conversations.

All three of us also thank
MSRI and the town of Berkeley, California, 
along with the organizers of the workshop
``Generalized McKay correspondences and representation theory''
(March 2006) where great progress was made.
For the same reason,
M.H. and K.H. thank 
KITP, Santa Barbara (August 2005),
ESI, Wien (June 2006), 
LMU,  Munchen (July 2006), 
IAS, Princeton (January 2007), 
EPFL, Lausanne (September 2007) 
and the
organizers of the workshops that were held there.

M.H. wants to thank the string theory group at DESY, 
Hamburg, including Yasuaki Hikida, Yuji Okawa, Ioannis Papadimitriou, 
Sylvain Ribault, Volker Schomerus, J\"org Teschner, for valuable discussions.
An essential part of his contribution to this work 
was done during the time at DESY.
M.H. further wants to thank MPRG, Philadelphia (September 2005), 
CERN, Geneva (January and October 2006), 
University of Toronto (March 2006), Trinity College, 
Dublin (March 2007), ETH, Z\"urich (March 2007), 
ASC, Munich (May 2007), CSF, Ascona (August 2007), 
University of Augsburg (February 2008) and MPI, 
Potsdam (March 2008) for hospitality during his visit.

K.H. thanks CEA Saclay, Paris
for warm hospitality during his visit (May-June 2006),
as well as other places, in particular,
KdV Institute, Amsterdam (July 2006),
DESY, Hamburg (July 2006),
ETH, Z\"urich (July 2006 and September 2007),
BIRS, Banff (March 2007 and October 2005),
Aspen Center for Physics, Aspen (Summer 2007)
and
RIMS, Kyoto (December 2007)
where crucial progress was made.
K.H. is supported also by NSERC and the Alfred P. Sloan Foundation.

\newpage

\begin{appendix}
\section{Supersymmetry}

\label{app:susy}

\newcommand{\tilx}{\widetilde{x}}
\newcommand{\tily}{\widetilde{y}}
\newcommand{\tilF}{\widetilde{F}}

Here we record the supersymmetry variation of the bulk action.

\subsection{Non-Linear Sigma Models And Landau-Ginzburg Models}

We consider the non-linear sigma model on a K\"ahler manifold
$(M,g)$
with a superpotential $W$. We suppose that the metric is given in terms of
a K\"ahler potential $K$ in a coordinate patch, 
$g_{i\bj}=\partial_i\partial_{\bj}K$.
The bulk Lagrangian is the sum of the following two terms
(no partial integration to be used):
\beqa
{\mathcal L}_K&=&\int K(\Phi,\bPhi)\dd^4\theta
+\partial_+\partial_-K(\phi,\bphi)\nn\\
&=&2g_{i\bj}\left(\partial_+\phi^i\partial_-\bphi^{\bj}
+\partial_-\phi^i\partial_+\bphi^{\bj}
+i\bpsi_-^{\bj}\lrD_{\!\!\!+}\psi_-^i
+i\bpsi_+^{\bj}\lrD_{\!\!\!-}\psi_+^i
\right)
+R_{i\bj k\bl}\psi_+^i\psi_-^k\bpsi_-^{\bj}\bpsi_+^{\bl}
\nn\\
&&+g_{i\bj}\tilF^i\overline{\tilF}^{\bj}
\\
&=&g_{i\bj}\left(\partial_0\phi^i\partial_0\bphi^{\bj}
-\partial_1\phi^i\partial_1\bphi^{\bj}
+i\bpsi_-^{\bj}(\lrD_{\!\!\! 0}+\lrD_{\!\!\! 1})\psi_-^i
+i\bpsi_+^{\bj}(\lrD_{\!\!\! 0}-\lrD_{\!\!\! 1})\psi_+^i
\right)
\nn\\
&&+R_{i\bj k\bl}\psi_+^i\psi_-^k\bpsi_-^{\bj}\bpsi_+^{\bl}
+g_{i\bj}\tilF^i\overline{\tilF}^{\bj}
\nn\\
{\mathcal L}_W&=&{1\over 2}\int W(\Phi)\dd^2\theta\,\,+\,\,c.c.
\nn\\
&=&{1\over 2}\left(\,
F^i\partial_iW(\phi)-\partial_i\partial_jW(\phi)\psi_+^i\psi_-^j
\,\,+\,\, c.c.\,\right)
\nn\\
&=&{1\over 2}\left(\,
\tilF^i\partial_iW(\phi)-D_i\partial_jW(\phi)\psi_+^i\psi_-^j
\,\,+ \,\, c.c.\, \right)
\eeqa
Here $\tilF^i$ are the covariantized auxiliary fields
\beq
\tilF^i=F^i-\Gamma^i_{jk}\psi_+^j\psi_-^k.
\eeq
Under $(2,2)$ supersymmetry transformation
$\delta=i\epsilon_+{\bf Q}_--i\epsilon_-{\bf Q}_+
-i\bepsilon_+\overline{\bf Q}_-
+i\bepsilon_-\overline{\bf Q}_+$,
\beqa
&&\delta\phi^i=\epsilon_+\psi_-^i-\epsilon_-\psi_+^i,\nn\\
&&\psi_{\pm}^i=\pm 2i\bepsilon_{\mp}\partial_{\pm}\phi^i
+\epsilon_{\pm}F^i,\\
&&\delta F^i=-2i\bepsilon_+\partial_-\psi_+^i
-2i\bepsilon_-\partial_+\psi_-^i,
\nn
\eeqa
${\mathcal L}_K$ and ${\mathcal L}_W$ transform as
\beqa
\delta {\mathcal L}_K&=&
\partial_-\left[\epsilon_+g_{i\bj}\left(
2\psi_-^i\partial_+\bphi^{\bj}-i\tilF^i\bpsi_+^{\bj}\right)\right]
-\partial_+\left[\epsilon_-g_{i\bj}\left(
2\psi_+^i\partial_-\bphi^{\bj}+i\tilF^i\bpsi_-^{\bj}\right)\right]
\nn\\
&&+\partial_+\epsilon_+\left[ 4 g_{ij}\psi_-^i\partial_-\bphi^{\bj}
\right]
-\partial_-\epsilon_-\left[ 4 g_{ij}\psi_+^i\partial_+\bphi^{\bj}
\right]\,\,+\,\, c.c.
\label{LKv}
\\
\delta {\mathcal L}_W&=&
-\epsilon_+\partial_-\left[i\bpsi_+^{\bi}\partial_{\bi}\overline{W}\right]
-\epsilon_-\partial_+\left[i\bpsi_-^{\bi}\partial_{\bi}\overline{W}\right]
\,\,+\,\, c.c.
\label{LWv}
\eeqa
We see that the supercurrent $G_{\pm}^{\,\mu},\overline{G}_{\pm}^{\,\mu}$
defined by
$$
\delta S=\int \dd^2x\left(\partial_{\mu}\epsilon_+G_-^{\,\mu}-
\partial_{\mu}\epsilon_-G_+^{\,\mu}\,\,+\,\, c.c.\right)
$$
is given by
\beqa
G_-^{\,\mu}:&& G_-^{\,\,+}=4g_{i\bj}\psi_-^i\partial_-\bphi^{\bj},\,\,\,
G_-^{\,\,-}=i\bpsi_+^{\bi}\partial_{\bi}\overline{W},
\\
G_+^{\,\mu}:&& G_+^{\,\,+}=
-i\bpsi_-^{\bi}\partial_{\bi}\overline{W},\,\,\,
G_+^{\,\,-}
=4g_{i\bj}\psi_+^i\partial_+\bphi^{\bj}.
\eeqa

Let us formulate this theory on the left half space $\surface=\{(t,\s)|
s\leq 0\}$ which has one timelike boundary line at $\s=0$.
We focus on the ${\mathcal N}=2_B$ supersymmetry generated by
${\bf Q}=\overline{\bf Q}_++\overline{\bf Q}$,
${\bf Q}^{\dag}={\bf Q}_++{\bf Q}$.
The corresponding variation $\delta=i\epsilon{\bf Q}^{\dag}-i\bepsilon{\bf Q}$
is obtained from the $(2,2)$ variation by
setting
$$
\epsilon_+=-\epsilon_-=\epsilon,\qquad
\bepsilon_+=-\bepsilon_-=\bepsilon.
$$
The bulk action variation is found from (\ref{LKv})-(\ref{LWv}):
\beq
\delta S_{\rm bulk}=
\Rp\int_{\partial\surface}\dd t\,
\bepsilon\left(
2g_{i\bj}\bpsi_-^{\bj}\partial_+\phi^i
-2g_{i\bj}\bpsi_+^{\bj}\partial_-\phi^i
+ig_{i\bj}\overline{\tilF}^{\bj}(\psi_+^i+\psi_-^i)
+i(\psi_+^i+\psi_-^i)\partial_iW
\right).
\label{Sbkkkv}
\eeq
The standard boundary term is defined by
\beq
L^{(0)}_{\rm bdry}
=-{i\over 2}g_{i\bj}\left(\psi_+^i\bpsi_-^{\bj}+\bpsi_+^{\bj}\psi_-^i
\right)
\eeq
and its ${\mathcal N}=2_B$ variation is
\beq
\delta L^{(0)}_{\rm bdry}
=\Rp\left[\bepsilon
\left(+2g_{i\bj}\bpsi_+^{\bj}\partial_-\phi^i
-2g_{i\bj}\bpsi_-^{\bj}\partial_+\phi^i
-ig_{i\bj}\overline{\widetilde{F}}^{\bj}(\psi_+^i+\psi_-^i)
\right)\right]
\label{Lstv}
\eeq
The sum of (\ref{Sbkkkv}) and (\ref{Lstv})
is simply
\beq
\delta S_{\rm bulk}+\delta\int_{\partial\surface}L^{(0)}_{\rm bdry}
=\Rp\int_{\partial\surface}\dd t\,\bepsilon\Bigl(i(\psi_+^i+\psi_-^i)
\partial_i W\Bigr).
\eeq
This is the Warner term \cite{Warner}.

\subsection{Linear Sigma Model}

The gauge kinetic term, the matter kinetic term and
the FI-theta term are given by
\beqa
{\mathcal L}_{\rm g}&=&\int\dd^4\theta\left(-{1\over 2e^2}
\overline{\Sigma}\Sigma\right)+\mbox{total derivative}\nn\\
&=&
{1\over 2e^2}\,\left[
|\partial_0\sigma|^2-|\partial_{1}\sigma|^2
+i\blambda_-(\lrd_{\!\! 0}+\lrd_{\!\! 1})\lambda_-
+i\blambda_+(\lrd_{\!\! 0}-\lrd_{\!\! 1})\lambda_+
+v_{01}^2+D^2
\right],\nn\\
\label{appLg}\\
{\mathcal L}_{\rm m}&=&
\int\dd^4\theta\,\,\bPhi\e^V\Phi
+\mbox{total derivative}\nn\\
&=&
|D_0\phi|^2-| D_1\phi|^2
+i\bpsi_-(\lrD_{\!\! 0}+\lrD_{\!\! 1})\psi_-
+i\bpsi_+(\lrD_{\!\! 0}-\lrD_{\!\! 1})\psi_+
+\bphi D\phi+|F|^2
\nn\\
&&\,
-|\sigma \phi|^2-\bpsi_-\sigma\psi_+-\bpsi_+\bsigma\psi_-
-i\bphi\lambda_-\psi_++i\bphi\lambda_+\psi_-+i\bpsi_+\blambda_-\phi
-i\bpsi_-\blambda_+\phi,
\nn\\
\label{appLm}\\
{\mathcal L}_{{\rm FI}\,\theta}&=&
{\rm Re}\int\dd^2\widetilde{\theta}\Bigl(-t\Sigma\Bigr)
=-rD+\theta v_{01}.
\label{appLFI}
\eeqa
Only the special case of $T=U(1)$ and with
just one charge $1$ matter field is presented, 
since the generalization is obvious.
The superpotential term is
\beqa
{\mathcal L}_W&=&{\rm Re}\int \dd^2\theta \,W(\Phi)
={\rm Re}\left[\,\,\sum_{i=1}^NF_i{\partial W\over\partial\phi_i}(\phi)
-\sum_{i,j=1}^N{\partial^2W\over\partial\phi_i\partial\phi_j}(\phi)
\psi_+^i\psi_-^j\,\,\right].
\label{appLW}
\eeqa

The $(2,2)$ supersymmetry transformation of the component fields
is given by (where $v_{\pm}:={1\over 2}(v_0\pm v_1)$)
\beqa
&&\delta v_{\pm}={i\over 2}\bepsilon_{\pm}\lambda_{\pm}
+{i\over 2}\epsilon_{\pm}\blambda_{\pm},
\nonumber\\
&&\delta\sigma=-i\bepsilon_+\lambda_-
-i\epsilon_-\blambda_+,
\nonumber\\
&&\delta D={1\over 2}\Bigl(
-\bepsilon_+(\partial_0-\partial_1)\lambda_+
-\bepsilon_-(\partial_0+\partial_1)\lambda_-
+\epsilon_+(\partial_0-\partial_1)\blambda_+
+\epsilon_-(\partial_0+\partial_1)\blambda_-\Bigr).
\nonumber\\
&&\delta\lambda_+=
i\epsilon_+(D+iv_{01})
+\epsilon_-(\partial_0+\partial_1)\bsigma,
\nonumber\\
&&\delta\lambda_-=
i\epsilon_-(D-iv_{01})
+\epsilon_+(\partial_0-\partial_1)\sigma,
\label{vctSUSY}\\
\mbox{and}&&
\nonumber\\
&&\delta\phi=\epsilon_+\psi_--\epsilon_-\psi_+,
\nonumber\\
&&\delta\psi_+=i\bepsilon_-(D_0+D_1)\phi+\epsilon_+F
-\bepsilon_+\bsigma\phi,
\nonumber\\
&&\delta\psi_-=-i\bepsilon_+(D_0-D_1)\phi+\epsilon_-F
+\bepsilon_-\sigma\phi,
\nonumber\\
&&\delta F=
-i\bepsilon_+(D_0-D_1)\psi_+-i\bepsilon_-(D_0+D_1)\psi_-
+(\bepsilon_+\bsigma\psi_-+\bepsilon_-\sigma\psi_+)
+i(\bepsilon_-\blambda_+-\bepsilon_+\blambda_-)\phi.
\nonumber\\
&&\label{chiSUSY}
\eeqa
Under this, the Lagrangians vary as follows:
\beqa
\delta{\mathcal L}_{\rm g}&=&
\partial_+\epsilon_+\left(-{2i\over e^2}\blambda_-\partial_-\sigma\right)
-\partial_-\epsilon_-\left({2i\over e^2}\blambda_+\partial_+\bsigma\right)
+\partial_-\epsilon_+\left({i \over e^2}v_{01}\blambda_+\right)
-\partial_+\epsilon_-\left({i\over e^2}v_{01}\blambda_-\right)
\nn\\&&
+\partial_-\left[{\epsilon_+\over 2e^2}
\left(-2i\blambda_-\partial_+\sigma
+\blambda_+(D+iv_{01})\right)\right]
-\partial_+\left[{\epsilon_-\over 2e^2}
\left(2i\blambda_+\partial_-\bsigma
-\blambda_-(D-iv_{01})\right)\right]
\nn\\
&&+c.c.
\label{Lg22SUSY}
\\
\delta{\mathcal L}_m&=&
\partial_+\epsilon_+4D_-\bphi\psi_--\partial_-\epsilon_-4D_+\bphi\psi_+
+\partial_-\epsilon_+(-\bphi\blambda_+\phi)
-\partial_+\epsilon_-\bphi\blambda_-\phi
\nn\\
&&+\partial_-\left[\epsilon_+\left(
2D_+\bphi\psi_-+\bphi\blambda_+\phi-i\bphi\sigma\psi_+-i\bpsi_+F\right)
\right]
\nn\\
&&-\partial_+\left[\epsilon_-\left(
2D_-\bphi\psi_+-\bphi\blambda_-\phi-i\bphi\bsigma\psi_-
+i\bpsi_-F\right)\right]
\nn\\&&
+c.c.
\label{Lm22SUSY}
\\
\delta{\mathcal L}_{{\rm FI}, \theta}&=&
\partial_-\epsilon_+(r\blambda_+)-\partial_+\epsilon_-(-r\blambda_-)
+\partial_-\left[-t\epsilon_+\blambda_+\right]
-\partial_+\left[\overline{t}\epsilon_-\blambda_-\right]
\nn\\
&&+c.c.
\label{LFI22SUSY}
\\
\delta{\mathcal L}_W&=&
\partial_-\epsilon_+\left(i\bpsi^{\bi}_+\partial_{\bi}\overline{W}\right)
-\partial_+\epsilon_-\left(-i\bpsi^{\bi}_-\partial_{\bi}\overline{W}\right)
\nn\\
&&
+\partial_-\left(-i\epsilon_+\bpsi^{\bi}_+\partial_{\bi}\overline{W}\right)
-\partial_+\left(i\epsilon_-\bpsi^{\bi}_-\partial_{\bi}\overline{W}\right)
\nn\\
&&+c.c.
\label{LW22SUSY}
\eeqa
This shows that the supercurrent is given by
\beqa
&&
G^+_-=-{2i\over e^2}\blambda_-\partial_-\sigma+4D_-\bphi\psi_-
\\&&
G^-_-=\left(r-|\phi|^2+{i\over e^2}v_{01}\right)\blambda_+
+i\bpsi_+^{\bi}\partial_{\bi}\overline{W},
\\&&
G^+_+=-\left(r-|\phi|^2-{i\over e^2}v_{01}\right)\blambda_-
-i\bpsi_-^{\bi}\partial_{\bi}\overline{W},
\\&&
G^-_+={2i\over e^2}\blambda_+\partial_+\bsigma+4D_+\bphi\psi_+.
\eeqa

Let us formulate the model on the left-half space
$\surface$ and consider the ${\mathcal N}=2_B$
supersymmetry $\epsilon_{\pm}=\pm\epsilon$, $\bepsilon_{\pm}=\pm\bepsilon$:
\beqa
\delta v_0&=&{i\over 2}\bepsilon(\lambda_+-\lambda_-)
+{i\over 2}\epsilon (\blambda_+-\blambda_-),\nn\\
\delta v_1&=&{i\over 2}\bepsilon(\lambda_++\lambda_-)
+{i\over 2}\epsilon (\blambda_++\blambda_-),\nn\\
\delta\sigma&=&-i\bepsilon\lambda_-+i\epsilon \blambda_+,\nn\\
\delta D&=&-\bepsilon \partial_-\lambda_++\bepsilon \partial_+\lambda_-
+\epsilon\partial_-\blambda_+-\epsilon\partial_+\blambda_-,\nn\\
\delta \lambda_+&=&i\epsilon (D+iv_{01}+2i\partial_+\bsigma),\nn\\
\delta \lambda_-&=&-i\epsilon (D-iv_{01}+2i\partial_-\sigma).\nn\\
\delta \phi&=&\epsilon(\psi_++\psi_-),\nn\\
\delta\psi_+&=&-i\bepsilon(D_0+D_1)\phi+\epsilon F-\bepsilon\bsigma\phi,\nn\\
\delta\psi_-&=&-i\bepsilon(D_0-D_1)\phi-\epsilon F-\bepsilon\sigma\phi,\nn\\
\delta F&=&-i\bepsilon(D_0-D_1)\psi_+
+i\bepsilon (D_0+D_1)\psi_-
+\bepsilon\bsigma\psi_--\bepsilon\sigma\psi_+
-i\bepsilon (\blambda_++\blambda_-)\phi,\nn\\
\label{2BvarLSM}
\eeqa
The variation of the terms of the action can be easily 
read from (\ref{Lg22SUSY})-(\ref{LW22SUSY}).

\noindent
\underline{Gauge kinetic term}~~
Variation of the gauge kinetic term can be written succinctly
in terms of the ${\mathcal N}=2_B$ boundary superspace \cite{blin,HKLM} as
\beq
\delta {1\over 2\pi}\int_{\surface}\dd^2\s\, {\mathcal L}_{\rm g}
=-{1\over 2\pi}
{\rm Re}\left[\,{i\over 2e^2}\int_{\partial_B\surface}\dd t\dd\theta\,
\bepsilon\,\bD\Sigma\bD\bSigma
\,\,\right].
\label{delLg}
\eeq
The following counter term can cancel it:
\beq
S^{\rm c.t.}_{\rm g}
=-{1\over 2\pi}{\rm Re}\int_{\partial_B\surface}\dd t\dd\theta\,
\left(\sum_ia_i\bD b_i\right)
\label{gaugect}
\eeq
where $a_i$ and $b_i$ are functions of
$\Sigma$ and $\bSigma$ such that
$$
\Theta:=\sum_ia_i\dd b_i\quad
\mbox{obeys}
\quad
\dd\Theta=
{i\over 2e^2}\dd\sigma\wedge\dd\bsigma.
$$
The boundary counter term in (\ref{Sgggg}) corresponds to the choice
$
\Theta={i\over 4e^2}\left(\sigma\dd\bsigma-\bsigma \dd \sigma\right).
$
Other choice of $\Theta$ differs from this by an exact 1-form and hence the
corresponding counter term differs from (\ref{Sgggg}) by
a boundary D-term.

\noindent
\underline{Matter kinetic term}~~
Variation of the matter kinetic term is
\beqa
\delta{1\over 2\pi}\int_{\surface}\dd^2\s\,{\mathcal L}_{\rm m}
\!\!&=&\!\!
{1\over 2\pi}{\rm Re}\int_{\partial\surface}\dd t\,\bepsilon\Bigl[
\,\bphi(\lambda_++\lambda_-)\phi
+\bpsi_-(D_0+D_1)\phi-\bpsi_+(D_0-D_1)\phi
\nn\\
&&\quad\quad\quad\quad\quad
+i\bpsi_+\bsigma\phi
-i\bpsi_-\sigma\phi
+i\overline{F}(\psi_++\psi_-)\Bigr]
\eeqa
This is canceled by a simple boundary Lagrangian
\beq
S^{\rm c.t.}_{\rm m}={1\over 2\pi}\int_{\partial\surface}\dd t\,\left[
{i\over 2}(\bpsi_-\psi_+-\bpsi_+\psi_-)-{i\over 2}
\bphi(\sigma-\bsigma)\phi\right].
\label{matterct}
\eeq

\noindent
\underline{FI-theta term}~~
Variation of the FI-theta term is 
\beq
\delta{1\over 2\pi}\int_{\surface}\dd^2\s\,{\mathcal L}_{{\rm FI}\,\theta}
={1\over 2\pi}\Rp\int_{\partial\surface}\dd t\,
\left[t\epsilon\blambda_+-t\bepsilon\lambda_-\right]
={1\over 2\pi}\Ip\int_{\partial\surface}\dd t\,
(t\delta\sigma)
\eeq
This is cancelled by the variation of the counter term
\beq
S^{\rm c.t.}_{{\rm FI}\,\theta}
={1\over 2\pi}\int_{\partial\surface}\dd t\,
 {\rm Im}\Bigl(\,-t\sigma\,\Bigr),
\label{FIct}
\eeq

\end{appendix}

\bigskip
\noindent
{\it E-mail addresses}:\\
\texttt{manfred.herbst@cern.ch}, \texttt{hori@physics.utoronto.ca},
\texttt{d.c.page@gmail.com}

\end{document}